\documentclass[a4paper, 11pt, twoside, onecolumn, openright, final]{memoir}

\usepackage{mdframed}
\usepackage{tikz}
\usepackage[force]{feynmp-auto}
\usepackage{afterpage, everypage}
\usepackage{fmtcount}
\usepackage{etoc}
\usepackage{ifthen}
\usepackage[labelfont={bf,sc},labelsep=endash]{caption}
\usepackage{subcaption}
\usepackage{xstring}
\usepackage[colorlinks=true, breaklinks, linkcolor=MPLblue,urlcolor=darkgray]{hyperref}
\usepackage{verbatim}
\usepackage{amsfonts, amssymb, mathtools, amsmath, amsthm}
\usepackage{aligned-overset}
\usepackage{enumitem}
\usepackage{etoolbox}
\usepackage{xcolor}
\usepackage{longtable}
\usepackage{lipsum}
\usepackage{minitoc}
\usepackage{slashed}
\usepackage{empheq}
\usepackage[version=4]{mhchem}
\usepackage{lettrine}
\usepackage[normalem]{ulem}

\usepackage{standalone}
\usepackage{import}

\setlrmarginsandblock{1.4in}{0.95in}{*} 
\setulmarginsandblock{3cm}{2.5cm}{*}
\setheadfoot{1cm}{1.5cm}
\setheaderspaces{*}{1cm}{*}

\usepackage[utf8]{inputenc}
\usepackage[T1]{fontenc}
\usepackage{newpxtext}
\usepackage{newpxmath}
\usepackage{blindtext}

\usepackage{thcover}
\setDisplayskipStretch{0.0} 

\midsloppy

\checkandfixthelayout
\sloppybottom
\feetatbottom

\allowdisplaybreaks

\usepackage[bb=dsserif, scr=rsfso]{mathalpha}
\usepackage{bm}
\newcommand{\Id}{\mathbb{1}}

\newcommand{\bnu}{\bar{\nu}}
\newcommand{\vrho}{\varrho}
\newcommand{\bvrho}{\bar{\varrho}}
\newcommand{\dd}{\mathrm{d}}
\newcommand{\Neff}{N_{\mathrm{eff}}}

\newcommand{\ddp}[1]{[\dd^3 \vec{p}_{#1}]}

\newcommand{\x}{\mathrm{x}}
\newcommand{\y}{\mathrm{y}}
\newcommand{\z}{\mathrm{z}}

\newcommand{\mW}{m_W}

\newcommand{\me}{m_e}

\newcommand{\ii}{{\mathrm i}}

\newcommand{\YP}{Y_{\mathrm{p}}}
\newcommand{\TFO}{T_{\mathrm{FO}}}
\newcommand{\TNuc}{T_{\mathrm{Nuc}}}
\newcommand{\Tcm}{T_{\mathrm{cm}}}
\newcommand{\He}[1]{{}^#1{\mathrm{He}}}
\newcommand{\Li}{{}^7{\mathrm{Li}}}
\newcommand{\Be}{{}^7{\mathrm{Be}}}

\newcommand{\had}{\hat{a}^\dagger}
\newcommand{\ha}{\hat{a}}
\newcommand{\hbd}{\hat{b}^\dagger}
\newcommand{\hb}{\hat{b}}
\newcommand{\hcd}{\hat{c}^\dagger}
\newcommand{\hc}{\hat{c}}
\newcommand{\tr}{\mathrm{tr}}
\newcommand{\Tr}{\mathrm{Tr}}

\newcommand{\vp}{\vec{p}}
\newcommand{\vpp}{\vec{p}'}
\newcommand{\bi}{{\bar{\imath}}}
\newcommand{\bj}{{\bar{\jmath}}}

\newcommand{\Hself}{\mathcal{J}}
\newcommand{\Anti}{\mathcal{A}}
\newcommand{\Hcal}{\mathcal{H}}
\newcommand{\Hamil}{\mathcal{V}}
\newcommand{\Hvac}{\mathcal{H}_0}
\newcommand{\vecHvac}{\vec{\mathcal{H}}_0}
\newcommand{\vecHlep}{\vec{\mathcal{H}}_{\mathrm{lep}}}
\newcommand{\Hlep}{\mathcal{H}_\mathrm{lep}}
\newcommand{\Hvacy}{\underline{\mathcal{H}}_0}
\newcommand{\Hlepy}{\underline{\mathcal{H}}_\mathrm{lep}}
\newcommand{\vecHvacy}{\vec{\underline{\mathcal{H}}}_0}
\newcommand{\vecHlepy}{\vec{\underline{\mathcal{H}}}_\mathrm{lep}}
\newcommand{\vecHeff}{\vec{\Hamil}_{\mathrm{eff}}}

\newcommand{\ATAOH}{{ATAO}-$\mathcal{V}\,$}
\newcommand{\ATAOJ}{{ATAO}-$\Hself\,$}
\newcommand{\ATAOJH}{{ATAO}-$(\Hself\pm\mathcal{V})\,$}

\newcommand{\LATAO}{\langle}
\newcommand{\RATAO}{\rangle}

\newcommand{\abs}[1]{\left\lvert#1\right\rvert}
\newcommand{\norm}[1]{\left\lVert#1\right\rVert}
\newcommand{\ket}[1]{\lvert #1 \rangle}
\newcommand{\bra}[1]{\langle #1 \rvert}

\DeclareMathOperator{\sign}{sgn}

\usepackage{scalerel}
\usepackage{stackengine,wasysym}

\newcommand\reallywidetilde[1]{\ThisStyle{%
  \setbox0=\hbox{$\SavedStyle#1$}%
  \stackengine{-.1\LMpt}{$\SavedStyle#1$}{%
    \stretchto{\scaleto{\SavedStyle\mkern.2mu\AC}{.5150\wd0}}{.6\ht0}%
  }{O}{c}{F}{T}{S}%
}}

\usepackage[style=alphabetic,maxcitenames=2,maxbibnames=10,labelnumber=true,doi=false,isbn=false,url=false,natbib=true,giveninits=true,hyperref=true,bibencoding=utf8,backend=biber,backref=true]{biblatex}
\AtEveryBibitem{%
  \clearfield{note}%
  \clearfield{issn}%
  \clearlist{language}%
  \clearfield{pages}%
  \clearfield{month}%
  \clearfield{isbn}%
  }
 
 \renewbibmacro{in:}{}

\DeclareSourcemap{
  \maps[datatype=bibtex]{
    \map[overwrite]{
      \step[fieldsource=author, match=\regexp{Froustey}, final]
      \step[fieldsource=keywords, match=\regexp{\A.+\Z}, final]
      \step[fieldset=keywords, fieldvalue={,hit}, append]
    }
    \map[overwrite]{
      \step[fieldsource=author, match=\regexp{Froustey}, final]
      \step[fieldsource=keywords, notmatch=\regexp{.+}, final]
      \step[fieldset=keywords, fieldvalue={hit}]
    }
    \map{
      \step[fieldsource=author, match=\regexp{Froustey}, final]
      \step[fieldset=keywords, fieldvalue={hit}]
    }
  }
}

\DeclareSourcemap{
  \maps[datatype=bibtex]{
    \map[overwrite]{
      \step[fieldsource=author, notmatch=\regexp{Froustey}, final]
      \step[fieldsource=keywords, match=\regexp{\A.+\Z}, final]
      \step[fieldset=keywords, fieldvalue={,nohit}, append]
    }
    \map[overwrite]{
      \step[fieldsource=author, notmatch=\regexp{Froustey}, final]
      \step[fieldsource=keywords, notmatch=\regexp{.+}, final]
      \step[fieldset=keywords, fieldvalue={nohit}]
    }
    \map{
      \step[fieldsource=author, notmatch=\regexp{Froustey}, final]
      \step[fieldset=keywords, fieldvalue={nohit}]
    }
  }
}

\DeclareSourcemap{
  \maps[datatype=bibtex]{
    \map{
      \step[nocited, final]
      \step[fieldsource=author, notmatch=\regexp{Froustey}, final]
      \step[entrynull]
    }
  }
}

\AtBeginBibliography{\small}

\definecolor{UBCgreen}{RGB}{154, 205, 50}
\definecolor{MPLblue}{RGB}{68, 119, 178}
\definecolor{firebrick}{RGB}{178,34,34}

\DeclareFieldFormat{keywordcite}{%
\ifkeyword{hit}{%
\hypersetup{citecolor=firebrick}{%
#1%
}%
\hypersetup{citecolor=black}%
}{%
\hypersetup{citecolor=MPLblue}{%
#1%
}%
\hypersetup{citecolor=black}%
}%
}%
\csletcs{old:cite}{abx@macro@cite}
\renewbibmacro{cite}{\printtext[keywordcite]{\csuse{old:cite}}}

\newbibmacro{string+doiurlisbn}[1]{%
	\iffieldundef{doi}{%
		\iffieldundef{url}{%
			#1}{%
			\href{\thefield{url}}{#1}%
		}%
	}{%
		\href{https://dx.doi.org/\thefield{doi}}{#1}%
	}%
}

\DeclareFieldFormat{title}{\usebibmacro{string+doiurlisbn}{\mkbibemph{#1}}}
\DeclareFieldFormat[article,incollection]{title}%
{\usebibmacro{string+doiurlisbn}{\mkbibquote{#1}}}

\makeatletter
\newrobustcmd{\mkbibfixedbrackets}[1]{%
	\begingroup
	\blx@blxinit
	\blx@setsfcodes
	\bibleftbracket#1\bibrightbracket
	\endgroup}

\renewbibmacro*{doi+eprint+url}{%
	\ifboolexpr{test {\iffieldequalstr{eprinttype}{arxiv}} or test {\iffieldequalstr{eprinttype}{arXiv}}}
	{\setunit{\addspace}\newblock}
	{\newunit\newblock}%
	\iftoggle{bbx:eprint}
	{\usebibmacro{eprint}}
	{}%
	\newunit\newblock
}

\usepackage{xpatch}
\xpatchbibdriver{online}
{\newunit\newblock
	\iftoggle{bbx:eprint}
	{\usebibmacro{eprint}}
	{}}
{\ifboolexpr{test {\iffieldequalstr{eprinttype}{arxiv}} or test {\iffieldequalstr{eprinttype}{arXiv}}}
	{\setunit{\addspace}\newblock}
	{\newunit\newblock}%
	\iftoggle{bbx:eprint}
	{\usebibmacro{eprint}}
	{}}
{}{}

\DeclareFieldFormat{eprint:arxiv}{%
	\mkbibbrackets{%
		\ifhyperref
		{\href{https://arxiv.org/\abx@arxivpath/#1}{%
				\UrlFont{arXiv\addcolon}%
				\nolinkurl{#1}%
			}}
		{arXiv\addcolon
			\nolinkurl{#1}%
			\iffieldundef{eprintclass}
			{}
			{\addspace\UrlFont{\mkbibfixedbrackets{\thefield{eprintclass}}}}}}}
\makeatother

\definecolor{shadecolor}{gray}{0.9}

\makeatletter
\newlength{\interruptlength}
\newrobustcmd\interruptrule[3]{%
\color{#1}%
\hspace*{\dimexpr\mdfboundingboxwidth+
\mdf@innerrightmargin@length\relax}%
\rule[\dimexpr-\mdfboundingboxdepth+
#2\interruptlength\relax]%
{\mdf@middlelinewidth@length}%
{\dimexpr\mdfboundingboxtotalheight-#3\interruptlength\relax}%
}
\newrobustcmd\overlaplines[2][white]{%
\mdfsetup{everyline=false}%
\setlength{\interruptlength}{#2}
\appto\mdf@frame@leftline@single{\llap{\interruptrule{#1}{1}{2}}}
\appto\mdf@frame@rightline@single{\rlap{\interruptrule{#1}{1}{2}}}
\appto\mdf@frame@leftline@first{\llap{\interruptrule{#1}{0}{1}}}
\appto\mdf@frame@rightline@first{\rlap{\interruptrule{#1}{0}{1}}}
\appto\mdf@frame@leftline@second{\llap{\interruptrule{#1}{1}{1}}}
\appto\mdf@frame@rightline@second{\rlap{\interruptrule{#1}{1}{1}}}
\appto\mdf@frame@leftline@middle{\llap{\interruptrule{#1}{0}{0}}}
\appto\mdf@frame@rightline@middle{\rlap{\interruptrule{#1}{0}{0}}}
}
\makeatother

\usetikzlibrary{shapes,shadows}
\tikzstyle{abstractbox} = [draw=black, fill=white, rectangle, inner sep=4pt] 
\tikzstyle{abstracttitle} = [fill=white]

\newcommand{\boxabstract}[2][fill=white]{
\begin{center}
  \begin{tikzpicture}
    \node [abstractbox, #1] (box)
    {
    \begin{minipage}{0.90\linewidth}
        \vspace{0.5em}
        \centering
        \begin{minipage}{0.90\linewidth}
           \setlength{\parindent}{0.2mm}
        \itshape #2
        \end{minipage}
        \vspace{0.5em}
      \end{minipage}};
  \end{tikzpicture}
\end{center}
}

\newcommand{\partmarkLOCAL}{}

\tikzstyle{partbox} = [draw=black, fill=white, rectangle, inner sep=10pt]
\tikzstyle{parttitle} = [fill=white]

\newcommand{\boxpart}[2][fill=white]{
\begin{center}
  \begin{tikzpicture}
    \node [partbox, #1] (box)
    {
    \begin{minipage}{0.90\linewidth}
        \vspace{1em}
        \centering
        \begin{minipage}{0.90\linewidth}
           \setlength{\parindent}{2mm}
        #2
        \end{minipage}
        \vspace{1em}
      \end{minipage}};
    \node[parttitle, right=12pt] at (box.north west) {\LARGE\scshape\MakeLowercase Contents};
  \end{tikzpicture}
\end{center}
}

\newlength\tocrulewidth
\renewcommand{\thepart}{\Alph{part}}

\renewcommand*{\parttitlefont}{\normalfont\HUGE\MakeUppercase}

\renewcommand*{\printparttitle}[1]{\parttitlefont #1%
\IfEqCase{\thepart}{%
{C}{
\addcontentsline{lof}{part}{#1}%
}
{A}{
\addcontentsline{lof}{part}{\numberline{\thepart} #1}%
\addcontentsline{lot}{part}{\numberline{\thepart} #1}
}
{B}{
\addcontentsline{lof}{part}{\numberline{\thepart} #1}%
\addcontentsline{lot}{part}{\numberline{\thepart} #1}
}}}

\renewcommand{\afterpartskip}{
\IfEqCase{\thepart}{%
{C}{
\renewcommand{\partmarkLOCAL}{Introduction}
\begin{figure}[hb!]
\begin{center}
\includegraphics[width=\textwidth]{Figures/Figure_Striatum_1.pdf}
\caption*{{\bf Coronal slice of mouse brain with cortex (C) and striatum (S).} [C.~Piette].}
\addtocontents{lof}{\protect\addvspace{1em}}
\addcontentsline{lof}{figure}{Coronal slice of mouse brain with cortex (C) and striatum (S) [C.~Piette]}
\end{center}
\end{figure}
}
{D}{
\renewcommand{\partmarkLOCAL}{Contributions}
\vfill*
\begin{figure}[hb!]
\begin{center}
\includegraphics[width=0.8\textwidth]{Figures/Figure_Striatum_3.pdf}
\caption*{{\bf Corticostriatal projection from secondary somatosensory cortex to striatum.}
Axio Zoom image of sensory cortical axon terminals expressing ChR2-YFP in the striatum [E.~Perrin].}
\end{center}
\addtocontents{lof}{\protect\addvspace{1em}}
\addcontentsline{lof}{figure}{Corticostriatal projection from secondary somatosensory cortex to striatum [E.~Perrin]}
\end{figure}
\markboth{}{}%
}
{A}{
\renewcommand{\partmarkLOCAL}{Stochastic neural networks and synaptic plasticity}%
\vspace*{3cm}%
\etocsettocstyle{}{}
\etocsetnexttocdepth{0}%
\boxpart{\localtableofcontents*}}
{B}{
\renewcommand{\partmarkLOCAL}{Influence of STDP in computational models of the striatum}%
\vspace*{3cm}%
\etocsettocstyle{}{}
\etocsetnexttocdepth{0}%
\boxpart{\localtableofcontents*}}
{E}{
\renewcommand{\partmarkLOCAL}{Discussion}
\vfill*
\begin{figure}[hb!]
\begin{center}
\includegraphics[width=\textwidth]{Figures/Figure_Striatum_2.png}
\caption*{{\bf Sagittal section of a mouse brain.} A retrograde AAV-GFP was injected in the striatum, revealing mesencephalic
projections known as the MFB (medial forebrain bundle), as well as deep-layer cortical pyramidal neurons. GFP in green;
Tyrosine Hydroxylase in red [S.~Valverde].}
\addtocontents{lof}{\protect\addvspace{1em}}
\addcontentsline{lof}{figure}{Sagittal section of a mouse brain [S.~Valverde]}
\end{center}
\end{figure}
\markboth{}{}%
}}%
\thispagestyle{empty}
\vspace*{\fill}}

\makeatletter
\@addtoreset{chapter}{part}
\makeatother

\chapterstyle{veelo}

\setsecheadstyle{\LARGE \raggedright}%

\setsubsecheadstyle{\Large\bfseries\raggedright}%

\setsecnumdepth{subsection}

\setsubsubsecheadstyle{\normalfont\large\itshape\raggedright}%

\makeatletter
\cftsetindents{figure}{0pt}{2em}

\cftsetindents{table}{0pt}{2em}
\makeatother

\createmark{chapter}{left}{nonumber}{}{}
\makepagestyle{myheadings}
\makeevenhead{myheadings}{\thepage}{}{\textsc{\small\MakeUppercase\partmarkLOCAL}}
\makeoddhead{myheadings}{\textsc{\small\leftmark}}{}{\thepage}

\makepagestyle{myheadings2}
\makeevenhead{myheadings2}{\thepage}{}{\textsc{\small\MakeUppercase\partmarkLOCAL}}
\makeoddhead{myheadings2}{\textsc{\small\leftmark}}{}{\thepage}

\newcolumntype{M}[1]{>{\centering\arraybackslash}m{#1}}
\newcolumntype{N}{@{}m{0pt}@{}}

\begin{document}

\begin{titlingpage}

\begin{center}\Huge
 \bfseries The Universe at the MeV era: neutrino evolution and cosmological observables
\par\end{center}\vskip 0.5em

\vspace{2cm}

\begin{center}
            \LARGE \lineskip 0.5em%
\begin{tabular}[t]{c}
             Julien Froustey
\end{tabular}\par
\end{center}

\vspace{1cm}
\begin{center}
    \Large June 10, 2022
\end{center}

\end{titlingpage}

\pagestyle{ruled}

\frontcover

\nobibintoc

\dominitoc

\frontmatter 
\pagestyle{ruled}
\clearpage

\openright
\newpage

{
\hypersetup{linkcolor=black}
    \tableofcontents*
}

\newpage

{
\hypersetup{linkcolor=black}
    \listoftables*
    \listoffigures*
}

\newpage

\chapter*{Table of Symbols}

\vspace{-1cm}

\renewcommand{\arraystretch}{1.3}

\begin{center}
	\begin{longtable}{|lll|}
  	\hline 
  Notation & Description & Definition, relation \\\hline\hline
  \endhead
  
 $g_{\mu \nu}$ & 4-dimensional spacetime metric & Signature $(+,-,-,-)$ \\
 $\gamma_{ij}$ & Spatial metric & $g_{ij} = - a^2 \gamma_{ij}$ \\
 $a(t)$ & Scale factor & \\
 $H$ & Hubble rate & $H = \dot{a}/a$ \\
 $\Tcm$ & Comoving temperature & $\Tcm \propto a^{-1}$ \\
 $x$ & Reduced scale factor & $x = m_e / \Tcm$ \\
 $y$ & Comoving momentum & $y = p / \Tcm$ \\
 $z$ or $z_\gamma$ & Dimensionless plasma temperature & $z = T_\gamma / \Tcm$ \\
  $f_{\nu_\alpha}(p,t)$ & Neutrino distribution function & \\
 $z_{\nu_\alpha}$ & Effective temperature of $\nu_\alpha$ & Equations~\eqref{eq:param_fnu} and~\eqref{eq:param_rho} \\
 $\delta g_{\nu_\alpha}$ & Non-thermal spectral distortions of $\nu_\alpha$ & \\
 $\rho_{\nu_\alpha}$ & Energy density of $\alpha-$flavour neutrinos & $\rho_{\nu_\alpha} = 7/8 \times \pi^2/30 \times (z_{\nu_\alpha} \Tcm)^4$ \\
 $\rho_\nu$ & Total neutrino energy density & $\rho_{\nu} = \rho_{\nu_e} + \rho_{\nu_\mu} + \rho_{\nu_\tau}$ \\
 $\Neff$ & Effective number of neutrino species & See Eq.~\eqref{eq:intro_def_Neff} and section~\ref{subsec:results_Neff} \\
 $\mathcal{N}$ & Heating rate & Equation~\eqref{eq:Nheating} \\ \hline
 \textbf{BBN} & & Section~\ref{subsec:intro_BBN} and chapter~\ref{chap:BBN} \\
 $n_b$ & Baryon density & \\
 $X_i$ & Number fraction of isotope $i$ & $X_i = n_i / n_b$ \\ 
 $Y_i$ & Mass fraction of isotope $i$ & $Y_i = A_i X_i$ \\
 $\YP$ & Helium-4 primordial abundance & $\YP = Y_{\He4}$ \\
 $i/\mathrm{H}$ & Density ratio & $i/\mathrm{H} =  {n_i}/{n_\mathrm{H}}$ \\ \hline 
 $\Delta m^2_{ij}$ & Mass-squared difference & $\Delta m^2_{ij} = m_{\nu_i}^2 - m_{\nu_j}^2$ \\
 $\mathbb{M}^2$ & Matrix of mass-squared differences & $\mathbb{M}^2 = \mathrm{diag}(0,\Delta m^2_{21}, \Delta m^2_{31})$ \\
 $\theta_{ij}$ & Mixing angle & \\
 $U$ & PMNS mixing matrix & See section~\ref{subsec:Values_Mixing} \\ \hline
 \textbf{BBGKY} & & Chapter~\ref{chap:QKE}  \\
 $\vrho^{i_1 \cdots i_s}_{j_1 \cdots j_s}$ & $s-$body reduced density matrix & $\vrho^{i_1 \cdots i_s}_{j_1 \cdots j_s} = \langle \had_{j_s} \cdots \had_{j_1} \ha_{i_1} \cdots \ha_{i_s} \rangle$ \\
 $C^{ik}_{jl}$ & Correlated part of $\vrho^{ik}_{jl}$ & $\vrho^{ik}_{jl} = \vrho^i_j \vrho^k_l - \vrho^i_l \vrho^k_j + C^{ik}_{jl}$ \\
 $\tilde{v}^{ik}_{jl}$ & Interaction matrix element & $\tilde{v}^{ik}_{jl} = \bra{ik} \hat{H}_\mathrm{int} \ket{jl}$ \\
 $\Gamma^i_j$ & Mean-field potential & $\Gamma^i_j = \sum_{k,l}{\tilde{v}^{ik}_{jl} \vrho^l_k}$ \\
 $\mathcal{I}$ & Collision integral (see Eq.~\eqref{eq:C11} for $\mathcal{C}$) & $\mathcal{C}^{\vp}_{\vpp} = (2 \pi)^3 \, 2 E \, \delta^{(3)}(\vp - \vpp) \mathcal{I}(p)$ \\ \hline
 $\Hvac$ & Vacuum Hamiltonian & Equation~\eqref{eq:Hvac} \\
 $\Hlep$ & Lepton mean-field Hamiltonian & Equation~\eqref{eq:Hlep} \\
 $\Hamil$ & Effective Hamiltonian (no asymmetry) & $\Hamil = \Hvac + \Hlep$ \\
 $\Anti$ & Asymmetry matrix & $\Anti = \int{(\vrho - \bvrho) y^2 \dd{y}/(2 \pi^2)}$ \\
 $\Hself$ & Self-interaction mean-field & Equation~\eqref{DefJ} \\
 $\mathcal{K}$ & Dimensionless collision term & $\mathcal{K} = \mathcal{I}/xH$ \\
 $\gamma$ & Adiabaticity parameter & Equation~\eqref{Defgammatr1} \\ \hline
\end{longtable}
\end{center}

\renewcommand{\arraystretch}{1}

\pagestyle{simple}

\pagestyle{ruled}



\chapter[\protect\numberline{}Remerciements][Remerciements]{Remerciements}

\lettrine[lines=3]{U}{ne} thèse est, avant qu'elle commence, une expérience tout à la fois excitante et effrayante. Excitante d'abord, pour la liberté qu'elle annonce, la possibilité de travailler sur un sujet passionnant, de rencontrer des chercheurs de tous horizons... mais effrayante aussi car il faut faire face à un problème a priori « sans réponse ». Ce manuscrit est le fruit de cette expérience, et il n'aurait pu voir le jour sans l'apport de tant de personnes que je vais essayer bien modestement de mentionner.

Qui dit soutenance de thèse dit jury, dont je remercie les membres pour avoir accepté un rôle si particulier : celui de me faire entrer dans la « communauté de la recherche ». Merci notamment à George Fuller et Pasquale Serpico d'avoir accepté d'être les rapporteurs de cette thèse. Plus généralement, je souhaite remercier ici tous les chercheurs et toutes les chercheuses que j'ai eu le plaisir et la chance de côtoyer ces dernières années, que ce soit à l'IAP ou ailleurs. J'ai notamment pu collaborer avec Cristina Volpe que je remercie pour tout ce qu'elle a pu m'apprendre sur les neutrinos et leur monde si mystérieux. Une mention particulière aux membres du GReCO et à mes parrains de thèse, Silvia Galli et Patrick Boissé, pour leur accompagnement précieux. L'IAP a été un excellent environnement de travail, et ce grâce à ses équipes administratives et techniques, ainsi qu'au travail de l'équipe de direction que je salue ici. À côté de la recherche, l'enseignement a constitué une part de ce doctorat très importante pour moi, et je suis reconnaissant à Arnaud Raoux, Agnès Maître, Christophe Balland, Quentin Grimal, Laurent Coolen et Arnaud Cassan de m'avoir permis d'enseigner dans leurs différentes UE.

Parmi tous ces chercheurs, il y en a évidemment un que je dois remercier tout particulièrement : mon directeur de thèse, Cyril. Plus qu'un simple encadrant, j'ai eu avec toi une relation de quasi-collègue, me donnant de formidables clés pour entrer dans ce monde de la recherche. Nous avons franchi tant bien que mal les différents confinements en gardant toujours un contact étroit, et je ne peux que souligner tout spécialement tes capacités exceptionnelles de relecture de manuscrit dans des lieux inhabituels. Je ne sais pas si je retrouverai ailleurs une telle disponibilité bienveillante, et je resterai longtemps marqué par ta vision géométrique (fort éclairante le plus souvent~!) des choses, ainsi que cette volonté systématique de chercher à comprendre la physique derrière des équations ou des résultats numériques — une attitude qui devrait être, j'en suis bien convaincu, systématique. Pour tout cela, merci.

Les amis, bien sûr. Celles et ceux sans qui trois années de thèse (d'autant plus en rajoutant une surcouche de confinement) seraient bien plus tristes. Merci à mes amis de lycée que je retrouve annuellement en rouge \& blanc, à mes amis de prépa devenus partenaires de brunch. Je pense aux moments passés avec mes amis de l'ENS, qu'ils soient musicaux\footnote{« Chevaleresques », dixit un certain groupe nommé BrassENS.}, pseudo-sportifs (une pensée pour cette brochette d'athlètes olympiques du BDS)... je ne peux pas citer tout le monde, mais tout le monde a joué un rôle dans cette thèse, et j'espère avoir soudé des amitiés qui dureront encore longtemps. Un groupe tout particulier m'aura beaucoup accompagné, car notre amitié a été forgée dans \sout{le sang} des préparations de plans pour l'agrégation (et bien plus que ça depuis~!). Je me permets de mentionner en particulier Hugo et Jules, partenaires d'un projet d'édition un peu fou, mais qui traduit notre amour de la physique.

La vie quotidienne à l'IAP serait bien moins agréable sans la présence d'un formidable groupe de doctorants, organisant (les regrettées) movie nights, nombreuses pauses et autres vendredis à des bars non loin du laboratoire. Mille mercis à ceux que j'ai vus devenir docteurs, ceux qui m'ont accompagné pendant trois ans et ceux qui resteront encore un peu après mon départ\footnote{Félicitations à Pierre qui entre dans ces trois catégories.} : Aline, Amaury, Axel, Clément, Daniel, Denis, Eduardo, Emilie, Emma, Etienne, François, Louis, Lucas, Marko, Pierre, Quentin, Shweta, Virginia, Warren et tous les autres.

Parmi tous mes amis, une mention particulière à celui avec qui je ne pensais pas passer autant de temps\footnote{Vous avez dit confinement ?}, mais qui se sera avéré être un merveilleux colocataire, compagnon de jeux, de boissons \& fromages, de visionnage de séries, de coinche, soutien indéfectible quand cela s'est avéré nécessaire, bref merci infiniment à toi Emilien.

J'en termine avec ceux sans qui, évidemment, je ne serai pas arrivé là tant ils m'ont accompagné depuis (c'est le cas de le dire) le début. Merci à toute ma famille, merci Alexandre, merci Papa, merci Maman.


\pagestyle{ruled}

\renewcommand{\partmarkLOCAL}{Résumé en français}


\chapter[\protect\numberline{}Résumé en français][Résumé]{Résumé}

\lettrine[lines=3]{C}{ette} thèse s'inscrit dans le cadre de la cosmologie moderne, qui est définitivement entrée dans une ère de précision. Par exemple, les dernières mesures du fond diffus cosmologique (CMB) par \emph{Planck} ont fourni une nouvelle validation du modèle standard de la cosmologie $\Lambda$CDM et de ses extensions directes. Un résultat notable de \emph{Planck} est la mesure d'une quantité clé, le \emph{nombre effectif d'espèces de neutrinos} $\Neff$. Ce paramètre quantifie l'excès de densité d'énergie dans le fond cosmique de neutrinos entre l'évolution réelle de l'Univers et l'approximation dite de découplage instantané : puisque le découplage des neutrinos du plasma électromagnétique de photons, d'électrons et de positrons n'est pas entièrement terminé lorsque les annihilations des électrons/positrons prennent place, la prise en compte de ce "chevauchement" conduit à une plus grande densité d'énergie. Plus généralement, une prédiction robuste et précise des conséquences de ce "découplage incomplet des neutrinos" est cruciale car les neutrinos ont un impact sur de nombreuses étapes cosmologiques, de la nucléosynthèse primordiale (BBN) à la formation des structures.

\subsection*{Étude du découplage "standard" des neutrinos}

\paragraph{Aspects formels} Les premières déterminations de $\Neff$ négligeaient les oscillations de saveur des neutrinos, ce qui réduisait le problème physique à la résolution d'une équation de Boltzmann pour les fonctions de distribution des neutrinos. Cependant, puisque les particules n'évoluent pas dans le vide mais dans un bain thermique, des corrections d'électrodynamique quantique (QED) à la thermodynamique du plasma doivent aussi être prises en compte. Enfin, la prise en compte des oscillations de saveur nécessite de remplacer l'ensemble des fonctions de distribution par un objet plus général, une \emph{matrice densité}, et d'introduire de même une généralisation adaptée de l'équation de Boltzmann. Jusqu'à présent, les méthodes permettant d'obtenir la dénommée “équation cinétique quantique" (\emph{Quantum Kinetic Equation}, QKE) étaient une approche opérationnelle reposant sur un développement perturbatif en l'interaction faible~\cite{SiglRaffelt}, ou l'approche fonctionnelle basée sur les fonctions de Green et le formalisme CTP (\emph{Closed-Time-Path})~\cite{BlaschkeCirigliano}.

Dans le chapitre~\ref{chap:QKE}, nous présentons une méthode alternative --- à savoir une \emph{hiérarchie BBGKY généralisée} ---, où le développement perturbatif de~\cite{SiglRaffelt} est remplacé par une séparation contrôlée des contributions (non-)corrélées à la matrice densité à $1-$, $2-$, ... $n-$ corps. Cette méthode avait été utilisée dans~\cite{Volpe_2013} pour obtenir les termes de champ moyen de la QKE dans l'approximation de Hartree-Fock. Nous avons été au-delà de cette approximation et inclus des corrélations d'ordre plus élevé (utilisant l'ansatz du chaos moléculaire) afin d'obtenir le terme de collision, c'est-à-dire les contributions de diffusion et d'annihilation entre $\nu$, $\bnu$ et avec $e^-$, $e^+$ --- et donc l'équation cinétique complète.

\begin{figure}[!ht]
	\centering
	\includegraphics{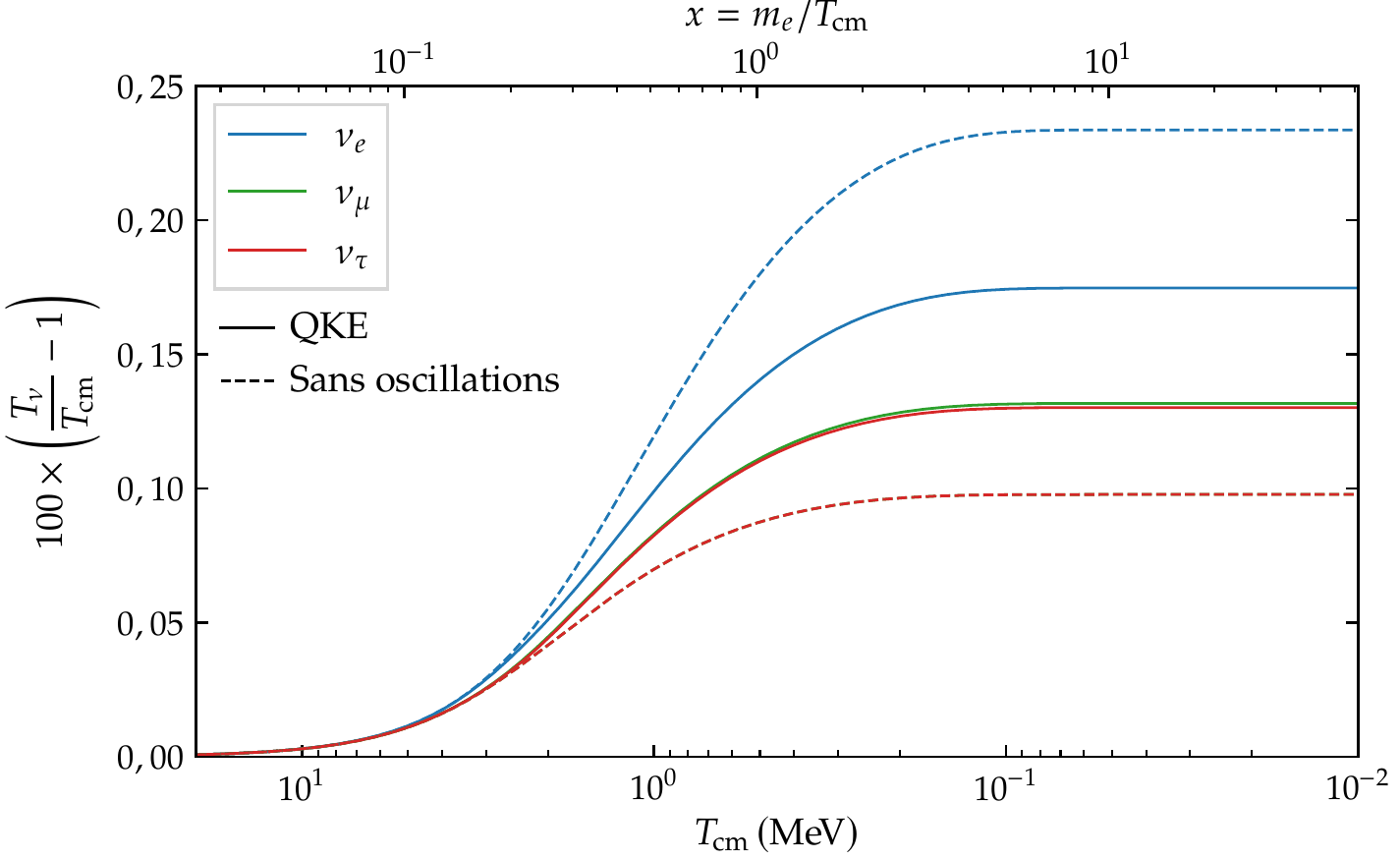}
	\caption[Évolution des températures effectives des neutrinos]{\label{fig:evolution_tnu_fr} Évolution de la température (effective) des neutrinos au cours du découplage, en prenant ou non en compte les oscillations de saveur. $\Tcm$ est la température comobile, proportionnelle à l'inverse du facteur d'échelle, qui correspond à la température des neutrinos dans l'approximation de découplage instantané.}
\end{figure}

\paragraph{Aspects numériques} Dans le chapitre~\ref{chap:Decoupling}, nous présentons un calcul du découplage des neutrinos avec, pour la première fois, le terme de collision complet (et les corrections QED susmentionnées). À titre d'exemple, l'évolution de la température\footnote{Il s'agit en réalité d'une température \emph{effective} car le processus de découplage est légèrement hors-équilibre, cela est précisé dans la section~\ref{subsec:results_Neff}.} des différentes saveurs de neutrinos est représentée Figure~\ref{fig:evolution_tnu_fr}. Avec ces résultats, nous avons obtenu la nouvelle valeur
\[ \Neff = 3, \! 0440 \, , \]
 avec une précision de quelques $10^{-4}$. Cette incertitude est due aux valeurs expérimentales actuelles des paramètres physiques (et notamment l'angle de mélange $\theta_{12}$), ainsi que la variabilité reliée aux paramètres numériques du code \texttt{NEVO} que nous avons développé.

Des calculs précédents, qui tenaient compte des oscillations de saveur et qui ont abouti à la valeur $\Neff \simeq 3, \! 045$~\cite{Relic2016_revisited}, ne tenaient pas compte de l'intégralité du terme de collision : ses composantes hors diagonale étaient approchées par un terme d'amortissement. L'inclusion de ce terme sans aucune approximation est un véritable défi numérique, en particulier à cause de la raideur qu'il apporte à l'équation différentielle et parce qu'il se calcule en un temps $\mathcal{O}(N^3)$ où $N$ est la taille de la grille d'impulsions. Nous assurons un temps de calcul raisonnable grâce à une amélioration majeure, à savoir le calcul direct du jacobien du système différentiel. Notre résultat sur $\Neff$ a été confirmé ultérieurement par~\cite{Bennett2020}.

\paragraph{Description approchée de l'évolution} Nous avons également introduit une description effective des oscillations de saveur, qui donne des résultats indiscernables de ceux obtenus en résolvant l'équation exacte. Elle permet également de réduire considérablement le temps de calcul, ce qui constitue une autre amélioration importante de notre code. Cette approximation repose sur l'existence d'une grande séparation d'échelles entre les fréquences d'oscillation et le taux de collision, ce qui permet de faire la moyenne de ces oscillations. En d'autres termes, la matrice densité reste toujours diagonale dans la base de la matière (la base des états propres du Hamiltonien prenant en compte les effets du vide et de champ moyen). Nous avons appelé cette description simplifiée l'approximation de \emph{Transfert Adiabatique d'Oscillations Moyennées} (ATAO pour \emph{Adiabatic Transfer of Averaged Oscillations}). De plus, nous avons utilisé cette approximation afin de mieux comprendre certains résultats comme l'absence d'effets de la phase CP dans le découplage standard des neutrinos.

\subsection*{Nucléosynthèse primordiale et découplage incomplet des neutrinos}

En résolvant la QKE, nous obtenons les distributions gelées des (anti)neutrinos, qui à leur tour donnent accès aux paramètres cosmologiques tels que $\Neff$ (cf.~ci-dessus) ou la densité d'énergie des neutrinos \emph{aujourd'hui} $\Omega_\nu$. Ainsi, nous sommes notamment en possession des deux paramètres qui fixent les différents effets du découplage incomplet des neutrinos sur la plus ancienne sonde de l'histoire de l'Univers dont nous disposons --- la BBN --- : la distribution de $\nu_e$, $\bnu_e$ et la densité d'énergie paramétrée par $\Neff$.

Le chapitre~\ref{chap:BBN} est dédié à l'évaluation des changements des abondances primordiales d'hélium, de deutérium et de lithium dus au découplage incomplet des neutrinos. Tout d'abord, les abondances des éléments légers dépendent du taux d'expansion de l'Univers (donc de $\Neff$, via l'effet dit d'horloge --- \emph{clock effect} ---). Ensuite, l'abondance des neutrons au début de la BBN est entre autres fixée par le rapport neutrons-protons qui varie si l'on change les distributions de $\nu_e$, $\bnu_e$. Nous avons étudié en détail comment ces effets interagissaient, en comparant leurs contributions relatives et en fournissant des estimations analytiques lorsque cela était possible. Ce travail théorique est mené conjointement à une étude numérique, en combinant notre code d'évolution des neutrinos et le code de BBN \texttt{PRIMAT}~\cite{Pitrou_2018PhysRept}. En particulier, nous avons pu résoudre un désaccord existant dans la littérature entre~\cite{Mangano2005} et~\cite{Grohs2015} concernant la variation de l'abondance du deutérium due à un découplage incomplet des neutrinos.

\subsection*{Évolution des asymétries primordiales}

Le cas "standard" du découplage des neutrinos suppose que l'asymétrie des leptons est nulle, une approximation justifiée pour les électrons et les positrons (dont la dégénérescence doit être de l'ordre du rapport baryon/photon $\eta \simeq 6 \times 10^{-10}$~\cite{Fields:2019pfx} par neutralité de charge), mais il n'existe pas de telle contrainte pour les neutrinos. Le CMB et, plus important encore, le BBN sont en fait les meilleures sources de limites sur les asymétries des neutrinos, puisque de telles asymétries affecteraient les abondances primordiales par les mêmes mécanismes que ceux présentés précédemment.

La présence d'asymétries de neutrinos non nulles ajoute une couche de complexité considérable à la physique de l'évolution des neutrinos. En effet, il faut désormais prendre en compte un terme supplémentaire de champ moyen d'auto-interaction dans la QKE, qui domine pendant une grande partie de l'ère du découplage des neutrinos pour des asymétries $\mu/T \in [10^{-3},10^{-1}]$. Dans la lignée de nos travaux sur la résolution des QKEs dans le cas standard avec le terme de collision complet, nous étendons notre code au cas asymétrique dans le chapitre~\ref{chap:Asymmetry}.

\begin{figure}[!h]
	\centering
	\includegraphics{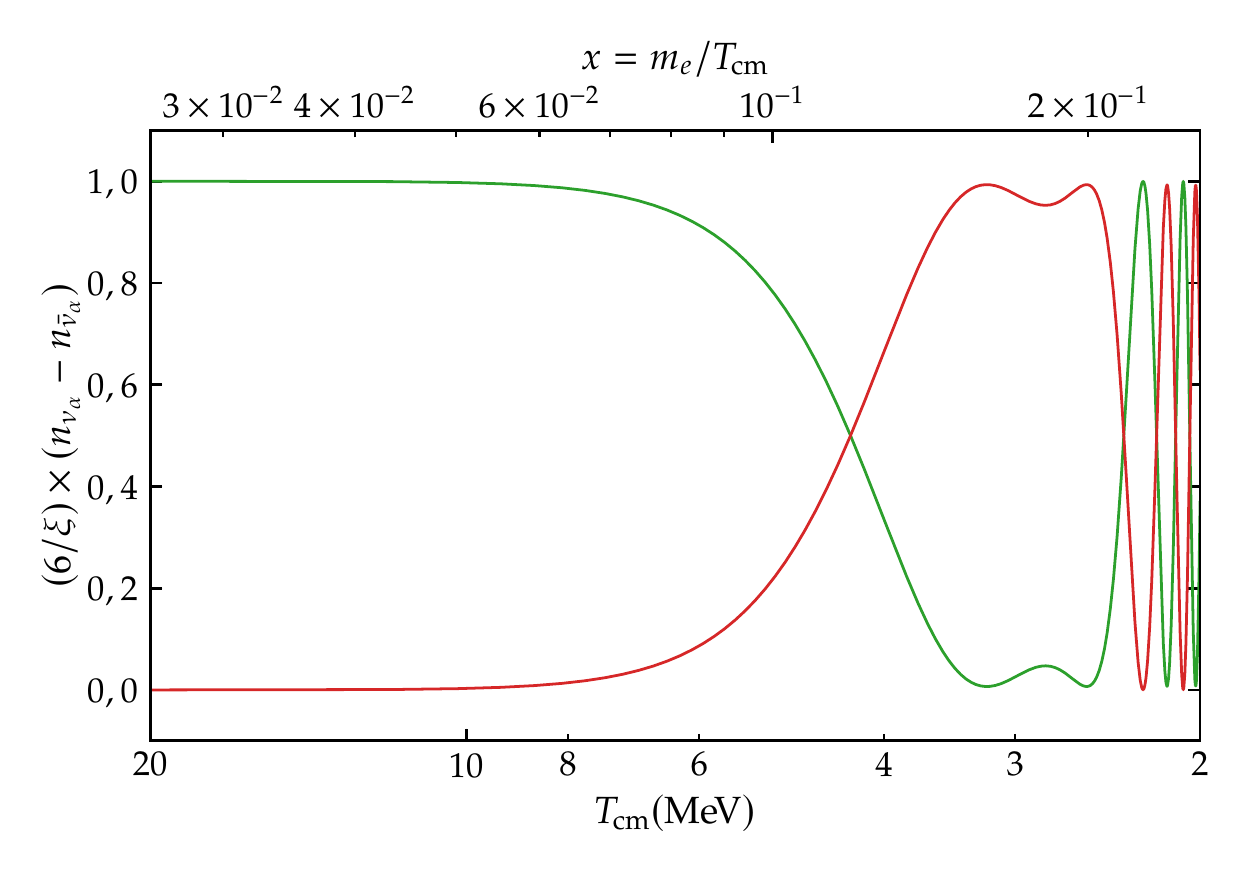}
	\caption[Exemple d'évolution des neutrinos dans un cas asymétrique à deux saveurs]{\label{fig:asym_resume_fr} Évolution de l'asymétrie dans un cas à deux saveurs $\nu_\mu$ (vert) - $\nu_\tau$ (rouge), avec une différence de masse $\Delta m^2 = 2,45 \times 10^{-3} \, \mathrm{eV}^2$ et un angle de mélange $\theta = 0,831$. Les résultats de la résolution directe de la QKE (trait plein) et via l'approximation ATAO (pointillés) sont indiscernables. \emph{Paramètres initiaux :} $\xi_1 = \mu_1/T = 0,001 \equiv \xi$, $\xi_2 = 0$.}
\end{figure}

De plus, nous avons généralisé l'approximation ATAO pour tenir compte des auto-interactions, qui rendent le Hamiltonien non linéaire. Cette description approchée nous a permis de retrouver analytiquement les résultats connus sur les oscillations collectives dites \emph{synchrones}, mais aussi de découvrir que ce régime est généralement suivi d'oscillations \emph{quasi-synchrones} de fréquences croissant plus rapidement. Nous avons fourni de nombreuses vérifications analytiques et numériques de ce nouveau résultat, dans le cas simplifié à deux saveurs mais aussi dans le cadre général à trois saveurs. À titre d'illustration, on représente Figure~\ref{fig:asym_resume_fr} un exemple d'évolution de l'asymétrie dans un cas simplifié à deux saveurs. Dans cet exemple, le régime d'oscillations synchrones prend place jusqu'à environ $2,8 \, \mathrm{MeV}$, et est suivi par le régime d'oscillations quasi-synchrones.

Nous avons exploré plus avant la dépendance de la configuration finale des neutrinos par rapport aux paramètres de mélange, et nous avons notamment montré que la phase CP de Dirac ne peut pas affecter substantiellement la valeur finale de $\Neff$ ni le spectre électronique final des (anti)neutrinos, et ne devrait donc pas affecter les observables cosmologiques.



\mainmatter 

\pagestyle{ruled}

\chapter*{Introduction}
\addcontentsline{toc}{chapter}{\protect\numberline{}Introduction}

\begin{quote}
\emph{I have done a terrible thing today, something which no theoretical physicist should ever do. I have suggested something that can never be verified experimentally.}

 \sourceatright{Wolfgang Pauli, 1930, quoted in~\cite{Hoyle1967}}
\end{quote}

W. Pauli is reported to have confessed this “terrible thing” to his friend Walter Baade, after having proposed the existence of a particle in order to save the principle of energy conservation: the \emph{neutrino}. Indeed, the measurement of the energy of electrons emitted in beta decays was in complete disagreement with predictions. For instance, the predicted decay $\ce{^{14}_6\mathrm{C} -> ^{14}_7\mathrm{N} + e^-}$ should lead to a very peaked electron energy $E_e = (m_{^{14}\mathrm{C}} - m_{^{14}\mathrm{N}})c^2$. On the contrary, scientists measured a continuous spectrum of energies... Pauli thus proposed that the final state actually contained a third particle, neutral and very light, that would take away part of the disintegration energy, such that the decay actually reads $\ce{_6^{14}\mathrm{C} -> _7^{14}\mathrm{N} + e^- + \bar{\nu}_e}$. However, with such properties, a neutrino (actually here, an antineutrino) should be very difficult to detect, hence the quotation at the start of this introduction. Yet, a few decades later, in 1956, F. Reines and C. L. Cowan sent a telegram informing Pauli of the discovery of the electronic antineutrino. Reines was awarded half the Nobel Prize in Physics “for the detection of the neutrino” in 1995, Cowan having passed away.

Today, we associate one neutrino to each charged lepton, bringing to $3$ the number of known neutrinos: $\nu_e$, $\nu_\mu$ and $\nu_\tau$. In the Standard Model of particle physics, they are \emph{massless} particles which only interact via the \emph{weak} interaction. However, there is now a large body of experimental evidence that neutrinos have properties that are not predicted by the Standard Model. In particular, they undergo \emph{flavour oscillations}, a property that cannot be  understood with massless species, whose discovery led to another Nobel Prize in 2015, awarded to T. Kajita (from the Super-Kamiokande experiment) and A. B. McDonald (from the Sudbury Neutrino Observatory).

These exciting developments in particle physics have a particular resonance at an incredibly wider scaler --- in cosmology. Indeed, neutrinos play a key role at various stages of the evolution of the Universe, and the imprints they leave on cosmological observables will be uncovered more and more precisely as new detectors are being developed. In this period of “precision cosmology”, there is therefore an important need for accurate theoretical predictions on the values of these cosmological observables, in order to be able to pinpoint potential hints for beyond-the-Standard-Model physics.

During this PhD, we have focused on the epoch when the temperature of the Universe was about $10^{10} \, \mathrm{K} \sim 1 \, \mathrm{MeV}$, the so-called “\emph{MeV age}”. As we will detail in the forthcoming chapters, this period is extremely rich in physical events involving neutrinos: they decouple from the electromagnetic plasma, electrons and positrons annihilate, and primordial nucleosynthesis starts. It is crucial to accurately predict the features of neutrino distributions at this epoch, as the MeV era is the neutrino “Grand Finale” before their behaviour is simply described by that of a free streaming particle bath, the Cosmic Neutrino Background. Our goal was twofold: develop new theoretical tools, regarding the derivation of the resolution of the evolution equations, and apply a numerical code to investigate neutrino evolution in various frameworks. In particular, we have studied in-depth the cases of “standard” neutrino decoupling, which involves solely Standard Model physics with the known results about flavour oscillations, and the situation of potentially large primordial neutrino/antineutrino asymmetry, which could largely affect the cosmological expansion.

\noindent This PhD is based on the following publications:
\begin{itemize}
	\item \cite{Froustey2019} focusing on the various ways in which incomplete neutrino decoupling (without taking into account flavour mixing) affects the abundances of light elements produced during Big Bang Nucleosynthesis, in order to understand semi-analytically the physics at play;
	\item \cite{Froustey2020} in which we performed the first calculation of neutrino decoupling including all known physical effects required to reach a few $10^{-4}$ accuracy on the parameter $\Neff$ prediction;
	\item \cite{FrousteyTAUP2021} Proceedings of the TAUP2021 conference where the results of the above paper were presented, and a discussion on the neutrino energy density parameter $\Omega_\nu$ was added;
	\item \cite{Froustey2021} extending the study of neutrino evolution to account for the possibility of non-zero neutrino/antineutrino asymmetry.
\end{itemize}

\noindent The manuscript is organised as follows. In chapter~\ref{chap:IntroCosmo}, we provide a wide introduction to cosmology and, in particular, neutrino physics in connection with cosmology. This gives the basis for the in-depth analyses of the next chapters. In order to perform the precision calculation of $\Neff$, we first present a new derivation of the evolution equation of neutrinos and antineutrinos (the “Quantum Kinetic Equations”) in chapter~\ref{chap:QKE}. Chapter~\ref{chap:Decoupling} is dedicated to the standard calculation of $\Neff$. We then use these results to study their consequences on Big Bang Nucleosynthesis in chapter~\ref{chap:BBN}. Finally, we extend the previous analytical and numerical tools to the case of non-zero asymmetries in chapter~\ref{chap:Asymmetry}.

\adjustmtc

\clearpage

\pagestyle{ruled}

\chapter{Neutrinos in cosmology: an overview}
\label{chap:IntroCosmo}

\setlength{\epigraphwidth}{0.63\textwidth}
\epigraph{Our whole universe was in a hot, dense state \\
Then nearly fourteen billion years ago expansion started [...]}{Barenaked Ladies, \emph{Big Bang Theory Theme}}

{
\hypersetup{linkcolor=black}
    \minitoc
}

Neutrinos are probably the most exciting particles in the Standard Model: being neutral fermions, they could be Majorana particles, and we \emph{know} (see section~\ref{sec:Intro_massive_nu}) that at least two neutrino states are massive --- a feature not predicted by the Standard Model. If many particle physics experiments have been able to determine the properties of these particles with increasing accuracy in the last decades, there is another promising laboratory that can be used: the Universe itself.
Cosmology indeed provides complementary results, as the imprints left by neutrino evolution on cosmological observables are directly dependent on neutrino properties. In this introductory chapter, we present the main elements of neutrino physics that are relevant to study their evolution in the early Universe. This presentation, which is necessarily limited, only scratches the surface of many topics that are developed in, e.g.~\cite{Neutrino_Cosmology,LesgourguesPastor}.

\section{Elements of standard cosmology}

We describe in this section the key features of the Standard Model of cosmology, introducing the necessary notations and equations for the forthcoming sections and chapters. It is intended as a concise and oriented presentation of cosmology, and we refer to many excellent references such as~\cite{KolbTurner,PeterUzan,WeinbergCosmology,ModernCosmology} for a more complete presentation of cosmology.
The reader familiar with standard cosmology can skip this first section and jump to section~\ref{sec:Intro_Neutrinos}, dedicated to an overview of neutrinos in the early Universe. 

\subsection{The homogeneous and isotropic universe}

The Standard Model of cosmology is based on two main assumptions~\cite{PeterUzan}:
\begin{itemize}
	\item General Relativity is an adequate theory of gravitation ;
	\item the \emph{cosmological principle}: on the largest scales, the Universe is spatially homogeneous and isotropic.
\end{itemize}
This second hypothesis must be understood as a statistical, averaged property on scales typically $\gtrsim 100 \, \mathrm{Mpc}$. Under these assumptions, it can be shown that space-time must be described by a Friedmann-Lemaître-Robertson-Walker (FLRW) geometry, that is with the metric (we use the same conventions as~\cite{KolbTurner,GiuntiKim}):
\begin{equation}
\dd{s}^2 = g_{\mu \nu} \dd{x^\mu} \dd{x^\nu} \equiv \dd{t}^2 - a^2(t) \left(\frac{\dd{r}^2}{1-Kr^2} + r^2 \dd{\theta}^2 + r^2 \sin{\theta}^2 \dd{\varphi}^2 \right) \, ,
\end{equation}
where $(t,r,\theta,\varphi)$ are the coordinates, $K=-1,0,+1$ for spaces with negative, zero and positive curvature, and $a(t)$ is the \emph{scale factor}. Such a cosmology is entirely determined by the evolution of $a(t)$, which is given by Einstein theory of General Relativity. It unveils the relationship between the geometry of spacetime (through the metric $g_{\mu \nu}$) and its energy content (through the stress-energy tensor $T_{\mu \nu}$). Einstein field equations read
\begin{equation}
\label{eq:Einstein}
R_{\mu \nu} - \frac12 R g_{\mu \nu} = 8 \pi \mathcal{G} T_{\mu \nu} + \Lambda g_{\mu \nu} \, .
\end{equation}
In this equation $R_{\mu \nu}$ is the Ricci tensor and $R = g^{\mu \nu} R_{\mu \nu}$ the Ricci scalar, $\mathcal{G}$ the gravitational constant, and $\Lambda$ the \emph{cosmological constant}.  Note that we use natural units in which $\hbar = c = k_B = 1$. We assume that basics of General Relativity are known to the reader, and refer for instance to~\cite{PeterUzan,Wald,Carroll_RG}.

\subsubsection{Dynamical equations}

Thanks to the cosmological principle, the Universe can be described as a collection of perfect fluids, for which the energy-momentum tensor reads
\begin{equation}
T^{\mu \nu} = (\rho + P) u^\mu u^\nu - P g^{\mu \nu} \, ,
\end{equation}
where $\rho$ is the energy density, $P$ is the pressure, and $u^\mu = \dd{x^\mu}/\dd s$ is the four-velocity of the fluid. In the comoving frame where the perfect fluid is at rest, $u^\mu = (1,0,0,0)$ and one has
\begin{equation}
{T^\mu}_\nu = \mathrm{diag}(\rho,-P,-P,-P) \, .
\end{equation}
We see then from~\eqref{eq:Einstein} that it is possible to interpret the cosmological constant as the energy density of the vacuum, through
\begin{equation}
\rho_\Lambda = \frac{\Lambda}{8 \pi \mathcal{G}} \qquad \text{and} \qquad P_\Lambda = - \rho_\Lambda \, .
\end{equation}
We detail below that this \emph{a priori} peculiar \emph{negative} pressure amounts to the fact that the cosmological constant corresponds to a constant energy density.

\noindent We can now obtain the equations governing the dynamics of expansion:\footnote{Although we only quote the results, let us give here the non-vanishing Christoffel symbols of the FLRW metric, which we write $\dd{s}^2 = \dd{t}^2 - a^2(t) \gamma_{ij} \dd{x^i}\dd{x^j}$:
\begin{equation*}
{\Gamma^{0}}_{ij} = \dot{a} a \gamma_{ij} \quad , \quad {\Gamma^{i}}_{0j} = \frac{\dot{a}}{a} {\delta^i}_j \quad , \quad {\Gamma^{i}}_{jk} = \frac12 \gamma^{il}\left(\partial_j \gamma_{kl} + \partial_k \gamma_{jl} - \partial_l \gamma{jk} \right) = ^{(3)}\!{\Gamma^i}_{jk} \, .
\end{equation*}
This allows to compute the non-zero components of the Ricci tensor,
\begin{equation*}
R_{00} = - 3 \frac{\ddot{a}}{a} \quad , \quad R_{ij} = \left(\frac{\ddot{a}}{a} + 2 H^2 + \frac{2 K}{a^2} \right) a^2 \gamma_{ij}  \quad \text{and the Ricci scalar} \quad R = - 6 \left( \frac{\ddot{a}}{a} + H^2 + \frac{K}{a^2} \right) \, .
\end{equation*}
}
\begin{itemize}
	\item from the $0-0$ component of~\eqref{eq:Einstein}, one gets the famous \emph{Friedmann equation}:
	\begin{equation}
		\label{eq:Friedmann}
		\boxed{\left(\frac{\dot{a}}{a}\right)^2 \equiv H^2 = \frac{8 \pi \mathcal{G}}{3} \rho - \frac{K}{a^2} } \, ,
	\end{equation}
	where we defined the \emph{Hubble parameter} $H \equiv \dot{a}/a$. Introducing the Planck mass $M_\mathrm{Pl} \equiv \mathcal{G}^{-1/2} \simeq 1.22 \times 10^{19} \, \mathrm{GeV}$, we have
	\[ H^2 = \frac{8 \pi}{3 M_\mathrm{Pl}^2} \rho - \frac{K}{a^2} \, . \]
	We will also sometimes use the reduced Planck mass $m_\mathrm{Pl} \equiv M_\mathrm{Pl}/\sqrt{8 \pi}$. The \emph{critical density} is the energy density corresponding to a flat ($K=0$) Universe today, $\rho_\mathrm{crit} \equiv 3 H_0^2 / 8 \pi \mathcal{G}$. The energy density parameter is then defined as $\Omega = \rho/\rho_\mathrm{crit}$, the different values in the standard model of cosmology being given below.
	\item from the $i-i$ component of~\eqref{eq:Einstein}, one gets
	\begin{equation}
	\label{eq:acceleration}
	\frac{\ddot{a}}{a} = - \frac{4 \pi \mathcal{G}}{3}(\rho + 3 P) \, .
	\end{equation}
	\item the energy-momentum conservation $\nabla_{\mu} T^{\mu \nu} = 0$ ($\nabla_\mu$ being the covariant derivative) reduces to
	\begin{equation}
		\label{eq:eq_conservation}
		\boxed{\dot{\rho} + 3 H (\rho + P) = 0} \, .
	\end{equation}
\end{itemize}
These three equations are not independent, which is a consequence of Bianchi identities. The most often used equations are then~\eqref{eq:Friedmann} and~\eqref{eq:eq_conservation}.

\subsubsection{Some solutions} 

\paragraph{Friedmann equation} Perfect fluids are characterized by their \emph{equation of state} $P = w \rho$, where $w$ is independent of time. With such a relation, we see from~\eqref{eq:eq_conservation} that the energy density evolves as $\rho \propto a^{-3(1+w)}$. We distinguish three examples of interest:
\begin{itemize}
	\item \emph{radiation} ($w = 1/3$), for which $\rho \propto a^{-4}$,
	\item pressureless \emph{matter} ($w = 0$), for which $\rho \propto a^{-3}$,
	\item \emph{dark energy} ($w = -1$), for which $\rho = \mathrm{const}$.
\end{itemize}

\paragraph{The $\bm{\Lambda}$CDM model} In the standard model of cosmology, which accounts extremely well for an incredible variety of observations,\footnote{There are of course some tensions, like the $H_0$ tension between “early” and “late” measurements of the Hubble constant. We do not discuss such limits of the $\Lambda$CDM model in this introduction.} the constituents of the Universe are~\cite{PDG}:
\begin{itemize}
	\item baryonic matter (the “usual” matter), which behaves as pressureless matter, amounting today to $\Omega_b^0 \simeq 0.049$,
	\item photons which behave as radiation and amounting to $\Omega_r^0 \sim 10^{-4}$,
	\item neutrinos, which are the only known particles in the Standard Model which were ultrarelativistic at early times (during Big Bang Nucleosynthesis and Cosmic Microwave Background formation) and thus behaving as radiation, and are non-relativistic today (at least for two eigenstates, cf.~section~\ref{subsec:Omeganu}). Their contribution $\Omega_\nu^0$ is thus split between $\Omega_r^0$ and the total matter part $\Omega_m^0$,
	\item “dark” components, which account for the missing energy (the baryonic and photon components which come from the Standard Model of particle physics only represent $5 \, \%$ of the total energy budget):
	\begin{itemize}
		\item a cosmological constant $\Lambda$ corresponding to \emph{dark energy}, which dominates the energy density today with $\Omega_\Lambda^0 \simeq 0.685$,
		\item \emph{cold dark matter}, necessarily non-baryonic and whose nature is still unknown today, which amounts for $\Omega_c^0 \simeq 0.265$.
	\end{itemize}
\end{itemize}
These last two components are at the origin of the name “$\Lambda$CDM” of the model.

\begin{figure}[!h]
	\centering
	\includegraphics{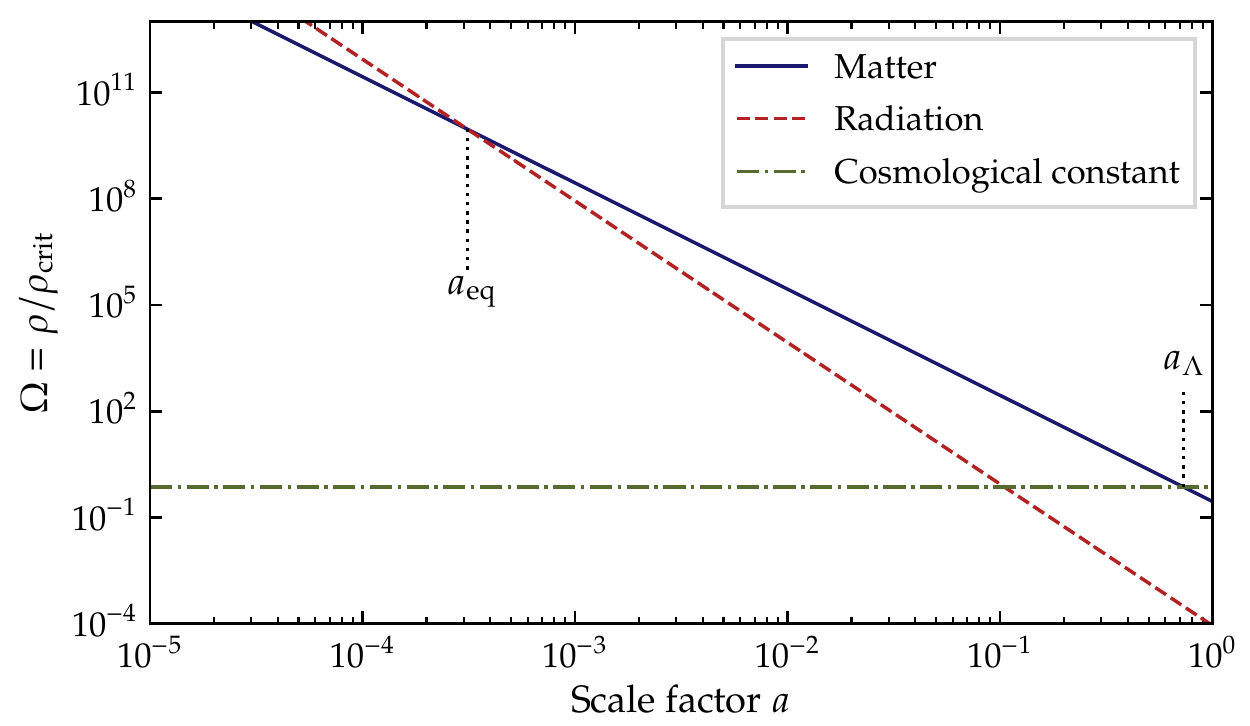}
	\caption[Energy density as a function of the scale factor for different constituents]{\label{fig:rho_scalefactor} Energy density as a function of the scale factor for nonrelativistic matter ($\propto a^{-3}$), radiation ($\propto a^{-4}$) and a cosmological constant ($= \mathrm{const.}$). At early times, the energy density of the Universe is dominated by the radiation component.}
\end{figure}

We plot on Figure~\ref{fig:rho_scalefactor} the evolution of the energy density for the different constituents of the Universe. Given the different scalings of $\rho$ with the scale factor, it appears that, although today the dark energy is dominating the Universe, this was not the case in the past. In the early Universe, we were in the so-called \emph{radiation-dominated} era, until the energy densities of radiation and matter became equal (“matter-radiation equality”) at the scale factor $a_\mathrm{eq}$. We then entered the \emph{matter-dominated} era, until the recent period of accelerated expansion driven by the cosmological constant, the transition taking place at the scale factor $a_\Lambda$ at which the acceleration of the expansion was zero.\footnote{Indeed, one can see from~\eqref{eq:acceleration} that the dominance of matter makes the Universe decelerate, while dark energy drives an accelerated expansion. The transition thus corresponds to a vanishing acceleration.}

\paragraph{Scale factor and time} Assuming that the Universe is flat ($K=0$), we can solve Friedmann equation~\eqref{eq:Friedmann} for a fluid of equation of state $P = w \rho$:
\begin{equation}
	\dot{a} \propto a^{-(1+3w)/2} \quad \implies \quad a(t) \propto t^{\frac{2}{3(1+w)}} \, .
\end{equation}
In particular, we find for radiation the important relationship between the Hubble rate and cosmic time:
\begin{equation}
	\label{eq:hubble_radiation}
	a(t) \propto \sqrt{t} \quad \text{hence} \quad H = \frac{1}{2 t} \, .
\end{equation}

\vspace{-0.5cm}

\subsubsection{Redshifting of momenta} 

We end this section with a very useful result valid in FLRW spacetime: the so-called “redshifting” of physical momentum as the Universe expands.

Let us show that the physical linear momentum of a free-falling particle decreases as $1/a$ as the Universe expands, following~\cite{Neutrino_Cosmology,ModernCosmology}. We start with the geodesic equation
\begin{equation}
\label{eq:geodesic}
\frac{\dd^2{x^\mu}}{\dd \lambda^2} + \Gamma^{\mu}_{\alpha \beta} \frac{\dd{x^\alpha}}{\dd \lambda} \frac{\dd{x^\beta}}{\dd \lambda} = 0 \, ,
\end{equation}
where the use of the affine parameter $\lambda$ instead of the proper time $\tau$ allows to treat at the same time massive and massless particles. It is implicitly defined such that the four-momentum $P^\mu = (E, P^i)$ reads
\begin{equation}
P^\mu \equiv \frac{\dd{x^\mu}}{\dd \lambda} \, .
\end{equation}
Note that the $0-$component of this definition gives $\dd / \dd{\lambda} = E \dd / \dd{t}$. The $0-$component of the geodesic equation~\eqref{eq:geodesic} can thus be written
\begin{equation}
E \frac{\dd E}{\dd t} + \Gamma^{0}_{ij} P^i P^j = 0 \, ,
\end{equation}
where we use the fact that only the spatial components of $\Gamma^0_{\alpha \beta}$ are non-zero. Since $\Gamma^0_{ij} = - (\dot{a}/a) g_{ij} = \dot{a} a \gamma_{ij}$, we have
\begin{equation}
\label{eq:geodesic_temp}
E \frac{\dd E}{\dd t} +\dot{a} a \gamma_{ij} P^i P^j = 0 \, .
\end{equation}

The norm of the four-momentum is $m^2 = P^\mu P_\mu = E^2 - a^2 \gamma_{ij} P^i P^j$. It is customary to introduce the \emph{physical} linear momentum $p^i \equiv a(t) P^i$, which allows to rewrite the norm of the four-momentum $E^2 - \lvert \vec{p} \rvert^2 = m^2$ with $\lvert \vec{p} \rvert^2 \equiv \gamma_{ij} p^i p^j$ (as in flat spacetime). Deriving this relation gives (note that the calculation is also valid if $m=0$)
\begin{equation}
\label{eq:norm_4momentum}
E \frac{\dd E}{\dd t} - \lvert \vec{p} \rvert \frac{\dd \lvert \vec{p} \rvert}{\dd t} = 0 \, .
\end{equation}
Rewriting~\eqref{eq:geodesic_temp} with the physical momentum finally leads to
\begin{equation}
E \frac{\dd E}{\dd t} +\frac{\dot{a}}{a} \lvert \vec{p} \rvert^2 = 0 \quad \xRightarrow[\text{using \eqref{eq:norm_4momentum}}]{} \quad \frac{1}{\lvert \vec{p} \rvert} \frac{\dd \lvert \vec{p} \rvert}{\dd t} = - \frac{\dot{a}}{a} \, .
\end{equation}
Therefore, we have proven that
\begin{equation}
\label{eq:scaling_p}
\boxed{\lvert \vec{p} \rvert \propto a^{-1}} \, .
\end{equation}

\subsection{Equilibrium thermodynamics}

Having described gravity in the homogeneous Universe, we must now turn to the equations governing matter and radiation. The statistical properties of the particles filling the Universe are described by their \emph{distribution functions} $f(p,t)$, which give the number of particles of momentum $p$ at time $t$ (there is no dependence on space nor on the direction of $\vec{p}$ thanks to homogeneity and isotropy):
\begin{equation}
	\dd N(p,t) = f(p,t) \frac{4 \pi p^2 \dd{p}}{(2\pi)^3} \, .
\end{equation}

The number density $n$, energy density $\rho$, and pressure $P$ of a dilute, weakly-interacting of a gas of particles with $g$ internal degrees of freedom (for example, $2$ for photons or charged leptons) and distribution function $f(p,t)$ read
\begin{subequations}
	\label{eq:thermo_intro}
	\begin{align}
		n &= \frac{g}{2 \pi^2} \int_{0}^{\infty}{p^2 \dd{p} f(p,t)} \, , \\
		\rho &= \frac{g}{2 \pi^2} \int_{0}^{\infty}{E(p) p^2 \dd{p} f(p,t)} \, , \\
		P &= \frac{g}{2 \pi^2} \int_{0}^{\infty}{\frac{p^2}{3 E(p)} p^2 \dd{p} f(p,t)} \, . \label{eq:pressure_general}
	\end{align}
\end{subequations}
If the reaction rates of a particle species are high enough (cf.~the discussion on decoupling below), it will be maintained in \emph{kinetic} equilibrium, such that its distribution function reads
\begin{equation}
f(p) = \frac{1}{\displaystyle e^{(E-\mu)/T} \pm 1} \, ,
\end{equation}
with $\mu$ the chemical potential and $T$ the temperature of the species. The $+$ sign is for fermions (Fermi-Dirac (FD) distribution), the $-$ sign for bosons (Bose-Einstein (BE) distribution). If this particle species is in \emph{chemical} equilibrium, there is an additional constraint on the chemical potentials, namely, the reaction $a + b \leftrightarrow c + d$ implies $\mu_a + \mu_b = \mu_c + \mu_d$. If this reaction is an elastic scattering, this is trivial since the incoming and outgoing particles are identical.\footnote{That is why we should rather talk about chemical equilibrium \emph{with} a given system. In contrast, kinetic equilibrium is an “intrinsic” property: self-interactions can maintain equilibrium spectra even if there is no external species to be at equilibrium with.} An interesting case is particle/antiparticle annihilation: for instance, annihilations into pairs of photons will impose $\mu = - \bar{\mu}$ since $\mu_\gamma = 0$.

Let us give the explicit results for the different thermodynamic quantities in the relativistic limit $T \gg m$, for non-degenerate particles $T \gg \mu$:
\begin{equation}
\label{eq:thermo_quantities_relat}
n = \left\{ \begin{aligned} &g \frac{\zeta(3)}{\pi^2} T^3 &\text{(BE)} \\
&g \frac34 \frac{\zeta(3)}{\pi^2} T^3 &\text{(FD)} \end{aligned} \right.
\qquad , \qquad 
\rho = \left\{ \begin{aligned} &g \frac{\pi^2}{30} T^4 &\text{(BE)} \\
&g \frac78 \frac{\pi^2}{30} T^4 &\text{(FD)} \end{aligned} \right. 
\qquad , \qquad
P = \frac{\rho}{3} \, ,
\end{equation}
while in the non-relativistic limit ($m \gg T$), the results are the same for fermions and  for bosons:
\begin{equation}
\label{eq:thermo_quantities_NR}
n = g \left(\frac{mT}{2 \pi}\right)^{3/2} e^{-(m-\mu)/T} \qquad , \qquad \rho = m n \qquad , \qquad P = n T \ll \rho \, .
\end{equation}
These results justify the equations of state $w=1/3$ for radiation and $w=0$ for non-relativistic matter.

\paragraph{Entropy} One can show, combining the conservation equation~\eqref{eq:eq_conservation} and the derivative of pressure~\eqref{eq:pressure_general} with respect to temperature (with an equilibrium distribution), that the \emph{entropy density}
\begin{equation}
	\label{eq:def_entropy}
	s \equiv \frac{\rho + P - \mu n}{T} \, ,
\end{equation}
satisfies
\begin{equation}
\label{eq:entropy_evo}
\dd{(sa^3)} = - \frac{\mu}{T} \dd{(na^3)} \, .
\end{equation}
Therefore, for non-degenerate matter ($\mu/T \ll 1$) or when it is neither destroyed nor created ($\dd{(n a^3)} = 0$), the product $s a^3$ is constant.

Note that these relations can be deduced from standard equilibrium thermodynamics. Indeed, the Gibbs free energy (or free enthalpy) $G = U + PV - TS$ is a function of $(T,P,N)$ thanks to the property of Legendre transforms: starting from $U(S,V,N)$, the two Legendre transforms via $+PV$ and $-TS$ change the natural variables to $(T,P,N)$, and the fundamental thermodynamic identity becomes
\begin{equation}
\label{eq:thermo_identity}
 \dd{U} = T \dd{S} - P \dd{V} + \mu \dd{N} \qquad \implies \qquad \dd{G} = - S \dd{T} + V \dd{P} + \mu \dd{N} \, .
\end{equation}
However, $G$ is an additive function, and its only additive variable is $N$, therefore $G(T,P,N) = \mu(T,P) N$. We thus deduce that the entropy is $S = (U + PV - \mu N)/T$, which gives~\eqref{eq:def_entropy} after dividing by the volume. Moreover, using the identity~\eqref{eq:thermo_identity} for a comoving volume $V \propto a^3$, we get $T \dd{(s a^3)} = \dd{(\rho a^3)} + P \dd{a^3} - \mu \dd{(n a^3)}$, but the first two terms on the right-hand side cancel according to the conservation equation~\eqref{eq:eq_conservation}.

In the following section, we discuss some consequences of the expansion of the Universe. The conservation of entropy plays an important role as it gives directly some information on the evolution of the temperature.

\subsection{Thermal history of the Universe}

In short, the history of the Universe in the hot Big Bang model is the history of its cooling as it expands. This cooling has several consequences, and we discuss here two crucial ones to understand neutrino evolution in the (early) Universe. 

\subsubsection{Nonrelativistic transition and entropy transfer}

It is useful to write the entropy density of the plasma as a function of the photon temperature $T_\gamma$:
\begin{equation}
s_\text{pl} \equiv \frac{2 \pi^2}{45} g_s(T_\gamma) T_\gamma^3 \, .
\end{equation}
Let us assume that we are in a non-degenerate case ($\mu =0$ for all species), such that the entropy is conserved according to~\eqref{eq:entropy_evo}: $s_\text{pl} a^3 = \text{const}$. We thus obtain the very important result:
\begin{equation}
\label{eq:gs}
\boxed{T_\gamma \propto g_s^{-1/3} a^{-1}} \, .
\end{equation}
Whenever $g_{s}$ is constant, the temperature decreases as $a^{-1}$. However, it is possible that particles in the plasma become non-relativistic, in which case their entropy is exponentially suppressed (in other words, they do not contribute anymore to $g_s$) --- cf.~the integrals given in~\eqref{eq:thermo_quantities_relat} and~\eqref{eq:thermo_quantities_NR}. The conservation of $s_\text{pl} a^3$ then shows that entropy is \emph{transferred} to the other species. The mechanism behind this transfer is the displacement of the equilibrium of the reaction $X + \bar{X} \leftrightarrow \gamma + \gamma $ towards the right when the temperature gets below $m_X$, since the average energy of photons is then too small to create $X-\bar{X}$ pairs. This temperature threshold is precisely the one of the non-relativistic transition, which is why we will equivalently talk about non-relativistic transition and entropy transfer, or particle/antiparticle annihilation.

We show a concrete example in section~\ref{sec:Intro_Neutrinos} when we discuss electron/positron annihilations and the associated reheating of photons.

\subsubsection{Decoupling} 

The second consequence of the cooling of the Universe is the \emph{decoupling} of species when their interaction rate becomes too small compared to the expansion rate. Below the decoupling temperature, they interact too little to remain in thermal contact with other species. As a rule of thumb, we say that decoupling occurs when\footnote{Another way to justify this is to say that the heat bath temperature varies as $T_\gamma \propto a^{-1}$ (we neglect a variation of $g_s$ for this argument), such that $\dot{T}_\gamma/T_\gamma = - H$. Therefore, the relation~\eqref{eq:condition_decoupling} corresponds to the moment when the interactions are not fast enough to adjust to the changing temperature.}
\begin{equation}
	\label{eq:condition_decoupling}
	\frac{\Gamma}{H} \sim 1 \, ,
\end{equation}
with $\Gamma = n \langle \sigma v \rangle$, where $n$ is the number density of target particles, $\sigma$ is the cross-section and $v \sim 1$ the relative velocity (in the ultrarelativistic case). The angle brackets denote thermal averaging. This expression shows why decoupling occurs when the temperature decreases: the interactions may become too weak, or the target density can be suppressed (this is the case after recombination and thus for photon decoupling at $T \sim 0.3 \, \mathrm{eV}$).

\paragraph{Temperature evolution of a decoupled species} Once they are decoupled from the plasma, particles are free-streaming and their distribution functions are \emph{frozen}: if we write with a subscript $_D$ the quantities at decoupling, we have
\begin{equation}
\label{eq:temp_decoup}
f(p,t) = f(p_D, t_D) = f\left(\frac{a(t)}{a_D}p,t_D\right) \, ,
\end{equation}
where we have used the scaling relation~\eqref{eq:scaling_p}. In general, we cannot define an effective temperature and an effective chemical potential,\footnote{Such quantities have to be \emph{effective}, since equilibrium is not maintained anymore by interactions. However this is not in contradiction with the fact that, in the ultra- and non-relativistic limits, equilibrium distributions are maintained.} except in the two following limits~\cite{GiuntiKim}.
\begin{itemize}
	\item If the particles are ultrarelativistic (and non-degenerate) at decoupling (which is the case for massless particles), we have from~\eqref{eq:temp_decoup} and using $E=p$,
	\begin{equation}
	\label{eq:decouple_massless}
	f(p,t) = \frac{1}{\displaystyle e^{\frac{a}{a_D}p/T_D} \pm 1} = \frac{1}{\displaystyle e^{p/T} \pm 1} \quad \text{with} \quad \boxed{T = T_D \frac{a_D}{a} \propto a^{-1}} \, .
	\end{equation}
	The particles keep a relativistic equilibrium distribution with an effective temperature scaling as $a^{-1}$. Note that even if, at some point, the particles become nonrelativistic ($m \sim T$), the spectrum keeps this shape.
	\item If the particles are nonrelativistic at decoupling, we can simplify $E_D = \sqrt{p_D^2 + m^2} \simeq m + (p^2/2 m)$, hence,
	\begin{align}
	f(p,t) = \frac{1}{\displaystyle e^{(E_D - \mu_D)/T_D} \pm 1} &\simeq e^{(\mu_D - m)/T_D} e^{-p_D^2/(2mT_D)} \nonumber \\
	&\equiv e^{(\mu -m)/T} e^{-p^2/(2mT)} \, ,
	\end{align}
	where, using once again the scaling relation $p = p_D a_D/a$, we define the effective temperature
	\begin{equation}
	\boxed{T = T_D \left(\frac{a_D}{a}\right)^2 \propto a^{-2}} \, ,
	\end{equation}
	and the effective chemical potential
		\begin{equation}
	\mu =m + (\mu_D -m)\frac{T}{T_D} = m + (\mu_D -m)\left(\frac{a_D}{a}\right)^2  \, .
	\end{equation}
	Even if at decoupling $\mu_D=0$, it cannot remain equal to zero later.
\end{itemize}

\subsubsection{A (very) brief thermal history of the Universe}

If the number of relativistic degrees of freedom $g_s$ is constant, \eqref{eq:gs} shows that the temperature of the plasma (what we usually call “the temperature of the Universe”) decreases as $a^{-1}$. As the Universe cools down, equilibrium between species can no longer be maintained, and massive particles become non-relativistic. There are also very high-energy phenomena that we did not discuss such as the electroweak phase transition. A summary of the major events that are predicted by the standard model of cosmology is presented in Table~\ref{Table:chrono_universe}. The earliest experimental probe of this model is Big Bang Nucleosynthesis (BBN), which we discuss in section~\ref{subsec:intro_BBN}. The observation of the Cosmic Microwave Background (CMB), that is photons that decoupled from electrons 380 000 years after the Big Bang, has been another decisive argument towards the validation of this standard model.

\renewcommand{\arraystretch}{1.2}

\begin{table}[!htb]
	\centering
	\begin{tabular}{|M{4cm} |M{3cm} | M{6cm} |}
  	\hline 
  $\sim$ Age of the Universe & Temperature $(\mathrm{K})$ &  Major event(s)  \\
  \hline \hline
 $< 10^{-43} \, \mathrm{s} $& $> 10^{32}$   & ???   \\ 
 $ 10^{-43} - 10^{-35} \, \mathrm{s} $& $ 10^{32} - 10^{28} $   & Period of inflation   \\ 
  $ 10^{-35} - 10^{-12} \, \mathrm{s} $& $ 10^{28} - 10^{16} $   & Generation of matter/antimatter asymmetry   \\ 
  $ 10^{-12} \, \mathrm{s}$ & $10^{16}$ & Electroweak phase transition \\
   $ 10^{-4} \, \mathrm{s} $& $ 10^{12} $   & Quark-hadron transition,  $\mu^+ \mu^-$ annihilation   \\ 
  $ \bm{1 - 10^2 \, \mathrm{s}} $ & $\bm{\sim 10^{10} \, (\sim 1 \, \mathrm{MeV})}$ & \textbf{Neutrino decoupling, $\bm{e^+ e^-}$ annihilation, BBN} \\
  $10^5 \, \mathrm{yr}$ & $4000$ & Recombination, formation of the Cosmic Microwave Background \\
  $2 \times 10^5 - 10^9 \, \mathrm{yr}$ &  & Galaxy formation \\
  $13.7 \times 10^9 \, \mathrm{yr} $ & $2.73$ & Today \\ \hline
\end{tabular}
	\caption[(Very) short history of the Universe]{Major events occurring during the expansion of the Universe in the hot Big Bang model (adapted from~\cite{MohapatraPal}). Note that the values given for the epoch of inflation are particularly model-dependent.
	\label{Table:chrono_universe}}
\end{table}

As emphasized in Table~\ref{Table:chrono_universe}, the period during which the temperature of the Universe was about $1 \, \mathrm{MeV}$ (the so-called \emph{“MeV era”}) is eventful. Indeed, we show in the next section that neutrinos decouple only shortly before electrons/positrons annihilate, suggesting a partial overlap between these two phenomena. Moreover, this era also marks the beginning of BBN, meaning that neutrino decoupling influences the primordial abundances.

\section{Neutrinos in the early Universe}
\label{sec:Intro_Neutrinos}

In this section, we give an introduction to the main events summarized in Table~\ref{Table:chrono_universe} that take place during the MeV age: neutrino decoupling, $e^\pm$ annihilations, and primordial nucleosynthesis. The results presented here are standard, and will be the basis of the work presented in chapters~\ref{chap:Decoupling} and~\ref{chap:BBN}.

\noindent During this period, the constituents of the Universe are:
\begin{itemize}
	\item a QED plasma of electrons, positrons and photons, tightly coupled by QED interactions (note that heavier charged leptons, $\mu^\mp$ and $\tau^\mp$, have annihilated),
	\item a bath of neutrinos and antineutrinos, coupled to the QED plasma via weak interactions with electrons and positrons,
	\item baryons (initially neutrons and protons which combine later on to form light elements during BBN) in negligible abundance compared to the leptons, as shown by the latest measurement of the baryon-to-photon ratio $\eta = n_b/n_\gamma \simeq 6.1 \times 10^{-10}$~\cite{Fields:2019pfx}.
\end{itemize}

We emphasize that the calculations presented in this section do not take into account all the physics known to play a role at this epoch, as neutrino oscillations (see section~\ref{sec:Intro_massive_nu}) are discarded. Full calculations including these phenomena and providing an in-depth analysis of their effects are the subject of this thesis.

\subsection{Instantaneous neutrino decoupling}

To roughly estimate the temperature of neutrino decoupling via~\eqref{eq:condition_decoupling}, we need to compare $\Gamma$ and $H$. In the early universe, neutrinos are kept in equilibrium via weak interaction processes like $\nu + e^{-} \to e^{-} + \nu$. The cross-section of such processes scales as $\sigma \sim G_F^2 T^2$, the particle density (for ultrarelativistic electrons) as $n_e \sim T^3$ (see~\eqref{eq:thermo_quantities_relat} above) and the relative velocity is $v \sim 1$. Hence the interaction rate \[\Gamma = n_e \langle \sigma v \rangle \sim G_F^2 T^5 \, .\] 
From Eq.~\eqref{eq:Friedmann}, the Hubble rate is, in the radiation era,
\begin{equation*}
H = \sqrt{\frac{8 \pi}{3 M_\mathrm{Pl}^2} \rho_\mathrm{rad}} = \sqrt{\frac{8 \pi}{3 M_\mathrm{Pl}^2} g_* \frac{\pi^2}{30} T^4} \sim \sqrt{g_*} \frac{T_\gamma^2}{M_\mathrm{Pl}} \, ,
\end{equation*}
where $g_{*}$ is the number of relativistic degrees of freedom, defined such that
\begin{equation}
\rho_\text{rad} = \frac{\pi^2}{30} g_*(T_\gamma) T_\gamma^4 \, .
\end{equation}
At this time, the relativistic species in the Universe are photons (with two helicity states), three species of neutrinos and antineutrinos (each with one helicity state), plus electrons and positrons (with two spin states), thus
\begin{equation}
\label{eq:gstar}
g_* = 2 + 3\times 2 \times \frac{7}{8} + 2 \times 2 \times \frac{7}{8} = \frac{43}{4} = 10.75 \, .
\end{equation}
The ratio of the interaction rate to the expansion rate is thus
\begin{equation}
\frac{\Gamma}{H} \sim \frac{G_F^2 T^5}{\sqrt{g_*} \, T^2/M_\mathrm{Pl}} \simeq \left(\frac{T}{1 \ \mathrm{MeV}}\right)^3 \, .
\end{equation}
Therefore, at $T_{\nu D} \sim 1 \, \mathrm{MeV}$, there are not enough collisions compared to the expansion of the Universe and neutrinos subsequently decouple from the electromagnetic plasma. Once it has decoupled, the fluid of neutrinos and antineutrinos is called the \emph{Cosmic Neutrino Background} (C$\nu$B): if it could be directly detected, it would give a snapshot of the Universe as it was a few dozens of seconds after the Big Bang, a tremendous jump in the past compared to the CMB, which was formed 380 000 years later...

In the \emph{instantaneous decoupling approximation}, one considers that at $T_{\nu D}$, all neutrinos suddenly decouple. From this moment onward, their distribution function remains a Fermi-Dirac distribution~\eqref{eq:decouple_massless}
\[f_\nu(p) = \frac{1}{e^{p/T_\nu} +1} \qquad ; \qquad T_\nu \propto a^{-1} \, .\]
In the electromagnetic plasma, the equilibrium of the reaction $e^+ + e^- \leftrightarrow \gamma + \gamma$ gets very displaced on the right when the temperature drops below $m_e \simeq 0.511 \ \mathrm{MeV}$.  Electrons and positrons annihilate and their entropy is transferred to the gas of photons, whose temperature will thus decrease slower than $a^{-1}$.

Let us estimate the final ratio of the photon-to-neutrino temperature once the $e^+ e^-$ annihilation is over. Using~\eqref{eq:gs}, the conservation of the entropy of the electromagnetic plasma reads\footnote{We put a superscript $^{\mathrm{(pl)}}$ to highlight the fact that neutrinos do not interact with the plasma anymore, and are thus not counted in $g_s^{(\mathrm{pl})}$.}
\[g_{s}^{(\mathrm{pl})}  \times T_\gamma^3 a^3 = \mathrm{cst} \implies g_{s}^{(\mathrm{pl})} \times \left(\frac{T_\gamma}{T_\nu}\right)^3 = \mathrm{cst} \, ,\]
since $T_\nu \propto a^{-1}$. Long before the annihilation of electrons and positrons (but after neutrino decoupling), $T_\gamma = T_\nu$ and the relativistic species are photons and $e^\pm$ pairs, so $g_{s}^{(\mathrm{pl})} = 2 +  2 \times 2 \times \frac78 = \frac{11}{2}$. After the annihilation, there are only photons left and $g_{s}^{(\mathrm{pl})} = 2$. Therefore, long after decoupling, the ratio of temperatures is
\begin{equation}
\label{eq:TgTnu}
\boxed{\frac{T_\gamma}{T_\nu} = \left(\frac{11}{4}\right)^{1/3} \simeq 1.40} \ .
\end{equation}

This is of course an approximate result, for various reasons. First, since the temperatures of neutrino decoupling and $e^+ e^-$ annihilation are close, neutrinos are not fully decoupled when the annihilation takes place. This leads to a small “reheating” of neutrinos compared to the instantaneous decoupling (ID) limit. Historically, this has been parameterized in the following way: should ID be true, then after $e^+ e^-$ annihilation the radiation energy density would read
\begin{equation}
\rho_\text{rad} = \left(1 + 3 \times \frac78 \left(\frac{4}{11}\right)^{4/3}\right) \rho_\gamma \, ,
\end{equation}
where we used the expressions~\eqref{eq:thermo_quantities_relat} and the ratio of temperatures~\eqref{eq:TgTnu}. The departure from this “ideal” picture is defined such that $\rho_\text{rad}$ reads
\begin{equation}
\label{eq:intro_def_Neff}
\rho_\text{rad} = \left(1 + \Neff \times \frac78 \left(\frac{4}{11}\right)^{4/3}\right) \rho_\gamma \, ,
\end{equation}
where the parameter $\Neff$ is called\footnote{The key is in “effective”: there are still exactly 3 active neutrinos, but their energetic contribution is equivalent to the one of not exactly 3 instantaneously decoupled neutrinos. In addition, note that any beyond-the-Standard-Model relativistic species that would contribute to $\rho_\text{rad}$ is taken into account in $\Neff$.} the \emph{effective number of neutrino species}. It represents the (non-integer) number of instantaneously decoupled neutrinos that would have the same energy density as the actual 3 active neutrinos. It is a convenient cosmological observable as it encapsulates all the information on the energy density during the radiation-dominated era. Due to the overlap between $e^+ e^-$ annihilation and neutrino decoupling, we expect $\Neff > 3$. It should be noted that $\Neff$ is one of the main cosmological quantities we will be interested in throughout this manuscript.

Another limitation to the instantaneous decoupling limit is the fact that the more energetic neutrinos will remain in thermal contact with the plasma longer than the low-energy ones, which should source \emph{spectral distortions} of the neutrino distribution functions. Finally, the different flavours of neutrinos do not have the same coupling with electrons/positrons. Indeed, electronic neutrinos can interact with $e^\pm$ via neutral \emph{and} charged current processes (i.e., exchanges of $Z$ and $W$ bosons), while the other flavours of neutrinos can only interact with the heat bath via neutral current processes. Assuming that neutrinos had Maxwell-Boltzmann distributions, Dolgov found~\cite{Dolgov_2002PhysRep}:
\begin{equation*}
T_{\nu_e D} \simeq 1.87 \ \mathrm{MeV} \quad ; \quad T_{\nu_{\mu,\tau} D} \simeq 3.12 \ \mathrm{MeV} \, .
\end{equation*}

These different arguments show that, in reality, neutrino decoupling is a much more complicated process than what is described in the instantaneous decoupling approximation. The true decoupling process is referred to as \emph{incomplete neutrino decoupling}, and describing it requires to solve the Boltzmann equations which drive the evolution of the neutrino distributions functions. The following subsection describes standard numerical results on this topic.

\subsection{Incomplete neutrino decoupling}
\label{subsec:intro_decoupling}

Previous works have tackled the problem of incomplete neutrino decoupling, with increasing precision and taking into account the distortions of the spectra:\footnote{A good summary of the existing literature as of 2015 can be found in~\cite{Grohs2015}.} \cite{Dolgov1992,Dodelson_Turner_PhRvD1992} approximated the distribution functions with Maxwell-Boltzmann statistics, an approximation overcome in~\cite{Hannestad_PhRvD1995,Dolgov_NuPhB1997,Dolgov_NuPhB1999,Esposito_NuPhB2000} which used various numerical methods. Corrections to the plasma thermodynamics (see section~\ref{subsec:QED}) were included in~\cite{Mangano2002,Mangano2005,Grohs2015,Relic2016_revisited,Gariazzo_2019}. We can also quote the recent approximate but accurate methods developed in~\cite{Escudero_2018,Escudero_2020}.

We present in this section the results of a typical calculation of incomplete neutrino decoupling which \emph{does not} take into account neutrino masses and mixings, features that we describe in the next section. The full calculation is done in chapter~\ref{chap:Decoupling} and requires a more complex formalism. Without flavour oscillations, calculating neutrino decoupling~\cite{Mangano2002,Grohs2015,Froustey2019} requires to solve the covariant Boltzmann kinetic equation which reads for neutrinos~\cite{ModernCosmology}
\begin{equation}
\label{eq:boltzmann_nu}
\left[\frac{\partial}{\partial t} - H p \frac{\partial}{\partial p}\right]f_{\nu_\alpha}(p,t) = C_{\nu_\alpha}[f_\nu,f_{e^\pm}] \, ,
\end{equation}
where $C_{\nu_\alpha}[f_j]$ is the collision term. This collision integral is dominated by two-body reactions $1+2 \to 3+4$ and is given by (the sum is over reactions)
\begin{multline}
\label{eq:collision_integral_intro}
C_{\nu_1} = \frac{1}{2E_1} \sum{\int{\frac{\dd^3 p_2}{2 E_2 (2\pi)^3}\frac{\dd^3 p_3}{2 E_3 (2\pi)^3}\frac{\dd^3 p_4}{2 E_4 (2\pi)^3} \times (2\pi)^4 \delta^{(4)}(p_1+p_2-p_3-p_4)}} \\
\times  S \langle \abs{\mathcal{M}}^2\rangle \times F[f^{(1)},f^{(2)},f^{(3)},f^{(4)}] \, ,
\end{multline}
where $S$ is the symmetrization factor, $\langle \abs{\mathcal{M}}^2\rangle$ the summed-squared matrix element (given for instance in~\cite{Grohs2015}), and \[F \equiv f^{(3)} f^{(4)}(1-f^{(1)})(1-f^{(2)}) - f^{(1)} f^{(2)} (1-f^{(3)})(1-f^{(4)}) \, ,\]
the notation $f_a^{(j)}$ meaning $f_a(p_j)$. The standard calculation assumes no neutrino asymmetry $f_{\nu_\alpha} = f_{\bar{\nu}_\alpha}$. Finally, the distribution functions are the same for $\nu_\mu$ and $\nu_\tau$, as at the energy scales of interest the muon and tau neutrinos have the same interactions (while the distribution of $\nu_e$ is different because of charged current interactions with the background medium). We are thus restricted to two unknown neutrino distributions, $f_{\nu_e}$ and $f_{\nu_\mu}$. In addition to the Boltzmann equation for neutrinos, the last necessary equation\footnote{All remaining species (electrons, positrons, photons) are kept at equilibrium by fast QED interactions, so their properties are summarized by a single parameter, the plasma temperature $T_\gamma$ --- whose evolution is thus given by energy conservation.} is the total energy conservation~\eqref{eq:eq_conservation}.

\paragraph{Comoving variables} We define the comoving temperature $\Tcm \propto a^{-1}$ \cite{Grohs2015}, which corresponds to the physical temperature of all species when they are strongly coupled, i.e. $T_\nu = T_\gamma = \Tcm$ when $\Tcm \gg 1 \, \mathrm{MeV}$, and is also the temperature of neutrinos at all times in the instantaneous decoupling approximation $T_\nu^\mathrm{ID} = \Tcm$. From this proxy for the scale factor, we define the comoving variables \cite{Esposito_NuPhB2000,Mangano2005}
\begin{equation}
\label{eq:comoving_variables}
x \equiv m_e/\Tcm\, , \qquad y \equiv p/\Tcm\, ,\quad \text{and} \quad z \equiv T_\gamma/\Tcm\, ,
\end{equation}
which are respectively the reduced scale factor, the comoving momentum, and the dimensionless photon temperature, such that $f(p,t)$ is now expressed $f(x,y)$. We also introduce the dimensionless thermodynamic quantities $\bar{\rho} \equiv (x/m_e)^4 \rho $ and $\bar{P} \equiv  (x/m_e)^4 P$. 

\paragraph{Neutrino spectral distortions due to $e^\pm$ annihilation} It is then possible to follow the evolution of neutrino distribution functions across the decoupling era. Since this is an out-of-equilibrium process, we cannot properly talk about neutrino “temperatures”, although such quantities are quite convenient to give a global picture of the results. Note that for instance in Refs.~\cite{Dodelson_Turner_PhRvD1992,Hannestad_PhRvD1995,Dolgov_NuPhB1997}, an “effective temperature” is defined as
\begin{equation*}
T_\mathrm{eff}(p) = \frac{p}{\ln{[1/f_\nu(p) -1]}}
\end{equation*}
and comparing $T_\mathrm{eff}$ with $T_\gamma$ along the evolution would seem to be a good indicator of decoupling. However, such a temperature is just the temperature of the only Fermi-Dirac spectrum which takes the value $f_\nu(p)$ at $p$, so it doesn't give the "global" information one is looking for when defining a temperature. Moreover, we are interested in the parameter $\Neff$ which depends on the energy density of neutrinos --- but $T_\mathrm{eff}$ is not a convenient parameter to compute the energy density.

Therefore, we rather introduce the following parameterization:
\begin{equation}
\label{eq:param_fnu}
f_{\nu_\alpha}(x,y) \equiv \frac{1}{e^{y/z_{\nu_\alpha}} + 1} \left[1 + \delta g_{\nu_\alpha}(x,y)\right] \, ,
\end{equation}
where the reduced effective temperature $z_{\nu_\alpha} \equiv T_{\nu_\alpha}/\Tcm$ is the reduced temperature of the Fermi-Dirac spectrum with zero chemical potential which has the same energy density as the real distribution:
\begin{equation}
\label{eq:def_znu_intro}
\bar{\rho}_{\nu_\alpha} \equiv \frac78 \frac{\pi^2}{30} z_{\nu_\alpha}^4 \,.
\end{equation}
Note that the effective distortions are constrained so that~\eqref{eq:def_znu_intro} holds:
\begin{equation}
\int_{0}^{\infty}{\dd{y} y^3\frac{\delta g_{\nu_\alpha}}{e^{y/z_{\nu_\alpha}}+1}} = 0 \, .
\end{equation}

\begin{figure}[!h]
	\centering
	\includegraphics{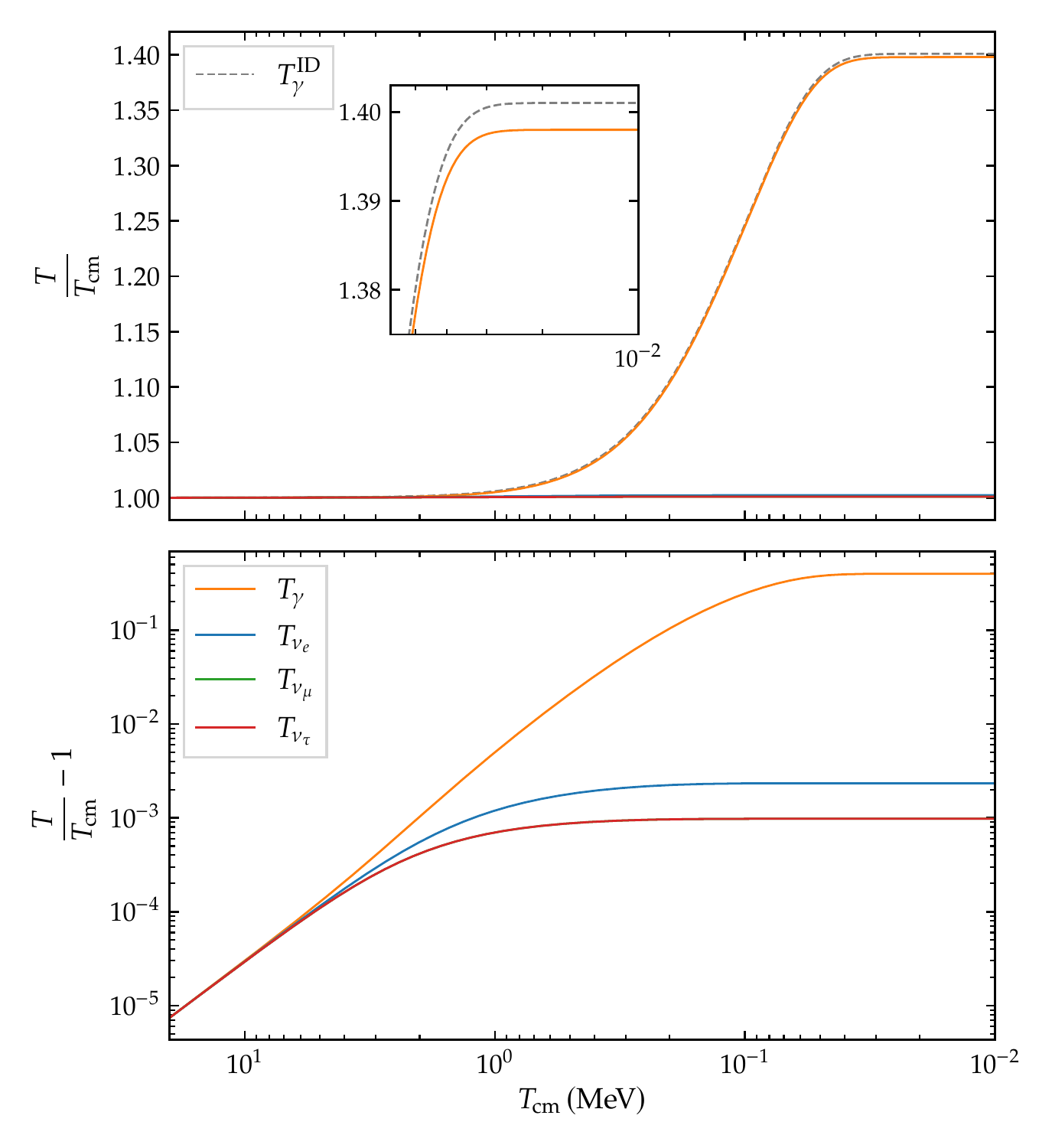}
	\caption[Temperature evolution during neutrino decoupling]{\label{fig:TnuNoMix} Evolution of the (effective) temperatures of the relativistic species (photons and neutrinos) across the decoupling era. \emph{Top panel:} Comoving photon temperature in the instantaneous decoupling approximation (dashed grey line) and taking into account incomplete neutrino decoupling (solid orange line). The asymptotic value of $T_\gamma^{\mathrm{ID}}/\Tcm$ is, as derived in~\eqref{eq:TgTnu}, $(11/4)^{1/3}\simeq 1.40102$. The neutrino effective temperatures cannot be distinguished by eye from $\Tcm$. \emph{Bottom panel:} Relative difference between the temperatures and $\Tcm$, showing the higher reheating of $\nu_e$ compared to $\nu_{\mu,\tau}$.}
\end{figure}

The evolution of the effective temperatures is shown on Figure~\ref{fig:TnuNoMix}. As expected, photons are less reheated by $e^+ e^-$ annihilation, hence the smaller final value of $T_\gamma$. Even though we draw two lines, $T_{\nu_\mu}$ and $T_{\nu_\tau}$ are exactly equal. The higher value of $T_{\nu_e}$ is due to the charged current processes: these additional interactions maintain thermal contact longer and make electronic (anti)neutrinos the main channel of entropy transfer between the QED plasma and the neutrino bath. The effective number of neutrino species is conveniently computed from the effective temperatures:
\begin{equation}
\label{eq:defNeff}
\rho_\nu + \rho_{\bnu} = \Neff \times \frac78 \left(\frac{4}{11}\right)^{4/3} \rho_\gamma  \ \iff \ \Neff \equiv \left[ \frac{(11/4)^{1/3}}{z} \right]^4 \times \left(z_{\nu_e}^4 + z_{\nu_\mu}^4 + z_{\nu_\tau}^4 \right) \, ,
\end{equation}
where we assumed no asymmetry between neutrinos and antineutrinos ($z_{\nu_\alpha} = z_{\bnu_\alpha}$). In the general case, one simply has to replace $z_{\nu_\alpha}^4 \to (z_{\nu_\alpha}^4+z_{\bnu_\alpha}^4)/2$.

\begin{figure}[!ht]
	\centering
	\includegraphics{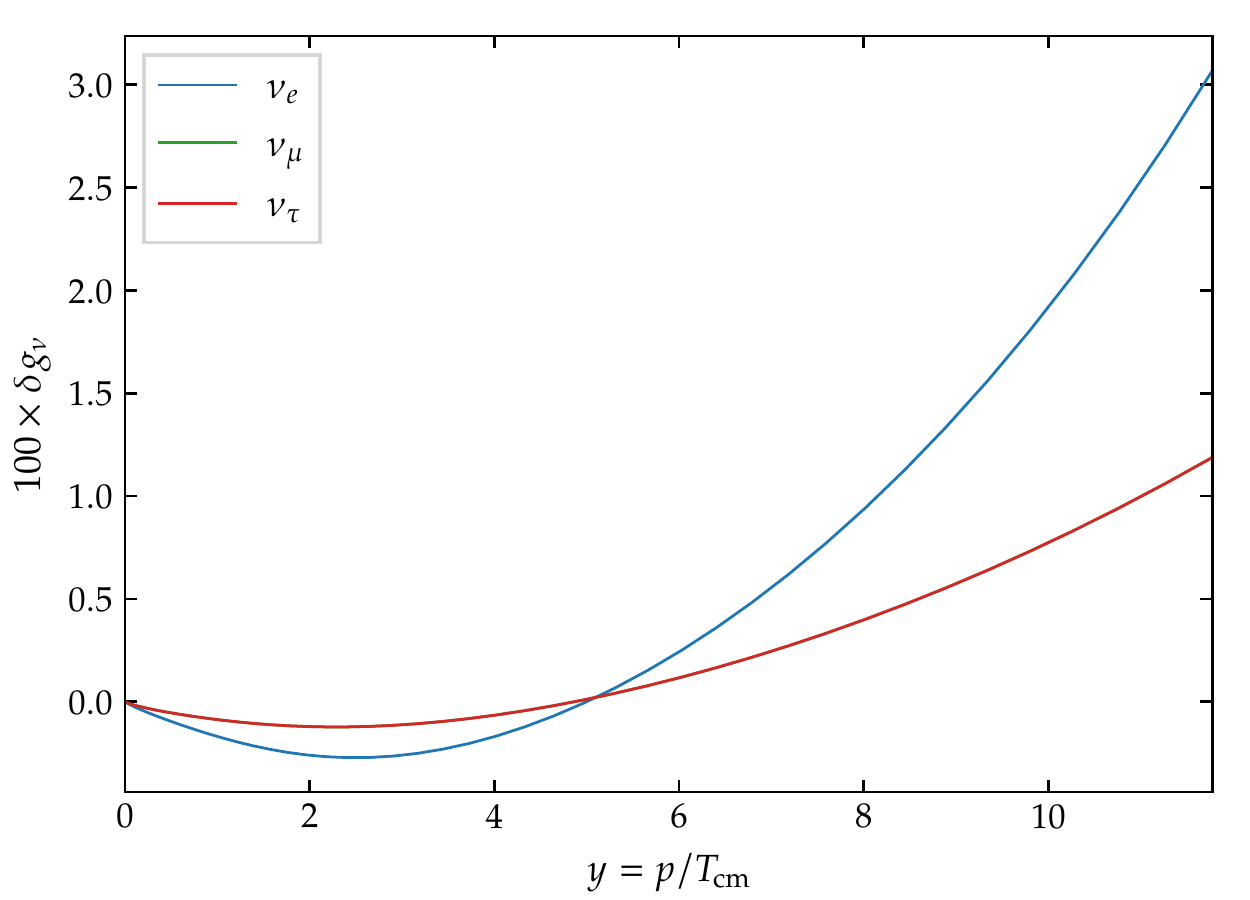}
	\caption[Final non-thermal distortions of the neutrino spectra]{\label{fig:gnuNoMix} Final non-thermal distortions of the neutrino spectra. Since they have exactly the same interactions, $\delta g_{\nu_\mu} = \delta g_{\nu_\tau}$. The larger amplitude of distortions for electronic neutrinos is due to the charged current processes.}
\end{figure}

We plot on Figure~\ref{fig:gnuNoMix} the final non-thermal distortions $\delta g_{\nu_\alpha}$. The same comments can be made as for the final effective temperatures. Note that the typical size of the corrections to the ID limit is $\sim 1 \, \%$.

For completeness, we give the value of $\Neff$ deduced from the results shown on Figure~\ref{fig:TnuNoMix}: $\Neff \simeq 3.043$. It is close to the values previously obtained in the literature, notably the previous reference $\Neff = 3.046$~\cite{Relic2016_revisited}. However, these values strongly depend on the inclusion of QED corrections, of a proper treatment of flavour mixing (which we haven't introduced yet), etc. We postpone the dedicated computation of $\Neff$ to chapter~\ref{chap:Decoupling}.

\subsubsection{Experimental constraints}

We gather in Table~\ref{Table:constraints_Neff} some of the latest bounds on $\Neff$ obtained from different CMB experiments~\cite{WMAP2012,Planck18,ACT:2020,SPT-3G:2021}.\footnote{A detailed discussion of the effects of neutrinos on CMB can be found in, e.g.,~\cite{Neutrino_Cosmology}.} These results notably confirm the existence of only 3 neutrino species being thermally populated close to decoupling, which constrains the properties of possible sterile states. For instance, light sterile neutrinos with a $\sim \mathrm{eV}$ mass, favoured to fit some experimental anomalies, would unavoidably bring $\Neff \simeq 4$~\cite{Gariazzo_2019} thanks to oscillation mechanisms introduced in section~\ref{sec:Intro_massive_nu}.

\begin{table}[!htb]
	\centering
	\begin{tabular}{|M{4cm} |M{2cm}  l |}
  	\hline 
  Experiment & $\Neff$ &   \\
  \hline \hline
 WMAP~\cite{WMAP2012} & $3.84 \pm 0.40$ & [WMAP + ACT + SPT + BAO + $H_0$]  \\ 
 Planck~\cite{Planck18}   & $2.99 \pm 0.17$ & [Planck + BAO]    \\ 
 ACT~\cite{ACT:2020} & $2.74 \pm 0.17$ & [ACT + Planck] \\
  SPT-3G~\cite{SPT-3G:2021}   & $2.95 \pm 0.17$ & [SPT + Planck]   \\  \hline
\end{tabular}
	\caption[Recent constraints on $\Neff$]{Recent constraints on the value of $\Neff$ obtained from different experiments and combined datasets. The uncertainties are 68 \% confidence intervals.
	\label{Table:constraints_Neff}}
\end{table}

One of the main results of this PhD is the reevaluation of the cosmological observable $\Neff$, including all relevant effects to reach a $10^{-4}$ precision, also including the effect of neutrino masses (to be introduced in the following section~\ref{sec:Intro_massive_nu}). To this aim, we first derive the neutrino evolution equations in chapter~\ref{chap:QKE}, extending the work of~\cite{Volpe_2013} for astrophysical environments, and implement two-body collisions in an isotropic and homogeneous environment, including neutrino self-interactions with their full matrix structure. Then, we numerically solve these equations, but also present an approximate solution where an adiabatic evolution is considered, exploiting the different timescales involved in the problem. This procedure allows to maintain the required precision while decreasing substantially the computation time, gaining some physical insight on the role of the phenomenon of flavour oscillations in neutrino decoupling. The numerical results we present correspond to the case of zero chemical potential. Finally we investigate the impact of neutrino masses and mixings on BBN predictions in chapter~\ref{chap:BBN}, going beyond works available in the literature~\cite{Mangano2005,Gava:2010kz,Gava_corr}.

\subsection{Big Bang Nucleosynthesis}
\label{subsec:intro_BBN}

Big Bang Nucleosynthesis (BBN) is one of the historical pillars of the Big Bang model, together with the expansion of the Universe and the Cosmic Microwave Background. It corresponds to the period when the temperature was small enough to enable the formation of light elements by combining neutrons and protons. The idea first appeared in the seminal paper by Alpher, Bethe and Gamow~\cite{AlpherBetheGamow}, and later studies showed that BBN was responsible for the primordial production of deuterium ($\mathrm{D}={}^2\mathrm{H}$), helium-3, helium-4 and lithium-7. 

\begin{figure}[!ht]
	\centering
	\includegraphics{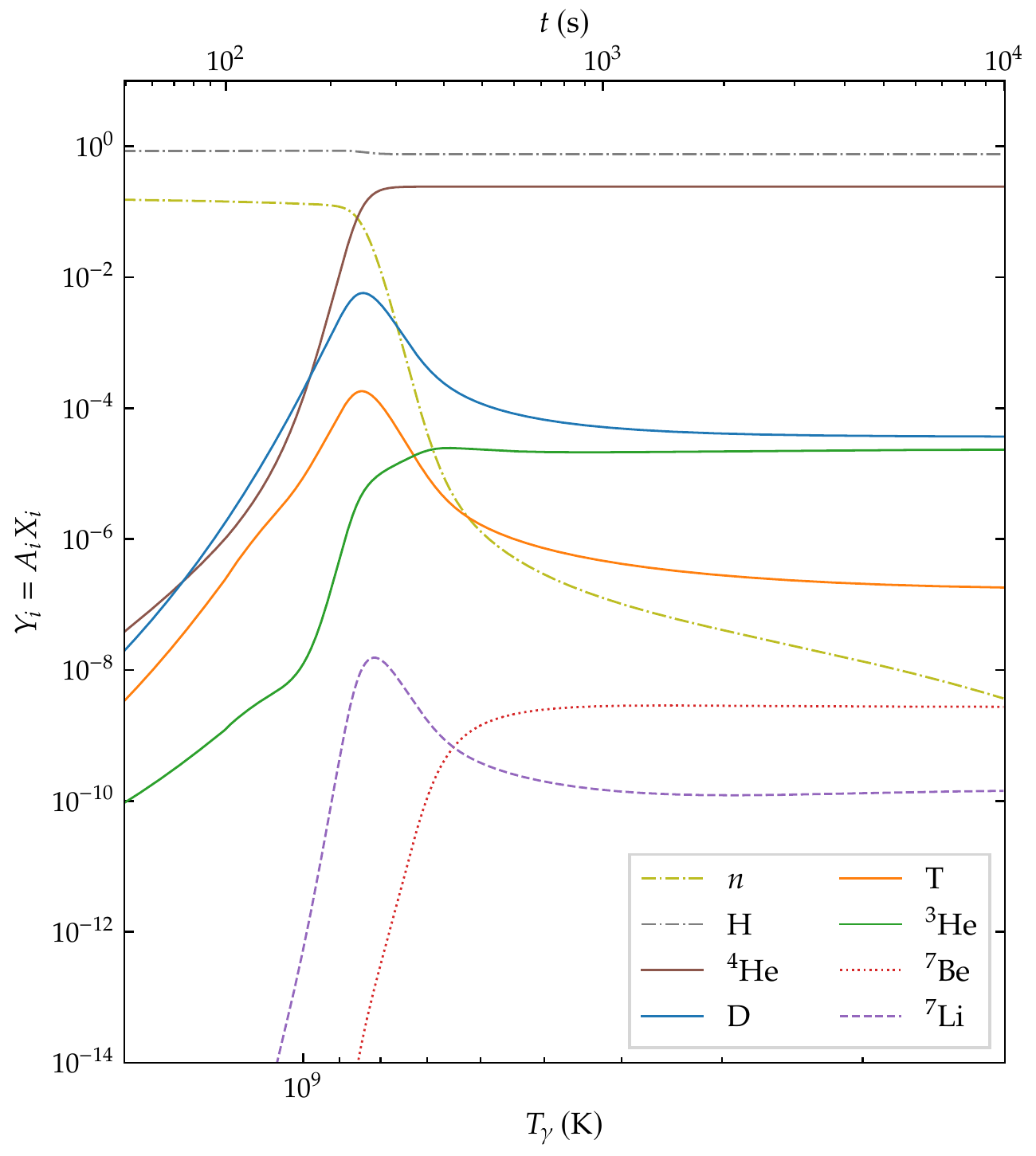}
	\caption[Evolution of light element abundances during BBN]{\label{fig:IntroBBN} Evolution of light element abundances, computed with the code \texttt{PRIMAT}. At the end of BBN, the baryonic content of the Universe is mainly made of hydrogen and helium-4. Deuterium is also an important cosmological probe (see text).}
\end{figure}

We do not present in-depth the different steps of BBN, referring for more details to books like~\cite{PeterUzan} or reviews like~\cite{Pitrou_2018PhysRept}, and to chapter~\ref{chap:BBN}. There are nevertheless three big steps to keep in mind:
\begin{enumerate}
	\item At high temperatures, neutrons and protons are kept in chemical equilibrium by weak interactions:
\begin{equation}
\label{eq:weakrates_BBN_intro}
\begin{aligned}
n + \nu_e &\leftrightarrow p + e^-  \\
n &\leftrightarrow p + e^- + \bar{\nu}_e  \\
n + e^+ &\leftrightarrow p + \bar{\nu}_e 
\end{aligned}
\end{equation}
Similarly to neutrino decoupling, these reactions freeze-out at a temperature $\TFO \sim 0.7 \, \mathrm{MeV}$.
	\item Below $\TFO$, the only reaction left\footnote{This is actually an oversimplification, as evidenced for instance in~\cite{Grohs:2016vef}. We discuss this point later in chapter~\ref{chap:BBN}.} is neutron decay $n \to p + e^- + \bar{\nu}_e$, until the beginning of nucleosynthesis at $\TNuc$, the first nuclear reaction being $n + p \to \mathrm{D} + \gamma$.
	\item A whole set of out-of-equilibrium nuclear reactions take place, producing heavier nuclei until BBN eventually stops when the temperature is too low to maintain high enough nuclear rates. This leads to the primordial abundances represented on Figure~\ref{fig:IntroBBN} (the results were obtained with a numerical code, \texttt{PRIMAT}, whose principle is presented in chapter~\ref{chap:BBN}).
\end{enumerate}
The standard notations are the following: $n_i$ is the number density of isotope $i$ and the \emph{number} fraction of isotope $i$ is $X_i \equiv n_i/n_b$, with $n_b$ the baryon density. The \emph{mass} fraction is $Y_i \equiv A_i X_i$, where $A_i$ is the nucleon number. It is customary to define:
\begin{equation}
 \YP \equiv Y_{\He4} = 4 \frac{n_{\He4}}{n_b} \qquad \text{and} \qquad i/\mathrm{H} \equiv \frac{X_i}{X_\mathrm{H}} = \frac{n_i}{n_\mathrm{H}} \, .
 \end{equation}
 Note that, in addition to the species previously mentioned, we also plot the evolution of the abundances of $\mathrm{T} = {}^3\mathrm{H}$ and $\Be$. These nuclei are actually unstable: tritium decays into helium-3 with a half-life of 12.32 years, and the half-life of the decay $\Be \to \Li$ is 53.22 days. Since these periods are much higher than the time-scale on Figure~\ref{fig:IntroBBN}, and that the next major event in the history of the Universe is the formation of the CMB 380 000 years later, we add the abundances $(\mathrm{T} + \He{3})$ and $(\Be + \Li)$ in the final abundances reported (cf.~Table~\ref{Table:General_BBN}).
 
 Neutrinos affect BBN at various levels: via $\Neff$, the Hubble expansion rate is modified, affecting the “clock” of BBN. Then, the neutron-to-proton ratio, set by~\eqref{eq:weakrates_BBN_intro}, is modified when the electronic neutrino distribution functions get distorted, which is parameterized by $z_{\nu_e}$ and $\delta g_{\nu_e}$.
 
 As of today, the experimental measurements of primordial abundances are precise enough to be compared with numerical predictions are $\YP$ and $\mathrm{D}/\mathrm{H}$, and the excellent agreement is seen as a decisive proof of the hot Big Bang model (see chapter~\ref{chap:BBN}). We can however make two remarks:
\begin{itemize}
	\item the predicted abundance of lithium-7, given the baryon density otherwise measured by Planck and in agreement with helium-4 and deuterium measurements, is three times larger than the observed one: this is called the \emph{cosmological lithium problem}. Many solutions have been proposed, but it is for instance very hard to change the abundance of $\Li$ without dramatically affecting the very well constrained $\mathrm{D}/\mathrm{H}$. For a review, see for instance~\cite{Fields2011} ;
	\item the uncertainty on the prediction of the deuterium abundance has recently been reduced thanks to the updated rates of the reaction $\mathrm{D} + p \leftrightarrow \He{3} + \gamma$ from the LUNA experiment~\cite{Mossa2020}. On the one hand, a series of works~\cite{Pisanti2020,Yeh2020} confirm the agreement between predictions and measurements, while the analysis of~\cite{Pitrou2020} hints for a possible tension. As discussed in~\cite{Pitrou2021}, the difference between these results is due to the values selected for the nuclear rates of the reactions $\mathrm{D} + \mathrm{D} \leftrightarrow n + \He{3}$ and $\mathrm{D} + \mathrm{D} \leftrightarrow p + \mathrm{T}$. When the same rates are used, the different numerical codes agree. However, there is no definitive argument in favour of one selection or the other, which calls for higher precision measurements in the future.
\end{itemize}
These issues show how crucial it is to understand the physics at play during BBN. In chapter~\ref{chap:BBN}, we study in detail how incomplete neutrino decoupling affects the primordial abundances, combining a numerical and a theoretical analyses.

\section{From massless to massive neutrinos}
\label{sec:Intro_massive_nu}

Up to this point in the discussion, we have considered that neutrinos were the particles predicted by the Standard Model of particle physics --- cf.~appendix~\ref{App:StandMod}. However, there is now a substantial amount of experimental evidence for the fact that neutrinos are not massless and can undergo \emph{neutrino oscillations}, that is the possible change of flavour as neutrinos propagate. The concept of neutrino oscillations was first proposed by Pontecorvo~\cite{Pontecorvo1957,Pontecorvo1958} but for $\nu \leftrightarrow \bar{\nu}$ mixing, as he had been misled by some experimental results. In 1962, Maki, Nakagawa and Sakata~\cite{MNS} proposed that the “weak neutrinos”\footnote{At that time, the $\tau$ lepton had not yet been discovered.} $\nu_e$ and $\nu_\mu$ were quantum superpositions of the “true neutrinos” $\nu_1$ and $\nu_2$:
\begin{equation}
	\label{eq:mixing_2D}
\begin{aligned}
	\nu_e &= \cos{\theta} \nu_1 + \sin{\theta} \nu_2 \\
	\nu_\mu &= - \sin{\theta} \nu_1 + \cos{\theta} \nu_2
\end{aligned} \, .
\end{equation}
However, it was about a decade later that the current theory of neutrino oscillations was truly developed, cf.~for instance~\cite{Eliezer1975}.

We shall not review the history neutrino oscillations, but we will present it through the prism of a particular topic: the so-called \emph{solar neutrino problem}. This will allow us to introduce all the elements we will later need to take into account neutrino mixing in our calculations.

Moreover, the theoretical description of neutrino masses and mixings is presented in appendix~\ref{App:StandMod}. We do not discuss models of neutrino masses, such as the see-saw mechanisms, and refer to, e.g.,~\cite{GiuntiKim,MohapatraPal,King2015}.

\subsection{Massive neutrinos: the example of the solar neutrino problem}

The Sun is a powerful source of electronic neutrinos with energy of the order of $1 \, \mathrm{MeV}$, produced in the thermonuclear reactions which generate solar energy. One of the two main chains of reactions, the so-called $pp$ chain, involves the following reactions:
\begin{equation}
\label{eq:ppchain}
\begin{aligned}
	p + p &\to \mathrm{D} + e^+ + \nu_e  &(pp) \\
	p + e^- + p &\to \mathrm{D} + \nu_e &(pep) \\
	\He{3} + p &\to \He{4} + e^+ + \nu_e &(hep) \\
	\Be + e^- &\to \Li + \nu_e &(\Be) \\
	{}^8\mathrm{B} &\to {}^8\mathrm{Be}^* + e^+ + \nu_e &({}^8\mathrm{B})
\end{aligned}
\end{equation}
Each of these reactions produces neutrinos with different energy distributions, and in particular different average energies (see for instance the spectrum of solar neutrino fluxes in~\cite{PDG} or~\cite{Vitagliano_Review}).

The flux of solar neutrinos on the Earth is about $6 \times 10^{10} \, \mathrm{cm^{-2}s^{-1}}$. Such neutrinos were detected for the first time in 1970 in the Homestake experiment. However, the observed number of neutrinos was about one third of what the Standard Solar Model\footnote{A SSM is a “solar model that is constructed with the best available physics and input data” and is “required to fit the observed luminosity and radius of the Sun at the present epoch, as well as the observed heavy-element-to-hydrogen ratio at the surface of the Sun”~\cite{Bahcall}.} (SSM) predicted. This discrepancy was called the \emph{solar neutrino problem}, in particular after the disagreement between the SSM and experimental counts had been confirmed by several experiments (Kamiokande, GALLEX/GNO, SAGE, Super-Kamiokande), which probed different parts of the solar neutrino energy spectrum.

All these experiments measured a smaller number of electronic neutrinos compared to the expected flux. A solution to this problem can be obtained with the phenomenon of neutrino oscillations. Assuming that the \emph{mass states} $\nu_1$ and $\nu_2$ introduced in~\eqref{eq:mixing_2D} have a squared mass-difference $\Delta m^2 = m_2^2 - m_1^2$, one can show that the probability that a neutrino of energy $E$, initially in a state $\ket{\nu_e}$, be measured in a state of the same flavour $\ket{\nu_e}$ (the \emph{survival probability}) after a distance $L$ is~\cite{GiuntiKim}\footnote{This formula is easily derived in standard quantum mechanics with the so-called “equal momentum” assumption, which states that neutrinos propagate with identical momenta but different energies due to their different masses. This assumption is highly questionable, but does not affect the result, see~\cite{GiuntiKim} for a thorough discussion.}
\begin{equation}
\mathcal{P}_{e \to e} = 1 - \sin^2(2 \theta) \sin^2\left(\frac{\Delta m^2 L}{4 E}\right) \, .
\end{equation}
For neutrinos coming from the sun, the “source-detector” distance is huge and it happens that $\Delta m^2$ is not too small, such that the measurable quantity over the energy resolution of the detector is the average probability
\begin{equation}
\label{eq:survival_vacuum}
\mathcal{P}_{e \to e}^\mathrm{vacuum} = 1 - \frac12 \sin^2(2 \theta) \, .
\end{equation}
The superscript $^\mathrm{vacuum}$ highlights the fact that we do not consider any \emph{matter effect}, which comes from the different behaviour of neutrinos propagating in matter and in vacuum, as evidenced in the following.

The first experiments on solar neutrinos were only able to measure a deficit in the number of detected electronic neutrinos. The observations of the Sudbury Neutrino Observatory (SNO)~\cite{SNO_2002} were crucial in proving the validity of the mechanism of neutrino oscillations. In this heavy-water Cherenkov detector located in the Creighton mine near Sudbury (Ontario, Canada), solar neutrinos are detected through both charged and neutral currents on deuterium (and also via elastic scattering):
\begin{equation*}
\begin{aligned}
 \nu_e + \mathrm{D} &\to p + p + e^-  &\mathrm{(CC)} \\
 \nu_\alpha + \mathrm{D} &\to p + n + \nu_\alpha &\mathrm{(NC)}
 \end{aligned}
 \end{equation*}
 The neutral-current reaction on deuterium is equally sensitive to all neutrinos, while the charged-current one is only sensitive to electronic neutrinos: this provides a direct way to check the flavour transformation and the survival probability by comparing the flux from CC to the total neutrino flux.
 
\begin{figure}[!ht]
	\centering
	\includegraphics{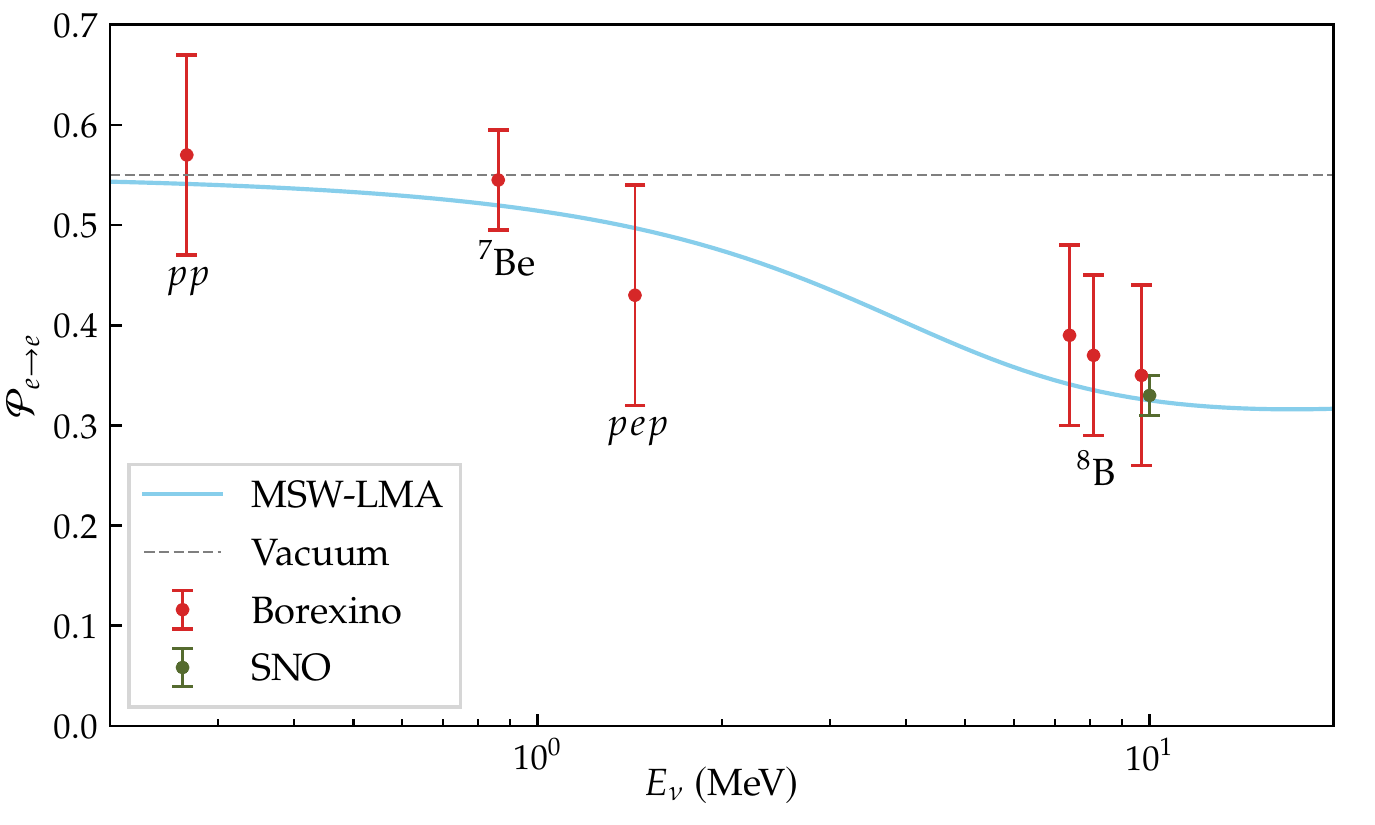}
	\caption[$\nu_e$ survival probability as a function of neutrino energy]{\label{fig:solar_neutrinos} Electron neutrino survival probability as a function of neutrino energy. The points represent, from left to right, the Borexino $pp$, ${}^7\mathrm{Be}$, $pep$, and ${}^8\mathrm{B}$ data (red points)~\cite{Borexino} and the SNO ${}^8\mathrm{B}$ data (green point)~\cite{SNO_2016}. The error bars represent the $\pm 1 \sigma$ experimental + theoretical uncertainties. The blue curve corresponds to the prediction of the MSW-LMA solution using the formulae given in~\cite{Vissani2017a,Vissani2017b,Borexino} and the parameters from~\cite{PDG}. The dashed grey line is the vacuum prediction (equation~\eqref{eq:survival_vacuum}, corrected to account for three-neutrino mixing).}
\end{figure}

 We show on Figure~\ref{fig:solar_neutrinos} the measured electron neutrino survival probability for different neutrino energies. These different energies correspond to different production channels of solar neutrinos~\eqref{eq:ppchain}, and the results plotted on this Figure come from two experiments: SNO, already mentioned, and Borexino, a liquid scintillator experiment based at Gran Sasso (Italy).
 
First and foremost, $\mathcal{P}_{e \to e} \neq 1$: there is a large conversion of electronic neutrinos into the other flavours. At low energies, the prediction\footnote{This formula is actually corrected in the three-neutrino case and reads (the mixing angles are introduced in appendix~\ref{App:StandMod})
\[ \mathcal{P}_{e \to e}^\mathrm{vacuum} = \cos^4{\theta_{13}} \left(1 - \frac12 \sin^2(2 \theta_{12})\right) + \sin^4{\theta_{13}} \, . \] } of the vacuum solution~\eqref{eq:survival_vacuum} is in agreement with experimental data. However, this solution does not explain the results at higher energies, notably the SNO one. The answer to this discrepancy requires to take into account \emph{matter effects}.

\paragraph{Matter effects and MSW resonance} Flavour oscillation is a consequence of the fact that flavour and mass eigenstates are not identical (flavour mixing) and that mass eigenstates propagate with different velocities (mass differences). The propagation in a medium different from vacuum is thus expected to change the oscillation mechanism, similarly to the refraction index of a dielectric medium for photons (the matter effects we discuss here are sometimes called \emph{refractive effects}). Wolfenstein~\cite{MSW_W} discovered in 1978 that neutrinos propagating in matter were subject to a mean-field potential which modifies their mixing.

The effective potential due to charged-current interactions, $V_\text{CC}$, is obtained by averaging the weak Hamiltonian~\eqref{eq:hcc_app} over the background particle distribution. For a homogeneous and isotropic gas of unpolarized electrons, we obtain an average effective Hamiltonian (we assume we are at sufficiently low energy to use the effective four-Fermi theory)
\begin{equation}
	\label{eq:VCC_intro}
	\langle H_\text{eff}^{\mathrm{(CC)}} \rangle = V_\text{CC} \overline{\nu_{eL}}(x) \gamma^0 \nu_{eL}(x) \quad \text{with} \quad \boxed{V_\text{CC} = \sqrt{2} G_F n_{e^-}} \, .
\end{equation}
A background of positrons leads to $V_\text{CC} = - \sqrt{2} G_F n_{e^+}$, because of the anticommutation of annihilation/creation operators in the Hamiltonian when dealing with antiparticles. The same procedure can be applied to neutral-current interactions, and the effective potential experienced by a neutrino of any flavour in a background of unpolarized fermions $f$ reads
\begin{equation}
V_\text{NC}^f = \sqrt{2} G_F n_f g_V^f \, ,
\end{equation}
with $g_V^f$ the weak vector coupling of fermion $f$. Given the values of the different $g_V^f$, in an environment of electrons, positrons, protons and neutrons such as the early Universe, and because of charge neutrality, all contributions cancel except that of neutrons:
\begin{equation}
V_\text{NC} = - \frac12 \sqrt{2} G_F n_n \, .
\end{equation}
In summary, the effective potential felt by a neutrino of flavour $\alpha$ is~\cite{Neutrino_Cosmology,GiuntiKim}
\begin{equation}
	V_\alpha = \sqrt{2} G_F \left[ (n_{e^-} - n_{e^+})\delta_{\alpha e} - \frac12 n_n \right] \, .
\end{equation}
Note that in astrophysical environments, there are usually no positrons and the effective potential is the one boxed in~\eqref{eq:VCC_intro}. Since $V_\text{NC}$ is flavour-independent, it does not affect mixing.\footnote{This is true as long as we only consider active-active oscillations. The presence of sterile neutrino species would break this simplification.}


Due to this extra term in the Hamiltonian, the effective mass states are modified --- in other words, the \emph{mass basis} becomes the \emph{matter basis}. In the two-neutrino simplified case, the eigenstates of the total Hamiltonian, $\nu_1^m$ and $\nu_2^m$, are related to the flavour eigenstates via the same relation as~\eqref{eq:mixing_2D}, but the effective mixing angle in matter
\begin{equation}
\label{eq:thetam}
\tan{2 \theta_m} = \frac{\tan{2 \theta}}{ 1 - \frac{2 E V_\text{CC}}{\Delta m^2 \cos{2 \theta}}} \, ,
\end{equation}
and the effective mass-squared difference is
\begin{equation}
\Delta m_m^2 = \sqrt{(\Delta m^2 \cos{2 \theta} - 2 E V_\text{CC})^2 + (\Delta m^2 \sin{2 \theta})^2} \, .
\end{equation}
It was first pointed out in 1985 by Mikheev and Smirnov~\cite{MSW_MS} that in a medium with varying density, there was a region in which the effective mixing angle passes through the maximal mixing value of $\pi/4$. From~\eqref{eq:thetam}, this \emph{resonance} occurs for
\begin{equation}
\Delta m^2 \cos{2 \theta} = 2 p V_\text{CC} \quad \iff \quad n_{e^-}^\mathrm{res} = \frac{\Delta m^2 \cos{2 \theta}}{2 \sqrt{2} G_F E} \, ,
\end{equation}
if we assume there are no positrons (standard case in astrophysical environments where this effect was first discussed). This mechanism is now commonly referred to as the \emph{MSW effect}.

The possibility of large flavour conversion will depend on the hierarchy between two timescales: the oscillation frequency $\Delta m^2/2E$ and the rate of variation of the mixing angle $\dd{\theta_m}/\dd \x$, where we parameterize the neutrino trajectory by a position in space $\x$. In particular, if $\Delta m^2/2E \gg \abs{\dd{\theta_m}/\dd \x}$, the evolution is \emph{adiabatic} and transitions between $\nu_1^m$ and $\nu_2^m$ are suppressed. The criterion of adiabaticity, which suppresses the transitions between matter eigenstates, will be at the core of the approximation developed to study neutrino decoupling including flavour oscillations in chapters~\ref{chap:Decoupling} and~\ref{chap:Asymmetry}.

For solar neutrinos, assuming $\Delta m^2 \sim 7 \times 10^{-5} \, \mathrm{eV^2}$ and $\tan^2{\theta} \simeq 0.4$, we obtain the survival probability represented by the blue curve on Figure~\ref{fig:solar_neutrinos}. This set of parameters, that is well-measured today in the 3-neutrino framework, was historically called the “Large Mixing Angle” (LMA) solution, and was for instance in competition with the “Small Mixing Angle” (SMA) solution, for which $\Delta m^2 \sim 5 \times 10^{-6} \, \mathrm{eV}^2$, $\tan^2{\theta} \sim 5 \times 10^{-4}$. The agreement with the experimental points is a proof of the validity of the flavour oscillation mechanism, with the additional complexity of matter effects.

\subsection{Three-neutrino mixing}

In our discussion of the solar neutrino problem, we have reduced the problem to the mixing of two neutrino states. After many other experiments (atmospheric neutrinos, reactors and accelerators), the “standard” model of neutrino mixing involves the three active flavour species and the associated three mass eigenstates. The parameterization of the mixing is presented in appendix~\ref{subsec:Values_Mixing}, and the values of all parameters are now well measured, except for two:
\begin{itemize}
	\item the \emph{mass ordering} (we also talk about the mass hierarchy) is unknown. Defining $\Delta m_{ji}^2 = m_{\nu_j}^2 - m_{\nu_i}^2$, there are two possibilities concerning the sign of $\Delta m_{31}^2$, represented on Figure~\ref{fig:hierarchies};

 \begin{figure}[!ht]
	\centering
	\includegraphics{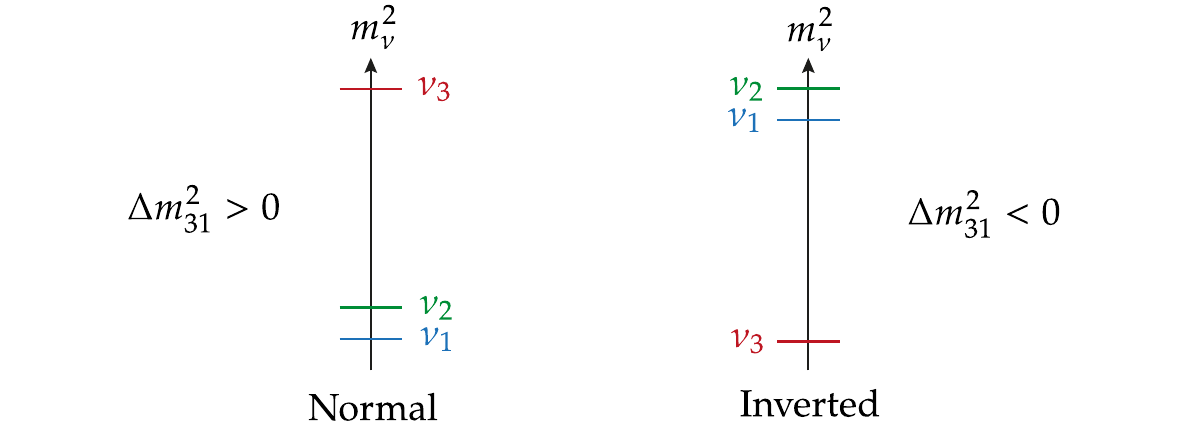}
	\caption[Normal and inverted neutrino mass orderings]{\label{fig:hierarchies} The two possible mass orderings, normal (\emph{left}) and inverted (\emph{right}). The small mass gap $\Delta m_{21}^2$ is the one involved in solar neutrino oscillations.}
\end{figure}

	\item the value of the Dirac CP phase: a non-vanishing phase would indicate a difference between the mixing of neutrinos and antineutrinos. An important result of this thesis is the strong independence of neutrino physics in the MeV era, as long as beyond-the-Standard-Model mechanisms (outside of neutrino masses and mixings) are not invoked.
\end{itemize}
This standard “three-neutrino mixing” model will be used throughout the manuscript.

\subsection{Massive neutrinos in cosmology}

A robust and precise prediction of the consequences of incomplete neutrino decoupling is crucial since neutrinos impact many cosmological stages.
\begin{enumerate}
	\item During Big Bang Nucleosynthesis (BBN), neutrinos control neutron/proton conversions as they participate to weak interactions, and the frozen neutron abundance subsequently affects nuclear reactions and light element relics --- see~\ref{subsec:intro_BBN}.
	\item During the Cosmic Microwave Background (CMB) formation, the free streaming of neutrinos is crucial to predict the CMB angular spectrum. Also, the value of $\Neff$ affects the cosmological expansion, and thus also the radiative transfer of CMB. From these effects, CMB alone can be used to place constraints on $\Neff$ or in combination with BBN constraints on primordial light elements.
	\item In the late universe, neutrino free streaming also affects structure formation, via its effect on the growth of perturbations. This is used to place the constraint $\sum_i m_{\nu_i} < 0.12 \, \mathrm{eV}$ (see e.g.~\cite{Planck18,eBOSS}) on the sum of neutrino masses.
\end{enumerate}
It is striking that neutrino masses play a key role in both the earliest stage 1 and the latest stage 3 for very different reasons.
In stage 1, neutrino oscillations, which are due to small neutrino mass-squared differences and mixing angles, affect the non-thermal part of the spectra, as they lead to less distortion in electron-type neutrinos and more distortion in other types than if there were no oscillations at all. Also, oscillations lead to a mild modification of $\Neff$ --- see chapter~\ref{chap:Decoupling}. In stage 3, and due to cosmological redshifting, all massive neutrinos undergo at some point a transition from being very relativistic (they behave gravitationally like decoupled photons) to being non-relativistic. This transition depends only on neutrino masses and not on mixing angles, since frozen neutrino spectra inherited from stage 1 are generated incoherently in the mass basis.
Finally, stage 2 would also be affected beyond the standard cosmological model, if we were to consider exotic physics with increased neutrino self-interactions, so that they would still behave effectively as a perfect fluid around CMB formation~\cite{Kreisch:2019yzn,Grohs:2020xxd}.

This interplay between the various cosmological eras implies that it is crucial to understand neutrino decoupling as precisely as possible, in order to use these predictions as initial conditions for the subsequent eras. For instance, current constraints from CMB on cosmological parameters~\cite{Planck18} were placed using $\Neff=3.046$  when solving numerically for the linear evolution of cosmological perturbations.


\pagestyle{ruled}

\chapter{The Quantum Kinetic Equations}
\label{chap:QKE}

\epigraph{You think this is hard? Try being waterboarded, \emph{that's} hard!}{Sue Sylvester, \emph{Glee} [S01E01]}

{
\hypersetup{linkcolor=black}
    \minitoc
}

\boxabstract{The material of this chapter was published in~\cite{Froustey2020}.}

A precision calculation of neutrino evolution requires to take into account the phenomenon of flavour oscillations. This means that the Boltzmann kinetic equation must be generalized to account for \emph{flavour coherence}, that is the possibility to have a non-vanishing statistical average of mixed-flavour states. A convenient formalism consists in promoting the set of distribution functions to a one-body density matrix, a strategy notably introduced in a seminal paper by Sigl and Raffelt~\cite{SiglRaffelt}. They obtained the so-called “Quantum Kinetic Equation” (QKE) through a perturbative expansion of this “matrix of densities” on the interaction parameter (i.e.~the Fermi constant $G_F$). In the following, we present a formalism that follows quite closely this historical approach, but in the more general framework of a hierarchy of equations. Alternatively to this operator approach, we can quote the functional approach\footnote{We do not compare here these approaches, but simply note that the functional formalism contains information about the spectrum of the theory (i.e., which states are available)~\cite{Berges:2004,Vlasenko_PhRevD2014,Berges:2015,Drewes:2017}. However, for neutrinos in the early Universe the spectral function is, at the order we are interested in, the same as the massless, free-field one, such that the two formalisms can be used interchangeably.} of~\cite{BlaschkeCirigliano} which uses the Closed-Time Path formalism.

After deriving the QKEs in the general case of a system subject only to two-body interactions, we apply the formalism to the specific case of neutrinos in the early Universe, which leads to many simplifications.

\section{Hierarchy of evolution equations}
\label{sec:BBGKY}

In this section, we derive from first principles the neutrino quantum kinetic equations, which generalize the Boltzmann kinetic equation for distribution functions~\eqref{eq:boltzmann_nu}, to account for neutrino masses and mixings. We present the Bogoliubov-Born-Green-Kirkwood-Yvon (BBGKY) hierarchy~\cite{Bogoliubov,BornGreen,Kirkwood,Yvon} that was historically derived for a non-relativistic $N-$body system and heavily used in nuclear physics~\cite{WangCassing1985,Cassing1990,Reinhard1994,Lac04,Simenel,Lac14}, but which can also be applied to a relativistic system such as the bath of neutrinos and antineutrinos in the early Universe. We extend the formalism of~\cite{Volpe_2013}, where the BBGKY formalism was applied to derive extended mean-field equations for astrophysical applications, and include the collision term. 

\subsection{BBGKY formalism}

The exact evolution of a $N-$body system under the Hamiltonian $\hat{H}$ is given by the Liouville-von Neumann equation for the many-body density matrix
\begin{equation}
\label{eq:vonneumann}
\ii \frac{\dd \hat{D}}{\dd t} = [\hat{H}, \hat{D}] \, ,
\end{equation}
where $\hat{D} = \ket{\Psi}\bra{\Psi}$, with $\ket{\Psi}$ the quantum state, from which we define the $s$-body reduced density matrices, 
\begin{equation}
\label{eq:defrhos}
\hat{\varrho}^{(1 \cdots s)} \equiv \frac{N !}{(N-s)!} \mathrm{Tr}_{s+1\dots N} \hat{D} \, .
\end{equation}
Its components (we drop the superscript $^{(1\cdots s)}$, redundant with the number of indices) read, as detailed in the appendix~\ref{app:vrho_components}:
\begin{equation}
\label{eq:defrhos2}
\varrho^{i_1 \cdots i_s}_{j_1 \cdots j_s} \equiv \langle \hat{a}_{j_s}^\dagger \cdots \hat{a}_{j_1}^\dagger \hat{a}_{i_1} \cdots \hat{a}_{i_s} \rangle \, ,
\end{equation}
where the indices $i,j$ label a set of quantum numbers (species $\phi_i$, momentum $\vp_i$, helicity $h_i$) which describe a one-particle quantum state, and $\langle \cdots \rangle$ is shorthand for $\bra{\Psi} \cdots \ket{\Psi}$. Let us give one example, where the summing rules are also made explicit:\footnote{Reminder: the factors of $2E$ in the denominator of $\ddp{}$ arise from relativistic phase-space constraints. More precisely, the phase space integrals should be four-dimensional, with the on-shell condition:
	\[\int{\mathrm{d}^3 \vec{p}}\int_{0}^{\infty}{\mathrm{d}E \, \delta(E^2 - p^2 - m^2)} = \int{\mathrm{d}^3 \vec{p}}\int_{0}^{\infty}{\mathrm{d} E \, \frac{\delta(E - \sqrt{p^2 + m^2})}{2 E}} \, ,\]
	and the additional $(2\pi)^3$ are actually the fundamental phase space volumes $(2 \pi \hbar)^3$ with $\hbar=1$.}
\begin{equation}
\label{eq:set_quantum_numbers}
\sum_i{\had_i}= \sum_{\phi_i} \sum_{h_i} \int{\ddp{i} \, \had_{\phi_i}(\vp_i,h_i)} \qquad \text{with} \qquad \ddp{i} \equiv \frac{\dd^3 \vp_i}{(2 \pi)^3 2 E_i}   \, .
\end{equation}
The creation and annihilation operators satisfy the fermionic anticommutation rules $\{ \had_i, \ha_j \} = \delta_{ij}$, with $\delta$ the Kronecker delta (generalized to the set of quantum numbers precised above)
\begin{equation}
\delta_{ij} \equiv (2 \pi)^3 \, 2 E_i \, \delta^{(3)}(\vp_i - \vp_j \,) \delta_{h_i h_j} \delta_{\phi_i \phi_j} \, .
\end{equation}

The central object is the one-body reduced density matrix \cite{SiglRaffelt},
\begin{equation}
\label{eq:defrho}
\vrho^i_j \equiv \langle \ha_j^\dagger \ha_i \rangle \, ,
\end{equation}
whose diagonal entries correspond to the standard occupation numbers. 

The Hamiltonian for this system is given by the sum of the kinetic and the two-body interaction terms (such an interaction Hamiltonian being adequate for neutrinos whose interactions are described by Fermi theory),
\begin{equation}
\hat{H} = \hat{H}_0 + \hat{H}_{\mathrm{int}} = \sum_{i,j}{t^{i}_{j} \, \had_i \ha_j} + \frac14 \sum_{i,j,k,l}{\tilde{v}^{ik}_{jl} \, \had_i \had_k \ha_l \ha_j}  \label{eq:defHint} \, .
\end{equation} 
The interaction matrix elements are fully anti-symmetrized by construction:
\begin{equation}
\label{eq:defvint}
\bra{ik} \hat{H}_{\mathrm{int}} \ket{jl} \equiv \tilde{v}^{ik}_{jl}= - \tilde{v}^{ki}_{jl} = \tilde{v}^{ki}_{lj} \, .
\end{equation}
The traditional presentation of the BBGKY formalism in nuclear physics is sometimes based on tensor products of one-particle states rather than fully antisymmetrized states (defined as $\ket{i_1 \cdots i_s} = \had_{i_1} \cdots \had_{i_s} \ket{0}$ with $\ket{0}$ the quantum vacuum state). The equivalence between both approaches is discussed for completeness in the appendix~\ref{app:BBGKY_Antisym}.

This set of definitions ensures proper transformation laws under a unitary transformation of the one-particle quantum state $\psi^i = \mathcal{U}^{i}_{a} \psi^a$: all lower indices are covariant while upper indices are contravariant, namely,
\begin{equation}
\label{eq:transfo}
\vrho^a_b = {\mathcal{U}^\dagger}^a_i \, \vrho^i_j \, \mathcal{U}^j_b \quad , \quad t^a_b = {\mathcal{U}^\dagger}^a_i \, t^i_j \, \mathcal{U}^j_b \quad , \quad \tilde{v}^{ac}_{bd} = {\mathcal{U}^\dagger}^a_i {\mathcal{U}^\dagger}^c_k \, \tilde{v}^{ik}_{jl} \, \mathcal{U}^j_b \mathcal{U}^l_d \, .
\end{equation}
One must keep in mind that the unitary transformation “matrix” $\mathcal{U}$ is extremely complicated a priori, all the complexity being hidden in the use of the generalized indices.

\paragraph{BBGKY hierarchy} The evolution equation for $\vrho$ can be obtained directly via the Ehrenfest theorem. One can also apply partial traces to \eqref{eq:vonneumann}, which leads to the well-known \emph{BBGKY hierarchy}~\cite{Bogoliubov,BornGreen,Kirkwood,Yvon}, whose first two equations read\footnote{We explicitly wrote the components of the tensors compared to the expressions found in~\cite{Volpe_2013} or~\cite{Lac04,Simenel}.} (Einstein summation convention implied): 

\begin{subequations}
\label{eq:hierarchy}
\begin{empheq}[left=\empheqlbrace]{align}
\ii  \frac{\dd \vrho^{i}_{j}}{\dd t} &= \left( t^{i}_{k} \vrho^{k}_{j} - \vrho^{i}_{k} t^{k}_{j} \right) + \frac12 \left(\tilde{v}^{ik}_{ml} \vrho^{ml}_{jk} - \vrho^{ik}_{ml} \tilde{v}^{ml}_{jk} \right)\,, \label{eq:hierarchy_1} \\
\ii  \frac{\dd \vrho^{ik}_{jl}}{\dd t} &= \left(t^{i}_{r} \vrho^{rk}_{jl} + t^{k}_{p} \vrho^{ip}_{jl} + \frac12 \tilde{v}^{ik}_{rp} \vrho^{rp}_{jl} - \vrho^{ik}_{rl} t^{r}_{j} - \vrho^{ik}_{jp} t^{p}_{l} - \frac12 \vrho^{ik}_{rp} \tilde{v}^{rp}_{jl} \right) \label{eq:hierarchy_2}  \\
&\qquad \qquad + \frac12 \left(\tilde{v}^{im}_{rn} \vrho^{rkn}_{jlm} + \tilde{v}^{km}_{pn} \vrho^{ipn}_{jlm} - \vrho^{ikm}_{rln} \tilde{v}^{rn}_{jm} - \vrho^{ikm}_{jpn} \tilde{v}^{pn}_{lm} \right)  \, . \nonumber
\end{empheq}
\end{subequations}

More than simply recasting in a less compact form the very complicated problem \eqref{eq:vonneumann}, this hierarchy furnishes a set of evolution equations which depend on higher-order reduced density matrices. In order to solve these equations, one necessarily needs to \emph{truncate} this hierarchy. Different truncation schemes exist, and we will only discuss the useful ones for neutrino evolution in the early Universe.

\subsection{Hartree-Fock approximation and mean-field terms}

It proves convenient to split the two-body density matrix into its uncorrelated (i.e., products of one-body density matrices) and correlated contributions \cite{Lac04,Simenel,Lac14}:
\begin{equation}
\label{eq:splitrho}
\vrho^{ik}_{jl} \equiv 2 \vrho^{i}_{[j} \vrho^{k}_{l]} + C^{ik}_{jl} \equiv \vrho^{i}_{j} \vrho^{k}_{l} - \vrho^{i}_{l} \vrho^{k}_{j} + C^{ik}_{jl} \, .
\end{equation}
Inserting this decomposition into \eqref{eq:hierarchy_1}, we get
\begin{equation}
\label{eq:eqvrho}
\ii  \frac{\dd \vrho^{i}_{j}}{\dd t} = \left[ \left(t^{i}_{k} + \Gamma^{i}_{k}\right) \vrho^{k}_{j} - \vrho^{i}_{k} \left(t^{k}_{j} + \Gamma^{k}_{j}\right) \right] + \frac12 \left(\tilde{v}^{ik}_{ml} C^{ml}_{jk} - C^{ik}_{ml} \tilde{v}^{ml}_{jk} \right) 
=  \left [ \hat{t} + \hat{\Gamma} , \hat{\vrho} \right]^{i}_{j} + \ii \,  \mathcal{C}^{i}_{j} \, ,
\end{equation}
which defines the collision term $\hat{C}$ (discussed later) and the \emph{mean-field potential} $\hat{\Gamma}$ (for once, we explicit the summation)
\begin{equation}
\label{eq:Gamma}
\Gamma^{i}_{j} = \sum_{k,l}{\tilde{v}^{ik}_{jl} \vrho^{l}_{k}} \, .
\end{equation}
It accounts for the effective potential “felt” by the particles when propagating in a non-vacuum background.

\subsubsection{Mean-field approximation}

The simplest non-trivial closure of the BBGKY hierarchy is the so-called Hartree-Fock or \emph{mean-field} approximation. It consists in neglecting $C^{ik}_{jl} \simeq 0$ and keeping only the commutator part in \eqref{eq:eqvrho}.

However, in the context of neutrino decoupling in the early Universe, one seeks a generalization of the Boltzmann equation for neutrino distribution functions \cite{Dolgov_NuPhB1997,Esposito_NuPhB2000,Mangano2002,Grohs2015,Froustey2019}, which describes the evolution of densities under two-body collisions. In other words, we need to truncate the hierarchy~\eqref{eq:hierarchy} assuming the \emph{molecular chaos} ansatz: correlations between the one-body density matrices arise from two-body interactions between uncorrelated matrices~\cite{Lac04}. This prescribes the form of $C^{ik}_{jl}(t)$, leading to a formal expression for $\hat{\mathcal{C}}$, which we establish in the following section.

\subsection{Derivation of the structure of the collision term}
\label{subsec:derivation_collision}

Compared to the Boltzmann treatment of neutrino evolution, which neglects flavour mixing, the QKE contains mean-field terms, and the collision term has a richer matrix structure with non-zero off-diagonal components. To derive this collision term, i.e., the contribution to the evolution of the one-body density matrix from two-body correlations, one needs an expression for the correlated part $\hat{C}$ in \eqref{eq:eqvrho}. It is obtained from the evolution equation for $\hat{\vrho}^{(12)}$, where we separate correlated and uncorrelated parts \cite{Lac04}.

To do so, we need a splitting similar to \eqref{eq:splitrho} for the three-body density matrix,
\begin{equation}
\label{eq:splitrho123}
\vrho^{ikm}_{jln} = 6 \vrho^{i}_{[j}\vrho^{k}_{l}\vrho^{m}_{n]} + 9 \vrho^{[i}_{[j} C^{km]}_{ln]} + C^{ikm}_{jln} \, .
\end{equation}
This allows to rewrite~\eqref{eq:hierarchy_2} as an equation for the two-body correlation function \cite{Volpe_2013}. In the \emph{molecular chaos} ansatz, correlations are built through collisions between uncorrelated particles. These correlations then evolve ``freely'', i.e., we do not take into account a mean-field background for $\hat{C}$. The evolution equation is thus greatly simplified and reads
\begin{equation}
\begin{aligned}
\ii  \frac{\dd C^{ik}_{jl}}{\dd t} &\simeq \left[t^{i}_{r} C^{rk}_{jl} + t^{k}_{p} C^{ip}_{jl} - C^{ik}_{rl} t^{r}_{j} - C^{ik}_{jp} t^{p}_{l} \right] \\
&\quad + \underbrace{(\hat{\Id} - \vrho)^i_r (\hat{\Id}-\vrho)^k_p \, \tilde{v}^{rp}_{sq} \, \vrho^{s}_{j} \vrho^{q}_{l} -  \vrho^i_r \vrho^k_p \, \tilde{v}^{rp}_{sq} \, (\hat{\Id} -\vrho)^{s}_{j} (\hat{\Id} - \vrho)^{q}_{l}}_{\displaystyle \equiv B^{ik}_{jl}} \, ,
\end{aligned}
\end{equation}
The commutator in the first row is a “vacuum term” which accounts for the evolution of correlations in the vacuum, hence depending only on the kinetic part of the Hamiltonian~\eqref{eq:defHint}. The second row is the “Born term” which only involves the uncorrelated part of~\eqref{eq:splitrho}. The other neglected terms can be found in e.g.~\cite{WangCassing1985,Lac04,Volpe_2013}.

We can solve this equation, starting from $C(t=0)=0$,
\begin{equation}
\label{eq:solveC}
C^{ik}_{jl}(t) = - \ii \int_{0}^{t}{\dd s \,  T^{ik}_{mp}(t,s) B^{mp}_{nq}(s) {T^\dagger}^{nq}_{jl}(t,s)} \, ,
\end{equation}
with the evolution operator
\begin{equation}
T^{ik}_{jl}(s,s') = \exp{\left(-\ii \int_{s'}^{s}{\dd \tau \, \hat{t}(\tau)}\right)}^{i}_{j} \exp{\left(-\ii \int_{s'}^{s}{\dd \tau \, \hat{t}(\tau)}\right)}^{k}_{l} \, .
\end{equation}
Now we consider that there is a clear separation of scales \cite{SiglRaffelt}, i.e.the duration of one collision is very small compared to the variation timescale of the density matrices (i.e., compared to the duration between two collisions, and the typical inverse oscillation frequency). Therefore, the argument inside the integral of \eqref{eq:solveC} is only non-zero for $s \simeq 0$: we can extend the integration domain to $+ \infty$, while the operators keep their $t=0$ value. Finally we symmetrize the integration domain\footnote{See section 6.1~\cite{FidlerPitrou} for a detailed discussion of this procedure, which amounts to separate the macroscopic evolution from the microphysics processes.} with respect to 0 (with an extra factor of $1/2$), which leads to the equation with collision term:\footnote{The product $\hat{T} \hat{B} \hat{T}^\dagger$ must be done from left to right, as can be seen with the components in~\eqref{eq:solveC}.}
\begin{subequations}
\begin{align}
\ii \frac{\dd \vrho^{i}_{j}}{\dd t} &= \left[ \hat{t} + \hat{\Gamma}, \hat{\varrho}\right]^{i}_{j} - \frac{\ii}{4} \int_{-\infty}^{+\infty}{\dd t\,  \left[ \tilde{v},\hat{T}(t,0)\hat{B}(0) \hat{T}^\dagger(t,0) \right]}^{ik}_{jk} \\ 
&= [(t^i_k + \Gamma^i_k)\varrho^k_j - \varrho^i_k (t^k_j + \Gamma^k_j)] \nonumber \\
&\qquad  - \frac{\ii}{4}  \underbrace{\int_{-\infty}^{+\infty}{\dd t \, e^{-\ii (E_m+E_l-E_j-E_k)t}}}_{(2\pi) \, \delta(E_m + E_l - E_j - E_k)}\left[\tilde{v}^{ik}_{rl} B^{rl}_{jk} - B^{ik}_{rl} \tilde{v}^{rl}_{jk} \right] \, , \\
&\equiv \left[ \hat{t} + \hat{\Gamma}, \hat{\varrho}\right]^{i}_{j} + \ii \, \mathcal{C}^{i}_{j}
\end{align}
\end{subequations}
The exponential of energies comes from the $\hat{T}$ terms, using that the density matrix for a given momentum $\hat{\vrho}(p)$ satisfies\footnote{We anticipate the simplification of the density matrices due to homogeneity and isotropy~\eqref{eq:homogeneity} in order to get a practical result. Note that we do not take into account the contribution of masses to neutrino energies in the collision term, since they are completely negligible compared to the ultrarelativistic part.}  $\hat{t}\hat{\vrho}(p) = p \,\hat{\varrho}(p)$. The general form of the collision term is then
\begin{multline}
\label{eq:C11}
\mathcal{C}^{i_1}_{i_1'} = \frac14 \sum_{i_2, i_3, i_4} \sum_{j_1, j_2, j_3, j_4} \left(\tilde{v}^{i_1 i_2}_{i_3 i_4} \vrho^{i_3}_{j_3} \vrho^{i_4}_{j_4} \tilde{v}^{j_3 j_4}_{j_1 j_2} (\hat{\Id} -\vrho)^{j_1}_{i_1'} (\hat{\Id} - \vrho)^{j_2}_{i_2} - \tilde{v}^{i_1 i_2}_{i_3 i_4} (\hat{\Id} -\vrho)^{i_3}_{j_3} (\hat{\Id} - \vrho)^{i_4}_{j_4} \tilde{v}^{j_3 j_4}_{j_1 j_2} \vrho^{j_1}_{i_1'} \vrho^{j_2}_{i_2} \right. \\ 
\left. + (\hat{\Id} -\vrho)^{i_1}_{j_1} (\hat{\Id} - \vrho)^{i_2}_{j_2} \tilde{v}^{j_1 j_2}_{j_3 j_4} \vrho^{j_3}_{i_3} \vrho^{j_4}_{i_4} \tilde{v}^{i_3 i_4}_{i_1' i_2}  - \vrho^{i_1}_{j_1} \vrho^{i_2}_{j_2} \tilde{v}^{j_1 j_2}_{j_3 j_4} (\hat{\Id} - \vrho)^{j_3}_{i_3} (\hat{\Id}-\vrho)^{j_4}_{i_4} \tilde{v}^{i_3 i_4}_{i_1' i_2}  \right) \\
\times  (2 \pi) \, \delta(E_{i_1} + E_{i_2} - E_{i_3} - E_{i_4}) \, .
\end{multline}
It has the standard structure ‘‘gain $-$ loss $+$ h.c.'', which will be made more explicit when we give the full expressions for a system of neutrinos and antineutrinos interacting with standard model weak interactions, cf.~section~\ref{subsec:collision_integral}. 

In the following sections, we will focus on the case of the early Universe and consider three active species of neutrinos in a background of electrons and positrons, muons and antimuons (in traces), and photons. The influence of baryons can be discarded given their negligible density compared to relativistic species (the baryon-to-photon ratio is $\eta \equiv n_b/n_\gamma \simeq 6.1\times 10^{-10}$ from the most recent measurement of the baryon density~\cite{Fields:2019pfx}). Therefore, we focus on the lepton sector evolution, and baryons will only be dealt with when we discuss BBN in chapter~\ref{chap:BBN}.

\section{Application to neutrinos in the early Universe}

Assuming the Universe to be homogeneous and isotropic in the period of interest, the density matrices read,\footnote{The annihilation and creation operators satisfy the equal time anticommutation rules
\[\{\ha_{\nu_\alpha}(\vp, h),\had_{\nu_\beta}(\vpp, h') \} = (2\pi)^3 \, 2E_p \, \delta^{(3)}(\vp - \vpp) \, \delta_{h h'} \, \delta_{\alpha \beta} \ ; \ \{\had_{\nu_\alpha}(\vp, h),\had_{\nu_\beta}(\vpp, h') \} = \{\ha_{\nu_\alpha}(\vp, h),\ha_{\nu_\beta}(\vpp, h') \} = 0 \]
Similar relations hold for the antiparticle operators.}
\begin{subequations}
\label{eq:homogeneity}
\begin{align}
 \langle \had_{\nu_\beta}(\vpp, h') \ha_{\nu_\alpha}(\vp, h)  \rangle &= (2 \pi)^3 \, 2 E_p \, \delta^{(3)}(\vp - \vpp) \delta_{h h'} \, \vrho^{\nu_\alpha}_{\nu_\beta}(p,t) \, \delta_{h-}  \, , \\
  \langle \hbd_{\nu_\alpha}(\vp, h) \hb_{\nu_\beta}(\vpp, h')  \rangle &= (2 \pi)^3 \, 2 E_p \, \delta^{(3)}(\vp - \vpp) \delta_{h h'}  \, \bvrho^{\nu_\alpha}_{\nu_\beta}(p,t) \, \delta_{h+} \, .
\end{align}
\end{subequations}
The Kronecker delta ensures that only left-handed neutrinos and right-handed antineutrinos are included. However, in anisotropic environments, “wrong-helicity“ densities or \emph{pairing} densities (like the non-lepton-number-violating correlator, in the Dirac neutrino case, $\langle \hb_{\nu_\alpha} \ha_{\nu_\beta} \rangle$) can be sourced~\cite{Volpe_2013,SerreauVolpe,Volpe_2015,KartavtsevRaffelt}. We do not consider such terms here. The energy function is $E_p = p$ for neutrinos,\footnote{We always neglect the small neutrino masses compared to their typical momentum, except for the vacuum term since the diagonal momentum contribution disappears from the evolution equation (section~\ref{subsec:vacuum}).} and $E_p = \sqrt{p^2 + m_{e,\mu}^2}$ for $e^\pm, \, \mu^\pm$.

In the following, we will apply the BBGKY formalism to a system of neutrinos, leaving the details of the inclusion of antineutrinos\footnote{Note that the antineutrino density matrix $\bvrho^i_j \equiv \langle \hb_i^\dagger \hb_j \rangle$ is defined with a transposed convention, compared to the neutrino density matrix, to have similar evolution equations and transformation properties.} to appendix~\ref{app:antiparticles}. Note that, for a relativistic system, the hierarchy is given by an infinite set of equations~\cite{CalzettaHu} (basically, $N \to \infty$ in \eqref{eq:defrhos}, but this does not affect the reduced equations for the one-body density matrix). 
We emphasize that we use the notation $\vrho$ for a slightly different object from the one defined in equation~\eqref{eq:defrho}. $\vrho^{\nu_\alpha}_{\nu_\beta}(p,t)$ is a reduced part of the full $\vrho$, namely the diagonal values in helicity and momentum space. It nevertheless possesses a matrix structure in flavour space. However, the previous formalism must be applied with the \emph{full} density matrix,\footnote{In practice, this is hardly a problem: one must just remember to sum over all possible species.} thus one must also look at the charged lepton indices. Since only neutrinos mix between themselves (i.e., the flavour structure is strictly diagonal except in the neutrino-neutrino subspace of $\vrho$), we can distinguish two “blocks” in $\vrho$: the lepton part, purely diagonal
\[\vrho_\text{lep}(p,t) = \mathrm{diag}(f_{e^-}(p,t), f_{\mu^-}(p,t),0) \qquad \text{and} \qquad \bvrho_\text{lep} = \mathrm{diag}(f_{e^+}(p,t), f_{\mu^+}(p,t),0) \, , \]
and the neutrino part for which we will use the simplified notation $\vrho^\alpha_\beta \equiv \vrho^{\nu_\alpha}_{\nu_\beta}$. In the following, we will only use the distribution functions when it comes to leptons. Furthermore, and as expected from homogeneity and isotropy assumptions, all quantities are diagonal in momentum space, such that in the end, only the diagonal values $A(p)$ of operators $A^{\vp}_{\vpp} = A(p) \bm{\delta}_{\vp \vpp}$ will be dealt with, where the ‘‘Kronecker symbol'' in momentum space is $\bm{\delta}_{\vp \vpp} = (2 \pi)^3 \, 2 E_p \, \delta^{(3)}(\vp - \vpp)$. This will be made explicit in the upcoming calculations.

To determine the equation of evolution of the statistical ensemble of neutrinos, we now have to calculate the relevant expressions of the vacuum, the mean-field~\eqref{eq:Gamma} and collision~\eqref{eq:C11} terms.

\paragraph{Remark: Majorana and Dirac neutrinos} In the case of Dirac neutrinos, the discussion above can be directly applied: right-handed (RH) neutrinos and left-handed (LH) antineutrinos are not present in the early Universe~\cite{Shapiro1980,LesgourguesPastor,Dolgov_2002PhysRep}, such that $\vrho$ and $\bvrho$ correspond respectively to the LH part of the full neutrino density matrix in helicity space, and the RH part of full antineutrino density matrix in helicity space.

In the Majorana case, what we call antineutrinos are actually the right-handed neutrinos. More precisely, since the energy scale is much higher than the mass of neutrinos, helicity-flip processes are suppressed and one can match these definitions safely. The interactions of the neutrino field have the same form regardless of the mass mechanism, see for instance Eq.~(14.23) in~\cite{GiuntiKim}. Therefore, all the following calculations, including the mean-field and collision terms, can be done without considering the nature of neutrinos. Note that this would not hold in an anisotropic setup: taking into account \emph{spin coherence} effects shows differences between the Majorana and Dirac cases~\cite{Vlasenko_PhRevD2014,SerreauVolpe,Cirigliano2014,BlaschkeCirigliano}. 

\subsection{Vacuum term}
\label{subsec:vacuum} 

The neutrino kinetic term is, by definition, diagonal in the mass basis (the basis elements being the eigenstates of the vacuum Hamiltonian $\hat{H}_0$):
\begin{equation}
\left. t^{a}_{b}(p)\right|_{\text{mass basis}} = \sqrt{p^2 + m_a^2} \, \delta^a_b \simeq p \delta^a_b + \frac{m_a^2}{2p} \delta^a_b  \, .
\end{equation}
Since terms proportional to the identity do not contribute to flavour evolution (their commutator with $\vrho$ vanishes), the first term will later disappear from the evolution equation. In the flavour basis, the vacuum term is obtained following the transformation laws \eqref{eq:transfo}:
\begin{equation}
t^i_j = p \delta^i_j + \left( U \frac{\mathbb{M}^2}{2p} U^\dagger \right)^i_j \, ,
\end{equation}
with $\mathbb{M}^2$ the matrix of mass-squared differences and $U$ the Pontercorvo-Maki-Nakagawa-Sakata (PMNS) mixing matrix \cite{GiuntiKim,PDG}.

\subsection{Weak interaction matrix elements} 
\label{subsec:weak_matrix_el_QKE}

Neutrinos and antineutrinos in the early Universe interact with each others and with the charged leptons forming the homogeneous and isotropic plasma. The interaction Hamiltonian is thus given by the charged- and neutral-current terms from the standard model of weak interactions, expanded at low energies compared to the gauge boson masses. All expressions and subsequent interaction matrix elements~\eqref{eq:defvint} are gathered in the appendix~\ref{app:matrix_el_MF}, while we give here the details of the calculation for the charged-current processes $\nu_e - e^\pm$ so as to illustrate how the formalism works.

\subsubsection{Example: matrix elements for charged-currents with $e^\pm$}

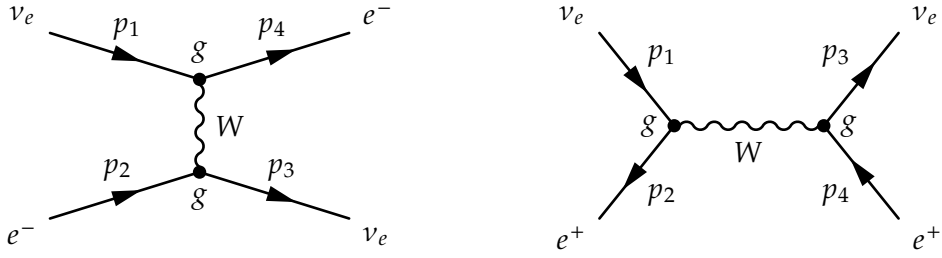
\begin{figure}[!ht]
	\vspace{0.3cm}
	\begin{fmffile}{CC-Feynman}
	\begin{equation*}
	\begin{gathered}
	\begin{fmfgraph*}(140,70)
		\fmfleft{i1,i2}
		\fmfright{o1,o2}
		\fmflabel{$e^-$}{i1}
		\fmflabel{$\nu_{e}$}{o1}
		\fmflabel{$\nu_{e}$}{i2}
		\fmflabel{$e^-$}{o2}
		\fmf{fermion,label=$p_2$,label.side=left}{i1,v1}
		\fmf{fermion,label=$p_3$,label.side=left}{v1,o1}
		\fmf{photon, label=$W$}{v1,v2}
		\fmf{fermion,label=$p_1$,label.side=left}{i2,v2}
		\fmf{fermion,label=$p_4$,label.side=left}{v2,o2}
		\fmfdot{v1,v2}
		\fmfv{label=$g$}{v1,v2}
	\end{fmfgraph*}
	\end{gathered}
	\qquad \qquad \qquad
		\begin{gathered}
	\begin{fmfgraph*}(140,70)
		\fmfleft{i1,i2}
		\fmfright{o1,o2}
		\fmflabel{$e^+$}{i1}
		\fmflabel{$e^+$}{o1}
		\fmflabel{$\nu_e$}{i2}
		\fmflabel{$\nu_e$}{o2}
		\fmf{fermion,label=$p_2$,label.side=left}{v1,i1}
		\fmf{fermion,label=$p_4$,label.side=left}{o1,v2}
		\fmf{photon, label=$W$}{v1,v2}
		\fmf{fermion,label=$p_1$,label.side=left}{i2,v1}
		\fmf{fermion,label=$p_3$,label.side=left}{v2,o2}
		\fmfdot{v1,v2}
		\fmfv{label=$g$}{v1,v2}
	\end{fmfgraph*}
	\end{gathered}
	\end{equation*}
\end{fmffile}
	\vspace{0.3cm}
	\caption[Charged-current processes]{\label{fig:Feynman-diagram-CC} Charged-current processes.}
\end{figure}

The part of the interaction Hamiltonian corresponding to charged-current processes with electrons and positrons is~\eqref{eq:hcc_app}, which we recall here (cf.~also equation (4.10) in~\cite{SiglRaffelt}):
\begin{multline}
\label{eq:hcc}
\hat{H}_{CC} = 2 \sqrt{2} G_F m_W^2 \int{\ddp{1} \ddp{2} \ddp{3} \ddp{4}} \ (2\pi)^3 \delta^{(3)}(\vp_1 + \vp_2 - \vp_3 - \vp_4) \\ \times  [\overline{\psi}_{\nu_e}(\vp_1)\gamma_\mu P_L\psi_e(\vp_4)] W^{\mu \nu}(\Delta) [\overline{\psi}_e(\vp_2) \gamma_\nu P_L \psi_{\nu_e}(\vp_3)] \, ,
\end{multline}
with $\psi(\vec{p}) = \sum_{h} \left[ \ha(\vec{p},h) u^h(\vec{p})+ \hbd(-\vec{p},h) v^h(-\vec{p}) \right]$ the Fourier transform of the quantum fields, $P_L = (1-\gamma^5)/2$ the left-handed projection operator, and the gauge boson propagator
\begin{equation}
\label{eq:propagator}
W^{\mu \nu}(\Delta) = \frac{\eta^{\mu \nu} - \frac{\Delta^\mu \Delta^\nu}{m_W^2}}{m_W^2 - \Delta^2} \simeq \frac{\eta^{\mu \nu}}{m_W^2} + \frac{1}{m_W^2}\left(\frac{\Delta^2 \eta^{\mu \nu}}{m_W^2} - \frac{\Delta^\mu \Delta^\nu}{m_W^2}\right) \, .
\end{equation}
The lowest order in the previous expansion is the Fermi four-fermion effective theory. The momentum transfer is $\Delta = p_1-p_4$ for a $t$-channel ($\nu_e-e^-$ scattering), and $\Delta= p_1 + p_2$ for the $s$-channel ($\nu_e-e^+$), see figure~\ref{fig:Feynman-diagram-CC}. More precisely, going from $t$-channel to $s$-channel corresponds to the change of variables $- p_2 \leftrightarrow p_4$ in the integral~\eqref{eq:hcc}, which will be important to keep track of the correct signs when computing the interaction matrix elements. Finally, note that we only discuss the interaction between $\nu_e - e^{\pm}$, but the exact same calculation can be carried out with $\nu_\mu - \mu^{\pm}$.

Our goal is to identify the $\tilde{v}$ coefficients by matching the expression~\eqref{eq:hcc} with the general definition~\eqref{eq:defHint}. Note that it is important to write the sums and integrals in the form~\eqref{eq:set_quantum_numbers} to ensure the proper identification.

To compute the coefficient $\tilde{v}^{\nu_e e}_{\nu_e e}$, we need to extract from~\eqref{eq:hcc} the terms involving $\hat{a}_e^{(\dagger)}$ and $\hat{a}_{\nu_e}^{(\dagger)}$. It reads:
\begin{multline}
\label{eq:hcc_matrixel}
\hat{H}_{CC} \supset 2 \sqrt{2} G_F m_W^2 \sum_{h_1, \cdots} \int{\ddp{1} \cdots} \ (2\pi)^3 \delta^{(3)}(\vp_1 + \vp_2 - \vp_3 - \vp_4) \times [\bar{u}_{\nu_e}^{h_1}(\vp_1) \gamma_\mu P_L u_e^{h_4}(\vp_4)] \\ \times  W^{\mu \nu}(p_4-p_1) [\bar{u}_e^{h_2}(\vp_2) \gamma_\nu P_L u_{\nu_e}^{h_3}(\vp_3)] \times \underbrace{\had_{\nu_e}(\vp_1,h_1)\ha_e(\vp_4,h_4)\had_e(\vp_2,h_2)\ha_{\nu_e}(\vp_3,h_3)}_{\displaystyle = - \had_{\nu_e}(1)\had_e(2) \ha_e(4) \ha_{\nu_e}(3)} \, .
\end{multline}
In anticommuting $\had_e$ and $\ha_e$ we were slightly careless regarding the delta-function that appears if $\vp_2 = \vp_4$ and $h_2 = h_4$, but it corresponds to disconnected parts in the Hamiltonian which only affect the ground-state energy. In other words, it disappears when prescribing the normal ordering of the Hamiltonian, which we have not mentioned here for brevity. Note that we explicitly replaced the gauge boson momentum $\Delta = p_4 - p_1$ since we look at the interaction between $\nu_e$ and $e^-$.

Finally, in order to fit with the general definition~\eqref{eq:set_quantum_numbers} so that we can directly read off the coefficients $\tilde{v}$ from its expression, the Hamiltonian~\eqref{eq:hcc_matrixel} must explicitly show a sum on the different species. This is a crucial step to get the numerical prefactor right: indeed, for now the coefficients appearing in the expression~\eqref{eq:hcc_matrixel} are not $\tilde{v}$, as it does not contain all the necessary (normal) orderings of the annihilation/creation operators, like for instance $\had_{\nu_e}(1)\had_e(2) \ha_{\nu_e}(3) \ha_{e}(4)$. In other words, as long as we do not properly antisymmetrize the ordering of $\ha, \had$ operators in the Hamiltonian, we would get coefficients like $\tilde{v}^{\nu_e(1) e(2)}_{\nu_e(3) e(4)} \neq 0$ but $\tilde{v}^{\nu_e(1) e(2)}_{e(4) \nu_e(3)} = 0$, which does not respect the antisymmetrization property~\eqref{eq:defvint}. We thus write (using $i$ instead of $(\vp_i,h_i)$ for brevity):
\begin{align*}
 \had_{\nu_e}(1)\had_e(2) \ha_e(4) \ha_{\nu_e}(3) = \frac14 \Big( &\had_{\nu_e}(1)\had_e(2) \ha_e(4) \ha_{\nu_e}(3) - \had_{\nu_e}(1)\had_e(2) \ha_{\nu_e}(3)\ha_e(4) \\
  + \, &\had_e(2) \had_{\nu_e}(1) \ha_{\nu_e}(3)\ha_e(4) - \had_e(2) \had_{\nu_e}(1)\ha_e(4) \ha_{\nu_e}(3)\Big) \, ,
 \end{align*}
Given the factor $1/4$ in the definition~\eqref{eq:defHint} we obtain
 \begin{multline}
\tilde{v}^{\nu_e(1) e(2)}_{\nu_e(3) e(4)} = - 2 \sqrt{2} G_F m_W^2 \, (2 \pi)^3 \delta^{(3)}(\vec{p}_1 + \vec{p}_2 - \vec{p}_3 - \vec{p}_4) \\
\times [\bar{u}_{\nu_e}^{h_1}(\vp_1) \gamma_\mu P_L u_e^{h_4}(\vp_4)] \ W^{\mu \nu}(p_4-p_1) \ [\bar{u}_e^{h_2}(\vp_2) \gamma_\nu P_L u_{\nu_e}^{h_3}(\vp_3)] \, .
\end{multline}
With the expression of the propagator~\eqref{eq:propagator}, we have three contributions to $\tilde{v}$.

\paragraph{Fermi order} Keeping only the lowest order in~\eqref{eq:propagator}, the exchange of a $W$ boson is reduced to a contact interaction. We can then perform a Fierz transformation~\cite{GiuntiKim,PeskinSchroeder} (it amounts to rewriting this charged-current process as a neutral-current one):
\begin{equation}
\label{eq:Fierz_CC}
[\bar{u}_{\nu_e}^{h_1} (\vec{p}_1) \gamma^\mu P_L u_{e}^{h_4} (\vec{p}_4)] \ [\bar{u}_{e}^{h_2} (\vec{p}_2)\gamma_\mu P_L ] = - [\bar{u}_{\nu_e}^{h_1} (\vec{p}_1) \gamma^\mu P_L u_{\nu_e}^{h_3} (\vec{p}_3)] \ [\bar{u}_{e}^{h_2} (\vec{p}_2)\gamma_\mu P_L u_{\nu_e}^{h_3} (\vec{p}_3)] \, .
\end{equation}
This leads us to the final expression of the weak interaction matrix element for the charged-current processes, at Fermi order:
\begin{multline}
\label{eq:vtildeCC_Fermiorder}
\tilde{v}^{\nu_e(1) e(2)}_{\nu_e(3) e(4)} \underset{\text{CC, Fermi}}{=} 2 \sqrt{2} G_F \, (2 \pi)^3 \delta^{(3)}(\vec{p}_1 + \vec{p}_2 - \vec{p}_3 - \vec{p}_4) \\
\times [\bar{u}_{\nu_e}^{h_1} (\vec{p}_1) \gamma^\mu P_L u_{\nu_e}^{h_3} (\vec{p}_3)] \ [\bar{u}_{e}^{h_2} (\vec{p}_2)\gamma_\mu P_L u_{e}^{h_4} (\vec{p}_4)] \, .
\end{multline}

\paragraph{First post-Fermi order} Note that we can perform a Fierz transformation with the part of the propagator $\propto \eta^{\mu \nu}$, but not with the last term. Therefore, we have:
\begin{multline}
\label{eq:vtildeCC_Delta2}
\tilde{v}^{\nu_e(1) e(2)}_{\nu_e(3) e(4)} \underset{\text{CC, $\Delta^2/m_W^2$}}{=} 2 \sqrt{2} G_F  \, (2 \pi)^3 \delta^{(3)}(\vec{p}_1 + \vec{p}_2 - \vec{p}_3 - \vec{p}_4) \\
\times [\bar{u}_{\nu_e}^{h_1} (\vec{p}_1) \gamma^\mu P_L u_{\nu_e}^{h_3} (\vec{p}_3)] \ \frac{(p_4 - p_1)^\nu (p_4-p_1)_\nu}{m_W^2} [\bar{u}_{e}^{h_2} (\vec{p}_2)\gamma_\mu P_L u_{e}^{h_4} (\vec{p}_4)] \, ,
\end{multline}
and
\begin{multline}
\label{eq:vtildeCC_DmuDnu}
\tilde{v}^{\nu_e(1) e(2)}_{\nu_e(3) e(4)} \underset{\text{CC, $-\Delta^\mu \Delta^\nu/m_W^2$}}{=} 2 \sqrt{2} G_F  \, (2 \pi)^3 \delta^{(3)}(\vec{p}_1 + \vec{p}_2 - \vec{p}_3 - \vec{p}_4) \\
\times [\bar{u}_{\nu_e}^{h_1} (\vec{p}_1) \gamma_\mu P_L u_{e}^{h_4} (\vec{p}_4)] \ \frac{(p_4 - p_1)^\mu (p_4 - p_1)^\nu}{m_W^2} [\bar{u}_{e}^{h_2} (\vec{p}_2)\gamma_\nu P_L u_{\nu_e}^{h_3} (\vec{p}_3)] \, .
\end{multline}

We chose to use Fierz identities to avoid remaining minus signs and to make the calculation of the mean-field potentials more straightforward (see next subsection).

\paragraph{Positron background} If we are interested in the interaction between electronic neutrinos and positrons, we extract the contributions in $\hat{H}_{CC}$ involving $\hb_e, \hbd_e$. It reads:\footnote{Recall the change of variables compared to~\eqref{eq:hcc_matrixel} corresponding to the crossing symmetry $p_2 \leftrightarrow - p_4$.}
\begin{multline}
\label{eq:hcc_matrixel_bar}
\hat{H}_{CC} \supset 2 \sqrt{2} G_F m_W^2 \sum_{h_1, \cdots} \int{\ddp{1} \cdots} \ (2\pi)^3 \delta^{(3)}(\vp_1 + \vp_2 - \vp_3 - \vp_4) \times [\bar{u}_{\nu_e}^{h_1}(\vp_1) \gamma_\mu P_L v_e^{h_2}(\vp_2)] \\ \times  W^{\mu \nu}(p_1+p_2) [\bar{v}_e^{h_4}(\vp_4) \gamma_\nu P_L u_{\nu_e}^{h_3}(\vp_3)] \times \had_{\nu_e}(\vp_1,h_1)\hbd_e(\vp_2,h_2)\hb_e(\vp_4,h_4)\ha_{\nu_e}(\vp_3,h_3) \, ,
\end{multline}
which has the opposite sign compared to~\eqref{eq:hcc_matrixel}, since there is no need to anticommute the antiparticle creation/annihilation operators which already appear “in the reference order”. Of course, the full antisymmetrization is still necessary to read the $\tilde{v}$ coefficients and get the prefactor right. For instance, at Fermi order, we get following the same steps as before (including a Fierz transformation):
\begin{multline}
\tilde{v}^{\nu_e(1) \bar{e}(2)}_{\nu_e(3) \bar{e}(4)} \underset{\text{CC, Fermi}}{=} - 2 \sqrt{2} G_F \, (2 \pi)^3 \delta^{(3)}(\vec{p}_1 + \vec{p}_2 - \vec{p}_3 - \vec{p}_4) \\
\times [\bar{u}_{\nu_e}^{h_1} (\vec{p}_1) \gamma^\mu P_L u_{\nu_e}^{h_3} (\vec{p}_3)] \ [\bar{v}_{e}^{h_4} (\vec{p}_4)\gamma_\mu P_L v_{e}^{h_2} (\vec{p}_2)] \, .
\end{multline}

\subsubsection{Set of matrix elements at Fermi order}

We show in table~\ref{Table:MatrixElements} the set of interaction matrix elements derived from the Hamiltonians~\ref{subsec:Hamiltonians}, at Fermi order. They are of special importance, since they give the first non-zero contribution to the collision term: in other words, we will only use these matrix elements to compute $\mathcal{C}$. Conversely, to compute the mean-field potentials at order $\Delta^2/m_{W,Z}^2$, one needs the matrix elements from the expansion of the propagator \eqref{eq:propagator}, which are obtained similarly (cf. the example detailed above) and not reproduced here for the sake of brevity.

\renewcommand{\arraystretch}{1.5}

\begin{table}[!h]
	\centering
	\begin{tabular}{|l|r|}
  	\hline 
 Interaction process &  $\tilde{v}^{12}_{34}/\left[\sqrt{2} G_F (2 \pi)^3 \delta^{(3)}(\vp_1+\vp_2-\vp_3-\vp_4)\right]$   \\
  \hline \hline
  $CC$ &   \\ \hline
  $\nu_e(1) e(2) \nu_e(3) e(4)$ & $2 \times [\bar{u}_{\nu_e}^{h_1} (\vec{p}_1) \gamma^\mu P_L u_{\nu_e}^{h_3} (\vec{p}_3)]  [\bar{u}_{e}^{h_2} (\vec{p}_2)\gamma_\mu P_L u_{e}^{h_4} (\vec{p}_4)]$  \\
    $\nu_e(1) \bar{e}(2) \nu_e(3) \bar{e}(4)$ & $- 2 \times [\bar{u}_{\nu_e}^{h_1} (\vec{p}_1) \gamma^\mu P_L u_{\nu_e}^{h_3} (\vec{p}_3)] [\bar{v}_{e}^{h_4} (\vec{p}_4)\gamma_\mu P_L v_{e}^{h_2} (\vec{p}_2)]$  \\
  $\nu_e(1) \bar{\nu}_e(2) e(3) \bar{e}(4)$ & $2 \times [\bar{u}_{\nu_e}^{h_1} (\vec{p}_1) \gamma^\mu P_L v_{\nu_e}^{h_2} (\vec{p}_2)]  [\bar{v}_{e}^{h_4} (\vec{p}_4)\gamma_\mu P_L u_{e}^{h_3} (\vec{p}_3)]$  \\ \hline \hline
  $NC, \text{matter}$ &   \\ \hline
  $\nu_e(1) e(2) \nu_e(3) e(4)$ & $2 \times [\bar{u}_{\nu_e}^{h_1} (\vec{p}_1) \gamma^\mu P_L u_{\nu_e}^{h_3} (\vec{p}_3)]  [\bar{u}_{e}^{h_2} (\vec{p}_2)\gamma_\mu (g_L P_L + g_R P_R) u_{e}^{h_4} (\vec{p}_4)]$  \\
    $\nu_e(1) \bar{e}(2) \nu_e(3) \bar{e}(4)$ & $- 2 \times [\bar{u}_{\nu_e}^{h_1} (\vec{p}_1) \gamma^\mu P_L u_{\nu_e}^{h_3} (\vec{p}_3)] [\bar{v}_{e}^{h_4} (\vec{p}_4)\gamma_\mu (g_L P_L + g_R P_R) v_{e}^{h_2} (\vec{p}_2)]$  \\
  $\nu_e(1) \bar{\nu}_e(2) e(3) \bar{e}(4)$ & $2 \times [\bar{u}_{\nu_e}^{h_1} (\vec{p}_1) \gamma^\mu P_L v_{\nu_e}^{h_2} (\vec{p}_2)]  [\bar{v}_{e}^{h_4} (\vec{p}_4)\gamma_\mu (g_L P_L + g_R P_R) u_{e}^{h_3} (\vec{p}_3)]$  \\ \hline  \hline
    $NC, \text{self-interactions}$ &   \\ \hline
  $\nu_\alpha(1) \nu_\beta(2) \nu_\alpha(3) \nu_\beta(4)$ & $(1+\delta_{\alpha \beta})
\times [\bar{u}_{\nu_\alpha}^{h_1} (\vec{p}_1) \gamma^\mu P_L u_{\nu_\alpha}^{h_3} (\vec{p}_3)]  [\bar{u}_{\nu_\beta}^{h_2} (\vec{p}_2)\gamma_\mu P_L u_{\nu_\beta}^{h_4} (\vec{p}_4)]$  \\
    $\nu_\alpha(1) \bnu_\beta(2) \nu_\alpha(3) \bnu_\beta(4)$ & $- (1 + \delta_{\alpha \beta}) \times [\bar{u}_{\nu_\alpha}^{h_1} (\vec{p}_1) \gamma^\mu P_L u_{\nu_\alpha}^{h_3} (\vec{p}_3)]  [\bar{v}_{\nu_\beta}^{h_4} (\vec{p}_4) \gamma_\mu P_L v_{\nu_\beta}^{h_2} (\vec{p}_2)]$  \\
  $\nu_\alpha(1) \bnu_\alpha(2) \nu_\beta(3) \bnu_\beta(4)$ & $(1+ \delta_{\alpha \beta}) \times [\bar{u}_{\nu_\alpha}^{h_1} (\vec{p}_1) \gamma^\mu P_L v_{\nu_\alpha}^{h_2} (\vec{p}_2)]   [\bar{v}_{\nu_\beta}^{h_4} (\vec{p}_4) \gamma_\mu P_L u_{\nu_\beta}^{h_3} (\vec{p}_3)]$  \\ \hline 
\end{tabular}
	\caption[Interaction matrix elements at Fermi order]{Interaction matrix elements at lowest order in the expansion of the gauge boson propagators (Fermi effective theory of weak interactions). We have not written the matrix elements with (anti)muons which are exactly similar to the ones with electrons/positrons with $\nu_e \leftrightarrow \nu_\mu$. The neutral-current couplings are $g_L = -1/2 + \sin^2{\theta_W}$ and $g_R = \sin^2{\theta_W}$, where $\sin^2{\theta_W} \simeq 0.231$ is the weak-mixing angle. These expressions are derived in the appendix~\ref{app:matrix_el_MF}.
	\label{Table:MatrixElements}}
\end{table}

At leading order, the charged-current processes have been written as neutral-current ones thanks to Fierz rearrangement identities (cf. example above). Therefore one can write the global expression\footnote{To be precise, we should call these coupling matrices $G_{L,(e)}$ and $G_{R,(e)}$, with similarly for interactions with muons $G_{L,(\mu)} = \mathrm{diag}(g_L,g_L+1,g_L)$ and $G_{R,(\mu)} = \mathrm{diag}(g_R,g_R,g_R)$. We omit this heavy notation since only interactions with $e^\pm$ are considered in the collision term.} for all interactions between $\nu_e$ and $e^-$:
\begin{multline}
\label{eq:vtilde_nue_full}
\tilde{v}^{\nu_\alpha(1) e(2)}_{\nu_\beta(3) e(4)} = 2 \sqrt{2} G_F \, (2 \pi)^3 \delta^{(3)}(\vec{p}_1 + \vec{p}_2 - \vec{p}_3 - \vec{p}_4) \\
\times [\bar{u}_{\nu_\alpha}^{h_1} (\vec{p}_1) \gamma^\mu P_L u_{\nu_\beta}^{h_3} (\vec{p}_3)] \ [\bar{u}_{e}^{h_2} (\vec{p}_2)\gamma_\mu (G_L^{\alpha \beta} P_L +  G_R^{\alpha \beta} P_R ) u_{e}^{h_4} (\vec{p}_4)] \, ,
\end{multline}
with, in the Standard Model,
\begin{equation}
\label{eq:couplingmatrices}
G_L = \mathrm{diag}(g_L+1,g_L,g_L) \quad , \quad G_R = \mathrm{diag}(g_R,g_R,g_R) \, .
\end{equation}
One can also introduce non-standard interactions which promote those couplings to non-diagonal matrices \cite{Relic2016_revisited}.

\subsection{Mean-field potential}

With the set of all relevant $\tilde{v}^{ik}_{jl}$, one can compute the mean-field potential from \eqref{eq:Gamma}. This procedure is outlined in \cite{Volpe_2013}, and we continue the example of the charged-current processes with electrons and positrons, leaving the other cases to the appendix~\ref{app:matrix_el_MF}.

\subsubsection{Example: mean-field due to charged-current interactions with $e^{\pm}$}

Following the definition~\eqref{eq:Gamma}, we have:
\begin{align*}
\Gamma^{\nu_e(\vp_1,h_1)}_{\nu_e(\vp_3,h_3)} &= \sum_{h_2,h_4} \int{\ddp{2} \ddp{4}} \tilde{v}^{\nu_e(1) e(2)}_{\nu_e(3) e(4)} \times \langle \had_e(\vp_2, h_2) \ha_e(\vp_4, h_4)\rangle \\
&= \sum_{h_2,h_4} \int{\ddp{2} \ddp{4}} \tilde{v}^{\nu_e(1) e(2)}_{\nu_e(3) e(4)} \times (2\pi)^3 \, 2E_{p_2} \, \delta^{(3)}(\vp_2 - \vp_4) \, \delta_{h_2 h_4} \, f_{e^-}(p_2) \, .
\end{align*}
Let us now calculate the potentials arising from the three contributions to $\tilde{v}$.

\paragraph{Fermi order} We use the matrix element~\eqref{eq:vtildeCC_Fermiorder}, and write $\vp = \vp_2 = \vp_4$ (which is enforced by the delta-function), such that
\begin{multline*}
\Gamma^{\nu_e(\vp_1,h_1)}_{\nu_e(\vp_3,h_3)} \underset{\text{CC, Fermi}}{=} 2 \sqrt{2} G_F \, (2 \pi)^3 \delta^{(3)}(\vec{p}_1 - \vec{p}_3) \sum_{h} \int{\ddp{}} [\bar{u}_{\nu_e}^{h_1} (\vec{p}_1) \gamma^\mu P_L u_{\nu_e}^{h_3} (\vec{p}_3)] \\
\times   [\bar{u}_{e}^{h} (\vec{p})\gamma_\mu P_L u_{e}^{h} (\vec{p})] \,  f_{e^-}(p) 
\end{multline*}
The spinor products can be simplified thanks to trace technology: 
\begin{align*}
\sum_{h} [\bar{u}_{e}^{h} (\vec{p})\gamma_\mu P_L u_{e}^{h} (\vec{p})] &= \sum_{h} [\bar{u}_{e,i}^{h} (\vec{p})\left(\gamma_\mu P_L\right)_{ij} u_{e,j}^{h} (\vec{p})] \\
&= \left(\sum_{h}{u_{e}^{h}(\vec{p}) \bar{u}_{e}^{h}(\vec{p})}\right)_{ji} \left(\gamma_\mu P_L \right)_{ij}
\end{align*}
Next, we use the spin sum~\cite{PeskinSchroeder} $\sum_{h}{u_e^h(\vec{p})\bar{u}_e^h(\vec{p})} = \slashed{p} + m_e$. Moreover, we are dealing with ultra-relativistic neutrinos which have only one possible helicity state ($-$), the useful formula being then the projection of the spin sum on spinors with definite helicity in the ultra-relativistic limit\footnote{See for instance equation~(38.31) in~\cite{Srednicki}, being careful that it uses the metric $(-,+,+,+)$ and the convention $\{\gamma^\mu,\gamma^\nu\} = - 2 \eta^{\mu \nu}$, hence the different sign for $\slashed{k}$.} $u_{\nu_e}^{(-)}(\vec{k}) \bar{u}_{\nu_e}^{(-)}(\vec{k}) = P_L \slashed{k}$. With this, we can rewrite
\begin{align*}
\sum_{h} [\bar{u}_{e}^{h} (\vec{p})\gamma_\mu P_L u_{e}^{h} (\vec{p})] = \tr \left[(\gamma_\nu p^\nu + m_e) \gamma_\mu P_L \right] &= 2 p_\mu \, , \\
\intertext{and}
[\bar{u}_{\nu_e}^{(-)} (\vec{p}_1) \gamma^\mu P_L u_{\nu_e}^{(-)} (\vec{p}_1)] = \tr [P_L \gamma^\nu \gamma^\mu P_L] p_{1, \nu} &= 2 p_1^\nu \, ,
\end{align*}
where we have used the identities recalled in equation~\eqref{eq:trace_identities}. All in all, we have

\begin{equation}
\label{eq:Gamma_CC_Fermi_temp}
\Gamma^{\nu_e(\vp_1,-)}_{\nu_e(\vp_3,-)}  \underset{\text{CC, Fermi}}{=} 2 \sqrt{2} G_F \, (2 \pi)^3 \delta^{(3)}(\vec{p}_1 - \vec{p}_3) \times 4 \times  \int{\frac{\dd^3 \vp}{(2 \pi)^3 \, 2 E_p} } \underbrace{(p_1 \cdot p)}_{p_1 E_p - \vp_1 \cdot \vp} f_{e^-}(p) \, .
\end{equation}
Because of isotropy, the integral $\int{\dd^3 \vp \cdots  (\vp_1 \cdot \vp)} = 0$,
\begin{equation}
\label{eq:meanfield_CC_Fermi}
\Gamma^{\nu_e(\vp_1,-)}_{\nu_e(\vp_3,-)}  \underset{\text{CC, Fermi}}{=} \sqrt{2} G_F \, (2 \pi)^3 \, 2 p_1 \, \delta^{(3)}(\vec{p}_1 - \vec{p}_3) \times 2   \int{\frac{\dd^3 \vp}{(2 \pi)^3} \, f_{e^-}(p)} = \sqrt{2} G_F n_{e^-} \, \bm{\delta}_{\vp_1 \vp_3} \, ,
\end{equation}
where we recognized the electron number density from~\eqref{eq:thermo_intro}. We find back the well-known mean-field potential~\eqref{eq:VCC_intro} responsible, for instance, for the MSW effect~\cite{MSW_W,MSW_MS} in stars.

\paragraph{Post-Fermi order, $\bm{\Delta^2/m_W^2}$ term} With the matrix element~\eqref{eq:vtildeCC_Delta2}, we can follow the same steps as before to get
\begin{equation*}
\Gamma^{\nu_e(\vp_1,-)}_{\nu_e(\vp_3,-)}  \underset{\text{CC, $\Delta^2/m_W^2$}}{=} 8 \sqrt{2} \frac{G_F}{m_W^2} \, (2 \pi)^3 \, \delta^{(3)}(\vec{p}_1 - \vec{p}_3) \times  \int{\frac{\dd^3 \vp}{(2 \pi)^3 \, 2 E_p} \, (p_1 \cdot p) \times (p_1-p)^2 \times f_{e^-}(p)} \, .
\end{equation*}
We have $(p_1 - p)^2 = m_e^2 - 2 (p_1 \cdot p)$. The first term is identical to the Fermi order calculation with an overall multiplication by $(m_e/m_W)^2$. To compute the second contribution we use spherical coordinates aligned with $\vec{p_1}$, such that $p_1 \cdot p = p_1 E_p - p_1 p \cos{\theta}$. Therefore we are left with the integral
\begin{equation*}
 - 2 p_1^2 \int{\frac{2 \pi p^2 \dd p}{(2 \pi)^3 \, 2 E_p}} \int_{0}^{\pi}{\sin{\theta} \dd \theta} \, (E_p - p \cos{\theta})^2  f_{e^-}(p) \, .
\end{equation*}
We need the three angular integrals:
\begin{equation*}
\int_{0}^{\pi}{\sin{\theta} \dd{\theta}} = 2 \ , \quad \int_{0}^{\pi}{\sin{\theta} \cos{\theta} \dd{\theta}} = 0 \ , \quad \int_{0}^{\pi}{\sin{\theta} \cos^2{\theta} \dd{\theta}} = \frac23 \, ,
\end{equation*}
with which we obtain, using the thermodynamic formulas~\eqref{eq:thermo_intro},
\begin{align*}
 - 2 p_1^2 \int{\frac{2 \pi p^2 \dd p}{(2 \pi)^3 \, 2 E_p}} \int_{0}^{\pi}{\sin{\theta} \dd \theta} \, (E_p - p \cos{\theta})^2  f_{e^-}(p) &= - p_1^2 \int{\frac{4 \pi p^2 \dd p}{(2 \pi)^3}} \left(E_p + \frac{p^2}{3 E_p}\right) f_{e^-}(p) \\
 &= - \frac12 p_1^2 \left(\rho_{e^-} + P_{e^-}\right) \, .
\end{align*}
Hence the second contribution to the mean-field potential 
\begin{equation}
\label{eq:meanfield_CC_Delta2}
\Gamma^{\nu_e(\vp_1,-)}_{\nu_e(\vp_3,-)}  \underset{\text{CC, $\Delta^2/m_W^2$}}{=} \sqrt{2} G_F n_{e^-} \left(\frac{m_e}{m_W}\right)^2  \, \bm{\delta}_{\vp_1 \vp_3} - \frac{2 \sqrt{2} G_F p_1}{m_W^2}(\rho_{e^-} + P_{e^-}) \, \bm{\delta}_{\vp_1 \vp_3} \, .
\end{equation}
The mean-field potentials up to first order in $\Delta^2/m_W^2$ do not usually take into account the non-relativistic nature of electrons and positrons \cite{SiglRaffelt,Mangano2005,Relic2016_revisited,Gariazzo_2019,Akita2020}. Instead, our expression involves both the energy density and the pressure of charged leptons, as mentioned for instance in \cite{NotzoldRaffelt_NuPhB1988}. As expected, we recover the more common expression in the ultra-relativistic limit $\rho_{e^-} + P_{e^-} \to (4/3) \rho_{e^-}$.

\paragraph{Post-Fermi order, $\bm{- \Delta^\mu \Delta^\nu/m_W^2}$ term} We finally include the matrix element~\eqref{eq:vtildeCC_DmuDnu} in the general expression of $\Gamma$. It reads
\begin{multline*}
\Gamma^{\nu_e(\vp_1,-)}_{\nu_e(\vp_3,-)} \underset{\text{CC, $-\Delta^\mu \Delta^\nu/m_W^2$}}{=} 2 \sqrt{2} \frac{G_F}{m_W^2} \, (2 \pi)^3 \delta^{(3)}(\vec{p}_1 - \vec{p}_3) \sum_{h} \int{\ddp{}} \, (p - p_1)_\mu (p - p_1)_\nu \\  \times [\bar{u}_{\nu_e}^{-} (\vec{p}_1) \gamma^\mu P_L u_{e}^{h} (\vec{p})] [\bar{u}_{e}^{h} (\vec{p})\gamma^\nu P_L u_{\nu_e}^{-} (\vec{p}_3)] \times  f_{e^-}(p) \, .
\end{multline*}
As before, we rewrite the spinor product (enforcing once again $\vp_1 = \vp_3$):
\begin{align*}
\sum_{h} [\bar{u}_{\nu_e}^{-} (\vec{p}_1) \gamma^\mu P_L u_{e}^{h} (\vec{p})] [\bar{u}_{e}^{h} (\vec{p})\gamma^\nu P_L u_{\nu_e}^{-} (\vec{p}_1)] &= \tr [ P_L \gamma^\sigma p_{1,\sigma} \gamma^\mu P_L (\gamma^\lambda p_\lambda + m_e) \gamma^\nu P_L ] \\
&= m_e p_{1, \sigma} \underbrace{\tr [\gamma^\sigma \gamma^\mu \gamma^\nu P_L]}_{=0} + \, p_{1, \sigma} p_\lambda \tr [\gamma^\sigma \gamma^\mu \gamma^\lambda \gamma^\nu P_L ] 
\end{align*}
With the trace identities~\eqref{eq:trace_identities}, we split the result into its imaginary and its real part.
\begin{itemize}
	\item The imaginary part of the mean-field contains the product:
	\[ \underbrace{(p-p_1)_{\mu} (p-p_1)_\nu}_{\text{sym. $\mu \leftrightarrow \nu$}} \ \ \times \underbrace{\epsilon^{\sigma \mu \lambda \nu}}_{\text{asym. $\mu \leftrightarrow \nu$}} = 0 \, . \]
	\item The real part reads after calculation:
	\begin{equation*}
	(p-p_1)_\mu (p-p_1)_\nu p_{1,\sigma} p_\lambda \times 2 \left(\eta^{\sigma \mu} \eta^{\lambda \nu} - \eta^{\sigma \lambda} \eta^{\mu \nu} + \eta^{\sigma \nu} \eta^{\mu \lambda} \right) = 2 m_e^2 (p_1 \cdot p) \, .
	\end{equation*}
\end{itemize}
Comparing with the expression~\eqref{eq:Gamma_CC_Fermi_temp} in the Fermi order calculation, we get
\begin{equation}
\label{eq:meanfield_CC_DmuDnu}
\Gamma^{\nu_e(\vp_1,-)}_{\nu_e(\vp_3,-)} \underset{\text{CC, $-\Delta^\mu \Delta^\nu/m_W^2$}}{=} \frac{G_F}{\sqrt{2}} n_{e^-} \left(\frac{m_e}{m_W}\right)^2 \bm{\delta}_{\vp_1 \vp_3} \, .
\end{equation}

\paragraph{Positron background} Up until now, we have only considered the mean-field due to the interaction with the bath of electrons. Without any additional calculation, we can obtain the potential due to the positron background, simply looking at the additional signs that appear along the derivation:
\begin{itemize}
	\item due to the minus sign coming from the anti-commutation of $\hb_e, \, \hbd_e$ (cf.~for instance the expression of the matrix element at Fermi order in table~\ref{Table:MatrixElements}), the Fermi order result for the $e^+$ background is exactly the opposite of the $e^-$ one, with $n_{e^-}$ replaced by $n_{e^+}$:
	\begin{equation*}
\Gamma^{\nu_e(\vp_1,-)}_{\nu_e(\vp_3,-)} \overset{[e^-]}{ \underset{\text{CC, Fermi}}{=}} - \sqrt{2} G_F n_{e^+} \, \bm{\delta}_{\vp_1 \vp_3} \, ;
\end{equation*}
	\item beyond Fermi order, the interaction being now an $s-$channel instead of a $t-$channel, $\Delta = (p_1 - p)$ is replaced by $(p_1 + p)$. This changes the sign inside $\Delta^2 = m_e^2 - 2 (p_1 \cdot p)$. Therefore the energy density and pressure contributions to the mean-field get an additional minus sign (and end up with the same sign whether it is an $e^-$ or $e^+$ background), but not the density terms.
\end{itemize}

\paragraph{Full CC mean-field} We can now gather all the previous contributions \eqref{eq:meanfield_CC_Fermi}, \eqref{eq:meanfield_CC_Delta2}, \eqref{eq:meanfield_CC_DmuDnu} (and the corresponding potentials due to the interactions with positrons), which leads to
\begin{equation}
\Gamma^{\nu_e(\vp_1, -)}_{\nu_e(\vp_3, -)} = \left\{ \sqrt{2} G_F (n_{e^-} - n_{e^+}) \left[1 + \mathcal{O}\left(\frac{m_e^2}{m_W^2}\right) \right] - \frac{2 \sqrt{2} G_F p_1}{m_W^2}(\rho_{e^-} + P_{e^-} + \rho_{e^+} + P_{e^+}) \right \} \bm{\delta}_{\vp_1 \vp_3} \, .
\end{equation}
The post-Fermi order contribution proportional to the charged lepton densities is negligible compared to the Fermi order one. Moreover, the asymmetry $(n_{e^-}-n_{e-+})/n_\gamma \simeq \eta \sim 6 \times 10^{-10}$ is of the order of the baryon-to-photon ratio, hence completely negligible compared to the (yet ”higher order") energy density/pressure term.

Since, as expected given the assumptions of homogeneity and isotropy, $\Gamma$ is diagonal in momentum space, we only deal from now on with its diagonal part $\Gamma(p)$ (such that $\Gamma^{\vp,-}_{\vpp,-} = \Gamma(p) \bm{\delta}_{\vp \vpp}$). Moreover, we restrict ourselves to the $(-,-)$ helicity subspace (otherwise we would just need to add some helicity Kronecker symbols).  Finally, as mentioned before, the exact same calculation can be made with $\nu_\mu - \mu^{\pm}$, which allows to display the full charged-current contribution, showing the flavour matrix structure:
\begin{multline}
\Gamma^{\nu_\alpha}_{\nu_\beta}(p) \underset{\text{CC}}{=} \sqrt{2} G_F (n_{e^-} - n_{e^+}) \delta^{\alpha}_{e} \delta^e_{\beta} - \frac{2 \sqrt{2} G_F p}{m_W^2} \left(\rho_{e^-} + P_{e^-} + \rho_{e^+} + P_{e^+}\right) \delta^{\alpha}_{e} \delta^e_{\beta} \\
+ \sqrt{2} G_F (n_{\mu^-} - n_{\mu^+}) \delta^{\alpha}_{\mu} \delta^\mu_{\beta} - \frac{2 \sqrt{2} G_F p}{m_W^2} \left(\rho_{\mu^-} + P_{\mu^-} + \rho_{\mu^+} + P_{\mu^+}\right) \delta^{\alpha}_{\mu} \delta^\mu_{\beta}  \, .
\end{multline}

\subsubsection{Complete mean-field expression}

In addition to the charged-current processes, one needs to take into account the neutral-current processes with the background leptons and with (anti)neutrinos (so-called self-interactions). Some elements of the calculation are outlined in the appendix~\ref{app:matrix_el_MF}, and lead to the following final expression:\footnote{The absence of extra complex conjugation on $\mathbb{N}_{\bnu}$, i.e.~on $\bvrho$, compared to \cite{Volpe_2013} is due to the transposed definition of the antineutrino density matrix~\eqref{eq:homogeneity}.}
\begin{multline}
\label{eq:Gamma_potential}
\Gamma^{\alpha}_{\beta} = \sqrt{2} G_F (\mathbb{N}_\mathrm{lep^-} - \mathbb{N}_\mathrm{lep^+})^\alpha_\beta + \sqrt{2} G_F \left(\mathbb{N}_\nu - \mathbb{N}_{\bnu}\right)^\alpha_\beta \\ - \frac{2 \sqrt{2} G_F p}{m_W^2}(\mathbb{E}_\mathrm{lep^-} + \mathbb{P}_\mathrm{lep^-} + \mathbb{E}_\mathrm{lep^+} + \mathbb{P}_\mathrm{lep^+})^\alpha_\beta- \frac{8 \sqrt{2} G_F p}{m_Z^2}\left(\mathbb{E}_\nu + \mathbb{E}_{\bnu} \right)^{\alpha}_{\beta} \, .
\end{multline}
The first two terms are the particle/antiparticle asymmetric mean-field potentials arising from the V$-$A Hamiltonian. There is no contribution from the neutral-current processes with the matter background as they are flavour-independent (see appendix~\ref{app:matrix_el_MF}). As shown in the example above, expanding the gauge boson propagators to next-to-leading order in the exchange momentum leads to the symmetric terms proportional to the neutrino momentum $p$. This expression is derived in the flavour basis in which $\delta^{\alpha}_{e}$ is the Kronecker symbol. However it can be directly read in any basis, through the contravariant (covariant) transformation of upper (lower) indices \eqref{eq:transfo}.

 The various thermodynamic quantities involved are, where we use the standard definitions~\eqref{eq:thermo_intro} for the charged leptons,
\begin{equation}
\begin{aligned}
\mathbb{N}_\mathrm{lep^-} &= \mathrm{diag}(n_{e^-} , n_{\mu^-}, 0) \\
\left. \mathbb{N}_\nu \right|^\alpha_\beta &= \int{\frac{\dd^3 \vp}{(2 \pi)^3} \vrho^\alpha_\beta(p)}
\end{aligned}  \qquad
\begin{aligned} \mathbb{E}_\mathrm{lep^-} &= \mathrm{diag}(\rho_{e^-} , \rho_{\mu^-}, 0) \, ,  \\
\left. \mathbb{E}_\nu \right|^\alpha_\beta &=  \int{\frac{\dd^3 \vp}{(2 \pi)^3} \, p \, \vrho^\alpha_\beta(p)} \, ,
\end{aligned}
\end{equation}
and the corresponding quantities for antiparticles are obtained by replacing $f_{e^-} \to f_{e^+}$ and $\vrho^\alpha_\beta \to \bvrho^\alpha_\beta$. We define the total charged lepton energy density matrix $\mathbb{E}_\mathrm{lep} \equiv \mathbb{E}_\mathrm{lep^-} + \mathbb{E}_\mathrm{lep^+}$, and likewise for the pressure. In the following, we will systematically neglect the very small asymmetry of electrons/positrons, which is constrained to be of the order of the baryon-to-photon ratio $\eta \sim 10^{-9}$. Since electrons and positrons undergo very efficient electromagnetic interactions with the photon background, ensuring that their distribution function remains a Fermi-Dirac one at the photon temperature $T_\gamma$ \cite{Grohs2019}., we will use as electron distribution function
\begin{equation}
\label{eq:distrib_electrons}
f_{e^-}(p) = f_{e^+}(p) = \dfrac{1}{e^{\sqrt{p^2 + m_e^2}/{T_\gamma}}+1} \equiv f_e(p) \, .
\end{equation}
Note that the total neutrino energy density, that is the sum of the diagonal contribution of $\mathbb{E}_\nu$, reads:
\begin{equation}
	\label{eq:rhonu_matrix}
	\rho_\nu = \Tr(\mathbb{E}_\nu) \, .
\end{equation}

\subsection{Collision integral}
\label{subsec:collision_integral}

The remaining part of the QKE is the collision term, which is derived by inserting all possible matrix elements in the general expression \eqref{eq:C11}.
This leads to collision integrals previously derived in \cite{SiglRaffelt,BlaschkeCirigliano}, and progressively included in numerical computations, except for the self-interactions, whose off-diagonal components were approximated by damping terms or discarded \cite{Mangano2005,Gava:2010kz,Gava_corr,Relic2016_revisited,Gariazzo_2019}. In the following, we illustrate how our formalism applies by carrying out an explicit derivation for neutrino-neutrino scattering, displaying the full matrix structure of the statistical factor. The other contributions to the collision term are discussed in the appendix~\ref{app:collision_term}.

\subsubsection{Neutrino self-interactions collision term}

As an illustration of the use of the BBGKY formalism to derive the collision integrals, we detail the steps to obtain the neutrino-neutrino scattering contribution to the expression to come~\eqref{eq:C_nn}.

Neutrino-neutrino scattering processes correspond to the terms in \eqref{eq:C11} for which the inner matrix elements are scattering ones $\tilde{v}^{\nu_\delta \nu_\sigma}_{\nu_\delta \nu_\sigma}$. For simplicity, we focus here on the first term in the expression of $\mathcal{C}^{i_1}_{i_1'}$ \eqref{eq:C11}. Here, the index $i_1$ will refer to $\nu_\alpha(\vp_1)$ and $i_1'$ to $\nu_\beta(\vp_{\underline{1}})$. We do not specify the helicity, which is necessarily $h=(-)$ for ultra-relativistic neutrinos. Finally, we impose $\vp_k = \vpp_k$ for all $k$, which is enforced by the assumption of homogeneity~\eqref{eq:homogeneity}. There are two non-zero contributions to this part of the collision matrix. 
\begin{itemize}
	\item when 1 and 3 have the same flavour, that is for the following term:
	\[ \frac14 \tilde{v}^{\nu_\alpha(1) \nu_\gamma(2)}_{\nu_\alpha(3) \nu_\gamma(4)}\times \varrho^{\alpha(3)}_{\delta(3)} \varrho^{\gamma(4)}_{\sigma(4)} \times \tilde{v}^{\nu_\delta(3') \nu_\sigma(4')}_{\nu_\delta(1') \nu_\sigma(2')} \times (\Id- \varrho)^{\delta(1)}_{\beta(1)} (\Id- \vrho)^{\sigma(2)}_{\gamma(2)} \, . \]
	The scattering amplitude is then
\begin{align}
&\tilde{v}^{\nu_\alpha(1) \nu_\gamma(2)}_{\nu_\alpha(3) \nu_\gamma(4)}\times \tilde{v}^{\nu_\delta(3') \nu_\sigma(4')}_{\nu_\delta(1') \nu_\sigma(2')} \nonumber \\ 
&\quad = 2 G_F^2 \times (2 \pi)^6 \delta^{(3)}(\vp_1 + \vp_2 - \vp_3 - \vp_4) \delta^{(3)}(\vp_1 - \vp_{\underline{1}})  \nonumber \\
&\quad \quad \times [\bar{u}_{\nu_\alpha}(1) \gamma^\mu P_L u_{\nu_\alpha}(3)][\bar{u}_{\nu_\delta}(3) \gamma^\nu P_L u_{\nu_\delta}(1)] \times  [\bar{u}_{\nu_\gamma}(2) \gamma_\mu P_L u_{\nu_\gamma}(4)][\bar{u}_{\nu_\sigma}(4) \gamma_\nu P_L u_{\nu_\sigma}(2)]   \nonumber\\ 
&\quad = 2 G_F^2  \times (2 \pi)^6 \delta^{(3)}(\vp_1 + \vp_2 - \vp_3 - \vp_4) \delta^{(3)}(\vp_1 - \vp_{\underline{1}})   \nonumber\\
&\quad \quad \times p_{3 \eta} p_{1 \rho} \tr[\gamma^\rho \gamma^\mu P_L \gamma^\eta \gamma^\nu P_L] \times p_4^\lambda p_2^\tau \tr[\gamma_\tau \gamma_\mu P_L \gamma_\lambda \gamma_\nu P_L]  \nonumber \\
&\quad = 2^5 G_F^2  \times (2 \pi)^6 \delta^{(3)}(\vp_1 + \vp_2 - \vp_3 - \vp_4) \delta^{(3)}(\vp_1 - \vp_{\underline{1}}) \times (p_1 \cdot p_2) (p_3 \cdot p_4) \, , \label{eq:scatt_ampl_13}
\end{align}
while the matrix product associated to this scattering amplitude is
\[\varrho^{\alpha(3)}_{\delta(3)} \varrho^{\gamma(4)}_{\sigma(4)} (\Id- \varrho)^{\delta(1)}_{\beta(1)} (\Id- \vrho)^{\sigma(2)}_{\gamma(2)} = \Big[ \Tr[ \vrho_4 \cdot (\Id-\vrho_2) ] \cdot \vrho_3 \cdot  (\Id-\vrho_1) \Big]^\alpha_\beta \, . \]
	\item when 1 and 4 have the same flavour, the scattering amplitude is
\begin{align}
&\tilde{v}^{\nu_\alpha(1) \nu_\gamma(2)}_{\nu_\gamma(3) \nu_\alpha(4)} \times \tilde{v}^{\nu_\delta(3') \nu_\sigma(4')}_{\nu_\delta(1') \nu_\sigma(2')}  \nonumber \\ &\quad = - 2 G_F^2 \times  (2 \pi)^6 \delta^{(3)}(\vp_1 + \vp_2 - \vp_3 - \vp_4) \delta^{(3)}(\vp_1 - \vp_{\underline{1}})  \nonumber \\
&\quad \quad \times [\bar{u}_{\nu_\alpha}(1) \gamma^\mu P_L u_{\nu_\alpha}(4)][\bar{u}_{\nu_\sigma}(4) \gamma^\nu P_L u_{\nu_\sigma}(2)] [\bar{u}_{\nu_\gamma}(2) \gamma_\mu P_L u_{\nu_\gamma}(3)][\bar{u}_{\nu_\delta}(3) \gamma_\nu P_L u_{\nu_\delta}(1)] \nonumber \\ 
&\quad = - 2 G_F^2  \times (2 \pi)^6 \delta^{(3)}(\vp_1 + \vp_2 - \vp_3 - \vp_4) \delta^{(3)}(\vp_1 - \vp_{\underline{1}}) \nonumber \\
&\quad \quad  \times p_{3 \lambda} p_{1 \rho} p_{4 \eta} p_{2 \tau} \tr[\gamma^\mu P_L \gamma^\eta \gamma^\nu P_L \gamma^\tau \gamma_\mu P_L \gamma^\lambda \gamma_\nu P_L \gamma^\rho] \nonumber \\
&\quad = 2^5 G_F^2  \times (2 \pi)^6 \delta^{(3)}(\vp_1 + \vp_2 - \vp_3 - \vp_4) \delta^{(3)}(\vp_1 - \vp_{\underline{1}}) \times (p_1 \cdot p_2) (p_3 \cdot p_4) \, , \label{eq:scatt_ampl_14}
\end{align}
and the matrix product reads
\[\varrho^{\gamma(3)}_{\delta(3)} \varrho^{\alpha(4)}_{\sigma(4)} (\Id- \varrho)^{\delta(1)}_{\beta(1)} (\Id- \vrho)^{\sigma(2)}_{\gamma(2)} = \Big[ \vrho_4 \cdot (\Id-\vrho_2) \cdot \vrho_3 \cdot  (\Id-\vrho_1) \Big]^\alpha_\beta \, . \]
\end{itemize}
We chose the compact notation $\vrho_k \equiv \vrho(p_k)$ for brevity, and used $\vrho_1 = \vrho_{\underline 1}$ thanks to the momentum-conserving function $\delta^{(3)}(\vp_1 - \vp_{\underline{1}})$.

The scattering amplitudes of the four terms in~\eqref{eq:C11} are identical, and their matrix products arrange such that the final result has the expected “gain $-$ loss $+$ h.c.” structure. Note that we considered here a particular ordering of the indices, while the full expression is symmetric through the exchange of the indices $(1',2')$. In other words, one must take twice the previous results~\eqref{eq:scatt_ampl_13} and \eqref{eq:scatt_ampl_14} to account for all non-zero combinations.\footnote{This symmetry vanishes if $\delta$ and $\sigma$ represent the same flavour. However, this is exactly compensated by the extra factor of $2$ in the matrix elements for identical flavour, cf.~table~\ref{Table:MatrixElements}. This extra factor of $2$ is already accounted for regarding the couple $(\alpha, \gamma)$ as it allows to treat the case $\alpha=\gamma$ like the others (i.e.~separating a “trace“ and a “non-trace“ contributions).} Therefore,
\begin{equation}
\label{eq:C_nnscatt}
\begin{aligned}
\mathcal{C}^{[\nu\nu \leftrightarrow \nu \nu]} = &(2 \pi)^3 \delta^{(3)}(\vec{p}_1-\vec{p}_{\underline{1}}) \frac{2^5 G_F^2}{2}\int{[\dd^3 \vec{p}_2] [\dd^3 \vec{p}_3] [\dd^3 \vec{p}_4] (2 \pi)^4 \delta^{(4)}(p_1 + p_2 - p_3 - p_4)} \\
&\times (p_1 \cdot p_2)(p_3 \cdot p_4) \times F_\mathrm{sc}(\nu^{(1)},\nu^{(2)},\nu^{(3)},\nu^{(4)})
\end{aligned}
\end{equation}
with the statistical factor:
\begin{multline}
F_\mathrm{sc}(\nu^{(1)},\nu^{(2)},\nu^{(3)},\nu^{(4)}) =  \left[ \varrho_4 (\Id - \varrho_2) + \Tr(\cdots) \right] \varrho_3 (\Id -\varrho_1) + (\Id - \varrho_1) \varrho_3 \left[ (\Id - \varrho_2) \varrho_4 + \Tr(\cdots)\right]  \\
- \left[ (\Id - \varrho_4) \varrho_2  + \Tr(\cdots)\right] (\Id -\varrho_3)  \varrho_1 - \varrho_1  (\Id -\varrho_3)  \left[\varrho_2(\Id -\varrho_4)  + \Tr(\cdots)\right]  \, ,
\end{multline}
where $\Tr(\cdots)$ means the trace of the term in front of it.

A useful check consists in neglecting flavour mixing, i.e., assuming that the neutrino density matrices are diagonal, the diagonal entries being the distribution functions. In this case, we have $\vrho_4 (\Id - \vrho_2) + \Tr(\cdots) \to 4 f_4 (1-f_2)$, hence a total amplitude for neutrino-neutrino scatterings $2^5 G_F^2 \times 4 \times (p_1 \cdot p_2)(p_3 \cdot p_4) = 2^7 G_F^2 (p_1 \cdot p_2)(p_3 \cdot p_4)$. This is in agreement with the results quoted in~\cite{Grohs2015} (Table I), \cite{Dolgov_NuPhB1997} (Tables 1 and 2) or \cite{FidlerPitrou}.

\subsubsection{Final form of the QKE}
The full expression of the collision integral is derived in the appendix~\ref{app:collision_term}. As for the mean-field term, it is diagonal in momentum space: $\mathcal{C}^{\vp_1}_{\vp_{\underline{1}}}$ is proportional to $\bm{\delta}_{\vp_1 \vp_{\underline{1}}}$. Therefore, all the terms in the QKE are momentum-diagonal, and the actual equation is obtained by removing the momentum-conserving delta-functions. Notably, we will denote as the collision integral the quantity $\mathcal{I}$, related to $\mathcal{C}$ via $\mathcal{C}^{\vp_1}_{\vp_{\underline{1}}} = (2 \pi)^3 \, 2 E_1\, \delta^{(3)}(\vp_1 - \vp_{\underline{1}}) \mathcal{I}[\vrho]$, and the QKE reads:
\begin{equation}
 \boxed{\ii \, \frac{\dd \vrho(p)}{\dd t} = \left[t + \Gamma, \vrho \right] + \ii \, \mathcal{I}} \, . 
 \end{equation}

\section{QKEs for neutrinos in the early Universe}

In this last section, we gather the previous elements of the QKE to introduce the precise equations that will be numerically solved in various cases.

\subsection{Set of Quantum Kinetic Equations}

We present here the QKE for $\vrho(p,t)$, obtained from \eqref{eq:eqvrho} after dividing each term by the momentum-conserving function $\bm{\delta}_{\vp \vpp}$ from \eqref{eq:homogeneity}. Moreover, the time derivative $\dd/\dd t$ becomes $\partial/\partial t - H p \, \partial/\partial p$ to account for the expansion of the Universe, $H\equiv \dot a/a$ being the Hubble rate, given by Friedmann's equation $H^2 = (8 \pi \mathcal{G}/3) \rho$. The QKEs read: 
\begin{multline}
\ii \left[ \frac{\partial}{\partial t} - H p \frac{\partial}{\partial p}\right] \vrho = \Big[ U \frac{\mathbb{M}^2}{2p}U^\dagger, \varrho \Big] + \sqrt{2} G_F \Big[ \mathbb{N}_\nu - \mathbb{N}_{\bnu}, \varrho \Big] \\
- 2 \sqrt{2} G_F p \Big[ \frac{\mathbb{E}_\text{lep} + \mathbb{P}_\text{lep}}{m_W^2} + \frac43 \frac{\mathbb{E}_{\nu} + \mathbb{E}_{\bnu}}{m_Z^2},\varrho \Big ] + \ii \mathcal{I} \label{eq:QKE_rho}
\end{multline}
where we recall the definitions in flavour space $\mathbb{E}_\text{lep} \equiv \mathrm{diag}(\rho_{e^-} + \rho_{e^+},\rho_{\mu^-} + \rho_{\mu^+},0)$ and likewise for $\mathbb{P}_\text{lep}$. Similarly, the QKEs for the antineutrino density matrix read (cf.~appendix~\ref{app:antiparticles}):
\begin{multline}
\ii \left[ \frac{\partial}{\partial t} - H p \frac{\partial}{\partial p}\right] \bvrho = - \Big[ U \frac{\mathbb{M}^2}{2p}U^\dagger, \bvrho \Big] + \sqrt{2} G_F \Big[ \mathbb{N}_\nu - \mathbb{N}_{\bnu}, \bvrho \Big] \\
+ 2 \sqrt{2} G_F p \Big[ \frac{\mathbb{E}_\text{lep} + \mathbb{P}_\text{lep}}{m_W^2} + \frac43 \frac{\mathbb{E}_{\nu} + \mathbb{E}_{\bnu}}{m_Z^2},\bvrho \Big ] + \ii \bar{\mathcal{I}} \label{eq:QKE_rhobar}
\end{multline}

The collision term is the sum of the contributions from different physical processes: scattering with charged leptons ($\nu e^\pm\leftrightarrow \nu e^\pm$), annihilation ($\nu \bnu \leftrightarrow e^+ e^-$) and self-interactions (involving only $\nu$ and $\bnu$). Note that in the collision integral, we do not take into account the interactions with muons whose number density is negligible in the range of temperatures of interest. The expressions for the processes involving charged leptons are exactly the same as the ones quoted in \cite{Relic2016_revisited} [eqs.~(2.4)--(2.10)], and we do not report them here for brevity. This reference, however, does not contain the full expressions for neutrino self-interactions, of which we derived explicitly a part in the section~\ref{subsec:collision_integral}. Our complete expression for the self-interactions contribution to the collision integral reads:\footnote{It is equivalent to eq.~(96) of~\cite{BlaschkeCirigliano} (one only needs to swap the variables $\vp_3 \leftrightarrow \vp_4$ in the second and fourth terms of \eqref{eq:F_sc_nn}). Our expression highlights the ‘‘gain $-$ loss $+$ h.c.'' structure of this collision term.}
\begin{equation}
\label{eq:C_nn}
\begin{aligned}
\mathcal{I}^{[\nu  \nu]} = &\frac12 \frac{2^5 G_F^2}{2 p_1} \int{[\dd^3 \vec{p}_2] [\dd^3 \vec{p}_3] [\dd^3 \vec{p}_4] (2 \pi)^4 \delta^{(4)}(p_1 + p_2 - p_3 - p_4)} \\
&\Big[ (p_1 \cdot p_2)(p_3 \cdot p_4) F_\mathrm{sc}(\nu^{(1)},\nu^{(2)},\nu^{(3)},\nu^{(4)})  \\
&+ (p_1 \cdot p_4)(p_2 \cdot p_3) \left( F_\mathrm{sc}(\nu^{(1)},\bar{\nu}^{(2)},\nu^{(3)},\bar{\nu}^{(4)}) + F_\mathrm{ann}(\nu^{(1)},\bar{\nu}^{(2)},\nu^{(3)},\bar{\nu}^{(4)}) \right) \Big] \, ,
\end{aligned} 
\end{equation}
with the statistical factors for scattering and annihilation processes:
\begin{multline}
\label{eq:F_sc_nn}
F_\mathrm{sc}(\nu^{(1)},\nu^{(2)},\nu^{(3)},\nu^{(4)}) =  \left[ \varrho_4 (\Id - \varrho_2) + \Tr(\cdots) \right] \varrho_3 (\Id -\varrho_1) + (\Id - \varrho_1) \varrho_3 \left[ (\Id - \varrho_2) \varrho_4 + \Tr(\cdots)\right]  \\
- \left[ (\Id - \varrho_4) \varrho_2  + \Tr(\cdots)\right] (\Id -\varrho_3)  \varrho_1 - \varrho_1  (\Id -\varrho_3)  \left[\varrho_2(\Id -\varrho_4)  + \Tr(\cdots)\right]  \, ,
\end{multline}
\begin{multline}
\label{eq:F_sc_nbn}
F_\mathrm{sc}(\nu^{(1)},\bar{\nu}^{(2)},\nu^{(3)},\bar{\nu}^{(4)}) = \left[ (\Id - \bar{\varrho}_2) \bar{\varrho}_4 + \Tr(\cdots) \right] \varrho_3 (\Id -\varrho_1) + (\Id - \varrho_1) \varrho_3 \left[ \bar{\varrho}_4 (\Id - \bar{\varrho}_2) + \Tr(\cdots)\right]  \\
- \left[ \bar{\varrho}_2 (\Id -\bar{\varrho}_4) + \Tr(\cdots) \right] (\Id -\varrho_3) \varrho_1 - \varrho_1 (\Id -\varrho_3) \left[ (\Id -\bar{\varrho}_4) \bar{\varrho}_2 + \Tr(\cdots)\right] \, ,
\end{multline}
\begin{multline}
\label{eq:F_ann_nn}
F_\mathrm{ann}(\nu^{(1)},\bar{\nu}^{(2)},\nu^{(3)},\bar{\nu}^{(4)}) = \left[ \varrho_3 \bar{\varrho}_4 + \Tr(\cdots) \right] (\Id -\bar{\varrho}_2) (\Id -\varrho_1) + (\Id - \varrho_1) (\Id -\bar{\varrho}_2) \left[ \bar{\varrho}_4 \varrho_3 + \Tr(\cdots)\right]  \\
- \left[ (\Id -\varrho_3) (\Id -\bar{\varrho}_4) + \Tr(\cdots) \right] \bar{\varrho}_2 \varrho_1 - \varrho_1 \bar{\varrho}_2 \left[ (\Id -\bar{\varrho}_4) (\Id -\varrho_3) + \Tr(\cdots)\right] \, ,
\end{multline}
where, as before, we chose the more compact notation $\vrho_k = \vrho(p_k)$, and $\Tr(\cdots)$ means the trace of the term in front of it.

\subsection{Reduced equations}
\label{subsec:reduced_equations_QKE}

In this final section, we transform the QKE~\eqref{eq:QKE_rho} and in particular the collision integral into a form suitable for numerical resolution, which will be the topic of the next chapter.

\paragraph{Reduction of the collision integral} The most time consuming part of the QKE is the computation of the collision term. Thanks to the homogeneity and isotropy of the early Universe, and the particular form of the scattering amplitudes, the nine-dimensional collision integrals can be reduced to two-dimensional ones \cite{Hannestad_PhRvD1995,Semikoz_Tkachev,Dolgov_NuPhB1997,Grohs2015}.  We follow here the reduction method of~\cite{Dolgov_NuPhB1997}. The first idea is to use the integral representation of the delta function:
\[ \delta^{(3)}(\vp_1 + \vp_2 - \vp_3 - \vp_4) = \int{\frac{\dd^3 \vec{\lambda}}{(2 \pi)^3} e^{\ii \vec{\lambda} \cdot (\vp_1 + \vp_2 - \vp_3 - \vp_4)}} \, , \]
and use spherical coordinates defined as follows: the “$\vec{e}_\z$ unit vector” for $\vec{\lambda}$ is aligned with $\vp_1$, while $\vec{\lambda}$ is the “$\vec{e}_\z$ unit vector” for $\vp_{i \geq 2}$, that is,
\[ \cos{\theta_\lambda} \equiv \frac{\vp_1 \cdot \vec{\lambda}}{p_1 \lambda} \qquad ; \qquad \cos{\theta_i} \equiv \frac{\vp_i \cdot \vec{\lambda}}{p_i \lambda} \ \ \text{for } i = 2,3,4 \, ,  \]
the associated azimuthal angles $\varphi_\lambda, \varphi_{i \geq 2}$ being defined as usual. Then, thanks to the very simple form of the scattering amplitudes in the four-fermion approximation  --- cf.~for instance~\eqref{eq:scatt_ampl_13} and~\eqref{eq:scatt_ampl_14} ---, we can perform all the $\varphi$ and $\theta$ integrations. The next step consists in performing the integration on $\lambda$, whose result can be analytically calculated~\cite{Dolgov_NuPhB1997,BlaschkeCirigliano} (these are the so-called “$D-$functions” of Dolgov, Hansen and Semikoz~\cite{Dolgov_NuPhB1997}, which are piecewise quadrivariate polynomials). Finally, we are left with a three-dimensional integral on the momentum moduli $p_2, p_3, p_4$, and one of them is done integrating out the energy delta function $\delta(E_1 + E_2 - E_3 - E_4)$, namely $\int{p_4 \dd p_4 \, \delta(E_1+E_2-E_3-E_4)} = E_1+E_2-E_3$, since $p_4 \dd p_4 = E_4 \dd E_4$. This procedure is outlined in the Appendix~\ref{app:reduc_collision_integral}.

\paragraph{Equation in comoving variables}  In view of a numerical implementation, let us use the comoving variables introduced in~\eqref{eq:comoving_variables}.

Therefore, the QKEs are rewritten:
\begin{multline}
\frac{\partial \vrho(x,y_1)}{\partial x} = - \frac{\ii}{xH} \left(\frac{x}{m_e}\right) \left[ U \frac{\mathbb{M}^2}{2y_1}U^\dagger, \varrho \right]  +  \ii \frac{2 \sqrt{2} G_F}{xH} y_1 \left(\frac{m_e}{x}\right)^5 \left[ \frac{\bar{\mathbb{E}}_\mathrm{lep} + \bar{\mathbb{P}}_\mathrm{lep}}{m_W^2} ,\varrho \right ] \\
- \ii \frac{\sqrt{2} G_F}{x H} \left(\frac{m_e}{x}\right)^3 \left[ \overline{\mathbb{N}}_\nu - \overline{\mathbb{N}}_{\bnu}, \vrho \right] + \ii \frac{8 \sqrt{2} G_F}{3 x H} y_1 \left( \frac{m_e}{x}\right)^5 \left[\frac{\bar{\mathbb{E}}_\nu + \bar{\mathbb{E}}_{\bar{\nu}}}{m_Z^2}, \vrho\right]  + \frac{1}{xH} \mathcal{I} \, , \label{eq:QKE_fullfinal}
\end{multline}
with the two-dimensional collision integral (recall that we assume $f_e = f_{\bar{e}}$, which regroups some terms):
\begin{equation}
\label{eq:I_full}
\begin{aligned}
\mathcal{I} = &\frac{G_F^2}{2 \pi^3 y_1} \left(\frac{m_e}{x}\right)^5 \int{y_2 \dd y_2 \, y_3 \dd y_3 \, \bar{E}_4 \times \frac12} \\
\times &\Big[ 4 \left[ 2 d_1 + 2 d_3 + d_2(1,2) + d_2(3,4) - d_2(1,4) - d_2(2,3) \right] \\
&\qquad \qquad  \times \left(F_\mathrm{sc}^{LL}(\nu^{(1)},e^{(2)},\nu^{(3)},e^{(4)}) + F_\mathrm{sc}^{RR}(\nu^{(1)},e^{(2)},\nu^{(3)},e^{(4)})\right) \\
&- 4 x^2 \left[d_1 - d_2(1,3) \right]/\bar{E}_2 \bar{E}_4 \times \left(F_\mathrm{sc}^{LR}(\nu^{(1)},e^{(2)},\nu^{(3)},e^{(4)}) + F_\mathrm{sc}^{RL}(\nu^{(1)},e^{(2)},\nu^{(3)},e^{(4)}) \right) \\
+ \, &4 \left[ d_1 + d_3 - d_2(1,4) - d_2(2,3) \right] \times  \left( F_\mathrm{ann}^{LL}(\nu^{(1)},\bar{\nu}^{(2)},e^{(3)},e^{(4)}) + F_\mathrm{ann}^{RR}(\nu^{(1)},\bar{\nu}^{(2)},e^{(3)},e^{(4)}) \right)  \\
&+ 2 x^2\left[d_1 + d_2(1,2) \right]/\bar{E}_3 \bar{E}_4 \times \left(F_\mathrm{ann}^{LR}(\nu^{(1)},\bar{\nu}^{(2)},e^{(3)},e^{(4)}) + F_\mathrm{ann}^{RL}(\nu^{(1)},\bar{\nu}^{(2)},e^{(3)},e^{(4)}) \right) \\
+ \, &  \left[ d_1 + d_3 + d_2(1,2) + d_2(3,4) \right] \times F_\mathrm{sc}(\nu^{(1)},\nu^{(2)},\nu^{(3)},\nu^{(4)})  \\
&+ \left[ d_1 + d_3 - d_2(1,4) - d_2(2,3) \right] \times \left( F_\mathrm{sc}(\nu^{(1)},\bar{\nu}^{(2)},\nu^{(3)},\bar{\nu}^{(4)}) + F_\mathrm{ann}(\nu^{(1)},\bar{\nu}^{(2)},\nu^{(3)},\bar{\nu}^{(4)}) \right) \Big]
\end{aligned}
\end{equation}
The $d-$functions are $d_i = (x/m_e) d_i^{\mathrm{DHS}}$, with $d_i^{\mathrm{DHS}}$ defined in\cite{Dolgov_NuPhB1997} as functions of the momenta $p$, hence the prefactor $x/m_e$. It should be noted that \cite{Relic2016_revisited,Bennett2021} use a different convention (4 times greater $D-$functions and opposite sign for $D_2$). $\bar{E} \equiv E/\Tcm$ is the comoving energy, and $E_4$ stands for $E_1 + E_2 - E_3$ by energy conservation. The full expressions can be found in appendix A of~\cite{Dolgov_NuPhB1997}, appendix D of~\cite{BlaschkeCirigliano} or in the appendix~\ref{app:reduc_collision_integral} of this manuscript.

\paragraph{QKE for $\bm{\bvrho}$} Similarly to the neutrino density matrix QKE, we rewrite~\eqref{eq:QKE_rhobar} using the comoving variables, which leads to:
\begin{multline}
\frac{\partial \bvrho(x,y_1)}{\partial x} = + \frac{\ii}{xH} \left(\frac{x}{m_e}\right) \left[ U \frac{\mathbb{M}^2}{2y_1}U^\dagger, \bvrho \right]  -  \ii \frac{2 \sqrt{2} G_F}{xH} y_1 \left(\frac{m_e}{x}\right)^5 \left[ \frac{\bar{\mathbb{E}}_\mathrm{lep} + \bar{\mathbb{P}}_\mathrm{lep}}{m_W^2} ,\bvrho \right ] \\
- \ii \frac{\sqrt{2} G_F}{x H} \left(\frac{m_e}{x}\right)^3 \left[ \overline{\mathbb{N}}_\nu - \overline{\mathbb{N}}_{\bnu}, \bvrho \right] - \ii \frac{8 \sqrt{2} G_F}{3 x H} y_1 \left( \frac{m_e}{x}\right)^5 \left[\frac{\bar{\mathbb{E}}_\nu + \bar{\mathbb{E}}_{\bar{\nu}}}{m_Z^2}, \bvrho\right]  + \frac{1}{xH} \bar{\mathcal{I}} \, . \label{eq:QKEbar_fullfinal}
\end{multline}

These equations are at the core of any study of (anti)neutrino evolution in the early Universe, and allow to take into account mixing (via the vacuum term), refractive matter effects (mean-field potentials, including the self-interaction one), and collisions which notably drive the transfer of entropy from electron/positron annihilations towards (anti)neutrinos throughout the decoupling era. In the following chapters, we will adapt these equations to particular setups of interest: the “standard” calculation of neutrino decoupling in chapter~\ref{chap:Decoupling}, and the evolution of primordial neutrino asymmetries in chapter~\ref{chap:Asymmetry}. We will systematically be focusing on the final neutrino spectra, which allow to compute the cosmological observables we are interested in (namely, $\Neff$), before exploring further the consequences on BBN in chapter~\ref{chap:BBN}.

\pagestyle{ruled}

\chapter[Standard neutrino decoupling including flavour oscillations][Standard neutrino decoupling]{Standard neutrino decoupling including flavour oscillations}
\label{chap:Decoupling}



\setlength{\epigraphwidth}{0.44\textwidth}
\epigraph{Sometimes I'll start a sentence and I don't even know where it's going. I just hope I find it along the way.}{Michael Scott, \emph{The Office} [S05E12]}

{
\hypersetup{linkcolor=black}
    \minitoc
}

\boxabstract{The material of this chapter has been partly published in~\cite{Froustey2020}.}

Several effects take place during neutrino decoupling, with some of them leaving signatures on cosmological observables. From the numerical resolution of the QKE that was derived in the previous chapter, we can obtain the evolution of the density matrix $\vrho$ and fully characterize the neutrino spectra during BBN and later cosmological stages. In particular, we present in this chapter the calculation of “standard” neutrino decoupling and notably the resulting value of $\Neff$. Its previous reference value was $\Neff = 3.045$~\cite{Relic2016_revisited}, but this calculation did not include some important finite-temperature QED corrections (see below) and approximated the off-diagonal components of the self-interaction collision term as damping factors. We present the first calculation relaxing completely the damping approximation and including all (known) QED corrections.

\noindent This standard calculation is thus based on the following assumptions.
\begin{itemize}
	\item The early Universe is considered homogeneous and isotropic. Hence, the dynamical evolution of spacetime is entirely described by the evolution of the scale factor through Friedmann equation. This hypothesis was made in the derivation of the QKEs in chapter~\ref{chap:QKE}, and is discussed in section~\ref{subsec:limitations_NuDec}.
	\item There is no asymmetry between neutrinos and antineutrinos. Therefore, we will only follow the evolution of the neutrino density matrix $\vrho$, which is justified in section~\ref{sec:set_of_equations}.\footnote{We have checked numerically that solving additionally the QKE on $\bvrho$ gives consistent results.} 
	\item We do not include a CP phase in the PMNS matrix (mostly because of the uncertainty on its value). However, as detailed in section~\ref{subsec:Decoupling_CP}, its value does not affect $\vrho^e_e$ nor $\Neff$, that is the key physical parameters for BBN. More generally, the cosmological observables are not affected by the value of the Dirac CP-violating phase --- a wider discussion of this property when including asymmetries is presented in chapter~\ref{chap:Asymmetry}.
	\item The QED plasma differs from an ideal gas because of finite-temperature corrections~\cite{Dicus1982,Heckler_PhRvD1994,Fornengo1997,LopezTurner1998,BrownSawyer,Mangano2002,Bennett2020}. The associated corrections to the equation of state are included in the energy conservation equation (see below). The modifications to the scattering rates are yet to be included, since they add a considerable layer of complexity to the computation of the collision integrals~\cite{Bennett2021}.
\end{itemize}
Under these assumptions, we can specify the set of equations we solve and the features of the numerical code, \texttt{NEVO}, developed for that purpose. Our main focus being the role of flavour oscillations in neutrino evolution, we additionally introduce a particular approximation of the evolution equations, based on the large separation of time scales in the problem. Its excellent accuracy provides a powerful framework to understand how the final neutrino distributions depend on physical parameters.

\section{Set of equations}
\label{sec:set_of_equations}

We present here the set of differential equations one needs to solve to determine the evolution of all relevant quantities (notably, the neutrino spectra) across the decoupling era. 

\subsection{Neutrino sector}

On the neutrino side, we are interested in determining the evolution of the density matrix $\vrho$, which is given by the QKE~\eqref{eq:QKE_fullfinal}. Given the above hypotheses, this equation can be simplified, reducing the number of terms in the oscillation Hamiltonian.

\paragraph{Neutrino asymmetry mean-field} Except in the subsection~\ref{subsec:Decoupling_CP} specifically dedicated to it, we neglect the Dirac CP phase in the PMNS matrix~\eqref{eq:PMNS}, therefore the Hamiltonian is a real symmetric matrix (instead of a hermitian matrix in the general case). We show in the appendix~\ref{subsec:QKE_consistency} that the structure of the QKEs preserves the equality $\vrho = \bvrho^*$ if it is true initially (note that $\bvrho^* = \bvrho^T$ where $^T$ stands for the transposed of the matrix). A key result of this chapter (cf.~section~\ref{sec:ATAO}) is that $\vrho$ is diagonal in the “matter” basis, that is the basis in which the Hamiltonian is diagonal. Since the Hamiltonian is a real symmetric matrix, the effective mixing matrix between the matter and flavour bases is orthogonal. Therefore, $\vrho$ and $\bvrho$ cannot have imaginary components and we can conclude that $\vrho = \bvrho$ at all times.\footnote{There may be a caveat in this reasoning in the inverted hierarchy case for which an instability might lead to the growing of the imaginary off-diagonal components of $\vrho - \bvrho$~\cite{Hansen_Isotropy} --- but this is only the case if $\vrho$ is not exactly diagonal in the matter basis. We discuss this case in section~\ref{subsec:Inverted_Hierarchy}. }

We will thus neglect the term $\overline{\mathbb{N}}_\nu - \overline{\mathbb{N}}_{\bnu}$ in the mean-field Hamiltonian, and solve only the QKE for $\vrho$ instead of $\vrho$ and $\bvrho$.

\paragraph{Neutrino energy density mean-field} We will also discard the mean-field term proportional to $\bar{\mathbb{E}}_\nu + \bar{\mathbb{E}}_{\bar{\nu}}$, as the deviations of $\vrho$ from the equilibrium distribution $\propto \Id$ are very small (cf. numerical results below). Thus, such a mean-field term will give a negligible contribution within the commutator compared to $\bar{\mathbb{E}}_\mathrm{lep}$.

\paragraph{Summary} Therefore, the QKE for standard neutrino decoupling in the early Universe reduces to
\begin{equation}
\frac{\partial \vrho(x,y_1)}{\partial x} = - \frac{\ii}{xH} \left(\frac{x}{m_e}\right) \left[ U \frac{\mathbb{M}^2}{2y_1}U^\dagger, \varrho \right]  +  \ii \frac{2 \sqrt{2} G_F}{xH} y_1 \left(\frac{m_e}{x}\right)^5 \left[ \frac{\bar{\mathbb{E}}_\mathrm{lep} + \bar{\mathbb{P}}_\mathrm{lep}}{m_W^2} ,\varrho \right ] + \frac{1}{xH} \mathcal{I} \, . \label{eq:QKE_final}
\end{equation}
For convenience in the forthcoming discussion, we write the effective Hamiltonian $\Hamil \equiv \Hvac + \Hlep$, with
\begin{itemize}
	\item the vacuum contribution
	\begin{equation}
	\label{eq:Hvac}
	 \Hvac \equiv  \frac{1}{xH} \left(\frac{x}{m_e}\right) U \frac{\mathbb{M}^2}{2 y_1} U^\dagger \, ,
	 \end{equation}
	which is inversely proportional to the momentum $y_1$ ;
	\item the lepton mean-field part
	\begin{equation}
	\label{eq:Hlep}
	 \Hlep \equiv - \frac{1}{xH} \left(\frac{m_e}{x}\right)^5 2 \sqrt{2} G_F y_1 \frac{\bar{\mathbb{E}}_\mathrm{lep} + \bar{\mathbb{P}}_\mathrm{lep}}{m_W^2} \, ,
	 \end{equation}
	which depends linearly on $y_1$.
\end{itemize}
Introducing the dimensionless collision term $\mathcal{K} \equiv \mathcal{I}/xH$, the QKE can be rewritten as
\begin{equation}
\label{eq:QKE_compact}
\boxed{\frac{\partial \vrho}{\partial x} = - \ii [\Hamil,\vrho] + \mathcal{K}} \, .
\end{equation}

The mass matrix $\mathbb{M}^2$ differs depending on the mass ordering considered. Except in the subsection~\ref{subsec:Inverted_Hierarchy}, we will systematically consider a normal mass ordering, favoured by current neutrino oscillation data~\cite{deSalas_Mixing,Esteban2020}.

\subsection{Electromagnetic plasma sector}
\label{subsec:QED}

In principle, one could also follow the evolution of the distribution functions of photons, electrons and positrons throughout the decoupling era via a Boltzmann equation. The collision term would then contain QED interactions, which are extremely efficient compared to weak ones. In other words, photons, electrons and positrons will always be kept at equilibrium by these interactions. This explains why we have taken the $e^\pm$ distribution functions to be equilibrium Fermi-Dirac ones in the derivation of the QKEs.

Therefore, all the information on the statistical distribution of these particles is contained in the plasma temperature $T_\gamma$, or equivalently the comoving temperature $z = T_\gamma / \Tcm$. Its evolution is most simply obtained from the continuity equation $\dot{\rho} = - 3 H (\rho + P)$ with $\rho$ and $P$ the \emph{total} energy density and pressure, that is $\rho = \rho_\gamma + \rho_\nu + \rho_{\bnu} + \rho_{e^\pm}$. Note that we do not include the (anti)muon energy density in this equation as it is negligible compared to $\rho_{e^\pm}$ in the decoupling era. This choice is consistent with the absence of interactions with muons and antimuons in the collision integral: the only role $\mu^\pm$ play in this problem is at the Hamiltonian level, ensuring the coincidence between the flavour and matter bases at high temperature. The reheating due to $\mu^- \mu^+$ annihilations affects equally neutrinos and the QED plasma (as it happens when all species are still coupled), so we do not take it into account --- it amounts to a different normalization of $\Tcm$.

\subsubsection{Energy conservation and QED equation of state}

We rewrite the continuity equation as an equation on the dimensionless photon temperature $z(x)$ \cite{Mangano2002,Bennett2020}:
\begin{equation}
\frac{\dd z}{\dd x} =  \frac{\displaystyle \frac{x}{z}J(x/z) - \frac{1}{2 \pi^2 z^3} \frac{1}{xH} \int_{0}^{\infty}{\dd y \, y^3 \, \Tr \left[\mathcal{I}\right]} + G_1(x/z)}{ \displaystyle \frac{x^2}{z^2}J(x/z) + Y(x/z) + \frac{2 \pi^2}{15} + G_2(x/z)} \, , \label{eq:zQED}
\end{equation}
with
\begin{align}
J(\tau) &\equiv \frac{1}{\pi^2} \int_{0}^{\infty}{\dd \omega \, \omega^2 \frac{\exp{(\sqrt{\omega^2 + \tau^2})}}{(\exp{(\sqrt{\omega^2 + \tau^2})}+1)^2}} \, , \\
Y(\tau) &\equiv \frac{1}{\pi^2} \int_{0}^{\infty}{\dd \omega \, \omega^4 \frac{\exp{(\sqrt{\omega^2 + \tau^2})}}{(\exp{(\sqrt{\omega^2 + \tau^2})}+1)^2}} \, .
\end{align}
In~\eqref{eq:zQED}, the integral involving the neutrino collision integral comes from $\dd \bar{\rho}_\nu / \dd x$, given that $\bar{\rho}_\nu = \frac{1}{2 \pi^2} \int{\dd{y} \, y^3 \, \Tr[\vrho]}$.

The $G_1$ and $G_2$ functions account for the modifications of the plasma equation of state (departure from an ideal gas) due to finite-temperature QED corrections \cite{Heckler_PhRvD1994,Mangano2002,Bennett2020}. Indeed, this is part of the features encountered in interacting quantum fields at finite temperature; often interpreted as a modification of the dispersion relation of electrons/positrons and photons which get extra “thermal masses”.\footnote{One must however be very careful with this interpretation which can lead to a missing $1/2$ factor in the pressure correction as in~\cite{Grohs2015}. This factor is emphasized in~\cite{Mangano2002} and notably discussed in~\cite{Bennett2020}.} They can be calculated order by order in an expansion in powers of $\alpha = e^2/4\pi$: starting from an expansion of the partition function $Z$ and therefore the free energy $F$~\cite{KapustaGale}, one gets the thermodynamical quantities $\rho$ and $P$ at the desired order, and the $G-$functions after implementing these modified energy density/pressure in the continuity equation. The expressions read
\begin{align}
G_1^{(2)}(\tau) &= 2 \pi \alpha \left[\frac{K'(\tau)}{3} + \frac{J'(\tau)}{6} + J'(\tau) K(\tau) + J(\tau) K'(\tau) \right] \, , \label{eq:G1e2} \\
G_2^{(2)}(\tau) &= - 8 \pi \alpha \left[\frac{K(\tau)}{6} + \frac{J(\tau)}{6} - \frac12 K(\tau)^2 + K(\tau)J(\tau)\right]  \nonumber \\
&\phantom{=}  + 2 \pi \alpha \tau \left[\frac{K'(\tau)}{6} - K(\tau)K'(\tau) + \frac{J'(\tau)}{6} + J'(\tau) K(\tau) + J(\tau) K'(\tau) \right] \, , \label{eq:G2e2} \\
G_1^{(3)}(\tau) &= - \sqrt{2 \pi} \alpha^{3/2} \sqrt{J(\tau)} \times \tau \left[2 j(\tau) - \tau j'(\tau) + \frac{\tau^2 j(\tau)^2}{2 J(\tau)} \right] \, , \label{eq:G1e3} \\
G_2^{(3)}(\tau) &= \sqrt{2 \pi} \alpha^{3/2} \sqrt{J(\tau)} \left[\frac{\left(2 J(\tau) + \tau^2 j(\tau)\right)^2}{2 J(\tau)} + 6 J(\tau) + \tau^2 \left( 3 j(\tau) - \tau j'(\tau)\right) \right] \, , \label{eq:G2e3}
\end{align}
where $(\cdots)' = \dd(\cdots)/\dd \tau$, and with the additional functions
\begin{align}
j(\tau) &\equiv \frac{1}{\pi^2} \int_{0}^{\infty}{\dd \omega \,  \frac{\exp{(\sqrt{\omega^2 + \tau^2})}}{(\exp{(\sqrt{\omega^2 + \tau^2})}+1)^2}} \, , \\
K(\tau) &\equiv \frac{1}{\pi^2} \int_{0}^{\infty}{\dd \omega \, \frac{\omega^2}{\sqrt{\omega^2 + \tau^2}} \frac{1}{\exp{(\sqrt{\omega^2 + \tau^2})}+1}} \, , \\
k(\tau) &\equiv \frac{1}{\pi^2} \int_{0}^{\infty}{\dd \omega \, \frac{1}{\sqrt{\omega^2 + \tau^2}} \frac{1}{\exp{(\sqrt{\omega^2 + \tau^2})}+1}} \, . 
\end{align}
We discarded a logarithmic contribution to $G_{1,2}^{(2)}$ that is insignificant compared to the dominant contribution to $G_{1,2}^{(2)}$ and even compared to $G_{1,2}^{(3)}$ \cite{Bennett2020}. Note that our expressions look formally different from those of previous literature. For instance \eqref{eq:G1e2} is formally different from the equivalent equation in \cite{Mangano2002,Bennett2020}, while \eqref{eq:G2e2} matches formally with \cite{Mangano2002}, but not with \cite{Bennett2020}. Finally, \eqref{eq:G1e3} and \eqref{eq:G2e3} slightly differ from expressions reported in \cite{Bennett2020}. All expressions are in fact identical, since one can prove (after integrations by parts and rearrangements) the following identities:
\begin{equation}
J'(\tau) = - \tau j(\tau) \; , \
K'(\tau) = - \tau k(\tau) \; , \
Y'(\tau) = - 3 \tau J(\tau) \; , \
2 K(\tau) + \tau^2 k(\tau) = J(\tau) \, .
\end{equation}

\section{Adiabatic transfer of averaged oscillations}
\label{sec:ATAO}

Solving the full QKE \eqref{eq:QKE_final} is \emph{a priori} a considerable numerical challenge because of the need to resolve numerically both the effect of the mean-field terms and of computationally expensive collision integrals. However, the previous numerical results including flavour mixing~\cite{Mangano2005,Relic2016_revisited,Akita2020} seem to indicate that the expected oscillations are somehow “averaged” while there is a comparatively slow evolution due to collisions. Indeed, these studies solved the full QKE (except some approximations in the collision term), hence the vacuum and matter effects must be fully included in their results.

\begin{figure}[!ht]
	\centering
	\includegraphics{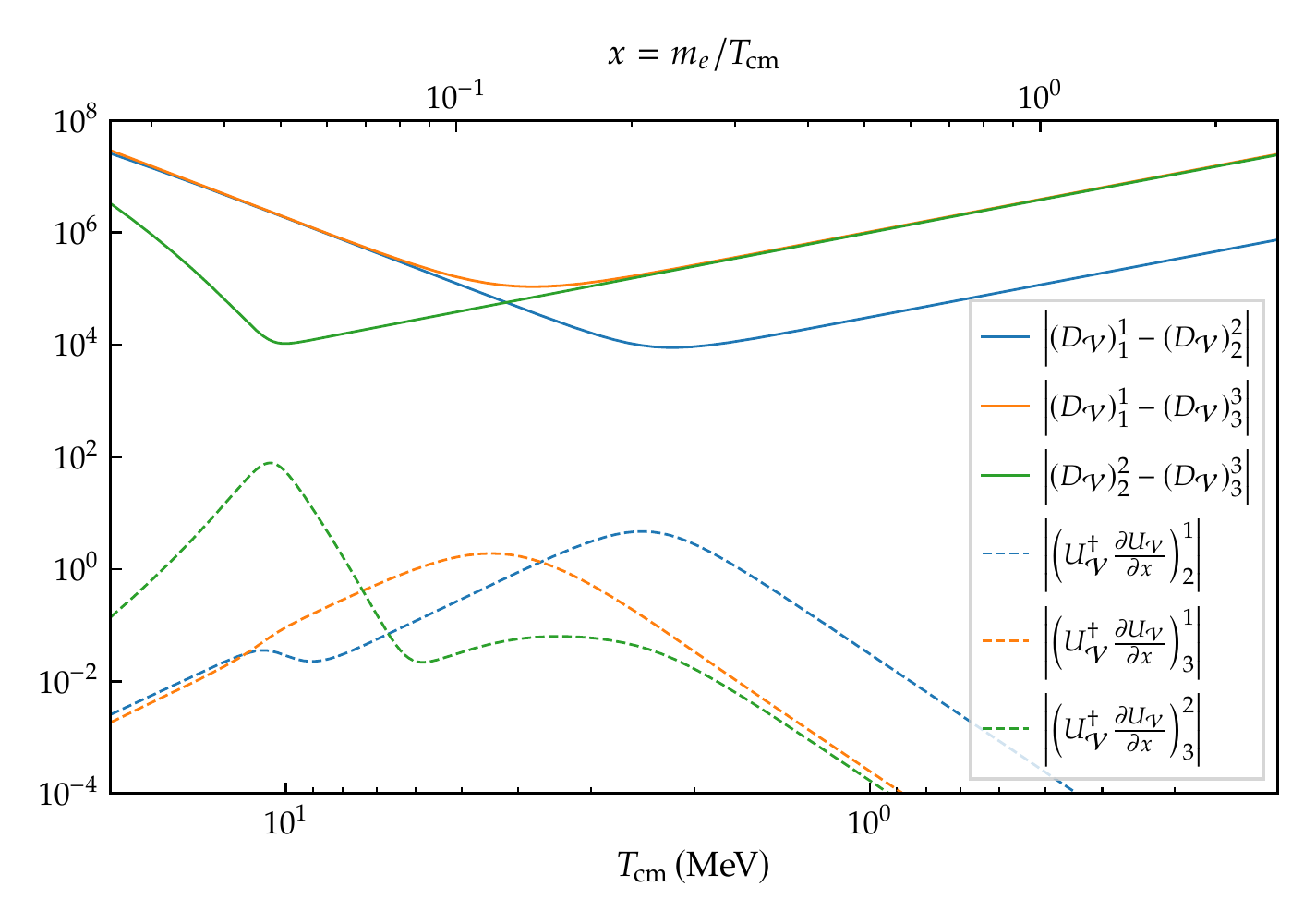}
	\caption[Checking the Hamiltonian adiabaticity condition]{\label{fig:ODG_adiab} Evolution of the different quantities appearing in \eqref{eq:QKEmatt} in the normal hierarchy of masses, for a comoving momentum $y=5$. The condition~\eqref{eq:adiab_cond} is satisfied throughout the evolution.}
\end{figure}
\begin{figure}[!ht]
	\centering
	\includegraphics{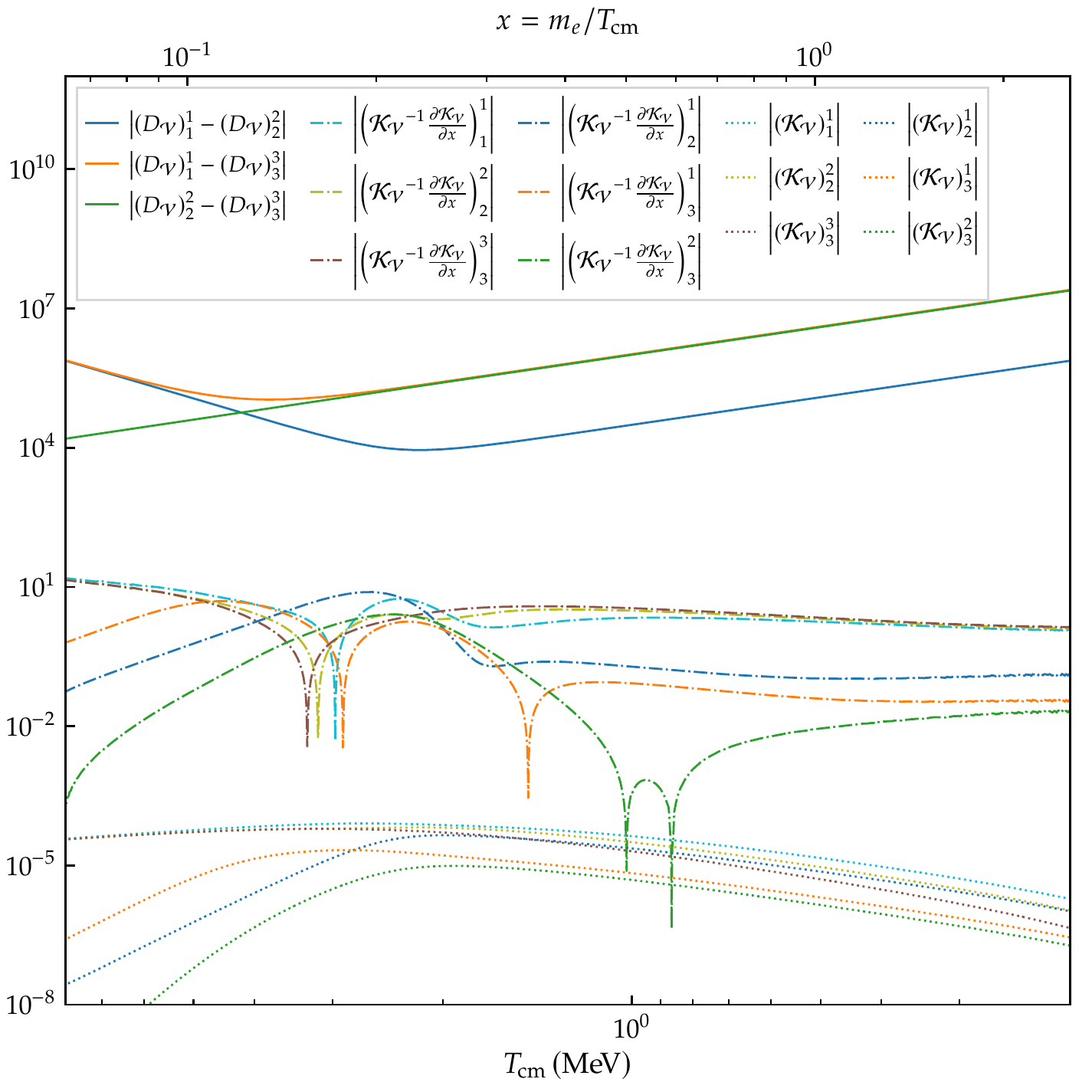}
	\caption[Checking the slowness of the rates associated to the collision term]{\label{fig:ODG_coll} Comparison of the evolution of the collision term, its relative variation and the effective oscillation frequencies in the normal hierarchy of masses, for a comoving momentum $y=5$. We check that the condition \eqref{eq:collis_cond} is satisfied with several orders of magnitude.}
\end{figure}

We expect a clear separation of time-scales to hold between the fast oscillations and the secular evolution due to the change of Hamiltonian and the collision term, which would allow for an effective description correctly capturing the salient features of the dynamical evolution. Let us start from the QKE written in its compact form~\eqref{eq:QKE_compact}. We treat the $y$ dependence of $\Hamil$ implicitly, as the following procedure must be applied for each $y$. Since the Hamiltonian $\Hamil$ is Hermitian, it can be diagonalized by the unitary transformation
\begin{equation}
\Hamil = U_\Hamil D_\Hamil  U_\Hamil^\dagger \qquad \text{with} \qquad (D_\Hamil)^j_k = (D_\Hamil)^j_j \, \delta^{j}_{k} \, .
\end{equation}
The density matrix in the matter basis reads $\vrho_\Hamil = U_\Hamil^\dagger \, \vrho \, U_\Hamil$, and evolves according to
\begin{equation}
\label{eq:QKEmatt}
\frac{\partial \vrho_\Hamil}{\partial x} = -\ii \left[ D_\Hamil,\vrho_\Hamil \right] -  \left[U_\Hamil^\dagger \frac{\partial U_\Hamil}{\partial x},\vrho_\Hamil \right] + U_\Hamil^\dagger \mathcal{K} U_\Hamil \, .
\end{equation}
The first approximation that we consider is the \emph{adiabatic approximation} \cite{HannestadTamborra,GiuntiKim} which consists in neglecting the time evolution of the matter PMNS matrix compared to the inverse effective oscillation frequency:
\begin{align}
&\text{\textsc{ Adiabatic approximation}}&   \norm{U_\Hamil^\dagger \frac{\partial U_\Hamil}{\partial x}} &\ll \norm{D_\Hamil} \, . \label{eq:adiab_cond}
\\
  \intertext{This condition means that the effective mixing matrix elements vary very slowly compared to the effective oscillation frequencies, so that the matter basis evolves adiabatically. More specifically, we need to check that $\abs{\left(U_\Hamil^\dagger \frac{\partial U_\Hamil}{\partial x}\right)^j_k} \ll \abs{(D_\Hamil)^j_j - (D_\Hamil)^k_k}$. Such adiabaticity condition is particularly important in presence of Mikheev-Smirnov-Wolfenstein (MSW) resonances \cite{MSW_MS,MSW_W}. Note that the sign of the mean-field contribution to $\Hamil$ \eqref{eq:Hlep} is opposite to the one encountered due to charged-current neutrino-electron scattering at lowest order, important for astrophysical environments (Sun, supernovae, binary neutron star mergers).
We numerically check (Figure~\ref{fig:ODG_adiab}) that the condition \eqref{eq:adiab_cond} is indeed satisfied throughout the range of temperatures of interest. \endgraf
If we now assume that many oscillations take place before the collision term varies substantially and write the collision term in matter basis $\mathcal{K}_\Hamil \equiv U_\Hamil^\dagger \mathcal{K} U_\Hamil$, its variation frequency $\sim \mathcal{K}_\Hamil^{-1} (\partial \mathcal{K}_\Hamil/\partial x)$ must be small compared to the effective oscillation frequency $D_\Hamil$. We also assume that the collision rate itself is small compared to the oscillation frequencies, namely} 
&\text{\textsc{ Averaged oscillations}}&  \norm{\mathcal{K}_\Hamil}, \norm{\mathcal{K}_\Hamil^{-1} \frac{\partial \mathcal{K}_\Hamil}{\partial x}} &\ll \norm{D_\Hamil} \label{eq:collis_cond}
\, .
\end{align} 
We check on Figure~\ref{fig:ODG_coll} that this separation of time-scales holds. Therefore, it is possible to \emph{average} the evolution of $\vrho_\Hamil$ over many oscillations (the collision term produces at constant rate neutrinos with random initial phases). The non-diagonal parts will then be washed out if the collision rate is not too strong. More precisely, we can write
\begin{equation}
\label{eq:solveapprox}
(\vrho_\Hamil)^j_k(x,y) \equiv e^{- \ii (D_\Hamil)^j_j x} R^{j}_{k}(x,y) e^{\ii (D_\Hamil)^k_k x} \ \implies \ \frac{\partial R^{j}_{k}}{\partial x} = e^{\ii (D_\Hamil)^j_j x}(\mathcal{K}_\Hamil)^j_k e^{- \ii (D_\Hamil)^k_k x} \, , 
\end{equation}
where we also assumed a slow variation of $D_\Hamil$, as a consequence of the adiabatic approximation. If \eqref{eq:collis_cond} holds, $\partial R^{j}_{k}/\partial x$ is integrated over many oscillations and the off-diagonal parts vanish. This leaves us with the effective equation in matter basis:
\begin{equation}
\label{eq:ATAO}
\text{\textsc{\bfseries Adiabatic Transfer of Averaged Oscillations}}  \qquad \left\{ \begin{aligned}
 \frac{\partial \tilde{\vrho}_\Hamil}{\partial x} &= \reallywidetilde{U_\Hamil^\dagger \mathcal{K} U_\Hamil} \\
 \vrho_\Hamil &= \tilde{\vrho}_\Hamil \end{aligned} \right. \, ,
\end{equation}
where the tilde means that we only keep the diagonal terms of $\vrho_\Hamil$, then convert it to the flavour basis to compute the collision term $\mathcal{K}$ and only keep the diagonal part of the collision term $U_\Hamil^\dagger \mathcal{K} U_\Hamil$ when transforming back to the matter basis. In the flavour basis, the density matrix $\vrho = U_\Hamil \tilde{\vrho}_\Hamil U_\Hamil^\dagger$ has off-diagonal components, while $\tilde{\vrho}_\Hamil$ is diagonal. Therefore the collision term destroys the coherence between these components (since it aims at a diagonal $\vrho$ in flavour space, with equilibrium distributions), which modifies in turn the diagonal values of $\vrho_\Hamil$ (whose off-diagonal terms average out). 

For clarity, we refer to this approximate numerical scheme to determine the neutrino evolution “Adiabatic Transfer of Averaged Oscillations'' (ATAO) and we then solve \eqref{eq:ATAO} instead of \eqref{eq:QKE_compact}.  Note that we explained the procedure with $\Hamil$, but it could be carried out with any Hamiltonian $\mathcal{H}$, which will be relevant when dealing with asymmetries (cf.~chapter~\ref{chap:Asymmetry}). The full designation of this scheme is thus rather \ATAOH but we will simply call it “ATAO” in this chapter since there cannot be any confusion.

In the following section, we will numerically solve the QKEs in both the full “exact” case and the ATAO approximation and discuss the validity of the approximate numerical solution.

\newpage

\section{Numerical implementation}
\label{sec:numeric_Dec}

We integrate numerically the QKE for neutrinos~\eqref{eq:QKE_final}, or \eqref{eq:ATAO} in the ATAO approximation, along with the energy conservation equation~\eqref{eq:zQED}. We developed the code \texttt{NEVO} (Neutrino EVOlver), written in Python with the \texttt{scipy} and \texttt{numpy} libraries.\footnote{Time consuming functions are compiled with the just-in-time compiler \texttt{numba}.}

\paragraph{Solver and initial conditions}

The collision term consists most of the time in nearly compensating gain and loss terms, and for temperatures larger than $0.1\, \text{MeV}$, the system is very stiff. Hence, one must rely on an implicit method. We chose the \texttt{LSODA} method which consists in a Backward Differentiation Formula (BDF) method (with adaptative order and adaptative step) when the system is stiff, which switches to an explicit method when not stiff (the Adams method). It was first distributed within the \texttt{ODEPACK} Fortran library~\cite{ODEPACK}, but we used the Python wrapper \texttt{solve\_ivp} distributed with the Python \texttt{scipy} module. We noticed that when setting the absolute and relative error tolerances to $10^{-n}$, the spectra are typically obtained with precision better than $10^{-n+2}$, in agreement with section B.5 of \cite{Gariazzo_2019}. Hence we fixed these error tolerances to $10^{-7}$ so as to obtain results with numerical errors below $10^{-5}$.

The initial common temperature of all species, that is all types of neutrinos and the electromagnetic plasma, is inferred from the conservation of total entropy. Choosing the initial comoving temperature $T_{\mathrm{cm},\mathrm{in}}  = 20 \, \text{MeV}$, the initial common temperature of all species is slightly larger because of early $e^\pm$ annihilations, and given by $T_\mathrm{in} = z_\mathrm{in} T_{\mathrm{cm},\mathrm{in}}$ with $z_\mathrm{in}-1 = 7.42\times 10^{-6}$. Had we chosen to start at $T_{\mathrm{cm},\mathrm{in}}  = 10 \, \text{MeV}$, the initial comoving temperature would have been $z_\mathrm{in}-1 = 2.98\times 10^{-5}$, in agreement with~\cite{Mangano2005,Dolgov_NuPhB1999}. 
As initial condition for the density matrix we take
\begin{equation}
\label{eq:initial_condition}
\vrho(x_{\mathrm{in}},y) = \begin{pmatrix}
f_\nu^{({\mathrm{in}})}(y) & 0 & 0 \\
0 & f_\nu^{(\mathrm{in})}(y) & 0 \\
0 & 0 & f_\nu^{(\mathrm{in})}(y)
\end{pmatrix} \quad , \quad \text{with} \quad f_\nu^{(\mathrm{in})}(y) \equiv \frac{1}{e^{y/z_\mathrm{in}} + 1} \, .
\end{equation}

\paragraph{Momentum grid} 

The neutrino spectra are sampled with $N$ points on a grid in the reduced momentum $y$. When choosing a linear grid, we use the range $0.01 \leq y\leq 16 + [N/20]$, and integrals are evaluated with the Simpson method. However, for functions which decay exponentially for large $y$, it is motivated to use the Gauss-Laguerre quadrature which was already proposed in~\cite{Gariazzo_2019}. We confirm that this method typically requires half of the grid points to reach the same precision as the one obtained with a linear spacing. In practice, when choosing the nodes and weights of the quadrature, we restrict to $y\leq 20+[N/5]$. When using $N=80$, we have thus restricted nodes to $y \leq 36$, and we used Laguerre polynomials of order $439$ to compute the weights with Eq.~(B.14) of \cite{Gariazzo_2019}. Since the tools provided in \texttt{numpy} are restricted to much lower polynomial orders, we used \emph{Mathematica} to precompute once and for all in a few hours the nodes and weights. The results reported in the following were performed with $N=80$ and the Gauss-Laguerre quadrature, checking that with $N=100$ the differences are smaller than the desired precision.

For each momentum $y_i$ of the grid, and with $N_{\nu}$ flavours, each density matrix has $N_{\nu}^2$ independent degrees of freedom ($N_\nu(N_{\nu} +1)$ real parts and $N_\nu(N_{\nu} -1)$ imaginary parts). In practice we reorganize these independent matrix entries into a vector $A^j(y_i)$ with $j=1,\dots,N_{\nu}^2$ and we concatenate them with the $y_i$ spanning the momentum grid. We thus solve for serialized variables, that is a giant vector of length $N N_\nu^2$. When using the ATAO approximation, one needs only to keep the diagonal part in the matter basis, and the giant vector is of size $N N_\nu$.\footnote{Results are then only converted at the very end in the flavour basis if desired.} Note that we do not store the binned density matrix components $\vrho^\alpha_\beta(y_i)$, which would be sub-optimal. Indeed, if neutrinos decoupled instantaneously, their distribution function would then be
\begin{equation}
\label{eq:fnueq}
f_{\nu}^{\mathrm{(eq)}}(x,y) \equiv \frac{1}{e^{y}+1} \, .
\end{equation}
Therefore, we can parametrize the density matrix $\vrho^\alpha_\beta(x,y) = \left[\delta^{\alpha}_{\beta}+a^\alpha_\beta(x,y)\right]\times f_\nu^{\mathrm{(eq)}}(x,y)$, and we store the values of $a^\alpha_\beta$, which encapsulate the deviation from instantaneous decoupling.

\paragraph{Mixing parameters} As specified at the end of section~\ref{sec:set_of_equations}, we will solve the QKEs in the normal ordering case and without a Dirac CP phase, therefore using the parameters given in~\ref{subsec:Values_Mixing}. We specifically include the CP phase in section~\ref{subsec:Decoupling_CP}, and discuss the case of the inverted hierarchy of masses in section~\ref{subsec:Inverted_Hierarchy}.

\paragraph{Numerical optimization via Jacobian computation.} The implicit method requires to solve algebraic equations and thus to obtain the Jacobian of the differential system. For the sake of this discussion, and to alleviate the notation, we ignore the different flavours and consider that we have only one neutrino flavour with spectrum $f(y)$. Noting the grid points $y_i$ and the values of the spectra $f_i = f(y_i)$ on the grid, the differential system is of the type $\partial_x f_i = \mathcal{K}_i(x, f_j)$. The implicit method requires the Jacobian $J_{ij} \equiv \partial \mathcal{K}_i/\partial f_j$. If no expression is provided, it is evaluated by finite differences in the $\{ f_i\}$ at a given $x$. Since the collision term involves a two-dimensional integral for each point of the grid, its computation on the whole grid is of order $\mathcal{O}(N^3)$. Hence, the computation of the Jacobian with finite differences is of order $\mathcal{O}(N^4)$. Since algebraic manipulations (mostly the LU decomposition) are at most of order $\mathcal{O}(N^3)$, reducing the cost of the Jacobian numerical evaluation is crucial to improve the speed of the implicit method. Fortunately, it is possible to compute the Jacobian with an $\mathcal{O}(N^3)$ complexity. To use a simple example, let us only consider the contribution from the loss part of the neutrino self-interactions, without including Pauli-blocking factors. This component of the collision term, once computed numerically with a quadrature, is of the form
\begin{equation}
\label{ContriC}
\mathcal{K}_i(x, f_j) = -\sum_{j,k} w_j w_k g(y_i,y_j,y_k) f_i f_j\,.
\end{equation}
In this expression $\sum_j w_j$ (resp. $\sum_k w_k$) accounts for the integration on $y_2$ (resp.~$y_3$) in~\eqref{eq:I_full} using a quadrature, and the function $g$ takes into account the specific form of the factor multiplying the statistical function (which is for the contribution considered $f_i f_j$).
Noting then that
\begin{equation}
\partial f_i /\partial f_j = \delta_{ij} \, ,\label{dfdf}
\end{equation}
the Jacobian associated with the contribution~\eqref{ContriC} is
\begin{equation}\label{Jimtwocontrib}
J_{i m} = \partial \mathcal{K}_i/\partial f_m = -\delta_{im} \sum_{j,k} w_j w_k g(y_i,y_j,y_k) f_j - \sum_{k} w_m w_k g(y_i,y_m,y_k) f_i \, . 
\end{equation}
The complexity of the second sum is of order $\mathcal{O}(N)$, and since the Jacobian has $N^2$ entries, it leads to a complexity of order $\mathcal{O}(N^3)$. The first term is not worse even though the double sum is of order $\mathcal{O}(N^2)$, because it concerns only the diagonal entries of the Jacobian due to the prefactor $\delta_{im}$. More generally for all contributions to the collision term, the complexity when computing the associated Jacobian is always of order $\mathcal{O}(N^3)$, even when taking into account Pauli-blocking factors which bring terms which are cubic or quartic in the density matrix. For instance, terms similar to~\eqref{ContriC}, but with factors $f_i f_j f_k$, are handled with the same method and would lead to three contributions instead of two in~\eqref{Jimtwocontrib}. As for terms with factor $f_i f_j f_l$, they would be handled using total energy conservation $y_i + y_j = y_k + y_l$, which allows for instance to replace the variables of summations (e.g. $\sum_{j,k} \to \sum_{j,l}$) when varying with respect to $f_l$. Following these arguments, one notices that the exponent of the complexity for both the collision term and its associated Jacobian is given by the number of independent momenta magnitudes, given that integrations on momenta directions have all been removed with the integration reduction method using the isotropy of momentum distribution. In the case at hand, we have only two-body collisions, for which total energy conservation implies that only three momenta magnitudes are independent, hence the complexity in $\mathcal{O}(N^3)$. 

When restoring the fact that we do not have a single flavour but density matrices, the discussion is similar when using the serialized variables described above, and again the complexity is of order $\mathcal{O}(N^3)$. In practice, we found that it takes roughly five times more time to compute a Jacobian than a collision term. Hence, when compared with the finite difference method, providing a numerical method for the Jacobian leads to a factor $N/5$ speed-up. Note that we must also integrate $z$ with eq.~\eqref{eq:zQED} jointly with the density matrices, so that we must pad the Jacobian obtained with the previous description with one extra line and one extra column. Again, the corresponding entries can be deduced using~\eqref{dfdf} and their computation is also of order $\mathcal{O}(N^3)$.

It is worth mentioning that providing a method for the Jacobian is not specific to the ATAO approximation. Indeed, when solving the full QKE one can also compute the Jacobian of the collision term, and one only needs to add the contribution from the vacuum and mean-field commutators whose complexity is simply of order $\mathcal{O}(N^2)$. The precise description of the Jacobian calculation, also adequate for the case of non-zero asymmetries (cf.~chapter~\ref{chap:Asymmetry}) is done in appendix~\ref{App:Numerics}.

When compared with the full QKE method, the ATAO numerical resolution allows to gain at least a factor 5 in time. Hence when using both a method for the Jacobian and the ATAO approximation, we gain typically a factor $N$ and computations that would otherwise last days on CPU clusters, are reduced to just few hours on a single CPU. Moreover, nothing prevents the computation of collision terms and Jacobians to be parallelized on the momentum grid, as we checked on the 4 or 8 CPUs of desktop machines, reducing even further the computation time.

\section{Neutrino temperature and spectra after decoupling}
\label{sec:results_nudec}

We use our code \texttt{NEVO} with the parameters given in the previous section to follow neutrino evolution across the MeV era. The quantities we are most interested in are the frozen-out spectra of neutrinos after decoupling, which allow to compute in particular their energy density (hence $\Neff$ as we also know the final photon temperature).

\subsection[Numerical results: benchmark value for $\Neff$]{Numerical results: benchmark value for $\bm{\Neff}$}
\label{subsec:results_Neff}

Since the diagonal entries of the neutrino density matrix correspond to the generalization of the distribution functions, we can use the same parameterization as~\eqref{eq:param_fnu} to separate effective temperatures and residual distortions, namely,
\begin{equation}
\label{eq:param_rho}
\vrho^\alpha_\alpha(x,y) \equiv \frac{1}{e^{y/z_{\nu_\alpha}} + 1} \left[1 + \delta g_{\nu_\alpha}(x,y)\right] \, ,
\end{equation}
where we recall that the reduced effective temperature $z_{\nu_\alpha} \equiv T_{\nu_\alpha}/\Tcm$ is the reduced temperature of the Fermi-Dirac spectrum with zero chemical potential which has the same energy density as the real distribution:
\begin{equation}
\frac{1}{2 \pi^2} \int{\dd{y} y^3 \vrho^\alpha_\alpha(y)} = \bar{\rho}_{\nu_\alpha} \equiv \frac78 \frac{\pi^2}{30} z_{\nu_\alpha}^4 \,.
\end{equation}
The appeal of this parameterization lies in the clear separation between energy changes compared to instantaneous decoupling (hence gravitational effects through the Hubble parameter), and non-thermal distortions which can only affect the reaction rates. This will be very useful to study the consequences of incomplete neutrino decoupling on BBN in chapter~\ref{chap:BBN}. 

\begin{figure}[!ht]
	\centering
	\includegraphics{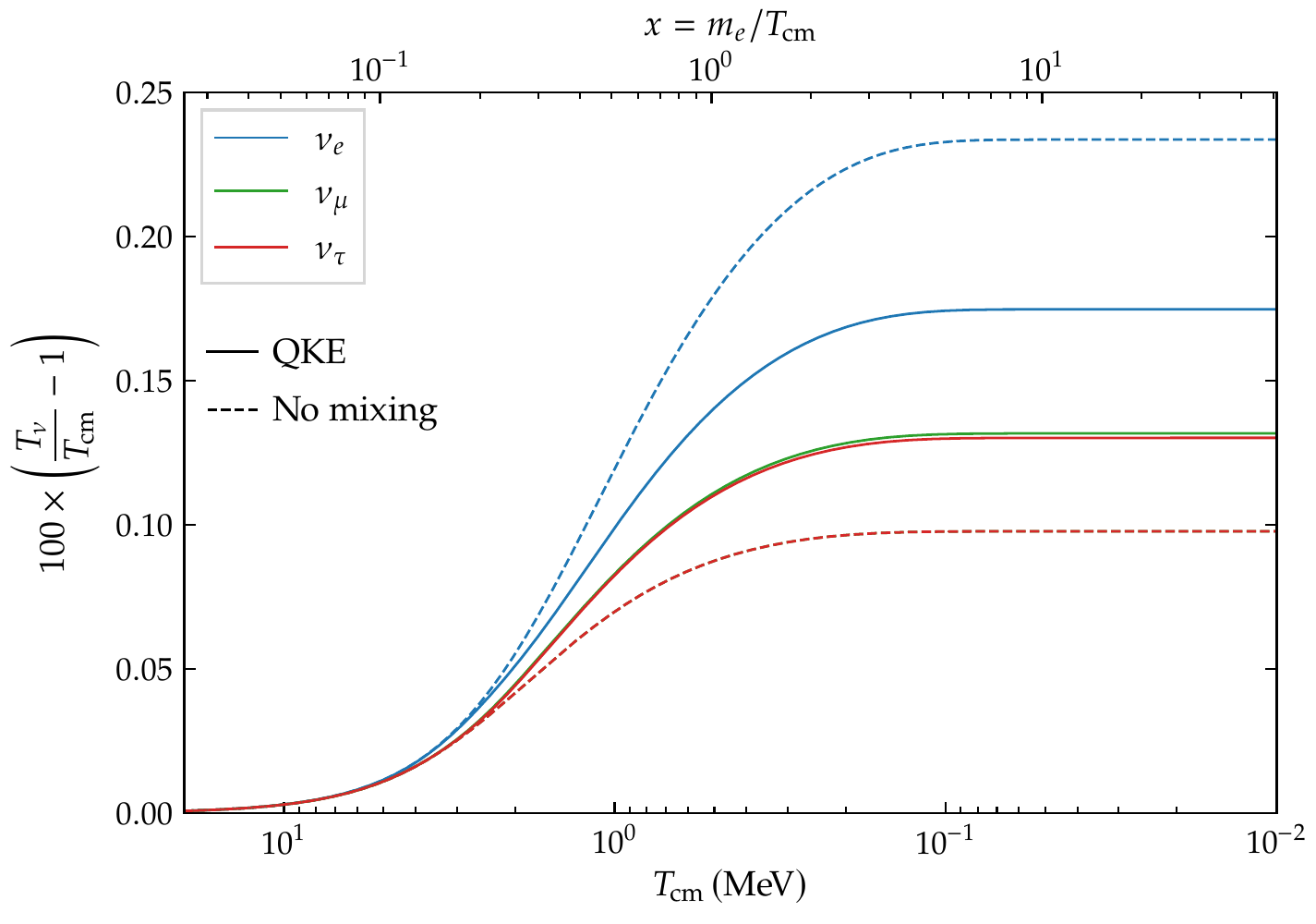}
	\caption[Evolution of the effective neutrino temperatures]{\label{fig:Tnu} Evolution of the effective neutrino temperatures, with and without oscillations. Long before decoupling, they remain equal to the photon temperature $z$, before freezing-out at different values depending on the interaction with the electromagnetic plasma. Without mixing, the distribution functions (and thus, the effective temperatures) are identical for $\nu_\mu$ and $\nu_\tau$.}
\end{figure}

\begin{figure}[!ht]
	\centering
	\includegraphics{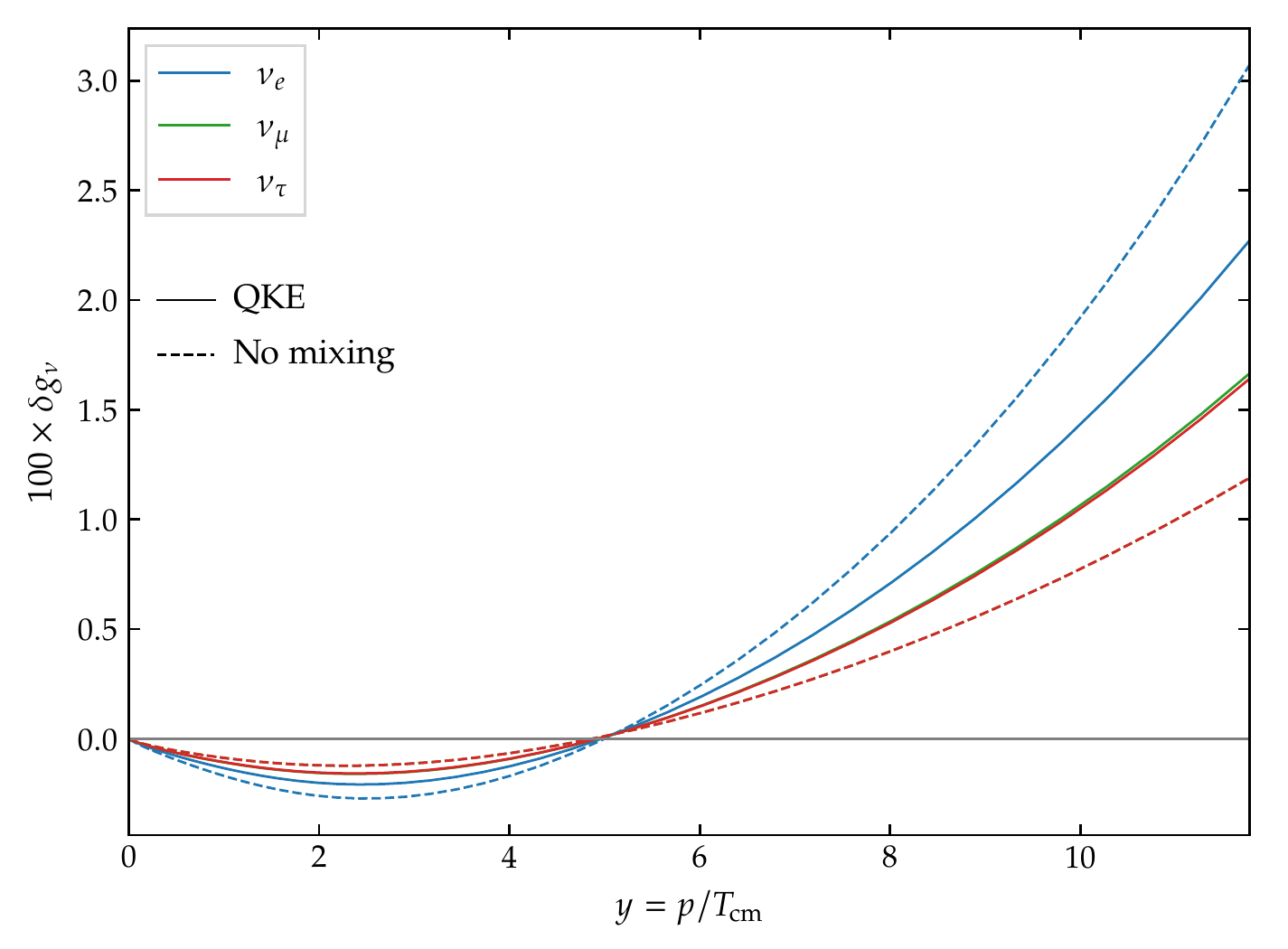}
	\caption[Frozen-out effective spectral distortions]{\label{fig:deltagnu} Frozen-out effective spectral distortions, with and without oscillations, for $x_f \simeq 51$ (corresponding to $T_{\mathrm{cm},f} = 0.01 \, \mathrm{MeV}$). The full QKE results are indistinguishable from the ATAO approximate ones.}
\end{figure}

We plot in Figure~\ref{fig:Tnu} the evolution of the neutrino effective temperatures, with and without flavour oscillations (including all QED corrections to the plasma thermodynamics), and in Figure~\ref{fig:deltagnu} the non-thermal residual distortions. In the absence of flavour mixing (dashed lines on Figures~\ref{fig:Tnu} and \ref{fig:deltagnu}), we cannot distinguish between $\nu_\mu$ and $\nu_\tau$ since they have exactly the same interactions with $e^\pm$ (neutral-currents only) and with (anti)neutrinos, thus nothing distinguishes these two flavours in the no-mixing case. On the contrary, the higher effective temperatures or non-thermal distortions for the electronic flavour are due to the charged-current processes, which increase the transfer of entropy from electrons and positrons.

The final values of the comoving temperatures and $\Neff$ are given in Table~\ref{Table:Res_NuDec}. We find that including the different corrections shifts $\Neff$ in agreement with the results quoted in Table 5 of~\cite{Bennett2021}: 
\begin{itemize}
	\item the $\mathcal{O}(e^2)$ finite-temperature QED correction to the plasma equation of state increases $\Neff$ by $\sim 0.01$ (fourth and fifth rows in Table~\ref{Table:Res_NuDec}) ;
	\item the next order in the QED corrections, $\mathcal{O}(e^3)$ reduces $\Neff$ by $\sim 0.001$ (fifth and sixth rows in Table~\ref{Table:Res_NuDec}), as predicted in~\cite{Bennett2020}, and also observed in~\cite{Akita2020} ;
	\item flavour oscillations have a subdominant contribution compared to QED corrections, namely $+ \, 6 \times 10^{-4}$ (third and sixth rows in Table~\ref{Table:Res_NuDec}).
\end{itemize}

\renewcommand{\arraystretch}{1.3}

\begin{table}[!htb]
	\centering
	\begin{tabular}{|l|ccccc|}
  	\hline 
  Final values & $z$ & $z_{\nu_e}$  & $z_{\nu_\mu}$ & $z_{\nu_\tau}$ &$\Neff$  \\
  \hline \hline
  Instantaneous decoupling, no QED & $1.40102$ &  $1.00000$ & $1.00000$ &$1.00000$ & $3.00000$ \\ \hline
  No oscillations (NO), QED $\mathcal{O}(e^3)$ &$1.39800$ & $1.00234$ &  $1.00098$ & $1.00098$ &$3.04338$ \\
  ATAO, no QED  & $1.39907$ & $1.00177$ & $1.00134$ & $1.00132$ & $3.03463$   \\
ATAO, QED $\mathcal{O}(e^2)$ & $1.39786$ & $1.00175$ & $1.00132$ & $1.00130$ & $3.04491$   \\ \hline 
    ATAO, QED $\mathcal{O}(e^3)$ & $1.39797$ & $1.00175$ & $1.00132$ & $1.00130$ & $3.04396$   \\
    Full QKE, QED $\mathcal{O}(e^3)$ & $1.39797$ & $1.00175$ & $1.00132$ & $1.00130$ & $3.04396$   \\ \hline
\end{tabular}
	\caption[Final (effective) temperatures after decoupling]{Frozen-out values of the dimensionless photon and neutrino temperatures, and the effective number of neutrino species. We detail the results of different implementations in order to assess the contribution of each correction w.r.t.~the instantaneous decoupling approximation. $\Neff$ differs between the ATAO and full QKE calculations by a few $10^{-6}$, which we attribute mainly to numerical errors. The values quoted here differ slightly from~\cite{Froustey2020} due to the updated value of $\mathcal{G}$~\cite{PDG}, which affects the Hubble expansion rate.
	\label{Table:Res_NuDec}}
\end{table}

Flavour oscillations reduce the discrepancy between the different flavours, thus $z_{\nu_e}$ is reduced while $z_{\nu_\mu}$ and $z_{\nu_\tau}$ are increased, with a very slightly higher value for $z_{\nu_\mu}$. This enhanced entropy transfer towards $\nu_\mu$ compared to $\nu_\tau$ is due to the more important $\nu_e-\nu_\mu$ mixing (cf.~Figure~\ref{fig:ATAO_Transfer} and the corresponding discussion).

\paragraph{Comparison with the literature}

The deviation of the dimensionless temperatures with respect to $1$ can be expressed as a relative change in the energy density, $\delta \bar{\rho}_{\nu_\alpha} \simeq 4 (z_{\nu_\alpha} - 1)$. Our values for the increase in the neutrino energy density are $\delta \bar{\rho}_{\nu_e} \simeq  0.70 \, \%$, $\delta \bar{\rho}_{\nu_\mu} \simeq  0.53 \, \%$ and $\delta \bar{\rho}_{\nu_e} \simeq  0.52 \, \%$. This is in agreement with the results of~\cite{Relic2016_revisited} (Table 1) or~\cite{Akita2020} (Table 2), except for the relative variation of muon and tau flavours: these works obtain a higher reheating of $\nu_\tau$ compared to $\nu_\mu$, while we find the opposite. This is due to a difference in the values of the mixing angles.\footnote{For instance, the older values used in \cite{Mangano2005} lead to higher distortions for $\nu_\mu$ than for $\nu_\tau$.} Nevertheless, if we use the mixing angles from \cite{Relic2016_revisited}, we obtain $\delta \bar{\rho}_{\nu_e} \simeq  0.694 \, \%$, $\delta \bar{\rho}_{\nu_\mu} \simeq  0.525 \, \%$ and $\delta \bar{\rho}_{\nu_\tau} \simeq  0.530 \, \%$. Furthermore, if  $\mathcal{O}(e^3)$ QED corrections are not included and only the diagonal components of the self-interaction collision term are kept, the spectra reach less flavour equilibration and the results of \cite{Relic2016_revisited} are recovered (at the level of a few $10^{-5}$): $\delta \bar{\rho}_{\nu_e} \simeq  0.706 \, \%$, $\delta \bar{\rho}_{\nu_\mu} \simeq  0.515 \, \%$ and $\delta \bar{\rho}_{\nu_\tau} \simeq  0.522 \, \%$.

We emphasize that our work is the first to include the full form of the self-interaction collision term and QED corrections to the plasma thermodynamics up to $\mathcal{O}(e^3)$ order. Compared to~\cite{Relic2016_revisited}, the closeness of our results is due to a compensation between the update in the value of physical parameters (namely $\mathcal{G}$, $G_F$) and the new ingredients of our calculation. Our results have been independently confirmed in~\cite{Bennett2021}, providing the new reference value for $\Neff$~\cite{PDG}.

\paragraph{Validity of the ATAO approximation} Finally, the results in Table~\ref{Table:Res_NuDec} show the striking accuracy of the ATAO approximation, as expected since the conditions \eqref{eq:adiab_cond} and \eqref{eq:collis_cond} are satisfied by several orders of magnitude (Figures~\ref{fig:ODG_adiab} and \ref{fig:ODG_coll}). The frozen-out values of the comoving temperatures and of $\Neff$ differ by $10^{-6}$, which is beyond our desired accuracy, and beyond the expected effect of neglected contributions.

In chapter~\ref{chap:Asymmetry}, we extend the ATAO approximation to the case of a non-zero neutrino/antineutrino asymmetry, showing once again its accuracy and how it provides a very efficient numerical way to tackle the problem of neutrino evolution.

\paragraph{Increase of $\bm{\Neff}$ due to mixing} The numerical solution of the QKE shows a larger $\Neff$ value (Table~\ref{Table:Res_NuDec}) compared to the no-oscillation case. To understand this slight increase of the total energy density of neutrinos,  one should keep in mind that electron-positron annihilations, which is the dominant energy-transferring process during decoupling~\cite{Grohs2015}, are more efficient in producing electronic type neutrinos (because of the existence of charged-current processes). Now the mixing and mean-field terms tend to depopulate $\nu_e$ and populate the other flavours, which frees some phase space for the reactions which create $\nu_e$, while increasing the effect of Pauli-blocking factors for reactions creating $\nu_{\mu,\tau}$. Since the former are the dominant reactions, the net effect is a larger entropy transfer from $e^\pm$, hence the larger value of $\Neff$. We further clarify the effect of mixing and mean-field terms in the light of the ATAO approximation in the section~\ref{sec:dependence_parameters}.

\vspace{0.5 cm}

Before discussing the various limitations to this calculation in the next section, we can quote the main result, namely the value of $\Neff$ in the Standard Model of particle physics and in the $\Lambda$CDM model of cosmology,
\begin{equation}
\label{eq:result_Neff}
\boxed{\Neff = 3.0440} \ ,
\end{equation}
read on the two last rows of Table~\ref{Table:Res_NuDec}. The uncertainty on this result will be discussed in section~\ref{sec:dependence_parameters}.

\subsection{Limitations of the calculation}
\label{subsec:limitations_NuDec}

Let us discuss the some physical features that are not taken into account in this calculation of neutrino decoupling.

\paragraph{Neglected corrections to the plasma thermodynamics} 
As explained in section~\ref{subsec:QED}, the energy density and pressure of $e^\pm$ and photons are modified at finite temperature, which is taken into account via the functions $G_1$ and $G_2$ in~\eqref{eq:zQED}. These functions are computed from the partition function of the QED plasma, in an expansion in powers of $e$. In our calculation, we included the contributions at order $\mathcal{O}(e^2)$ and $\mathcal{O}(e^3)$.

However, we discarded at order $\mathcal{O}(e^2)$ a “log-dependent” term, that is a contribution to the energy density and pressure involving a double integral with a logarithmic dependence on momenta. It was indeed estimated that it should lead to a negligible variation $\Delta \Neff \sim - 5 \times 10^{-5}$ (see Eq.~(4.21) in~\cite{Bennett2020}), a result confirmed in~\cite{Bennett2021} (Table 3).

We have shown that the $\mathcal{O}(e^3)$ contribution leads to a reduction of $\Neff$ by $10^{-3}$, which naturally asks the question of the higher order contributions. We can safely ignore them according to the estimate of~\cite{Bennett2020}, which found for the $\mathcal{O}(e^4)$ correction a contribution $\Delta \Neff \sim 3.5 \times 10^{-6}$ in the ultra-relativistic limit. This departure from $\Neff = 3.044$ is way beyond our uncertainty goal.

Therefore, the finite-temperature QED corrections to the plasma thermodynamics can be considered as fully taken into account at the level of $10^{-4}$ for the value of $\Neff$.

\paragraph{Finite-temperature corrections to the scattering rates} In the problem we consider, finite-temperature corrections do not appear only in bulk thermodynamic quantities, but also in weak scattering rates. Therefore, they should be accounted for in the collision integral $\mathcal{I}$, in reactions involving electrons and positrons. There are four types of such corrections to the weak rates~\cite{Tomalak2019,Bennett2021}: modification to the dispersion relation (which also leads to the modifications of the energy density and pressure), vertex corrections, real emission or absorption of photons, and closed fermion loops. Including them consistently in the collision term is a very tedious task that has not been done yet.

The only estimate of these effects in the literature has been done using the value of the difference in the energy loss rate in $e^- + e^+ \to \nu + \bar{\nu}$ due to the rate corrections in~\cite{Esposito_QED}, which would lead to $\Delta \Neff \sim - 0.001$ according to~\cite{Escudero_2020}. However, the estimate of~\cite{Esposito_QED} uses “temperature-dependent wave-function renormalization” techniques, which have led to disagreements in the literature (notably with detailed balance requirements being missed~\cite{BrownSawyer}). We thus conclude that a specific study of finite-temperature corrections to the scattering rates and their effect on neutrino decoupling must be undertaken.

\paragraph{Hypothesis of isotropy} There is, finally, a global underlying assumption that simplifies considerably the problem: isotropy. Indeed, without this hypothesis the density matrices would depend on both the direction and the magnitude of $\vec{p}$. This leads, for instance, to non-vanishing angular integrals in the self-interaction mean-field (that is important in the asymmetric case, see chapter~\ref{chap:Asymmetry}). Moreover, isotropy is also used in the reduction of the collision integral down to two dimensions. For these reasons, the calculation would be considerably more involved if one released this assumption.

We can quote the work~\cite{Hansen_Isotropy}, which tried to tackle this problem with a simplified framework: two bins of neutrinos (left-moving and right-moving, as a very crude anisotropic model), and a simplified collision term. Their results seem to indicate that the mean-field asymmetric neutrino contribution that we discarded might sometimes play a role. However, their treatment of anisotropies remains incomplete: even though a simple anisotropic model might be necessary at this stage, one should also consistently consider the anisotropic degrees of freedom in the metric, whose dynamics is governed by Einstein equations.

\subsection{Neutrinos today}

The neutrino spectra obtained after decoupling remain frozen from that point onwards, so our numerical results allow us to calculate the thermodynamic quantities expected for neutrinos today.

\subsubsection{Neutrino energy density}
\label{subsec:Omeganu}

Given the value of the CMB temperature today $T_\mathrm{CMB}=2.7255 \pm 0.0006 \, \mathrm{K} = (2.3487 \pm 0.0005) \times 10^{-4} \, \mathrm{eV}$~\cite{Fixsen:2009,PDG}, we can estimate the temperature of the C$\nu$B in the instantaneous decoupling limit:\footnote{We do not transfer the uncertainties from $T_\mathrm{CMB}$ since they are below the $\sim 0.1 \, \%$ variation of $T_\nu$ due to incomplete neutrino decoupling, and thus meaningless.}
\begin{equation}
T_\mathrm{C\nu B} = \left(\frac{4}{11}\right)^{1/3} T_\mathrm{CMB} = 1.945 \, \mathrm{K} = 1.676 \times 10^{-4} \, \mathrm{eV} \, .
\end{equation}
Given the values of the mass-squared differences~\eqref{ValuesStandard} $\Delta m_{21}^2 \simeq 7.5 \times 10^{-5} \, \mathrm{eV}^2$ and $\lvert \Delta m_{31}^2 \rvert \simeq 2.5 \times 10^{-3} \, \mathrm{eV}^2$ , the two heaviest mass eigenstates have necessarily masses $m_{\nu_i} \gg T_\mathrm{C\nu B}$ today. We could be more precise by using the results from section~\ref{subsec:results_Neff} and compare $m_{\nu_i}$ and the effective $T_{\nu_i}$ today, but the conclusion would be identical since $T_{\nu_i}$ and $T_\mathrm{C\nu B}$ differ by $\simeq 0.1 \, \%$ (cf.~Table~\ref{Table:Res_NuDec}).

We can then write the total neutrino energy density parameter as
\begin{equation}
\label{eq:calcul_Omeganu}
\Omega_\nu = \frac{\rho_\nu + \rho_{\bnu}}{\rho_\mathrm{crit}} \simeq \frac{2 \sum_{i}{(m_{\nu_i} n_{\nu_i})}}{\rho_\mathrm{crit}} \, ,
\end{equation}
where we assume a zero asymmetry and that all mass species are non-relativistic today. Should this be wrong, the error made would be completely negligible since $\rho_\nu$ is strongly dominated, in the matter-dominated era, by the contribution from the non-relativistic neutrinos.

The critical density today is \cite{PDG}
\begin{equation*}
 \rho_\text{crit} = \frac{3 H_0^2}{8 \pi \mathcal{G}} = 8.0959 \times 10^{-11} \times h^2 \, \mathrm{eV^4} \equiv \rho_\text{crit}^{100} \times h^2
 \end{equation*}
This value corresponds to the updated value for Newton's constant of gravitation $\mathcal{G} = 6.67430 \times 10^{-11} \, \mathrm{m^3 \cdot kg^{-1} \cdot s^{-2}}$. $h$ is the present value of the Hubble parameter in units of $100 \, \mathrm{km \cdot s^{-1} \cdot Mpc^{-1}}$.

The conversion between the comoving quantities and the physical ones is done in the following way:
\begin{equation*}
n_{\nu_i} = \bar{n}_{\nu_i} \times \Tcm^3 = \bar{n}_{\nu_i} \times \left(\frac{T_\gamma}{z_\gamma}\right)^3
\end{equation*}
Today $T_\gamma = T_\mathrm{CMB}$, so we rewrite
\begin{equation}
\label{eq:Omeganu}
\Omega_\nu = \frac{2 \sum_{i}{m_{\nu_i} \bar{n}_{\nu_i}}}{\rho_\text{crit}^{100} \times h^2} \times \frac{T_\mathrm{CMB}^3}{z_\gamma^3} \, .
\end{equation}

\paragraph{Instantaneous decoupling}

In the instantaneous decoupling limit,
\begin{equation*}
\boxed{\text{Inst. Dec.}} \quad \left\{ 
\begin{aligned}
z_\gamma &= \left(\frac{11}{4}\right)^{1/3} \\
\bar{n}_\mathrm{ID} &= \frac{3 \zeta(3)}{4 \pi^2}
\end{aligned} \right. \, .
\end{equation*}
From that, we can compute
\[ \frac{\rho_\text{crit}^{100} z_\gamma^3}{2 \bar{n}_\mathrm{ID} T_\mathrm{CMB}^3} = \frac{\rho_\text{crit}^{100} \times \frac{11}{4}}{2 \times \frac{3 \zeta(3)}{4 \pi^2} \times T_\mathrm{CMB}^3} \simeq 94.06 \, \mathrm{eV} \, . \]
Therefore \eqref{eq:Omeganu} becomes:
\begin{equation}
\label{eq:Omeganu_ID}
\boxed{\Omega_\nu^\mathrm{ID} = \frac{\sum_{i}{m_{\nu_i}}}{94.06 \, \mathrm{eV} \times h^2}} \, .
\end{equation}

\paragraph{Incomplete neutrino decoupling} Thanks to our results, rewritten in the mass basis in Table~\ref{Table:Resmass_NuDec}, we can actually make a precise prediction for $\Omega_\nu$.

\begin{table}[!htb]
	\centering
	\begin{tabular}{|l|ccccc|}
  	\hline 
  Final values & $z$ & $z_{\nu_1}$  & $z_{\nu_2}$ & $z_{\nu_3}$ &$\Neff$  \\
  \hline \hline
    ATAO, QED $\mathcal{O}(e^3)$ & $1.39797$ & $1.00191$ & $1.00143$ & $1.00102$ & $3.04396$   \\ \hline
\end{tabular}
	\caption[Frozen-out values of the comoving temperatures for the mass eigenstates]{Frozen-out values of the dimensionless photon and neutrino effective temperatures, defined for the \emph{mass} eigenstates. We only report the ATAO values, which differ from the full QKE results at the level of $10^{-6}$.
	\label{Table:Resmass_NuDec}}
\end{table}

As the differences between the different $z_{\nu_i}$ are very small (see Table~\ref{Table:Resmass_NuDec}), the approximation $n_{\nu_1} \simeq n_{\nu_2} \simeq n_{\nu_3}$ is often made, allowing to factorize out the sum of neutrino masses $\sum{m_{\nu_i}}$ in~\eqref{eq:calcul_Omeganu}. Alternatively, one can assume the neutrino masses to be quasidegenerate, i.e.~$m_0 \equiv m_{\nu_1} \simeq m_{\nu_2} \simeq m_{\nu_3} \gg T_\nu^0$. In this case, the differences of masses can be neglected and we can once again factorize them. We then have
\[ \frac{\rho_\text{crit}^{100} z_\gamma^3}{2(\bar{n}_{\nu_1} + \bar{n}_{\nu_2} + \bar{n}_{\nu_3}) T_\mathrm{CMB}^3} \simeq 31.04 \, \mathrm{eV} \, . \]
Thus we can write the result as
\begin{equation}
\label{eq:Omeganu_degenerate}
\boxed{\Omega_\nu = \frac{3 m_0}{93.12 \, \mathrm{eV} \times h^2} = \frac{\sum_{i}{m_{\nu_i}}}{93.12 \, \mathrm{eV} \times h^2}} \, ,
\end{equation}
as long as we do not make a difference between the masses today. We can compare this to the previous result~\cite{Mangano2005}, whose denominator value was $93.14 \, \mathrm{eV}$. Although this is extremely close to what our much more precise study of neutrino decoupling gives, one must remember that some differences are hidden in the new values of the physical constants. For instance, the instantaneous decoupling value derived in Eq.~\eqref{eq:Omeganu_ID} above ($94.06 \, \mathrm{eV}$) was quoted to be $94.12 \, \mathrm{eV}$ at the time of the study~\cite{Mangano2005}.

\begin{figure}[!ht]
	\centering
	\includegraphics{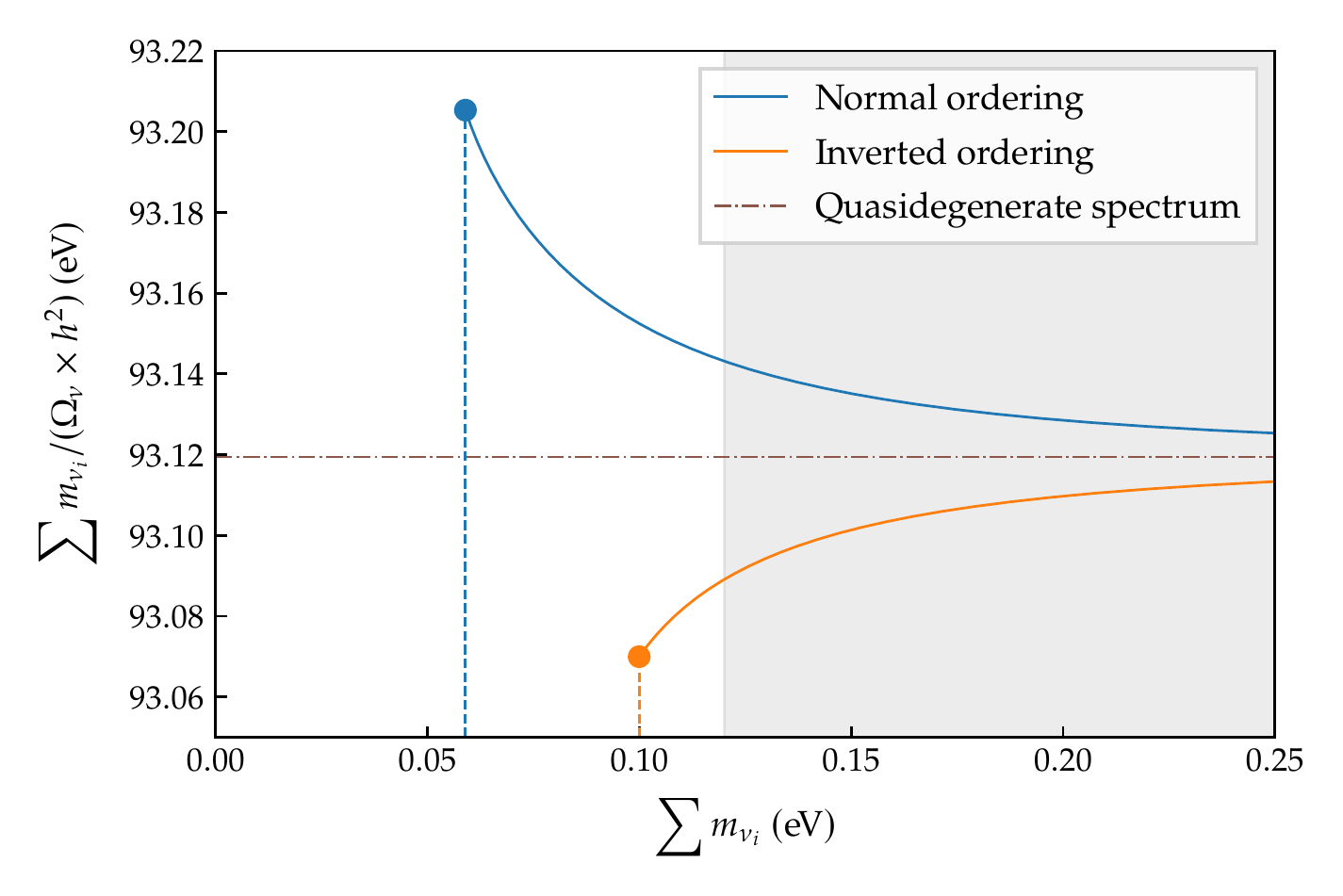}
	\caption[Dependence of the neutrino energy density parameter today with the sum of neutrino masses]{\label{fig:Omeganu} Dependence of the neutrino energy density parameter today with the sum of neutrino masses. The grey area is the 95 \% excluded zone by~\cite{Planck18}, $\sum{m_{\nu_i}} < 0.12 \, \mathrm{eV}$. The brown dash-dotted line corresponds to the value quoted in~\cite{Mangano2005,PDG}, for which $m_{\nu_1} \simeq m_{\nu_2} \simeq m_{\nu_3} \gg T_{\nu}^0$, cf.~equation~\eqref{eq:Omeganu_degenerate}.}
\end{figure}

In order to fully exploit the results of our study, we plot in Figure~\ref{fig:Omeganu} the generalization of this coefficient to any value of the masses. We vary the minimal neutrino mass and deduce the other two given the values~\eqref{ValuesStandard}, depending on the choice of mass ordering. The endpoints of each line correspond to the minimal sum of masses, respectively reached for $m_{\nu_1} = 0$ (normal ordering) and $m_{\nu_3} = 0$ (inverted ordering). We also show the exclusion zone obtained in~\cite{Planck18}, showing the current range of variation of this coefficient, and thus, of $\Omega_\nu$.

\subsubsection{Neutrino number density}

To obtain the neutrino energy density today, we have actually computed their number density for each mass eigenstate. Summing over all states, the total number density reads
\begin{equation} 
\sum_{i}{(n_{\nu_i} + n_{\bar{\nu}_i})} \simeq 339.5 \, \mathrm{cm^{-3}} \, .
\end{equation}
This is, among all astrophysical and cosmological sources, the largest neutrino density at Earth (cf.~the Grand unified neutrino spectrum in~\cite{Vitagliano_Review}). However, due to the very small energy of these neutrinos today, their direct detection is a considerable task~\cite{Neutrino_Cosmology,PTOLEMY2018,PTOLEMY2019}.

\section{Dependence on the physical parameters}
\label{sec:dependence_parameters}

In this section, we discuss how the results obtained in section~\ref{sec:results_nudec} depend on the various physical parameters that enter in the calculation. This allows us to get a nominal uncertainty on the main result~\eqref{eq:result_Neff}. Moreover, we also explore the physical role played by flavour oscillations, aided by the ATAO approximation.

\subsection{Effect of the CP phase}
\label{subsec:Decoupling_CP}

The standard calculation we presented does not include a non-zero Dirac CP phase, while this is not excluded by oscillation data. Indeed, the analysis of the appearance channels $\nu_\mu \to \nu_e$ and $\bnu_\mu \to \bnu_e$ (whose oscillation probabilities get opposite shifts due to the CP phase) in long-baseline accelerator experiments (T2K, NO$\nu$A) and in neutrino atmospheric data (Super-Kamiokande) show a preference\footnote{Note that there is a relative tension between the determinations of $\delta$ obtained from T2K and NO$\nu$A data in the normal mass ordering case, while there is an excellent agreement in the inverted ordering scenario~\cite{Esteban2020,Kelly2020}.} for values $\delta \neq 0^{\circ}, \, 180^{\circ}$.  We show in this section why this is nevertheless a safe choice, as such a phase does not affect our results.

The generalized parameterization of the PMNS matrix~\eqref{eq:PMNS} when including a CP violating phase\footnote{We do not include possible Majorana phases that have no effect on neutrino oscillations.} reads
\begin{equation}
\label{eq:PMNS_CP}
U = R_{23} S R_{13} S^\dagger R_{12} = \begin{pmatrix} 
c_{12} c_{13} & s_{12} c_{13} & s_{13} e^{-\ii \delta} \\
- s_{12}c_{23} - c_{12}s_{23}s_{13} e^{\ii \delta} & c_{12} c_{23} - s_{12}s_{23}s_{13} e^{\ii \delta} & s_{23} c_{13} \\
s_{12}s_{23} - c_{12}c_{23}s_{13} e^{\ii \delta} & -c_{12}s_{23} - s_{12}c_{23}s_{13} e^{\ii \delta} & c_{23} c_{13}
\end{pmatrix} \, ,
\end{equation}
where $S = \mathrm{diag}(1,1,e^{\ii \delta})$.  The best-fit value quoted in~\cite{PDG} is $\delta = 1.36 \, \pi \, \mathrm{rad}$, while~\cite{deSalas_Mixing} quotes $\delta_\mathrm{NO} = 1.08 \, \pi \, \mathrm{rad}$ and $\delta_\mathrm{IO} = 1.58 \, \pi \, \mathrm{rad}$. We will use, as always in this manuscript, the value from~\cite{PDG}, but the conclusions remain the same regardless of the value chosen.

Although this new phase affects the vacuum oscillation term in the QKEs, it is actually possible to factorize this dependence and reduce the problem to the case $\delta = 0$, in some limits that we expose below. We revisit the derivation of \cite{Balantekin:2007es,Gava:2008rp,Gava:2010kz,Gava_corr}, where conditions under which the CP phase has an impact on the evolution of $\vrho$ in matter were first uncovered.

In this section, we will note with a superscript $^0$ the quantities in the $\delta =0$ case. We introduce a convenient unitary transformation 
\[ \check{S} \equiv R_{23} S R_{23}^\dagger \, , \]
and define $\check{\vrho} \equiv \check{S}^\dagger \vrho \check{S}$ (likewise for $\bvrho$). Let us now prove that $\check{\vrho} = \vrho^0$. First, we need to show that $\check{\vrho}$ has the same evolution equation as $\vrho^0$. Let us rewrite the QKE~\eqref{eq:QKE_final} in a very compact way:
\begin{equation}
\label{eq:QKE_compact_CP}
\ii \frac{\partial \vrho}{\partial x} =  \lambda [U \mathbb{M}^2 U^\dagger,\vrho] + \mu [\bar{\mathbb{E}}_\mathrm{lep} + \bar{\mathbb{P}}_\mathrm{lep}, \vrho]  + \ii \mathcal{K}(\vrho,\bvrho) \, ,
\end{equation}
with coefficients $\lambda, \mu$ which can be read from \eqref{eq:QKE_final}. Applying $\check{S}^\dagger (\cdots) \check{S}$ on both sides of the QKE gives the evolution equation for $\check{\vrho}$. 

First, using that $\check{S}^\dagger U = U^0 S^\dagger$ (we recall that $U^0$ is the PMNS matrix without CP phase) and that $\mathbb{M}^2$ and $S$ commute since they are diagonal, the vacuum term reads $\check{S}^\dagger [U \mathbb{M}^2 U^\dagger,\vrho] \check{S} = [U^0 \mathbb{M}^2 {U^0}^\dagger, \check{\vrho}] $. 

Then, the mean-field term satisfies 
\begin{equation}
\label{eq:meanfield_Sdagger}
\check{S}^\dagger [\bar{\mathbb{E}}_\mathrm{lep} + \bar{\mathbb{P}}_\mathrm{lep}, \vrho] \check{S} \simeq [\bar{\mathbb{E}}_\mathrm{lep} + \bar{\mathbb{P}}_\mathrm{lep}, \check{\vrho}] \, .
\end{equation}
This equality would be exact if we completely neglected the mean-field contribution of the background muons, as the energy density would read $\bar{\mathbb{E}}_\mathrm{lep} \simeq \mathrm{diag}(\rho_{e^-} + \rho_{e^+}, 0, 0)$ (likewise for the pressure). This is justified since the energy density of muons is negligible compared to the electron one across the decoupling era, and it results in muon and tau neutrinos having the very same interactions, a condition evidenced in~\cite{Gava:2010kz}. Moreover, in the region of high temperatures where the muon energy density get closer to the electron one (even though, at $20 \, \mathrm{MeV}$, muons are still largely non-relativistic), the commutator of $\bar{\mathbb{E}}_\mathrm{lep}$ with $\vrho$ vanishes as $\vrho \propto \Id$. The density matrix differs from $\Id$ only when $e^\pm$ annihilations are effective, which is “too late” for the muon mean-field to have an effect. Therefore, we can safely take~\eqref{eq:meanfield_Sdagger} as exact.

Finally, the collision term contains products of density matrices and $G_{L,R}$ coupling matrices for the scattering/annihilation terms with electrons and positrons. Since $[G_{L,R}, \check{S}^{(\dagger)}]=0$ (as we only consider standard interactions), we can write $\check{S}^\dagger \mathcal{K}(\vrho, \bvrho) \check{S} = \mathcal{K}(\check{\vrho},\check{\bvrho})$. Once again, the fact that $\nu_\mu$ and $\nu_\tau$ have identical interactions is key to this factorization, as pointed out in~\cite{Gava:2010kz} and previously in~\cite{Balantekin:2007es,Gava:2008rp} in the astrophysical context. In~\cite{Gava:2010kz}, the collision term is approximated by a damping factor ; the factorization then holds since the damping coefficients are identical whether they involve $\nu_\mu$ or $\nu_\tau$.

Therefore, the QKE for $\check{\vrho}$ reads:
\begin{equation}
\label{eq:QKE_compact_CP_S}
\ii \frac{\partial \check{\vrho}}{\partial x} =  \lambda [U^0 \mathbb{M}^2 {U^0}^\dagger,\check{\vrho}] + \mu [\bar{\mathbb{E}}_\mathrm{lep} + \bar{\mathbb{P}}_\mathrm{lep}, \check{\vrho}]  + \ii \mathcal{K}(\check{\vrho},\check{\bvrho}) \, ,
\end{equation}
which is exactly the QKE for $\vrho^0$, i.e.the QKE without CP phase. Moreover, the initial condition~\eqref{eq:initial_condition} is unaffected by the $\check{S}$ transformation: $\check{\vrho}(x_\mathrm{in},y) = \vrho^0(x_\mathrm{in},y)$. Since the initial conditions and the evolution equations are identical for $\check{\vrho}$ and $\vrho^0$, then at all times $\vrho^0(x,y) = \check{\vrho}(x,y)$. We can therefore write the relation between the density matrices with and without CP phase,
\begin{equation}
\label{eq:CP_flavour}
\vrho(x,y) = \check{S} \vrho^0(x,y) \check{S}^\dagger \, .
\end{equation}
This relation has two major consequences:
\begin{enumerate}
	\item The trace of $\vrho$ is unaffected by $\delta$, therefore $\Neff = \Neff(\delta = 0)$ ;
	\item The first diagonal component is unchanged $\vrho^e_e = (\vrho^0)^e_e$. Equivalently with the parameterization~\eqref{eq:param_rho}, $z_{\nu_e} = z_{\nu_e}^0$ and $\delta g_{\nu_e} = \delta g_{\nu_e}^0$.
\end{enumerate}
Therefore, under the assumptions made above (in particular, the initial distribution has no chemical potentials), the CP phase will have no effect on BBN, since light element abundances are only sensitive to $\Neff$, $z_{\nu_e}$ and $\delta g_{\nu_e}$ (see chapter~\ref{chap:BBN}). Note that in presence of initial degeneracies, the initial conditions do not necessarily coincide $\check{\vrho}(x_\mathrm{in},y) \neq \vrho^0(x_\mathrm{in},y)$ and signatures of a CP phase could in principle be found in the primordial abundances~\cite{Gava:2010kz,Gava_corr}. We discuss this topic in section~\ref{SecDiracPhase}.

A useful rewriting of~\eqref{eq:CP_flavour} can be made with the final distributions ($x = x_f$), when mean-field effects are negligible. The correspondence between the $\delta =0$ and $\delta \neq 0$ cases reads in the matter basis (which is then the mass basis):
\begin{equation}
 \vrho_{\Hvac}(x_f,y) = S \vrho_{\Hvac}^0(x_f,y) S^\dagger \ .   \label{eq:CP_mass}
\end{equation}
Note that the transformation involves now $S$ instead of $\check{S}$ (this is linked to the fact that the transformation between $\vrho$ and $\vrho_{\Hvac}$ is made through $U$, while the transformation between $\vrho^0$ and $\vrho_{\Hvac}^0$ involves $U^0$). We can go further using the ATAO approximation, which constrains the form of $\vrho$ and allows to analytically estimate the effect of the CP phase. \emph{In the ATAO approximation}, $\vrho_{\Hvac}^0$ is diagonal, such that we get the result:
\begin{equation}
 \text{\textsc{\bfseries ATAO}} \qquad \qquad   \vrho_{\Hvac}(x_f,y) = \vrho_{\Hvac}^0(x_f,y) \, . \qquad \qquad
\end{equation}
Defining effective temperatures $z_{\nu_i}$ for the mass states ($i=1,2,3$), we have then $z_{\nu_i} = z_{\nu_i}^0$. Using the PMNS matrix to express the results in the flavour basis, the effective temperatures read:\footnote{These expressions are rigorously exact for the energy densities, and they can be rewritten for the effective temperatures since $z_\nu - 1 \ll 1$.}
\begin{equation}
\label{eq:resultsCP}
\begin{aligned}
z_{\nu_e} &= z_{\nu_e}^0 \ , \\
z_{\nu_\mu} &= z_{\nu_\mu}^0 - \frac12 (z_{\nu_1} - z_{\nu_2}) \sin{(2 \theta_{12})} \sin{(\theta_{13})}\sin{(2 \theta_{23})} \left[ 1 - \cos{(\delta)}\right] \ , \\
z_{\nu_\tau} &= z_{\nu_\tau}^0 + \frac12 (z_{\nu_1} - z_{\nu_2}) \sin{(2 \theta_{12})} \sin{(\theta_{13})}\sin{(2 \theta_{23})} \left[ 1 - \cos{(\delta)}\right] \ .
\end{aligned}
\end{equation}
These relations show that the CP phase only affects the muon and tau neutrino distribution functions, with a $[\cos{(\delta)} - 1]$ dependence. For the preferred values of $\delta = 1.36 \, \pi \, \mathrm{rad}$ and the mixing angles~\cite{PDG}, and with the results for $\delta=0$ from section~\ref{sec:results_nudec}, we expect $\lvert z_{\nu_\mu} - z_{\nu_\mu}^0 \rvert =  \lvert z_{\nu_\tau} - z_{\nu_\tau}^0 \rvert \simeq 4.7 \times 10^{-5}$.  This is in excellent agreement with the numerical results obtained solving the QKE with a CP phase (see Table~\ref{Table:Res_NuDec_CP}).

\begin{table}[!htb]
	\centering
	\begin{tabular}{|l|ccccc|}
  	\hline 
  Final values & $z$ & $z_{\nu_e}$  & $z_{\nu_\mu}$ & $z_{\nu_\tau}$ &$\Neff$  \\
  \hline \hline
    $\delta = 0$ & $1.39797$ & $1.00175$ & $1.00132$ & $1.00130$ & $3.04396$   \\
    $\delta = 1.36 \, \pi \, \mathrm{rad}$  & $1.39797$ & $1.00175$ & $1.00127$ & $1.00135$ & $3.04396$   \\ \hline
\end{tabular}
	\caption[Final effective temperatures with and without CP phase]{Frozen-out values of the dimensionless photon and neutrino temperatures, and the effective number of neutrino species. We compare the results without CP phase (see also Table~\ref{Table:Res_NuDec}) and with the average value for $\delta$ from~\cite{PDG}.
	\label{Table:Res_NuDec_CP}}
\end{table}

Finally, the antineutrino density matrices satisfy the same relation as for neutrinos~\eqref{eq:CP_flavour} $\bvrho(x,y) = \check{S} \bvrho^0(x,y) \check{S}^\dagger$. The QKEs in the absence of CP phase preserve the property $\vrho^0 = {(\bvrho^0)}^{*}$ if it is true initially. The asymmetry with $\delta \neq 0$ would then read $\vrho - \bvrho = \check{S}(\vrho^0 - {\vrho^0}^*)\check{S}^\dagger$. Therefore, CP violation effects in the $\nu_\mu$ and $\nu_\tau$ distributions (which would be contributions $\propto \sin{\delta}$) can arise from the complex components of $\vrho^0$, thus requiring the ATAO approximation to break down. Since in the cosmological context without initial degeneracies the approximation is very well satisfied, there can be no additional CP violation and the formulae~\eqref{eq:resultsCP} are equally valid for antineutrinos.

Let us mention that we also performed a calculation solving the full QKEs for both neutrinos and antineutrinos, with a non-zero CP phase, and the results were once again the same as the ATAO approximate ones and consistent with~\eqref{eq:CP_mass}.

\subsection{Role of flavour oscillations}

As we have shown in section~\ref{subsec:results_Neff}, flavour oscillations do not significantly affect the value of $\Neff$ compared to, for instance, the inclusion of $\mathcal{O}(e^3)$ finite-temperature QED corrections to the plasma thermodynamics, although they affect much more importantly the neutrino spectra (see Figures~\ref{fig:Tnu} and~\ref{fig:deltagnu}). Therefore, if one is interested in understanding how flavour oscillations modify the physics at play during decoupling --- for instance because $\Neff$ is not the only relevant quantity (for BBN, for instance) --- a deeper study is called for. Notably, we will study precisely the changes brought by the vacuum + mean-field terms in the QKE, and how the mixing parameters play a role (through their values or their sign, for instance considering the inverted hierarchy of masses).

\paragraph{ATAO transfer functions}

The ATAO approximation allows to get some insight on the impact of the mixings and mean-field terms, as its extreme accuracy shows that flavour oscillations “only act as” changes of matter basis across the evolution.

\begin{figure}[ht]
	\centering
     \includegraphics{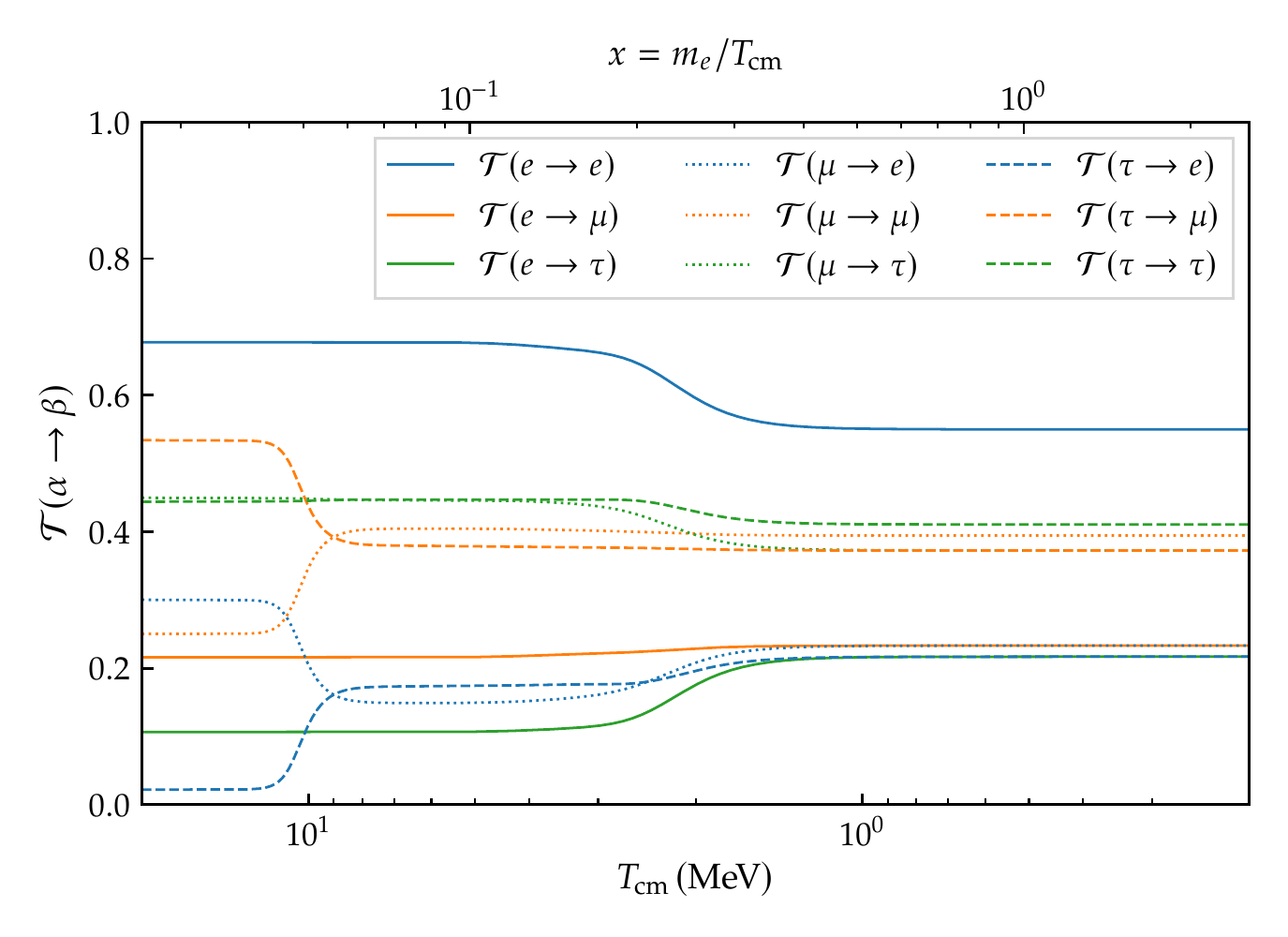}
	\caption[Asymptotic ATAO transfer functions (normal ordering)]{\label{fig:ATAO_Transfer} Asymptotic ATAO transfer function $\mathcal{T}(\alpha \to \beta, x, y)$ for $y = 5$. The asymptotic values for large $x$ correspond to the vacuum oscillation averages~\eqref{VacuumAverage}.}
      \end{figure}

Let us define the \emph{“ATAO transfer function”}
\begin{equation}
\label{eq:defATAO}
\mathcal{T}(\alpha \to \beta, x \to x', y) = \left[U_\Hamil(x',y)  \,\reallywidetilde{\left(U_\Hamil^\dagger(x,y)  D(\alpha) U_\Hamil(x,y)\right)} \,U_\Hamil^\dagger(x',y)\right]^{\beta}_{\beta} \, ,
\end{equation}
where $D(\alpha)$ is a diagonal matrix with a non-vanishing (unit) component, that is $\left[D(\alpha)\right]^\beta_{\gamma} \equiv \delta_\alpha^\beta \delta^\alpha_\gamma$ (no summation). Equation \eqref{eq:defATAO} corresponds to the probability for a state of flavour $\alpha$ and momentum $y$ generated at a pseudo scale factor $x$, ``averaged'' according to the ATAO approximation, to re-emerge as a flavour $\beta$ at later $x'$, if it is not affected by collisions in the meantime.
When evaluated at $x' \to \infty$, the asymptotic $\mathcal{T}(\alpha \to \beta, x, y) \equiv \mathcal{T}(\alpha \to \beta, x \to \infty, y)$ provide information on neutrino flavour conversion from their last scattering with other species, until all neutrino spectra are frozen since mean-field and collisions are then negligible. These asymptotic functions are shown on Figure~\ref{fig:ATAO_Transfer}.

If mean-field effects can be ignored, the asymptotic ATAO transfer function converges to the following expression
\begin{equation}
\label{VacuumAverage}
\mathcal{T}^\mathrm{vac}(\alpha \to \beta) \equiv \left[U \,\reallywidetilde{\left(U^\dagger D(\alpha) U \right)}\,U^\dagger\right]^{\beta}_{\beta}\,,
\end{equation}
which is independent of $y$ and where the PMNS matter matrix is replaced by the vacuum one. Note that $\mathcal{T}^\mathrm{vac}(\alpha \to \beta) = \mathcal{T}^\mathrm{vac}(\beta \to \alpha)$, as can be seen on Figure~\ref{fig:ATAO_Transfer} at small temperatures.

\subsubsection{Proof that (only) mixing in vacuum matters}

Due to the particular features of this standard calculation (no asymmetries, very small corrections compared to the instantaneous decoupling limit), we will show that the role of flavour mixing on neutrino spectra is essentially captured by considering solely vacuum mixing. To gain this insight on the impact of the mixing and mean-field terms, we have performed two schematic calculations, including either the neutrino probabilities at the end of the evolution, i.e. $T_{\mathrm{cm}, f} = 0.01 \, \mathrm{MeV}$ (``No osc., post-aver.''), or keeping only the mixing and collision terms during the evolution (``Without mean-field''). The corresponding results are shown in Table~\ref{Table:Res_NuDec_Transfer}.

\begin{table}[!htb]
	\centering
	\begin{tabular}{|l|ccccc|}
  	\hline 
  Final values & $z$ & $z_{\nu_e}$  & $z_{\nu_\mu}$ & $z_{\nu_\tau}$ &$\Neff$  \\
  \hline \hline
    No oscillations, QED $\mathcal{O}(e^3)$ &$1.39800$ & $1.00234$ &  $1.00098$ & $1.00098$ &$3.04338$ \\ \hline
  No osc., post-averaging, QED $\mathcal{O}(e^3)$ & $1.39800$ & $1.00173$ & $1.00130$ & $1.00127$ & $3.04340$   \\
W/o mean-field, QED $\mathcal{O}(e^3)$ & $1.39796$ & $1.00175$ & $1.00132$ & $1.00131$ & $3.04405$   \\ \hline 
    ATAO, QED $\mathcal{O}(e^3)$ & $1.39797$ & $1.00175$ & $1.00132$ & $1.00130$ & $3.04396$   \\ \hline
\end{tabular}
	\caption[Results of schematic calculations with an approximate inclusion of flavour oscillation physics]{\label{Table:Res_NuDec_Transfer} Frozen-out values of the dimensionless photon and neutrino temperatures, and the effective number of neutrino species. The no oscillations result is presented here to facilitate the discussion. The post-averaging result corresponds to Eq.~\eqref{eq:DefPost}.}
\end{table}

\paragraph{A crude treatment of flavour mixing: post-averaging method} The goal of the first schematic calculation is to start from the no-mixing results (Boltzmann equation, second row in Table~\ref{Table:Res_NuDec}), and “post-average” them (in the ATAO sense), namely, 
\begin{equation}
\label{eq:DefPost}
(\vrho^\mathrm{post})^\beta_{\beta} \equiv \sum_\alpha (\vrho^\mathrm{NO})^\alpha_\alpha \, \mathcal{T}^\mathrm{vac}(\alpha \to \beta)  \, .
\end{equation}
From Table~\ref{Table:Res_NuDec_Transfer} one can see that the electronic spectra are suppressed and other neutrino types spectra are enhanced by this vacuum averaging procedure. Since one can nearly recover the exact oscillation case results by averaging the final results found without oscillations, it proves that the different values of the effective neutrino temperatures between the no-oscillation case and the full oscillation case are likely to be essentially due to the effect of the mixings. However, the post-averaging of the no-oscillation case preserves, by construction, the trace of $\vrho$ hence the energy density $\rho_\nu$. Therefore, it cannot capture the enhancement of $\Neff$ discussed at the end of section~\ref{subsec:results_Neff}.

\paragraph{Role of the mean-field term}

In the second schematic calculation we have solved the QKEs \eqref{eq:QKE_final} without the mean-field term, i.e., keeping only the vacuum and collision terms.\footnote{We thus have, at all times, $U_\Hamil = U$ and the matter basis coincides with the mass basis.} This is somehow an improvement of the ``post averaging'' procedure, since it neglects the variation of the transfer functions (which always have their asymptotic vacuum values), but accounts correctly for the effect of collisions. The accuracy of the results compared to the full treatment shows once more that the effect of the mean-field is very mild in this case. Indeed, the mean-field contribution becomes effective when $\vrho$ deviates from a matrix proportional to the identity, which only happens when $x \sim 3\times10^{-1}$: however at this point the mean-field contribution is becoming negligible compared to the vacuum one (cf.~Figure~\ref{fig:ATAO_Transfer}). Note that this would not hold if we introduced chemical potentials \cite{Bell98,Dolgov_NuPhB2002,Gava:2010kz,Mirizzi2012,Saviano2013,HannestadTamborra}. The higher value obtained for $\Neff$ in this case can be qualitatively understood. Since $\mathcal{T}^\mathrm{vac}(e \to e) < \mathcal{T}(x \ll1, e \to e)$, $\nu_e$ produced by collisions will be more converted into other flavours (in particular $\nu_\tau$) at early times compared to the full calculation. This frees some phase space for the reheating of $\nu_e$, which is the dominant process. More entropy is transferred from $e^\pm$ annihilations, which increases slightly $\Neff$.

\subsection{Dependence on the mixing angles}
\label{subsec:dependence_mixing}

The transfer functions introduced in the previous section also shed some light on the importance of the precise value of the mixing angles, which explain some discrepancy with previous results (see section~\ref{subsec:results_Neff}). Indeed, varying $\theta_{ij}$ within their uncertainty ranges slightly modify the $\mathcal{T}(\alpha \to \beta)$ curves, which can cross each other. For instance, with the set of parameters used in \cite{Relic2016_revisited}, the asymptotic value $\mathcal{T}^\mathrm{vac}(e \to \tau)$ is higher than $\mathcal{T}^\mathrm{vac}(e \to \mu)$, contrary to Figure~\ref{fig:ATAO_Transfer}. This higher conversion of electron neutrinos into tau neutrinos explains why their final temperatures are $z_{\nu_\tau} \gtrsim z_{\nu_\mu}$ (the values remaining very close).

The experimental uncertainties on the values of the mixing angles~\cite{PDG} lead to small variations of the neutrino distribution functions and $\Neff$. The numerical sensitivity of $\Neff$ to the variation of the mixing angles around their preferred values is found to be:
\begin{equation}
\frac{\partial \Neff}{\partial \theta_{12}} \simeq 8 \times 10^{-4} \ \mathrm{rad^{-1}} \quad ; \quad  
\frac{\partial \Neff}{\partial \theta_{13}} \simeq 9 \times 10^{-4} \ \mathrm{rad^{-1}} \quad ; \quad
 \abs{\frac{\partial \Neff}{\partial \theta_{23}}} \ll  \abs{\frac{\partial \Neff}{\partial \theta_{12}}},  \abs{\frac{\partial \Neff}{\partial \theta_{13}}} \, .
\end{equation} 
The sensitivity with respect to $\theta_{23}$ is much smaller than for the other mixing angles, and cannot be separated from numerical noise. Given the uncertainties at $\pm 1 \sigma$ on the mixing angles~\cite{PDG}, we estimate the associated variation of $\Neff$ to be $\Delta \Neff \simeq 10^{-5}$, beyond our accuracy goal.

\subsection{Inverted mass ordering case}
\label{subsec:Inverted_Hierarchy}

\begin{figure}[!h]
	\centering
     \includegraphics{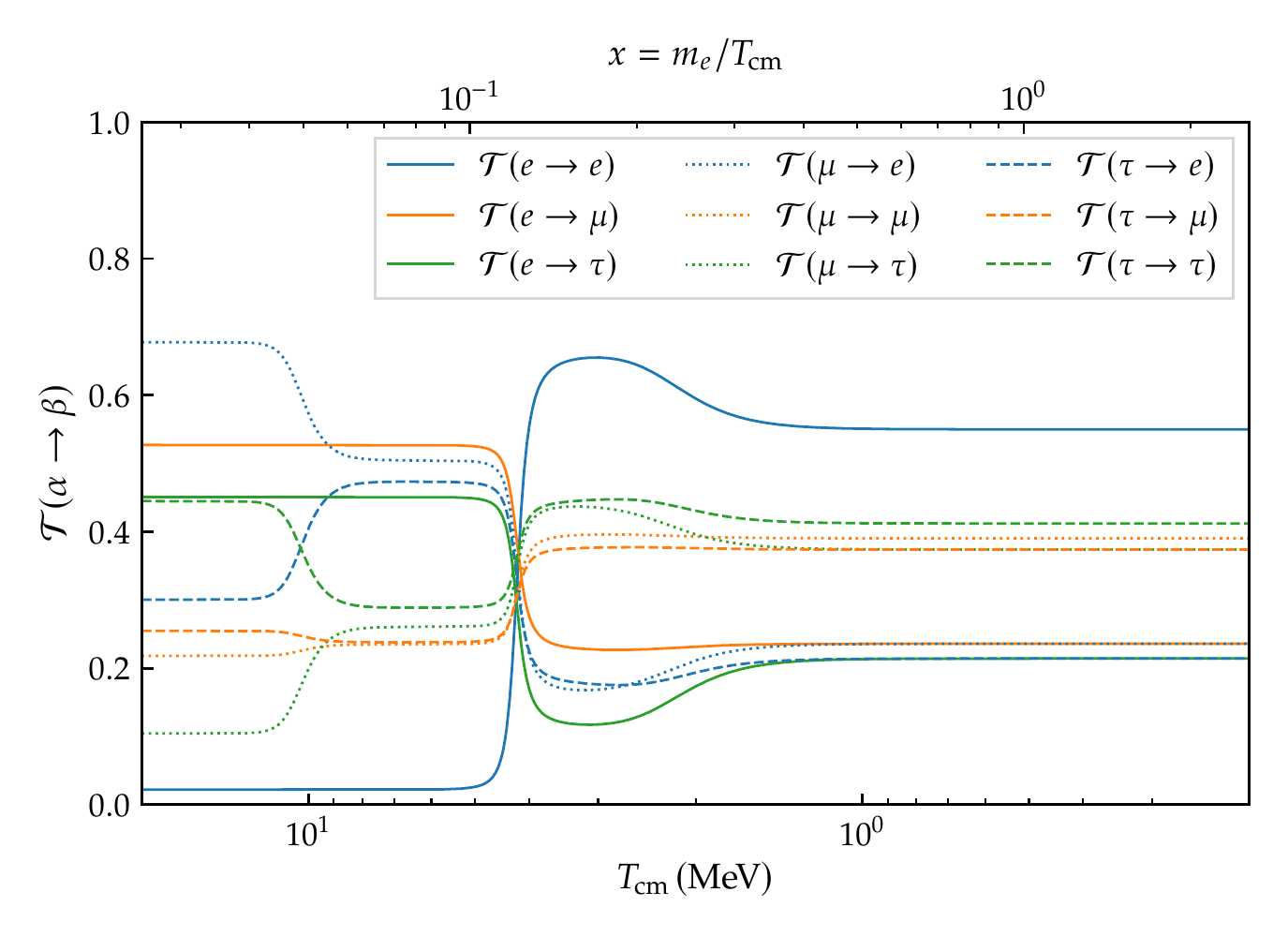}
	\caption[Asymptotic ATAO transfer functions (inverted ordering)]{\label{fig:ATAO_Transfer_IH} Asymptotic ATAO transfer function $\mathcal{T}(\alpha \to \beta, x, y)$ for $y = 5$ in the inverted hierarchy of neutrino masses.}
      \end{figure}

In the inverted mass ordering, for which $\Delta m_{31}^2 < 0$,  we obtain an increase of $\Neff$ by $5\times10^{-6}$ when solving the QKE~\eqref{eq:QKE_final}. To understand this, we plot on Figure~\ref{fig:ATAO_Transfer_IH} the transfer functions $\mathcal{T}(\alpha \to \beta, x, y)$ in the inverted hierarchy case.  Comparing this plot with Figure~\ref{fig:ATAO_Transfer}, we see that electronic neutrinos can be generated above an MSW resonance (e.g.~at about $4\, \mathrm{MeV}$ for $y=5$), and are converted nearly entirely into $\nu_\mu$ and $\nu_\tau$ (solid lines on Figure~\ref{fig:ATAO_Transfer_IH}). Again, this impacts subsequent collisions because it frees some phase space for $\nu_e$, which is beneficial for the total production of neutrinos. However, since neutrino decoupling occurs mainly at temperatures which are below the MSW resonance,\footnote{This is not the case for very large $y$ but they are subdominant in the total energy density budget.} the differences between normal and inverted hierarchies are extremely small. 

Given the discussion of the previous section, it appears that neutrino decoupling is mostly sensitive to the neutrino mixings, whereas it has little sensitivity to the mass-squared differences and therefore to the neutrino mass hierarchy.

However, a recent work by \emph{Hansen et al.}~\cite{Hansen_Isotropy} uncovered potential instabilities that could occur during neutrino evolution. In particular, in the inverted ordering case with symmetric initial conditions, the number densities of neutrinos and antineutrinos are the same, but there can be an effect of different off-diagonal terms that build up and trigger collective effects.  Indeed, if there are non-zero off-diagonal imaginary terms in $\vrho$, then $\vrho-\bvrho$ will be non-zero even though the number densities would be equal. Therefore, including the asymmetric neutrino term in the QKE (which is not the case in our calculation and in other standard studies of neutrino decoupling~\cite{Relic2016_revisited,Bennett2021}), it could be possible to see such imaginary parts grow exponentially. However, in the “standard” calculation of $\Neff$ we performed there should be no such imaginary parts: they are zero initially, and the validity of our ATAO approximation (which amounts to say that $\vrho$ is diagonal in the matter basis, hence notably real in the flavour basis) shows that they should keep vanishing. 

Imaginary parts in the flavour basis can physically only appear from the oscillatory phases of $\vrho$ (that is, from departures from the ATAO approximation). Given the values of the oscillation frequencies and the fact that they are $y$-dependent, the oscillations of the different $y$-modes will be dephased so one would need extremely many grid points to follow the evolution and capture the cancellation --- or not --- of the imaginary parts in $\mathbb{N}_\nu - \mathbb{N}_{\bnu}$.

The results of~\cite{Hansen_Isotropy} nevertheless indicate that, even when the instability occurs, $\Neff$ is changed at most by $5 \times 10^{-4}$. This is due to the very small difference between the distributions of $\nu_e$, $\nu_\mu$ and $\nu_\tau$: even if there are large flavour conversions due to this instability, it will remain a higher order effect compared to the global reheating of the three neutrino flavours.

\pagestyle{ruled}

\chapter{Consequences for Big Bang Nucleosynthesis}
\label{chap:BBN}

\setlength{\epigraphwidth}{0.45\textwidth}
\epigraph{I'm too young to die! And too old to eat off the kids' menu! What a stupid age I am!}{Jason Mendoza, \emph{The Good Place} [S02E02]}

{
\hypersetup{linkcolor=black}
    \minitoc
}

\boxabstract{The material of this chapter was partly published in~\cite{Froustey2019} and~\cite{Froustey2020}.}

During the MeV era, electron-positron annihilations and the subsequent distortions of neutrino spectra throughout their decoupling leave some imprints that we dicussed in the previous chapter: an increased energy density parameterized by $\Neff$, and some non-thermal distortions. However, we only focused on the lepton sector, completely disregarding the baryons that are also present in the early Universe. There are indeed in negligible number, as the baryon-to-photon ratio is estimated at $\eta = n_b/n_\gamma \simeq 6.1\times 10^{-10}$~\cite{Fields:2019pfx}. Yet, the cooling of the Universe allows for the formation of light nuclides during the so-called Big Bang nucleosynthesis (BBN), topic of this chapter. Due to the value of $\eta$, it is justified to take neutrino evolution as a background result and to study how it affects BBN.

We show in Table~\ref{Table:General_BBN} the latest experimental values of the primordial abundances, and the ones predicted numerically. BBN numerical codes have indeed been developed over the past decades since the pioneering work of Wagoner~\cite{Wagoner}, the most widely used being \texttt{PRIMAT}~\cite{Pitrou_2018PhysRept}, \texttt{AlterBBN}~\cite{AlterBBN,AlterBBNv2} or \texttt{PArthENoPE}~\cite{Parthenope,Parthenope_reloaded,Parthenope_revolutions}. In this chapter, the values presented are updated from the works~\cite{Froustey2019,Froustey2020} and obtained with \texttt{PRIMAT}. We recall the notations introduced in~\ref{subsec:intro_BBN}: we call $n_i$ the number density of isotope $i$ and $n_b$ is the baryon density, which allow to define the \emph{number} fraction of isotope $i$, $X_i \equiv n_i/n_b$. The \emph{mass} fraction is therefore $Y_i \equiv A_i X_i$, where $A_i$ is the nucleon number. It is customary to define\footnote{This notation originates from the old astrophysical practice to call the mass fraction of hydrogen $X$, $Y$ for helium and $Z$ for heavier elements (“metals”), while the index $\mathrm{p}$ stands for “primordial”.} $\YP \equiv Y_{\He4}$ and $i/\mathrm{H} \equiv X_i/X_\mathrm{H}$.

\renewcommand{\arraystretch}{1.3}

\begin{table}[!htb]
	\centering
	\begin{tabular}{|l|M{2.8cm} M{2.2cm}  M{2.9cm}  M{2.3cm} |}
  	\hline 
  Abundances & $\YP$  &  $\mathrm{D}/\mathrm{H} \,  (\times 10^{-5})$ &  $\He3/\mathrm{H} \, (\times 10^{-5})$ &  $\Li/\mathrm{H} \,( \times 10^{-10})$ \\
  \hline \hline
 Observations & $0.2453 \pm 0.0034$  \cite{Aver2020}  & $2.527 \pm 0.030$  \cite{Cooke2017} & $\leq 1.1 \pm 0.2$  \cite{Bania2002,Cooke2022}  & $1.6 \pm 0.3$  \cite{Sbordone2010} \\ \hline
 This work & $0.24721 \pm 0.00014$  & $2.438 \pm 0.037$ & $1.039 \pm 0.014$  & $5.505 \pm 0.220$\\ \hline 
\end{tabular}
	\caption[State-of-the-art data on light element abundances]{Light element abundances: latest observations and results from this work.  $\He3$ stands for $\He3 + \mathrm{T}$, and $\Li$ stands for $\Li + \Be$ to account for slow radioactive decays. The uncertainties on the predicted values are obtained assuming that the baryon density is determined from CMB+BAO~\cite{Planck18}, and performing a Monte-Carlo method on the posterior of this baryon abundance, but also on the uncertainties of nuclear rates and the neutron lifetime, see~\cite{Pitrou_2018PhysRept,Pitrou2020}.
	\label{Table:General_BBN}}
\end{table}

The broad agreement between predictions and observations, spanning about nine orders of magnitude from helium-4 to lithium-7, has been a decisive argument towards the validation of the hot Big Bang model. However, problems persist such as the famous \emph{lithium problem}, with no good explanation to this day~\cite{Fields2011,Fields2022}.

This chapter is a synthesis and update of results partly published in two papers. In~\cite{Froustey2019}, we studied the various ways in which incomplete neutrino decoupling affects BBN, without taking into account flavour oscillations. Yet we compared the results depending on the corrections that we included in the BBN code. In~\cite{Froustey2020}, all corrections were included while we focused on the difference between the oscillation and no-oscillation cases. In this chapter, we will always use the results with flavour mixing, but we will study the interplay between neutrino decoupling and BBN without weak rates corrections to keep the discussion simple, before giving the “full” results in section~\ref{subsec:full_BBN}.

Comparisons with respect to a fiducial cosmology, where neutrinos are artificially decoupled instantaneously prior to electron-positron annihilations, require the ability to map different homogeneous cosmologies. There is no unique way to perform this cosmology mapping, that is, to compute variations, similarly to the gauge freedom that exists when comparing a perturbed cosmology with a background cosmology. For instance, we can compare the fiducial instantaneous decoupling with the full neutrino decoupling physics, either using the same cosmological times or the same cosmological factors, or even the same plasma temperatures. The fact that there is no unique choice complicates the discussion of the physical effects at play, but the physical observables, e.g., the final BBN abundances, do not depend on it. We will systematically specify which variable is left constant (cosmic time, scale factor, or photon temperature) when comparing the true Universe to the fiducial one. Quantities written with a superscript $^{(0)}$ correspond to the fiducial (instantaneous decoupling) cosmology, and the variation of a quantity $\psi$ will be written as
\begin{equation*}
\delta \psi \equiv \frac{\Delta \psi}{\psi^{(0)}} \equiv \frac{\psi - \psi^{(0)}}{\psi^{(0)}} \, .
\end{equation*}

\section{Incomplete neutrino decoupling and BBN}
\label{sec:overview_BBN}

By modifying the expansion rate of the Universe and affecting the neutron/proton weak reaction rates, incomplete neutrino decoupling will slightly modify the BBN abundances of light elements \cite{Pitrou_2018PhysRept,Mangano2005,Grohs2015}.

To get a clear understanding of the physics at play, it is useful to recall the standard picture of BBN \cite{KolbTurner,PeterUzan}.
\begin{enumerate}
\item Neutrons and protons track their equilibrium abundances, 
\begin{equation}
\label{eq:nse}
\left. \frac{n_n}{n_p}\right|_\mathrm{eq} = \exp{(-\Delta/T_\gamma)} \,,
\end{equation}
where $\Delta = m_n - m_p \simeq 1.293 \, \mathrm{MeV}$ is the difference of nucleon masses, until the so-called ``weak freeze-out," when the rates of $n \leftrightarrow p$ reactions drop below the expansion rate,
	\begin{equation}
	\label{eq:defofFO}
	\lambda \equiv \left. \frac{\Lambda_{n \to p} + \Lambda_{p \to n}}{H}\right|_{\TFO} \simeq 1\,.
	\end{equation}
\item After the freeze-out, neutrons only undergo beta decay until the beginning of nucleosynthesis, and a good approximation is
	\begin{equation}
	\label{eq:modelFO}
	X_n(\TNuc) = X_n(\TFO) \times \exp{\left[- \frac{t_\mathrm{Nuc} - t_\mathrm{FO}}{\tau_n}\right]}\, ,
	\end{equation}
	where $\tau_n \simeq 879.4 \, \mathrm{s}$~\cite{PDG} is the neutron mean lifetime. The nucleosynthesis temperature is usually defined when the \emph{deuterium bottleneck} is overcome, with the criterion $n_D/n_b \sim 1$ \cite{PeterUzan,Neutrino_Cosmology}. It can also be associated with the maximum in the evolution of the deuterium abundance~\cite{bernstein1989}, which coincides with the drop in the density of neutrons (converted into heavier elements). We will adopt this definition, which is very close to the other criterion. Note that $t_\mathrm{Nuc} - t_\mathrm{FO} \simeq t_\mathrm{Nuc}$, since $t_\mathrm{FO} \ll t_\mathrm{Nuc}$. Indeed, we have numerically $\TFO \simeq 0.67 \, \mathrm{MeV}$ and $t_\mathrm{FO} \simeq 1.7 \, \mathrm{s}$ for the freeze-out, and $\TNuc \simeq 73 \, \mathrm{keV}$ and $t_\mathrm{Nuc} \simeq 245 \, \mathrm{s}$ for the start of nucleosynthesis.
	
There is a caveat in this oversimplified description: as shown in~\cite{Grohs:2016vef}, the evolution of the neutron abundance is not ruled only by beta decay below $\TFO$. On the contrary, one needs to take into account all weak interactions even for smaller temperatures to avoid making potentially large mistakes on $\YP$ (see for instance Figure~8 in~\cite{Grohs:2016vef}). However, the approximation leading to~\eqref{eq:modelFO} is eventually valid for large enough times (below $T \simeq 0.28 \, \mathrm{MeV}$ according to~\cite{Pitrou_2018PhysRept}): since $t_\mathrm{Nuc} \gg t_\mathrm{FO}$, using this “beta decay” model will provide good results for the upcoming semi-analytical analysis, namely the calculation of $\delta X_n^{[\Delta t]}$ (see equations~\eqref{eq:deltaxn} and~\eqref{eq:dxnclock} below).

\item Almost all free neutrons are then converted into $\He4$, leading to
	\begin{equation}
	\YP \simeq 2 X_n(\TNuc)\, .
	\end{equation}
\end{enumerate}
This indicates very precisely where incomplete neutrino decoupling will intervene. Weak rates, and thus the freeze-out temperature, are modified through the changes in the distribution functions (different temperatures and spectral distortions $\delta g_{\nu_e}$). But the changes in the energy density will also modify the relation $t(T_\gamma)$, leaving more or less time for neutron beta decay and light element production. This is the so-called \emph{clock effect}, originally discussed in~\cite{Dodelson_Turner_PhRvD1992,Fields_PhRvD1993}. In summary, the neutron fraction at the onset of nucleosynthesis is modified as 
\begin{align}
\delta X_n^{[\mathrm{Nuc}]} \equiv \frac{\Delta X_n(\TNuc)}{X_n^{(0)}(\TNuc)} 
&= \frac{\Delta X_n(\TFO)}{X_n^{(0)}(\TFO)} - \frac{\Delta t_\mathrm{Nuc}}{\tau_n}  \nonumber \\
&\equiv \delta X_n^{[\mathrm{FO}]} + \delta X_n^{[\Delta t]} \, , \label{eq:deltaxn}
\end{align}
with $\Delta t_\mathrm{Nuc} \equiv t_\mathrm{Nuc} - t_\mathrm{Nuc}^{(0)}$ (we neglected the variation of $t_\mathrm{FO}$). For freeze-out ($\delta X_n^{[\mathrm{FO}]}$), it is a variation at constant $\lambda = 1$, which we take as our definition of freeze-out. $\delta X_n^{[\mathrm{Nuc}]}$ is the neutron abundance variation between the onset of nucleosynthesis in the ``actual" Universe and the one in the reference universe. Given our definition of $\TNuc$, the constant quantity here is $\dd X_D/\dd t = 0$.

Note that this model of freeze-out is quite similar to the instantaneous decoupling approximation for neutrinos, i.e., we condense a gradual process into a snapshot. Actually, in the range $4 \gtrsim \lambda \gtrsim 0.2$, there is a smooth transition between nuclear statistical equilibrium [Eq.~\eqref{eq:nse}] and pure beta decay. For the sake of argument, we keep the criterion $\lambda \simeq 1$, and we will point out the limits of this model in the following discussions when necessary.

\section{Detailed analysis with \texttt{PRIMAT}}

In order to check the qualitative predictions of section~\ref{sec:overview_BBN}, we incorporate the results of neutrino decoupling from chapter~\ref{chap:Decoupling} into the BBN code \texttt{PRIMAT} and investigate the associated modification of abundances.

\subsection{Overview of the BBN code}

\texttt{PRIMAT} is a \emph{Mathematica} code developed from the \emph{Fortran} code used for instance in~\cite{Coc2006,CocVangioni2010,Coc2015}, designed to compute as precisely as possible the primordial abundances by including the various corrections to the weak and nuclear reaction rates. It is presented in~\cite{Pitrou_2018PhysRept,Pitrou:2019nub}, and is broadly designed as follows:
\begin{itemize}
	\item it solves the dynamics of the background, first obtaining the scale factor as a function of the temperature $a(T_\gamma)$ (either from entropy conservation in the approximation of instantaneous neutrino decoupling, or taking into account the entropy transfer --- see below), then $a(t)$ via Friedmann equation;
	\item once the thermodynamics and cosmological expansion are known, the weak rates are computed on a grid of plasma temperatures so as to interpolate them. Different corrections can be included in order to determine these rates with great precision;
	\item finally, it builds and solves the system of differential equations that accounts for the nuclear and weak reactions.
\end{itemize}

We already mentioned the two levels at which incomplete neutrino decoupling intervenes: changing the relation $t(T_\gamma)$ via the different energy density and Friedmann equation, and affecting the weak rates via the different electronic (anti)neutrino distribution functions. Let us thus focus on the weak interaction reactions and how they are implemented in \texttt{PRIMAT}.

\subsubsection{Weak interaction reactions}

The reactions which determine the neutron-to-proton ratio are:
\begin{subequations}
\label{eq:weakrates_BBN}
\begin{align}
n + \nu_e &\longleftrightarrow p + e^- \label{eq:weakrates_BBN_1} \\
n &\longleftrightarrow p + e^- + \bar{\nu}_e \label{eq:weakrates_BBN_2} \\
n + e^+ &\longleftrightarrow p + \bar{\nu}_e \label{eq:weakrates_BBN_3}
\end{align}
\end{subequations}
The neutron and proton densities evolve according to
\begin{equation*}
\dot{n}_n + 3 H n_n = - n_n \Lambda_{n \to p} + n_p \Lambda_{p \to n} \quad \text{and} \quad \dot{n}_p + 3 H n_p = - n_p \Lambda_{p \to n} + n_n \Lambda_{n \to p} \, ,
\end{equation*}
where the rates $\Lambda_{n \to p}$ correspond to~\eqref{eq:weakrates_BBN} from left to right, and conversely for $\Lambda_{p \to n}$. In the \emph{Born approximation} (also called \emph{infinite nucleon mass approximation}), the scattering matrix elements take a simple form and we can write schematically (the bar shows that we are at the Born approximation level)
\begin{equation}
\label{eq:weakrate_schematic}
\overline{\Lambda} = K \int_{0}^{\infty}{p^2 \dd{p} \, E_\nu^2 \times \left[\textit{Stat. fact.} \right]} \, ,
\end{equation}
with $p$ the electron or positron momentum. The prefactor reads $K = 4 G_F^2 \abs{V_{ud}}^2 (1+3 g_A^2)/(2 \pi)^3$ with $V_{ud}$ the first entry of the Cabibbo-Kobayashi-Maskawa (CKM) matrix~\cite{Cabibbo,KobayashiMaskawa} and $g_A = 1.2753(13)$ the axial-vector constant\footnote{Several conventions exist regarding this constant. In~\cite{PDG}, the vector and axial weak coupling constants $c_V$ and $c_A$ are defined such that the matrix elements include the term $[\gamma_\mu (c_V + c_A \gamma_5)]$, while $c_A$ is often defined with an opposite sign. With this convention however, we have $g_A = - c_V/c_A > 0$.} of nucleons~\cite{PDG}. The neutrino energy must satisfy the conservation condition, which actually limits the domain of integration. The statistical factor part contains only the product of electron/positron and (anti)neutrino distribution functions (that is the entries $\vrho^{e}_e$ and $\bvrho^{e}_e$ of the density matrix when we take into account mixing), since we can neglect the baryon Pauli-blocking factors due to the very small baryon-to-photon ratio, and the other neutron or proton distribution function is integrated upon, giving the density which gets factored out.

Let us consider reaction~\eqref{eq:weakrates_BBN_1}. Energy conservation requires, in the infinite nucleon mass approximation, $E_\nu = m_p + E - m_n = E - \Delta$, with $E$ the electron energy. Since we must have $E_\nu > 0$, the integral~\eqref{eq:weakrate_schematic} is thus limited to the domain $E > \Delta$, that is $p > \sqrt{\Delta^2 - m_e^2}$. The statistical factor is
\begin{equation}
\label{eq:statfact_BBN_1}
[\textit{Stat. fact.}] = [1-f_e(E)]f_{\nu_e}(E - \Delta) = f_e(-E) \times \frac{1 + \delta g_{\nu_e}(E-\Delta)}{e^{(E-\Delta)/T_{\nu_e}}+1} \, ,
\end{equation}
where we used the functional property of equilibrium Fermi-Dirac spectra $1 - f_e(E) = f_e(-E)$, and the parameterization of neutrino spectral distortions~\eqref{eq:param_rho}. If we assume instantaneous neutrino decoupling, the neutrino distribution function reads $f_\nu^{(\mathrm{eq})}$ (i.e. $T_{\nu_e} = \Tcm$ and $\delta g_{\nu_e}=0$), and the reaction rate can be written:
\begin{equation}
\overline{\Lambda}_{n+\nu_e \to p + e^-} = K \int_{E > \Delta}{p^2 \dd{p} \, (E - \Delta)^2 f_e(-E) f_\nu^{(\mathrm{eq})}(E - \Delta)} \, ,
\end{equation}

A similar procedure can be applied to reaction~\eqref{eq:weakrates_BBN_2}. Energy conservation requires $E_{\bnu} = \Delta - E > 0$ hence $E < \Delta$. The statistical factor reads
\begin{equation}
\label{eq:statfact_BBN_2}
[\textit{Stat. fact.}] = [1-f_e(E)][1-f_{\bnu_e}(\Delta-E) ]= f_e(-E) \times \left[1- \frac{1 + \delta g_{\bnu_e}(\Delta-E)}{e^{(\Delta-E)/T_{\bnu_e}}+1}\right] \, ,
\end{equation}
which can also be simplified in the instantaneous decoupling limit, using $1 - f_\nu^{(\mathrm{eq})}(\Delta-E) = f_\nu^{(\mathrm{eq})}(E-\Delta)$, such that
\begin{equation}
\overline{\Lambda}_{n \to p + e^- + \bnu_e} = K \int_{E < \Delta}{p^2 \dd{p} \, (E-\Delta)^2 f_e(-E) f_\nu^{(\mathrm{eq})}(E-\Delta)} \, ,
\end{equation}

Finally, for the reaction~\eqref{eq:weakrates_BBN_3}, energy conservation gives $E_{\bnu} = E + \Delta$ which does not put any constraint on $p$. The statistical factor reads
\begin{equation}
\label{eq:statfact_BBN_3}
[\textit{Stat. fact.}] = f_e(E)[1-f_{\bnu_e}(E + \Delta) ]= f_e(E) \times \left[1- \frac{1 + \delta g_{\bnu_e}(E+\Delta)}{e^{(E+\Delta)/T_{\bnu_e}}+1}\right] \, ,
\end{equation}
which we simplify in the instantaneous decoupling limit as
\begin{equation}
\overline{\Lambda}_{n + e^+ \to p + \bnu_e} = K \int_{0}^{\infty}{p^2 \dd{p} \, (-E - \Delta)^2 f_e(E) f_\nu^{(\mathrm{eq})}(-E - \Delta)} \, .
\end{equation}
We therefore introduce the functions:
\begin{equation}
E_\nu^\mp(E) \equiv E \mp \Delta \qquad \text{and} \qquad \chi_\pm(E) \equiv \left[E_\nu^\mp(E)\right]^2 f_e(-E) \, f_\nu^{(\mathrm{eq})}\left(E_\nu^\mp(E)\right) \, .
\end{equation}
We can then gather all the contributions in a single expression for the Born rates:
\begin{equation}
\label{eq:Lambda_nTOp}
\overline{\Lambda}_{n \to p} = K \int_{0}^{\infty}{p^2 \dd{p} \left[\chi_+(E) + \chi_+(-E)\right]} \, .
\end{equation}
The first term in the integrand comes from the sum of $\overline{\Lambda}_{n+\nu_e \to p + e^-}$ and $\overline{\Lambda}_{n \to p + e^- + \bnu_e}$, and the second term is $\overline{\Lambda}_{n + e^+ \to p + \bnu_e}$. This expression coincides with Eq.~(77) in~\cite{Pitrou_2018PhysRept}.

The reaction rate for protons is obtained by replacing $\Delta \to - \Delta$, that is $\chi_+ \to \chi_-$, which reads for completeness
\begin{equation}
\label{eq:Lambda_pTOn}
\overline{\Lambda}_{p \to n} = K \int_{0}^{\infty}{p^2 \dd{p} \left[\chi_-(E) + \chi_-(-E)\right]} \, .
\end{equation}

In addition to these Born rates, several corrections are implemented: radiative corrections at zero and finite temperature, finite nucleon mass corrections, weak magnetism... and incomplete neutrino decoupling. We now focus on this latter particular feature, the extensive derivation above showing directly where to modify the rates to include the results from chapter~\ref{chap:Decoupling}. Note that, unless stated otherwise, QED corrections to the plasma thermodynamics are included in the calculation.

\subsection{Implementation of incomplete neutrino decoupling in \texttt{PRIMAT}}

In the version of \texttt{PRIMAT} used in~\cite{Pitrou_2018PhysRept}, the lack of effective temperatures and spectral distortion values across the nucleosynthesis era required an approximate strategy to include incomplete neutrino decoupling. It consisted in neglecting spectral distortions $\delta g_{\nu} = 0$ while considering that all neutrinos shared the same temperature, that is an effective average temperature $\widehat{T}_\nu$ consistent with the energy transfer between the QED plasma and neutrinos. We can define it from the effective temperatures $T_{\nu_\alpha}$:
\begin{equation}
\label{eq:def_Taverage}
\bar{\rho}_{\nu} = \frac78 \frac{\pi^2}{30} \left(z_{\nu_e}^4 + z_{\nu_\mu}^4 + z_{\nu_\tau}^4\right) \equiv 3 \times \frac78 \frac{\pi^2}{30} \times \hat{z}_\nu^4 \qquad \text{with} \qquad \hat{z}_\nu = \frac{\widehat{T}_\nu}{\Tcm} \, .
\end{equation}
It is not necessary to have the individual values of $T_{\nu_\alpha}$ to compute $\widehat{T}_{\nu}$, as one can use another key quantity: the \emph{heating rate}~\cite{Parthenope}
\begin{equation}
\label{eq:Nheating}
\mathcal{N} \equiv  \frac{1}{z^4} \left(x \frac{\dd (\bar{\rho}_\nu + \bar{\rho}_{\bnu})}{\dd x} \right)_{x=x(z)} = \frac{1}{z^4} \frac{1}{2 \pi^2 H} \int{\dd{y} y^3 \, \Tr[\mathcal{I} + \bar{\mathcal{I}}]} \, .
\end{equation}
which is obtained from~\eqref{eq:rhonu_matrix} and the QKE~\eqref{eq:QKE_final}, noting that the trace of a commutator is zero. $T_\gamma^4 \mathcal{N}$ can be viewed as the volume heating rate of the neutrino bath in units of the Hubble rate~\cite{Pitrou_2018PhysRept}. The values of $\mathcal{N}$ were obtained from a fit given in \texttt{PArthENoPE} \cite{Parthenope}  [Eqs.~(A23)--(A25)], computed by Pisanti \emph{et al.}~from the results of~\cite{Mangano2002,Mangano2005}.

\begin{figure}[!ht]
	\centering
	\includegraphics{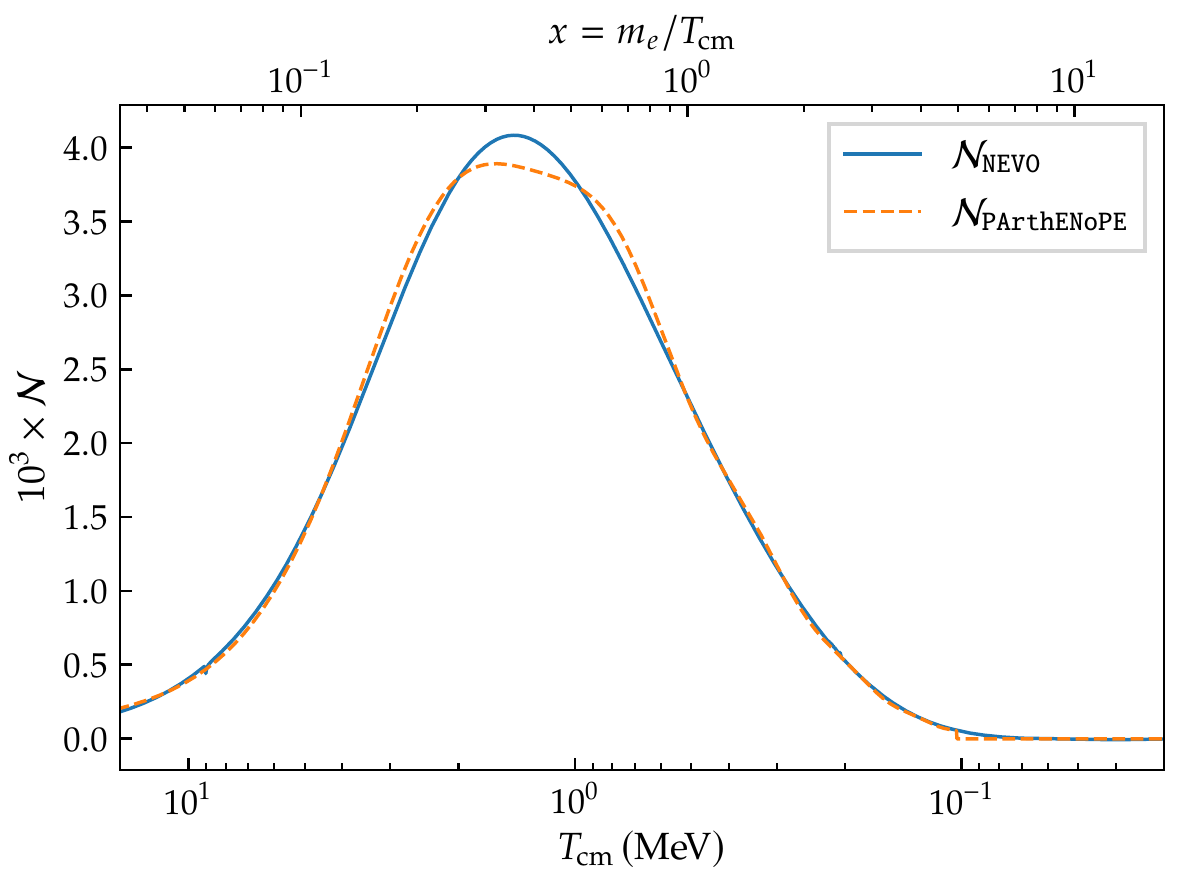}
	\caption[Dimensionless heating rate from \texttt{NEVO} and \texttt{PArthENoPE}]{\label{fig:Compare_N} Comparison of the dimensionless heating rate $\mathcal{N}$ computed with the results from \texttt{NEVO} and equation~\eqref{eq:compute_N}, and the fit given in \texttt{PArthENoPE}~\cite{Parthenope}. For consistency with the results at the time, we did not include the $\mathcal{O}(e^3)$ finite-temperature corrections to the plasma thermodynamics in this calculation.}
\end{figure}

Note that we can also compute $\mathcal{N}$ from the variation on the comoving photon temperature $z$, which is more convenient if we want to treat our numerical results a posteriori, since we keep the values of $z(x)$ across the decoupling era. Starting from~\eqref{eq:zQED}, we have:
\begin{equation}
\label{eq:compute_N}
\mathcal{N}(x) = \frac{2 x}{z} \left\{\frac{x}{z} J\left(\frac{x}{z}\right) + G_1\left(\frac{x}{z}\right) - \left[ \frac{x^2}{z^2} J\left(\frac{x}{z}\right) + Y\left(\frac{x}{z}\right) + \frac{2 \pi^2}{15} + G_2\left(\frac{x}{z}\right)\right] \times  \frac{\dd z}{\dd x} \right\} \, .
\end{equation}
We plot on Figure~\ref{fig:Compare_N} the quantity $\mathcal{N}$ deduced from the neutrino decoupling results of chapter~\ref{chap:Decoupling}, and the fit provided in \texttt{PArthENoPE}.

This approximate handling of incomplete neutrino decoupling in the earlier version of \texttt{PRIMAT} correctly captures the changes in the expansion rate (since~\eqref{eq:def_Taverage} shows that the energy density is well computed from $\widehat{T}_\nu$), but \emph{a priori} it handles the weak rates poorly: electron neutrinos are too cold ($T_{\nu_e} > \widehat{T}_\nu$), and their spectrum is not distorted. This should in principle have consequences for the neutron-to-proton ratio at freeze-out, and thus on the final abundances.

We modified \texttt{PRIMAT} to introduce the results from neutrino transport analysis. Since the useful variable in nucleosynthesis is the plasma temperature $T_\gamma$, all other quantities\footnote{The non-thermal distortions depend both on the momentum (sampling on a grid with points $y_i$) and time, hence on the plasma temperature $T_\gamma$.} ($x$, $T_{\nu_\alpha}$, $\delta g_{\nu_\alpha}(y_i)$) are interpolated. Depending on the options chosen, one can then use the ``real" effective neutrino temperatures or the average temperature for comparison with the previous approach (keeping the total energy density unchanged in each case).

\begin{figure}[!ht]
	\centering
	\includegraphics{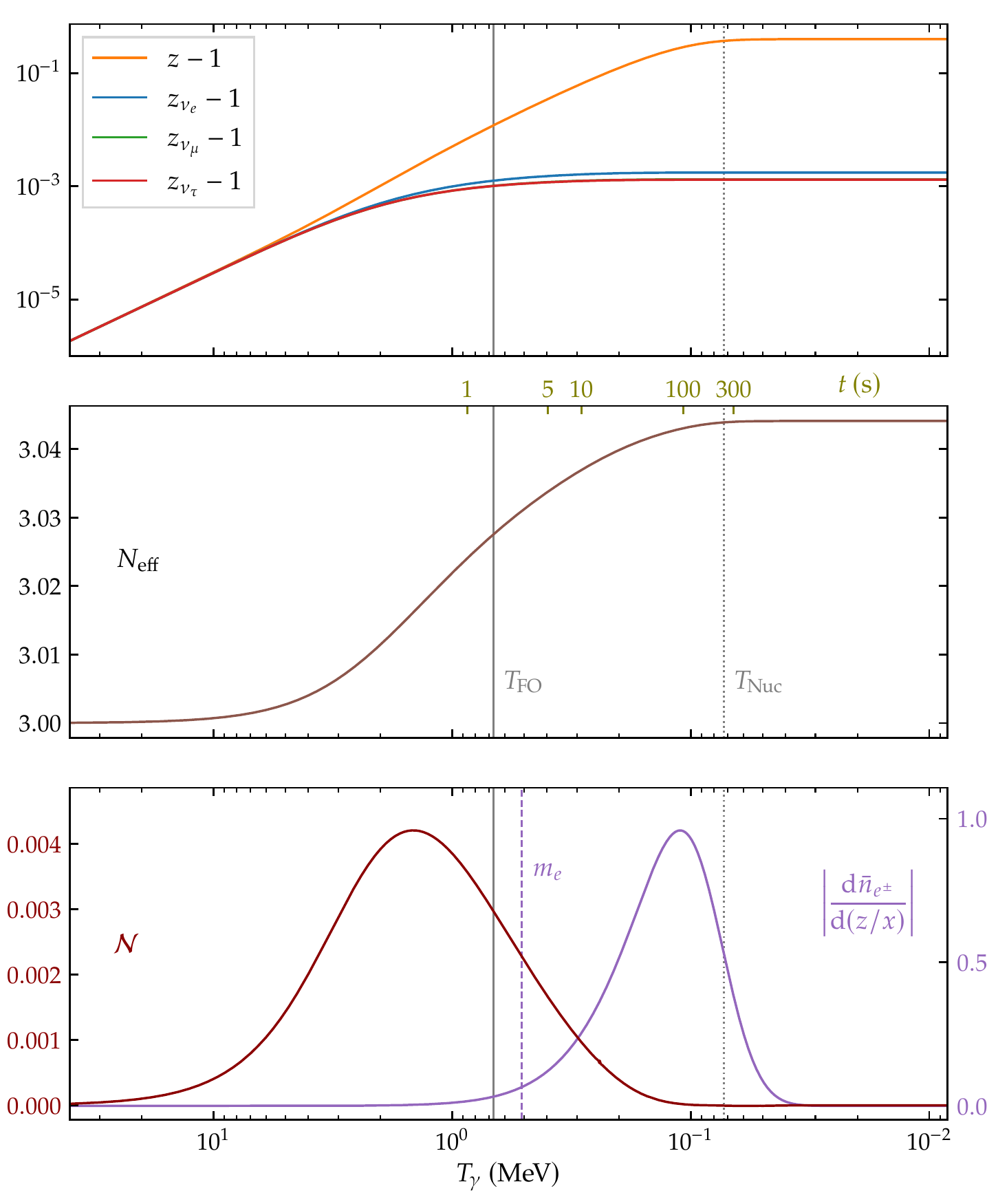}
	\caption[Evolution of relevant quantities for neutrino decoupling, as a function of $T_\gamma$]{\label{fig:summary_nubbn} Evolution of relevant quantities for neutrino decoupling, as a function of the plasma temperature. \emph{Top}: Comoving (effective) temperatures of the plasma and neutrinos. \emph{Middle}: Effective number of neutrinos, as defined in Eq.~\eqref{eq:defNeff_bbn}. \emph{Bottom}: Neutrino heating rate and variation of the comoving electron+positron density (derivative taken with respect to $z/x = T_\gamma/m_e$).}
\end{figure}

\paragraph{Summary of neutrino decoupling results} We plot on Figure~\ref{fig:summary_nubbn} the evolution of different quantities that play a significant role for BBN, obtained through the numerical resolution presented in chapter~\ref{chap:Decoupling}. The evolution of the comoving temperatures has already been discussed in the previous chapter. The reheating of the different species is due to the entropy transfer from electrons and positrons, which is visualized by plotting the variation of their number density. For $T_\gamma \gg m_e$, electrons are relativistic and $\bar{n}_{e^\pm} \equiv (n_{e^-} + n_{e^+})\times (x/m_e)^3$ is constant, while for $T_\gamma \ll m_e$ the density drops to zero. The variation between those two constants corresponds to the annihilation period, which indeed starts around $T_\gamma \sim m_e$ and is over for $T_\gamma \sim 30 \, \mathrm{keV}$. At the beginning of this period, neutrinos progressively decouple and there is a heat transfer from the plasma, visualized through the dimensionless heating rate $\mathcal{N}$ defined in~\eqref{eq:Nheating}. The slight overlap between the two curves in the bottom panel of Fig.~\ref{fig:summary_nubbn} is the very reason why neutrinos are partly reheated. 

Finally, we plot the evolution of $N_\mathrm{eff}$, from $3$ before the MeV age to its frozen value $3.044$. To do so, we define it such that we can compute its value across the decoupling era and not only long after decoupling, via
\begin{equation}
\label{eq:defNeff_bbn}
\rho_\nu + \rho_{\bnu} = \frac78 \left(\frac{\Tcm}{T_\gamma}\right)^4  \Neff \times \rho_\gamma \quad \text{hence} \quad \boxed{\Neff = 3 \left(\frac{\hat{z}_\nu z^{(0)}}{z}\right)^4} \, .
\end{equation}
In this expression, $z^{(0)}$ is the photon temperature in the instantaneous decoupling limit. All these quantities are taken as functions of either the comoving temperature $\Tcm$ or the photon temperature $T_\gamma$. Comparing with Fig.~5 in~\cite{Grohs2015}, we note that there is no ``plateau" before the freeze-out. This behavior can be considered as an artifact due to plotting $\Neff$ as a function of $x=m_e/\Tcm$: the plateau is due to the difference between $\Tcm$ and $\Tcm^{(0)}$ for a given $T_\gamma$, and does not represent a meaningful physical effect (see also Fig.~7 in~\cite{Esposito_NuPhB2000}). In other words, the asymptotic values of $\Neff$ are meaningful, while the intermediate ones depend on the reference chosen, which is less significant.

\subsubsection{Corrections to the Born rates} The weak rates are modified at the Born level by including the effective temperatures and distortions of electronic (anti)neutrinos in the rates~\eqref{eq:Lambda_nTOp} and~\eqref{eq:Lambda_pTOn}. First, we have the same expressions where the neutrino distribution functions are replaced in $\chi_{\pm}$
\[ f_{\nu}^{(\mathrm{eq})}(E_\nu^\mp,\Tcm) \to f_{FD}(E_\nu^\mp,T_{\nu_e}) \qquad \text{i.e.} \qquad \frac{1}{e^{E_\nu^\mp/\Tcm}+1} \to \frac{1}{e^{E_\nu^\mp/T_{\nu_e}}+1} \, . \]
Then, we derive the contribution from the non-thermal distortions starting from the expressions of the statistical factors.
\begin{itemize}
	\item[(a)] The case of eq.~\eqref{eq:statfact_BBN_1} is the simplest since the neutrino is in the initial state, we just need to add the extra contribution 
	\[(E-\Delta)^2 f_e(-E) \frac{\delta g_{\nu_e}(E-\Delta)}{e^{(E-\Delta)/T_{\nu_e}}+1} = (E_\nu^-)^2 f_e(-E) \frac{\delta g_{\nu_e}(E_\nu^-)}{e^{E_\nu^-/T_{\nu_e}}+1} \, , \]
	and note that $E_\nu^- = \abs{E_\nu^-} > 0$.
	\item[(b)] For the reaction $n \to p + e^- + \bnu_e$, i.e. the statistical factor~\eqref{eq:statfact_BBN_2}, note that $\Delta - E = - E_\nu^-(E) = \abs{E_\nu^-(E)}$, such that the extra contribution reads
		\[(E-\Delta)^2 f_e(-E) \times (-1) \times \frac{\delta g_{\bnu_e}(\Delta-E)}{e^{(\Delta-E)/T_{\bnu_e}}+1} = - (E_\nu^-)^2 f_e(-E) \frac{\delta g_{\bnu_e}(\abs{E_\nu^-})}{e^{\abs{E_\nu^-}/T_{\bnu_e}}+1} \, . \]
	\item[(c)] Finally, for~\eqref{eq:statfact_BBN_3}, the extra contribution is
		\[(E+\Delta)^2 f_e(E) \times (-1) \times \frac{\delta g_{\bnu_e}(E+\Delta)}{e^{(E+\Delta)/T_{\nu_e}}+1} = - (E_\nu^-)^2 f_e(E) \frac{\delta g_{\nu_e}(-E_\nu^-)}{e^{-E_\nu^-/T_{\nu_e}}+1} \, , \]
		where $E_\nu^- = -E - \Delta$ is evaluated at $-E$. Since $E_\nu^- < 0$, we can replace $- E_\nu^- = \abs{E_\nu^-}$.
\end{itemize}
If, as assumed in standard neutrino decoupling, we consider that neutrinos and antineutrinos have the same distributions, we can gather the corrections under the common equation (the case of $p \to n$ rates is treated in the same way):
\begin{subequations}
\label{eq:deltagamma}
\begin{align}
\Delta \Lambda_{n \to p} &=  K \int_{0}^{\infty}{p^2 \mathrm{d} p \left[\delta \chi_+(E) + \delta \chi_+(-E)\right]} \, ,  \label{subeq:np} \\
\Delta \Lambda_{p \to n} &= K \int_{0}^{\infty}{p^2 \mathrm{d} p \left[\delta \chi_-(E) + \delta \chi_-(-E)\right]} \, , \label{subeq:pn}
\end{align}
\end{subequations}
with
\begin{equation}
\delta \chi_{\pm} (E) = \abs{E_\nu^\mp(E)}^2 f_{e}(-E) \, \frac{\sign{(E_\nu^\mp)} \times \delta g_{\nu_e}(\abs{E_\nu^\mp})}{e^{\abs{E_\nu^\mp}/T_{\nu_e}}+1} \, .
\end{equation}
As evidenced above, the $\sign$ function accounts for the fact that $f_{\nu_e}(\abs{E_\nu^\mp})$ appears as part of a Pauli blocking factor if $E_\nu^\mp < 0$, i.e., the neutrino is in a final state.

\subsubsection{Analysis via different implementations}

We consider three different implementations:
\begin{itemize}
\item[\textit{(i)}] The earlier \texttt{PRIMAT} approach (no distortions and an average neutrino temperature), where $\widehat{T}_\nu$ is computed from the effective temperatures via~\eqref{eq:def_Taverage}, although we checked that computing it from the heating rate $\mathcal{N}$ deduced from \texttt{NEVO} gave the same results. We call this approach ``$\widehat{T}_\nu$" in Tables~\ref{Table:Corrections} and \ref{Table:Full_Corrections} and Figs.~\ref{fig:analyzevariations} and \ref{fig:analyzevariationsfull}.
\item[\textit{(ii)}] The weak rates including the real electron neutrino temperature, but still without spectral distortions. We call this approach ``$T_{\nu_e},$ no distortions."
\item[\textit{(iii)}] Full results from neutrino evolution. We call this approach ``$T_{\nu_e},$ with distortions."
\end{itemize}
Note that these three scenarios take place in identical cosmologies, with the \emph{same} energy density; using the proper $\nu_e$ temperature and including distortions only affect the weak rates. 

\renewcommand{\arraystretch}{1.3}

\begin{table}[!htb]
	\centering
	\begin{tabular}{|l|cccc|}
  	\hline 
  & $\YP$  &  $\mathrm{D}/\mathrm{H}  \times 10^5$ &  $\He3/\mathrm{H} \times 10^5$ &  $\Li/\mathrm{H} \times 10^{10}$ \\
  \hline \hline
  Inst. decoupling, no QED & $0.24268$  & $2.4027$ & $1.0339$  & $5.4690$ \\ \hline
  $\widehat{T}_\nu$ & $0.24281$  & $2.4122$ & $1.0353$  & $5.4470$ \\
    $T_{\nu_e},$ no distortions & $0.24278$  & $2.4120$  & $1.0353$  & $5.4466$ \\
  $T_{\nu_e},$ with distortions & $0.24278$  & $2.4125$ & $1.0354$  & $5.4479$ \\ \hline \hline
   Inst. decoupling, with QED & $0.24268$  & $2.4052$ & $1.0343$  & $5.4620$ \\ \hline
  $\widehat{T}_\nu$ & $0.24280$  & $2.4146$ & $1.0357$  & $5.4403$ \\
    $T_{\nu_e},$ no distortions & $0.24278$  & $2.4145$ & $1.0356$  & $5.4399$ \\
  $T_{\nu_e},$ with distortions & $0.24286$  & $2.4149$ & $1.0357$  & $5.4412$\\ \hline 
\end{tabular}
	\caption[Light element abundances at the Born approximation level]{Light element abundances, at the Born approximation level, for various implementations of neutrino-induced corrections. See section~\ref{subsec:full_BBN} for results with the full corrections derived in~\cite{Pitrou_2018PhysRept}. The number of digits is larger than the nominal uncertainty but is chosen here so as to show the variations.
	\label{Table:Corrections}}
\end{table}

\renewcommand{\arraystretch}{1.2}

We show in Table~\ref{Table:Previous} the relative variations of abundances when taking into account incomplete neutrino decoupling, i.e., when going from the instantaneous decoupling to the “$T_{\nu_e}$, with distortions” row in Table~\ref{Table:Corrections}. We also compare these variations with previous results in the literature. We check that our results are in close agreement with Grohs \emph{et al.}~\cite{Grohs2015}, but with opposite signs of variation (except for ${}^4 \mathrm{He}$) compared to the results of Mangano \emph{et al.}~\cite{Mangano2005}. The extensive study in the next section sheds a new light on the different phenomena involved: our aim is to justify physically these results.

\begin{table}[!ht]
	\centering
	\begin{tabular}{|l|rrrr|}
  	\hline 
 Variation of abundances & \multicolumn{1}{c}{$\delta \YP$ } & \multicolumn{1}{c}{$\delta \left(\mathrm{D}/\mathrm{H}\right) $}& \multicolumn{1}{c}{$ \delta \left(\He3/\mathrm{H}\right) $} &  \multicolumn{1}{c|}{$\delta \left(\Li/\mathrm{H}\right) $ }\\
  \hline \hline
  \emph{No QED corrections} & & & &   \\ 
  Grohs \emph{et al.}~\cite{Grohs2015} & $4.636 \times 10^{-4}$ &  $3.686 \times 10^{-3}$ & $1.209 \times 10^{-3}$ &$-3.916 \times 10^{-3}$ \\
  This work & $7.707 \times 10^{-4}$ & $4.086 \times 10^{-3}$ & $1.369 \times 10^{-3}$ & $-3.855 \times 10^{-3}$  \\ \hline \hline
    \emph{QED corrections included} & & & & \\ 
  Naples group \cite{Mangano2005} & $7.05 \times 10^{-4}$ &  $-2.8 \times 10^{-3}$ & $-1.1 \times 10^{-3}$ & $3.92 \times 10^{-3}$  \\
  This work & $7.648 \times 10^{-4}$ & $4.043 \times 10^{-3}$ & $1.355 \times 10^{-3}$ & $-3.814 \times 10^{-3}$  \\ \hline 
\end{tabular}
	\caption[Comparison with previous results]{Comparison with previous results. Note that baseline values are different in the cases that do or do not include QED corrections (see Table~\ref{Table:Corrections}). The values given by the Naples group in~\cite{Mangano2005} are absolute variations, and we need the baseline values to compute relative variations; as these were not given, we use our own baseline values. Besides, our results with QED corrections include the $\mathcal{O}(e^3)$ corrections which were absent in~\cite{Mangano2005} ; however, we checked that the relative variations were the same if we restricted to $\mathcal{O}(e^2)$ corrections. 
	\label{Table:Previous}}
\end{table}

\subsection{Neutron fraction at the onset of nucleosynthesis}

We now review the physics that allows us to understand the numerical results of Table~\ref{Table:Corrections}. We first detail the physics affecting the helium abundance, which is directly related to the neutron fraction at the onset of nucleosynthesis, before turning to the production of other light elements, for which the clock effect dominates.

\subsubsection{Neutron/proton freeze-out}

Previous articles~\cite{Dodelson_Turner_PhRvD1992,Fields_PhRvD1993,Mangano2005} studied the variation of $n \leftrightarrow p$ rates due to incomplete neutrino decoupling at constant scale factor, claiming that the Hubble rate $H$ was left unchanged at a given $x$. This argument of constant total energy density, namely $\Delta (\rho_\nu + \rho_{\bnu}) = - \Delta \rho_\mathrm{em}$ with $\rho_{\mathrm{em}}$ the energy density of the QED plasma, requires $T_\gamma \simeq T_\nu$, as proven in the Appendix 3 of~\cite{Dodelson_Turner_PhRvD1992}. However, by looking at the top panel of Fig.~\ref{fig:summary_nubbn} it appears that at freeze-out $T_\gamma$ and $T_{\nu_{\alpha}}$ differ by $\sim 1 \, \%$, which is the typical order of magnitude of variations we are interested in. Moreover, the analysis of~\cite{Fields_PhRvD1993} used thermal-equivalent distortions of neutrinos spectra (i.e., only effective temperatures, no $\delta g_{\nu}$), calling for a more precise study making full use of our numerical results.

Due to the rich interplay of the processes involved, an analytical estimate of $\delta X_n^{[\mathrm{FO}]}$ is particularly challenging. Since our goal is to provide a satisfactory physical picture of the role of neutrinos in BBN, and thus in particular to check Eq.~\eqref{eq:deltaxn}, we perform a numerical evaluation. 

\begin{figure}[!ht]
	\centering
	\includegraphics{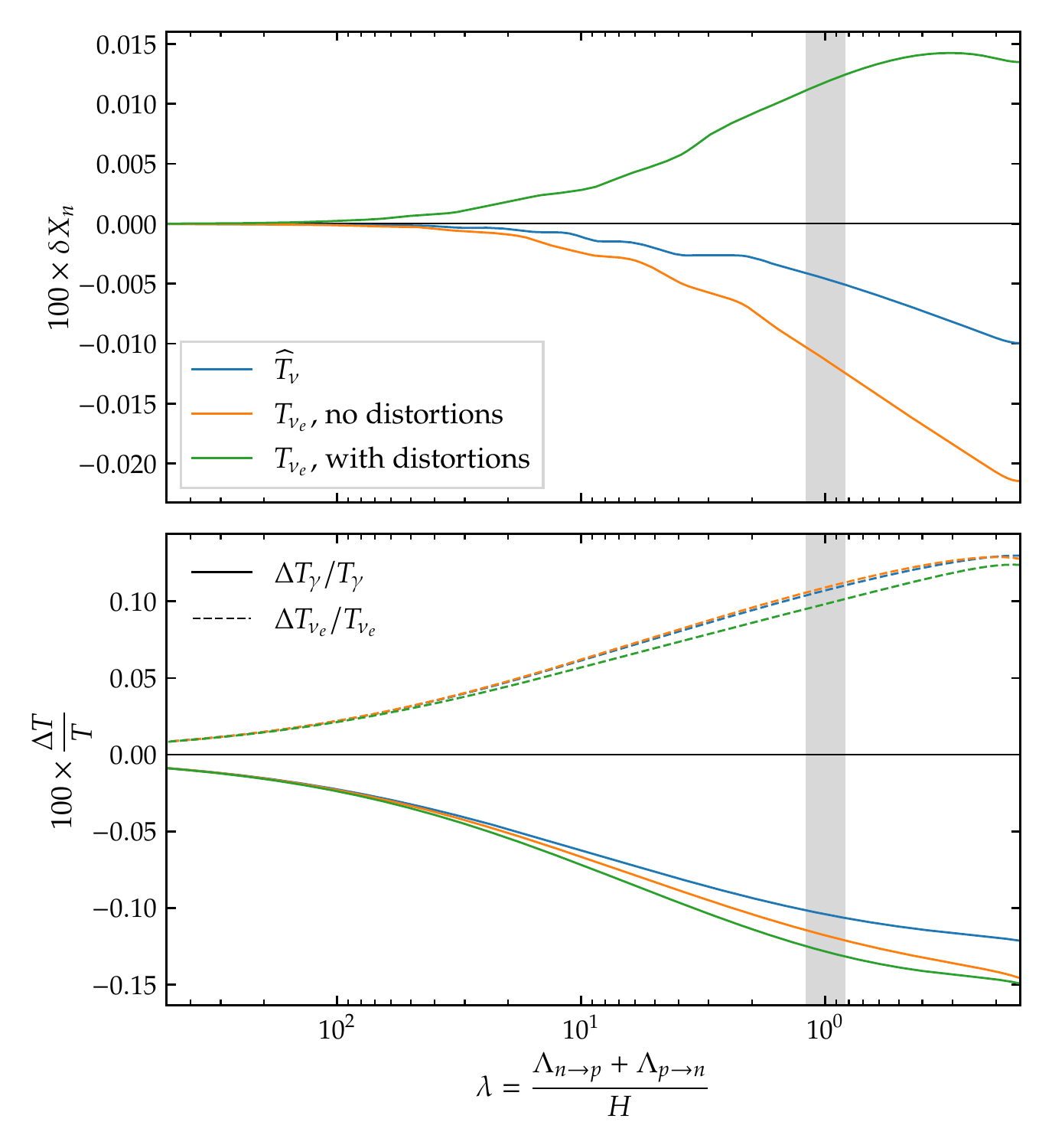}
	\caption[Neutron fraction and temperature variations for various implementations of neutrino-induced corrections]{\label{fig:analyzevariations} Neutron fraction (\emph{top}) and temperature (\emph{bottom}) variations with respect to instantaneous decoupling, in the different implementations of neutrino-induced corrections. These quantities are plotted as a function of the ratio of the total $n \leftrightarrow p$ reaction rate and the Hubble expansion rate, which is approximately equal to $1$ at weak freeze-out.}
\end{figure}

Figure~\ref{fig:analyzevariations} shows the variation of $X_n$ and $T_{\nu_e,\gamma}$ for the different implementations of neutrino-induced corrections around the time of freeze-out. In each case, incomplete neutrino decoupling leads to a decrease of $X_n$ at freeze-out.

For each implementation of neutrino-induced corrections the evolution of the photon temperature $z(x)$ is the same; the difference lies in whether or not we include $z_{\nu_e}$ and $\delta g_{\nu_e}$. But the quantities in Fig.~\ref{fig:analyzevariations} are plotted with respect to $\lambda$, which is a different function of $x$ in each case. For instance, when including the real $\nu_e$ temperature, weak rates increase and freeze-out is delayed, leading to a smaller $T_\gamma(\lambda \simeq 1) \equiv \TFO$: the orange curve is below the blue one in the bottom panel of Fig.~\ref{fig:analyzevariations}. Adding the distortions increases the rates even more, and slightly decreases $\TFO$ (green curve). One would then expect a reduction of $X_n$, which would track its equilibrium value longer. While this is true for thermal corrections (orange curve below the blue one in the top panel of Fig.~\ref{fig:analyzevariations}), adding the distortions disrupts this picture.

\begin{figure}[!ht]
	\centering
	\includegraphics{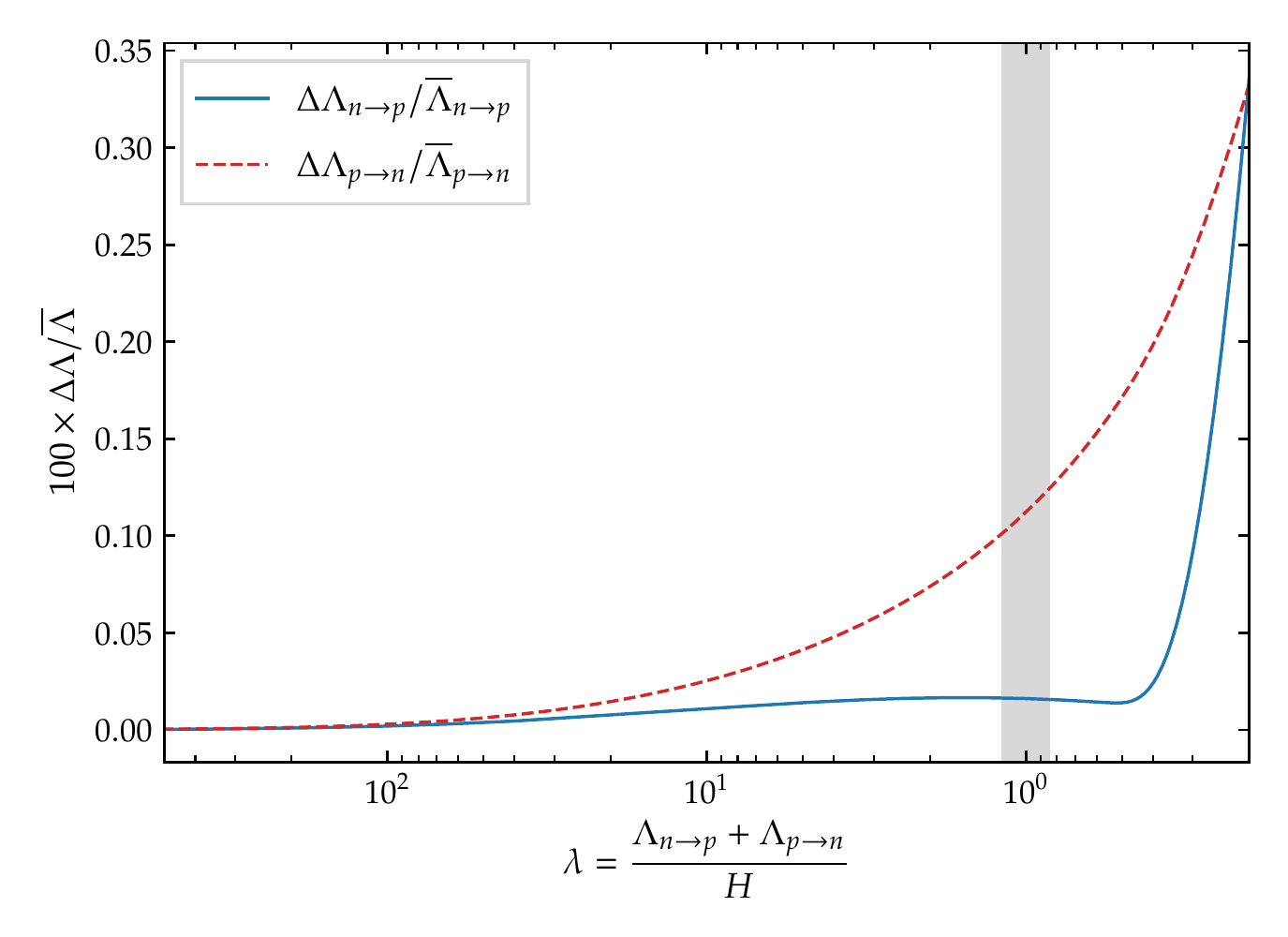}
	\caption[Relative corrective to $n \leftrightarrow p$ weak rates]{\label{fig:detailedbalance} Relative corrections to $n \leftrightarrow p$ weak rates, with $\Delta \Lambda_{n \leftrightarrow p}$ defined in Eq.~\eqref{eq:deltagamma}. To ensure detailed balance requirements, we enforce $T_{\nu_e} = T_\gamma$.}
\end{figure}

\paragraph{Disruption of detailed balance} Indeed, the main effect of including neutrino spectral distortions is to alter the detailed balance relation $\overline{\Lambda}_{p \to n} = e^{-\Delta/T} \overline{\Lambda}_{n \to p}$. Let us parameterize this deviation from detailed balance as\footnote{Note that a similar deviation from detailed balance would arise if electronic neutrinos had a chemical potential ($\sigma_\nu$ would then be $\mu_{\nu_e}/T$). However, this is a coincidence: here, neutrinos and antineutrinos have the same distributions, but the existence of non-thermal distortions lead to a disruption of detailed balance that we parameterize, \emph{for convenience}, like a chemical potential-like deviation.}
\begin{equation}
\label{eq:detailedbalance}
\Lambda_{p \to n} = \exp{ \left(-\frac{\Delta}{T} + \sigma_\nu\right)} \Lambda_{n \to p} \, ,
\end{equation}
with $\sigma_\nu \ll 1$. Writing this in terms of the Born rates $\overline{\Lambda}$ (which satisfy the detailed balance equation), we get
\begin{equation}
\sigma_\nu = \frac{\Delta \Lambda_{p \to n}}{\overline{\Lambda}_{p \to n}} - \frac{\Delta \Lambda_{n \to p}}{\overline{\Lambda}_{n \to p}} \, ,
\end{equation}
leading to a change in the equilibrium neutron abundance,
\begin{equation}
\label{eq:dxnsigma}
\delta X_n^{(\mathrm{eq})} = (1 - X_n) \sigma_\nu \, ,
\end{equation}
since $X_n/(1-X_n) = n_n/n_p$ and $(n_n/n_p)_{\mathrm{eq}} = \Lambda_{p \to n}/\Lambda_{n \to p}$. Corrections to the Born rates are shown in Fig.~\ref{fig:detailedbalance}. Equations \eqref{eq:detailedbalance} and thus \eqref{eq:dxnsigma} are not absolutely valid for $\lambda \simeq 1$ because deviations from detailed balance start earlier, but we can nonetheless estimate from this plot that $\sigma_\nu(\lambda \simeq 1) \simeq 0.001$. With $X_n(\lambda \simeq 1) \simeq 0.2$, we find from Eq.~\eqref{eq:dxnsigma} that including the spectral distortions increases the neutron fraction at freeze-out by
\begin{equation}
\label{eq:shift_disto}
\delta X_n^{[\mathrm{FO}],\delta g_{\nu_e}}  \lesssim 0.08 \, \% \, ,
\end{equation}
where the definition of this value corresponds to the shift from the orange curve to the green curve in the top panel of Fig.~\ref{fig:analyzevariations}:
\begin{equation}
\delta X_n^{[\mathrm{FO}],\delta g_{\nu_e}}  \equiv \delta X_{n,[T_{\nu_e},\text{with dist.}]}^{[\mathrm{FO}]} - \delta X_{n,[T_{\nu_e},\text{no dist.}]}^{[\mathrm{FO}]} \, .
\end{equation}
The value \eqref{eq:shift_disto} is overestimated because at $\lambda =1$, the neutron-to-proton ratio has already deviated from nuclear statistical equilibrium. In fact, one can reasonably consider that the shift in $\delta X_n^{[\mathrm{FO}]}$ is due to the deviation from detailed balance at higher temperatures, when nuclear statistical equilibrium was actually verified (namely, for $\lambda \sim 4$). Indeed, using Eq.~\eqref{eq:dxnsigma} for $\lambda \sim 4$, we obtain the observed shift $\delta X_n^{[\mathrm{FO}],\delta g_{\nu_e}} = 0.02 \, \%$.

We conclude this detailed analysis of neutron/proton freeze-out by stating the obtained value for $\delta X_n^{[\mathrm{FO}]}$, which can be read from Fig.~\ref{fig:analyzevariations} at $\lambda \sim 1$ in the ``$T_{\nu_e},$ with distortions" case:
\begin{equation}
\label{eq:dxnfo}
\delta X_n^{[\mathrm{FO}]} \simeq + 0.014 \, \% \, .
\end{equation}

\subsubsection{Clock effect}

The \emph{clock effect} is due to the higher radiation energy density for a given plasma temperature, which reduces the time necessary to go from $\TFO$ to $\TNuc$. This leads to less neutron beta decay, and thus a higher $X_n(\TNuc)$ and consequently a higher $\YP$. To estimate this contribution we will make several assumptions, justified by observing Fig.~\ref{fig:summary_nubbn}. Since $t_\mathrm{Nuc} \sim 245 \, \mathrm{s} \gg t_\mathrm{FO}$, the freeze-out modification discussed previously will only result in a very small change in duration; indeed, we find numerically that $\Delta t_\mathrm{FO} \simeq 0.002 \ \mathrm{s}$. We also checked that $\TNuc$ is almost not modified ($\delta \TNuc \simeq - 0.01 \ \%$), which is expected since the onset of nucleosynthesis is essentially determined only by $T_\gamma$. Therefore, the clock effect is mainly described by the change of duration between $\TFO^{(0)}$ and $\TNuc \simeq \TNuc^{(0)}$.

An additional assumption is made by observing the time scale in Fig.~\ref{fig:summary_nubbn}: most of the neutron beta decay takes place when neutrinos have decoupled and electrons and positrons have annihilated. We will thus consider that between the freeze-out and the beginning of nucleosynthesis, neutrinos are decoupled and $N_\mathrm{eff} \simeq N_\mathrm{eff}^\mathrm{fin}$ is constant.

Therefore, we can write $H \propto 1/2t$ as we are in the radiation era, cf.~\eqref{eq:hubble_radiation}. Using Friedmann equation $H^2 \propto \rho$, we get
\begin{equation}
\frac{\Delta t_\mathrm{Nuc}}{t_\mathrm{Nuc}^{(0)}} = - \frac12 \left. \frac{\Delta \rho}{\rho^{(0)}}\right|_{T_\gamma = \TNuc} = - \left. \frac{\Delta \rho_\nu}{\rho_\nu^{(0)}}\right|_{\TNuc}  \times \frac{\rho_\nu^{(0)}}{\rho^{(0)}}\,.
\end{equation}
The factor $1/2$ disappears since the variation of the energy density is $\Delta \rho = \Delta \rho_\nu + \Delta \rho_{\bnu} = 2 \Delta \rho_\nu$ (we consider the standard case without asymmetry). This shift in the neutrino energy density is parameterized by $N_\mathrm{eff}$, while the ratio of instantaneously decoupled energy densities is, at $\TNuc$, 
\begin{equation}
\frac{\rho_\nu^{(0)}}{\rho^{(0)}} = \frac{\rho_\nu^{(0)}}{\rho_\gamma^{(0)} + \rho_\nu^{(0)} + \rho_{\bnu}^{(0)}} = \frac{\frac78 \frac{\pi^2}{30} \times 3 \times \left(\frac{4}{11}\right)^{4/3} \TNuc^4}{2 \times \frac{\pi^2}{30} \TNuc^4 + 2 \times \frac78 \frac{\pi^2}{30} \times 3 \times \left(\frac{4}{11}\right)^{4/3} \TNuc^4} \simeq 0.203 \, .
\end{equation}
This gives
\begin{equation*}
\frac{\Delta t_\mathrm{Nuc}}{t_\mathrm{Nuc}^{(0)}} \simeq - 0.203 \times \frac{\Delta N_\mathrm{eff}}{3} \simeq -3.0\times 10^{-3} \, ,
\end{equation*}
but since we included the QED corrections to the plasma thermodynamics at all stages of the calculation, the reference value for $\Neff$ is not $3$ but $\Neff^{(0),\text{QED}} = 3.00965$, which leads to
\begin{equation}
\frac{\Delta t_\mathrm{Nuc}}{t_\mathrm{Nuc}^{(0)}} \simeq - 0.203 \times \frac{\Delta N_\mathrm{eff}}{\Neff^{(0),\text{QED}}} \simeq -2.3\times 10^{-3} \, ,
\end{equation}
This estimate is actually in very good agreement with the numerical result
\begin{equation}
\left. \frac{\Delta t_\mathrm{Nuc}}{t_\mathrm{Nuc}^{(0)}}\right|_{\texttt{PRIMAT}} \simeq -2.1 \times 10^{-3} \, .
\end{equation}
 Hence, the estimate for the clock effect contribution is, from Eq.~\eqref{eq:deltaxn},
\begin{equation}
\label{eq:dxnclock}
\delta X_n^{[\Delta t]} = - \frac{\Delta t_\mathrm{Nuc}}{t_\mathrm{Nuc}^{(0)}} \times \frac{t_\mathrm{Nuc}^{(0)}}{\tau_n} \simeq 0.064 \, \% \, ,
\end{equation}
where we recall that $t_\mathrm{Nuc}^{(0)} \simeq 245 \, \mathrm{s}$ and $\tau_n = 879.4 \, \mathrm{s}$.

\subsection{Primordial abundances}

The previous results allow to estimate the changes to the primordial abundances. We separate the discussion between the $\He4$ abundance, which is essentially set by the neutron fraction at freeze-out, and the other light element abundances, for which the clock effect affects the nuclear reactions.

\subsubsection{Helium abundance}

The previous study allows us to estimate the change in the $\He4$ abundance. Since most neutrons are converted into $\He4$, by combining Eqs.~\eqref{eq:dxnfo} and \eqref{eq:dxnclock} (``$T_{\nu_e},$ with distortions" case) we get
\begin{equation}
\delta \YP = \delta X_n^{[\mathrm{Nuc}]} = \delta X_n^{[\mathrm{FO}]} + \delta X_n^{[\Delta t]} \simeq 0.078 \, \% \, ,
\label{eq:deltaYp}
\end{equation}
which is in excellent agreement with the result in Table~\ref{Table:Previous}. 

The different values of $\YP$ depending on the implementations (see Table~\ref{Table:Corrections}) are very well described by this explanation: since the energy density is always the same, $\delta X_n^{[\Delta t]}$ remains identical, while the varying $\delta X_n^{[\mathrm{FO}]}$ (Fig.~\ref{fig:analyzevariations}) controls $\delta \YP$.

\subsubsection{Other abundances}

We now focus on the other light elements produced during BBN, up to $\Be$. To understand the individual variations of abundances due to incomplete neutrino decoupling, in Table~\ref{Table:LightElements} we separate the final abundances of $\He3$, $\mathrm{T}$, $\Be$, and $\Li$.

\begin{table}[h]
	\centering
	\begin{tabular}{|l|rrrrr|}
  	\hline
  & \multicolumn{1}{c}{$\mathrm{D}/\mathrm{H}$} & \multicolumn{1}{c}{${\He3}/\mathrm{H}$}&  \multicolumn{1}{c}{$\mathrm{T}/\mathrm{H}$} &  \multicolumn{1}{c}{${\Be}/\mathrm{H}$} &  \multicolumn{1}{c|}{${\Li}/\mathrm{H} $} \\
  \hline \hline
  $(i/\mathrm{H})^{(0),\infty}$ & $2.41 \times 10^{-5}$ & $1.03 \times 10^{-5}$ & $7.69 \times 10^{-8}$ & $5.17 \times 10^{-10}$ & $2.76 \times 10^{-11}$ \\
  $\Delta (i/\mathrm{H})^\infty$ & $9.7 \times 10^{-8}$ & $1.4 \times 10^{-8}$  & $3.3 \times 10^{-10}$ & $- 2.2 \times 10^{-12}$ & $1.2 \times 10^{-13}$ \\
   $\delta (i/\mathrm{H})^\infty$ & $0.40 \, \% $ &  $0.13 \, \%$ & $0.43 \, \%$ &$- 0.42 \, \%$ & $0.43 \, \%$   \\ \hline 
\end{tabular}
	\caption[Variation of primordial abundances due to incomplete neutrino decoupling, at the Born level]{Neutrino-induced corrections to the primordial production of light elements other than $\He4$. QED corrections to the plasma thermodynamics are included up to order $\mathcal{O}(e^3)$, and the weak rates are computed at the Born level.
	\label{Table:LightElements}}
\end{table}

There are two contributions to the change in the final abundance of an element:
\begin{equation}
\delta (i/\mathrm{H})^{\infty} = \delta X_i^{\infty} - \delta X_\mathrm{H}^{\infty} \simeq \delta X_i^{[\Delta t]} + \delta X_n^{[\mathrm{Nuc}]} \, .
\label{eq:abund_light}
\end{equation}
The variation of the proton final abundance is directly related to $\delta X_n^{[\mathrm{Nuc}]}$ given in Eq.~\eqref{eq:deltaxn}, because an increase of $X_n^{[\mathrm{Nuc}]}$ corresponds to a higher neutron-to-proton ratio and/or less beta decay, and thus less protons. On the other hand, the variation of $X_i^{\infty}$ is entirely encapsulated in the clock effect contribution $\delta X_i^{[\Delta t]}$ (it does not depend on $X_n(\TNuc)$ at first order, since all light elements except $\He4$ only appear at trace level). Indeed, nucleosynthesis consists in elements being produced/destroyed until the reaction rates (which depend only on $T_\gamma$) become too small compared to the Hubble rate~\cite{SmithBBN}. Because of incomplete neutrino decoupling, a given value of $T_\gamma$ is reached sooner and the nuclear reactions have had less time to be efficient. In other words, there is less time to produce or destroy the different elements.\footnote{This argument does not apply to $\He4$ since it is the most stable light element: for such small variations of the expansion rate, almost all neutrons still end up in $\He4$, so $\YP$ is only affected by $\delta X_n^{[\mathrm{Nuc}]}$.}

\begin{figure}[!ht]
	\centering
	\includegraphics{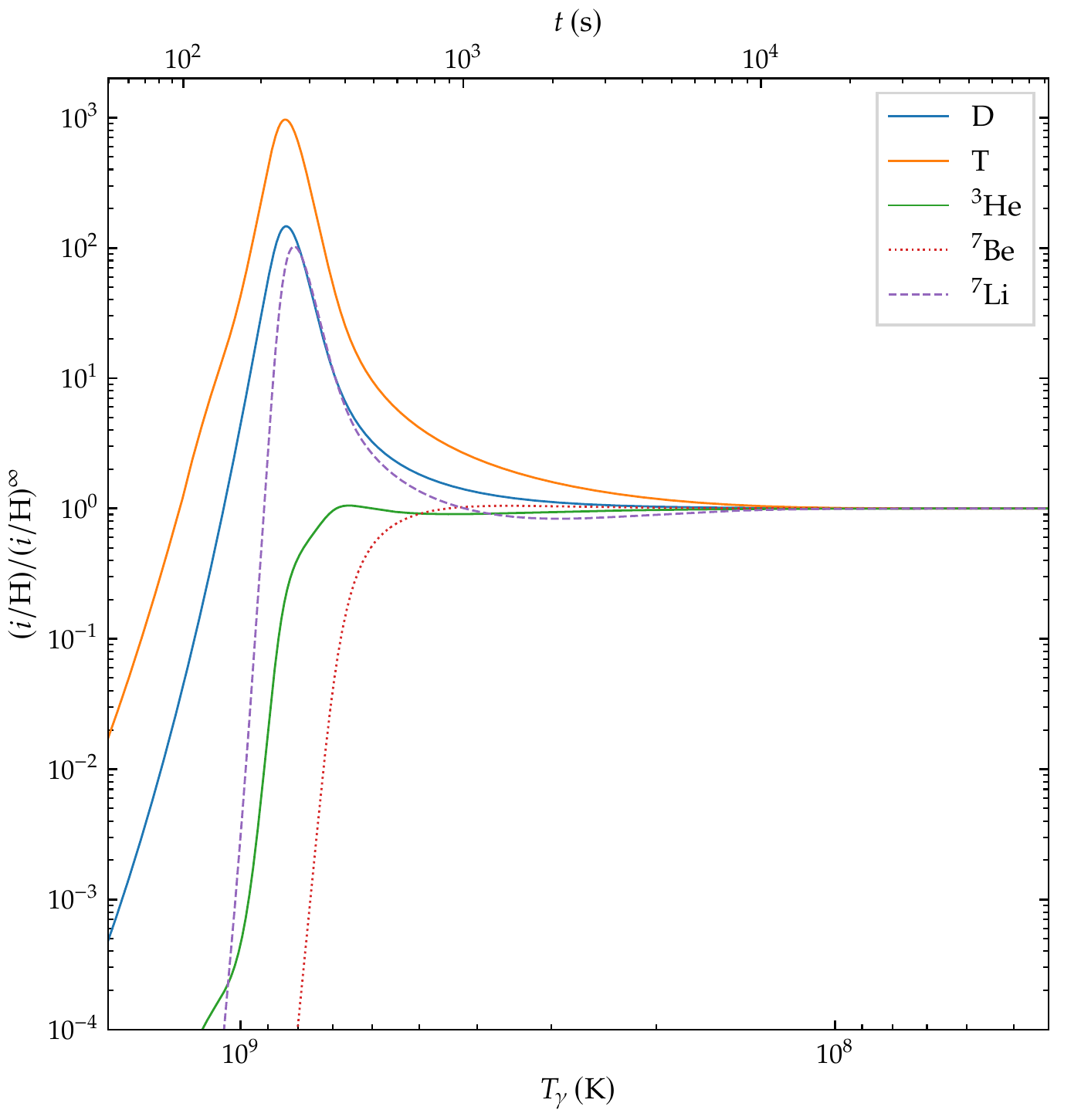}
	\caption[Evolution of light element abundances, rescaled by their frozen-out value]{\label{fig:evolutionabundances} Evolution of light element abundances computed with \texttt{PRIMAT}, including incomplete neutrino decoupling corrections at the Born approximation level. To compare the evolution for different elements, all abundances are rescaled by their frozen-out value.}
\end{figure}

We can thus understand the values of Table~\ref{Table:LightElements} by looking at the evolution of abundances at the end of nucleosynthesis, shown in Fig.~\ref{fig:evolutionabundances}. All elements except $\Be$ are mainly destroyed when the temperature drops below $\TNuc$. The very similar evolutions of $\mathrm{D}$, $\mathrm{T}$, and $\Li$ explain their similar values of $\delta X_i^\infty$: their destruction rates go to zero more quickly, resulting in a higher final abundance value. For $\Be$ it is the opposite: it is more efficiently produced than destroyed, and the clock effect reduces the possible amount formed (hence, the negative $\delta X_{\Be}^{\infty}$). Moreover, its evolution is even sharper than that of tritium, and thus we expect $\abs{\delta X_{\Be}^{\infty}} > \delta X_\mathrm{T}^{\infty}$. Finally, $\He3$ has much smaller variations, with a small amplitude of abundance reduction from $\TNuc$. This explains the comparatively small value of $\delta X_{\He3}^{\infty}$.

To recover the aggregated variations of Table~\ref{Table:Previous} (for $\He3$ and $\mathrm{T}$, and $\Be$ and $\Li$), one performs the weighted average of individual variations. Since $(\He3/\mathrm{H})^\infty \gg (\mathrm{T}/\mathrm{H})^\infty$, the contribution of $\He3$ dominates, and this argument can be immediately applied to $\Be$ and $\Li$.

\subsection{Precision nucleosynthesis}
\label{subsec:full_BBN}

Having thoroughly studied the physics at play by focusing on the Born approximation level, we can now present the results incorporating all weak rates corrections derived in~\cite{Pitrou_2018PhysRept}. These additional contributions (radiative corrections, finite nucleon mass, and weak magnetism) cannot in principle be added linearly, due to nonlinear feedback between them. Concerning incomplete neutrino decoupling, this means that we also include radiative corrections inside the spectral distortion part of the rates: we modify Eq.~\eqref{eq:deltagamma}, following Eqs.~(100) and (103) in~\cite{Pitrou_2018PhysRept}. Since the neutrino sector physics was already “accurate” (given that we used results with QED corrections and flavour mixing), the corrections to the weak rates are the last missing ingredient at the nucleosynthesis level to have the most precise predictions of primordial abundances.

The results, once again for the three implementations of neutrino-induced corrections, are given in Table~\ref{Table:Full_Corrections}.

\begin{table}[!htb]
	\centering
	\begin{tabular}{|l|cccc|}
  	\hline 
  & $\YP$  &  $\mathrm{D}/\mathrm{H}  \times 10^5$ &  $\He3/\mathrm{H} \times 10^5$ &  $\Li/\mathrm{H} \times 10^{10}$ \\
  \hline \hline
   Inst. decoupling, all corr. & $0.24711$  & $2.4291$ & $1.0379$  & $5.5270$ \\ \hline
  $\widehat{T}_\nu$ & $0.24716$  & $2.4381$ & $1.0392$  & $5.5038$ \\
    $T_{\nu_e},$ no distortions & $0.24713$  & $2.4380$ & $1.0392$  & $5.5033$ \\
  $\bm{T_{\nu_e},}$\textbf{ with distortions} & $\mathbf{0.24721}$  & $\mathbf{2.4384}$ & $\mathbf{1.0393}$  & $\mathbf{5.5045}$\\ \hline 
\end{tabular}
	\caption[Light element abundances with all corrections included]{Light element abundances, including all weak rate corrections and QED corrections to plasma thermodynamics, for various implementations of neutrino-induced corrections. See Table~\ref{Table:Corrections} for results at the Born approximation level. The final row (in boldface) gives the prediction for primordial abundances in the most accurate framework ; the values were reported in Table~\ref{Table:General_BBN}.
	\label{Table:Full_Corrections}}
\end{table}

Compared to the Born approximation level (Table~\ref{Table:Corrections}), the additional corrections result in higher final abundances, as discussed in~\cite{Pitrou_2018PhysRept}. Starting then from a baseline where all of these corrections are included except for incomplete neutrino decoupling, the shift in abundances due to neutrinos is slightly reduced by roughly $- \, 0.03 \, \%$; for instance $\delta \YP = + \, 0.043 \, \%$ instead of $+ \, 0.076 \, \%$. The other conclusions of the previous sections remain valid: the average temperature implementation is close to the complete one, we explain $\YP$ through $X_n(\TNuc)$, and the clock effect sources the variations of light elements other than $\He4$.

\begin{figure}[!ht]
	\centering
	\includegraphics{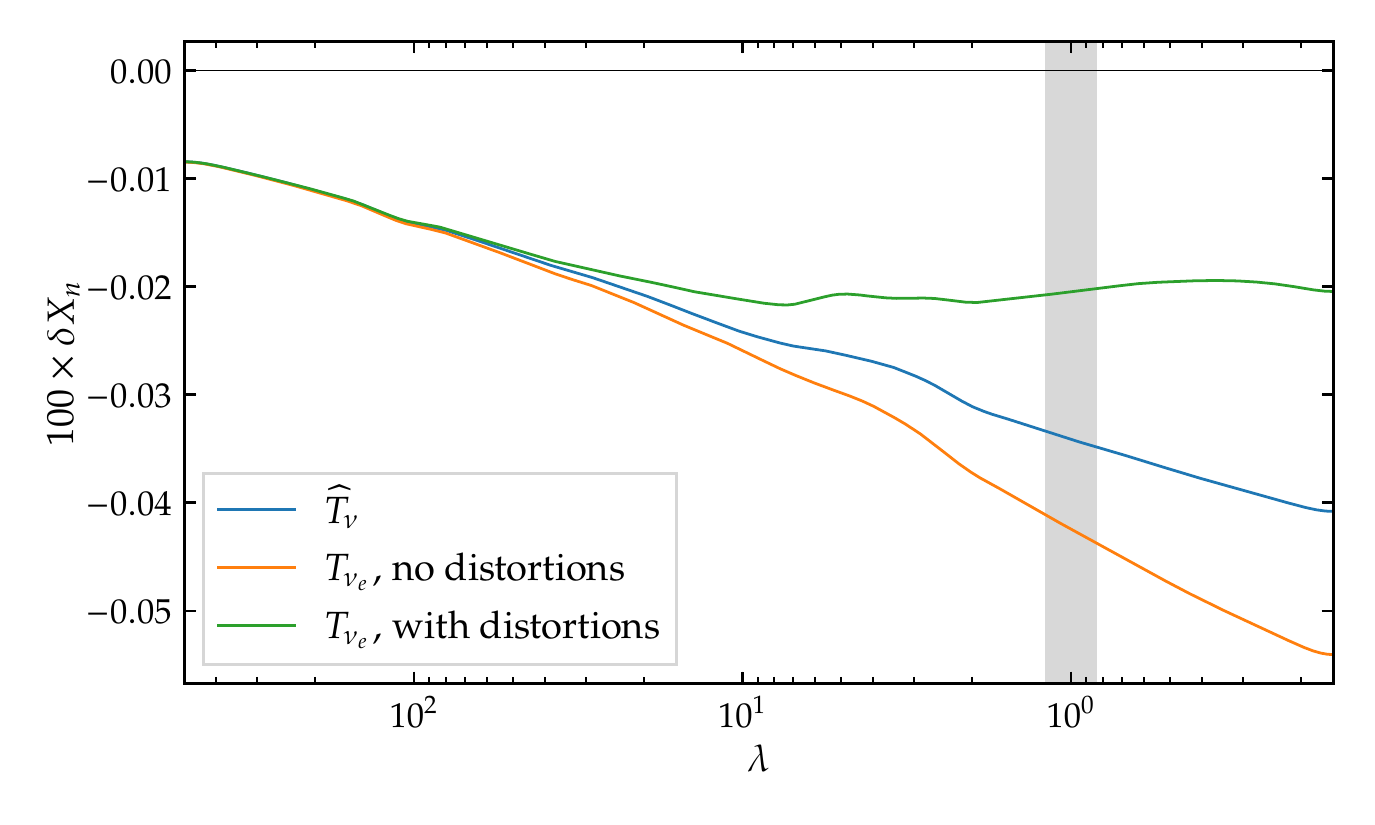}
	\caption[Neutron fraction variation for different implementations of neutrino-induced corrections (all other weak rate corrections included)]{\label{fig:analyzevariationsfull} Neutron fraction around freeze-out, in the different implementations of neutrino-induced corrections. Compared to Fig.~\ref{fig:analyzevariations}, all weak rate corrections are included.}
\end{figure}

Since the additional corrections like finite nucleon mass contributions only affect the weak rates and not the energy density, we expect that the only difference compared to the picture at the Born level will lie in $\delta X_n^{[\mathrm{FO}]}$, while $\sigma_\nu$ and $\delta X_i^{[\Delta t]}$ will remain unchanged. This is indeed what we observe in Fig.~\ref{fig:analyzevariationsfull}: the reduction of the neutron fraction at freeze-out due to incomplete neutrino decoupling is enhanced when including all weak rates corrections. Moreover, by comparing Figs.~\ref{fig:analyzevariationsfull} and \ref{fig:analyzevariations} we find
\begin{equation}
\delta X_{n, \mathrm{All}}^{[\mathrm{FO}]} - \delta X_{n, \mathrm{Born}}^{[\mathrm{FO}]} \simeq - 0.03 \, \% \, ,
\end{equation}
which, by inserting this difference into Eqs.~\eqref{eq:deltaYp} and \eqref{eq:abund_light}, explains the results of Table~\ref{Table:Full_Corrections}.

\section*{Concluding remarks} Neutrino decoupling is now included in the standard BBN codes, at least via the $\Neff$ parameter~\cite{AlterBBN,AlterBBNv2} or through the $\mathcal{N}$ heating function~\cite{Parthenope,Parthenope_reloaded,Parthenope_revolutions}. Note that the very small difference between the “$\widehat{T}_\nu$” and “$T_{\nu_e}$, with distortions” (way below the experimental uncertainties) justifies that the $\mathcal{N}$ method can give satisfactory results. However, no change to the weak rates seems to be taken into account in \texttt{AlterBBN}. 

Our detailed analysis has evidenced that including incomplete neutrino decoupling leads to an increase of helium-4, deuterium and helium-3 abundances, and a reduction of lithium-7 abundance, in agreement with~\cite{Grohs2015} but disagreeing with~\cite{Mangano2005}.

Can we be satisfied with an approximate treatment of neutrino decoupling in a BBN calculation? Looking at Tables~\ref{Table:General_BBN} and~\ref{Table:Previous}, one can see that the scale of the variations due to incomplete neutrino decoupling (compared to instantaneous decoupling) is below the experimental uncertainties on the primordial abundances --- except for deuterium where the precision has reached the percent level. A possible tension concerning the abundance of deuterium has been suggested~\cite{Pitrou2020}, which shows that an accurate treatment of neutrino decoupling is crucial for precision nucleosynthesis and to make conclusions regarding the cosmological model. Nevertheless, the very small variation of abundances \emph{between the various implementations} shows that a treatment that is not completely exact, for instance via the heating rate $\mathcal{N}$, can be satisfactory. But, given our results that can be provided on demand, and the existence of other neutrino decoupling public codes~\cite{Bennett2021}, one might as well be as precise as possible on this point.

\pagestyle{ruled}

\chapter[Primordial neutrino asymmetry evolution][Primordial neutrino asymmetry evolution]{Primordial neutrino asymmetry evolution}
\label{chap:Asymmetry}




\setlength{\epigraphwidth}{0.5\textwidth}
\epigraph{It's supposed to be a little asymmetrical. Apparently a small flaw somehow improves it.}{Sheldon Cooper, \emph{The Big Bang Theory} [S11E24]}

{
\hypersetup{linkcolor=black}
    \minitoc
}

\boxabstract{The material of this chapter was published in~\cite{Froustey2021}.}

The specific features of the cosmic neutrino background and its effects on primordial nucleosynthesis have been studied in the previous chapters, assuming systematically that neutrinos and antineutrinos kept the same distributions. This absence of asymmetry --- or rather, the extremely small asymmetry --- between matter and antimatter is measured for baryons via the parameter $\eta \sim 10^{-9}$~\cite{Fields:2019pfx}. Since, by charge conservation, $n_{e^-} - n_{e^+} = n_p$ during BBN ($\mathrm{H}$ being by far the most numerous species), there can be no sizable asymmetry in the charged lepton sector. However, thanks to their electric neutrality, no such constraint exists for neutrinos.

In thermal and chemical equilibrium, the chemical potential of a given neutrino flavour $\alpha$ and the corresponding antineutrino chemical potential are related through $\mu_\alpha =- \bar \mu_\alpha $. The initial asymmetry in a given flavour $\alpha$, defined as the difference between the neutrino and antineutrino comoving densities, is related to $\mu_\alpha$, which is not constrained \emph{a priori} for the reason explained above. One thus hopes to constrain them from their impact on cosmological observables. More specifically, to take into account the effects of momenta redshifting due to cosmological expansion, we aim at constraining the degeneracy parameters $\xi_\alpha \equiv \mu_\alpha/\Tcm$ which are conserved by expansion. There is first an effect of $\xi_e$ on the neutron/proton freeze-out, which affects Big-Bang Nucleosynthesis (BBN). Indeed, the detailed balance relation~\eqref{eq:nse} is shifted to
\begin{equation}
	\label{eq:nse_xie}
	\frac{n_n}{n_p} = \left. \frac{n_n}{n_p}\right\rvert_{\xi_e = 0} \times e^{- \xi_e} \, .
\end{equation}
Also the total energy density of a given neutrino flavour and its corresponding antineutrino is supplemented by a term $\propto \xi_\alpha^2$, leading to a modification of $\Neff$, and this has an impact on cosmological expansion which affects both BBN and the cosmic microwave background (CMB) anisotropies. Hence, assuming a full equilibration of neutrino asymmetries with a common $\xi$, a constraint can be obtained from BBN alone~\cite{Simha:2008mt,Fields:2019pfx}, from CMB alone~\cite{Oldengott:2017tzj,Planck18}, or using a combination of both~\cite{Pitrou_2018PhysRept} to give 
\begin{equation}
    \label{eq:xi_BBN_CMB}
\xi = 0.001 \pm 0.016\,.
\end{equation}
If standard baryogenesis models involving sphalerons suggest that $\xi$ should be of the order of the baryon asymmetry $\eta = n_b/n_\gamma \simeq 6.1\times 10^{-10}$~\cite{Neutrino_Cosmology,Davidson_Leptogenesis}, other proposed models like~\cite{McDonald:1999in,March-Russell:1999hpw,Gu:2010dg} manage to combine a large lepton asymmetry with the value of $\eta$. Therefore potentially “high” values of $\xi$ are not forbidden and \eqref{eq:xi_BBN_CMB} motivates why we focus in this chapter on degeneracy parameters in the range $[10^{-3},10^{-1}]$. 

The total asymmetry, that is the sum over each flavour asymmetry, is preserved by the physical processes at play. However, individual asymmetries can evolve towards the average, in which case we can talk about “\emph{flavour equilibration}”. The goal of this chapter is to review the physics of this equilibration, that is the evolution of the degeneracy parameters, accounting for all relevant physical effects at play during neutrino decoupling. To that end, it is necessary to solve the Quantum Kinetic Equations (QKEs) that dictate the evolution of (anti)neutrino density matrices, taking into account vacuum oscillations, mean-field effects with leptons and neutrinos (the latter being referred to as self-interaction mean-field), and collision processes.

First, self-interactions have a crucial effect in delaying equilibration as they are responsible for the so-called \emph{synchronous oscillations}~\cite{Pastor:2001iu,Dolgov_NuPhB2002,Abazajian2002,Wong2002}, and we find that in general there is also a second regime with \emph{quasi-synchronous oscillations} having much larger frequencies. Furthermore we find that the complete form of the neutrino collision term must be used, including the full matrix structure of both reactions among neutrinos and with electrons/positrons.

Unless the chemical potential differences are very small, there is always a period when self-interactions dominate over the lepton mean-field contribution and the vacuum Hamiltonian contribution. One of the dramatic consequences is that solving the exact evolution of neutrino number densities involves very short time scales compared to the cosmological time scale, which implies that it is numerically very difficult to treat them exactly. So far, the main approach when considering non-vanishing degeneracies consisted in using a damping approximation for the collision term~\cite{Bell98,Dolgov_NuPhB2002,Pastor:2008ti,Gava:2010kz,Gava_corr,Mangano:2010ei,Johns:2016enc,Barenboim:2016shh}, either for all its components or only for its off-diagonal components. Indeed the computation of the collision term is the time-consuming step with a $\mathcal{O}(N^3)$ complexity, where $N$ is the number of points used to sample the neutrino spectra. We use none of these approximations, and we find from the structure of the full collision term that it cannot efficiently damp all types of synchronous oscillations, a feature that is lost when relying on damping approximations. 

We have shown in chapter~\ref{chap:Decoupling} that the numerical resolution could be considerably improved, altering only subdominantly the precision of results, by using an approximate scheme which consisted in averaging over neutrino oscillations in the adiabatically evolving matter basis. In this chapter, we extend this method with initial degeneracies, that is taking into account the effect of the self-interaction mean-field. In section~\ref{SecTheory} we summarize the formalism used to describe the evolution of neutrino and antineutrino density matrices in the context of non-zero asymmetries, and in section~\ref{SecResScheme} we detail the various numerical schemes we developed, notably the extension of the ATAO scheme when considering self-interactions. Restricting to oscillations with only two neutrinos in section~\ref{Sec2Neutrinos}, we derive analytic expressions for synchronous and quasi-synchronous oscillations. Two physically motivated cases with two-neutrino flavours are then investigated in details in section~\ref{SecRelevant2Neutrinos}. They allow to understand the evolution of neutrino asymmetry in the general case with three neutrinos, which is presented in section~\ref{Sec:3neutrinos}, along with an assessment of the dependence on the main mixing parameters (mass ordering, mixing angles, Dirac phase). Finally we discuss the main differences with existing results in the literature in section~\ref{SecDiscussion}.

\section{Neutrino evolution in the primordial Universe with degeneracies}\label{SecTheory}

In order to determine neutrino evolution in the early Universe, one must solve a set of quantum kinetic equations in the expanding Universe, involving both neutrino oscillations and collisions. We present in this section all the variables relevant to this problem, with a particular emphasis on the physical quantities related to the presence of a neutrino/antineutrino asymmetry.

\subsection{QKE with a non-zero neutrino/antineutrino asymmetry}

Let us recall the general form of the QKE~\eqref{eq:QKE_fullfinal}, valid in the early Universe:
\begin{multline}
\frac{\partial \vrho(x,y_1)}{\partial x} = - \frac{\ii}{xH} \left(\frac{x}{m_e}\right) \left[ U \frac{\mathbb{M}^2}{2y_1}U^\dagger, \varrho \right]  +  \ii \frac{2 \sqrt{2} G_F}{xH} y_1 \left(\frac{m_e}{x}\right)^5 \left[ \frac{\bar{\mathbb{E}}_\mathrm{lep} + \bar{\mathbb{P}}_\mathrm{lep}}{m_W^2} ,\varrho \right ] \\
- \ii \frac{\sqrt{2} G_F}{x H} \left(\frac{m_e}{x}\right)^3 \left[ \overline{\mathbb{N}}_\nu - \overline{\mathbb{N}}_{\bnu}, \vrho \right] + \ii \frac{8 \sqrt{2} G_F}{3 x H} y_1 \left( \frac{m_e}{x}\right)^5 \left[\frac{\bar{\mathbb{E}}_\nu + \bar{\mathbb{E}}_{\bar{\nu}}}{m_Z^2}, \vrho\right]  + \frac{1}{xH} \mathcal{I} \, ,
\end{multline}
We will neglect the symmetric term proportional to (anti)neutrino energy densities, as it is always negligible compared to the same term proportional to charged lepton energy densities. Indeed, for initial Fermi-Dirac distributions at the same temperature and without degeneracies, this term is purely proportional to the identity matrix, hence it does not contribute to the dynamics of density matrices, which was the reason we discarded this term in chapter~\ref{chap:Decoupling}. Considering initial degeneracies, we have $\bar{\rho}_{\nu_\alpha}+\bar{\rho}_{\bar{\nu}_\alpha} \propto \xi_\alpha^2$, therefore this contribution is typically smaller than $\bar{\rho}_\mathrm{lep}$ (for relativistic leptons) by a factor which is of the order of the $\xi_\alpha^2$ differences.

We rewrite the QKE for $\vrho$, along with the equation for $\bvrho$, in a more concise way:
\begin{subequations}
  \label{eq:QKE_compact_asym}
\begin{align}  
\frac{\partial \vrho}{\partial x} &= - \ii [\Hamil +
\Hself,\vrho] + \mathcal{K} \, ,\\
\frac{\partial \bvrho}{\partial x} &= + \ii [\Hamil -\Hself,\bvrho] + \overline{\mathcal{K}} \, ,
\end{align}
\end{subequations}
with $\Hamil = \Hvac + \Hlep$, where the different Hamiltonians have been defined in~\eqref{eq:Hvac} and~\eqref{eq:Hlep}. For convenience, we recall here the expressions:
\begin{equation}
\label{eq:Hamil_general}
\Hvac \equiv  \frac{1}{xH} \left(\frac{x}{m_e}\right) U \frac{\mathbb{M}^2}{2 y_1} U^\dagger \quad , \qquad 
\Hlep \equiv - \frac{1}{xH} \left(\frac{m_e}{x}\right)^5 2 \sqrt{2} G_F y_1 \frac{\bar{\mathbb{E}}_\mathrm{lep} + \bar{\mathbb{P}}_\mathrm{lep}}{m_W^2} \, .
\end{equation}
 It proves convenient to separate the $y-$dependence in the Hamiltonian --- see section \ref{Sec2Neutrinos}. Hence we define 
\begin{equation}
    \label{eq:Hamil_y}
    \mathcal{H}_0 \equiv \Hvacy / y \, , \qquad  \mathcal{H}_\mathrm{lep} \equiv \Hlepy y \, .
\end{equation}

Contrary to the standard case studied in chapter~\ref{chap:Decoupling}, the self-interaction Hamiltonian $\Hself$ must be included when considering neutrino asymmetries. We introduce the notation
\begin{equation}
\label{DefJ}
\Hself = \frac{1}{xH} \left(\frac{\me}{x}\right)^3  \sqrt{2} G_F \Anti \ ,  \ \  \text{where} \quad \Anti \equiv (\overline{\mathbb{N}}_\nu - \overline{\mathbb{N}}_{\bnu}) = \int (\vrho-\bvrho)\mathcal{D}y\,.
\end{equation}
Note that $\Anti$, referred to as the “(integrated) neutrino asymmetry", is simply proportional to the lepton number matrix $\eta_\nu \equiv \Anti/n_\gamma = \pi^2/[2 \zeta(3) z^3] \times \Anti$. We introduced the convenient notation $\mathcal{D}y \equiv y^2 \dd{y}/(2 \pi^2)$.

The Hubble rate $H$ is given by the Friedmann equation, that we recall here to highlight its dependence on $x$,
\begin{equation}
    \label{eq:scaling_Hubble_x}
    H = \frac{\me}{m_\mathrm{Pl}} \times \frac{\me}{x^2} \times \sqrt{\frac{\bar{\rho}}{3}} \qquad \text{where} \quad \bar{\rho} =\bar{\rho}_\gamma + \bar{\rho}_{\nu, \bar{\nu}} + \bar{\rho}_{e^\pm} + \bar{\rho}_{\mu^\pm}  \, ,
\end{equation}
where we stress again that the “barred” energy densities are the comoving ones, differing by a factor $(\me/x)^4$ from the physical ones. $m_\mathrm{Pl} \simeq 2.435 \times 10^{18} \, \mathrm{GeV}$ is the reduced Planck mass.

\paragraph{Mixing parameters} We use the standard parameterization of the PMNS matrix~\eqref{eq:PMNS_bis} and the values~\eqref{ValuesStandard}, unless otherwise specified. We do not take into account the CP phase except in the dedicated subsection~\ref{SecDiracPhase}.

\paragraph{Dirac or Majorana neutrinos} It could be natural to believe that an asymmetry between neutrinos and antineutrinos is not possible if neutrinos are Majorana particles\footnote{For instance, the argument was raised in~\cite{Bernstein1982} before being corrected in~\cite{Langacker1982a,Langacker1982b} --- the original mistake was acknowledged in~\cite{Bernstein1984}.}. Indeed, in that case “neutrinos are their own antiparticles”, which should lead necessarily to $\mathbb{N}_\nu = \mathbb{N}_{\bnu}$. But this overlooks the fact that there are \emph{helicity} degrees of freedom to take into account. Due to the Majorana condition $\nu = \nu^C$ (where $^C$ denote charge conjugation), there are twice as many degrees of freedom for Dirac neutrinos (left-handed\footnote{Left-handed (resp. right-handed) referring to a negative (resp. positive) \emph{helicity}.}  and right-handed neutrinos, left-handed and right-handed antineutrinos), while there are only left-handed and right-handed neutrinos in the Majorana case. It is then common to refer to right-handed neutrinos in the Majorana case as “antineutrinos”. However, all these states are not in thermal equilibrium, since helicity-flip rates are suppressed by a factor $\mathcal{O}(m_\nu^2/E_\nu^2) \ll 1$ compared to helicity-conserving reactions.

In summary, a positive asymmetry is interpreted:
\begin{itemize}
	\item for Majorana neutrinos, as an excess of left-handed over right-handed neutrinos,
	\item for Dirac neutrinos, as an excess of left-handed neutrinos over right-handed antineutrinos. 
\end{itemize}

The counting of degrees of freedom with details about $CPT$ transformations is explained in~\cite{GiuntiKim}, Section~6.2.2. The thermal population of the \emph{a priori} two extra degrees of freedom in the Dirac case is discussed in~\cite{LesgourguesPastor}: the “wrong-helicity” states cannot be populated except if they had a mass at the $\mathrm{keV}$ scale~\cite{Dolgov_2002PhysRep}, which is excluded for active neutrinos.

\subsubsection{Evolution of the plasma temperature}

For large temperatures, that is before we start the numerical resolution ($x<x_\mathrm{init}$), the evolution of the comoving plasma temperature is estimated assuming that neutrino spectra are thermal with the same temperature ($z_{\nu}=z$). Afterwards, the evolution of $z$ is computed using the full (anti)neutrino density matrices and the exact form for collisions between neutrinos and electrons/positrons. Including QED corrections~\cite{Heckler_PhRvD1994,Mangano2002,Bennett2020}, we use
\begin{equation}\label{D1Froustey2020}
\frac{\dd z}{\dd x} =  \frac{\displaystyle \frac{x}{z}J(x/z) - {S}_\nu + G_1(x/z)}{ \displaystyle \frac{x^2}{z^2}J(x/z) + Y(x/z) + \frac{1}{4}\sum_\alpha Y_{\nu}(\xi_\alpha/z) +  \frac{2 \pi^2}{15} + G_2(x/z)} \, ,
\end{equation}
where we defined
\begin{equation}
\begin{aligned}
& S_\nu = 0\,, \quad Y_{\nu}(\zeta_\alpha) \equiv \frac{1}{\pi^2} \int_{0}^{\infty}{\dd \omega \, \omega^3 (\omega-\zeta_\alpha) \frac{\exp{(\omega-\zeta_\alpha)}}{[\exp{(\omega-\zeta_\alpha)}+1]^2}}\quad &\text{for}\quad x \leq x_\mathrm{init}\,, \\
&Y_\nu = 0 \,,\quad S_\nu \equiv \frac{1}{4 \pi^2 z^3} \int_{0}^{\infty}{\dd y \, y^3 \left( \Tr [\mathcal{K}] + \Tr [\overline{\mathcal{K}}] \right)} \ \quad &\text{for}\quad x > x_\mathrm{init}\,,
\end{aligned}
\end{equation}
the functions $J,\, Y, \, G_1, \, G_2$ having been introduced in~\eqref{eq:zQED}. The sum on $\alpha$ in the denominator of \eqref{D1Froustey2020} runs on $2 N_\nu$ elements, being all neutrinos and antineutrinos species. 

The starting condition $z_\mathrm{init}$ at $x_\mathrm{init}$ is found by solving the differential equation~\eqref{D1Froustey2020}, with the initial condition $z=1$ at $x=0$.\footnote{In principle $z$ increases at each species annihilation, and in particular we should consider $\mu^\pm$ annihilations since these leptons appear in the mean-field effects. This choice of initial conditions is however consistent with neglecting the interactions with $\mu^\pm$ in the collision term, which therefore do not reheat the plasma of neutrinos, photons and $e^\pm$.} When there are no neutrino degeneracies, it matches the condition found by all coupled species entropy conservation. However, it gives a slightly different $z_\mathrm{init}$ in the presence of initial degeneracies since entropy conservation is then violated. 

\subsection{Neutrino asymmetry matrix $\Anti$}
\label{subsec:Anti}

Long before neutrino decoupling, that is for temperatures much larger than $2\,\mathrm{MeV}$, neutrinos and antineutrinos are maintained at kinetic and chemical equilibrium, thus generally following Fermi-Dirac (FD) distributions with a chemical potential 
\[g(T_\nu, \mu, p) \equiv \left[e^{(p - \mu)/T_\nu}+1\right]^{-1} \, .\]
Introducing the reduced variables $z_\nu = T_\nu/\Tcm$ and $\xi = \mu/\Tcm$, we rewrite this FD distribution 
\[g(z_\nu, \xi, y) = \left[e^{(y - \xi)/z_\nu}+1\right]^{-1} \, .\]
In most of the temperature range of interest, since electrons and positrons have not annihilated and all species are coupled, we can consider\footnote{We only take $z_\nu = 1$ for the analytical discussion in order to simplify the presentation. In the numerical resolution, the spectra evolve following the QKEs and $e^\pm$ annihilations increase the neutrino temperatures.} $z_\nu = 1$. We thus define $g(\xi, y) \equiv g(1, \xi, y)$, and the initial conditions read
\begin{equation}\label{rhoinit}
\vrho_\mathrm{init} = \mathrm{diag}\left[g(\xi_\alpha,y)\right] \, , \quad\bvrho_\mathrm{init} = \mathrm{diag}\left[g(-\xi_\alpha,y)\right]\,.
\end{equation}

\paragraph{Useful properties of Fermi-Dirac spectra} In this chapter, we will often use the following relations:\footnote{They respectively intervene in the calculation of the asymmetry, the sum of energy densities, the leading and next-to-leading orders of the asymmetry oscillation frequency.} 
\begin{subequations}
\label{IntegralsFD}
\begin{align}
\int [g(\xi,y) - g(-\xi,y)] \mathcal{D} y &= \frac{\xi}{6} + \frac{\xi^3}{6\pi^2}\label{Intg1}\\
\int y \, [g(\xi,y) + g(-\xi,y)]  \mathcal{D} y &= \frac{7 \pi^2}{120}+ \frac{\xi^2}{4} + \frac{\xi^4}{8\pi^2}\label{Intg2}\\
\int y^{-1} \, [g(\xi,y) + g(-\xi,y)]  \mathcal{D} y &= \frac{1}{12} + \frac{\xi^2}{4\pi^2}\label{Intg3}\\
\int y^{-2} \, [g(\xi,y) - g(-\xi,y)] \mathcal{D} y &= \frac{\xi}{2 \pi^2} \label{Intg4}
\end{align}
\end{subequations}

From equation~\eqref{Intg1}, the asymmetry matrix is initially
\begin{equation}
\label{eq:init_Anti}
\Anti_\mathrm{init} = \frac{1}{6}\mathrm{diag}\left[\xi_\alpha+\frac{\xi_\alpha^3}{\pi^2}\right] \, .
\end{equation}

For reasons detailed in section~\ref{SecResScheme}, we also introduce the evolution equation for $\Anti$. It is obtained by combining the QKE \eqref{eq:QKE_compact_asym} with the definition~\eqref{DefJ},
\begin{equation}\label{dJdx}
\frac{\dd \Anti}{\dd x} = -\ii \int \left[\Hamil,
\vrho + \bvrho\right] \mathcal{D}y+ \int \left(\mathcal{K} -
  \overline{\mathcal{K}}\right) \mathcal{D}y\,.
\end{equation}
In principle there is no need to solve this equation for $\Anti$ because it is a simple consequence of the definition~\eqref{DefJ} with equations~\eqref{eq:QKE_compact_asym}. However, some approximate resolution schemes promote $\Anti$ to an independent variable, thus requiring this additional equation to ensure the overall consistency.

\subsection{MSW transitions}

\begin{figure}[!ht]
  \centering
  \includegraphics[]{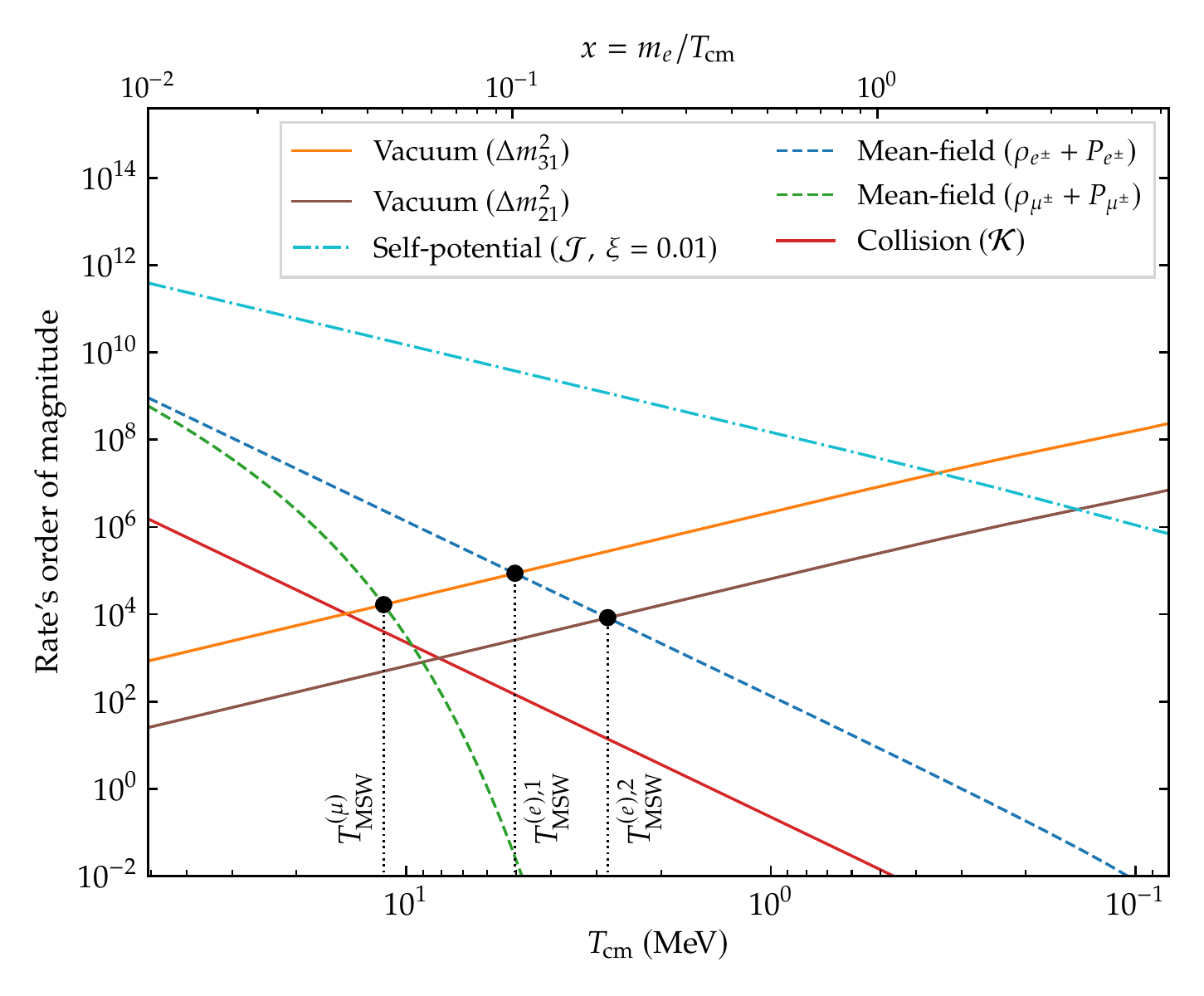}
	\caption[Orders of magnitude of the different rates involved in the QKE]{\label{fig:ODG_QKE} Orders of magnitude of the different rates involved in the QKE, for $y=y_\mathrm{eff}=3.15$ (this averaged value will be justified in section~\ref{FreqSyncOsc}). $\Hself$ is plotted with $\Anti$ given by \eqref{eq:init_Anti} and $\xi = 0.01$. The oscillation frequencies, set by the Hamiltonian eigenvalues, are very large compared to the collision rate and its variation (see section~\ref{sec:ATAO} for this discussion). As the temperature decreases, the dominant contribution in the Hamiltonian changes from $\Hself$ to $\Hamil$, $\Hamil$ itself being dominated first by $\Hlep$ and then by $\Hvac$. We estimate the magnitude of the collision rate as in Figure 1 of~\cite{Mirizzi2012}.}
\end{figure}

Schematically, the lepton mean-field term scales as $\Tcm^5$, whereas the vacuum oscillation Hamiltonian scales as $1/\Tcm$ (discarding the common $1/xH$ scaling). Hence there is always a Mikheev-Smirnov-Wolfenstein (MSW) transition~\cite{MSW_W,MSW_MS} from lepton mean-field domination to vacuum domination, which can be resonant or not depending on the mixing angles and the mass ordering. There are two differences with the MSW transition in stars. First, the lepton mean-field term in stellar environments is $\sqrt{2}G_F n_{e^-}$, but it is cancelled here by the positron contribution $-\sqrt{2}G_F n_{e^+}$ since the electron/positron asymmetry is negligible. Hence, in the cosmological case the dominant lepton mean-field contribution is given by~\eqref{eq:Hamil_general}. Second, the role of the electron density profile crossed by emitted neutrinos in a star is now played by the thermal evolution of the Universe. In the cosmological context, there are three transitions which are illustrated in Figure~\ref{fig:ODG_QKE}.

\begin{enumerate}
\item Since $m_\mu/\me \simeq 207$, the first MSW transition, that we call the muon-driven MSW transition, occurs when the $\mu^\pm$ mean-field effects become of the same order as the vacuum Hamiltonian associated with the large mass gap $\Delta m_{31}^2$ (or equivalently $\Delta m_{32}^2$), and this occurs around $T_\mathrm{MSW}^{(\mu)} \simeq 12\,\mathrm{MeV}$ (see section~\ref{SecDescriptionMuonTransition}), when muons are not relativistic.
\item When the $e^\pm$ mean-field effects also become of the same order as the vacuum Hamiltonian associated with the large mass gap $\Delta m_{31}^2$, we encounter the first electron-driven MSW transition around $T_\mathrm{MSW}^{(e), 1} \simeq 5\,\mathrm{MeV}$ (see section~\ref{SecElectronMSW}).
\item Finally, when the same mean-field term becomes of the same order as the vacuum Hamiltonian associated with the small mass gap $\Delta m_{21}^2$, we reach the second electron-driven MSW transition around $T_\mathrm{MSW}^{(e), 2} \simeq 2.8\,\mathrm{MeV}$ (see section~\ref{SecElectronMSW}).
\end{enumerate}
The presence of a neutrino asymmetry modifies this picture because self-interaction mean-field effects (abbreviated as \emph{self-interactions} when it is clear that we do not refer to collisions between (anti)neutrinos) scale as $\Tcm^3$ and the traceless part of the neutrino asymmetry is proportional to $|\xi_\alpha-\xi_\beta|$. Unless the degeneracy differences are very small, there is always a period when self-interactions dominate over the lepton mean-field contribution until they become smaller than the vacuum contribution (see Figure~\ref{fig:ODG_QKE}). At the beginning of this period of self-interaction mean-field domination we can encounter a Matter Neutrino Resonance (MNR)~\cite{Malkus:2014iqa,Johns:2016enc}, when lepton mean-field effects become smaller than self-interaction effects. However in that early phase all matrix densities and all mean-field contributions (save the negligible vacuum one), are diagonal in flavour space, therefore no conversion can occur. Conversely, describing the end of the self-interaction domination, when the vacuum Hamiltonian takes over the self-interaction effects, is rather complicated owing to the physics of synchronous oscillations which takes place, and which depends on the lepton-driven MSW transitions. One of the goals of this chapter is precisely to revisit the physics of these oscillations and their consequences for the equilibration of asymmetries.

Finally, note that cases where degeneracies are so small that self-interactions are at most of the order of the vacuum or lepton mean-field contributions around the MSW transition, lead to rather different physical effects since this condition is largely dependent on the magnitude of neutrino momenta. These low degeneracy regimes have been investigated in~\cite{Johns:2016enc}, but we will not explore such small values, motivated by the fact that BBN constraints are of the order of $10^{-2}$ on $\xi_e$, see equation~\eqref{eq:xi_BBN_CMB}.

\section{Resolution schemes}\label{SecResScheme}

The QKEs~\eqref{eq:QKE_compact_asym} are challenging to solve for various reasons, the main one being the coexistence of multiple time scales: the different terms in the Hamiltonian correspond to different oscillation frequencies, that need to be compared to the collision rate---the latter being in addition particularly computationally expensive. The orders of magnitude of the different terms involved in the QKE~\eqref{eq:QKE_compact_asym} are shown on Figure~\ref{fig:ODG_QKE}.

\paragraph{ATAO approximation}

The separation of these time scales allows for the use of effective resolution schemes. 
In general, for a given Hamiltonian $\mathcal{H}$ governing the evolution of a density matrix $\vrho$, i.e., if $\partial_x \vrho = - \ii [\mathcal{H},\vrho]$, the eigenvalues of $\mathcal{H}$ give the oscillation frequencies of $\vrho$. More precisely, noting $U_\mathcal{H}$ the unitary matrix which diagonalizes $\mathcal{H}$ (that is $\mathcal{H} = U_\Hcal D_\Hcal U_\Hcal^\dagger$ with $D_\Hcal$ diagonal), the density matrix in the “$\Hcal$-basis” is $U_\Hcal^\dagger \vrho U_\Hcal$. The off-diagonal components of this matrix have oscillatory phases equal to the differences of the diagonal components of $D_\Hcal$.

If $U_\Hcal$ evolves slowly enough\footnote{As explained in section~\ref{sec:ATAO}, $U_\Hcal$ must evolve slowly compared to the inverse oscillation frequency, that is schematically $\abs{(U_\Hcal^\dagger \partial_x U_\Hcal)^i_j} \ll \abs{(D_\Hcal)^i_i - (D_\Hcal)^j_j}$, which corresponds roughly to comparing the Hubble rate to the oscillation frequencies.}, the oscillation frequencies are so large that the off-diagonal components of $U_\Hcal^\dagger \vrho U_\Hcal$ are averaged out. Therefore, transforming back to the flavour basis, we define the averaged matrix $\LATAO \vrho \RATAO_\Hcal $ by
\begin{equation}
\label{DefAverage}
\LATAO\vrho \RATAO_\Hcal \equiv U_\Hcal \reallywidetilde{\left( U^\dagger_\Hcal \vrho U_\Hcal \right)} U_\Hcal^\dagger  \, .
\end{equation}
The wide overtilde notation means that we keep only the diagonal part---thus neglecting the fast off-diagonal oscillatory evolution which averages to zero. This procedure requires that the diagonalizing basis changes slowly relative to the oscillations, which is a standard case of adiabatic approximation. Since oscillations are averaged throughout the adiabatic evolution of the Hamiltonian, the \emph{adiabatic transfer of averaged oscillations} (ATAO) consists in the approximation
\begin{equation}
\label{eq:def_ATAO}
 \vrho \simeq \LATAO \vrho \RATAO_\Hcal  \quad \text{i.e.} \quad \vrho = U_\Hcal \widetilde{\vrho}_\Hcal U_\Hcal^\dagger \, ,
 \end{equation}
with $\widetilde{\vrho}_\Hcal$ diagonal. When including collisions, we account for their effects on time scales much larger than the one set by $\mathcal{H}$, which leads to the evolution equations
\begin{equation}\label{ATAOrhodot}
 \partial_x \widetilde{\vrho}_\Hcal = U^\dagger_\mathcal{H} \LATAO\mathcal{K}\RATAO_\mathcal{H}  U_\mathcal{H} = \widetilde{\mathcal{K}}_\Hcal\,. 
\end{equation}
Note that the collision term depends on $\vrho$, which is evaluated with the approximation~\eqref{eq:def_ATAO}. 

Such a situation is encountered by neutrinos in the early universe: the results of Figure~\ref{fig:ODG_QKE} show that the Hamiltonian governing the evolution of $\vrho$ is progressively dominated, as the temperature decreases, by the self-potential (and the lepton mean-field), then by the vacuum contribution, and we now detail the associated approximation schemes.

\subsection{\ATAOH}

If the self-potential can be ignored (for instance if we consider a case without neutrino asymmetries), the fast scale is set by the Hamiltonian $\Hamil$ and we will call this situation the \ATAOH approximation, which was used in chapter~\ref{chap:Decoupling}. As previously explained, we thus approximate\footnote{Concerning $\bvrho$, it is equivalent to average it around $\pm \Hamil$, hence our choice to use $\Hamil$ for both $\vrho$ and $\bvrho$.} $\vrho \simeq \LATAO \vrho \RATAO_{\Hamil}$ and $\bvrho \simeq \LATAO \bvrho \RATAO_{\Hamil}$, such that
\begin{equation}\label{rtildetorhoATAOH}
\vrho = U_{\Hamil} \, \widetilde{\vrho}_{\Hamil} \,U^\dagger_{ \Hamil}\,,\qquad \bvrho = U_{\Hamil}\, \widetilde{\bvrho}_{\Hamil} \,U^\dagger_{\Hamil}\,,
\end{equation}
with $\widetilde{\vrho}_{\Hamil}$ and $\widetilde{\bvrho}_{\Hamil}$ being diagonal. Therefore, it is convenient to solve for the $N_\nu$ diagonal components of these variables instead of the $N_\nu^2$ variables of the density matrices in flavour basis (which, in this approximation, are not independent). The evolution equation~\eqref{ATAOrhodot} leads to
\begin{equation}\label{BasicATAOH}
\partial_x \widetilde{\vrho}_\Hamil = \widetilde{\mathcal{K}}_\Hamil[\vrho,\bvrho]\,,\qquad \partial_x \widetilde{\bvrho}_\Hamil = \widetilde{\overline{\mathcal{K}}}_\Hamil[\vrho,\bvrho]\,.
\end{equation}
Since the collision term depends on $\vrho,\bvrho$, this means that the evolved variables $\widetilde{\vrho}_\Hamil$ and $\widetilde{\bvrho}_\Hamil$ are transformed to the flavour basis with \eqref{rtildetorhoATAOH}, so as to evaluate the collision term whose values  in flavour space are eventually transformed back into the matter basis. We then keep only their diagonal components through 
\begin{equation}
\widetilde{\mathcal{K}}_\Hamil \equiv \reallywidetilde{\left( U^\dagger_\Hamil \mathcal{K} U_\Hamil \right)}\,,\qquad \widetilde{\overline{\mathcal{K}}}_\Hamil \equiv \reallywidetilde{\left( U^\dagger_\Hamil \overline{\mathcal{K}} U_\Hamil \right)}\,.
\end{equation}

Actually, $\Hamil$ depends on both $x$ and $y$, and so does $U_\Hamil$. Hence this averaging scheme is momentum-dependent, which is a central feature to understand the evolution of density matrices. When lepton mean-field effects can be ignored, then the $y$ dependence is the same for all momenta (a $1/y$ prefactor in $\Hamil \simeq \Hvac$) and the unitary matrices $U_{\Hamil}$ do not depend on $y$ anymore since they all reduce to the PMNS matrix. 

\subsection{\ATAOJH}

When neutrino asymmetries cannot be ignored, we see on Figure~\ref{fig:ODG_QKE} that there is a range of temperatures for which $\Hself$ must necessarily be included in the Hamiltonian. As can be seen in the QKEs~\eqref{eq:QKE_compact_asym}, the Hamiltonian for $\vrho$ is then $\Hself + \Hamil$ while it is $\Hself - \Hamil$ for $\bvrho$. Therefore, the \ATAOJH approximation reads $\vrho \simeq \LATAO\vrho \RATAO_{\Hself+\Hamil}$ and $\bvrho \simeq \LATAO\bvrho \RATAO_{\Hself-\Hamil}$, such that 
\begin{equation}
\vrho = U_{\Hself + \Hamil} \, \widetilde{\vrho}_{\Hself + \Hamil} \,U^\dagger_{\Hself + \Hamil}\,,\qquad \bvrho = U_{\Hself - \Hamil}\, \widetilde{\bvrho}_{\Hself - \Hamil} \,U^\dagger_{\Hself - \Hamil}\, ,
\end{equation}
where $\widetilde{\vrho}_{\Hself + \Hamil}$ and $\widetilde{\bvrho}_{\Hself - \Hamil}$ are diagonal. We solve the evolution of $\vrho, \bvrho$ on timescales much larger than the one set by $\Hself \pm \Hamil$, on which oscillations are averaged, hence the evolution equation is given by~\eqref{ATAOrhodot}
\begin{equation}\label{BasicATAOJH}
\partial_x \widetilde{\vrho}_{\Hself +\Hamil} = \widetilde{\mathcal{K}}_{\Hself +\Hamil}[\vrho,\bvrho]\,,\qquad \partial_x \widetilde{\bvrho}_{\Hself - \Hamil} = \widetilde{\overline{\mathcal{K}}}_{\Hself - \Hamil}[\vrho,\bvrho]\,.
\end{equation}
The method is similar to the \ATAOH case, but we need to handle the fact that the Hamiltonian itself depends on $\vrho$, through the self-potential $\Hself$. In order to compute it at each time step, we would need to keep track of the $N_\nu^2$ entries of each density matrix in the flavour basis. A better possibility, which we choose, consists in promoting $\Hself$ (actually, $\Anti$) to be an independent variable with its own evolution equation~\eqref{dJdx}. Equation~\eqref{DefJ} is then only used to set the initial value of $\Hself$ from the initial conditions on $\vrho$, $\bvrho$. In doing so, we go from $2 \times N\times N_\nu^2$ to $2 \times N \times N_\nu + N_\nu^2$ variables, with $N$ the number of momentum nodes (cf.~section~\ref{subsec:numerics}). We stress that the evolution of $\Anti$ depends on the full collision terms in flavour space, and not just on the diagonal components in the matter basis $\widetilde{\mathcal{K}}_{\Hself \pm\Hamil}$, as is the case for $\widetilde{\vrho}_{\Hself +\Hamil}$ and $\widetilde{\bvrho}_{\Hself -\Hamil}$.

\paragraph{High temperatures: \ATAOJ} It is clear from Figure~\ref{fig:ODG_QKE} that at large temperatures, $\Hself$ largely dominates $\Hamil \simeq \Hlep$ (except for very small $\xi$ that lie outside the range of values we span here). That is why one could consider an even simpler \ATAOJ approximation, where $\Hself \pm \Hamil$ is replaced by $\Hself$. In that case, the changes of basis for $\vrho$ and $\bvrho$ are achieved with the same matrix $U_\Hself$. In section~\ref{FreqSyncOsc}, we show that this “leading order” Hamiltonian leads to theoretical estimates of synchronous oscillations frequencies in agreement with the existing literature, while using the full \ATAOJH allows to get an important correction which is responsible for quasi-synchronous oscillations.
The weight of $\Hamil$ in the \ATAOJH scheme becomes more important when the temperature decreases (i.e., $x$ increases), since $\Hself \propto x^{-2}$ and $\Hvac \propto x^2$. 

\subsection{QKE}

The QKE method is not an approximation scheme, but consists instead in solving exactly the neutrino and antineutrino evolutions, that is equations~\eqref{eq:QKE_compact_asym}. However
these equations are very stiff at early times given that all terms except the vacuum one increase for
large temperatures. Therefore, integration times are typically much longer, in addition to the fact that we need to keep track of the $N_\nu^2$ entries of each density matrix in the flavour basis, contrary to the $N_\nu$ diagonal ones in the matter basis when using an ATAO framework.

\subsection{Numerical methods}
\label{subsec:numerics}

The general method used to solve for the time evolution of density matrices is described in section~\ref{sec:numeric_Dec}. The neutrino spectra are sampled on a grid and we have several possible choices for the spacing of the reduced momenta $y$ in this grid. We found that in the context of asymmetry equilibration, a linear spacing is much more adequate than the Gauss-Laguerre quadrature. All numerical results presented in this chapter are performed with an extension of the code \texttt{NEVO}, using a linear grid with $N=40$ points, the minimum and maximum momenta being chosen as described in section~\ref{sec:numeric_Dec}. We start the numerical resolution at $\Tcm = 20\,\mathrm{MeV}$, the final temperature depending on the particular configuration investigated. For initial conditions, we set $z_\mathrm{init}$ using that photons, $e^\pm$ and neutrinos are fully thermalized with a common temperature, see equation~\eqref{D1Froustey2020}. In the case of vanishing degeneracies, this determines $z_\mathrm{init}-1 \simeq 7.42 \times 10^{-6}$.

In the general QKE method, the only difference in the code is the contribution of commutators of the type $[\Anti, \vrho]$ and  $[\Anti, \bvrho]$ in \eqref{eq:QKE_compact_asym}. However, when using the \ATAOJH method, one needs to add $N_\nu^2$ variables corresponding to the degrees of freedom of $\Anti$ whose evolution is determined by \eqref{dJdx}. 

When equations are stiff, we must rely on implicit methods that
require the computation of the Jacobian of the system of differential
equations. The default method consists in using a finite difference
estimation. The complexity of the calculation of the collision term is $\mathcal{O}(N^3)$ since for each momentum one must compute on a two-dimensional integral \cite{Dolgov_NuPhB1997}. Hence with finite differences the complexity for the Jacobian is $\mathcal{O}(N^4)$. However we can provide its explicit form to the solver and it reduces its evaluation to $\mathcal{O}(N^3)$. This method was used in chapter~\ref{chap:Decoupling} in both the QKE and the \ATAOH schemes. 

This powerful numerical technique can be extended to the \ATAOJH scheme, and the essential steps are described in appendix~\ref{App:Numerics}.
Since we only add $N_\nu^2$ variables, the complexity remains $\mathcal{O}(N^3)$. All in all, we found that the code was at least ten times faster with the \ATAOJH scheme, and even more at low temperatures where the fast oscillations (see  next section) slow even more the QKE algorithm.

\section{Synchronous oscillations with two neutrinos}\label{Sec2Neutrinos}

The presence (and domination) of the self-interaction mean-field in the QKEs radically changes the phenomenology of neutrino evolution. This non-linear term notably leads to oscillations of all momentum-modes at a common frequency, a phenomenon named \emph{synchronous oscillations}, studied both numerically \cite{Pastor:2001iu,Dolgov_NuPhB2002,Mangano:2010ei} and analytically \cite{Abazajian2002,Wong2002}. In this section, we extend this theoretical work in the framework of the ATAO approximations we developed: this allows to explicitly calculate the next-to-leading order contribution to the oscillation frequency that was not considered in previous works, and that we check numerically in the next section.

We restrict to a two-flavour case, which allows to easily perform the following calculations thanks to the vector representation of $2 \times 2$ Hermitian matrices. We do not specify yet the values of the mixing parameters, as they will be set for different physical setups in section~\ref{SecRelevant2Neutrinos}.

Let us thus consider in this section the vacuum Hamiltonian of the form
\begin{equation}
\label{eq:Hvac_2nu}
    \Hvac = \frac{1}{xH} \left( \frac{x}{\me} \right) U \begin{pmatrix} 0 & 0 \\ 0 & \Delta m^2/2y
    \end{pmatrix} U^\dagger \quad \text{with} \quad U = \begin{pmatrix} \cos{\theta} & \sin{\theta} \\ 
    - \sin{\theta} & \cos{\theta} \end{pmatrix} \, ,
\end{equation}
along with the lepton mean-field contribution of the type
\begin{equation}
\label{eq:Hlep_2nu}
    \Hlep = - \frac{1}{xH} \left(\frac{\me}{x}\right)^5 \frac{2 \sqrt{2} G_F y}{\mW^2}
    \begin{pmatrix}
    \bar{\rho}_{l^\pm} + \bar{P}_{l^\pm} & 0 \\
    0 & 0
    \end{pmatrix} \, .
\end{equation}
In order to maintain a similar expansion history as in the case of three neutrinos, we add one fully decoupled thermalised neutrino flavour to the energy content of the Universe when studying the case of only two neutrino oscillations.

\subsection{Transformation to vectors}

It is customary to rephrase the density matrix evolution as an evolution for vectors using the relation between a Hermitian $2\times2$ matrix $P$, and a vector of $\mathbb{R}^3$ $\vec{P}$
\begin{equation}\label{MatrixToVector}
P = \frac{1}{2}P^0 \Id  +\frac{1}{2} \vec{P} \cdot \vec{\sigma}  \, ,
\end{equation}
where $\vec{\sigma} = \left(\sigma_\x, \sigma_\y, \sigma_z \right)$ is the “vector” of Pauli matrices. Commutators of matrices are then handled using $[\sigma_i,\sigma_j] = 2 \ii
\epsilon_{ijk} \sigma_k$ as we obtain
\begin{equation}
-\ii [P, Q] =  \frac{1}{2} \left( \vec{P} \wedge \vec{Q} \right) \cdot \vec{\sigma} \,.
\end{equation}
The evolution of the neutrino and antineutrino density matrices in vector notations\footnote{For consistency, we write the “vector part” of the two-neutrino density matrix $\vec{\vrho}$, while it is common in the literature to call this the \emph{polarization vector} $\vec{P}$ \cite{SiglRaffelt,Dolgov_NuPhB2002,Johns:2016enc}.} are immediately obtained to be
\begin{equation}
\label{eq:QKE_2nu}
\partial_x \vec{\vrho} =  \left(\vec{\Hamil}  + \vec{\Hself}\right) \wedge \vec{\vrho} +
\vec{\mathcal{K}}\,,\qquad 
\partial_x \vec{\bvrho} = \left( -\vec{\Hamil}  +
\vec{\Hself}\right) \wedge \vec{\bvrho} + \vec{\overline{\mathcal{K}}}\,,
\end{equation}
which we must supplement by
\begin{equation}
\partial_x \vrho^0 = \mathcal{K}^0\,,\qquad \partial_x \bvrho^0 = \overline{\mathcal{K}}^0\,,
\end{equation}
to account for the evolution of the trace part of density matrices. 

In the QKE~\eqref{eq:QKE_2nu}, the vector form of the Hamiltonian $\vec{\Hamil}= \vec{\mathcal{H}}_0 + \vec{\mathcal{H}}_\mathrm{lep}$ is the sum of the vacuum contribution obtained from \eqref{eq:Hvac_2nu}
\begin{equation}
\label{eq:Hvec}
\vec{\mathcal{H}}_0 = \frac{1}{xH} \left(\frac{x}{\me}\right) \frac{\Delta m^2}{2 y} \begin{pmatrix} \sin(2 \theta) \\ 0 \\ - \cos(2 \theta) \end{pmatrix} \, ,
\end{equation}
and the lepton mean-field one, derived from \eqref{eq:Hlep_2nu},
\begin{equation}
\label{eq:Hlepvec}
\vec{\mathcal{H}}_\mathrm{lep} = - \frac{1}{xH} \left(\frac{\me}{x}\right)^5 \frac{2 \sqrt{2} G_F y}{\mW^2} \begin{pmatrix} 0 \\ 0 \\ \bar{\rho}_{l^\pm} + \bar{P}_{l^\pm} \end{pmatrix} \, .
\end{equation}
Finally, the asymmetry vector evolves as
\begin{equation}\label{ddxvecJ}
\frac{\dd \vec{\Anti}}{\dd x} = \int{\left(\vec{\Hamil} \wedge [\vec{\vrho } + \vec{\bvrho}] \right) \mathcal{D}y} + \int{\left(\vec{\mathcal{K}} - \vec{\overline{\mathcal{K}}}\right) \mathcal{D} y}\,.
\end{equation}

This vector formalism allows for a more visual representation of the ATAO schemes. Averaging $\vrho$ with respect to an Hamiltonian $\Hcal$ corresponds to \emph{projecting $\vec{\vrho}$ onto $\vec{\Hcal}$}. To see this, we first note that the restriction to the diagonal part of an Hermitian two-by-two matrix corresponds to a projection along $\vec{e}_\z$ in vector notation. Hence when applying the averaging definition~\eqref{DefAverage}, the first step is the rotation which aligns $\vec{\Hcal}$ with $\vec{e}_\z$, then the diagonal part restriction selects only the $\z$-component of this rotation $\vec{\vrho}_\Hcal$, and finally it is rotated back into the initial frame. As a result one has, in the case of two neutrinos,
\begin{equation}
\overrightarrow{\LATAO \vrho \RATAO}_{\Hcal} = (\vec{\vrho} \cdot
\hat{\Hcal}) \hat{\Hcal} \, ,
\end{equation}
where $\hat{\mathcal{H}}$ is the unit vector in the direction of $\vec{\mathcal{H}}$. Since the equations of motion~\eqref{eq:QKE_2nu} correspond to instantaneous precessions set by $\vec{\mathcal{H}}$ (up to the collision term), the averaging procedure corresponds to projecting along that precession vector, i.e. removing the fast rotating part that is orthogonal to it.

\subsection{Frequency of synchronous oscillations}
\label{FreqSyncOsc}

\begin{figure}[!ht]
  \centering
  \includegraphics[]{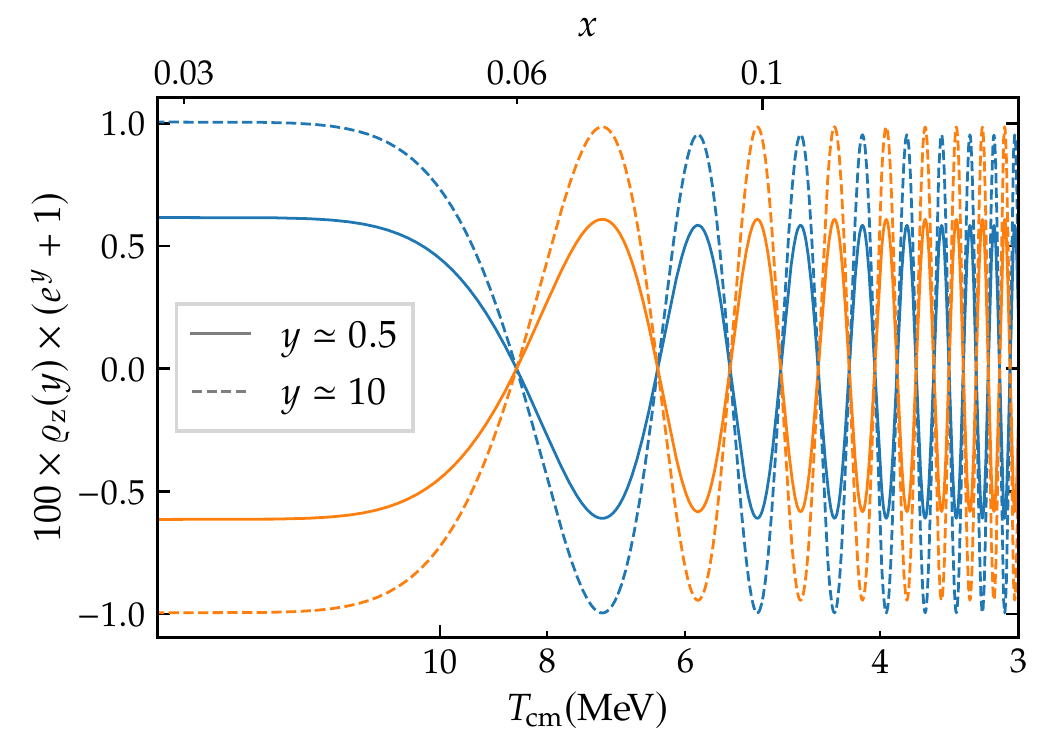}
	\caption[Synchronous oscillations in a two-neutrino $\nu_\mu - \nu_\tau$ case]{\label{fig:SyncOsc} Synchronous oscillations in a two-neutrino $\nu_\mu - \nu_\tau$ case with $\Delta m^2 = 2.45\times10^{-3} \, \mathrm{eV}^2$, $\theta=0.831$, without collisions. In blue $\vrho_\z = \vrho^{\mu}_{\mu} - \vrho^{\tau}_{\tau}$ and in orange $\bvrho_\z$. The initial degeneracy parameters are $\xi_\mu = 0.01$ and $\xi_\tau = 0$.}
\end{figure}

In some setups where the non-linear self-potential term in the QKEs dominates, such as dense neutrino gases or the early universe (for not too small asymmetries), it has been shown in references~\cite{Samuel:1993uw,Kostelecky:1993yt,Kostelecky:1993dm,Kostelecky:1993ys,Pastor:2001iu,Dolgov_NuPhB2002} that neutrinos develop so-called momentum-independent \emph{synchronous oscillations}, with all $y-$modes being “locked” on the asymmetry vector $\vec{\Anti}$. This is shown on Figure~\ref{fig:SyncOsc}, where the physical parameters are the same as in the upcoming section~\ref{SecMuonMSW}.

To understand this phenomenon and make quantitative predictions regarding the behaviour of the system of neutrinos and antineutrinos in different setups, we will first ignore the effect of collisions. The initial density matrices are given by equations~\eqref{rhoinit}.
Hence the initial vector components are $\vrho_\z(y) = g(\xi_1,y) - g(\xi_2,y)$ and $\bvrho_\z(y) = g(-\xi_1,y) - g(-\xi_2,y)$, and $\vrho_{\x, \y}(y) = \bvrho_{\x, \y}(y) = 0$. Since we neglect collisions in this section, it is clear from~\eqref{eq:QKE_2nu} that the norms of $\vec{\vrho}$ and $\vec{\bvrho}$ are conserved. The adiabatic evolution of these vectors thus consists in a rotation so as to follow the direction of their Hamiltonian. Therefore, in the \ATAOJH approximation, we can write the density matrix vectors
\begin{equation}
\label{eq:vrho_ATAOJV}
\begin{aligned}
\vec{{\vrho}} &= \abs{g(\xi_1,y)- g(\xi_2,y)} \widehat{\Hself+\Hamil} \, , \\
\vec{{\bvrho}}  &= - \abs{g(-\xi_1,y)- g(-\xi_2,y)} \widehat{\Hself-\Hamil} \, ,
\end{aligned}
\end{equation}
where $\widehat{\Hself+\Hamil}$ is the unit vector in the direction of $\vec{\Hself}+\vec{\Hamil}$. One must remember that at the initial temperatures we consider ($\Tcm \sim 20 \, \mathrm{MeV}$), the Hamiltonian is largely dominated by $\Hself$. The \ATAOJ approximation then corresponds to discarding $\Hamil$ in the above expressions, and this will give the leading order behaviour of the asymmetry.

\subsubsection{Evolution of the asymmetry vector}

\paragraph{Leading order} Let us then focus on this high temperature region first, when the misalignment between $\vec{\vrho}$ and $\vec{\bvrho}$ is negligible, i.e.,
\begin{equation}
    \vec{\vrho} =  \abs{g(\xi_1,y)- g(\xi_2,y)} \widehat{\Hself} \, , \qquad
\vec{{\bvrho}}  = - \abs{g(-\xi_1,y)- g(-\xi_2,y)} \widehat{\Hself} \, .
\end{equation}
Hence, the asymmetry vector is obtained from~\eqref{DefJ} and~\eqref{IntegralsFD} and reads
\begin{equation}
\label{JtohatJ}
\vec{\Anti} = \frac{1}{6}\left\lvert \xi_1-\xi_2 \right\rvert \left(1+\frac{\xi_1^2+\xi_2^2+\xi_1\xi_2}{\pi^2}\right)\widehat{\Anti} \, ,
\end{equation}
with the unit vector definition $\widehat{\Anti} = \widehat{\Hself}$, which is equal initially to $\mathrm{sgn}(\xi_1 - \xi_2) \vec{e}_\z$. We can use the expressions of $\vec{\vrho}, \, \vec{\bvrho}$ to explicitly compute the $y-$integral appearing in~\eqref{ddxvecJ}. It is then particularly convenient to use the quantities~\eqref{eq:Hamil_y} which isolate the momentum dependence of the Hamiltonian. Therefore, using the integrals given in \eqref{IntegralsFD}, we can rewrite~\eqref{ddxvecJ} as
\begin{equation}
\label{eq:derivA_leading}
\frac{\dd \vec{\Anti} }{\dd x}= F(\xi_1,\xi_2)
\vec{\underline{\mathcal{V}}}_\mathrm{eff}\wedge \vec{\Anti}\qquad \text{where} \qquad
\vec{\underline{\mathcal{V}}}_\mathrm{eff} \equiv \left(\vecHvacy + y^2_\mathrm{eff}
\vecHlepy \right) \, ,
\end{equation}
where we defined the slowness factor
\begin{equation}
\label{eq:slowness}
F(\xi_1,\xi_2) \equiv \frac{3}{2} \frac{\xi_1+\xi_2}{\pi^2 +\xi_1^2 +
  \xi_2^2+\xi_1 \xi_2}\,,
\end{equation}
in agreement with \cite{Abazajian2002,Wong2002}. The typical “average” momentum is
\begin{equation} 
y_\mathrm{eff} \equiv \pi\sqrt{1+\frac{\xi_1^2+\xi^2_2 }{2\pi^2}} \simeq \pi \, ,
\end{equation}
in agreement with equation~(33) in \cite{Wong2002} or equation~(2.19) in \cite{Abazajian2002} (derived in the particular case $\xi_1 = 0$).

This standard result allows to recover the key features of synchronous oscillations. The evolution of all $y-$modes is locked on the evolution of $\vec{\Anti}$, which precesses around the effective Hamiltonian computed for $y = y_\mathrm{eff}$. However, the oscillation frequency is greatly reduced compared to standard oscillations set by the Hamiltonian $\Hamil(y_\mathrm{eff})$, since for small degeneracies $F \propto (\xi_1 + \xi_2) \ll 1$. Let us provide a numerical evaluation. After muons and antimuons have annihilated (their remaining asymmetry is completely negligible here), and before electrons and positrons did so, that is in the range $200 \,\mathrm{MeV}\ge \Tcm \ge 0.5 \,\mathrm{MeV}$, the Hubble parameter~\eqref{eq:scaling_Hubble_x} reads
\begin{equation}\label{Htox2}
H \simeq \frac{\me}{M_\mathrm{Pl}} \times \frac{\me}{x^2} \sqrt{\frac{\pi^2}{45} \times \left[1 +
    (N_\nu+2)\frac{7}{8}\right]}\simeq \frac{\me}{x^2}\times 2.278 \cdot 10^{-22}\,,
\end{equation}
where in the last step we have taken $N_\nu=3$. When entering the correct numbers and approximating the slowness
factor by its lowest order in the $\xi_\alpha$, that is $F(\xi_1,\xi_2)\simeq 3(\xi_1+\xi_2)/(2\pi^2)$, we estimate the precession frequency to be
\begin{equation}\label{EvaluationOmega}
\Omega(x) \simeq 1.28\times 10^6 \,x^2\, \abs{\xi_1+\xi_2} \, \frac{\Delta m^2}{10^{-3}\,\mathrm{eV}^2} \, .
\end{equation}

Initially, all unit vectors are aligned $\widehat{\Anti} \parallel \hat{\Hamil} \simeq \hat{\mathcal{H}}_\mathrm{lep} \parallel \vec{e}_\z$. Then, as the temperature decreases, $\Hself$ dominates less compared to $\Hamil$ and the vectors $\vec{\vrho}$ and $\vec{\bvrho}$ become aligned with different directions (namely, $\widehat{\Hself + \Hamil}$ and $\widehat{\Hself - \Hamil}$), leading to~\eqref{eq:vrho_ATAOJV}.

\paragraph{Next-to-leading order} Let us therefore now account for the effect of $\Hamil$, in that $\vrho$ and $\bvrho$ do not get projected on the exact same directions. We will assume that $\lvert \vec{\Hamil} \rvert \ll \lvert \vec{\Hself} \rvert$, such that we can perform an expansion of the unit vector
\begin{equation}
\label{ExpandSH}
\widehat{\Hself+\Hamil} \simeq \widehat{\Hself} + \frac{\vec{\Hamil} }{ |\vec{\Hself}| } - \left(\frac{\vec{\Hamil}\cdot \vec{\Hself}}{|\vec{\Hself}|^2}\right) \widehat{\Hself}+\cdots
\end{equation}  
For $\widehat{\Hself-\Hamil} $ the expression is identical up to $\vec{\Hamil} \to -\vec{\Hamil}$. This expansion gives, in the \ATAOJH approximation, the next-to-leading order (NLO) terms that were not explicited in previous works (cf., for instance, equation~(25) in~\cite{Wong2002}). Including this expansion in equation~\eqref{ddxvecJ}, we can once again recast the evolution of the asymmetry as a precession equation
\begin{equation}
\label{eq:precession}
    \frac{\dd \vec{\Anti}}{\dd x} = \vec{\Omega} \wedge \vec{\Anti} \, ,
\end{equation}
where the oscillation frequency (in $x$ variable) reads, retaining only the vacuum contribution $\Hamil = \Hvac$ for simplicity,\footnote{This is not an oversimplification. Indeed, as long as $\Hlep$ dominates over $\Hvac$, all vectors are aligned along $\vec{e}_\z$ and no precession takes place.}
\begin{equation}
\label{eq:Omega_JH}
\vec\Omega = F(\xi_1, \xi_2 ) \vecHvacy \left[1
  -\frac{12}{\sqrt{2} G_F}\left(\frac{x}{\me}\right)^3 \frac{x H \vecHvacy \cdot
    \widehat{\Anti}}{|\xi_1-\xi_2|(\xi_1 + \xi_2)} \right]\,.
\end{equation}
The second term between brackets is the NLO term, which accounts for the difference between purely synchronous and \emph{quasi}-synchronous oscillations, given that its origin is rooted in the orientation differences between $\vec{\vrho}$ and $\vec{\bvrho}$. Note that we also took the lowest order contribution $\lvert \vec{\Anti} \rvert \simeq \abs{\xi_1 - \xi_2}/6$, valid for small degeneracy parameters (we can use the leading order expression~\eqref{JtohatJ} since the evolution of $\vec{\Anti}$ is a precession, hence its norm is unchanged). The expression~\eqref{eq:Omega_JH} leaves a priori the possibility of divide-by-zero if $\xi_1 = \pm \, \xi_2$. We discuss these special cases at the end of this section.

To estimate the precession frequency, we consider as before that initially $\widehat{\Anti} = \mathrm{sgn}(\xi_1 - \xi_2) \vec{e}_z$ and assume the transition between $\Hlep$ and $\Hvac$ to be abrupt enough such that we can estimate, using \eqref{eq:Hvac_2nu}, $x H \vecHvacy \cdot \widehat{\Anti} = - \mathrm{sgn}(\xi_1 - \xi_2) (x/\me) (\Delta m^2 / 2) \cos(2\theta)$. Therefore we get
\begin{equation}
\label{eq:Omega_norm}
    \lvert \vec\Omega \rvert = \frac{\abs{F(\xi_1, \xi_2 ) \Delta m^2}}{2 \me H} \times \left\lvert 1
  + \left(\frac{x}{x_\mathrm{tr}}\right)^4 \frac{\mathrm{sgn}(\Delta m^2 \cos(2 \theta))}{\xi_1^2 - \xi_2^2} \right \rvert \,,
\end{equation}
where we defined
\begin{equation}
x_\mathrm{tr} \equiv \me \left(\frac{\sqrt{2} G_F}{6 |\Delta m^2
    \cos(2\theta)|}\right)^{1/4}\simeq 3.7 \left(\frac{10^{-3}\,\mathrm{eV}^2}{|\Delta m^2
    \cos(2\theta)|}\right)^{1/4}\,.
\end{equation}
Given the scaling $H \propto x^{-2}$ recalled in~\eqref{eq:scaling_Hubble_x}, the frequency of synchronous oscillations keeps increasing as the Universe expands, first as $\Omega \propto x^2$ (leading order) and then as $\Omega \propto x^6$ (NLO domination).

This second behaviour is a novel result of this thesis. Although one might expect that the effect of $\Hamil$ would be completely subdominant compared to $\Hself$, the particular form of the equation of motion changes this picture. Keeping the dominant term in the expansion~\eqref{ExpandSH} (i.e., $\widehat{\Hself}$) leads in the evolution equation~\eqref{ddxvecJ} to an integral symmetric under $\xi_1 \to - \xi_1$ and similarly for $\xi_2$. Therefore, the associated contribution is proportional to $\xi_1^2 - \xi_2^2$, which after dividing by the norm of $\vec{\Anti}$ accounts for the precession frequency $\propto \xi_1 + \xi_2$ obtained in \eqref{eq:slowness}. However, the first order correction in~\eqref{ExpandSH} is odd with respect to $\vec{\Hamil}$, hence an antisymmetric integral with respect to $\xi_{1,2} \to - \xi_{1,2}$. The corresponding result is $\propto \xi_1 - \xi_2$, which is enhanced compared to the leading order term. 

The transition from leading to next-to-leading order is then found to be around
\begin{equation}
\label{eq:xNLO}
x_\mathrm{NLO} \equiv x_\mathrm{tr}|\xi_1^2-\xi_2^2|^{1/4}\,.
\end{equation}
Note that depending on the sign of $\Delta m^2 \cos(2\theta)/(\xi_1^2-\xi_2^2)$ the frequency can go through zero, which means that $\vec{\Anti}$ can precess in one direction, slow down, and then precess in the opposite direction with a frequency increasing as $\propto x^6$.

\paragraph{Particular cases} While the previous calculation seemed fairly general, there are two specific cases that deserve to be discussed.
\begin{itemize}
    \item \emph{Equal asymmetries ---} if $\xi_1 = \xi_2$ the asymmetry vector $\vec{\Anti}$ is strictly zero and the previous formalism is inadequate (namely, \eqref{eq:derivA_leading} cannot be obtained anymore since initially $\vec{\vrho} = \vec{\bvrho} = \vec{0}$). 
    \item \emph{Equal but opposite asymmetries ---} if $\xi_1 = - \xi_2$, the leading order term in~\eqref{eq:Omega_JH} vanishes (since $F(\xi_1,\xi_2) \propto \xi_1 + \xi_2$), but not the next-to-leading order contribution, a special case investigated at the end of section~\ref{SecMuonMSW}.
\end{itemize}

\paragraph{Summary: evolution of $\vec{\Anti}$} Initially, at high temperatures (typically $\Tcm \sim 20 \, \mathrm{MeV}$), the lepton term dominates over the vacuum one and $\vec{\Anti} \propto \vecHlep \parallel \vec{e}_z$. All vectors are aligned, and this situation does not change until the MSW transition between $\Hlep$-domination to $\Hvac$-domination. If this transition is slow enough compared to the precession frequency, then $\vec{\Anti}$ keeps following $\vec{\underline{\mathcal{V}}}_\mathrm{eff}$  and ends up aligned with $\vecHvacy$. This corresponds to an \emph{adiabatic} evolution of the asymmetry vector itself. Conversely, if the transition is too abrupt (that is much shorter than the precession timescale), $\vec{\Anti}$ gets brutally misaligned with $\vecHvac$ and oscillations develop. Let us stress that the evolution of $\vrho$ and $\bvrho$ is in general adiabatic, but it is the evolution of the vector that they track, namely $\vec{\Hself}$, which can be non-adiabatic depending on the value of the slowness factor~\eqref{eq:slowness}.

If oscillations do develop, initially at the frequency~\eqref{EvaluationOmega}, the increasing influence of $\Hamil$ compared to $\Hself$ leads to a new behaviour: beyond $x_\mathrm{NLO}$ given by~\eqref{eq:xNLO}, the frequency increases faster and $\Omega \propto x^6$. These features are illustrated in section~\ref{SecRelevant2Neutrinos}. Note that the calculation of the NLO assumes $\lvert \vec{\Hamil} \rvert \ll \lvert \vec{\Anti} \rvert$, but at some point the vacuum term becomes dominant over the self-potential one (cf.~Figure~\ref{fig:ODG_QKE}) and the \ATAOJH regime breaks down. This is discussed in more detail in section~\ref{SecTransitionToATAOV}.

\subsubsection{Adiabaticity parameter}

\begin{figure}[!ht]
	\centering
	\includegraphics{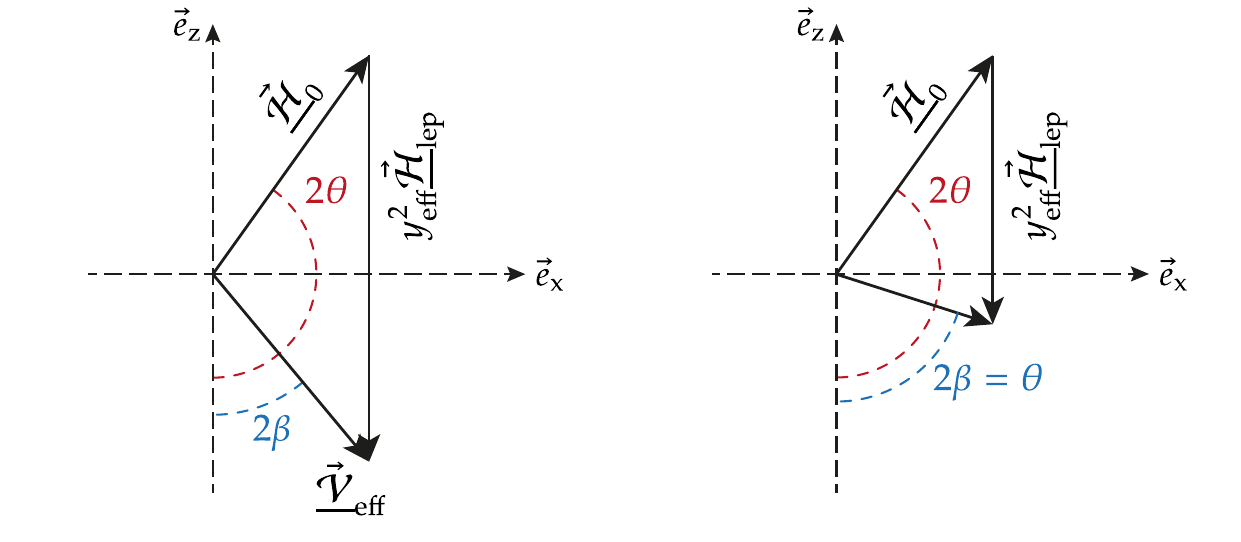}
	\caption[Parameterization of the adiabaticity of the evolution of $\vec{\Anti}$]{\label{FigAngles} Definition of the tipping angle $\beta$ on the left. The condition $\beta = \theta/2$ shown on the right corresponds to our definition of the MSW transition.}
\end{figure}

In the case with only two neutrinos, let us consider the evolution of the asymmetry vector $\vec{\Anti}$ without collisions and at leading order, which is dictated by equation~\eqref{eq:derivA_leading}. If the transition from a mean-field dominated to a vacuum dominated Hamiltonian, that is the MSW transition, is slow enough, then $\vec{\Anti}$ evolves adiabatically and follows $\vec{\underline{\mathcal{V}}}_\mathrm{eff} $. In order to assess the degree of (non-)adiabaticity, we thus need to quantify the speed at which the transition takes place. Let us first define a tipping angle $\beta$, illustrated in Figure~\ref{FigAngles}, by
\begin{equation}
\cos(2\beta) = -\hat{\underline{\mathcal{V}}}_\mathrm{eff} \cdot \vec{e}_\z \quad \text{since }\quad \hat{\underline{\mathcal{V}}}_\mathrm{eff}(x \ll x_\mathrm{MSW})= -\vec{e}_\z\,.
\end{equation}

Initially the tipping angle vanishes, and long after the transition it reaches $\theta$. We define the location of the MSW transition $x_\mathrm{MSW}$ as the moment when the tipping angle takes half of its final value, $\beta_\mathrm{MSW} = \theta/2$, which corresponds to $|\vecHvacy| = y^2_\mathrm{eff}|\vecHlepy|$ (as can be checked by trigonometric manipulations). Let us then define an adiabaticity parameter as 
\begin{equation}\label{Defgammatr1}
\gamma \equiv \frac{|\vec{\Omega}|}{\partial_x (2 \beta)}\, ,
\end{equation}
with $\vec{\Omega} = F(\xi_1,\xi_2)\vec{\underline{\mathcal{V}}}_\mathrm{eff}$, whose value is estimated from\footnote{The next-to-leading order contribution to $\vec{\Omega}$ is not relevant here as we focus on cases where $x_\mathrm{MSW} \ll x_\mathrm{NLO}$.} \eqref{eq:derivA_leading}. A large $\gamma$ corresponds to a rate of change of the Hamiltonian direction ($2 \partial_x \beta$) much smaller than the instantaneous precession frequency ($|\vec{\Omega}|$), that is to a very adiabatic evolution. We find
\begin{equation}
\gamma^{-1} = -\frac{(\underline{\mathcal{H}}_0^\perp)^2 y_\mathrm{eff}^2}{\abs{F(\xi_1,\xi_2)}|\vec{\underline{\mathcal{V}}}_\mathrm{eff}|^3}  \partial_x\left(\frac{|\vecHlepy|}{\underline{\mathcal{H}}_0^\perp}\right)\,,\quad \text{where}\quad \underline{\mathcal{H}}_0^\perp \equiv |\vecHvacy| \sin(2 \theta)\,.
\end{equation}
We have used $\partial_x (2\beta) = -\sin^2(2 \beta)y^2_\mathrm{eff}\partial_x\left(|\vecHlepy|/\underline{\mathcal{H}}_0^\perp \right)$ and $\sin(2 \beta) = \underline{\mathcal{H}}_0^\perp/|\vec{\underline{\mathcal{V}}}_\mathrm{eff}|$.
In order to assess the adiabaticity of the transition, we must estimate how large $\gamma$ is at the transition. Since $|\vecHlepy|/ \underline{\mathcal{H}}_0^\perp \propto 1/x^6$, and using that at the transition we have $|\vec{\underline{\mathcal{V}}}_\mathrm{eff}| = 2 |\vecHvacy| \cos \theta$ (see the geometry on the right plot of Figure~\ref{FigAngles}), the value of the adiabaticity parameter at the transition reduces to
\begin{equation}\label{Defgammatr2}
\gamma_\mathrm{MSW}  = \left.\abs{F(\xi_1,\xi_2)}\frac{2}{3} \frac{x |\vecHvacy| \cos^2 \theta}{\sin\theta}\right|_{x=x_\mathrm{MSW}}\,.
\end{equation}
Let us define the degree of non-adiabaticity through
\begin{equation}
P_\mathrm{n.a} \equiv \frac{1}{2}\left(1 \mp \widehat{\Anti} \cdot \widehat{\underline{\mathcal{V}}}_\mathrm{eff}\right)\,,
\end{equation}
with a $-$ sign (resp.~$+$ sign) if initially $\vec{\Anti}$ is aligned (resp.~anti-aligned) with $\vecHeff$ (i.e., $- \vec{e}_\z$). Its asymptotic value $P^\infty_\mathrm{n.a}$ when $x \gg x_\mathrm{MSW}$ estimates the misalignment of the final asymmetry vector due to lack of adiabaticity. Indeed if the transition is perfectly adiabatic, $\vec{\Anti}$ keeps tracking $\vec{\underline{\mathcal{V}}}_\mathrm{eff}$ and we always have $P_\mathrm{n.a}=0$. In general, the degree of non-adiabaticity needs not be much larger than unity to lead to a small $P^\infty_\mathrm{n.a}$, that is to a very non-adiabatic transition---see for instance the Landau-Zener approximation~\eqref{LZ}.

We note from equation~\eqref{Defgammatr2} that the adiabaticity parameter is of order $x \abs{F(\xi_1,\xi_2)}|\vecHvacy|$ evaluated at the transition, modulated by a geometric factor $(2/3) \cos^2 \theta/\sin \theta$. A transition is resonant when at some point the tipping angle goes through $2\beta = \pi/2$, that is through $\vec{\underline{\mathcal{V}}}_\mathrm{eff}$ having no component along $\vec{e}_\z$. Hence, a very non-resonant transition corresponds to $\theta \ll 1$, and in that case the geometric factor is enhanced by $1/\sin(\theta)$. It is less likely to encounter a small adiabaticity parameter because the tipping angle is small, and $|\vec{\Omega}|$ at the transition is larger than its final value (since $|\vec{\underline{\mathcal{V}}}_\mathrm{eff}|$ keeps decreasing). Conversely for a very resonant transition, $\pi/2-\theta \ll 1$, and the geometric factor is reduced by $\cos^2 \theta $ which is small, that is leads to a smaller adiabaticity parameter. This is partly because of the large tipping angle, but also because $|\vec{\Omega}|$ at the transition is much smaller than its final value (recall that at the transition $|\vec{\underline{\mathcal{V}}}_\mathrm{eff}| = 2 |\vecHvacy| \cos \theta$).

In the very resonant configuration ($\pi/2-\theta \ll 1$), the adiabaticity parameter for the dynamics of $\vec{\Anti}$ takes the simpler form
\begin{equation}\label{EqVeryResonant}
\gamma^{(\pi/2-\theta \ll 1)}_\mathrm{MSW} = \abs{F(\xi_1,\xi_2)}\underline{\mathcal{H}}_0^\perp \Delta x_\mathrm{MSW} \, , \ \ \text{with}\qquad \frac{1}{\Delta x_\mathrm{MSW}} \equiv y_\mathrm{eff}^2\left.\partial_x \left(\frac{|\vecHlepy|}{\underline{\mathcal{H}}_0^\perp}\right) \right|_{x=x_\mathrm{MSW}}\,.
\end{equation}

\paragraph{Landau-Zener formula} The Landau-Zener~\cite{Landau1932,Zener1932,Haxton:1986bc,Abazajian:2001nj,Johns:2016enc,Wittig} formula is an approximation of the degree of non-adiabaticity in this very resonant situation, using the approximation that the diagonal components of $\Hamil$ are linear in $x$ and that off-diagonal ones are constant, which reads
\begin{equation}\label{LZ}
P^\infty_\mathrm{n.a} \simeq \exp(-\pi \gamma_\mathrm{MSW}/2)\,.
\end{equation}

Note that the factors $F(\xi_1,\xi_2)$ and $y_\mathrm{eff}^2$ in \eqref{EqVeryResonant} are  specific to the fact that we consider the evolution of $\vec{\Anti}$. If we had considered the evolution of $\vec{\vrho}$ given by equation~\eqref{eq:QKE_2nu} without self-interactions nor collisions, that is $\partial_x \vec{\vrho} =  \vec{\Hamil}   \wedge \vec{\vrho} $, we would have obtained with a similar analysis (all quantities are written here for a given momentum $y$) the usual expression for the Landau-Zener adiabatic parameter
\begin{equation}\label{GammaLZ}
\gamma^{(\pi/2-\theta \ll 1)}_\mathrm{MSW} =\mathcal{H}_0^\perp \Delta x_\mathrm{MSW}\, , \quad \text{with} \qquad \frac{1}{\Delta x_\mathrm{MSW}} \equiv \left.\partial_x \left(\frac{|\vecHlep|}{\mathcal{H}_0^\perp}\right)\right|_{x=x_\mathrm{MSW}}
\end{equation}
where $\Hvac^\perp = \Hvac \sin(2 \theta)$. It is similar to equation~(9b) of \cite{Haxton:1986bc} and equation~(7.8) of \cite{Abazajian:2001nj}, 
the only difference being that $\Hvac^\perp$ is considered constant when dealing with solar neutrinos, whereas in the cosmological context it scales as $\propto x$, hence explaining why the expression \eqref{GammaLZ} for the transition width, $\Delta x_\mathrm{MSW}$, takes into account this evolution. However, note that this necessary modification for the expression of the adiabaticity parameter is absent from equation (28) of \cite{Johns:2016enc}, although this impacts only marginally the estimation of adiabaticity.

\section{Relevant two-neutrino cases for the primordial Universe}\label{SecRelevant2Neutrinos}

The previous results derived with only two neutrinos can shed some light on the physics at play in the standard case with three neutrinos and a general PMNS matrix. After the muon-driven MSW transition and before the electron-driven one, the oscillations only take place in the $\nu_\mu-\nu_\tau$ subspace since the unitary matrix $U_{\mathrm{eff}}$ that diagonalizes $\Hamil$ is approximately
\begin{equation}\label{UtoU}
U_{\mathrm{eff}} =  R_{23}(\theta_{23}^{\mathrm{eff}}) = \begin{pmatrix}
1 & 0 & 0\\
0 & \cos \theta_{23}^{\mathrm{eff}} &\sin \theta_{23}^{\mathrm{eff}} \\
0 & -\sin \theta_{23}^{\mathrm{eff}} & \cos \theta_{23}^{\mathrm{eff}}
\end{pmatrix} \,,
\end{equation}
this form being rigorously valid in the limit $m_\mu/m_e \to \infty$. Expanding in the ratio $\epsilon = \Delta m_{21}^2  / \Delta m_{32}^2$, we find 
\begin{equation}\label{th23th23}
\tan(2 \theta_{23}^{\mathrm{eff}}) = \tan(2 \theta_{23}) - \epsilon
\frac{ \sin (\theta_{13}) \sin(2 \theta_{12})}{\cos^2 (\theta_{13}) \cos^2(2 \theta_{23})} + \mathcal{O}(\epsilon^2)\,.
\end{equation}
Given the values~\eqref{ValuesStandard} we find $\theta_{23}^{\mathrm{eff}}  \simeq \theta_{23}$ with a difference of order $0.25\,\%$. Hence, we investigate the case $\Delta m^2 = \Delta m_{32}^2$ and $\theta=\theta_{23}$ in section~\ref{SecMuonMSW} to study the evolution of density matrices after the muon-driven transition.
 
Unfortunately, the system is not so easily reduced to a two-neutrino system when it comes to the description of the subsequent electron-driven MSW transitions. For simplicity, we choose to consider a fictitious configuration where $\theta_{13}=\theta_{23}=0$ such that oscillations take only place in the $\nu_e-\nu_\mu$ subspace, with the electrons/positrons being the relevant contribution to the lepton mean-field effects \eqref{eq:Hlep_2nu}. This configuration is detailed in section~\ref{SecElectronMSW}. For numerical applications we consider $\Delta m^2 = \Delta m_{21}^2$ and $\theta=\theta_{12}$. Although it is an ideal setup, it will provide important insight for the full three-neutrino case in section~\ref{Sec:3neutrinos}.

\subsection{Muon-driven MSW transition}\label{SecMuonMSW}

Let us consider a muon-driven MSW transition with $\theta = \theta_{23} \simeq 0.831$ and $\Delta m^2 = \Delta m_{32}^2 \simeq 2.453 \times 10^{-3} \, \mathrm{eV}^2$ for numerics \cite{PDG}. We restrict to normal ordering for simplicity, and do not include collisions yet. This means that electrons and positrons are absent from this description, except for their contribution to the energy density and thus the Hubble parameter~\eqref{eq:scaling_Hubble_x}.

\subsubsection{Description of the transition}\label{SecDescriptionMuonTransition}

As outlined before, synchronous oscillations of the neutrino ensemble can develop when the MSW transition occurs, provided this transition is abrupt enough for $\vec{\Anti}$ to get suddenly misaligned with $\vecHeff$ and precess around it. Let us first estimate the location of this transition. The energy density of muons/antimuons drops very rapidly once they become non-relativistic. In this limit, we get
\begin{equation}
\rho_{\mu^\pm}+ P_{\mu^\pm} = 4 \sigma^{5/2}
\left(\frac{x}{2\pi}\right)^{3/2} \mathrm{e}^{-\sigma x} \times \frac{m_e^4}{x^3}
\end{equation}
with $\sigma = m_\mu/m_e\simeq 206.77$. Approximating $y_{\mathrm{eff}} \simeq \pi$, we find that the vacuum term is equal in magnitude to the lepton term --- which is our definition of the transition, --- for $x_\mathrm{MSW}\simeq 0.043$, that is for $\Tcm \simeq 12\,\mathrm{MeV}$. Given the exponential drop $\exp(-\sigma x)$ of muons energy density, one can note that $x_\mathrm{MSW}$ is very mildly sensitive to the value of $\Delta m^2$.

The adiabaticity parameter given by \eqref{Defgammatr2} is then
\begin{equation}
\label{eq:gammaMSW_muon}
\gamma_\mathrm{MSW} \simeq 100 \times \abs{\xi_1 + \xi_2} \,.
\end{equation}
For $\xi_1+\xi_2$ of a few percent or smaller, we find $\gamma_\mathrm{MSW}<1$ and the transition is abrupt, that is the evolution of $\vec\Anti$ during the transition is very non-adiabatic. Hence we expect that as the direction of the effective Hamiltonian moves away from the vertical axis, $\vec{\Anti}$ will develop oscillations at the frequency $\Omega$. For much larger $\xi_1+\xi_2$ such that $\gamma_\mathrm{MSW} > 1$, and considering the Landau-Zener estimation for the degree of adiabaticity~\eqref{LZ}, $\vec{\Anti}$ should tend to follow adiabatically the transition to the vacuum Hamiltonian.

\begin{figure}[!ht]
  \centering
  \includegraphics[]{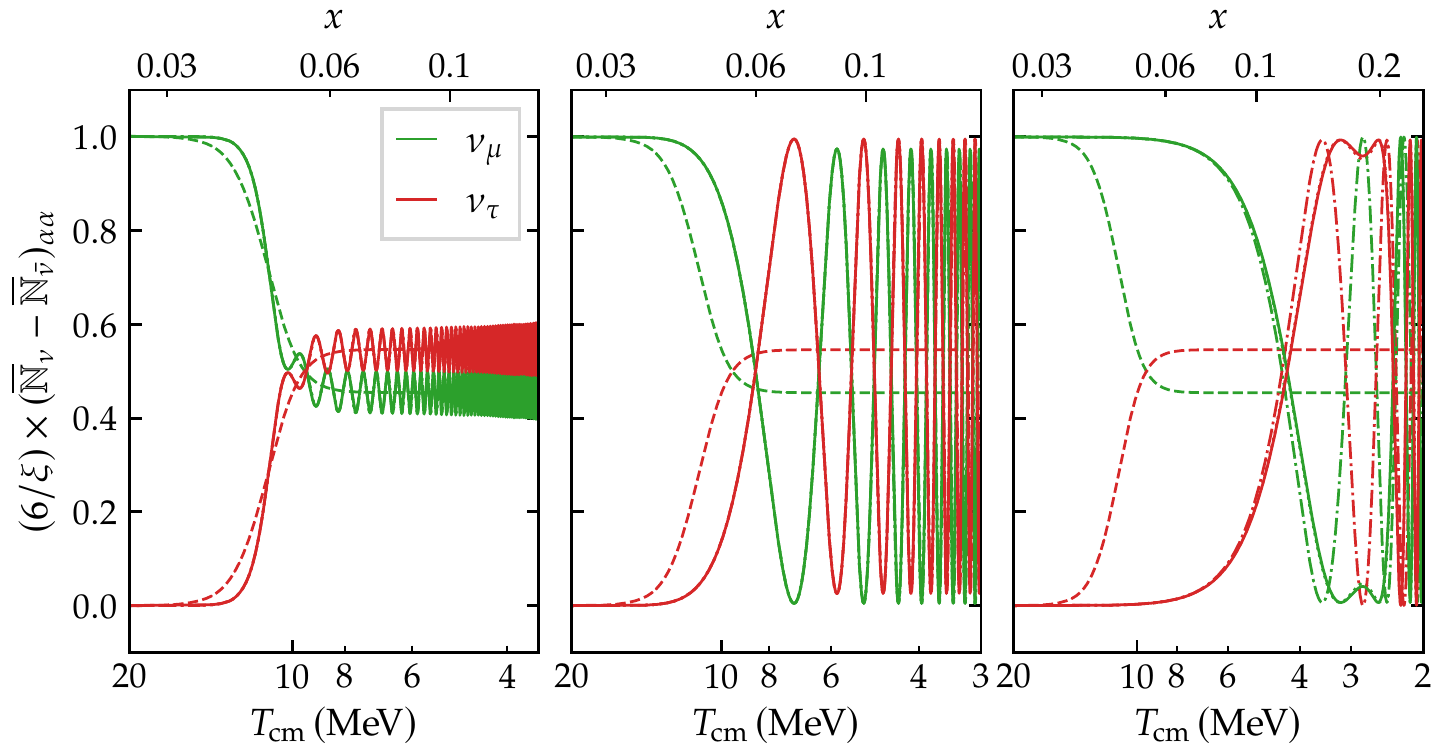} \\
  \includegraphics[]{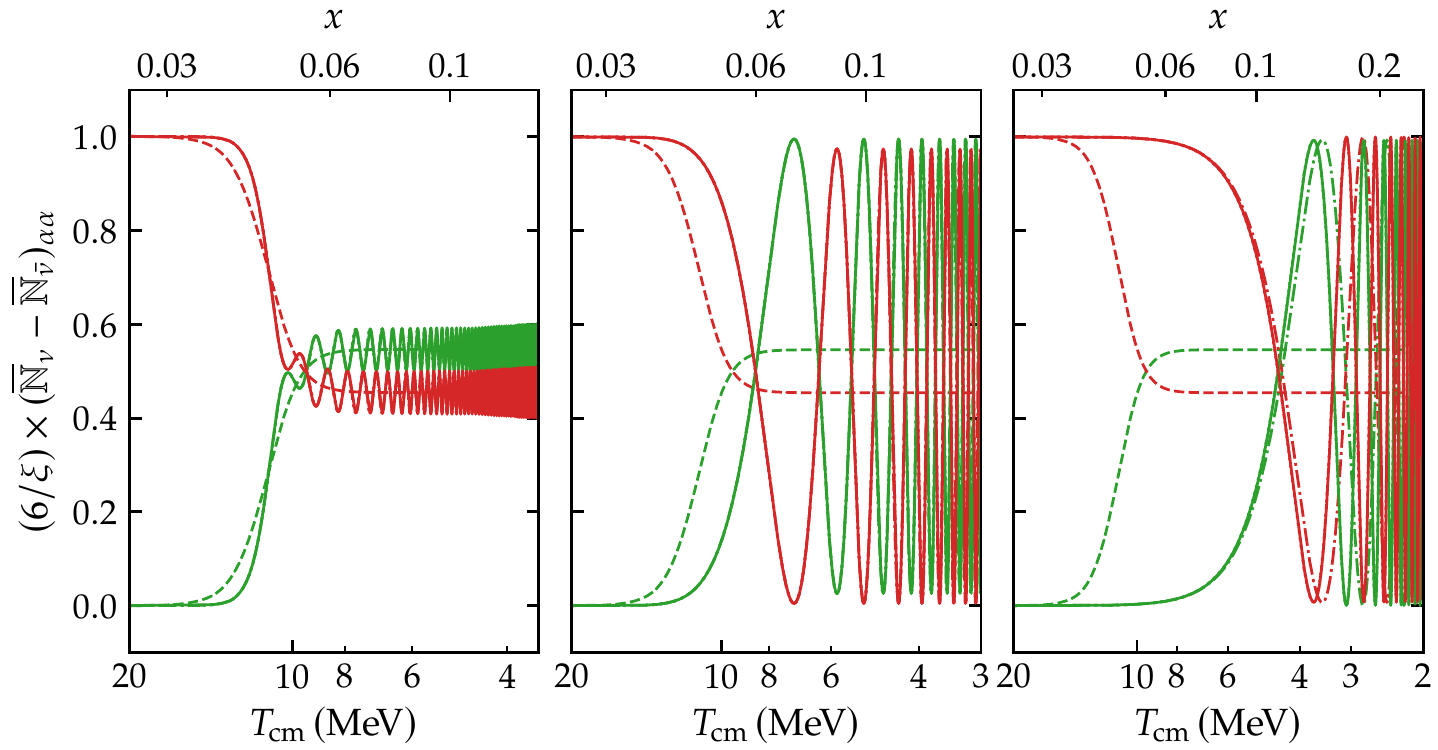}
	\caption[Evolution of the flavour asymmetries for a $\nu_\mu - \nu_\tau$ system without collisions]{\label{fig:2nu_mu-tau} Evolution of the flavour asymmetries for a two-neutrino $\nu_\mu$ (green) - $\nu_\tau$ (red) system without collisions, with $\Delta m^2 = 2.45 \times 10^{-3} \, \mathrm{eV}^2$ and $\theta = 0.831$. We compare different numerical schemes: in solid line QKE, in dots \ATAOJH (hidden behind QKE), in dot-dashes \ATAOJ, and in dashes \ATAOH. The initial degeneracy parameters are on the first row $\xi_2 =0$ and $\xi_1=0.1,0.01,0.001$ from left to right ; on the second row $\xi_1 =0$ and $\xi_2=0.1,0.01,0.001$. On the $\y$-axis label, $\xi$ stands for the non-zero initial $\xi_i$.}
\end{figure}

The evolution of asymmetry is illustrated in Figure~\ref{fig:2nu_mu-tau}. It is clear that the evolution with the self-interaction mean-field is completely different from the evolution where this has been ignored and which corresponds to the \ATAOH line: no synchronous oscillations take place in this scheme. These oscillations, in agreement with our adiabaticity estimate, do develop significantly for initial degeneracies smaller than one percent. On the contrary, for $\xi_1 + \xi_2 = 0.1$, the transition is quasi-adiabatic and $\vec{\Anti}$ follows the direction set by $\vecHeff$ with oscillations of much smaller amplitude compared to the smaller $\xi$ cases (right plots). Furthermore, it appears that at small degeneracies, the \ATAOJ results differ from the more accurate \ATAOJH scheme, the latter matching perfectly the QKE method. The difference between \ATAOJ and \ATAOJH can be understood by considering the NLO contribution to the precession frequency: this extra contribution explains the “wrong” frequency in the \ATAOJ case (see for instance the bottom right plot on Figure~\ref{fig:2nu_mu-tau}), or even the wrong qualitative behaviour of the asymmetry (top right plot).

\paragraph{Synchronous oscillation frequency} 

\begin{figure}[!ht]
	\centering
	\includegraphics[]{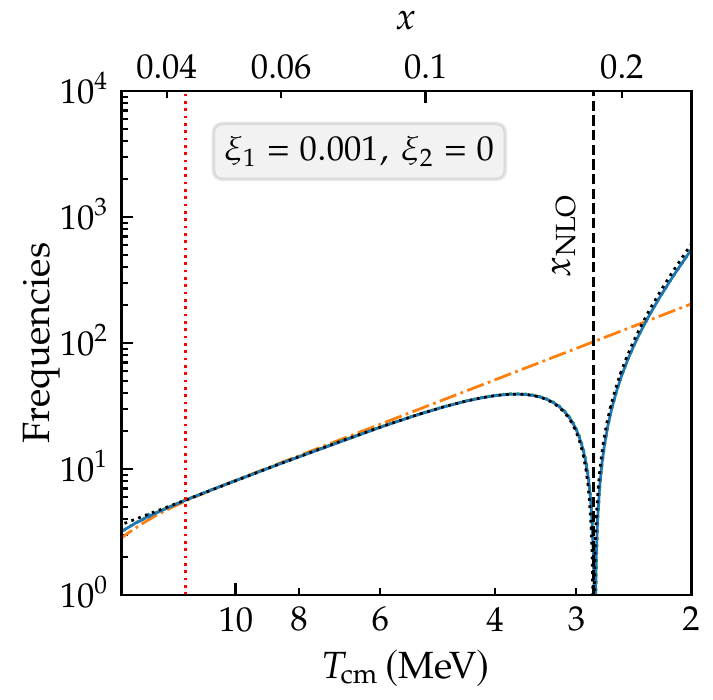}
        \includegraphics[]{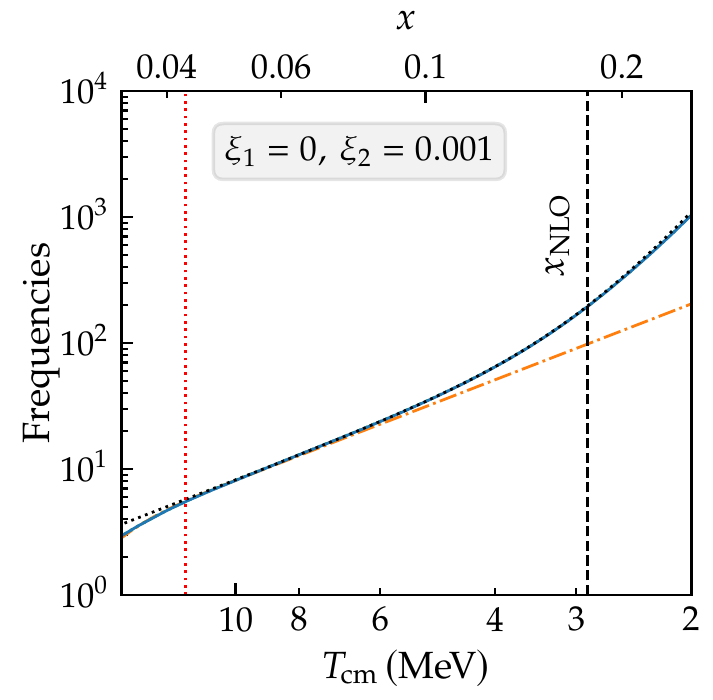}
	\caption[Frequency of synchronous oscillations for a $\nu_\mu - \nu_\tau$ system]{\label{fig:Omega_2nu_mu-tau} Frequency of synchronous oscillations in the case $\xi_1=0.001$, $\xi_2=0$
          (\emph{left}) and the case $\xi_1=0$, $\xi_2=0.001$ (\emph{right}). We consider a $\nu_\mu - \nu_\tau$ system with $\Delta m^2 = 2.45 \times 10^{-3}\,\mathrm{eV}^2$ and $\theta = 0.831$. The vertical red line is the location of the MSW
          transition. The dashed black line is $|\Omega \sin(2\theta)|$, that is the analytic approximation~\eqref{eq:Omega_norm}. The coloured lines correspond to $\lvert (\partial_x \vec{\Anti} \wedge
          \vec{\Anti}) \rvert /|\vec{\Anti}|^2$, in the QKE method (blue), the
          \ATAOJ scheme in orange and the \ATAOJH in green (hidden behind QKE).}
\end{figure}

To estimate the frequency from our runs we compute $(\partial_x \vec{\Anti} \wedge \vec{\Anti})/|\vec{\Anti}|^2$ which gives, given the precession equation~\eqref{eq:precession}, the projection of the rotation vector $\vec{\Omega}$ orthogonally to $\vec{\Anti}$. If the MSW transition is abrupt, the precession takes place around $\vecHvac$ with an angle $2 \theta$, hence the former quantity should be equal to $\lvert \sin(2 \theta) \vec{\Omega} \rvert$. Both frequencies are shown on Figure~\ref{fig:Omega_2nu_mu-tau}. We clearly see the transition from the regime $\Omega \propto x^2$ to
$\Omega \propto x^6$, that is the transition to the NLO regime, and in particular how the \ATAOJH scheme fits the QKE results, while (as expected by construction) the \ATAOJ scheme completely misses this change of regime. In the region $x_\mathrm{MSW} < x \ll x_\mathrm{NLO}$, $\Hself$ largely dominates over $\Hamil$ and all three schemes coincide.

Also, since $\theta=0.831 > \pi/4$, $\cos(2\theta)<0$ and according to~\eqref{eq:Omega_norm} the frequency can go through zero for normal ordering ($\Delta m^2>0$) with $|\xi_1|>|\xi_2|$ or inverted ordering ($\Delta m^2<0$) with $|\xi_2|>|\xi_1|$. That is the case in the top right plot of Figure~\ref{fig:2nu_mu-tau} ($\xi_1 = 0.001$, $\xi_2 = 0$ and normal ordering): we observe a back and forth motion of $\Anti$ for\footnote{The value predicted using \eqref{eq:xNLO} is slightly different from the one obtained numerically, because \eqref{eq:xNLO} assumes zero adiabaticity, such that the angle between $\vec{\Anti}$ and $\vecHvac$ is exactly $2 \theta$, whereas in reality $\vec{\Anti}$ partially follows the direction of $\vec{\Omega}$ during the MSW transition (see the bottom right plot on Figure~\ref{fig:Coll_e-mu} for a similar behaviour in an electron-driven transition).} $T_\mathrm{NLO} \simeq 2.8 \, \mathrm{MeV}$, which corresponds to $\Omega = 0$ as visible on Figure~\ref{fig:Omega_2nu_mu-tau}, left plot. This transition between two frequency regimes with a change of rotation direction is a feature also seen in Figures 9 and 10 of \cite{Johns:2016enc}.

If we consider smaller degeneracies, such that
\begin{equation}
|\xi_1^2-\xi_2^2| < \frac{x^4_\mathrm{MSW}}{x^4_\mathrm{tr}} \simeq
1.8\times 10^{-8}\left(\frac{|\Delta m^2
    \cos(2\theta)|}{10^{-3}\,\mathrm{eV}^2}\right)
\end{equation}
then the NLO contribution to the precession frequency will dominate already when the MSW transition occurs. The adiabaticity parameter $\gamma_\mathrm{MSW}$ must be rescaled by multiplying it by the factor in square brackets in equation~\eqref{eq:Omega_JH}, and we get approximately 
\begin{equation}
\gamma_\mathrm{MSW} =\frac{4.1 \times 10^{-7}}{\abs{\xi_1 - \xi_2}} \, ,
\end{equation}
assuming the NLO contribution does dominate in~\eqref{eq:Omega_JH}, which amounts to multiplying~\eqref{eq:gammaMSW_muon} by $(x_\mathrm{MSW}/x_\mathrm{tr})^4 /\abs{\xi_1^2 - \xi_2^2}$. Therefore, if we satisfy the condition $|\xi_1-\xi_2| \gg 4.1 \times10^{-7}$, the transition is still abrupt and oscillations do develop. For even smaller degeneracies, there is no clear region where $|\vec{\Hself}| \gg |\vec{\Hamil}|$, and the subsequent phenomenology can only be captured by a full QKE resolution as in~\cite{Johns:2016enc}. 

\paragraph{Beginning of oscillations} Provided the MSW transition is non-adiabatic, oscillations of $\vec{\Anti}$ appear, driving each individual mode. However, one can see on Figure~\ref{fig:2nu_mu-tau} that depending on the value of $(\xi_1, \xi_2)$, the apparent “start” of these oscillations looks shifted while $x_\mathrm{MSW}$ is the same. We can estimate how oscillations develop, which provides an additional check of our analytical developments.

The asymmetry evolves with a frequency $\Omega(x)$, therefore the phase of the oscillations is at any time given by
\begin{equation}
    \frac{\dd \Phi}{\dd x} = \Omega(x)  \qquad \text{hence} \qquad \Phi(x) = \frac{1}{3} \Omega(x) x \, ,
\end{equation}
where we used the fact that $\Omega \propto x^2$, keeping only the leading order contribution~\eqref{EvaluationOmega}. Half a period of oscillation is reached when $\Phi(x_\pi) = \pi$, which happens for
\begin{equation} 
\label{eq:x_startosc}
x_\pi = \left( \frac{1}{1.28 \times 10^6} \times \frac{10^{-3} \, \mathrm{eV}^2}{\abs{\Delta m^2}} \times \frac{1}{\abs{\xi_1 + \xi_2}} \times 3 \pi \right)^{1/3} \simeq 0.067 \, ,
\end{equation}
for $\xi_1 = 0.01$ and $\xi_2=0$, which agrees with Figure~\ref{fig:2nu_mu-tau}, top middle plot. 

\subsubsection{Particular cases}\label{SecParticular}

In this subsection, we use the $\nu_\mu - \nu_\tau$ framework to discuss the particular cases of equal and equal but opposite asymmetries, for which the calculations of section~\ref{FreqSyncOsc} are no longer valid. 

If $\xi_1 = \xi_2$, the vector parts of $\vrho(y)$, $\bvrho(y)$ and $\Anti$ are all equal to zero, and will therefore remain so. The self-potential term cancels in the QKE and the \ATAOH scheme describes accurately the neutrino evolution.

The case $\xi_1 = - \xi_2$ would correspond to a vanishing total lepton number density, while each flavour could display large asymmetries. This would result in a possibly significant contribution to the total energy density, hence the interest for this particular case. It was shown in \cite{Pastor:2001iu,Dolgov_NuPhB2002} that in this scenario synchronous oscillations are hampered as long as $\Hself$ dominates. This is in perfect agreement with our theoretical analysis: at leading order, as $F(\xi, -\xi) = 0$ the first term in \eqref{eq:Omega_JH} vanishes. However, our calculation of the NLO contribution shows that oscillations can still take place, but directly with a frequency $\propto x^6$. 

\begin{figure}[!ht]
  \centering
  \includegraphics[]{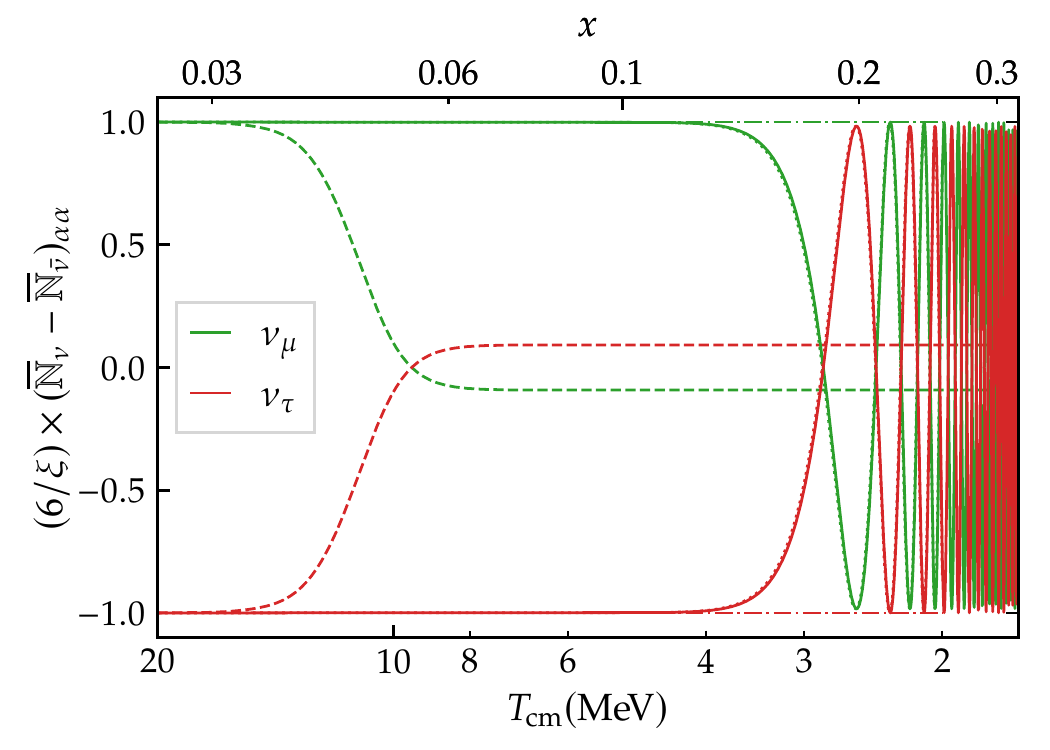}
	\caption[Equal but opposite asymmetries $\xi_1 = -\xi_2 = \xi = 0.001$]{\label{fig:xi1mxi2} Equal but opposite asymmetries $\xi_1 = -\xi_2 = \xi = 0.001$. We clearly see that, if we discarded the next-to-leading order contribution in~\eqref{eq:vrho_ATAOJV} (\ATAOJ curve, in dashed-dots), oscillations would be switched off. However, the lowest order term in the frequency expression is now $\propto x^6$ and gives rise to quasi-synchronous oscillations beyond $x\simeq 0.2$, see~equation~\eqref{eq:x_xi1mxi2}.}
\end{figure}

To check this prediction, we plot the evolution of asymmetries for $\xi_1 = -\xi_2 = \xi = 0.001$ on Figure~\ref{fig:xi1mxi2}. In the \ATAOJ scheme (which ignores the NLO contribution), oscillations never appear, contrary to the actual QKE evolution, correctly captured by the \ATAOJH scheme. The onset of synchronous oscillations is delayed compared for instance to the right plots of Figure~\ref{fig:2nu_mu-tau}, and we can estimate the location of this starting point exactly as in the previous section, the only difference being that we use the NLO part of~\eqref{eq:Omega_norm} $\Omega \propto x^6$. We find that the location of the first half-oscillation is
\begin{equation} 
\label{eq:x_xi1mxi2}
x_\pi = \left( \frac{8 \pi ^2}{3} \times 2.278 \times 10^{-22} \times \frac{\me^2}{\abs{\Delta m^2}} \times x_\mathrm{tr}^4 \times \xi \times 7 \pi \right)^{1/7} \simeq 0.20 \, ,
\end{equation}
for $\xi = 0.001$, in excellent agreement with Figure~\ref{fig:xi1mxi2}.

\subsection{Electron-driven MSW transition}\label{SecElectronMSW}

We now consider an electron/positron driven transition in the (fictitious) $\nu_e - \nu_\mu$ subspace, with the mixing angle $\theta = \theta_{12} \simeq 0.587$ and the small mass gap $\Delta m^2 = \Delta m_{21}^2 \simeq 7.53 \times 10^{-5} \, \mathrm{eV}^2$ for numerics. The difference with the previous case comes from the fact that the MSW transition now takes place when electrons are still relativistic. Moreover, we show in the following section how the collision term is very different from the one in the $\nu_\mu - \nu_\tau$ subspace. 

\begin{figure}[!ht]
  \centering
  \includegraphics[]{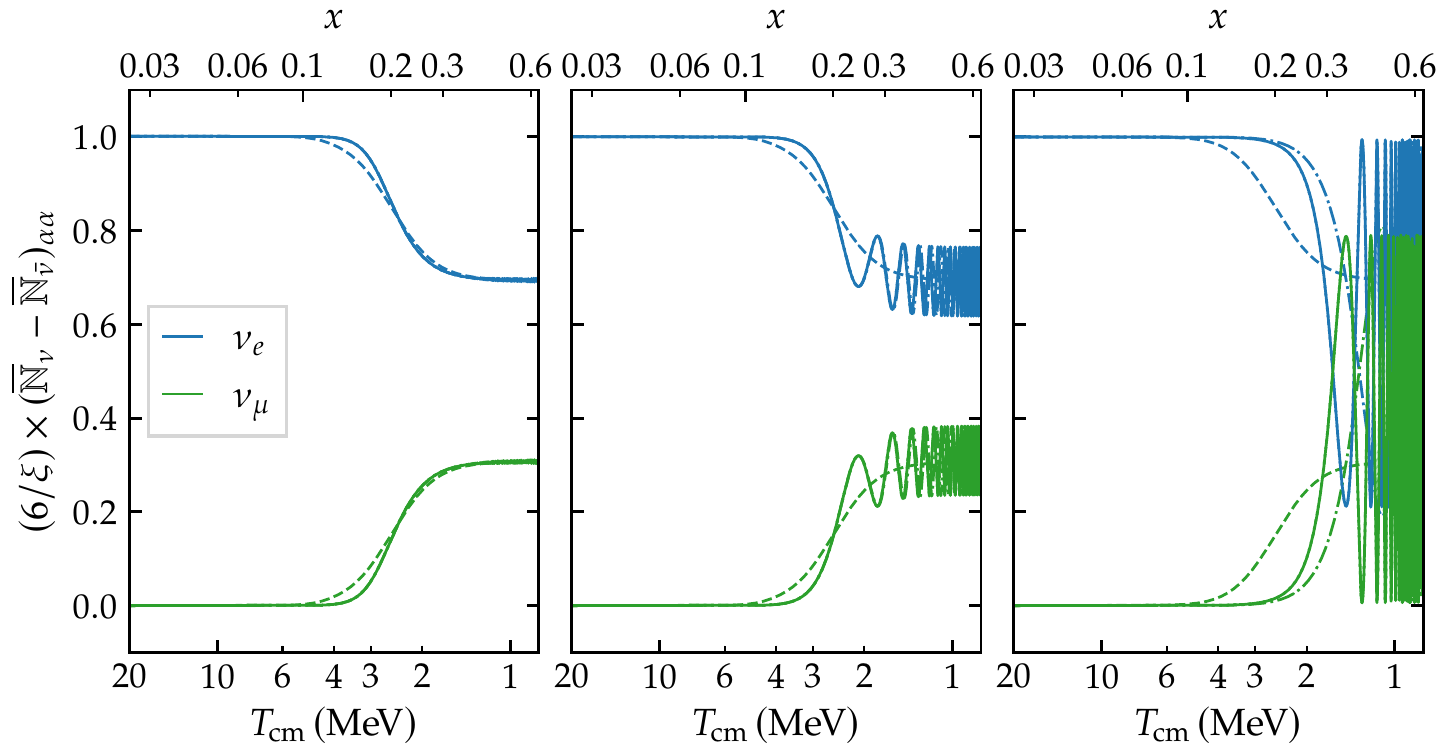} \\
  \includegraphics[]{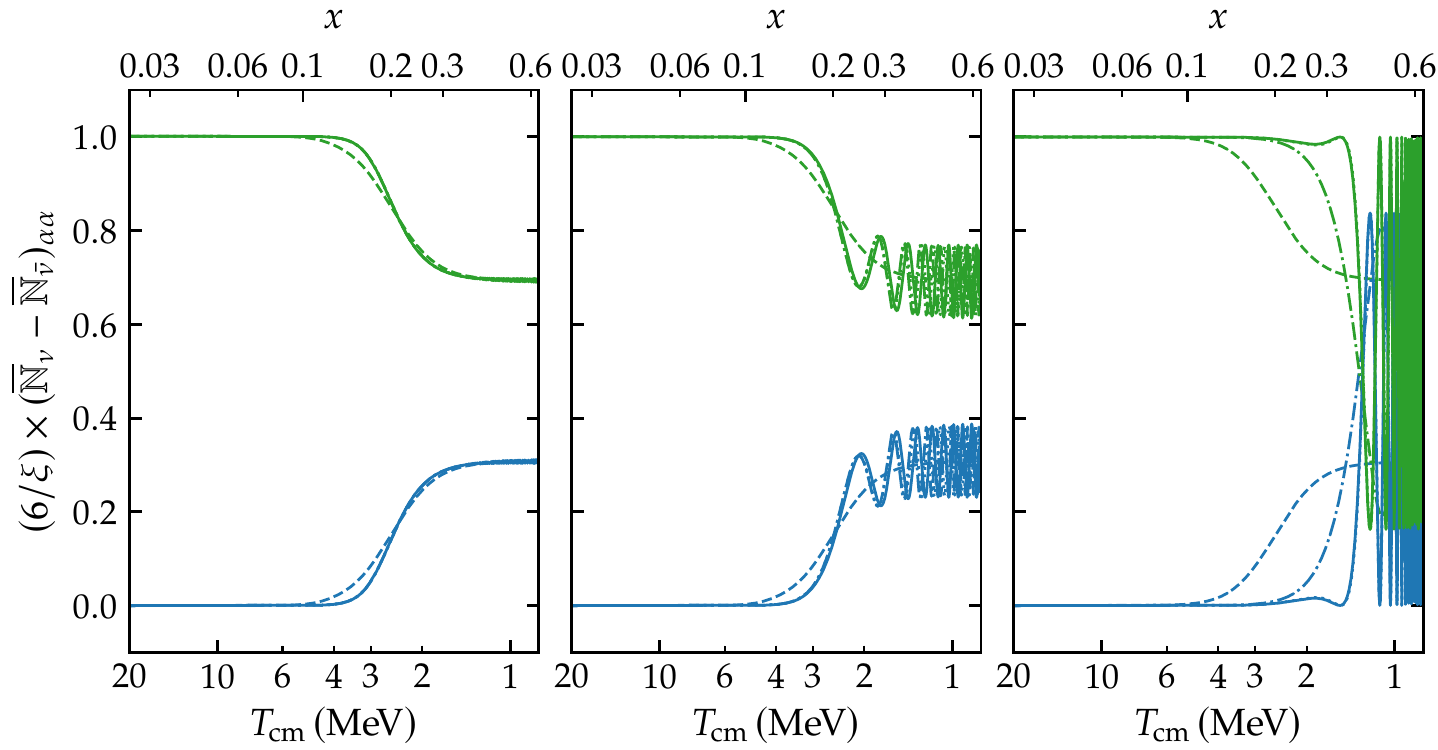}
	\caption[Evolution of the flavour asymmetries for a  $\nu_e - \nu_\mu$ system without collisions]{\label{fig:2nu_e-mu}Evolution of the flavour asymmetries for a two-neutrino $\nu_e$ (blue) - $\nu_\mu$ (green) system without collisions, with $\Delta m^2 = 7.53 \times 10^{-5} \, \mathrm{eV}^2$ and $\theta = 0.587$. We compare different numerical schemes: in solid line QKE, in dots \ATAOJH (hidden behind QKE), in dot-dashes \ATAOJ, and in dashes \ATAOH. The initial degeneracy parameters are on the first row $\xi_2 =0$ and $\xi_1=0.1,0.01,0.001$ from left to right ; on the second row $\xi_1 =0$ and $\xi_2=0.1,0.01,0.001$.}
\end{figure}

The relativistic limit is sufficient to estimate the location of the MSW transition, therefore we use (we take the comoving plasma temperature $z=1$ for simplicity, which is justified since $e^\pm$ annihilations are just beginning at this stage):
\begin{equation}
\rho_{e^\pm}+P_{e^\pm}  = \frac{7 \pi^2}{45} (\me/x)^4\, ,
\end{equation}
to deduce that the transition takes place for
\begin{equation}
x_\mathrm{MSW} = \left(\frac{\me^6 G_F}{\mW^2 \abs{\Delta
    m^2}}\frac{28\sqrt{2}\pi^2 y^2_\mathrm{eff}}{45}\right)^{1/6}=0.118\,\left(\frac{10^{-3} \, \mathrm{eV}^2}{\abs{\Delta m^2}}\right)^{1/6}\,.
\end{equation}
For the numerical values chosen, we find $x_\mathrm{MSW} \simeq 0.18$, that is $\Tcm \simeq 2.8\,\mathrm{MeV}$\footnote{Note also that the first electron-driven transition associated with the large mass gap should be around $\Tcm = 5\,\mathrm{MeV}$ by application of this estimate with $\Delta m^2 = \Delta m_{31}^2$.}. We estimate the adiabaticity of this transition with~\eqref{Defgammatr2}, and find
\begin{equation}
\gamma_\mathrm{MSW} \simeq 485 \times \abs{\xi_1 + \xi_2} \, .
\end{equation}
The larger prefactor compared to the estimate~\eqref{eq:gammaMSW_muon} in the muon-driven case makes the transition adiabatic up to smaller degeneracies. This is in agreement with the results of Figure~\ref{fig:2nu_e-mu}: for instance, the transition is much more adiabatic (small amplitude of synchronous oscillations) for $\xi_1 + \xi_2 = 0.01$ compared to Figure~\ref{fig:2nu_mu-tau}.

\begin{figure}[!ht]
	\centering
	\includegraphics{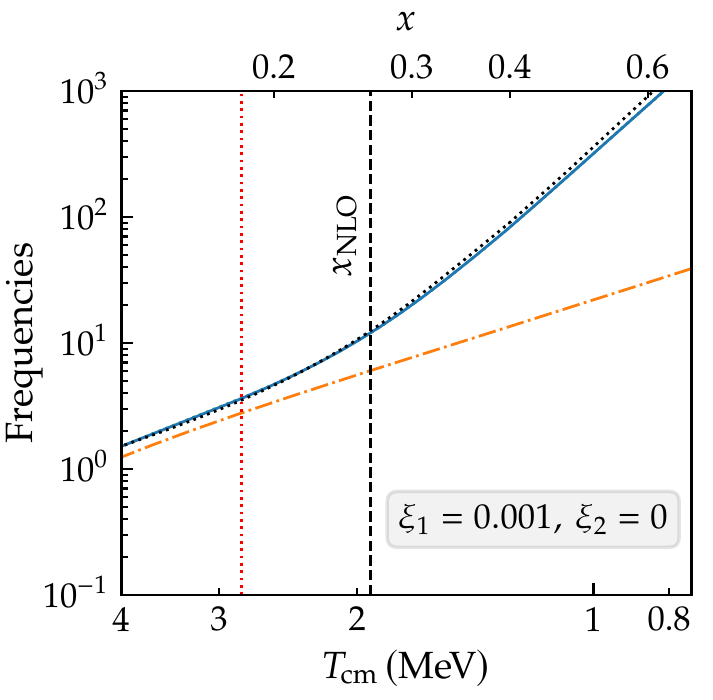}
        \includegraphics{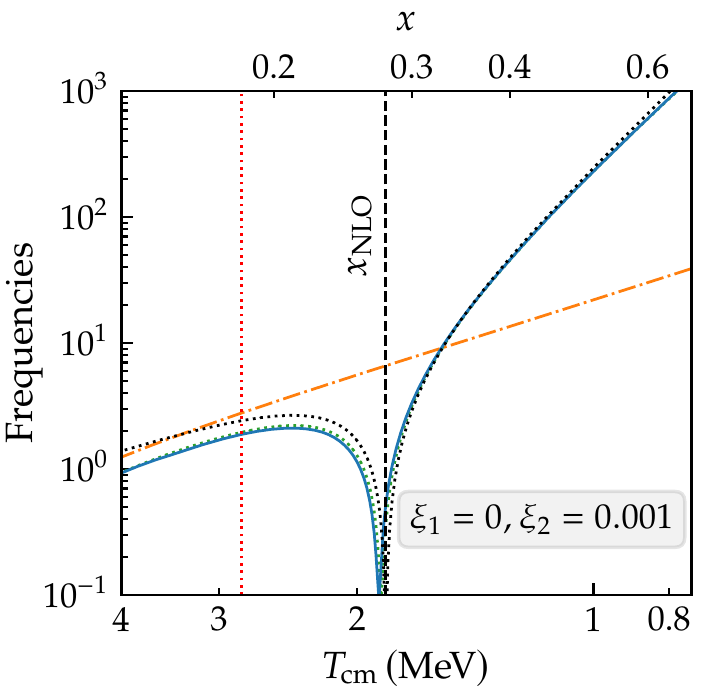}
	\caption[Frequency of synchronous oscillations for a $\nu_e - \nu_\mu$ system]{\label{fig:Omega_2nu_e-mu} Same plot as Figure~\ref{fig:Omega_2nu_mu-tau} for the $\nu_e - \nu_\mu$ system with mixing parameters $\theta=0.587$ and $\Delta m^2 = 7.53\times10^{-5}\,\mathrm{eV}^2$.}
\end{figure}

The frequency regimes outlined in section~\ref{FreqSyncOsc} are once again observed on Figure~\ref{fig:Omega_2nu_e-mu}, where we plot the quantities $|\Omega \sin(2\theta)|$ and $\lvert (\partial_x \vec{\Anti} \wedge \vec{\Anti}) \rvert /|\vec{\Anti}|^2$. We see the transition from $\Omega \propto x^2$ to $\Omega \propto x^6$ and the possible cancellation of the frequency at this transition. Contrary to the case studied in section~\ref{SecMuonMSW}, it happens now in normal ordering for $\abs{\xi_2}>\abs{\xi_1}$ because $\cos(2 \theta) > 0$.

\subsection{Effect of collisions}\label{SecCollisions}

In the previous sections, we systematically discarded the collision term in the QKEs in order to focus on the synchronous oscillation phenomenon and how approximate numerical schemes (namely, the \ATAOJH procedure) accurately capture the physics at play.

Taking into account the scattering and annihilation processes is nevertheless crucial for a precision calculation, not only since these processes will determine neutrino decoupling and partial reheating \cite{Dolgov_NuPhB1997,Esposito_NuPhB2000,Mangano2002,Mangano2005,Relic2016_revisited,Grohs2015,Grohs:2016cuu,Froustey2019,Froustey2020,Bennett2021}, but also because they can reduce flavour asymmetry differences. This second effect was notably shown in references~\cite{Dolgov_NuPhB2002,Johns:2016enc}, but these works used approximate expressions for the collision term (so-called damping approximation). We aim at showing the effect of the exact collision term, whose expression was derived for instance in~\cite{SiglRaffelt,BlaschkeCirigliano} and in chapter~\ref{chap:QKE} of this manuscript.

For the following discussion, we only recall that the collision term $\mathcal{K}[\vrho,\bvrho]$ is an integral whose matrix structure is determined by the statistical factors associated to two-body reactions $(1) + (2) \to (3) + (4)$. They read typically (we write $\vrho_i = \vrho(y_i)$ for particle $i$):
\begin{equation}
\label{eq:stat_fact1}
    \left[\vrho_4 (1- \vrho_2) + \mathrm{Tr}\left(\vrho_4 (1-\vrho_2) \right) \right] \vrho_3 (1- \vrho_1) - \{ \text{loss} \} + \mathrm{h.c.} \, ,
\end{equation}
for neutrino elastic scattering (this term corresponds to the process $\nu^{(1)} + \nu^{(2)} \to \nu^{(3)} + \nu^{(4)}$) and
\begin{equation}
\label{eq:stat_fact2}
    f_4^{(e)} (1- f_2^{(e)})G^{L/R} \vrho_3 G^{L/R} (1- \vrho_1) - \{ \text{loss} \} + \mathrm{h.c.} \, ,
\end{equation}
for reactions with electrons and positrons (this particular terms stands for a scattering $\nu^{(1)} + e^{(2)} \to \nu^{(3)} + e^{(4)}$). In the above expressions, the loss part corresponds to the exchange $\left\{ \vrho_i \leftrightarrow (1 - \vrho_i) \right\}$ for all distributions, and $\mathrm{h.c.}$ stands for “hermitian conjugate”. The coupling matrices $G^L$ and $G^R$ are diagonal in flavour space, and read in the full three-neutrino framework
\begin{equation}
    \label{eq:coupling_matrices}
    G^L = \begin{pmatrix} g_L +1 & 0 & 0 \\ 0 & g_L & 0 \\ 0 & 0 & g_L \end{pmatrix} \, , \quad
    G^R = \begin{pmatrix} g_R & 0 & 0 \\ 0 & g_R & 0 \\ 0 & 0 & g_R \end{pmatrix} \, ,
\end{equation}
with $g_L = -\frac12 + \sin^2{\theta_W}$, $g_R = \sin^2{\theta_W}$ where $\theta_W$ is Weinberg's angle. In the $ee$ entry of $G^L$, the extra factor of $1$ accounts for the charged currents between $e^\pm$ and $\nu_e$. Since we do not consider collisions with other charged leptons (due to their negligible density in the range of temperatures of interest), this is the only additional factor in $G^L$. Therefore the collision terms satisfy the general property
\begin{equation}\label{MagicKFundamental}
\mathcal{K}[U_\mathrm{s} \vrho U^\dagger_\mathrm{s}, U_\mathrm{s} \bvrho U^\dagger_\mathrm{s}] = U_\mathrm{s}\mathcal{K}[\vrho,\bvrho]  U^\dagger_\mathrm{s}\,,\qquad \overline{\mathcal{K}}[U_\mathrm{s} \vrho U^\dagger_\mathrm{s}, U_\mathrm{s} \bvrho U^\dagger_\mathrm{s}] = U_\mathrm{s}\overline{\mathcal{K}}[\vrho,\bvrho]  U^\dagger_\mathrm{s}
\end{equation}
for constant unitary matrices of the type
\begin{equation}\label{UtoU2}
U_\mathrm{s}  = \begin{pmatrix}
1 & 0 \\
0 & \mathcal{U}
\end{pmatrix} \,,\qquad \mathcal{U} \in \mathrm{U}(2) \,.
\end{equation}

In general, the collision term $\mathcal{K}[\vrho, \bvrho]$, being made of statistical factors like \eqref{eq:stat_fact1} and \eqref{eq:stat_fact2}, tends to make the density matrices in flavour basis diagonal, with entries being Fermi-Dirac distributions --- or $\vrho$ and $\bvrho$ must be obtained from conjugation of such matrices with a unitary matrix of the type~\eqref{UtoU2}. The degeneracies are not constrained by processes like $\nu_\alpha + \nu_\beta \to \nu_\alpha + \nu_\beta$ or $\nu_\alpha + \bar{\nu}_\alpha \to \nu_\beta + \bar{\nu}_\beta$. The only constraint is due to the processes $\nu_\alpha + \bar{\nu}_\alpha \to e^- + e^+$, which impose $\xi_\alpha = - \bar{\xi}_\alpha$ at equilibrium. Therefore, if collisions are strong enough, the density matrices are pushed towards
\begin{equation}
\label{eq:vrho_coll0}
\vrho \sim \mathrm{diag}[ g(\xi_\alpha,y) ]\,, \qquad  \bvrho \sim \mathrm{diag}[ g(-\xi_\alpha,y) ]\, ,
\end{equation}
where $\sim$ stands for the possible conjugation by a matrix of the form~\eqref{UtoU2}.

\paragraph{Muon-driven transition} In the framework of section~\ref{SecMuonMSW}, we considered a two-neutrino case with only $\nu_\mu$ and $\nu_\tau$. When focusing on the $\nu_\mu-\nu_\tau$ subspace of \eqref{eq:coupling_matrices}, the $G^L$ and $G^R$ matrices are proportional to the identity matrix.

Initially, the collision term vanishes since $\vrho$ and $\bvrho$ are in the form~\eqref{eq:vrho_coll0}. What may come as a surprise is the fact that it keeps vanishing even though $\vrho$ and $\bvrho$ evolve. Indeed, at high temperature the \ATAOJ scheme is valid and both $\vrho$ and $\bvrho$ are diagonalized by the same matrix $U_{\Hself}$, that is furthermore identical for all momenta $y$. This means that we have\footnote{For clarity, we omit the subscript $\Hself$ for the matter density matrix $\widetilde{\vrho}_{\Hself}$. More generally in this section, $\widetilde{\vrho}$ will be the diagonal density matrix, whether the Hamiltonian is $\Hself$, $\Hself + \Hamil$, \dots} $\vrho = U_{\Hself} \widetilde{\vrho} U_{\Hself}^\dagger$ (and similarly for $\bvrho$), so from the restriction of the general property~\eqref{MagicKFundamental} we deduce the relations 
\begin{equation}\label{MagicKUJ}
\mathcal{K}[\vrho,\bvrho] = U_{\Hself}
\mathcal{K}\left[\widetilde\vrho,\widetilde\bvrho\right]  U^\dagger_{\Hself} \ , \qquad 
\overline{\mathcal{K}}(\vrho,\bvrho) = U_{\Hself} \bar {\mathcal{K}}\left[\widetilde\vrho,\widetilde\bvrho\right]  U^\dagger_{\Hself}  \, .
\end{equation}
Thanks to this peculiar “factorization”, the collision term keeps vanishing as long as $\widetilde{\vrho}$ and $\widetilde{\bvrho}$ are diagonal matrices (which they are by definition) of Fermi-Dirac distributions. This is much less restrictive, and remains satisfied as long as $\Hself \gg \Hamil$ since $\widetilde{\mathcal{K}} = 0$ leads to $\partial_x \widetilde{\vrho} = 0$, hence the collision term keeps vanishing, and so on. With or without collisions, the evolution of $\vrho, \bvrho$ is purely due to the change of direction of $\vec{\Anti}$, which oscillates more or less around $\vec{\Hamil}$ depending on the adiabaticity of the MSW transition.

Note that the previous argument is only exact for the part of $\mathcal{K}$ corresponding to neutrino self-interactions. It extends to the scattering with electrons/positrons as long as all particles share the same temperature. But even beyond this, when $e^\pm$ annihilations populate the neutrinos, they do so in creating pairs of neutrinos/antineutrinos, so the collision term acts to maintain thermal distributions, but not to equilibrate asymmetries.

\begin{figure}[!ht]
	\centering
	\includegraphics[]{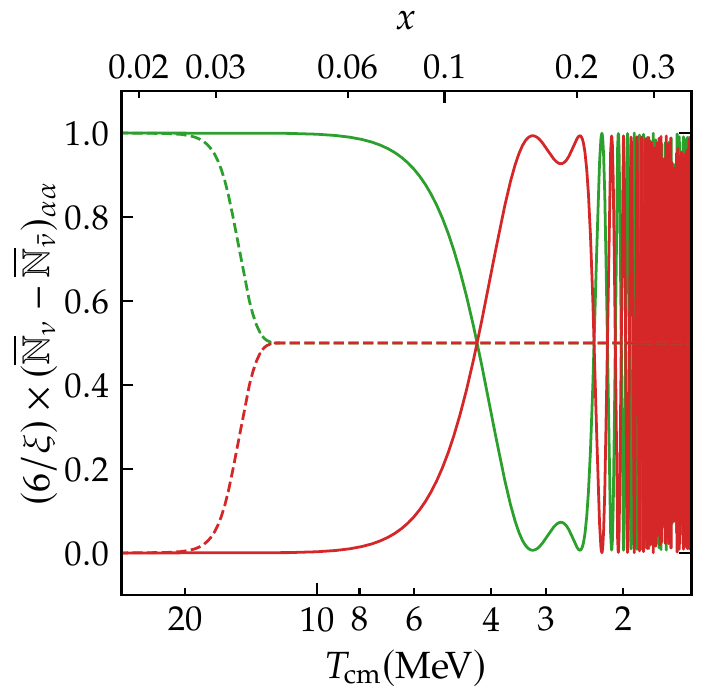}
    \includegraphics[]{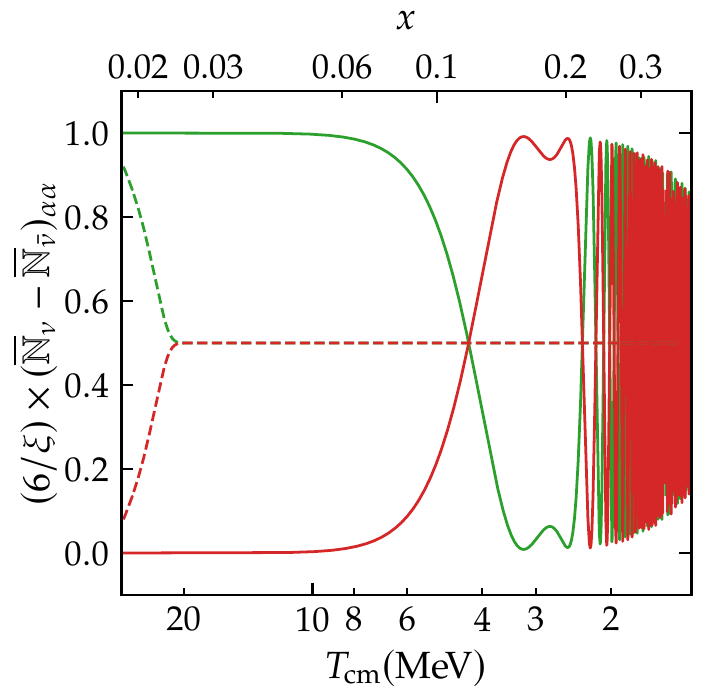}
	\caption[Effect of collisions on the muon-driven transition for $\xi_1=0.001$, $\xi_2=0$]{\label{fig:Collisions_mu} Effect of collisions on the muon-driven transition for $\xi_1=0.001$, $\xi_2=0$, with the collision term set to its actual value (\emph{left}) and artificially multiplied by $1000$ (\emph{right}). We only plot the result of two numerical schemes: \ATAOH (dashes) and \ATAOJH (solid, equivalent to QKE).}
\end{figure}

All in all, the asymmetries are not affected at all by the collision term as long as the \ATAOJ scheme is a good description of neutrino evolution. However, we have shown that below $\sim 10 \, \mathrm{MeV}$ the refined \ATAOJH scheme is necessary to capture the physics. The very fact that $\vrho$ and $\bvrho$ are not diagonalized with the same unitary matrix (either $U_{\Hself+\Hamil}$ or $U_{\Hself-\Hamil}$), and furthermore the $y$-dependence of these matrices, means that we lose the property \eqref{MagicKUJ}, that is
\begin{equation}\label{MagicKUJLost}
\mathcal{K}[\vrho,\bvrho] \neq U_{\Hself+\Hamil}(y)
\mathcal{K}[\widetilde\vrho,\widetilde\bvrho]  U^\dagger_{\Hself+\Hamil}(y) \ , \qquad
\overline{\mathcal{K}}[\vrho,\bvrho] \neq U_{\Hself-\Hamil}(y) \bar {\mathcal{K}}[\widetilde\vrho,\widetilde\bvrho]  U^\dagger_{\Hself-\Hamil}(y) \,. 
\end{equation}
When this non-equality is not meaningless --- it is necessarily suppressed by a factor $|
\vec{\Hamil}|/|\vec{\mathcal{\Hself}}| \propto x^4 /|\xi_1-\xi_2|$ --- the collision term starts to have a mild effect, which is even smaller for large $\xi_\alpha$ differences. Therefore, only for rather small $\xi_\alpha$ differences can a slight equilibration effect due to collisions be expected. However, since the frequency $\Omega$ of synchronous oscillations is then smaller, the actual start of oscillations is delayed until a moment when collisions are inefficient. This is why we expect collisions to have a negligible effect throughout the evolution in this $\nu_\mu - \nu_\tau$ system. We check this on Figure~\ref{fig:Collisions_mu}, left plot, where the evolution is indistinguishable from the one without collisions (Figure~\ref{fig:2nu_mu-tau}, top right plot). On the right plot, we artificially multiplied the collision term by $1000$, and we do observe in that case the damping of quasi-synchronous oscillations when $\abs{\Hamil} \sim \abs{\Hself}$, which corresponds to a reduction of asymmetry differences between the two flavours.

In principle, taking into account scattering and annihilations with muons and antimuons invalidates this picture since the charged currents with the $\nu_\mu$ and $\bar \nu_\mu$ make $G^L$ non-proportional to the identity, and the general property \eqref{MagicKUJ} would be lost. However their density is so suppressed in this range of temperature that we were able to check that the previous results are not affected.

Note also that the behaviour is very different if we ignore the self-interaction mean-field and rely on the pure \ATAOH scheme. In that case, the system is made of pure mass states of $\Hvac$ after the MSW transition (since each $\vrho$ is diagonal in the mass basis). However, given the $y$-dependence of $\Hamil$, this transition does not happen at the same time for all momenta. Thus there cannot be a property of the type~\eqref{MagicKUJ} with $U_{\Hself} \to U_{\Hamil}$ because there is no unique $\Hamil$, and furthermore $U_{\Hamil}(y)$ depends on $y$ which prevents its factorization out of the collision integral. Therefore the collision term will tend to restore diagonality in flavour space (that is reduce flavour coherence), and this can only be compatible with pure mass states (a requirement of the \ATAOH approximation) when all flavours have reached the same distributions, that is when the asymmetry matrix $\Anti$ is proportional to the identity and thus $\vec{\Anti}=\vec{0}$. In other words, the collision term is strongly incompatible with the evolution of asymmetries dictated by the \ATAOH scheme, and thus damps them. We observe this behaviour on Figure~\ref{fig:Collisions_mu}: $\Anti_\z \to 0$ in the presence of collisions, which was not the case on Figure~\ref{fig:2nu_mu-tau}.

To conclude, if we ignore the self-interaction mean-field, the collision term efficiently damps the asymmetry differences, because the unitary adiabatic evolution is not the same for density matrices at various momenta. When including the additional self-interaction potential, as long as it dominates the vacuum and lepton mean-fields, the density matrices at various momenta evolve adiabatically with the common
unitary matrix $U_{\Hself}$ and this preserves the initial absence of
effect of the collision term. It is only when \ATAOJ is insufficient and one has to rely on \ATAOJH that one starts to see the effect of the unitary evolution differing between momenta, but also between neutrinos and antineutrinos. This allows the collision term to damp the asymmetry vector. But this comes with a very large delay and the collision term is only able to barely damp $\Anti_\z$.

\paragraph{Electron-driven transition} In the framework of section~\ref{SecElectronMSW}, the difference with the $\nu_\mu - \nu_\tau$ case is the fact that $G^L$ is no longer proportional to the identity: $G^L = \mathrm{diag}(g_L +1, g_L)$. Once oscillations develop and $U_\Hself \neq \mathbb{I}$, there is no property like~\eqref{MagicKUJ}. In other words, the matrix $G^L$ sets the direction $\vec{e}_\z$ towards which the collision term now unavoidably attracts $\vec{\vrho}$ and $\vec{\bvrho}$ (whereas before, $\mathcal{K}$ was blind to any global rotation of axes). Therefore, the collision term tends to erase flavour coherence much more efficiently: it damps oscillations but does not necessarily allow to fully reach a state where the two neutrino flavours have identical distributions, because the collision term becomes too weak at temperatures below the MSW transition.

\begin{figure}[!ht]
	\centering
	    \includegraphics[]{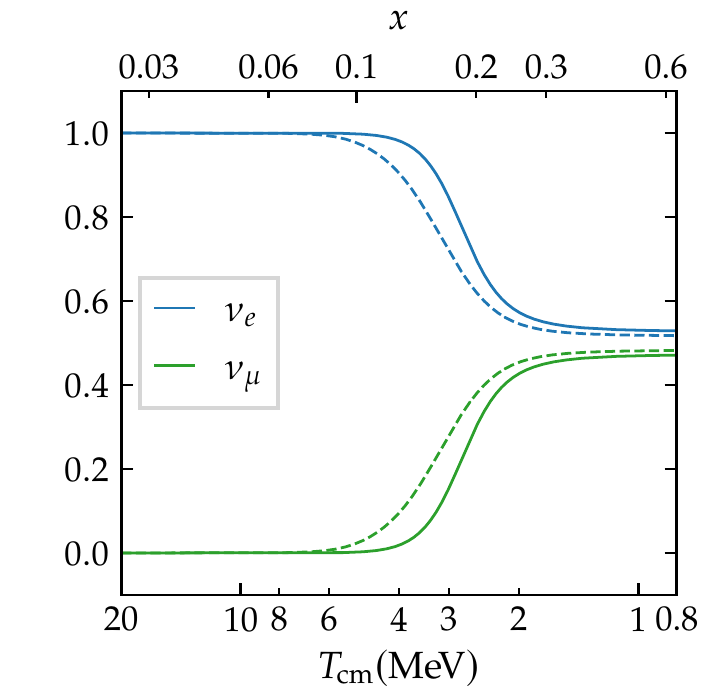}
	    \includegraphics[]{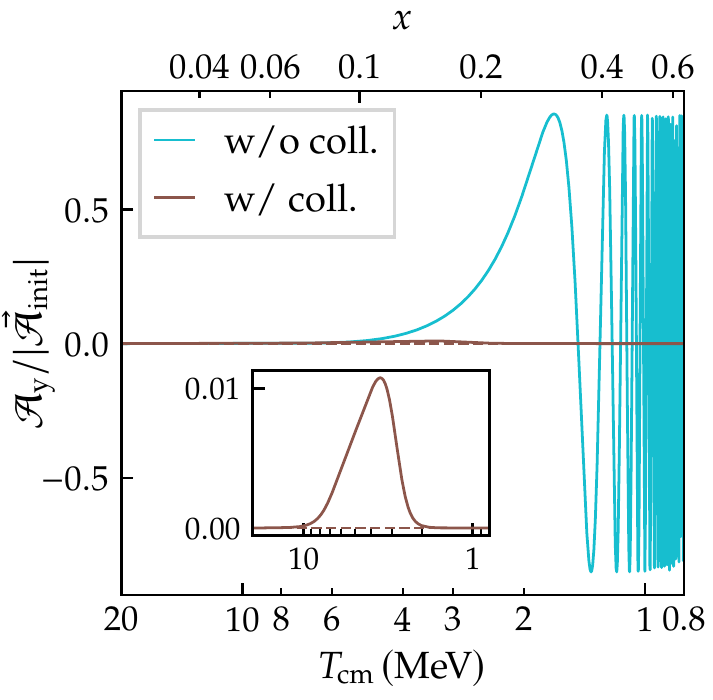}
	    \includegraphics[]{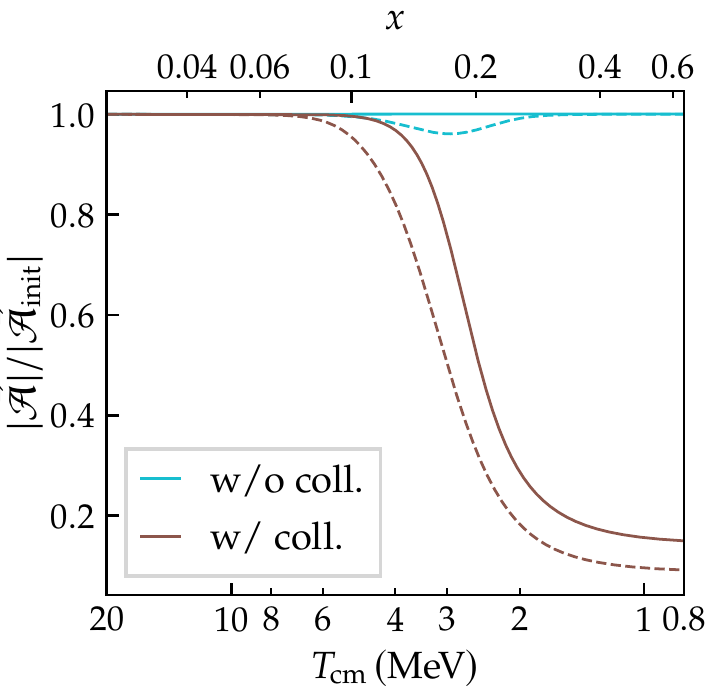}
        \includegraphics[]{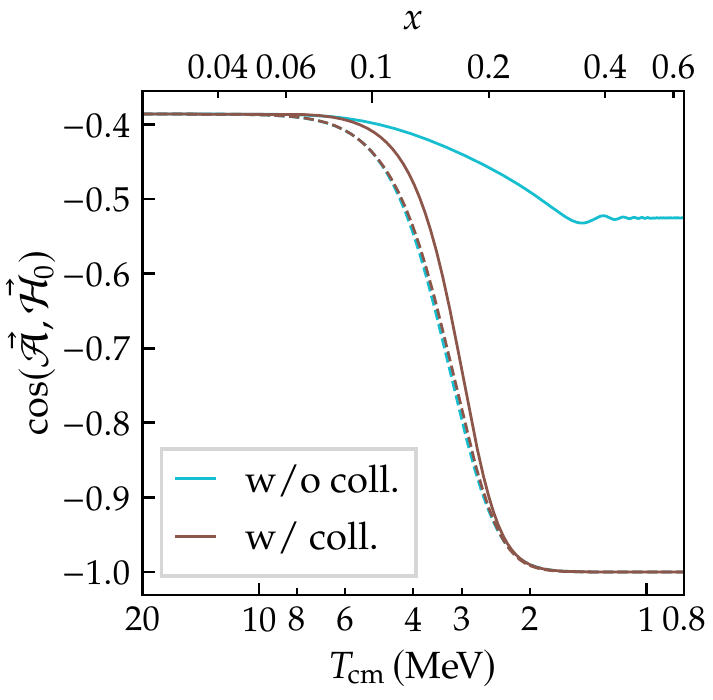}
	\caption[Effect of collisions on the evolution of the $\nu_e - \nu_\mu$ system for $\xi_1 = 0.001, \, \xi_2 = 0$]{\label{fig:Coll_e-mu} Effect of collisions on the evolution of the $\nu_e - \nu_\mu$ system, with $(\xi_1 = 0.001, \, \xi_2 = 0)$. The dashed lines correspond to the \ATAOH scheme (no self-interactions in the mean-field), and the solid lines to the \ATAOJH scheme (equivalent to the full QKE resolution). \emph{Top left plot:} evolution of electron and muon flavour asymmetries. \emph{Top right plot:} $\y-$component of the asymmetry vector $\vec{\Anti}$. \emph{Bottom left plot:} norm of $\vec{\Anti}$. \emph{Bottom right plot:} angle between $\vec{\Anti}$ and the final precession direction $\vecHvac$.}
\end{figure}

The top left plot of Figure~\ref{fig:Coll_e-mu} is equivalent to the top right plot of Figure~\ref{fig:2nu_e-mu} ($\xi_1 = 0.001$, $\xi_2 = 0$), but including collisions. As expected, the asymmetry is damped by $\mathcal{K}$ in both the \ATAOH and \ATAOJH schemes, and the evolution looks quite similar, suggesting that one could be satisfied with the simpler \ATAOH resolution scheme. However, this misses some important physical features, as the other plots on Figure~\ref{fig:Coll_e-mu} show. First, if one neglects the self-potential there is no precession of $\vec{\Anti}$ around $\vec{\Hamil}$, but simply the alignment of all $\vec{\vrho}, \, \vec{\bvrho}$ with $\vec{\Hamil}$ which evolves from $\Hlep$ domination to $\Hvac$ domination. This is clearly seen on the top right plot of Figure~\ref{fig:Coll_e-mu} (dashed lines): the $\y$-component of $\vec{\Anti}$ is constantly equal to zero, which is expected as $\widehat{\Anti}$ evolves from $\vec{e}_\z$ to $\hat{\Hcal}_0$ that lies in the $(\x-\z)$ plane. In the correct \ATAOJH scheme however, synchronous oscillations do develop when collisions are discarded. When one takes them into account, $\Anti_\y$ does not take sizeable values but starts an oscillation (that is strongly damped by $\mathcal{K}$), see the insert plot.

The bottom plots of Figure~\ref{fig:Coll_e-mu} show respectively the norm of $\vec{\Anti}$ and its alignment with $\vecHvac$. In the no-collision case (blue curves), the final angle between $\vec{\Anti}$ and $\vecHvac$ is non-zero (bottom right plot), but different from its initial value due to the very small adiabaticity of the MSW transition: $\vec{\Anti}$ slightly rotates towards $\vec{\Hamil}_\mathrm{eff}$, and precesses with an angle slightly different from $2 \theta$. Concerning its norm, $\lvert \vec{\Anti} \rvert$ is conserved without collisions.\footnote{The small “trough” in the \ATAOH case (dashed blue line) around $3 \, \mathrm{MeV}$ is not a numerical artefact. Since in that case each individual $\vec{\vrho}(y)$ changes its direction from $\vecHlep(y)$ to $\vecHvac(y)$ at different times (instead of being all locked on $\vec{\Anti}$), the norm of $\vec{\Anti}$ can only be compared at early and late times.} In contrast, we observe that including collisions (brown curves) $\vec{\Anti}$ gets aligned with $\vecHvac$ (but in the opposite direction due to the value of $\theta$), while the asymmetry differences are damped---a result of the competition between precession (which sets the preferred direction $\hat{\Hcal}_0$) and collisions (with the preferred direction $\vec{e}_\z$).

\section{Evolution with three flavours of neutrinos}\label{Sec:3neutrinos}

Having presented in the previous sections the salient features of two-neutrino evolution in the presence of flavour asymmetries, we can now turn to the full three-neutrino framework. Our goal is not to provide a thorough exploration of the parameter space, but instead to highlight the main physical characteristics of neutrino evolution with non-zero asymmetries.

\subsection{Method}

We have shown the accuracy of the \ATAOJH scheme that we can confidently use instead of a full QKE resolution, more computationally expensive. Therefore, the results are here obtained with this method and are compared with the \ATAOH scheme where we recall that the self-interactions are ignored in the mean-field, so as to highlight how the self-potential changes the dynamics.

However, it is impossible to integrate correctly the evolution at low temperatures. First, oscillations become too fast as their frequency grows as $x^6$ when the NLO dominates. Then, we reach the point where the \ATAOJH scheme starts to fail and oscillations must become gradually not synchronized. Eventually the system must converge to a state where fast oscillations disappear, that is an \ATAOH scheme. We chose to switch to an \ATAOH scheme at low temperature to effectively capture this transition from \ATAOJH to \ATAOH. In principle one should use the full QKE scheme to integrate numerically this phase, but for the same reasons it is numerically daunting. We chose to switch to the \ATAOH scheme around $2\,\mathrm{MeV}$, since the final MSW transition is over and collisions become rather inefficient (except for electron/positron annihilations). This method of instantaneous switching to the \ATAOH scheme necessarily misses some physics since it hides the complexity of the transition, but as we discuss in section~\ref{SecTransitionToATAOV} we expect that this does not significantly affect the results obtained for the evolution of asymmetry. 

\subsection{Results with standard mixing parameters}

Given the numerous possibilities for the values of the initial degeneracy parameters, we chose to restrict to two types of initial conditions. First, we consider the case where electronic flavour neutrinos have the largest degeneracy ($\xi_e=\xi_\mathrm{max}$), with $\xi_\tau=\xi_e/10$. Second we consider the case where muonic neutrinos have the largest non vanishing initial degeneracy ($\xi_\mu=\xi_\mathrm{max}$), with $\xi_\tau=\xi_\mu/10$. We do not report results where $\xi_\tau$ is the largest, because it is qualitatively very similar to the case where $\xi_\mu$ is the largest potential, since oscillations develop in exactly the same way. In any case, the third initial degeneracy parameter is set to zero.

\begin{figure}[!ht]
	\centering
\includegraphics[]{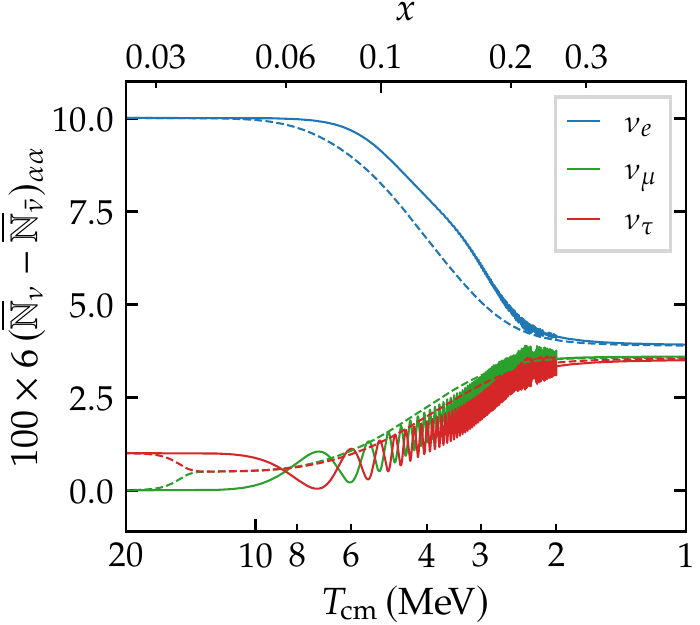}
\includegraphics[]{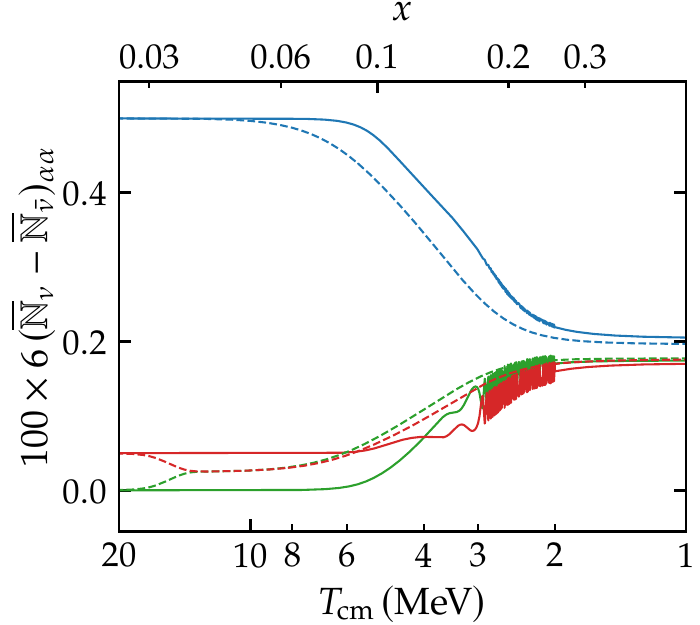}\\ \vspace{0.2cm}
	\includegraphics[]{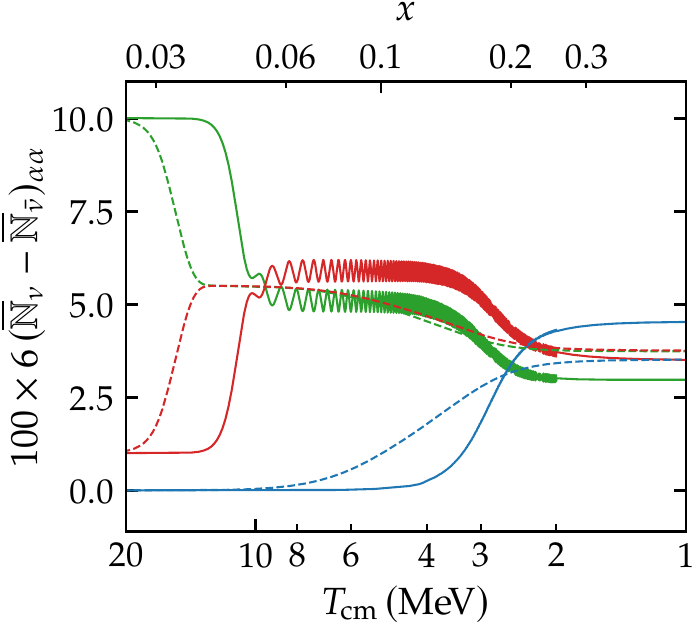}
	\includegraphics[]{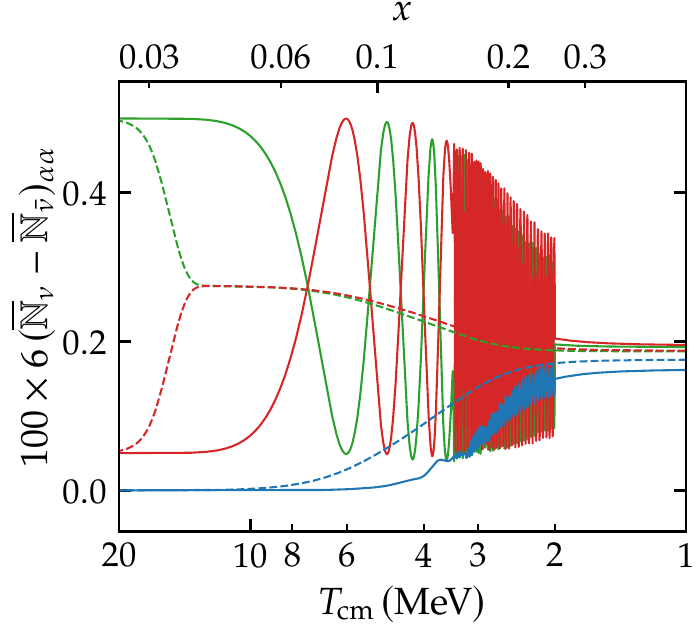}
	\caption[Evolution of flavour asymmetries in the standard three-neutrino case, for various initial conditions]{\label{figStandard} Initial conditions are $\xi_\alpha=(0.1,0,0.01)$ (top left), $\xi_\alpha=(0.005,0,0.0005)$ (top right), $\xi_\alpha=(0,0.1,0.01)$ (bottom left) and $\xi_\alpha=(0,0.005,0.0005)$ (bottom right). 
	The solid lines are the \ATAOJH schemes (extended into a simple \ATAOH below $2\,\mathrm{MeV}$), and the dashed lines are \ATAOH schemes throughout.}
\end{figure}

Results are depicted in Figure~\ref{figStandard} for both typically large ($\xi_\mathrm{max}=0.1$) and typically small ($\xi_\mathrm{max}=0.005$) potentials. We can observe how synchronous oscillations develop in the $\nu_\mu-\nu_\tau$ space after the muon-driven transition. Their amplitude is reduced for large initial potentials since the transition is then more adiabatic in that case, as detailed in section~\ref{SecMuonMSW}. In the case of small initial potentials, the transition from leading order to NLO oscillations is also clearly visible (right plots of Figure~\ref{figStandard}). Note that even though the general behaviour is that asymmetries tend to converge, this trend stops before equilibration is complete, and is in general less complete than in the case where the self-interaction mean-field is ignored. Furthermore in some cases, the ordering of final asymmetry is not the same as the ordering of initial ones.

If we consider cases with much larger initial asymmetries in the muonic and tauic neutrinos, typically such that $\xi_\mu+\xi_\tau \gg 0.1$, the muon-driven transition is very adiabatic, and oscillations do not develop at that transition as the asymmetry vector closely follows the evolution of $\Hamil(y_\mathrm{eff})$. This is illustrated in the left plot of Figure~\ref{figComparison}.

\begin{figure}[!ht]
	\centering
    \includegraphics[]{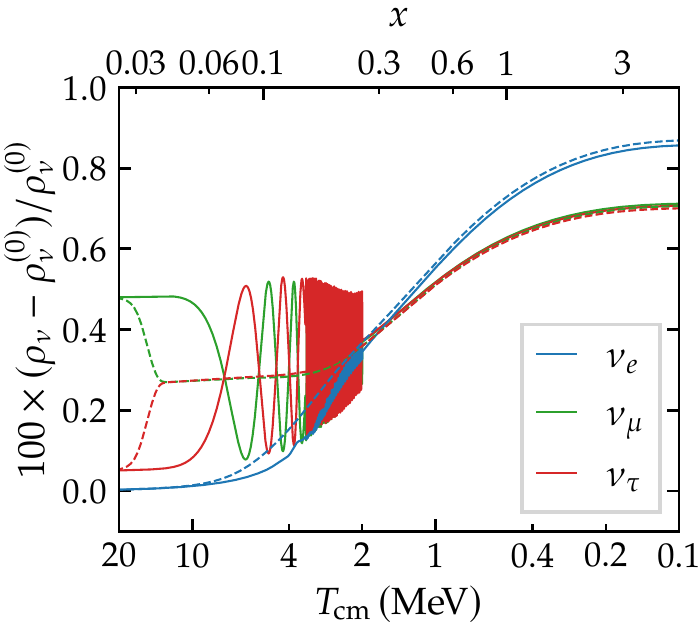}
    \includegraphics[]{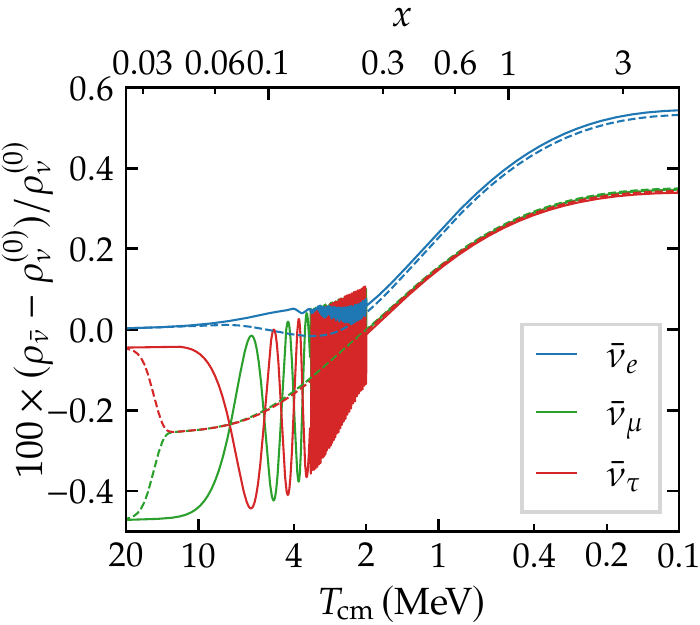}
	\caption[Fractional difference of the energy density with respect to a decoupled neutrino without chemical potential]{\label{figrho} Fractional difference of the energy density with respect to a decoupled neutrino without chemical potential. The solid lines are the \ATAOJH schemes (extended into a simple \ATAOH below $2\,\mathrm{MeV}$), and the dashed lines are \ATAOH schemes throughout. The left plot is for neutrinos and the right plot for antineutrinos. Initial conditions are $\xi_\alpha=(0,0.005,0.0005)$.}
\end{figure}

We stress that in this section we always consider the full collision term, and even though we show the results for the evolution of the asymmetry (since it is the relevant quantity to discuss synchronized oscillations), neutrinos are also partially reheated by electron/positron annihilations, which preserve the asymmetry. In Figure~\ref{figrho} we show the energy density fractional difference with respect to the one of a completely decoupled neutrino species with vanishing chemical potential $\bar{\rho}_\nu^{(0)} = 7 \pi^2/240$. In general the final effective temperature and distortions of electronic (anti-)neutrinos, which have a direct effect on neutron/proton freeze-out and thus BBN, depend on initial degeneracy parameters. Hence the impact on BBN predictions is not straightforward and one should perform a full BBN analysis for each set of initial conditions~\cite{Grohs:2016cuu}, as was done for the standard case of vanishing potentials in chapter~\ref{chap:BBN}.

It is beyond the scope of this chapter to perform a full exploration of parameters with all initial degeneracy parameters and all mixing parameters. However we aim here at highlighting how results are qualitatively modified when considering different mixing parameters.

\subsection{Dependence on neutrino mass ordering}

\begin{figure}[!ht]
	\centering
	\includegraphics[]{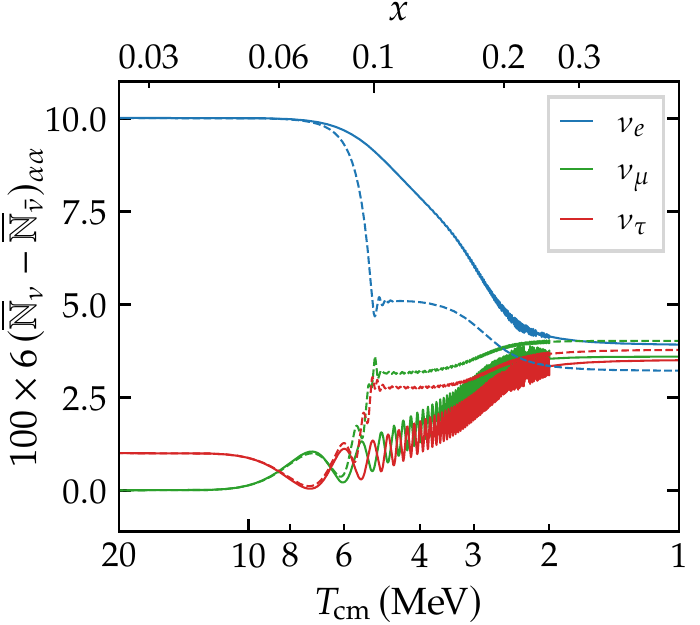}
     \includegraphics[]{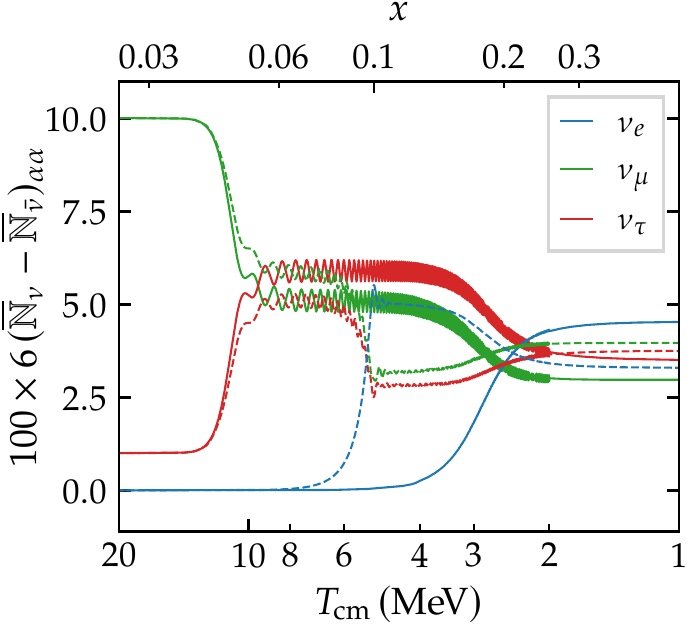}
	\caption[Comparison of normal and inverted ordering]{\label{figcomp4} The normal ordering case in solid lines, is compared to the inverted ordering case in dashed lines. Initial conditions are $\xi_\alpha=(0.1,0,0.01)$ on the left, and $\xi_\alpha=(0,0.1,0.01)$ on the right.}
\end{figure}

In Figure~\ref{figcomp4} we show the dependence on the neutrino mass ordering on two examples. The main difference is the resonant nature of the first electron-driven MSW transition at $T_\mathrm{MSW}^{(e),1} \simeq 5\,\mathrm{MeV}$ in IO. It leads to a much faster evolution, a feature also observed in Figure 1 of~\cite{Mangano:2011ip}. On the examples of Figure~\ref{figcomp4}, we see two consequences of this resonant transition: the ordering of degeneracy parameters can be modified, and collisions are much more efficient in damping the synchronous oscillations right after the transition.

\subsection{Dependence on mixing angles}\label{sec:mixing_angles}

In the two-neutrino case we have shown that $\vecHvac$ sets the precession direction of $\vec{\Anti}$; in other words, the values of the mixing angles are key parameters to determine the final asymmetry differences. Even though they are now better and better constrained~\cite{PDG}, let us explore qualitatively in this section their influence on the equilibration process.

\begin{figure}[!ht]
	\centering
	\includegraphics[]{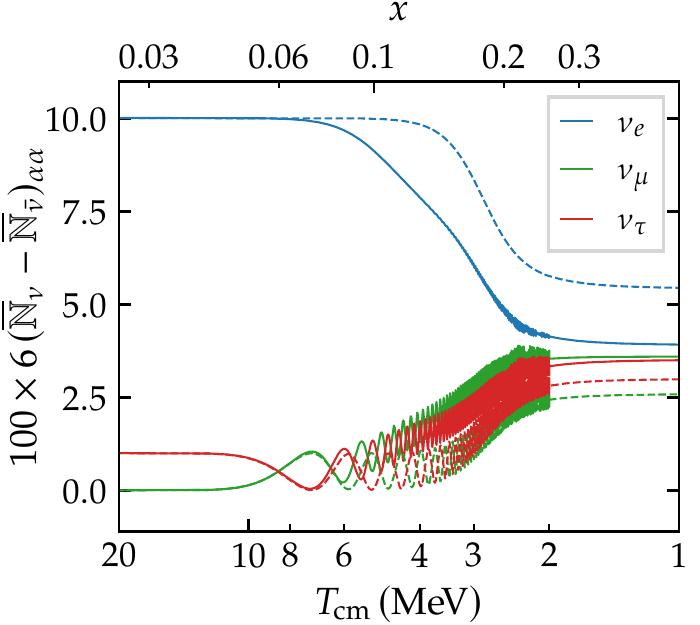}
     \includegraphics[]{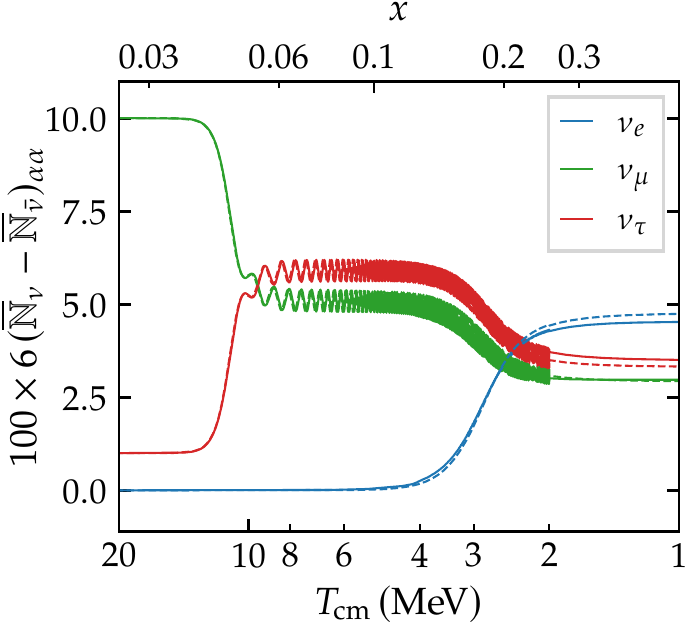}\\ \vspace{0.2cm}
     \includegraphics[]{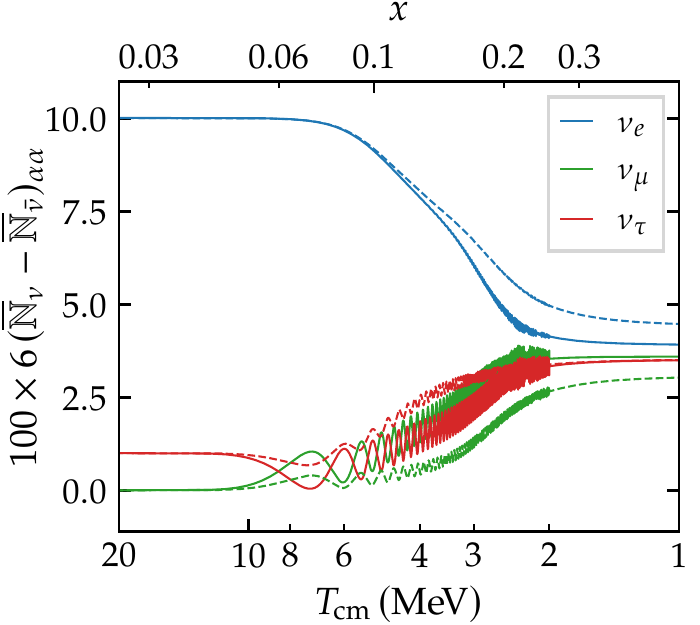}
     \includegraphics[]{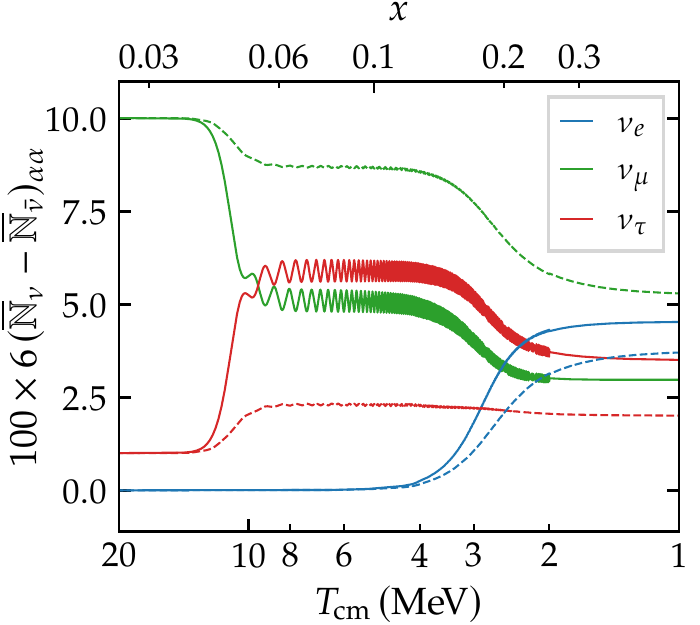}
	\caption[Evolution of asymmetries with different mixing angles]{\label{figcompamgles} Comparison of the standard case (solid lines) with a modified setup (in dashed lines). In the upper plots, the modification is  $\theta_{13}=0$, and in the lower plots $\theta_{12}=\theta_{23}=\pi/8$, with everything else unchanged. Initial conditions are $\xi_\alpha=(0.1,0,0.01)$ on the left, and $\xi_\alpha=(0,0.1,0.01)$ on the right. At low temperatures, the green lines overlap on the top right subplot, and similarly for the red lines on the bottom left subplot.}
      \end{figure}

To that purpose, we compare in Figure~\ref{figcompamgles} (upper plots) the standard case discussed above with modified setups. First, when $\theta_{13}=0$ (the rest being unchanged), we notice that equilibration is less efficient. Furthermore setting $\theta_{12} = \theta_{23} = \pi/8$, with $\theta_{13}$ at its standard value, the equilibration is also much less efficient as depicted in Figure~\ref{figcompamgles} (lower plots). Therefore, the general result that asymmetries mostly tend to equilibrate crucially depends on the values of the mixing angles. Typically, for small values of the mixing angles $\theta_{12}$ and $\theta_{23}$, that is far away from $\pi/4$, equilibration is less efficient, and a non-vanishing value for $\theta_{13}$ also significantly helps the equilibration process as highlighted in~\cite{Dolgov_NuPhB2002,Mangano:2010ei,Mangano:2011ip}. Also, using $\theta_{23} = \pi/8$ instead of the larger standard value~\eqref{ValuesStandard} increases the geometric factor $\cos^2 \theta/\sin \theta$ of the adiabatic parameter~\eqref{Defgammatr2} by a factor $\simeq 3.6$. Therefore, the muon-driven transition is much more adiabatic and the resulting oscillations are suppressed (see the discussion in section~\ref{SecDescriptionMuonTransition}), as can be checked on the bottom plots of Figure~\ref{figcompamgles}.

Evidently, even though we do not report it here, when all mixing angles vanish, equilibration entirely disappears. This highlights the importance of a consistent treatment of neutrino mixing when studying flavour equilibration in the early Universe. 

\subsection{Dependence on the Dirac phase}\label{SecDiracPhase}

We now examine the effect of the Dirac CP-violating phase $\delta$, that we discarded up until now in the PMNS matrix. Unless stated otherwise, all quantities in this section are considered in the case $\delta \neq 0$. $\vrho^{(\delta=0)}$ and $\bvrho^{(\delta=0)}$ refer to the solutions with vanishing Dirac phase, and we shall show how the general case with a non-zero phase can be deduced from it. It has been shown in \cite{Balantekin:2007es,Gava:2008rp,Gava:2010kz,Gava_corr} (see also section~\ref{subsec:Decoupling_CP}) that the evolution with a non-vanishing Dirac phase can be obtained from a transformation of the result obtained with a vanishing phase. More precisely, defining
$\check{S} \equiv R_{23} S R_{23}^\dagger$ (cf.~notations in section~\ref{subsec:Values_Mixing}), we can define
\begin{equation}\label{CheckAll}
\check{\mathcal{H}}_\mathrm{lep} = \check{S}^\dagger \mathcal{H}_\mathrm{lep} \check{S} \, , \qquad \check{\mathcal{H}}_0 = \check{S}^\dagger \mathcal{H}_0 \check{S} \, , \qquad \check{\vrho} \equiv \check{S}^\dagger \vrho \check{S} \, , \qquad \check{\bvrho} \equiv \check{S}^\dagger \bvrho \check{S}\,,
\end{equation}
and similar transformations for the collision terms. Since $\check{S}$ is of the type~\eqref{UtoU2} (see equation~\eqref{eq:Umat_SS} below), we infer from the property~\eqref{MagicKFundamental} that 
\begin{equation}\label{MagiccheckK}
\check{\mathcal{K}}(\vrho,\bvrho) = \mathcal{K}(\check{\vrho},\check{\bvrho}) \quad \text{and} \quad \check{\overline{\mathcal{K}}}(\vrho,\bvrho) = \overline{\mathcal{K}}(\check{\vrho},\check{\bvrho}) \, . 
\end{equation}
Furthermore, given that 
\begin{equation}\label{MagicUU}
\check{S}^\dagger U = U^{(\delta=0)}S^\dagger\,,
\end{equation}
and $[\mathbb{M}^2,S]=0$, we deduce that 
\begin{equation}
\label{eq:equalityofH0}
\check{\mathcal{H}}_0 = \mathcal{H}_0^{(\delta=0)} \, .
\end{equation}
Therefore, the evolution of $\check{\vrho}$ (resp. of $\check{\bvrho}$) is the same as the evolution of $\vrho^{(\delta=0)}$ (resp. of $\bvrho^{(\delta=0)}$) when the replacements $\mathcal{H}_\mathrm{lep} \to \check{\mathcal{H}}_\mathrm{lep}$ and $\Hself \to \check{\Hself}$ have been performed, that is 
\begin{equation}
\frac{\partial \check{\vrho} }{\partial x} = - \ii [\Hvac^{(\delta=0)}+\check{\mathcal{H}}_\mathrm{lep} +
\check{\Hself},\check{\vrho}] + \mathcal{K}(\check{\vrho},\check{\bvrho})\,,\qquad \frac{\partial \check{\bvrho} }{\partial x} = + \ii [\Hvac^{(\delta=0)}+\check{\mathcal{H}}_\mathrm{lep} -
\check{\Hself},\check{\bvrho}] + \overline{\mathcal{K}}(\check{\vrho},\check{\bvrho})\,. 
\end{equation}
In the standard case, that is with vanishing initial chemical potentials, $\check{\vrho}$ and $\vrho$ have the same initial conditions. If we further neglect the mean-field effects of muons/antimuons, then $[\check{S},\Hlep]=0$, hence $\check{\mathcal{H}}_\mathrm{lep} = \Hlep$ and $\check{\vrho}=\vrho^{(\delta=0)}$ (likewise for antineutrinos) at all times, as shown in \cite{Gava:2010kz,Gava_corr} and section~\ref{subsec:Decoupling_CP}. From this property, we obtain $\vrho$ from $\vrho^{(\delta=0)}$ using the inverse transformation, that is we get $\vrho = \check{S}\vrho^{(\delta =0)} \check{S}^\dagger$, with a similar relation for antineutrinos. It is equivalent to saying that both results are
exactly equal in their respective mass basis.

\begin{figure}[!ht]
	\centering
	\includegraphics[]{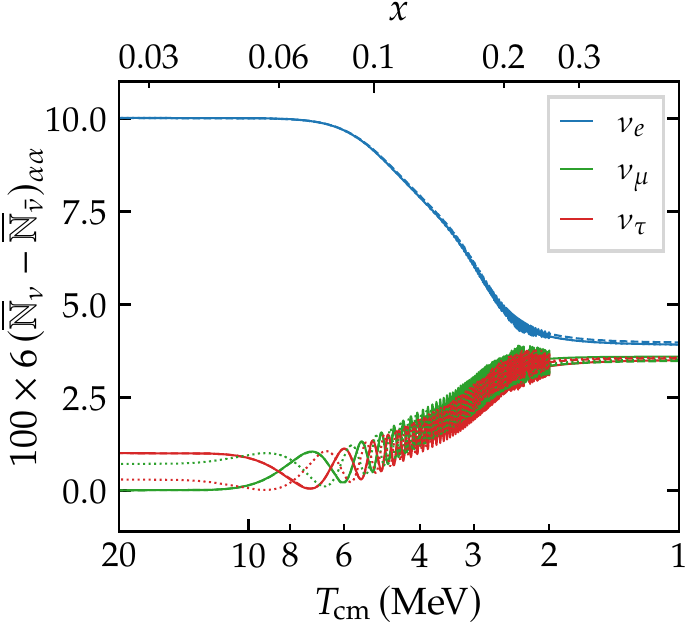}
	\includegraphics[]{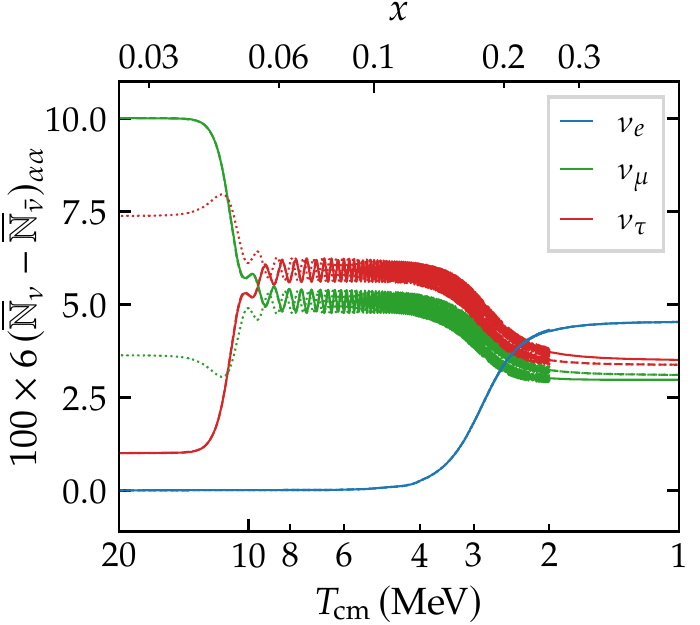}
	\caption[Effect of $\delta \neq 0$ on the evolution of asymmetries]{\label{figCP1} Effect of $\delta \neq 0$ on the evolution of asymmetries. The solid lines correspond to the standard case with $\delta=0$ (that is $\vrho^{(\delta=0)}$ and $\bvrho^{(\delta=0)}$). The dashed lines are the case with $\delta =245^\circ$ (central value in the most recent constraints~\cite{PDG}). The dotted lines correspond to the standard case results on which the transformation $\check{S}\vrho^{(\delta=0)} \check{S}^\dagger$ (and similarly for antineutrinos) has been applied. Initial conditions are $\xi_\alpha=(0.1,0,0.01)$ on the left and $\xi_\alpha=(0,0.1,0.01)$ on the right. Initially, the dashed lines are hidden behind the solid ones (see text). In the final stages the dotted lines are hidden behind the dashed lines hence showing that asymptotically $\vrho\simeq \check{S} \vrho^{(\delta=0)} \check{S}^\dagger$.
	}
\end{figure}

However in the presence of initial degeneracies, the initial conditions for $\vrho$ are not necessarily equal to those of $\check{\vrho}$. It is interesting to note that going from $\vrho$ to $\check{\vrho}$
amounts to a rotation in the vector description of the $\nu_\mu-\nu_\tau$
subspace. Indeed, both $S$  and $\check{S}$ are of the \eqref{UtoU2} type, and the associated $\mathcal{U}_S$ and $\mathcal{U}_{\check{S}}$ are expressed in terms of rotations as
\begin{equation}
\label{eq:Umat_SS}
\mathcal{U}_S = \mathrm{e}^{\ii \delta/2} \mathcal{R}_\z(\delta) \ , \qquad
\mathcal{U}_{\check{S}} = \mathrm{e}^{\ii \delta/2}
\mathcal{R}_\y (-2 \theta_{23}) \cdot
\mathcal{R}_\z(\delta) \cdot \mathcal{R}_\y^\dagger(-2 \theta_{23})\,.
\end{equation}
Forgetting the global $\mathrm{e}^{\ii \delta/2}$ factor which plays no role, we can use the property~\eqref{eq:SU2SO3} to interpret $\mathcal{U}_{\check{S}} $ as a rotation, when using the vector representation~\eqref{MatrixToVector}. It corresponds to a rotation of angle $\delta$ around an axis whose direction is obtained from the rotation $\mathcal{R}_\y (-2 \theta_{23})$ of the $\z$-axis. However, using that $\theta_{23}^\mathrm{eff} \simeq \theta_{23}$ from \eqref{th23th23}, this axis is approximately the one subtended by the $\nu_\mu-\nu_\tau$ restriction of the vacuum Hamiltonian. This has interesting consequences.
 \begin{itemize}
 \item Before the electron-driven transitions, using~\eqref{CheckAll} with $U$ of the effective form~\eqref{UtoU}, and~\eqref{eq:Umat_SS} taking $\theta_{23}^\mathrm{eff} \simeq \theta_{23}$, we infer the approximate relations
  \begin{equation}
  \left(\check{\mathcal{H}}_0\right)_{\alpha\beta} \simeq \left(\mathcal{H}_0\right)_{\alpha\beta} \quad  \xRightarrow[\text{using \eqref{eq:equalityofH0}}]{}  \quad \left(\mathcal{H}_0\right)_{\alpha\beta}  \simeq \left(\mathcal{H}^{(\delta=0)}_0\right)_{\alpha\beta}\quad \text{for}\quad \alpha,\beta \in\{\mu,\tau \}\,.
  \end{equation}
  Therefore at early times the evolution of $\vrho$ and $\vrho^{(\delta=0)}$ are nearly completely similar (likewise for antineutrinos) as can be checked by comparing the solid and dashed lines of Figure~\ref{figCP1} at high temperatures.
 \item Since the asymmetry vector precesses around that precise direction after the muon-driven transition, this amounts to the fact that after the muon-driven MSW transition and before the electron-driven ones, the oscillations of $\check{\vrho}$ (similarly for $\check{\bvrho}$) are simply phase shifted with respect to the ones of $\vrho^{(\delta=0)}$  (resp. $\bvrho^{(\delta=0)}$), by an angle $\delta$, as can be checked by comparing the dashed and dotted lines on Figure~\ref{figCP1} (see also the left plot of Figure~\ref{figComparison}).
\end{itemize}

Later, when the electron-driven transitions occur, the $\delta$-phase difference between $\check{\vrho}$ and $\vrho^{(\delta=0)}$ (likewise for antineutrinos) can only have an extremely marginal
effect because the oscillation frequency keeps increasing, and the amount of damping incurred in the magnitude of (the traceless part of) $\Anti$ is essentially only sensitive to the amplitude and axes of oscillations. Oscillations keep accelerating until synchronous oscillations disappear as we reach an average set by $\mathcal{H}_0$, which is captured by the \ATAOH scheme. Eventually the initial dephasing is lost, and it is impossible to distinguish between the final values of $\check{\vrho}$ and $\vrho^{(\delta=0)}$ (likewise for antineutrinos), therefore we can relate the final results to the case without Dirac phase by
\begin{equation}\label{FinalOperation}
\vrho \simeq \check{S}\vrho^{(\delta =0)} \check{S}^\dagger\,,\qquad \bvrho \simeq \check{S}\bvrho^{(\delta =0)} \check{S}^\dagger\,.
\end{equation}
There are two differences with the standard case without initial chemical potentials. On the one hand, \eqref{FinalOperation} is an approximate result and not an equality, based on the following approximations:
\begin{enumerate}
\item $\theta_{23}^\mathrm{eff} \simeq \theta_{23}$, which is guaranteed from equation~\eqref{th23th23} by $|\Delta m_{21}^2/\Delta m_{31}^2| \ll 1$ ;
\item the muon-driven MSW transition takes place well before the first electron-driven transition ($T_\mathrm{MSW}^{(\mu)} \gg T_\mathrm{MSW}^{(e),1}$), which is the case since $\me/m_\mu \ll 1$ ;
\item the amplitude and directions of oscillations are not meaningfully affected by the $\delta$-dephasing, and eventually this dephasing should not be observable as oscillations are averaged at large $x$.
\end{enumerate}
On the other hand, it is only valid at late times, whereas in the standard case it is valid at all times. It can be checked on Figure~\ref{figCP1} that it is very accurate, as the dashed lines ($\vrho$) and dotted lines ($\check{S}\vrho^{(\delta =0)} \check{S}^\dagger$) are nearly indistinguishable at late times, and are only $\delta$-dephased at early times if oscillations develop.

\noindent Finally the property \eqref{FinalOperation} allows to understand the physical effects of the Dirac phase. 
\begin{itemize}
\item When converted into mass basis components, the property~\eqref{MagicUU} and the relation~\eqref{FinalOperation} imply the (approximate) relation $\widetilde{\vrho} \simeq S \widetilde{\vrho}^{(\delta=0)} S^\dagger$. This means that in the \ATAOH approximation which holds at late times, we have $\widetilde \vrho = \widetilde \vrho^{(\delta=0)}$ and likewise for antineutrinos given that off-diagonal components in the matter basis vanish --- see section~\ref{subsec:Decoupling_CP}. Hence the difference in the final state, when interpreted in the flavour basis, is (approximately) only due to the different mass bases depending on the value of $\delta$. Structure formation, being sensitive to mass bases spectra, is therefore not affected by the Dirac phase. In addition, the trace of density matrices is conserved by~\eqref{FinalOperation} such that $N_\mathrm{eff}$ is preserved, and cosmological expansion is not modified. We conclude that there is no sizeable gravitational signature of the Dirac phase.
\item Since $\check{S}$ is of the~\eqref{UtoU2} type, the transformation~\eqref{FinalOperation} affects the number densities of $\nu_\mu$ and $\nu_\tau$, that is $\vrho^{\mu}_{\mu}$ and $\vrho^{\tau}_{\tau}$, but the number density for $\nu_{e}$, that is $\vrho^{e}_{e}$, is left invariant (likewise for antineutrinos). Only the coherence between the $e$ states and the $\mu$ and $\tau$ states, that is the $\vrho^{e}_{\mu}$ and $\vrho^{e}_{\tau}$ components, is affected. Therefore, there is also no perceptible effect on BBN because the neutron/proton freeze-out is sensitive only to the spectrum of electronic (anti)neutrinos.
\end{itemize}

\subsection{Equal but opposite asymmetries}

\begin{figure}[!ht]
\centering
\includegraphics[]{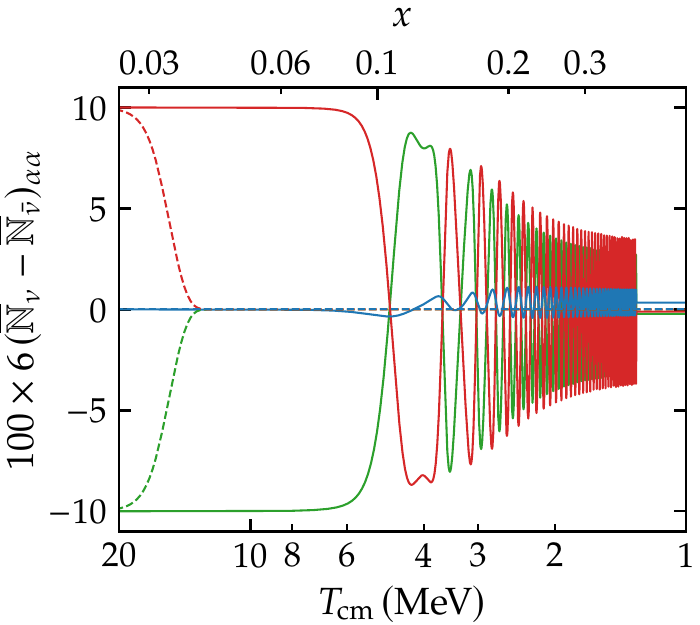}
\includegraphics[]{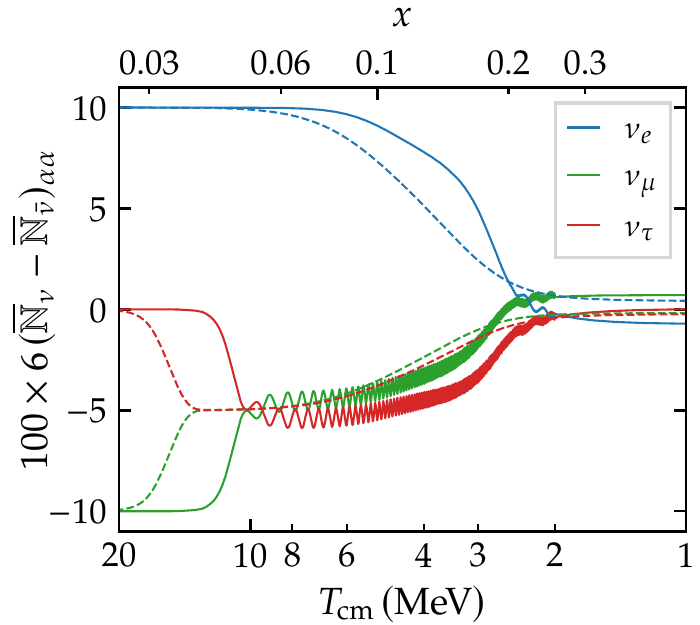}
	\caption[Special case of equal but opposite non-vanishing degeneracy parameters]{\label{figEBO} Special case of equal but opposite non-vanishing degeneracy parameters. The initial conditions on the left are $\xi_\alpha=(0,-0.1,0.1)$, and $\xi_\alpha=(0.1,-0.1,0)$ on the right. \emph{Numerical schemes:} \ATAOJH in solid lines, \ATAOH in dashed lines. On the left plot, we extended the \ATAOJH integration until $1.3 \, \mathrm{MeV}$ to see further the damping of the oscillations.}
\end{figure}

As noted in section~\ref{FreqSyncOsc}, the case of equal but opposite asymmetries is special because the leading order of synchronous oscillations vanishes --- meaning that the asymmetry should remain locked in its original configuration, --- but not the next-to-leading order. The results obtained in three-neutrino cases are depicted in Figure~\ref{figEBO}. When $\xi_\mu+\xi_\tau=0$, we expect from the analysis of section~\ref{SecParticular} that oscillations in the $\nu_\mu-\nu_\tau$ space should start at $1.4 \,\mathrm{MeV}$, while we observe the first half-oscillation at $4\,\mathrm{MeV}$. These oscillations must therefore be triggered by the first electron-driven transition.

One can compare the left plot of Figure~\ref{figEBO} with Figure~9 of~\cite{Dolgov_NuPhB2002}: the damping of synchronous oscillations is considerably reduced in our calculation, which we attribute to the use of the exact collision term, instead of a damping approximation (see also the discussion in section~\ref{SecDiscussion}).

If the opposite initial degeneracy parameters are the electronic and muonic (or tauic) ones, there is no substantial difference with the “standard” case (Figure~\ref{figStandard}). Indeed, the $\nu_\mu-\nu_\tau$ oscillations equilibrate partially (at least in the case of large $\xi_\tau$ taken in the Figure) the asymmetries in the $\nu_\mu - \nu_\tau$ subspace, and the common asymmetry is then not the opposite of $\xi_e$.

\section{Discussion}\label{SecDiscussion}

\subsection{Transition from \ATAOJH to \ATAOH}\label{SecTransitionToATAOV}

We have shown that the \ATAOJH scheme works very well as long as $|\Hamil| \ll |\Hself|$. Let us discuss here in more details the end of the \ATAOJH regime, in a simplified case with only two flavours so as to use the vector formalism. When $\lvert \vec{\Hamil} \rvert$ and $\lvert \vec{\Hself} \rvert$ are of the same order, quasi-synchronous oscillations cease to exist. Therefore, in principle, only the QKE scheme can handle this regime. However, the individual $\vec{\vrho}(y)$ then evolve independently rather than collectively, so that extremely rapid precessions around $\vec{\Hamil}$ should take place for all momenta, and given the $y$-dependence of $\vec{\Hamil}$, they tend to have different frequencies. 
Given that $|\vec{\Hamil}| \propto  (\Delta m^2/(2 y \me H) \simeq (\Delta m^2/10^{-3}\mathrm{eV}^2)\times 8 \cdot 10^6 \times x^2/y$, the oscillating part in the spectrum is typically a trigonometric function whose phase $\phi \propto x^3/y$. This implies extremely fast oscillations in the spectrum (i.e. in the variable $y$) whenever $x \gg 1$, that is at cosmological times. It is expected that even a mild collision term can average them out. Even if this is not the case, we only aim at describing the average of this incredibly fast oscillating spectrum, since this is the only part that will survive any measurement or physical process. This means that after a transitory regime, the \ATAOH scheme must become a good approximation. Sadly, given the $\mathcal{O}(N^3)$ complexity for computing the collision term, it becomes numerically impossible to integrate this transitory regime. Furthermore the $y$-grid becomes necessarily too sparse to account for these spectral oscillations. Hence one must rely on a certain approximation to handle the transition from a period where the \ATAOJH scheme applies to a regime where \ATAOH is sufficient.

We chose to push the \ATAOJH scheme as far as possible, typically down to $2\,\mathrm{MeV}$ and then to switch immediately to a \ATAOH scheme. In doing so we necessarily miss some features of the transitory regime. If $\vec{\Anti}$ is already well aligned with $\vecHeff$, it is nonetheless expected to be a very good approximation. Hence it misses some physics essentially due to the oscillations in the $\nu_\mu-\nu_\tau$ space which have developed right after the muon-driven transition, and for which we have seen that the collision term is not very effective in dampening these oscillations towards $\vecHeff$. We can however estimate the nature of the error made. If we focus on the two-neutrino case describing the $\nu_\mu-\nu_\tau$ space after the muon-driven transition, the density matrices are in the \ATAOH scheme in the form given by \eqref{eq:vrho_ATAOJV}. If we neglect terms of order $\mathcal{O}(\xi^2)$, then $\abs{g(-\xi_1,y)- g(-\xi_2,y)}  \simeq \abs{g(\xi_1,y)- g(\xi_2,y)} $, that is we have essentially $\vec{{\vrho}} \propto \widehat{\Hamil+\Hself}$ and $\vec{{\bvrho}}  \propto \widehat{\Hamil-\Hself} $ with the same prefactor. When the ratio $|\Hamil|/|\Hself|$ grows this tends to displace both $\vec{{\vrho}}$ and $\vec{{\bvrho}}$ in the same direction, namely, the projection of $\vec{\Hamil}$ in a plane orthogonal to $\vec{\Hself}$ (in agreement with the next-to-leading order term of the expansion \eqref{ExpandSH} of $\widehat{\Hself}$). The net result is that around the end of validity of the \ATAOJH regime, neutrinos and antineutrinos of one flavour are converted to neutrinos and antineutrinos of the other flavour, but in similar proportions for neutrinos and antineutrinos, hence preserving the asymmetry. Of course, since the ratio $|\Hamil| / |\Hself|$ reaches unity earlier for small $y$, it is expected that this concerns more the small momenta $y$. Also, the more  $\vec{\Hamil}$ and $\vec{\Hself}$ are misaligned in the end of the \ATAOJH regime, the more this phenomenon takes place. This is nicely seen in Figure~9 of \cite{Johns:2016enc}, a configuration solved numerically without collisions, where we observe that both the number of neutrinos and antineutrinos of a given flavour increase while the opposite takes place for the other flavour. By construction, our approach based on an instantaneous switching from \ATAOJH to \ATAOH cannot capture this phenomenon. 
Finding a method to handle this transitory regime when including the collision term is an upcoming numerical challenge for the computation of equilibration in the early universe.

\subsection{Comparison with the literature}

Our numerical results differ from the literature in several aspects, which we review here.

\paragraph{Collision term} It is clear that the muon-driven oscillations are less damped in our case, compared to the results for instance reported in \cite{Dolgov_NuPhB2002}. We explain this by the fact that the general collision term satisfies the “factorization” property \eqref{MagicKFundamental}, and thus \eqref{MagicKUJ} in the $\nu_\mu-\nu_\tau$ subspace.
This implies that oscillations developing in this subspace are only very mildly damped, as detailed in section~\ref{SecCollisions}. When relying on an approximate collision term, this property is lost. For computations where all entries of the collision term are based on a damping approximation, it is the repopulation term which fails to satisfy the property \eqref{MagicKFundamental}. On the other hand, for computations which use the full collision term for on-diagonal components, but a damping approximation for off-diagonal components (as is the case in \cite{Dolgov_NuPhB2002}), the property is lost precisely because not all components are computed in the same method and this introduces preferred directions in the collision term. In all cases, using an approximation for the collision term results in much more damping of the oscillations in the $\nu_\mu-\nu_\tau$ space compared to our results. We already mentioned the case where $\xi_\mu=-\xi_\tau$  of \cite{Dolgov_NuPhB2002} (Figure 9), compared to our results in Figure~\ref{figEBO}. One of the consequences is that it is not possible to consider that by $10\,\mathrm{MeV}$ we would have generically achieved $\xi_\mu=\xi_\tau$, as is assumed for instance in \cite{Pastor:2008ti,Mangano:2010ei,Castorina2012}.

\begin{figure}[!h]
	\centering
	\includegraphics[]{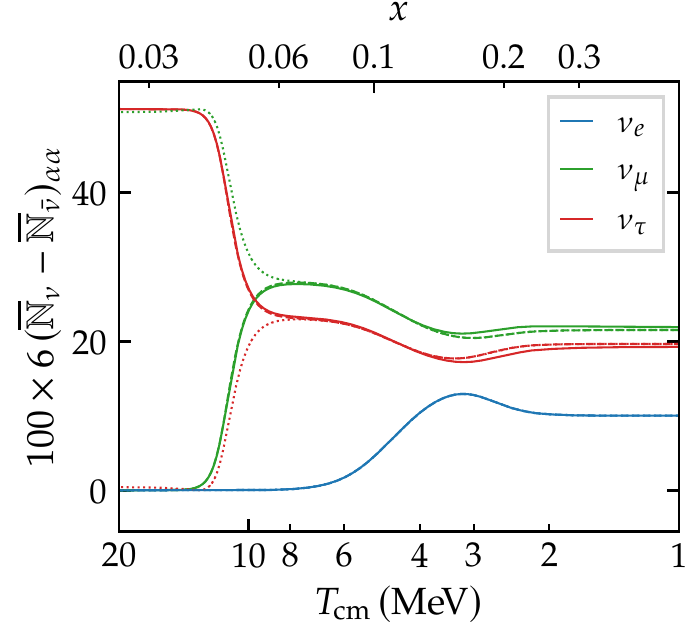}
	\includegraphics[]{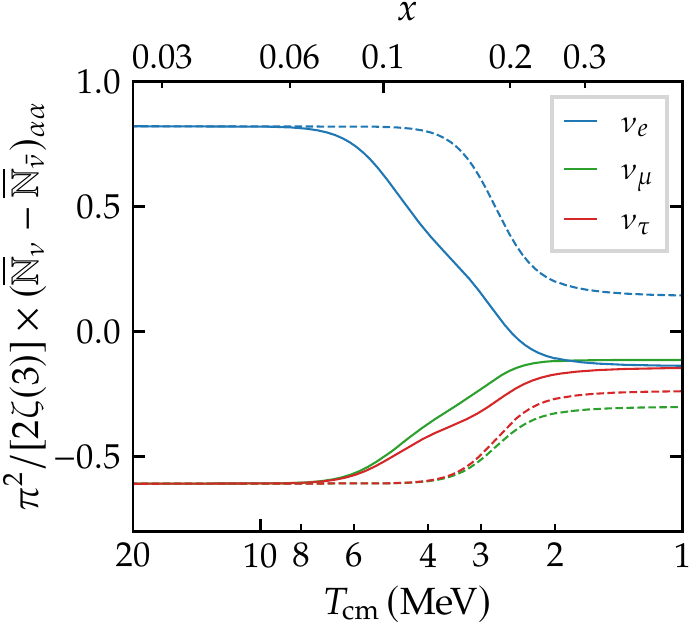}
	\caption[Comparison with previous results from the literature]{\label{figComparison} \emph{Left plot:} $\xi_\alpha=(0,0,0.5)$, with $\delta=0$ in solid line and $\delta=\pi$ in dashed line. Dotted lines correspond to the case $\delta=0$ on which the transformation $\check{S}\vrho^{(\delta=0)} \check{S}^\dagger$ has been applied, and are hidden behind dashed lines at late times. It can be compared with Figure~3  of \cite{Gava:2010kz}. \emph{Right plot:} $\xi_\alpha = (1.0732,-0.833,-0.833)$, with $\theta_{13}=0.20$ in solid line and $\theta_{13}=0$ in dashed line. It corresponds to Figure~1 of \cite{Mangano:2010ei}, noting that the initial conditions are $\eta_\mu=\eta_\nu = -0.61$ instead of the stated values $\eta_\mu=\eta_\nu = -0.41$.}
\end{figure}

\paragraph{Large mixing angle} Note that in references \cite{Dolgov_NuPhB2002,Gava:2010kz}, the large mixing angle value $\theta_{23}=\pi/4$ was used, which adds extra properties. Since $\cos(2 \theta_{23})=0$ we cannot use~\eqref{th23th23} to estimate $\theta_{23}^\mathrm{eff}$ after the muon-driven transition. Fortunately, in that case we get exactly 
\begin{equation}
\tan(2 \theta_{23}^\mathrm{eff}) =\frac{(\epsilon^{-1} + 1 )\cos^2 (\theta_{13}) +\sin^2 \theta_{12} \sin^2 \theta_{13} - \cos^2 \theta_{12}}{\sin(\theta_{13}) \sin(2 \theta_{12})} \,,
\end{equation}
hence we also find from $\epsilon = \Delta m_{21}^2  / \Delta m_{32}^2 \ll 1$ that $\theta_{23}^\mathrm{eff} \simeq \pi/4 $, and the vector representation of the vacuum Hamiltonian restricted to the $\nu_\mu-\nu_\tau$ space is approximately along the $\x$-axis. Therefore the precession direction corresponds to a state where the asymmetry is the same for $\nu_\mu$ and $\nu_\tau$. In that case, and given the extra damping incurred by the approximations in the collision term, it is expected that the equilibration of the degeneracy parameters $\xi_\mu$ and $\xi_\tau$ is very efficient right after the muon-driven transition. This is seen for instance in Figures (7-10) of \cite{Dolgov_NuPhB2002}, or Figure 3 of \cite{Gava:2010kz} whereas in its counterpart here (the left plot of \ref{figComparison}), $\xi_\mu \neq \xi_\tau$ after the muon-driven transition. Also this special choice of mixing angle explains why $\xi_\mu$ and $\xi_\tau$ remain equal on Figure 1 of \cite{Mangano:2010ei} or \cite{Mangano:2011ip}, whereas in the right plot of Figure~\ref{figComparison} they differ once the electron-driven MSW transitions are crossed, which results in a full equilibration being never achieved. Note that we find nonetheless the same influence of $\theta_{13}$, as discussed in section~\ref{sec:mixing_angles}.

\paragraph{CP phase} Furthermore our results about the effect of the Dirac phase differ with respect to \cite{Gava:2010kz,Gava_corr}. Although we confirm that the effect of the Dirac phase must be maximal when $\delta=\pi$ (strictly speaking there is no CP-violation in that specific case, as it is equivalent to $\delta=0$ and a change of sign in $\theta_{13}$), we find that it must necessarily be negligibly small given the structure of the equations (see section \ref{SecDiracPhase}) whereas it is found small but not negligible in \cite{Gava:2010kz}. To be specific, in Figure 3 of \cite{Gava:2010kz} there is a small effect of the Dirac phase, whereas in the left plot of Figure \ref{figComparison} no perceptible effect is found. Again these differences must find their origin in the differences for the treatment of the collision term, given that the property~\eqref{MagiccheckK} is not satisfied by an approximate repopulation term.

\section*{Concluding remarks}

The complexity of the physics of neutrino evolution in the early Universe considerably increases when including initial degeneracies, a problem studied analytically and numerically in the last two decades. The \ATAOH scheme, presented in chapter~\ref{chap:Asymmetry}, which relied on the adiabaticity of the evolution of the Hamiltonian governing the dynamics of $\vrho, \bvrho$ and the very fast scale of oscillations, was extended to the \ATAOJH scheme to account for non-vanishing chemical potentials.

A restriction to two-flavour systems showed the excellent accuracy of this method compared to a much longer QKE resolution, which we found to be at least ten times slower and even more when dealing with low temperatures. Even though our code can perform this “exact” QKE resolution, it is thus sufficient, notably if one wants to explore a wide range of parameters, to rely on the \ATAOJH scheme. Thanks to the \ATAOJH approximation, we recover the famous synchronous oscillations, but also predict and understand new results such as the existence of a phase of quasi-synchronous oscillations, that is an increased frequency regime ($\Omega(x) \propto x^2 \to x^6$) when the vacuum + mean-field Hamiltonian $\Hamil$ contribution becomes substantial compared to $\Hself$. The (non-)adiabaticity of the evolution of $\Anti$ during the lepton-driven transitions --- which depends on the degeneracies via the slowness factor~\eqref{eq:slowness} --- also allows to understand its qualitative behaviour, namely the (non-)efficiency of its alignment towards the vacuum Hamiltonian. In addition to their frequency, we can thus also estimate the beginning and the amplitude of the synchronous oscillations which develop afterwards.

We have shown that it is crucial to rely on the exact form of the collision term to fully take into account the physics of these oscillations --- approximate expressions previously overdamped degeneracy differences and led to a too rapid flavour equilibration. The $\mathcal{O}(N^3)$ complexity of the full collision term is the price to pay. Therefore, we argue that it is crucial to rely on a direct computation of the Jacobian to avoid worsening the problem. The method developed in the zero degeneracy case is extended to the situation with initial chemical potentials. Although it requires many more steps to implement it, as summarized in appendix~\ref{App:Numerics}, it keeps the appealing $\mathcal{O}(N^3)$ complexity.

The \ATAOJH scheme fails when the vacuum potential is of the order of the self-interaction potential. In principle, the transition from the \ATAOJH to the \ATAOH regime should be solved numerically with the full QKE method, but in practice this is numerically daunting. Nevertheless, by switching directly at sufficiently low temperature from \ATAOJH to \ATAOH, we argued that errors incurred on the neutrino asymmetry should be minimized. 

In the standard case with three neutrinos, we have highlighted the influence of the various mixing parameters. Notably, the Dirac phase is found to have no perceptible effect as it essentially changes only the phase of synchronous oscillations, and its residual effect is accurately captured by the transformation~\eqref{FinalOperation}. In general, degeneracies tend to equilibrate partially, and this is due to the fact that the mixing angles $\theta_{12}$ and $\theta_{23}$ are not so different from maximal mixing ($\pi/4$), but this statement strongly depends on the non-vanishing of $\theta_{13}$. Given the complexity of the physics involved during the evolution of density matrices, the degree of equilibration depends non-trivially on the values of the initial degeneracies (e.g. equilibration is far from being achieved in the left plot of Figure~\ref{figComparison}), and requires further systematic study. 


\pagestyle{ruled}

\chapter*{Conclusion}
\addcontentsline{toc}{chapter}{\protect\numberline{}Conclusion}

\setlength{\epigraphwidth}{0.48\textwidth}
\epigraph{Mozart, Beethoven and Chopin never died. They simply became music.}{Dr. Robert Ford, \emph{Westworld} [S01E10]}

Cosmology has entered exciting times, with the launch of terrestrial and spatial telescopes that will push always further our understanding of the Universe, and maybe unveil in the coming decade some of the enduring mysteries that plague the $\Lambda$CDM model: the nature of dark matter and dark energy, the $H_0$ tension, etc. In this era of “precision cosmology”, the physics of neutrinos is absolutely essential. Indeed, neutrinos intervene at all stages of cosmological expansion: relativistic in the early Universe, they decouple from the plasma of photons, electrons and positrons precisely when primordial nucleosynthesis begins --- the initial conditions of BBN being dependent on the neutrino distributions! In the late Universe, massive neutrinos can become non-relativistic and affect the formation of large-scale structure. Moreover, the neutrino sector is an immense room for new physics: sterile states, non-standard interactions, ...

The aim of this PhD was to achieve new levels of precision in the study of neutrino evolution in the early Universe, assuming no beyond-the-Standard-Model physics --- except for the crucial phenomenon of flavour oscillations.

We developed a new method to derive the QKE which drives neutrino evolution, namely an \emph{extended-BBGKY hierarchy}: the perturbative expansion of~\cite{SiglRaffelt} is replaced by a well-controlled hierarchy of (un)correlated contributions to the $1-$, $2-$, $\cdots$ $n-$body density matrix. This method had been used to derive the mean-field terms of the QKE in~\cite{Volpe_2013}, in the so-called Hartree-Fock approximation. We went beyond this approximation and included higher order correlations in the molecular chaos ansatz to obtain the collision term (that is all contributions from scattering and annihilations between $\nu$, $\bar{\nu}$ and with $e^-$, $e^+$) and thus the full QKE.

This allowed us to perform a calculation of neutrino decoupling with, for the first time, the \emph{full} collision term (and the aforementioned QED corrections). Thus, we set a new recommended value of the cosmological observable $\Neff$: $\Neff = 3.0440$ with a precision of a few $10^{-4}$. This precision is partly due to the experimental uncertainty on the physical parameters (notably the mixing angle $\theta_{12}$), but mostly the numerical variability depending on the settings of our algorithm. The previous calculations taking into account flavour oscillations, which led to the value $\Neff \simeq 3.045$~\cite{Relic2016_revisited}, did not consider the full collision term: its off-diagonal components were evaluated with a damping approximation. Including this term without any approximation is a real numerical challenge, in particular due to its stiffness and because it scales as $\mathcal{O}(N^3)$ with $N$ the size of the momentum grid. We ensured a reasonable computation time through a major improvement, namely the direct calculation of the Jacobian of the differential system. Our result on $\Neff$ was later confirmed by~\cite{Bennett2021}.

We also introduced an effective description of flavour oscillations that gives results indistinguishable from the ones obtained solving the exact equation. It also substantially reduces the computation time, another key improvement of our code. This approximation relies on the existence of a large separation of scales between the oscillation frequencies and the collision rate, which allows to average over these oscillations. In other terms, the density matrix always remains diagonal in the matter basis (the basis of eigenstates of the Hamiltonian taking into account vacuum and mean-field effects). We named this simplified description the \emph{Adiabatic Transfer of Averaged Oscillations} (ATAO) approximation. Moreover, we used this approximation to get a deeper understanding of some results like the (absence of) effects of the CP phase in standard neutrino decoupling.

Solving the QKE, we obtain the frozen-out distributions of (anti)neutrinos, which in turn give access to parameters like $\Neff$ (cf.~above) or the neutrino energy density parameter today $\Omega_\nu$. Thus, we are in possession of the two parameters that set the various effects of incomplete neutrino decoupling on the earliest probe of the history of the Universe we dispose of --- BBN ---: the distribution of $\nu_e, \, \bar{\nu}_e$ and the energy density parametrized by $\Neff$.

We have also assessed the changes in the primordial abundances of helium, deuterium and lithium due to incomplete neutrino decoupling. First, the light element abundances depend on the expansion rate of the Universe (hence on $\Neff$, via the so-called \emph{clock effect}). Then, the neutron abundance at the beginning of BBN is among other things set by the neutron-to-proton ratio which varies if one changes the distributions of $\nu_e, \bar{\nu}_e$.  We have studied in detail how those effects interplayed, comparing their relative contributions and providing analytical estimates when possible. This theoretical work was conducted hand-in-hand with a numerical study, combining our neutrino evolution code and the BBN code \texttt{PRIMAT}. In particular, we were able to resolve an existing discrepancy in the literature between~\cite{Grohs2015} and~\cite{Mangano2005} regarding the variation of deuterium abundance due to incomplete neutrino decoupling.

The presence of non-zero neutrino asymmetries, \emph{a priori} allowed, adds considerable complexity to the physics of neutrino evolution. Indeed, there is now an additional self-interaction mean-field term in the QKE, which dominates throughout a large part of the neutrino decoupling era for asymmetries $\mu/T \in [10^{-3},10^{-1}]$. In line with our work on the resolution of the QKE in the standard case with the full collision term, we extended our code to the asymmetric case. Moreover, we generalized the ATAO approximation to account for self-interactions, which make the Hamiltonian non-linear. This ATAO framework allowed us to analytically recover known results about the collective \emph{synchronous oscillations}, but also to discover that this regime is generally followed by \emph{quasi}-synchronous oscillations with larger frequencies. We have provided numerous analytical and numerical checks of this new result, in the simplified two-flavour case but also in the general three-flavour framework. We further explored the dependency of the final neutrino configuration on the mixing parameters, and notably showed that the CP-violating Dirac phase cannot substantially affect the final $\Neff$ nor the final electronic (anti)neutrino spectrum, and thus should not affect cosmological observables.

\paragraph{Prospects} The coming years will be exciting on the theoretical and experimental levels~\cite{Snowmass_Abazajian}. Neutrino properties will be constrained by cosmology and laboratory searches, these complementary results allowing to build a complete picture of the neutrino sector. On the cosmological side, the PTOLEMY experiment proposal~\cite{Long_PTOLEMY,PTOLEMY2018} which aims at observing directly the C$\nu$B is particularly exciting, as it would provide the first direct observation of the physics of neutrino decoupling, instead of all the “secondary” ones (BBN, effects on CMB, etc.). However, there has been some recent debate on the possibility of using such a setup (based on the measure of the beta decay and absorption processes of tritium bound to graphene), as this binding would lead to fundamental quantum uncertainties on the spectrum of the emitted $e^-$, well above the required energy resolution to detect the C$\nu$B~\cite{Cheipesh:2021fmg,PTOLEMY2022}. On the theoretical side, the results obtained during this PhD pave the way to many more applications. First, in the early Universe, we can explore some new physics, at the edge of the SM or including new mechanisms. For example, it is very important to provide precise constraints on the effect of sterile neutrinos of given masses and mixtures on cosmological observables, as they are often proposed as solutions to anomalies in laboratory experiments. Second, our thorough description of the evolution of primordial asymmetries is only the first step towards establishing new refined limits on the chemical potentials of neutrinos. Finally, the tools we have developed are not \emph{a priori} limited to cosmology, and the study of astrophysical environments such as binary neutron star mergers or core-collapse supernovae, where anisotropies and collective behaviours provide very rich and complex physics, is a clear path forward for research.

\adjustmtc

\clearpage

\appendix

\pagestyle{ruled}

\chapter{Elements of neutrino physics}
\label{App:StandMod}

\setlength{\epigraphwidth}{0.45\textwidth}
\epigraph{The only thing I'm not good at is modesty. Because I'm great at it.}{Gina Linetti, \emph{Brooklyn Nine-Nine} [S05E17]}

{
\hypersetup{linkcolor=black}
    \minitoc
}

In this appendix, we summarize some useful results regarding the description of neutrinos and their interactions. We first discuss the Standard Model case, before giving the parameters that are commonly used to describe massive neutrino mixings.

\section{Neutrinos in the Standard Model of particle physics}

The Standard Model (SM) is a gauge theory based on the local symmetry group $\mathrm{SU(3)}_C \times \mathrm{SU(2)}_L \times \mathrm{U(1)}_Y$, where the subscripts $C$, $L$, $Y$ denote respectively colour, left-handed chirality and hypercharge. The interaction of neutrinos is determined by the electroweak part of the SM, based on the gauge group $\mathrm{SU(2)}_L \times \mathrm{U(1)}_Y$.

\subsection{Tools for the Clifford algebra} 

We refer the reader to, e.g.~\cite{PeskinSchroeder,SchwartzQFT,Srednicki} for a detailed introduction to quantum field theory. Here, we simply recall the useful relations involving the $\gamma$ matrices, which as a reminder are defined such that
\begin{equation}
\label{eq:Clifford}
\{ \gamma^\mu, \gamma^\nu \} = 2 \eta^{\mu \nu} \, ,
\end{equation}
where $\eta^{\mu \nu}$ is the Minkowski metric (here taken with signature (+---), for consistency with the FLRW metric).

The left and right chiral parts of a Dirac spinor $\psi = \psi_L + \psi_R$ are obtained from the projectors
\begin{equation}
P_L \equiv \frac{1-\gamma^5}{2} \qquad , \qquad P_R \equiv \frac{1+ \gamma^5}{2} \, ,
\end{equation}
with the fifth gamma-matrix $\gamma^5 \equiv \ii \gamma^0 \gamma^1 \gamma^2 \gamma^3$.

The useful identities read:
\begin{equation*}
\label{eq:trace_identities}
\boxed{ \begin{aligned}
\eta^{\mu \nu} \eta_{\mu \nu} &= 4 \\
\tr \left[ \gamma^\mu \gamma^\nu P_{L,R} \right] &= 2 \eta^{\mu \nu} \\
\tr\left[\gamma^\sigma  \gamma^\mu \gamma^\lambda \gamma^\nu P_{L,R} \right] &= 2 \left(\eta^{\sigma \mu} \eta^{\lambda \nu} - \eta^{\sigma \lambda} \eta^{\mu \nu} + \eta^{\sigma \nu} \eta^{\mu \lambda} \right) \pm 2 i \epsilon^{\sigma \mu \lambda \nu} \\
\epsilon^{\mu \nu \rho \sigma} \epsilon_{\mu \nu \tau \lambda} &= - 2 ({\delta^\rho}_\tau {\delta^\sigma}_\lambda - {\delta^\rho}_\lambda {\delta^\sigma}_\tau) \\
\gamma^\mu \gamma^\nu \gamma^\rho \gamma_\mu &= 4 \eta^{\nu \rho} \\
\gamma^\mu \gamma^\nu \gamma^\rho \gamma^\sigma \gamma_\mu &= - 2 \gamma^\sigma \gamma^\rho \gamma^\nu
\end{aligned}}
\end{equation*}

Note that any specific choice of matrices satisfying the fundamental anticommutation relations~\eqref{eq:Clifford} constitutes a \emph{representation} of the $\gamma$ matrices. Two are often used:
\begin{itemize}
	\item the Dirac basis, in which
	\[ \gamma^0_D = \begin{pmatrix} \Id & 0 \\ 0 & \Id \end{pmatrix} \, , \quad \gamma^i_D = \begin{pmatrix} 0 & \sigma^i \\ - \sigma^i & 0 \end{pmatrix} \, , \quad \gamma^5_D = \begin{pmatrix} 0 & \Id \\ \Id & 0 \end{pmatrix} \, , \]
	with $\sigma^i$ the Pauli matrices,
	\item the Weyl (chiral) basis, for which one choice is \cite{PeskinSchroeder}
		\[ \gamma^0_C = \begin{pmatrix}  0 & \Id \\  \Id & 0 \end{pmatrix} \, , \quad \gamma^i_C = \begin{pmatrix} 0 & \sigma^i \\ - \sigma^i & 0 \end{pmatrix} \, , \quad \gamma^5_C = \begin{pmatrix}  - \Id & 0 \\ 0 & \Id \end{pmatrix} \, . \]
		This is the choice consistent\footnote{By consistent, we mean “which makes clear”: we can always define $\psi_{L,R}$ based on the actions of the left and right projectors on $\psi$, but they do not separate into two Weyl spinors.} with the decomposition of the 4-component spinor field $\psi(x)$ into the left-handed and right-handed two-component Weyl spinors:
		\[ \psi_C = \begin{pmatrix} \psi_L \\ \psi_R \end{pmatrix} \, , \quad P_L \psi_C = \frac{1 - \gamma^5_C}{2} = \begin{pmatrix} \Id & 0 \\ 0 & 0 \end{pmatrix} \psi_C = \psi_L \, . \]
\end{itemize}
%
%
%
%
%
%

\subsection{Neutrino interactions and Fermi theory}
\label{subsec:Hamiltonians}

The relevant two-body interactions correspond to Standard Model interactions involving neutrinos and antineutrinos. In the early universe, they interact through weak processes with electrons, positrons (also muons and antimuons) and other (anti)neutrinos. Therefore, we must take as interaction Hamiltonian \eqref{eq:defHint} the useful part of the SM Hamiltonian of weak interactions, that is given by
\begin{equation}
\label{eq:defHsm}
\hat{H}_{\mathrm{int}} = \hat{H}_{CC} + \hat{H}_{NC}^{\mathrm{mat}} + \hat{H}_{NC}^{\nu \nu} \, ,
\end{equation}
where we separate three contributions:
\begin{itemize}
	\item the charged-current hamiltonian,
\begin{multline}
\label{eq:hcc_app}
\hat{H}_{CC} = 2 \sqrt{2} G_F m_W^2 \int{\ddp{1} \ddp{2} \ddp{3} \ddp{4}} \ (2\pi)^3 \delta^{(3)}(\vp_1 + \vp_2 - \vp_3 - \vp_4) \\ \times  [\overline{\psi}_{\nu_e}(\vp_1)\gamma_\mu P_L\psi_e(\vp_4)] W^{\mu \nu}(\Delta) [\overline{\psi}_e(\vp_2) \gamma_\nu P_L \psi_{\nu_e}(\vp_3)] \, ,
\end{multline}
with $\psi(\vec{p}) = \sum_{h} \left[ \ha(\vec{p},h) u^h(\vec{p})+ \hbd(-\vec{p},h) v^h(-\vec{p}) \right]$ the Fourier transform of the quantum fields,  and the gauge boson propagator
\begin{equation}
\label{eq:app_propagator}
W^{\mu \nu}(\Delta) = \frac{\eta^{\mu \nu} - \frac{\Delta^\mu \Delta^\nu}{m_W^2}}{m_W^2 - \Delta^2} \simeq \frac{\eta^{\mu \nu}}{m_W^2} + \frac{1}{m_W^2}\left(\frac{\Delta^2 \eta^{\mu \nu}}{m_W^2} - \frac{\Delta^\mu \Delta^\nu}{m_W^2}\right) \, .
\end{equation}
The lowest order in this expansion is the usual 4-Fermi effective theory. The momentum transfer is $\Delta = p_1-p_4$ for a $t$-channel ($\nu_e-e^-$ scattering), and $\Delta= p_1 + p_2$ for the $s$-channel ($\nu_e-e^+$).

	\item the neutral-current interactions with the matter background (electrons and positrons, we would write the same term for $\mu^\pm$),
\begin{multline}
\label{eq:hncmat}
\hat{H}_{NC}^{\mathrm{mat}} = 2 \sqrt{2} G_F m_Z^2 \sum_{\alpha} \int{\ddp{1} \ddp{2} \ddp{3} \ddp{4}} \ (2\pi)^3 \delta^{(3)}(\vp_1 + \vp_2 - \vp_3 - \vp_4) \\ \times  [\overline{\psi}_{\nu_\alpha}(\vp_1)\gamma_\mu P_L\psi_{\nu_\alpha}(\vp_3)] Z^{\mu \nu}(\Delta) [\overline{\psi}_e(\vp_2) \gamma_\nu (g_L P_L + g_R P_R) \psi_e(\vp_4)] \, ,
\end{multline}
where $Z^{\mu \nu}$ is identical to $W^{\mu \nu}$ with the replacement $m_W \to m_Z$. The neutral-current couplings are $g_L = -1/2 + \sin^2{\theta_W}$ and $g_R = \sin^2{\theta_W}$, where $\sin^2{\theta_W} \simeq 0.231$ is the weak-mixing angle.
	\item the self-interactions of neutrinos,\footnote{To understand the different prefactor from $\hat{H}_{NC}^{\mathrm{mat}}$, start from the general neutral-current Hamiltonian:
\[
\hat{H}_{NC} = 2 \sqrt{2} G_F m_Z^2 \sum_{f,f'} \int{\cdots \ \left[\overline{\psi}_{f} \gamma_\mu (g_L^f P_L + g_R^f P_R) \psi_f\right]Z^{\mu \nu}(\Delta)\left[\overline{\psi}_{f'}\gamma_\nu (g_L^{f'} P_L + g_R^{f'} P_R) \psi_{f'}\right]} \]
Now the multiplicity of each term and the use of $g_L^\nu = 1/2$, $g_R^\nu=0$ lead to the Hamiltonians above.
}
\begin{multline}
\label{eq:hncself}
\hat{H}_{NC}^{\nu \nu} = \frac{G_F}{\sqrt{2}} m_Z^2 \sum_{\alpha, \beta} \int{\ddp{1} \ddp{2} \ddp{3} \ddp{4}} \ (2\pi)^3 \delta^{(3)}(\vp_1 + \vp_2 - \vp_3 - \vp_4) \\ \times  [\overline{\psi}_{\nu_\alpha}(\vp_1)\gamma_\mu P_L\psi_{\nu_\alpha}(\vp_3)] Z^{\mu \nu}(\Delta) [\overline{\psi}_{\nu_\beta}(\vp_2) \gamma_\nu P_L \psi_{\nu_\beta}(\vp_4)] \, .
\end{multline}
\end{itemize}

In order to calculate the collision term, we will restrict to the low-energy 4-Fermi theory (cf. section~\ref{app:collision_term}), while the next-to-leading order must be used to obtain the relevant mean-field terms in the Quantum Kinetic Equations.

\section{Neutrino masses and mixings}



\label{subsec:Values_Mixing}

\paragraph{PMNS mixing matrix}

For all numerical calculations in this work, we employ the standard parameterization of the PMNS matrix which reads \cite{Gariazzo_2019,GiuntiKim,PDG}
\begin{equation}
\label{eq:PMNS}
U = R_{23} R_{13} R_{12} = \begin{pmatrix} 
c_{12} c_{13} & s_{12} c_{13} & s_{13} \\
- s_{12}c_{23} - c_{12}s_{23}s_{13} & c_{12} c_{23} - s_{12}s_{23}s_{13} & s_{23} c_{13} \\
s_{12}s_{23} - c_{12}c_{23}s_{13} & -c_{12}s_{23} - s_{12}c_{23}s_{13} & c_{23} c_{13}
\end{pmatrix} \, ,
\end{equation}
with $c_{ij} = \cos{\theta_{ij}}$, $s_{ij}=\sin{\theta_{ij}}$ and $\theta_{ij}$ the mixing angles. $R_{ij}$ is the real rotation matrix of angle $\theta_{ij}$ in the $i$-$j$ plane, namely, $(R_{ij})^i_i = (R_{ij})^j_j = c_{ij}$, $(R_{ij})^k_k = 1$ where $k \neq i,j$, $(R_{ij})^i_j = - (R_{ij})^j_i = s_{ij}$ and the other components are zero. Note that we do not introduce here a CP-violating phase, postponing its treatment to specific sections of this thesis, namely~\ref{subsec:Decoupling_CP} and~\ref{SecDiracPhase}. We use the most recent values from the Particle Data Group \cite{PDG}:
\begin{align}
\left(\frac{\Delta m_{21}^2}{10^{-5} \, \mathrm{eV}^2},\frac{\Delta m_{31}^2}{10^{-3} \, \mathrm{eV}^2},s_{12}^2,s_{23}^2,s_{13}^2\right)_\mathrm{ NH} &= \left(7.53, 2.53, 0.307, 0.546, 0.0220 \right) \, , \label{ValuesStandard}  \\
\left(\frac{\Delta m_{21}^2}{10^{-5} \, \mathrm{eV}^2},\frac{\Delta m_{31}^2}{10^{-3} \, \mathrm{eV}^2},s_{12}^2,s_{23}^2,s_{13}^2\right)_\mathrm{ IH} &= \left(7.53, -2.46, 0.307, 0.539, 0.0220 \right) \, ,
\end{align}
where $\Delta m_{ij}^2 \equiv m_i^2 - m_j^2$ is the difference of the squared masses of the mass eigenstates $i$ and $j$. The associated values of the mixing angles are $\theta_{12} = 0.587$, $\theta_{13}=0.149$ and $\theta_{23}=0.831$ in normal ordering, the only different value is $\theta_{23} = 0.824$ in inverted ordering.

For completeness, we also give the most recent values of the physical constants used \cite{PDG}: the Fermi constant $G_F = 1.1663787 \times 10^{-5} \, \mathrm{GeV^{-2}}$ and the gravitational constant $\mathcal{G} = 6.70883 \times 10^{-39} \, \mathrm{GeV^{-2}}$.

Note that we could also use the values from the global fit of neutrino oscillation data~\cite{deSalas_Mixing}, a choice made in~\cite{Bennett2021}. The results we get for the standard value $\Neff$ are identical at the level of a few $10^{-6}$.

\paragraph{Parameterizations of $\bm{\mathrm{SU(2)}}$ and $\bm{\mathrm{SO(3)}}$} In chapter~\ref{chap:Asymmetry}, we study the case of two-flavour mixing, for which a vector representation of density matrices is possible. This allows for a more intuitive representation of their time evolution, namely in terms of precession. Let us precise here some definitions concerning the useful matrix groups $\mathrm{SU(2)}$ and $\mathrm{SO(3)}$.

Any matrix of $\mathrm{SU(2)}$ can be expressed in terms of Euler angles as
\begin{equation}
\mathcal{R}_2(\alpha,\beta,\gamma) = \mathcal{R}_\z(\alpha)\cdot
\mathcal{R}_\y(\beta) \cdot  \mathcal{R}_\z(\gamma) = \begin{pmatrix}
\mathrm{e}^{-\ii(\alpha+\gamma)/2} \cos(\beta/2) & -\mathrm{e}^{-\ii(\alpha-\gamma)/2}\sin(\beta/2) \\
\mathrm{e}^{\ii(\alpha-\gamma)/2} \sin(\beta/2)& \mathrm{e}^{\ii(\alpha+\gamma)/2}\cos(\beta/2)
\end{pmatrix}
\end{equation}
with $\mathcal{R}_i(\theta) \equiv \exp(- \ii \theta \sigma_i/2)$. Similarly a $\mathrm{SO(3)}$ matrix is also expressed with Euler angles as
\begin{equation}
R_3(\alpha,\beta,\gamma) = R_\z(\alpha)\cdot R_\y(\beta) \cdot  R_\z(\gamma) 
\end{equation}
where $R_j(\theta) \equiv \exp(-\ii \theta \mathcal{J}^j)$ and
$(\mathcal{J}^i)_{jk} = -\ii \epsilon_{ijk}$. Both sets of matrices are related since they share the same Lie algebra, thanks to 
\begin{equation}
\label{eq:SU2SO3}
\mathcal{R}_2\cdot  \sigma_i \cdot \mathcal{R}_2^\dagger = \sigma_j\left(R_3\right)^j_{\,\,\,i}
\,,
\end{equation}
where it is implied that the Euler angles defining $\mathcal{R}_2$ and $R_3$ are the same. Therefore, the conjugation of the traceless part of a two-neutrino density matrix by an element $\mathcal{R}_2$ of $\mathrm{SU(2)}$ is equivalent to the associated rotation $R_3$ applied on its vector representation defined in equation~\eqref{MatrixToVector}. 

Note that the PMNS matrix is thus defined as
\begin{equation}
\label{eq:PMNS_bis}
U =  R_\x(-\theta_{23})R_\y(\theta_{13})  R_\z(-\theta_{12}) \,.
\end{equation}

\pagestyle{ruled}

\chapter{On the BBGKY formalism}
\label{App:BBGKY}

\setlength{\epigraphwidth}{0.5\textwidth}
\epigraph{But, for better or worse, the Crown has landed on my head. And I say we go.}{Queen Elizabeth II, \emph{The Crown} [S01E08]}

{
\hypersetup{linkcolor=black}
    \minitoc
}

In this Appendix, we give some technical details on the BBGKY formalism and how it applies to (anti)neutrinos in the early Universe, in addition to the elements introduced in chapter~\ref{chap:QKE}. First, we explicit the components of the density matrix and the interaction potential, discussing some different presentations existing in the literature. Then, we explain how antiparticles can be included without much effort in the framework we presented in chapter~\ref{chap:QKE}.

\section{Further details on the formalism}
\label{app:BBGKY_Antisym}

The BBGKY hierarchy is often used in nuclear physics~\cite{Cassing1990, Reinhard1994, Lac04,Lac14}, which corresponds to a slightly different physical context than neutrinos and antineutrinos in the early Universe. In particular, we have chosen to use systematically a second-quantized approach (compared to the first-quantized formalism sometimes used), and we give in this section the equivalence between the various formalisms.


\subsection{Components of the density matrix}
\label{app:vrho_components}

Let's prove that the components of the $s$-body operator defined in \eqref{eq:defrhos} are given by \eqref{eq:defrhos2}. We roughly follow \cite{Simenel}, Eqs.~(12)--(15). For simplicity, we derive the result for $\hat{\varrho}^{(1)}$ only, but it can be readily generalized.

We need the closure relation which reads
\begin{equation}
\label{eq:closure_relation}
\hat{\Id}_N = \frac{1}{N!} \sum_{k_1 \cdots k_N}{\ket{k_1 \cdots k_N} \bra{k_1 \cdots k_N}} \, ,
\end{equation}
following~\cite{Simenel}, Eq.~(A28). Moreover, the trace of a $N-$particle operator $\hat{A}$ reads
\begin{equation}
\label{eq:def_trace}
\Tr \, \hat{A} = \sum_{k_1 \cdots k_N} \frac{1}{\sqrt{N!}}  \bra{k_1 \cdots k_N} \hat{A} \ket{k_1 \cdots k_N} \frac{1}{\sqrt{N!}} \, ,
\end{equation}
consistently with (A22) and (C3) in~\cite{Simenel}. The \emph{partial traces} are given by:
\begin{equation}
\label{eq:def_partial_trace}
\left. \Tr_{s+1 \dots N}(\hat{A}) \right \rvert^{i_1 \cdots i_s}_{j_i \cdots j_s} = \sum_{k_{s+1}\cdots k_N}  \frac{1}{\sqrt{N!}} \bra{i_1 \cdots i_s k_{s+1} \cdots k_N} \hat{A} \ket{j_1 \cdots j_s k_{s+1} \cdots k_N} \frac{1}{\sqrt{N!}} \, .
\end{equation}
Note that we have evidently $\Tr_{1 \dots N} \hat{A} = \Tr \,  \hat{A}$.

Starting from~\eqref{eq:defrho}, and inserting twice the closure relation~\eqref{eq:closure_relation},
\begin{align}
\langle \Psi | \hat{a}_j^\dagger \hat{a}_i | \Psi \rangle 
&= \langle \Psi | \frac{1}{N!}\sum_{k_1 \cdots k_N}{|k_1 \cdots k_N \rangle \langle k_1 \cdots k_N |} \,  \hat{a}_j^\dagger \hat{a}_i \, \frac{1}{N!}\sum_{k'_1 \cdots k'_N}{|k'_1 \cdots k'_N \rangle \langle k'_1 \cdots k'_N |} \Psi \rangle \nonumber
\\ &= \frac{1}{(N!)^2} \sum_{\substack{k_1 \cdots k_N \\  k'_1 \cdots k'_N}}{\langle k'_1 \cdots k'_N | \Psi \rangle \langle \Psi | k_1 \cdots k_N \rangle \times \langle k_1 \cdots k_N |  \hat{a}_j^\dagger  \hat{a}_i |k'_1 \cdots k'_N \rangle} \nonumber
\\ &= \frac{N^2}{(N!)^2} \sum_{\substack{k_2 \cdots k_N \\  k'_2 \cdots k'_N}}{\langle i \, k'_2 \cdots k'_N | \Psi \rangle \langle \Psi | j \, k_2 \cdots k_N \rangle \times \langle k_2 \cdots k_N |  k'_2 \cdots k'_N \rangle} \nonumber
\\ \intertext{Indeed, $\sum_{k'_2 \cdots k'_N}{\langle i \, k'_2 \cdots k'_N | \Psi \rangle \ha_i |k'_1 \cdots k'_N \rangle} = N \langle i \, k'_2 \cdots k'_N | \Psi \rangle |  k'_2 \cdots k'_N \rangle$ since, due to antisymmetry, only one of the $k'_p$ can be equal to $i$, and the minus signs appearing when anticommuting $\ha_i$ and $\had_{k'_1}\cdots \had_{k'_p}$ are compensated when $k'_p = i$ is moved at the beginning of the bra $\bra{i k'_2 \cdots k'_N}$.} &= \frac{1}{[(N-1)!]^2} \sum_{\substack{k_2 \cdots k_N \\  k'_2 \cdots k'_N}}{\langle i \, k'_2 \cdots k'_N | \Psi \rangle \langle \Psi | j \, k_2 \cdots k_N \rangle \times (N-1)! \cdot \delta_{k_2 k'_2}\cdots\delta_{k_N k'_N}} \nonumber
\\ &= N \sum_{k_2 \cdots k_N}{\frac{1}{\sqrt{N!}}\langle i \, k_2 \cdots k_N | \Psi \rangle \langle \Psi | j \, k_2 \cdots k_N \rangle \frac{1}{\sqrt{N!}}} \nonumber
\\ &= N \left. \mathrm{Tr}_{2 \dots N} \hat{D} \right |_{ij} \, ,
\end{align}
where we use the definition~\eqref{eq:def_partial_trace} of partial traces. We thus recover the expression~\eqref{eq:defrhos}.


\subsection{“Labelled particles” notation} 

It is very common in the literature~\cite{Lac04,Simenel,Lac14} to adopt a notation that overlooks --- temporarily --- the fact that the particles are indistinguishable. In other words, one “labels” particles and introduces tensor product states. For instance, for two particles, the ket $\ket{1:i,2:j}$ allow to define the antisymmetrized version (valid for fermions):\footnote{The ket $\ket{ij}$ is, in second quantization, directly defined from the vacuum via $\ket{ij} \equiv \had_i \had_j \ket{0}$, from which the antisymmetry properties follow.}
\begin{equation}
 \ket{ij} \equiv \frac{1}{\sqrt{2}} \left(\ket{1:i,2:j} - \ket{1:j,2:i}\right) \, . 
\end{equation}

\paragraph{Consistency of definitions} The interaction matrix elements are then usually defined as $\tilde{v}^{ik}_{jl} = v^{ik}_{jl} - v^{ik}_{lj}$, where \[v^{ik}_{jl} \equiv \bra{1:i,2:k}\hat{H}_{\mathrm{int}}\ket{1:j,2:l} \, . \] This definition is consistent with~\eqref{eq:defvint}. Indeed, one has:
\begin{align}
\bra{ik}\hat{H}_{\mathrm{int}}\ket{jl} &= \frac{1}{\sqrt{2}} \left( \bra{1:i,2:k} - \bra{1:k,2:i}\right) \hat{H}_{\mathrm{int}} \left( \ket{1:j, 2:l} - \ket{1:l,2:j} \right) \frac{1}{\sqrt{2}} \nonumber \\
&= \frac12 \bra{1:i,2:k} \hat{H}_{\mathrm{int}} \left( \ket{1:j, 2:l} - \ket{1:l,2:j} \right) \nonumber  \\
&\qquad \qquad \qquad \qquad + \frac12 \underbrace{\bra{1:k, 2:i} \hat{H}_{\mathrm{int}} \left( \ket{1:l, 2:j} - \ket{1:j,2:l} \right)}_{\text{first term with } 1 \leftrightarrow 2} \nonumber \\
&=  \bra{1:i,2:k} \hat{H}_{\mathrm{int}} \left( \ket{1:j, 2:l} - \ket{1:l,2:j} \right) \nonumber \\
&= v^{ik}_{jl} - v^{ik}_{lj} \label{eq:check_defHint}
\end{align}
We used the fact that nothing changes if we rename $1 \leftrightarrow 2$ (i.e. $v^{ik}_{jl} = v^{ki}_{lj}$).

We prefer to use solely the definition~\eqref{eq:defvint} which uses only creation/annihilation operators and no nonphysical labeling. We will however refer to it in this Appendix when necessary, to connect our equations with the corresponding ones in the literature.

\paragraph{Two-body operators} In this new “language”, the interaction Hamiltonian is not written directly in second quantization, but in terms of two-body operators:
\begin{equation}
\hat{H}_\mathrm{int} = \frac12 \sum_{q \neq q'}{\hat{V}(q,q')} = \sum_{q<q'}{\hat{V}(q,q')} \, ,
\end{equation}
where the indices $q$ and $q'$ label particles, and noting that $\hat{V}(q,q') = \hat{V}(q',q)$. The matrix elements of $\hat{V}$ read
\[v^{ik}_{jl} \equiv \bra{1:i,2:k}\hat{V}(1,2) \ket{1:j,2:l} \, ,\]
hence the same definition is valid with $\hat{H}_\mathrm{int}$, i.e. $v^{ik}_{jl} = \bra{1:i,2:k}\hat{H}_{\mathrm{int}}\ket{1:j,2:l}$. Note that this overlooks the fact that $\hat{H}_\mathrm{int}$ is a $N$-body operator and $\hat{V}(1,2)$ a 2-body one, such that they don't act on the same spaces.

Finally, we emphasize that the expression~\eqref{eq:defHint}, which involves annihilation and creation operators, is valid for any number of particles, hence in the entire Fock space~\cite{CohenIII}.

\paragraph{Traces and notation of the BBGKY hierarchy} Finally, let us discuss the expression of traces in this “labelled particles” notation --- an important point as traces appear in all equations of the BBGKY hierarchy.

An important relation is (we prove its consistency with the above definitions later, and admit it for now):
\begin{equation}
\label{eq:tr_v_rho}
\Tr_2 \left(\hat{V}(1,2) \hat{\vrho}^{(12)}\right)^{\textcolor{firebrick}{i}}_{\textcolor{firebrick}{j}} \equiv \sum_{k,l,r}{v^{\textcolor{firebrick}{i}k}_{rl} \vrho^{rl}_{\textcolor{firebrick}{j}k}} \, .
\end{equation}
The summations over $l$, $r$ come from the product operation, and the sum over $k$ is due to the trace. Given the antisymmetry properties of $\hat{\vrho}^{(12)}$, namely $\vrho^{rl}_{jk} = - \vrho^{lr}_{jk}$, we get (with a relabeling of the indices $l$, $r$ for one term):
\begin{align}
\Tr_2 \left(\hat{V}(1,2) \hat{\vrho}^{(12)}\right)^i_j &=\frac12 \sum_{k,l,r}{\left(v^{ik}_{rl} - v^{ik}_{lr}\right) \vrho^{rl}_{jk}} \nonumber \\
&= \frac12 \sum_{k,l,r}{\tilde{v}^{ik}_{rl} \vrho^{rl}_{jk}} \nonumber \\
&= \frac12 \Tr_2 \left( \hat{\tilde{v}}^{(12)} \hat{\vrho}^{(12)}\right)^i_j
\end{align}
This relations explain why the BBGKY hierarchy \emph{looks} a priori different depending on the references, cf.~for instance Eq.~(88) in~\cite{Simenel}, Eq.~(3) in~\cite{Lac04} or Eq.~(9) in~\cite{Volpe_2013}.

Let us now prove~\eqref{eq:tr_v_rho}. From~\eqref{eq:def_trace}, we have
\begin{align*}
\Tr_2 \left(\hat{V}(1,2) \hat{\vrho}^{(12)}\right)^i_j &=\frac{1}{2!} \sum_{k}{\bra{ik} \hat{V}(1,2) \hat{\vrho}^{(12)} \ket{jk}}  \\
&= \frac{1}{2!} \frac{1}{2!} \sum_{k,l,r} \underbrace{\bra{ik} \hat{V}(1,2) \ket{rl}}_{\displaystyle v^{ik}_{rl} - v^{ik}_{lr}}  \underbrace{\bra{rl} \hat{\vrho}^{(12)} \ket{jk}}_{\displaystyle \vrho^{rl}_{jk} - \vrho^{rl}_{kj}} \, ,  \\
\end{align*}
where the matrix elements are consistent with~\eqref{eq:check_defHint}. With the antisymmetry property of $\hat{\vrho}^{(12)}$, we have
\begin{align*}
\Tr_2 \left(\hat{V}(1,2) \hat{\vrho}^{(12)}\right)^i_j &= \frac{1}{2!} \frac{1}{2!} \sum_{k,l,r}{(v^{ik}_{rl}-v^{ik}_{lr}) \times 2 \vrho^{rl}_{jk}}  \\
&= \sum_{k,l,r}{v^{ik}_{rl} \vrho^{rl}_{jk}}
\end{align*}
thanks to the relabeling $l \leftrightarrow r$ in one term.

%
%

\section{Quantum kinetic equations with antiparticles}
\label{app:antiparticles}

\subsection{QKE for $\bm{\bvrho}$}

We present in this section the inclusion of antiparticles to the BBGKY formalism.

\paragraph{Generalized definitions}

One must adapt the definitions \eqref{eq:defrhos2} and \eqref{eq:defHint} to include the annihilation and creation operators $\hb,\hbd$. Throughout this appendix, we will emphasize the indices which are associated to antiparticles with a barred notation $(\bar{\imath},\bar{\jmath})$. Therefore, with capital indices $I$ being either $i$ or $\bi$, we have:
\begin{align}
\label{eq:defrho_anti}
\varrho^{I_1 \cdots I_s}_{J_1 \cdots J_s} &\equiv \langle \hat{c}_{J_s}^\dagger \cdots \hat{c}_{J_1}^\dagger \hat{c}_{I_1} \cdots \hat{c}_{I_s} \rangle \, , \\
\hat{H}_0 &= \sum_{I,J}{t^{I}_{J} \, \hcd_I \hc_J} \, , \\
\hat{H}_\mathrm{int} &= \frac14 \sum_{I,J,K,L}{\tilde{v}^{IK}_{JL} \, \hcd_I \hcd_K \hc_L \hc_J} \, , \label{eq:vint_anti}
\end{align}
where $\hc_I = \ha_i$ or $\hb_{\bar{\imath}}$ depending on the index $I$ labelling a particle or an antiparticle.

The evolution equations \eqref{eq:hierarchy} and \eqref{eq:eqvrho} are naturally extended to the antiparticle case thanks to the global indices. The downside of this strategy is that the transformation law of tensors is now implicit: since $\ha$ transforms like $\hbd$ under a unitary transformation $\psi^a = \mathcal{U}^a_i \psi^i$, the behaviour of upper and lower indices is inverted whenever they label an antiparticle degree of freedom, for instance:
\begin{equation}
\label{eq:transfo_unit}
t^{i}_{j} = {\mathcal{U}^\dagger}^{i}_{a} \, t^{a}_{b} \, \mathcal{U}^{b}_{j} \qquad ; \qquad t^{\bi}_{\bj} = \mathcal{U}^{a}_{i} \, t^{\bar{a}}_{\bar{b}} \, {\mathcal{U}^\dagger}^{j}_b \, .
\end{equation}

Since we assume an isotropic medium, there are no “abnormal” or “pairing” densities \cite{Volpe_2013,SerreauVolpe,Volpe_2015} such as $\langle \hb \ha \rangle$, which ensures the separation of the two-body density matrix between the neutrino density matrix (for which we keep the notation $\vrho$) and the antineutrino one $\bvrho$. In order for $\bvrho$ to have the same transformation properties as $\vrho$, we need to take a transposed convention for its components:
\begin{equation}
\bvrho^\bi_\bj = \vrho^{\{J=\bar{\jmath}\}}_{\{I=\bar{\imath}\}} = \langle \hcd_\bi \hc_\bj \rangle = \langle \hbd_i \hb_j \rangle \, .
\end{equation}
One could further take transposed conventions for the antiparticle indices in $t$ and $\tilde{v}$, which would ensure a clear correspondence between index position and transformation law --- contrary to \eqref{eq:transfo_unit}. For instance, $\bar{t}^\bi_\bj \equiv t^\bj_\bi$ transforms as $t^i_j$. However, in order to keep a unique expression for the mean-field potential or the collision term, we stick to the general definitions above. For instance, we have:
\begin{equation}
\label{eq:Gamma_full}
\Gamma^{i}_{j} = \sum_{K,L}{\tilde{v}^{iK}_{jL} \vrho^{L}_{K}} = \sum_{k,l}{\tilde{v}^{ik}_{jl} \vrho^{l}_{k}} + \sum_{\bar{k},\bar{l}}{\tilde{v}^{i \bar{k}}_{j \bar{l}} \bvrho^{\bar{k}}_{\bar{l}}} \, .
\end{equation}
Since the annihilation and creation operators do not appear naturally in normal order in the Hamiltonian \eqref{eq:defHsm}, recasting it in the form~\eqref{eq:vint_anti} leads to extra minus signs in $\tilde{v}$ involving antiparticles (cf.~table~\ref{Table:MatrixElements}).

These conventions being settled, we can include the full set of interaction matrix elements and compute all relevant contributions to the neutrino QKEs \eqref{eq:QKE_rho}. In the following, we derive the QKE for $\bvrho$.

\paragraph{QKE for antineutrinos}

Thanks to our conventions, the evolution equation for the antineutrino density matrix $\bvrho$ is similarly obtained within the BBGKY formalism, 
with some differences compared to the neutrino case. First and foremost, the evolution equation for $\bvrho^\bi_\bj$ correspond in the general formalism to the equation for $\vrho^\bj_\bi$:
	\begin{equation}
	\label{eq:drhobjbi}
	\ii  \frac{\dd \bvrho^{\bi}_{\bj}}{\dd t}  = \ii  \frac{\dd \vrho^{\bj}_{\bi}}{\dd t} = \left( \left[t^{\bj}_{K} + \Gamma^{\bj}_{K}\right] \vrho^{K}_{\bi} - \vrho^{\bj}_{K} \left[t^{K}_{\bi} + \Gamma^{K}_{\bi}\right] \right) + \ii \, \hat{\mathcal{C}}^\bj_\bi \, ,
	\end{equation}
showing that taking the commutator with a transposed convention leads to a minus sign. Moreover,
\begin{itemize}
	\item we express the kinetic terms $t^\bj_\bi$, starting from the mass basis:
	\begin{equation}
	 t^\bj_\bi = U^a_j  \left. \frac{\mathbb{M}^2}{2p}\right|^{\bar{a}}_{\bar{b}} {U^\dagger}^i_b = {U^\dagger}^i_b  \left. \frac{\mathbb{M}^2}{2p}\right|^b_a U^a_j = t^i_j \, ;
	 \end{equation}
	\item $\tilde{v}^{\bar{\jmath} k}_{\bar{\imath} l}$ is the coefficient in front of $\hbd_j \had_k \ha_l \hb_i$, so it will have the same expression (apart from the interchange of $u$ and $v$ spinors for neutrinos, which leaves the result identical) as the coefficient in front of $\ha_j \had_k \ha_l \had_i = - \had_i \had_k \ha_l \ha_j$, that is $- \tilde{v}^{ik}_{jl}$. Therefore, $\Gamma^{\bar{\jmath}}_{\bar{\imath}} = - \Gamma^{i}_{j}$.
\end{itemize}
Including these two results in~\eqref{eq:drhobjbi} show that, compared to the neutrino case, the vacuum term gets a minus sign (from the reversed commutator), but not the mean-field. Formally,
\begin{equation}
\ii \frac{\dd \bvrho^i_j}{\dd t} = \left[- \hat{t} + \hat{\Gamma}, \hat{\bvrho}\right]^{i}_{j} + \ii \, \hat{\mathcal{C}}^\bj_\bi \, .
\end{equation} 
Two additional remarks:
\begin{itemize}
	\item $s$ and $t$ channels are inverted when the particle $1$ is an antineutrino ($2$ and $4$ left unchanged). For instance, the scattering between $\bnu_e$ and $e^-$ is a $s-$channel (exchanged momentum $\Delta = p_1+p_2$), contrary to the scattering between $\nu_e$ and $e^-$ ($\Delta = p_1 - p_2$). This changes the sign of $\Delta^2$, leading to another minus sign for $\Gamma$ at order $1/m_{W,Z}^2$;
	\item the collision integral $\bar{\mathcal{C}}$ is obtained from $\mathcal{C}$ through the replacements $\vrho \leftrightarrow \bvrho$ and $g_L \leftrightarrow g_R$, as detailed in~\ref{app:collision_term}.
\end{itemize}
Considering all these remarks, we obtained the QKE for $\bvrho$ \eqref{eq:QKE_rhobar}.

\subsection{Particle/antiparticle symmetry and consistency of the QKEs}
\label{subsec:QKE_consistency}

If there is no asymmetry, we expect that $\vrho = \bvrho^T$ where $^T$ represents the transposed of the matrix. Indeed, we recall that the definitions of $\vrho$ and $\bvrho$ are transposed: $\vrho^\alpha_\beta \propto \langle \had_{\nu_\beta} \ha_{\nu_\alpha} \rangle$ and $\bvrho^\alpha_\beta \propto \langle \hbd_{\nu_\alpha} \hb_{\nu_\beta} \rangle$. Note that the density matrices being Hermitian, $\bvrho^T = \bvrho^*$, but talking about transposition will be more convenient here (essentially because of the commutators in the QKEs).

A consistency check of the QKEs consists in verifying that, assuming $\vrho = \bvrho^T$, equations~\eqref{eq:QKE_rho} and~\eqref{eq:QKE_rhobar} are equivalent. To do so, we transpose~\eqref{eq:QKE_rhobar}. We assume that there is no CP-phase in the PMNS matrix, as its effect would naturally be to break this equivalence.

\subsubsection{Mean-field consistency}

The key relation is $[A,B]^T = (AB-BA)^T = - [A^T,B^T]$. Then, assuming $\vrho = \bvrho^T$,
\begin{align*}
\left(U \frac{\mathbb{M}^2}{2p}U^\dagger\right)^T &= U^* \frac{\mathbb{M}^2}{2p} U^T = U\frac{\mathbb{M}^2}{2p} U^\dagger &\text{(the absence of CP phase is crucial),} \\
(\mathbb{N}_\nu - \mathbb{N}_{\bnu})^T &= \mathbb{N}_{\bnu} - \mathbb{N}_\nu = - (\mathbb{N}_\nu - \mathbb{N}_{\bnu}) \, , \\
(\mathbb{E}_\mathrm{lep} + \mathbb{P}_\mathrm{lep})^T &= \mathbb{E}_\mathrm{lep} + \mathbb{P}_\mathrm{lep} &\text{because these matrices are diagonal,} \\
(\mathbb{E}_\nu + \mathbb{E}_{\bnu})^T &= \mathbb{E}_{\bnu} + \mathbb{E}_\nu \, .
\end{align*}
Therefore,
\begin{align*}
\ii \left[ \frac{\partial}{\partial t} - H p \frac{\partial}{\partial p}\right] \bvrho^T &=  \Big[ \left(U \frac{\mathbb{M}^2}{2p}U^\dagger\right)^T, \bvrho^T \Big] - \sqrt{2} G_F \Big[ (\mathbb{N}_\nu - \mathbb{N}_{\bnu})^T, \bvrho^T \Big] \\
&\qquad \qquad \qquad + 2 \sqrt{2} G_F p \Big[ \frac{(\mathbb{E}_\text{lep} + \mathbb{P}_\text{lep})^T}{m_W^2} + \frac43 \frac{(\mathbb{E}_{\nu} + \mathbb{E}_{\bnu})^T}{m_Z^2},\bvrho^T \Big ] + \ii \bar{\mathcal{I}}^T  \\
&=  \Big[ U \frac{\mathbb{M}^2}{2p}U^\dagger , \vrho \Big] + \sqrt{2} G_F \Big[(\mathbb{N}_\nu - \mathbb{N}_{\bnu}, \vrho \Big] \\
&\qquad \qquad \qquad + 2 \sqrt{2} G_F p \Big[ \frac{\mathbb{E}_\text{lep} + \mathbb{P}_\text{lep}}{m_W^2} + \frac43 \frac{\mathbb{E}_{\nu} + \mathbb{E}_{\bnu}}{m_Z^2},\vrho \Big ] + \ii \bar{\mathcal{I}}^T  \\
\end{align*}
This coincides with~\eqref{eq:QKE_rho} if $\bar{\mathcal{I}}^T=\mathcal{I}$, which we check now.

\subsubsection{Collision term consistency}

Let us show that $\bar{\mathcal{I}}^T = \mathcal{I}$ if there is no asymmetry (i.e.~if we assume that $\bm{{\varrho = \bar{\varrho}^T}}$). We will systematically observe that the processes in $\mathcal{I}$ correspond to the ones in $\bar{\mathcal{I}}^T$ where particles are exchanged with their associated antiparticles. Let us show it on a few processes:

\paragraph*{Annihilation into charged leptons} The process $\nu + \bar{\nu} \leftrightarrow e^- + e^+$ indeed coincides between both collision integrals:
\begin{align*}
\bar{\mathcal{I}}^T_{[\bar{\nu} \nu \to e^- e^+]} &\propto \int{\ddp{2} \cdots \, 4 (p_1\cdot p_3)(p_2 \cdot p_4) \times \underbrace{f_3 \bar{f}_4}_{f_3 f_4} \left[ G_L (\Id -\varrho_2)G_L(\Id -\bar{\varrho}_1) + \mathrm{h.c.}\right]^T} + \cdots \\
&= \int{\ddp{2} \cdots \, 4 (p_1\cdot p_3)(p_2 \cdot p_4) \times f_3 f_4 \left[ (\Id -\bar{\varrho}_1^T) G_L (\Id -\varrho_2^T)G_L + \mathrm{h.c.}\right]} + \cdots \\
&= \int{\ddp{2} \cdots\, 4 (p_1\cdot p_4)(p_2 \cdot p_3) \times f_4 f_3 \left[ (\Id -\varrho_1) G_L (\Id -\bar{\varrho}_2)G_L + \mathrm{h.c.}\right]} + \cdots \\
&= \mathcal{I}_{[\nu \bar{\nu} \to e^- e^+]}
\end{align*}
We recognize the second term of the statistical factor \eqref{eq:F_ann_cl}.

\paragraph*{Scattering with charged leptons} The process $\nu + e^+ \to \nu + e^+$ in $\mathcal{I}$ is “exchanged” with $\bar{\nu} + e^- \to \bar{\nu} + e^-$ in $\bar{\mathcal{I}}$. Indeed, we have:
\begin{align*}
F_\mathrm{sc}^{LL}(\bar{\nu}^{(1)},e^{(2)},\bar{\nu}^{(3)},e^{(4)})^T &= f_4(\Id -f_2)\left[G_L \bar{\varrho}_3 G_L (\Id -\bar{\varrho}_1) + \mathrm{h.c.} \right]^T - \{\mathrm{loss}\} \\
&= \bar{f}_4(\Id -\bar{f}_2)\left[(\Id - \varrho_1) G_L \varrho_3 G_L + \mathrm{h.c}. \right]^T - \{\mathrm{loss}\} \\
&= F_\mathrm{sc}^{LL}(\nu^{(1)},\bar{e}^{(2)},\nu^{(3)},\bar{e}^{(4)})
\end{align*}
and the prefactor of these statistical factors is identical in both collision integrals, namely $(p_1 \cdot p_4)(p_2 \cdot p_3)$.

\paragraph*{(Anti)neutrino scattering} The correspondence is now between $\nu + \nu \to \nu + \nu$ and $\bar{\nu} + \bar{\nu} \to \bar{\nu} + \bar{\nu}$. Indeed, let's compare the statistical factors \eqref{eq:F_sc_nn} and \eqref{eq:F_sc_bnbn}. Since
\[ \left[ (\Id - \bar{\varrho}_1) \bar{\varrho}_3  (\Id - \bar{\varrho}_2) \bar{\varrho}_4 \right]^T = \varrho_4 (\Id -\varrho_2) \varrho_3 (\Id -\varrho_1) \, , \]
we do have $\mathcal{I}_{[\nu \nu \to \nu \nu]} = \bar{\mathcal{I}}^T_{[\bar{\nu}\bar{\nu}\to\bar{\nu}\bar{\nu}]}$.

This proves the consistency of the QKEs. If one adopts the opposite point of view, this shows that if $\vrho = \bvrho^*$ initially, then this symmetry is preserved along the evolution.


\chapter{Computing the terms of the QKEs}

\setlength{\epigraphwidth}{0.3\textwidth}
\epigraph{The Game is On!}{Sherlock Holmes, \emph{Sherlock}}

{
\hypersetup{linkcolor=black}
    \minitoc
}

We provide here the derivation of the terms of the QKE that were not discussed in chapter~\ref{chap:QKE}, that is the neutral-current mean-field potentia, the collision integral with charged leptons, the antineutrino collision integral. We also detail the dimensional reduction of the collision integral.

\section{Interaction matrix elements and mean-field potential}
\label{app:matrix_el_MF}

We have detailed the example of the calculation of the interaction matrix elements $\tilde{v}$ for charged-current processes in the chapter~\ref{chap:QKE}. Let us show briefly how the other contributions are computed and how they lead to the different terms in the full mean-field potential~\eqref{eq:Gamma_potential}. For brevity, we only treat the interactions in the four-Fermi approximation, but the contributions at order $\Delta^2/m_{W,Z}^2$ can be computed following the same procedure as in section~\ref{subsec:weak_matrix_el_QKE}. Moreover, once the matrix element is known for the interaction with a given particle, the result for the interaction for the associated antiparticle is obtained exactly as in section~\ref{subsec:weak_matrix_el_QKE}.

\subsection{Charged-current mean-field}

The mean-field potential due to the forward coherent scattering with $e^\pm$ is discussed in chapter~\ref{chap:QKE}.

\subsection{Neutral-current mean-field}

For neutral-current processes, the procedure is exactly similar to the charged-current case, with the Hamiltonian
\begin{subequations}
\label{eq:hnc}
\begin{align}
\hat{H}_{NC} &= \frac{G_F}{\sqrt{2}} \sum_{f,f'} \int{{\dd^3 \vec{x} \ \left[\bar{\psi}_{f} \gamma^\mu (g_V^f -g_A^{f} \gamma^5) \psi_f\right]\left[\bar{\psi}_{f'}\gamma_\mu (g_V^{f'} - g_A^{f'} \gamma^5) \psi_{f'}\right]}} \, , \\
&= 2 \sqrt{2} G_F \sum_{f,f'} \int{{\dd^3 \vec{x} \ \left[\bar{\psi}_{f} \gamma^\mu (g_L^f P_L + g_R^{f} P_R) \psi_f\right]\left[\bar{\psi}_{f'}\gamma_\mu (g_L^{f'} P_L + g_R^{f'} P_R) \psi_{f'}\right]}} \, ,
\end{align}
\end{subequations}
where the different couplings are related via (remember that $P_L = (1-\gamma^5)/2$ and $P_R = (1+\gamma^5)/2$)
\begin{equation}
g_L = \frac{g_V + g_A}{2} \quad , \quad g_R = \frac{g_V - g_A}{2} \, .
\end{equation}

We split the Hamiltonian in the contributions involving neutrinos and the matter background and only (anti)neutrinos, \eqref{eq:hncmat} and~\eqref{eq:hncself}.

\subsubsection{Matter background}

We can follow the exact same steps as in the charged-current case. In the following we deal with particles, but as show before considering antiparticles basically only require to add a minus sign.

We find:
\begin{multline}
\label{eq:vNCmat}
\tilde{v}^{\nu_\alpha(1) f(2)}_{\nu_\alpha(3) f(4)} = \frac{G_F}{\sqrt{2}} \ (2 \pi)^3 \delta^{(3)}(\vp_1+\vp_2- \vp_3 - \vp_4) \\ 
\times [\bar{u}_{\nu_\alpha}^{h_1}(\vp_1) \gamma^\mu (1 - \gamma^5) u_{\nu_\alpha}^{h_3}(\vp_3)] \ [\bar{u}_{f}^{h_2}(\vp_2) \gamma_\mu (g_V^{f} - g_A^{f} \gamma^5) u_{f}^{h_f}(\vp_4)] \, .
\end{multline}
Assuming a homogeneous and unpolarized background for the fermion $f$, we can use the trace technology on the $u_f$ spinor product:
\begin{equation}
\sum_{h_f}{[\bar{u}_{f}^{h_f}(\vp) \gamma_\mu (g_V^f - g_A^f \gamma^5) u_{f}^{h_f}(\vp)]} = \tr [(\gamma^\alpha p_\alpha + m_f) \gamma_\mu (g_V^f - g_A^f \gamma^5)] = g_V^f p_\alpha \tr [\gamma^\alpha \gamma_\mu] = 4 g_V^f p_\mu \, ,
\end{equation}
which leads to, writing $f_f(p)$ the distribution function of the fermion $f$,
 \begin{align}
\Gamma^{\nu_\alpha(\vp_1,-)}_{\nu_\alpha(\vp_3,-)} \underset{\text{NC, $f$}}{=} &\frac{G_F}{\sqrt{2}} \times (2 \pi)^3 \delta^{(3)}(\vp_1 - \vp_3) \times \int{\frac{\mathrm{d}^3\vec{p}}{(2 \pi)^3 2 E_p} 4 \, g_V^f \, p_\mu \times f_f(p)} \nonumber  \\
&\qquad \qquad \qquad \qquad \qquad \qquad \times  \underbrace{[\bar{u}_{\nu_\alpha}^{(-)}(\vp_1) \gamma^\mu (1 - \gamma^5) u_{\nu_\alpha}^{(-)}(\vp_3)]}_{4 {p_1}^\mu} \nonumber \\
= \ \ &\frac{G_F}{\sqrt{2}} \times (2 \pi)^3 \delta^{(3)}(\vp_1 - \vp_3) \times 4 p_1 \times g_V^f \times 2 \int{\frac{\mathrm{d}^3\vec{p}}{(2 \pi)^3} f_f(p)} \nonumber \\
= \ \ &(2 \pi)^3 2 p_1 \delta^{(3)}(\vp_1 - \vp_3) \times \sqrt{2} G_F g_V^f n_f  \nonumber\\
= \ \ &\sqrt{2} G_F g_V^f n_f \ \bm{\delta}_{\vp_1 \vp_3} \, .
\end{align} 
We assumed that the fermions were spin-1/2 fermions, with 2 helicity states.

Therefore, in the early universe, including for completeness protons and neutrons, the full mean-field potential due to neutral-current interactions with matter reads
\begin{equation}
\label{eq:Gamma_NC_temp}
\Gamma^{\nu_\alpha(\vp_1,-)}_{\nu_\alpha(\vp_3,-)} \underset{\text{NC, mat}}{=}  \sqrt{2} G_F \, \bm{\delta}_{\vp_1 \vp_3} \, \left[g_V^e (n_e - \bar{n}_e) + g_V^p n_p + g_V^n n_n \right]
\end{equation}
This term is the same for all three neutrino flavors, hence it is proportional to $\Id$ and does not contribute to the mean-field dynamics.\footnote{This is true only if there are no sterile species. Otherwise the mean-field matrix would be, noting $\Gamma_\text{NC}$ the value~\eqref{eq:Gamma_NC_temp}, $\mathrm{diag}(\Gamma_\text{NC},\Gamma_\text{NC},\Gamma_\text{NC},0,\cdots,0)$ with as many zero diagonal entries as there are sterile neutrino species.} Moreover, using $g_V^p = - g_V^e$, $g_V^n = - 1/2$ and the electric neutrality of the Universe $n_e - \bar{n}_e - n_p = 0$, we can write the final result:
\begin{equation}
\label{eq:Gamma_NC_mat_final}
\boxed{\Gamma^{\nu_\alpha(\vp_1,-)}_{\nu_\alpha(\vp_3,-)} \underset{\text{NC, mat}}{=}  - \frac{G_F}{\sqrt{2}} \, \delta_{\alpha \beta} \, \bm{\delta}_{\vp_1 \vp_3} \, n_n} \, .
\end{equation}

\subsubsection{Neutrino self-interactions}

To derive $\tilde{v}^{\nu_\alpha \nu_\beta}_{\nu_\alpha \nu_\beta}$, we rewrite the Hamiltonian \eqref{eq:hncself} in the form~\eqref{eq:defHint}. For convenience, we distinguish the diagonal and non-diagonal terms, and restrict to Fermi order.

\paragraph{Non-diagonal terms}

The Hamiltonian contains the terms
	\begin{multline}
	\hat{H}_{NC}^{\nu \nu} \supset \frac{G_F}{4 \sqrt{2}} \sum_{h_1 \dots} \int{[\mathrm{d}^3 \vp_2 \,] \cdots} \ (2 \pi)^3 \delta^{(3)}(\vp_1+\vp_2-\vp_3-\vp_4) \times [\bar{u}_{\nu_\alpha}^{h_1}(\vp_1) \gamma^\mu (1 - \gamma^5) u_{\nu_\alpha}^{h_4}(\vp_4))] \\ \times [\bar{u}_{\nu_\beta}^{h_2}(\vec{p}_2) \gamma_\mu (1 - \gamma^5) u_{\nu_\beta}^{h_3}(\vec{p}_3)]
	\times \had_{\nu_\alpha}(\vp_1,h_1) \ha_{\nu_\alpha}(\vp_4,h_4) \had_{\nu_\beta}(\vp_2,h_2) \ha_{\nu_\beta}(\vp_3,h_3)
	\end{multline}
where all the possible contributions in normal ordering, so as to be in the form~\eqref{eq:defHint}, read
	\begin{multline}
	\label{eq:factor14offdiag}
\had_{\nu_\alpha}(1) \ha_{\nu_\alpha}(4) \had_{\nu_\beta}(2) \ha_{\nu_\beta}(3)	= \frac14  \Big(- \had_{\nu_\alpha}(1) \had_{\nu_\beta}(2) \ha_{\nu_\alpha}(4) \ha_{\nu_\beta}(3) + \had_{\nu_\alpha}(1) \had_{\nu_\beta}(2) \ha_{\nu_\beta}(3) \ha_{\nu_\alpha}(4) \\ + \had_{\nu_\beta}(2) \had_{\nu_\alpha}(1) \ha_{\nu_\alpha}(4) \ha_{\nu_\beta}(3)  - \had_{\nu_\beta}(2) \had_{\nu_\alpha}(1) \ha_{\nu_\beta}(3) \ha_{\nu_\alpha(4)}\Big)
\end{multline}
This term appears \emph{twice} in $\hat{H}_{NC}^{\nu \nu}$, therefore we can identify
\begin{multline}
\label{eq:vNCnunu}
\tilde{v}^{\nu_\alpha(1) \nu_\beta(2)}_{\nu_\beta(3) \nu_\alpha(4)} = \frac{G_F}{2\sqrt{2}} \ (2 \pi)^3 \delta^{(3)}(\vp_1 + \vp_2 - \vp_3 - \vp_4) \\ 
\times [\bar{u}_{\nu_\alpha}^{h_1}(\vp_1) \gamma^\mu (1 - \gamma^5) u_{\nu_\beta}{h_3}(\vp_3)] \ [\bar{u}_{\nu_\beta}^{h_2}(\vp_2) \gamma_\mu (1 - \gamma^5) u_{\nu_\alpha}^{h_4}(\vp_4)]
\end{multline}
Note that we used a Fierz identity to get this result (cancellation of two minus signs). To get the mean-field, we follow the standard procedure,
\begin{equation}
\Gamma^{\nu_\alpha(\vp_1,h_1)}_{\nu_\beta(\vp_3,h_3)}  \underset{\text{NC, $\nu$}}{=} \sum_{h_2,h_4} \int{[\mathrm{d}^3 \vp_2] [\mathrm{d}^3 \vp_4]} \ \tilde{v}^{\nu_\alpha(1) \nu_\beta(2)}_{\nu_\beta(3) \nu_\alpha(4)} \times \underbrace{\vrho^{\nu_\alpha(\vp_4,h_4)}_{\nu_\beta(\vp_2,h_2)}}_{(2\pi)^3 \, \delta_{h_2 h_4} \, 2p_2 \, \delta^{(3)}(\vp_2 - \vp_4) \vrho^\alpha_\beta(p_2)} \, .
\end{equation}
Using trace technology, we get the result:
\begin{align}
\Gamma^{\nu_\alpha(\vp_1,-)}_{\nu_\beta(\vp_3,-)} &\underset{\text{NC, $\nu$}}{=} \frac{G_F}{2 \sqrt{2}} \times (2 \pi)^3 \delta^{(3)}(\vp_1 - \vp_3) \times \int{\frac{\mathrm{d}^3\vec{p}_2}{(2 \pi)^3 2 p_2} 16 p_2^\mu p_{1,\mu} \times \vrho^\alpha_\beta(p_2)} \nonumber \\
&\ = \sqrt{2} G_F \times (2 \pi)^3 \, 2 p_1 \, \delta^{(3)}(\vp_1 - \vp_3) \times \int{\frac{\mathrm{d}^3\vec{p}_2}{(2 \pi)^3} \, \vrho^\alpha_\beta(p_2)} \nonumber \\
&\ = \sqrt{2} G_F \, \left. \mathbb{N}_\nu \right\rvert^\alpha_\beta \, \bm{\delta}_{\vp_1 \vp_3} \, .
\end{align}
where we have eliminated angular integrals involving $\vec{p}_2 \cdot \vec{p}_1$, which vanish thanks to isotropy.

There is yet another term to consider: the propagation $\nu_\alpha \to \nu_\beta$ in a background of antineutrinos, namely, the matrix elements $\bra{\nu_\alpha \bar{\nu}_\alpha} \hat{H}_{NC}^{\nu \nu} \ket{\nu_\beta \bar{\nu}_\beta}$. It is equivalently the coefficient in front of $\had_{\nu_\alpha}(1) \hbd_{\nu_\alpha}(2) \hb_{\nu_\beta}(4) \ha_{\nu_\beta}(3)$ in the expansion of the Hamiltonian, namely,
\begin{multline}
\tilde{v}^{\nu_\alpha(1) \bnu_\beta(2)}_{\nu_\beta(3) \bnu_\alpha(4)} = \frac{G_F}{2\sqrt{2}} \ (2 \pi)^3 \delta^{(3)}(\vp_1 + \vp_2 - \vp_3 - \vp_4) \\ 
\times [\bar{u}_{\nu_\alpha}^{h_1}(\vp_1) \gamma^\mu (1 - \gamma^5)v_{\nu_\alpha}^{h_4}(\vp_4)] \ [\bar{v}_{\nu_\beta}^{h_2}(\vp_2) \gamma_\mu (1 - \gamma^5) u_{\nu_\beta}^{h_3}(\vp_3)]
\end{multline}
We then multiply this term by
\begin{equation}
\langle \hbd_{\nu_\alpha}(\vec{p}_4,h_4)  \hb_{\nu_\beta}(\vp_2, h_2) \rangle = (2 \pi)^3 \delta_{h_2 h_4} 2 p_2 \delta^{(3)}(\vp_2 - \vp_4) \bvrho^{\alpha}_{\beta}(p_2) \, .
\end{equation}
Note that it is $\bvrho$, and not $\bvrho^*$ that appears, thanks to the transposed definition of $\bvrho$ compared to $\vrho$. Using again a Fierz identity, we get an extra minus sign such that
\begin{align}
\Gamma^{\nu_\alpha(\vp_1,-)}_{\nu_\beta(\vp_3,-)} &\underset{\text{NC, $\bnu$}}{=} - \frac{G_F}{2 \sqrt{2}} \times (2 \pi)^3 \delta^{(3)}(\vp_1 - \vp_3) \times \int{\frac{\mathrm{d}^3\vec{p}_2}{(2 \pi)^3 2 p_2} 16 p_2^\mu p_{1,\mu} \times \bvrho^\alpha_\beta(p_2)} \nonumber \\
&\ = - \sqrt{2} G_F \times (2 \pi)^3 \, 2 p_1 \, \delta^{(3)}(\vp_1 - \vp_3) \times \int{\frac{\mathrm{d}^3\vec{p}_2}{(2 \pi)^3} \, \bvrho^\alpha_\beta(p_2)} \nonumber \\
&\ = - \sqrt{2} G_F \, \left. \mathbb{N}_{\bnu} \right\rvert^\alpha_\beta \, \bm{\delta}_{\vp_1 \vp_3} \, .
\end{align}

\paragraph{Diagonal terms} In order to compute $\Gamma^{\nu_\alpha}_{\nu_\alpha}$, two different interaction matrix elements are needed:
\begin{itemize}
	\item $\tilde{v}^{\nu_\alpha \nu_\beta}_{\nu_\alpha \nu_\beta}$ with $\alpha \neq \beta$, which are given in~\ref{eq:vNCnunu},
	\item $\tilde{v}^{\nu_\alpha \nu_\alpha}_{\nu_\alpha \nu_\alpha}$, which are twice as great as the former matrix elements. Indeed, they consider to only \emph{one} term in the $\hat{H}_{NC}^{\nu \nu}$, but since they only involve one species, there is no rewriting~\eqref{eq:factor14offdiag}, which leads to an extra factor of \emph{four} --- all in all, there is an extra factor of 2.\footnote{In terms of Feynman diagrams, this corresponds to the possibility of coherent forward scattering with a $t-$ and a $u-$ channel (or $s-$channel for the interaction with $\bnu_\alpha$), contrary to the off-diagonal case where only one channel is available.}
\end{itemize}
This allows to write the matrix elements gathered in Table~\ref{Table:MatrixElements}.

Finally, the diagonal part of the self-interaction mean-field reads (we do not detail the interaction with antiparticles, which as usual gives an extra minus sign)
\begin{align}
\Gamma^{\nu_\alpha(\vp_1,-)}_{\nu_\alpha(\vp_3,-)} &\underset{\text{NC}}{=} \sqrt{2} G_F \times (2 \pi)^3 \, 2 p_1 \, \delta^{(3)}(\vp_1 - \vp_3) \sum_{\beta} \int{\frac{\mathrm{d}^3\vec{p}_2}{(2 \pi)^3} \left(\vrho^{\beta}_\beta(p_2) - \bvrho^\beta_\beta(p_2)\right) \times [1 + \delta_{\alpha \beta}]} \nonumber \\
&\, = \sqrt{2} G_F \, \left(\mathbb{N} - \mathbb{N}_{\bnu}\right)^\alpha_\alpha \, \bm{\delta}_{\vp_1 \vp_3}  + \sqrt{2} G_F \, \Tr \left(\mathbb{N} - \mathbb{N}_{\bnu}\right) \, \bm{\delta}_{\vp_1 \vp_3}  \, .
\end{align}
The second term being flavour-independent, it does not contribute to flavour evolution (in other words, it is a contribution $\propto \Id$ to the mean-field, hence its contribution vanishes inside the commutators).

\section{Collision term}
\label{app:collision_term}

We follow the same procedure as in section~\ref{subsec:collision_integral} to compute the remaining contributions to the collision term, that is the parts involving charged leptons. We then present the calculation of the antineutrino collision integral.

\subsection{Neutrino-electron scattering}
\label{subsec:Nu_e_scatt}

We use the general interaction matrix element~\eqref{eq:vtilde_nue_full}, such that our expression will be valid even for non-standard interactions (which are not considered in our numerical calculations):
\begin{multline}
\label{eq:vtilde_nu_e_app}
\tilde{v}^{\nu_\alpha(1) e(2)}_{\nu_\beta(3) e(4)} = 2 \sqrt{2} G_F \, (2 \pi)^3 \delta^{(3)}(\vec{p}_1 + \vec{p}_2 - \vec{p}_3 - \vec{p}_4) \\
\times [\bar{u}_{\nu_\alpha}^{h_1} (\vec{p}_1) \gamma^\mu P_L u_{\nu_\beta}^{h_3} (\vec{p}_3)] \ [\bar{u}_{e}^{h_2} (\vec{p}_2)\gamma_\mu (G_L^{\alpha \beta} P_L +  G_R^{\alpha \beta} P_R ) u_{e}^{h_4} (\vec{p}_4)] \, .
\end{multline}
Let us compute the contribution to the scattering kernel for which $2=e^-,3=\nu_\gamma,3'=\nu_\delta,1'=\nu_\sigma$. We have:
\begin{align*}
 \tilde{v}^{\nu_\alpha(1) e(2)}_{\nu_\gamma(3)e(4)} \times \tilde{v}^{\nu_\delta(3') e(4')}_{\nu_\sigma(1') e(2')} &= 8 G_F^2 (2 \pi)^3 \delta^{(3)}(\vec{p}_1 + \vec{p}_2 - \vec{p}_3 - \vec{p}_4) (2 \pi)^3 \delta^{(3)}(\vec{p}_{\underline{1}}+\vec{p}_2 - \vec{p}_3 - \vec{p}_4) \\ &\qquad \qquad \sum_{h_2,h_3,h_4}  [\bar{u}_{\nu_\alpha}^{h_1} (\vec{p}_1) \gamma^\mu P_L u_{\nu_\gamma}^{h_3} (\vec{p}_3)] [\bar{u}_{\nu_\delta}^{h_3} (\vec{p}_3) \gamma^\nu P_L u_{\nu_\sigma}^{h_1} (\vec{p}_1)] \\ \times  [\bar{u}_{e}^{h_2} (\vec{p}_2)\gamma_\mu &(G_L^{\alpha \gamma} P_L +  G_R^{\alpha \gamma} P_R ) u_{e}^{h_4} (\vec{p}_4)][\bar{u}_{e}^{h_4} (\vec{p}_4)\gamma_\nu (G_L^{\delta \sigma} P_L +  G_R^{\delta \sigma} P_R ) u_{e}^{h_2} (\vec{p}_2)] \\
&= 8 G_F^2 \,  (2\pi)^6 \delta^{(3)}(\vec{p}_1 + \vec{p}_2 - \vec{p}_3 - \vec{p}_4) \delta^{(3)}(\vec{p}_1-\vec{p}_{\underline{1}}) \\
&\times \tr\left[\gamma^\rho p_{1\rho} \gamma^\mu P_L \gamma^\lambda p_{3 \lambda} \gamma^\nu P_L \right] \\ &\times \tr\left[(\gamma_\tau p_2^\tau + m_e) \gamma_\mu (G_L^{\alpha \gamma} P_L +  G_R^{\alpha \gamma} P_R ) (\gamma_\eta p_4^\eta + m_e) \gamma_\nu (G_L^{\delta \sigma} P_L +  G_R^{\delta \sigma} P_R ) \right]
\end{align*}
The first trace reads (we use $\{P_L,\gamma^\mu\} = 0$):
\[p_{1\rho} p_{3 \lambda} \tr\left[\gamma^\rho  \gamma^\mu P_L \gamma^\lambda  \gamma^\nu P_L \right] = 2 \left[ p_1^\mu p_3^\nu + p_1^\nu p_3^\mu - (p_1 \cdot p_3) g^{\mu \nu} \right] + 2 \ii p_{1\rho} p_{3 \lambda} \epsilon^{\rho \mu \lambda \nu} \, . \]
In the second, only even powers of $m_e$ survive:
\begin{align*}
\tr&\left[(\gamma_\tau p_2^\tau + m_e) \gamma_\mu (G_L^{\alpha \gamma} P_L +  G_R^{\alpha \gamma} P_R ) (\gamma_\eta p_4^\eta + m_e) \gamma_\nu (G_L^{\delta \sigma} P_L +  G_R^{\delta \sigma} P_R ) \right] \\ 
&= p_2^\tau p_4^\eta \tr \left[ \gamma_\tau \gamma_\mu \gamma_\eta \gamma_\nu (G_L^{\alpha \gamma} G_L^{\delta \sigma} P_L +  G_R^{\alpha \gamma} G_R^{\delta \sigma} P_R) \right] + m_e^2 \tr \left[\gamma_\mu \gamma_\nu  (G_R^{\alpha \gamma} G_L^{\delta \sigma} P_L +  G_L^{\alpha \gamma} G_R^{\delta \sigma} P_R) \right]
\\ &= 2 \left[ p_{2\mu} p_{4\nu} + p_{2\nu} p_{4\mu} - (p_2 \cdot p_4) g_{\mu \nu} \right] (G_L^{\alpha \gamma} G_L^{\delta \sigma} +  G_R^{\alpha \gamma} G_R^{\delta \sigma}) \\
&\qquad \qquad + 2 \ii p_2^\tau p_4^\eta \epsilon_{\tau \mu \eta \nu} (G_L^{\alpha \gamma} G_L^{\delta \sigma} -  G_R^{\alpha \gamma} G_R^{\delta \sigma}) + 2 m_e^2 g_{\mu \nu}   (G_R^{\alpha \gamma} G_L^{\delta \sigma} +  G_L^{\alpha \gamma} G_R^{\delta \sigma})
\end{align*}
The product of both terms reads:
\begin{itemize}
	\item product of imaginary parts
	\begin{multline*}
	-4  p_{1\rho} p_{3 \lambda} p_2^\tau p_4^\eta (G_L^{\alpha \gamma} G_L^{\delta \sigma} -  G_R^{\alpha \gamma} G_R^{\delta \sigma})  \epsilon^{\rho \mu \lambda \nu} \epsilon_{\tau \mu \eta \nu} \\
	= 8  (G_L^{\alpha \gamma} G_L^{\delta \sigma} -  G_R^{\alpha \gamma} G_R^{\delta \sigma}) \left[ (p_1 \cdot p_2) (p_3 \cdot p_4) - (p_1 \cdot p_4)(p_2 \cdot p_3) \right] \, ,
	\end{multline*}
	\item product of real parts, $\mathcal{O}(m_e^0)$
	\begin{multline*}4 (G_L^{\alpha \gamma} G_L^{\delta \sigma} +  G_R^{\alpha \gamma} G_R^{\delta \sigma})  \left[ p_1^\mu p_3^\nu + p_1^\nu p_3^\mu - (p_1 \cdot p_3) g^{\mu \nu} \right] \left[ p_{2\mu} p_{4\nu} + p_{2\nu} p_{4\mu} - (p_2 \cdot p_4) g_{\mu \nu} \right] \\
	= 8 (G_L^{\alpha \gamma} G_L^{\delta \sigma} +  G_R^{\alpha \gamma} G_R^{\delta \sigma}) \left[(p_1 \cdot p_2)(p_3 \cdot p_4) + (p_1 \cdot p_4)(p_2 \cdot p_3) \right] \, ,
	\end{multline*}
	\item product of real parts, $\mathcal{O}(m_e^2)$
	\begin{multline*} 4 (G_R^{\alpha \gamma} G_L^{\delta \sigma} +  G_L^{\alpha \gamma} G_R^{\delta \sigma}) \left[ p_1^\mu p_3^\nu + p_1^\nu p_3^\mu - (p_1 \cdot p_3) g^{\mu \nu} \right] m_e^2 g_{\mu \nu} \\
	 = - 8 (G_R^{\alpha \gamma} G_L^{\delta \sigma} +  G_L^{\alpha \gamma} G_R^{\delta \sigma}) (p_1 \cdot p_3) m_e^2 \, .
	\end{multline*}
\end{itemize}
Therefore,
\begin{align*}
 \tilde{v}^{\nu_\alpha(1) e(2)}_{\nu_\gamma(3)e(4)} \times \tilde{v}^{\nu_\delta(3') e(4')}_{\nu_\sigma(1') e(2')} = 2^5 G_F^2 \,  &(2\pi)^6 \, \delta^{(3)}(\vec{p}_1 + \vec{p}_2 - \vec{p}_3 - \vec{p}_4) \, \delta^{(3)}(\vec{p}_1-\vec{p}_{\underline{1}}) \\
\qquad \times \Big[ &4 (p_1 \cdot p_2)(p_3 \cdot p_4) (G_L^{\alpha \gamma} G_L^{\delta \sigma} )+ 4 (p_1 \cdot p_4)(p_2 \cdot p_3) (G_R^{\alpha \gamma} G_R^{\delta \sigma}) \\
- \, &2 (p_1 \cdot p_3) m_e^2  (G_R^{\alpha \gamma} G_L^{\delta \sigma} +  G_L^{\alpha \gamma} G_R^{\delta \sigma}) \Big]
\end{align*}
All $\tilde{v}\tilde{v}^*$ products in~\eqref{eq:C11} are equal to this, the only difference being the indices $\alpha,\gamma,\delta,\sigma$ which must respect the matrix structure. This expression is in perfect agreement with, for instance, Eq.~(2.10) of~\cite{Relic2016_revisited}. Schematically,
\[\underbrace{\tilde{v}^{\nu_\alpha e}_{\nu_\gamma e}}_{\to G_{\alpha \gamma}^a} \varrho^\gamma_\delta (3)f_e^{(4)} \underbrace{\tilde{v}^{\nu_\delta e}_{\nu_\sigma e}}_{\to G_{\delta \sigma}^b} (\Id -\varrho^{(1)})^{\sigma}_{\beta} (1 -f_e^{(2)}) = f_e^{(4)}(1-f_e^{(2)}) \left[G^a \cdot \varrho^{(3)}\cdot G^b \cdot (\Id -\varrho^{(1)})\right]^\alpha_\beta \, . \]

Remember that we get the same contribution exchanging $3$ and $4$. Therefore :
\begin{equation}
\label{eq:C_sc_cl}
\begin{aligned}
\mathcal{C}^{[\nu  e^- \to \nu  e^-]} = &(2 \pi)^3 \delta^{(3)}(\vec{p}_1-\vec{p}_{\underline{1}}) \frac{2^5 G_F^2}{2}\int{[\dd^3 \vec{p}_2] [\dd^3 \vec{p}_3] [\dd^3 \vec{p}_4] (2 \pi)^4 \delta^{(4)}(p_1 + p_2 - p_3 - p_4)} \\
&\Big[ 4 (p_1 \cdot p_2)(p_3 \cdot p_4) F_\mathrm{sc}^{LL}(\nu^{(1)},e^{(2)},\nu^{(3)},e^{(4)})  \\
&+ 4 (p_1 \cdot p_4)(p_2 \cdot p_3) F_\mathrm{sc}^{RR}(\nu^{(1)},e^{(2)},\nu^{(3)},e^{(4)}) \\
&- 2 (p_1 \cdot p_3) m_e^2 \left(F_\mathrm{sc}^{LR}(\nu^{(1)},e^{(2)},\nu^{(3)},e^{(4)}) + F_\mathrm{sc}^{RL}(\nu^{(1)},e^{(2)},\nu^{(3)},e^{(4)}) \right) \Big] \, .
\end{aligned}
\end{equation}

\noindent The statistical factors read:
\begin{multline}
\label{eq:F_sc_cl}
F_\mathrm{sc}^{AB}(\nu^{(1)},e^{(2)},\nu^{(3)},e^{(4)}) = f_4 (1-f_2) \left [ G^A \varrho_3 G^B (\Id -\varrho_1) + (\Id - \varrho_1) G^B \varrho_3 G^A \right] \\  - (1-f_4)f_2 \left [ G^A (\Id -\varrho_3) G^B \varrho_1 +  \varrho_1 G^B (\Id -\varrho_3) G^A \right] \, ,
\end{multline}
with the compact notations $f_i =f_e^{(i)} = f_e(p_i)$ and $\vrho_i = \vrho^{(i)} = \vrho(p_i)$.

\vspace{0.5cm}

The other scattering amplitudes can all be obtained via “crossing symmetry” methods, as will be shown in the next two sections.

\subsection{Neutrino-positron scattering}

Now note that the relevant matrix elements are (see Table~\ref{Table:MatrixElements})
\begin{multline}
\tilde{v}^{\nu_\alpha(1) \bar{e}(2)}_{ \nu_\beta(3) \bar{e}(4)} = - 2 \sqrt{2} G_F \ (2 \pi)^3 \, \delta^{(3)}(\vec{p}_1 + \vec{p}_2 - \vec{p}_3 - \vec{p}_4) \\
\times [\bar{u}_{\nu_\alpha}^{h_1} (\vec{p}_1) \gamma^\mu P_L u_{\nu_\beta}^{h_3} (\vec{p}_3)] \ [\bar{v}_{e}^{h_4} (\vec{p}_4)\gamma_\mu (G_L^{\alpha \beta} P_L +  G_R^{\alpha \beta} P_R ) v_{e}^{h_2} (\vec{p}_2)]
\end{multline}
Therefore, the matrix structure is exactly identical, but we have to interchange in the matrix elements $p_2 \leftrightarrow p_4$.

All in all, the scattering with charged leptons reads:
\begin{equation}
\label{eq:C_sc_cl_full}
\begin{aligned}
\mathcal{C}^{[\nu  e^\pm \to \nu  e^\pm]} = &(2 \pi)^3 \delta^{(3)}(\vec{p}_1-\vec{p}_{\underline{1}}) \frac{2^5 G_F^2}{2}\int{[\dd^3 \vec{p}_2] [\dd^3 \vec{p}_3] [\dd^3 \vec{p}_4] (2 \pi)^4 \delta^{(4)}(p_1 + p_2 - p_3 - p_4)} \\
&\Big[ 4 (p_1 \cdot p_2)(p_3 \cdot p_4) \left(F_\mathrm{sc}^{LL}(\nu^{(1)},e^{(2)},\nu^{(3)},e^{(4)}) + F_\mathrm{sc}^{RR}(\nu^{(1)},\bar{e}^{(2)},\nu^{(3)},\bar{e}^{(4)})\right) \\
&+ 4 (p_1 \cdot p_4)(p_2 \cdot p_3) \left(F_\mathrm{sc}^{RR}(\nu^{(1)},e^{(2)},\nu^{(3)},e^{(4)}) + F_\mathrm{sc}^{LL}(\nu^{(1)},\bar{e}^{(2)},\nu^{(3)},\bar{e}^{(4)}) \right) \\
&- 2 (p_1 \cdot p_3) m_e^2 \left(F_\mathrm{sc}^{LR}(\nu^{(1)},e^{(2)},\nu^{(3)},e^{(4)}) + F_\mathrm{sc}^{LR}(\nu^{(1)},\bar{e}^{(2)},\nu^{(3)},\bar{e}^{(4)}) + \{L \leftrightarrow R \}\right) \Big] \, .
\end{aligned}
\end{equation}

\subsection{Neutrino-antineutrino annihilation}

Now, the relevant matrix element is:
\begin{multline}
\tilde{v}^{\nu_\alpha(1) \bar{\nu}_\beta(2)}_{ e(3) \bar{e}(4)} = - 2 \sqrt{2} G_F \ (2 \pi)^3 \delta^{(3)}(\vec{p}_1 + \vec{p}_2 - \vec{p}_3 - \vec{p}_4) \\
\times [\bar{u}_{\nu_\alpha}^{h_1} (\vec{p}_1) \gamma^\mu P_L v_{\nu_\beta}^{h_2} (\vec{p}_2)] \ [\bar{v}_{e}^{h_4} (\vec{p}_4)\gamma_\mu (G_L^{\alpha \beta} P_L +  G_R^{\alpha \beta} P_R ) u_{e}^{h_3} (\vec{p}_3)] \, .
\end{multline}

\noindent Two remarks must be made:
\begin{itemize}
	\item Thanks to the transposed definition of $\bvrho$, the statistical factor keeps a simple expression not involving extra transpositions. Indeed, we have:
\[\underbrace{\tilde{v}^{\nu_\alpha \bar{\nu}_\gamma}_{e \bar{e}}}_{\to G_{\alpha \gamma}^a} f_e^{(3)} \bar{f}_e^{(4)}  \underbrace{\tilde{v}^{e \bar{e}}_{\nu_\delta \bar{\nu}_\sigma}}_{\to G_{\sigma \delta}^b} (\Id-\varrho^{(1)})^\delta_\beta \underbrace{(\Id-\varrho^{(2)})^{\bar{\sigma}}_{\bar{\gamma}}}_{=(\Id-\bar{\varrho}^{(2)})^\gamma_\sigma} = f_e^{(3)}\bar{f}_e^{(4)} \left[G^a \cdot (\Id-\bar{\varrho}^{(2)})\cdot G^b \cdot (\Id-\varrho^{(1)})\right]^\alpha_\beta \, .\]
	\item As can be seen with the associated Feynman diagrams, the amplitudes are obtained from $\nu + e^-$ scattering through:
\[p_1 \to p_1 \ ; \ p_2 \to - p_4 \ ; \ p_3 \to - p_2 \ ; \ p_4 \to p_3 \, . \]
If we stay in our formalism, the product of $\tilde{v}^{\nu \bar{\nu}}_{e \bar{e}} \times \tilde{v}^{e \bar{e}}_{\nu \bar{\nu}}$ involves the combinations $u_e^{h_3}(\vec{p}_3) \bar{u}_e^{h_3}(\vec{p}_3)$ and $v_e^{h_4}(\vec{p}_4) \bar{v}_e^{h_4}(\vec{p}_4)$. Therefore the $m_e^2$ term reverses sign [product $(\gamma_\rho p_3^\rho + m_e)(\gamma_\omega p_4^\omega - m_e)$].
\end{itemize}
This leads to the following statistical factor:
\begin{multline}
\label{eq:F_ann_cl}
F_\mathrm{ann}^{AB}(\nu^{(1)},\bar{\nu}^{(2)},e^{(3)},\bar{e}^{(4)}) = f_3 \bar{f}_4 \left[ G^A (1-\bar{\varrho}_2) G^B (1-\varrho_1) + (1- \varrho_1) G^B (1-\bar{\varrho}_2) G^A \right] \\  - (1-f_3)(1-\bar{f}_4) \left[ G^A \bar{\varrho}_2 G^B \varrho_1 +  \varrho_1 G^B \bar{\varrho}_2 G^A \right] \, ,
\end{multline}
and the collision term contribution
\begin{equation}
\label{eq:C_ann_cl}
\begin{aligned}
\mathcal{C}^{[\nu  \bar{\nu} \to e^- e^+]} = &(2 \pi)^3 \delta^{(3)}(\vec{p}_1-\vec{p}_{\underline{1}}) \frac{2^5 G_F^2}{2}\int{[\dd^3 \vec{p}_2] [\dd^3 \vec{p}_3] [\dd^3 \vec{p}_4] (2 \pi)^4 \delta^{(4)}(p_1 + p_2 - p_3 - p_4)} \\
&\Big[ 4 (p_1 \cdot p_4)(p_2 \cdot p_3) F_\mathrm{ann}^{LL}(\nu^{(1)},\bar{\nu}^{(2)},e^{(3)},\bar{e}^{(4)})  \\
&+ 4 (p_1 \cdot p_3)(p_2 \cdot p_4) F_\mathrm{ann}^{RR}(\nu^{(1)},\bar{\nu}^{(2)},e^{(3)},\bar{e}^{(4)}) \\
&+ 2 (p_1 \cdot p_2) m_e^2 \left(F_\mathrm{ann}^{LR}(\nu^{(1)},\bar{\nu}^{(2)},e^{(3)},\bar{e}^{(4)}) + F_\mathrm{ann}^{RL}(\nu^{(1)},\bar{\nu}^{(2)},e^{(3)},\bar{e}^{(4)}) \right) \Big] \, .
\end{aligned}
\end{equation}

\subsection{Antineutrino collision term}

The antineutrino collision term $\bar{\mathcal{C}}$ for $\bvrho$ can be deduced from $\mathcal{C}$ based on the transformations $\vrho \leftrightarrow \bvrho$ and $L \leftrightarrow R$. Let us prove it explicitly.

As explained in section~\ref{app:collision_term}, the evolution equation for $\bvrho^\alpha_\beta$ is obtained similarly to the one for $\vrho^\beta_\alpha$. In other words, $\bar{\mathcal{C}}^\alpha_\beta$ is calculated with the general formula~\eqref{eq:C11}, where we take $i_1 = \bnu_\beta(\vp_1)$ and $i_1' = \bnu_\alpha(\vp_1)$.

\subsubsection{Antineutrino - charged lepton scattering}

The relevant matrix element is now (note the minus sign with respect to~\eqref{eq:vtilde_nu_e_app}, as explained in section~\ref{app:antiparticles}):
\begin{multline}
\tilde{v}^{\bar{\nu}_\beta(1) e(2)}_{\bar{\nu}_\alpha(3) e(4)} = - 2 \sqrt{2} G_F \, (2 \pi)^3 \delta^{(3)}(\vec{p}_1 + \vec{p}_2 - \vec{p}_3 - \vec{p}_4) \\
\times [\bar{v}_{\nu_\alpha}^{h_3} (\vec{p}_3) \gamma^\mu P_L v_{\nu_\beta}^{h_1} (\vec{p}_1)] \ [\bar{u}_{e}^{h_2} (\vec{p}_2)\gamma_\mu (G_L^{\alpha \beta} P_L +  G_R^{\alpha \beta} P_R ) u_{e}^{h_4} (\vec{p}_4)] \, .
\end{multline}
When inserting it in the collision term formula~\eqref{eq:C11}, the only differences with respect to the derivation of subsection~\ref{subsec:Nu_e_scatt} are the transposition of the $G$ matrices --- which actually corresponds to the exchange of the first and second lines in~\eqref{eq:C11} ---, and most importantly the exchange $p_1 \leftrightarrow p_3$. Therefore, the collision term reads
\begin{equation}
\label{eq:Cbar_sc_cl}
\begin{aligned}
\bar{\mathcal{C}}^{[\bnu  e^- \to \bnu  e^-]} = &(2 \pi)^3 \delta^{(3)}(\vec{p}_1-\vec{p}_{\underline{1}}) \frac{2^5 G_F^2}{2}\int{[\dd^3 \vec{p}_2] [\dd^3 \vec{p}_3] [\dd^3 \vec{p}_4] (2 \pi)^4 \delta^{(4)}(p_1 + p_2 - p_3 - p_4)} \\
&\Big[ 4 (p_1 \cdot p_4)(p_2 \cdot p_3) F_\mathrm{sc}^{LL}(\bnu^{(1)},e^{(2)},\bnu^{(3)},e^{(4)})  \\
&+ 4 (p_1 \cdot p_2)(p_3 \cdot p_4) F_\mathrm{sc}^{RR}(\bnu^{(1)},e^{(2)},\bnu^{(3)},e^{(4)}) \\
&- 2 (p_1 \cdot p_3) m_e^2 \left(F_\mathrm{sc}^{LR}(\bnu^{(1)},e^{(2)},\bnu^{(3)},e^{(4)}) + F_\mathrm{sc}^{RL}(\bnu^{(1)},e^{(2)},\bnu^{(3)},e^{(4)}) \right) \Big] \, ,
\end{aligned}
\end{equation}
which is indeed the neutrino collision term~\eqref{eq:C_sc_cl}, with the replacement $\vrho \to \bvrho$ and $L \leftrightarrow R$. The statistical factor is given by~\eqref{eq:F_sc_cl} with antineutrino density matrices, that is
\begin{multline}
\label{eq:F_sc_bar_cl}
F_\mathrm{sc}^{AB}(\bnu^{(1)},e^{(2)},\bnu^{(3)},e^{(4)}) = f_4 (1-f_2) \left [ G^A \bvrho_3 G^B (\Id -\bvrho_1) + (\Id - \bvrho_1) G^B \bvrho_3 G^A \right] \\  - (1-f_4)f_2 \left [ G^A (\Id -\bvrho_3) G^B \bvrho_1 +  \bvrho_1 G^B (\Id -\bvrho_3) G^A \right] \, .
\end{multline}

To avoid any confusion with the possible transpositions, let us justify one term of the statistical factor. The first term in~\eqref{eq:C11} reads
\[\underbrace{\tilde{v}^{\bnu_\beta e}_{\bnu_\gamma e}}_{\to G_{\gamma \beta}^A} \overbrace{\vrho^{\bar{\gamma}}_{\bar{\delta}}(3)}^{\bvrho^\delta_\gamma(3)} f_e^{(4)} \underbrace{\tilde{v}^{\bnu_\delta e}_{\bnu_\sigma e}}_{\to G_{\sigma \delta}^B} \overbrace{(\Id -\vrho^{(1)})^{\bar{\sigma}}_{\bar{\alpha}}}^{(\Id - \bvrho^{(1)})^\alpha_\sigma} (1 -f_e^{(2)}) = f_e^{(4)}(1-f_e^{(2)}) \left[(\Id -\bvrho^{(1)}) \cdot G^B \cdot \bvrho^{(3)} \cdot G^A \right]^\alpha_\beta \, , \]
which is the second term in~\eqref{eq:F_sc_bar_cl}.

The scattering with positrons is treated in the same fashion.

\subsubsection{Antineutrino - neutrino annihilation}

In this case, the appropriate exchange is $p_1 \leftrightarrow p_2$. The results then read
\begin{multline}
F_\mathrm{ann}^{AB}(\bnu^{(1)},\nu^{(2)},e^{(3)},\bar{e}^{(4)}) = f_3 \bar{f}_4 \left[ G^A (1-\vrho_2) G^B (1-\bvrho_1) + (1- \bvrho_1) G^B (1-\vrho_2) G^A \right] \\  - (1-f_3)(1-\bar{f}_4) \left[ G^A \vrho_2 G^B \bvrho_1 +  \bvrho_1 G^B \vrho_2 G^A \right] \, ,
\end{multline}
and
\begin{equation}
\begin{aligned}
\bar{\mathcal{C}}^{[\bnu  \nu \to e^- e^+]} = &(2 \pi)^3 \delta^{(3)}(\vec{p}_1-\vec{p}_{\underline{1}}) \frac{2^5 G_F^2}{2}\int{[\dd^3 \vec{p}_2] [\dd^3 \vec{p}_3] [\dd^3 \vec{p}_4] (2 \pi)^4 \delta^{(4)}(p_1 + p_2 - p_3 - p_4)} \\
&\Big[ 4 (p_1 \cdot p_3)(p_2 \cdot p_4) F_\mathrm{ann}^{LL}(\bnu^{(1)},\nu^{(2)},e^{(3)},e^{(4)})  \\
&+ 4 (p_1 \cdot p_4)(p_2 \cdot p_3) F_\mathrm{ann}^{RR}(\bnu^{(1)},\nu^{(2)}, e^{(3)},e^{(4)}) \\
&+ 2 (p_1 \cdot p_2) m_e^2 \left(F_\mathrm{ann}^{LR}(\bnu^{(1)},\nu^{(2)},e^{(3)},e^{(4)}) + F_\mathrm{ann}^{RL}(\bnu^{(1)},\nu^{(2)},e^{(3)},e^{(4)}) \right) \Big] \, .
\end{aligned}
\end{equation}
Once again, they correspond to Eqs.~\eqref{eq:F_ann_cl} and~\eqref{eq:C_ann_cl} with the changes $\vrho \leftrightarrow \bvrho$ and $L \leftrightarrow R$.

\subsubsection{Antineutrino self-interactions}

\paragraph{Antineutrino-antineutrino scattering}

We can compare the relevant matrix element:
\begin{multline}
\tilde{v}^{\bnu_\beta(1) \bnu_\alpha(2)}_{ \bnu_\beta(3) \bnu_\alpha(4)} = (1 + \delta_{\alpha \beta}) \times \sqrt{2} G_F \ (2 \pi)^3 \, \delta^{(3)}(\vec{p}_1 + \vec{p}_2 - \vec{p}_3 - \vec{p}_4) \\
\times [\bar{v}_{\nu_\beta}^{h_3} (\vec{p}_3) \gamma^\mu P_L v_{\nu_\beta}^{h_1} (\vec{p}_1)] \ [\bar{v}_{\nu_\alpha}^{h_4} (\vec{p}_4)\gamma_\mu  P_L v_{\nu_\alpha}^{h_2} (\vec{p}_2)]
\end{multline}
with the one for neutrino-neutrino scattering given in Table~\ref{Table:MatrixElements}. Apart from the replacement of $u$ spinors by $v$ spinors, the appropriate exchanges are $1 \leftrightarrow 3$ and $2 \leftrightarrow 4$, which leaves the scattering amplitude unchanged compared to neutrino-neutrino scattering (see the calculation in section~\ref{subsec:collision_integral}), i.e. it will still be $(p_1 \cdot p_2)(p_3 \cdot p_4)$.

Concerning the statistical factor, we have for instance:
\[\tilde{v}^{\bnu_\beta \bnu_\gamma}_{\bnu_\gamma \bnu_\beta} \vrho^{\bar{\gamma}}_{\bar{\sigma}}(3) \vrho^{\bar{\beta}}_{\bar{\delta}}(4) \tilde{v}^{\bnu_\sigma \bnu_\delta}_{\bnu_\sigma \bnu_\delta} (\Id -\vrho^{(1)})^{\bar{\sigma}}_{\bar{\alpha}} (\Id -\vrho^{(2)})^{\bar{\delta}}_{\bar{\gamma}} \propto \left[(\Id -\bvrho^{(1)})  \cdot \bvrho^{(3)} \cdot (\Id - \bvrho^{(2)}) \cdot \bvrho^{(4)} \right]^\alpha_\beta \, , \]
such that the full statistical factor is identical to~\eqref{eq:F_sc_nn} with the replacement $\vrho \to \bvrho$, that is
\begin{multline}
\label{eq:F_sc_bnbn}
F_\mathrm{sc}(\bnu^{(1)},\bnu^{(2)},\bnu^{(3)},\bnu^{(4)}) =  \left[ \bvrho_4 (\Id - \bvrho_4) + \Tr(\cdots) \right] \bvrho_3 (\Id -\bvrho_1) + (\Id - \bvrho_1) \bvrho_3 \left[ (\Id - \bvrho_2) \bvrho_4 + \Tr(\cdots)\right]  \\
- \left[ (\Id - \bvrho_4) \bvrho_2  + \Tr(\cdots)\right] (\Id -\bvrho_3)  \bvrho_1 - \bvrho_1  (\Id -\bvrho_3)  \left[\bvrho_2(\Id -\bvrho_4)  + \Tr(\cdots)\right]  \, .
\end{multline}

\paragraph{Antineutrino-neutrino scattering/annihilation} Now the appropriate exchange from the neutrino-antineutrino matrix elements is $p_1 \leftrightarrow p_2$ and $p_3 \leftrightarrow p_4$, such that the prefactor $(p_1 \cdot p_2)(p_3 \cdot p_4)$ is left invariant.

The expressions for the statistical factors are Eqs.~\eqref{eq:F_sc_nbn} and~\eqref{eq:F_ann_nn} with $\vrho \leftrightarrow \bvrho$:
\begin{multline}
\label{eq:F_sc_bnn}
F_\mathrm{sc}(\bnu^{(1)},{\nu}^{(2)},\bnu^{(3)},{\nu}^{(4)}) = \left[ (\Id - {\varrho}_2) {\varrho}_4 + \Tr(\cdots) \right] \bvrho_3 (\Id -\bvrho_1) + (\Id - \bvrho_1) \bvrho_3 \left[ {\varrho}_4 (\Id - {\varrho}_2) + \Tr(\cdots)\right]  \\
- \left[ {\varrho}_2 (\Id -{\varrho}_4) + \Tr(\cdots) \right] (\Id -\bvrho_3) \bvrho_1 - \bvrho_1 (\Id -\bvrho_3) \left[ (\Id -{\varrho}_4) {\varrho}_2 + \Tr(\cdots)\right] \, ,
\end{multline}
\begin{multline}
\label{eq:F_ann_bnbn}
F_\mathrm{ann}(\bnu^{(1)},{\nu}^{(2)},\bnu^{(3)},{\nu}^{(4)}) = \left[ \bvrho_3 {\varrho}_4 + \Tr(\cdots) \right] (\Id -{\varrho}_2) (\Id -\bvrho_1) + (\Id - \bvrho_1) (\Id -{\varrho}_2) \left[ {\varrho}_4 \bvrho_3 + \Tr(\cdots)\right]  \\
- \left[ (\Id -\bvrho_3) (\Id -{\varrho}_4) + \Tr(\cdots) \right] {\varrho}_2 \bvrho_1 - \bvrho_1 {\varrho}_2 \left[ (\Id -{\varrho}_4) (\Id -\bvrho_3) + \Tr(\cdots)\right] \, .
\end{multline}

\vspace{0.5cm}

\noindent Finally, the full self-interaction antineutrino collision term reads
\begin{equation}
\begin{aligned}
\bar{\mathcal{C}}^{[\bnu  \bnu]} = &(2 \pi)^3 \delta^{(3)}(\vec{p}_1-\vec{p}_{\underline{1}}) \frac{2^5 G_F^2}{2}\int{[\dd^3 \vec{p}_2] [\dd^3 \vec{p}_3] [\dd^3 \vec{p}_4] (2 \pi)^4 \delta^{(4)}(p_1 + p_2 - p_3 - p_4)} \\
&\Big[ 4 (p_1 \cdot p_2)(p_3 \cdot p_4) F_\mathrm{sc}(\bnu^{(1)},\bnu^{(2)},\bnu^{(3)},\bnu^{(4)})  \\
&+ (p_1 \cdot p_4)(p_2 \cdot p_3)\left( F_\mathrm{sc}(\bnu^{(1)},\nu^{(2)}, \bnu^{(3)}, \nu^{(4)} + F_\mathrm{ann}(\bnu^{(1)},\nu^{(2)}, \bnu^{(3)}, \nu^{(4)}) \right)  \Big] \, .
\end{aligned}
\end{equation}

\section{Reduction of the collision integral}
\label{app:reduc_collision_integral}

For completeness, we detail the reduction of the collision integral from nine to two dimensions, following~\cite{Dolgov_NuPhB1997}. 

\subsection{Method}

The collision integral for each reaction reads generally, as shown in the calculations of the previous sections (recall that $\mathcal{C} = (2 \pi)^3 \, 2 E_1\, \delta^{(3)}(\vp_1 - \vp_{\underline{1}}) \mathcal{I}[\vrho]$):
\begin{equation}
\label{eq:general_I_app}
\mathcal{I}= \frac{1}{2 E_1} \int{[\dd^3 \vec{p}_2] [\dd^3 \vec{p}_3] [\dd^3 \vec{p}_4] \, (2 \pi)^4 \delta^{(4)}(p_1 + p_2 - p_3 - p_4)} \times S \langle \abs{\mathcal{M}}^2\rangle \times F[\vrho] \, ,
\end{equation}
with $F[\vrho]$ the statistical factor, $\langle \abs{\mathcal{M}}^2 \rangle$ the reaction matrix element and $S$ the symmetrization factor.\footnote{The symmetrization factor appears as part of the usual Feynman rules in diagrammatic Quantum Field Theory. Within the BBGKY formalism, these numerical prefactors arise from the combination of the interaction matrix elements $\tilde{v}$ — which is absolutely equivalent.} As explained in section~\ref{subsec:reduced_equations_QKE}, the key trick then consists in using the integral representation of the Dirac delta function:
\[ \delta^{(3)}(\vp_1 + \vp_2 - \vp_3 - \vp_4) = \int{\frac{\dd^3 \vec{\lambda}}{(2 \pi)^3} \, e^{\ii \vec{\lambda} \cdot (\vp_1 + \vp_2 - \vp_3 - \vp_4)}} \, , \]
and decompose the entire collision integral with spherical coordinates. The “$\vec{e}_\z$ unit vector” for $\vec{\lambda}$ is aligned with $\vp_1$, while $\vec{\lambda}$ is the “$\vec{e}_\z$ unit vector” for $\vp_{i \geq 2}$, that is,
\[ \cos{\theta_\lambda} \equiv \frac{\vp_1 \cdot \vec{\lambda}}{p_1 \lambda} \qquad ; \qquad \cos{\theta_i} \equiv \frac{\vp_i \cdot \vec{\lambda}}{p_i \lambda} \ \ \text{for } i = 2,3,4 \, ,  \]
the associated azimuthal angles $\varphi_\lambda, \varphi_{i \geq 2}$ being defined as usual. Recalling that $[\dd^3 \vec{p}] = \dd^3 \vec{p}/(2 \pi)^3 2 E = p^2 \dd{p} \dd{\Omega} / (2 \pi)^3 2 E$, with $\dd{\Omega} = \sin{\theta} \dd{\theta} \dd{\varphi}$ for the solid angles, we rewrite~\eqref{eq:general_I_app}
\begin{align*}
\mathcal{I} &= \frac{1}{2^4 E_1} \frac{1}{(2 \pi)^8} \int{\lambda^2 \dd{\lambda} \dd{\Omega_\lambda}}\prod_{i=2}^{4}\frac{p_i^2 \dd{p_i} \dd{\Omega_i}}{E_i} \, e^{\ii \vec{\lambda}\cdot(\vp_1 + \vp_2 - \vp_3 - \vp_4)} \\ 
&\qquad \qquad \qquad \times \delta(E_1 + E_2 - E_3 - E_4) \, S \langle \abs{\mathcal{M}}^2\rangle \, F[\vrho] \\
&= \frac{1}{2^6 \pi^3 E_1 p_1} \int{\prod_{i=2}^{4}{\frac{p_i \dd{p_i}}{E_i}} \, \delta(E_1 + E_2 - E_3 - E_4) \, F[\vrho] \times D(p_1, p_2, p_3, p_4)} \, ,
\end{align*}
where we defined
\begin{multline}
\label{eq:definition_Dfunction}
D(p_1,p_2,p_3,p_4) \equiv \frac{p_1 p_2 p_3 p_4}{2^6 \pi^5} \int_{0}^{\infty}{\lambda^2 \dd{\lambda}} \int{e^{\ii \vp_1 \cdot \vec{\lambda}} \dd{\Omega_\lambda}} \int{e^{\ii \vp_2 \cdot \vec{\lambda}} \dd{\Omega_2}} \\ \int{e^{-\ii \vp_3 \cdot \vec{\lambda}} \dd{\Omega_3}} \int{e^{-\ii \vp_4 \cdot \vec{\lambda}} \dd{\Omega_4}} \ S \langle \abs{\mathcal{M}}^2\rangle \, .
\end{multline}
It is finally the particular form of the matrix elements $S \langle \abs{\mathcal{M}}^2\rangle$ which allows for further simplifications. Indeed, in the Fermi approximation of weak interactions, there are only two kinds of matrix elements (see equations~\eqref{eq:C_nnscatt}, \eqref{eq:C_sc_cl_full}, \eqref{eq:C_ann_cl} and similarly for antineutrinos):
\begin{equation}
\label{eq:app_matrix_el}
\begin{aligned}
 K (q_1 \cdot q_2)(q_3 \cdot q_4) &= K (E_1 E_2 - \vec{q}_1 \cdot \vec{q}_2)(E_3 E_4 - \vec{q}_3 \cdot \vec{q}_4) \\
  \text{and} \qquad \qquad K' m_e^2 (q_3 \cdot q_4) &= K' m_e^2 (E_3 E_4 - \vec{q}_3 \cdot \vec{q}_4) \, ,
\end{aligned}
\end{equation}
where each $q_i$ corresponds to one of the $p_i$. The scalar products appearing in the matrix elements are thus explicitly
\[ \vec{q}_i \cdot \vec{q}_j = q_i q_j (\sin{\theta_i} \sin{\theta_j} \cos(\varphi_i - \varphi_j) + \cos{\theta_i} \cos{\theta_j}) \, , \]
but the first term vanishes after the $\varphi$ integration thanks to the isotropy of the system.\footnote{Simply stated, it comes from $\int_{0}^{2 \pi}{\cos{\varphi} \dd{\varphi}} = 0$.} With the second term, the $\varphi$ integrations give factors of $(2 \pi)$, which leads to two possible kinds of integrals appearing in $D$ depending on the matrix element:
\begin{subequations}
\label{eq:plus_ou_moins}
\begin{align}
\int{e^{\pm \ii \lambda q \cos{\theta}} \sin{\theta} \, \dd{\theta}} &= \frac{2}{\lambda q} \sin{(\lambda q)}  \, , \\
\int{e^{\pm \ii \lambda q \cos{\theta}} \cos{\theta}  \sin{\theta} \, \dd{\theta}} &= \mp \frac{2 \ii}{\lambda q} \left[\cos{(\lambda q)} - \frac{\sin{(\lambda q)}}{\lambda q}\right] \, .
\end{align}
\end{subequations}

With this intermediate result, we can perform all integrations except the $\lambda$ one, and obtain the expressions for the $D-$functions depending on the matrix element gathered in Table~\ref{tab:integral_D_functions}, where we define $D_1$, $D_2$ and $D_3$.

\renewcommand{\arraystretch}{0.1}

\begin{table}[!htb]
	\centering
	\begin{tabular}{|M{2.1cm} |M{12cm} |}
  	\hline 
	\vspace{0.15cm} $S \langle \abs{\mathcal{M}}^2\rangle$ \vspace{0.15cm} &  $D(q_1,q_2,q_3,q_4)$ \\ \hline \hline
	\[1\] & \[D_1 = \frac{4}{\pi} \int_{0}^{\infty}{\frac{\dd \lambda}{\lambda^2} \sin{(\lambda q_1)} \sin{(\lambda q_2)} \sin{(\lambda q_3)} \sin{(\lambda q_4)}}\] \\
	\[- \vec{q}_3 \cdot \vec{q}_4\] & \begin{multline*} D_2(3,4) = \frac{4 q_3 q_4}{\pi} \int_{0}^{\infty}\frac{\dd \lambda}{\lambda^2} \sin{(\lambda q_1)} \sin{(\lambda q_2)} \\ \times \left[\cos{(\lambda q_3)} - \frac{\sin{(\lambda q_3)}}{\lambda q_3}\right] \left[\cos{(\lambda q_4)} - \frac{\sin{(\lambda q_4)}}{\lambda q_3}\right] \end{multline*} \\
	\[(\vec{q}_1 \cdot \vec{q}_2)(\vec{q}_3 \cdot \vec{q}_4)\] & \[D_3 = \frac{4 q_1 q_2 q_3 q_4}{\pi} \int_{0}^{\infty}\frac{\dd \lambda}{\lambda^2} \left[\cos{(\lambda q_1)} - \frac{\sin{(\lambda q_1)}}{\lambda q_1}\right] \! \cdots \! \left[\cos{(\lambda q_4)} - \frac{\sin{(\lambda q_4)}}{\lambda q_3}\right]\] \\ \hline
\end{tabular}
	\caption[Integral expression of the $D-$functions]{Integral expression of the $D-$functions. It is important to emphasize that if the two arguments of $D_2$ do not correspond to both incoming or both outgoing particles, it changes sign — see equation~\eqref{eq:plus_ou_moins}.
	\label{tab:integral_D_functions}}
\end{table}

\renewcommand{\arraystretch}{1}

When inserting the matrix elements~\eqref{eq:app_matrix_el} in the definition~\eqref{eq:definition_Dfunction}, we thus get\footnote{There is a typo in equation (A.14) of~\cite{Dolgov_NuPhB1997}.}
\begin{equation}
\begin{aligned}
 D &= K \left[ E_1 E_2 E_3 E_4 D_1 + D_3 + E_1 E_2 D_2(3,4) + E_3 E_4 D_2(1,2) \right]  \\
  \text{and} \qquad \quad D &= K' m_e^2 \left[E_3 E_4 D_1 + D_2(3,4) \right] \, .
\end{aligned}
\end{equation}

\subsection{Expressions of the $D-$functions}

As shown in the expressions of Table~\ref{tab:integral_D_functions}, $D_1$ and $D_3$ are symmetric with respect to permutations of any variables, while $D_2$ is symmetric under the permutations $1 \leftrightarrow 2$ and $3 \leftrightarrow 4$. Therefore, we give in the following the expressions in the case $q_1 > q_2$ and $q_3 > q_4$, without loss of generality.

Distinguishing the following four physical cases, the integrals of Table~\ref{tab:integral_D_functions} can be computed and simplified using, e.g., \emph{Mathematica}.

\paragraph{Case (a)} $q_1 + q_2 > q_3 + q_4$, $q_1 + q_4 > q_2 + q_3$ and $q_1 \leq q_2 + q_3 + q_4$
\begin{subequations}
\begin{align}
	D_1 &= \frac12 (q_2 + q_3 + q_4 - q_1) \, , \\
	D_2 &= \frac{1}{12} \left( (q_1 - q_2)^3 + 2 (q_3^3 + q_4^3) - 3 (q_1 - q_2)(q_3^2 + q_4^2) \right) \, , \\
	D_3 &= \frac{1}{60} \left(q_1^5 - 5 q_1^3 q_2^2 + 5 q_1^2 q_2^3 - q_2^5 - 5 q_1^3 q_3^2 + 5 q_2^3 q_3^2 + 5 q_1^2 q_3^3 + 5 q_2^2 q_3^3 \right. \nonumber \\
	&\qquad \quad \left. - q_3^5 - 5 q_1^3q_4^2 + 5 q_2^3 q_4^2 + 5 q_3^3 q_4^2 + 5 q_1^2 q_3^3 + 5 q_2^2 q_4^3 + 5 q_3^2 q_4^3 - q_4^5 \right) \, .
\end{align}
\end{subequations}
Having $q_1 > q_2 + q_3 + q_4$ would be unphysical, and yields $D_1 = D_2 = D_3 = 0$.

\paragraph{Case (b)} $q_1 + q_2 > q_3 + q_4$ and $q_1 + q_4 < q_2 + q_3$
\begin{subequations}
\begin{align}
	D_1 &= q_4 \, , \\
	D_2 &= \frac13 q_4^3 \, , \\
	D_3 &= \frac{1}{30} q_4^3 \left(5 q_1^2 + 5 q_2^2 + 5 q_3^2 - q_4^2 \right) \, .
\end{align}
\end{subequations}

\paragraph{Case (c)} $q_1 + q_2 < q_3 + q_4$, $q_1 + q_4 < q_2 + q_3$ and $q_3 \leq q_1 + q_2 + q_4$
\begin{subequations}
\begin{align}
	D_1 &= \frac12 (q_1 + q_2 + q_3 - q_4) \, , \\
	D_2 &=  \frac{1}{12} \left( - (q_1 + q_2)^3 - 2 q_3^3 + 2 q_4^3 + 3 (q_1 + q_2)(q_3^2 + q_4^2) \right) \, , \\
	D_3 &= \frac{1}{60} \left(- q_1^5 + 5 q_1^3 q_2^2 + 5 q_1^2 q_2^3 - q_2^5 + 5 q_1^3 q_3^2 + 5 q_2^3 q_3^2 - 5 q_1^2 q_3^3 - 5 q_2^2 q_3^3  \right. \nonumber \\
	&\qquad \quad \left. + q_3^5 + 5 q_1^3 q_4^2 + 5 q_2^3 q_4^2 - 5 q_3^3 q_4^2 + 5 q_1^2 q_4^3 + 5 q_2^2 q_4^3 + 5 q_3^2 q_4^3 - q_4^5 \right) \, .
\end{align}
\end{subequations}
The expression for $D_3$ corresponds to case (a) with the exchanges $q_1 \leftrightarrow q_3$ and $q_2 \leftrightarrow q_4$. The case $q_3 > q_1 + q_2 + q_4$ would be unphysical, and yields $D_1 = D_2 = D_3 = 0$.

\paragraph{Case (d)} $q_1 + q_2 < q_3 + q_4$ and $q_1 + q_4 > q_2 + q_3$
\begin{subequations}
\begin{align}
	D_1 &= q_2 \, , \\
	D_2 &= \frac16 q_2 \left(3 q_3^2 + 3 q_4^2 - 3 q_1^2 - q_2^2 \right) \, , \\
	D_3 &= \frac{1}{30} q_2^3 \left(5 q_1^2 + 5 q_3^2 + 5 q_4^2 - q_2^2 \right) \, .
\end{align}
\end{subequations}

\pagestyle{ruled}

\chapter[The numerical code \texttt{NEVO}][The numerical code NEVO]{The numerical code \texttt{NEVO}}
\label{App:Numerics}

\setlength{\epigraphwidth}{0.53\textwidth}
\epigraph{Is this an instrument of communication or torture?}{Lady Violet Crawley, \emph{Downton Abbey} [S02E05]}

{
\hypersetup{linkcolor=black}
    \minitoc
}

During this PhD, we have developed the numerical code \texttt{NEVO} (\texttt{N}eutrino \texttt{EVO}lver) to follow neutrino evolution in the early Universe. We presented its main features in chapter~\ref{chap:Decoupling}, and give some additional information in this Appendix. First, we review in more details how the degrees of freedom in density matrices are serialized.  We then provide an extensive description of our method of calculation of the Jacobian of the differential system of equations, in particular how it is extended from the \ATAOH (chapter~\ref{chap:Decoupling}) to the \ATAOJH scheme (chapter~\ref{chap:Asymmetry}). Indeed, we show that this implementation allows to gain an order of magnitude in computation time, whether we consider the standard calculation or the asymmetric case.

%

\section{Serialization of density matrices in flavour space}

Since the spectra of density matrices are sampled on a grid of $N$ comoving momenta, the $\{y_n\}$, the variables which need to be solved for are the $\vrho_{\alpha\beta}(y_n)$, and $\bvrho_{\alpha\beta}(y_n)$. In the discretized numerical resolution, integrals are replaced by a quadrature method, that is by a weighted sum on the $y_n$, the weights depending on the chosen grid points. 

In order to alleviate the explanations, we will ignore the presence of antineutrinos and we shall consider that the variables are just the $\vrho_{\alpha\beta}(y_n)$, for which we use the short notation  $\vrho_{\alpha\beta,n}$.  In the \ATAOJH method, this is clearly wrong since an asymmetry matrix $\Anti_{\alpha\beta}$ requires that neutrinos and antineutrinos should have different distributions and one must always evolve both neutrinos and antineutrinos --- but that does not change the arguments presented here.

For each $n$, we serialize the matrix components.  This requires to define a basis ${P^a}$ for hermitian matrices with $N_\nu^2$ elements. These are divided into $N_\nu(N_\nu+1)/2$ basis matrices for the real components and $N_\nu(N_\nu-1)/2$ for the imaginary components. For instance when $N_\nu=2$ the matrices are 
\begin{equation}
P^1 = \begin{pmatrix} 
1&0\\
0&0\end{pmatrix} \, , \quad
P^2 = \begin{pmatrix} 
0&0\\
0&1\end{pmatrix} \, , \quad
P^3 = \begin{pmatrix} 
0&1\\
1&0\end{pmatrix} \, , \quad
P^4 = \begin{pmatrix} 
0& - \ii\\
\ii&0\end{pmatrix}\,.
\end{equation}
An inner product between two hermitian matrices is $(A,B) \equiv \Tr(A \cdot B^\dagger)$, hence the norms of the basis matrices are
\begin{equation}
\norm{P^a}^2 = \sum_{\alpha\beta} |{P^a}_{\alpha\beta}| ^2= \Tr(P^a \cdot P^{a\dagger})\,.
\end{equation}
Any density matrix is decomposed on serialized components as
\begin{equation}
\vrho_{\alpha \beta,n} \equiv \vrho_{a,n} {P^a}_{\alpha\beta}\,.
\end{equation}
The serialized components are also related to the components in the matter basis through
\begin{equation}\label{DefPaij}
\widetilde \vrho_{ij,n} = \vrho_{a,n} {P^{a}}_{ij}(n)\quad \text{with} \quad {P^{a}}_{ij}(n) \equiv [U^\dagger(n) \cdot P^a \cdot U(n)]_{ij} \, ,
\end{equation}
where $U(n)$ stands for $U_{\Hcal}(y_n)$, $\Hcal$ being the appropriate Hamiltonian (it will depend on the numerical scheme chosen). Note that we use the same notation $P^a$ in the matter ($P^a_{ij}$) and flavour ($P^a_{\alpha \beta}$) bases, the difference being identified through the indices. Conversely the serialized components $\vrho_{a,n}$ are obtained from
\begin{equation}\label{Convertrhoarhoi}
\vrho_{a,n} =  \vrho_{\alpha\beta,n} {P^{\alpha\beta}}_{a}(n)=\frac{\widetilde{\vrho}_{ij,n} {P^{a}}_{ji}(n)}{\norm{P^a}^2 }\quad \text{with} \quad {P^{\alpha\beta}}_{a}(n) \equiv \frac{ {P^{a\star}}_{\alpha\beta}(n)}{\norm{P^a}^2 } =\frac{ {P^{a}}_{\beta \alpha}(n) }{\norm{P^a}^2 }\,.
\end{equation}
In any ATAO scheme, we are only interested in the diagonal components of the matter basis, since by construction all off-diagonal components vanish, hence we define $\widetilde \vrho_{i,n} \equiv \widetilde \vrho_{ii,n}$ and obtain the following relations between the serialized (flavour) basis and the (diagonal) matter components
\begin{equation}
\begin{aligned}
\widetilde \vrho_{i,n} &= \vrho_{a,n} T^{a}_i(n)\quad &&\text{with}\quad T^{a}_i(n) \equiv {P^{a}}_{ii}(n) \, ,  \\
\vrho_{a,n}&= \widetilde \vrho_{i,n} T^i_{a}(n) \quad &&\text{with}\quad T^i_{a}(n) \equiv \frac{1}{\norm{P^a}^2} T^{a}_i(n)  \,.
\end{aligned}
\end{equation}
Since the $U(n)$ depend both on $x$ and on $y_n$, the $T^a_i(n)$ and $T^i_a(n)$ also depend on these variables, that is for each time step they must be computed for all points of the momentum grid.

\section{Direct computation of the Jacobian}

A key feature of the code \texttt{NEVO} is the calculation of the Jacobian of the system of differential equations, which is performed directly instead of relying on the extremely time-consuming default finite difference method.

Throughout this section we use a prime to denote a derivative with respect to $x$, and we stress again that we do not mention antineutrinos for the sake of clarity, but all developments must be carried out taking them into account.

\subsection{QKE scheme}

We must solve for the evolution of $z$ and of the flavour space serialized variables $\vrho_{a,n}$. Equation~\eqref{D1Froustey2020} dictates the evolution of $z$. The evolution of the $\vrho_{a,n}$ is governed by~\eqref{eq:QKE_compact_asym}
\begin{equation}\label{eq:QKE_compactserialized}
\vrho_{a,n}' = M_{a,n}^c \vrho_{c,n}  + \mathcal{K}_{a,n} \, ,
\end{equation}
with 
\begin{equation}
M_{a,n}^c \equiv (\Hamil_{b,n} + \Hself_b) {C^{bc}}_a\quad\text{and}\quad {C^{bc}}_a \equiv -\ii[P^b,P^c]_{\alpha \beta} {P^{\alpha\beta}}_a\,.
\end{equation}
The first term in \eqref{eq:QKE_compactserialized} comes  from mean-field effects and the second term from collisions.
The associated Jacobian has the general structure
\begin{equation}\label{JacobNumQKE}
\begin{pmatrix}
0&0\\ \\
\displaystyle \left(\frac{\partial M_{a,n}^c }{\partial z} \vrho_{c,n} \right)& 
\displaystyle \left(\frac{\partial M_{a,n}^c }{\partial \vrho_{b,m}} \vrho_{c,n} +  M_{a,n}^b \delta^n_m \right)
 \end{pmatrix}+
 \begin{pmatrix}
\displaystyle \frac{\partial z'}{\partial z}& \displaystyle \frac{\partial z'}{\partial \vrho_{b,m}}\\ \\
\displaystyle \frac{\partial {\mathcal{K}}_{a,n}}{\partial z}& \displaystyle \frac{\partial {\mathcal{K}}_{a,n}}{\partial  \vrho_{b,m}}
 \end{pmatrix}
\end{equation}
where, as before, the first matrix is due to mean-field effects, and the second to collisions. The contributions from the mean-field effects require to calculate
\begin{equation}
 \frac{\partial M_{a,n}^c }{\partial z} =\frac{\partial \Hamil_{b,n}}{\partial z}{C^{bc}}_a\,, \qquad
 \frac{\partial M_{a,n}^c }{\partial \vrho_{b,m}} = \frac{\partial \Hself_d}{\partial \vrho_{b,m}} {C^{cd}}_a\,,
\end{equation}
and the quantity $\partial \Hself_d/\partial \vrho_{b,m}$ is read on the integral definition \eqref{DefJ}. The computation of the Jacobian associated with mean-field effects is at most $\mathcal{O}(N^2)$ when self-interactions are taken into account, and only $\mathcal{O}(N)$ when they are ignored or when there is no asymmetry. Let us now review the complexity of the remaining terms. 
\begin{itemize}
\item $\partial \mathcal{K}_{a,n}/ \partial \vrho_{b,m}$ is the time-consuming part. Since the complexity for computing the collision term is $\mathcal{O}(N^3)$, using a finite difference method would scale as $\mathcal{O}(N^4)$. A method to reduce the complexity to $\mathcal{O}(N_\nu^2 N^3)$ is detailed in \cite{Froustey2020}, hence considerably speeding the numerical resolution.
\item $\partial \mathcal{K}_{a,n}/\partial z$ is just the collision term where the contribution coming from the distributions of electrons/positrons is varied with respect to $z$. Hence it has the same $\mathcal{O}(N^3)$ complexity as the collision term.
\item $\partial z'/\partial z$ can be obtained from equation~\eqref{D1Froustey2020}. In practice we simply use a finite difference method.
\item $\partial z'/\partial \vrho_{b,m}$ is obtained from the chain rule as
\begin{equation}\label{dzpdrho}
\frac{\partial z'}{\partial \vrho_{b,m}} = \frac{\partial z'}{\partial \mathcal{K}_{a,n}} \frac{\partial \mathcal{K}_{a,n}}{\partial \vrho_{b,m}}
\end{equation}
and we only need the variation $\partial z'/\partial \mathcal{K}_{a,n}$ which is easily read on equation~\eqref{D1Froustey2020}. Indeed, since only the trace of the collision term sources $z'$, the only serialized components $a$ leading to a non-vanishing $\partial z'/\partial \mathcal{K}_{a,n}$ are those for which $\Tr(P^a) \neq 0$.
\end{itemize}

\subsection{\ATAOH scheme}

In the \ATAOH scheme we integrate $z$ with \eqref{D1Froustey2020}, and the diagonal components $\widetilde \vrho_i$ with \eqref{BasicATAOH}, that is
\begin{equation}\label{QKEcompact}
\widetilde{\vrho}'_{i,n}= \widetilde{\mathcal{K}}_{i,n}\,.
\end{equation}
Note that this is a very compact notation which hides the fact that what is known in general are the $\mathcal{K}_{\alpha \beta}(y_n)$ which depend on the $\vrho_{\alpha\beta}(y_n)$. Hence at each step one must transform the matter basis components $\widetilde \vrho_{i,n}$ to the flavour basis, compute the collision terms, and convert back into the matter basis, and keep only the diagonal terms. The relation between the diagonal matter basis components and the (flavour basis) serialized components reads
\begin{equation}
\label{eq:relation_K_Ktilde}
\widetilde{\mathcal{K}}_{i,n} = \mathcal{K}_{a,n} T^a_i(n).
\end{equation}

The general form of the Jacobian is then
\begin{equation}\label{JacobNumATAOH}
 \begin{pmatrix}
\displaystyle \frac{\partial z'}{\partial z}& \displaystyle \frac{\partial z'}{\partial \widetilde\vrho_{j,m}}\\ \\
\displaystyle \frac{\partial \widetilde{\mathcal{K}}_{i,n}}{\partial z}& \displaystyle \frac{\partial \widetilde{\mathcal{K}}_{i,n}}{\partial \widetilde \vrho_{j,m}}
 \end{pmatrix}
\end{equation}
since in this scheme there are no mean-field effects to solve for, as they are hidden in the evolution of the matter basis. Again a finite difference method to compute $\partial \widetilde{\mathcal{K}}_{i,n}/\partial \widetilde \vrho_{j,m}$ would be of complexity $\mathcal{O}(N^4)$, but using the method detailed in \cite{Froustey2020} it is reduced to a complexity $\mathcal{O}(N_\nu N^3)$. This is even slightly faster (reduced by a factor $N_\nu$) than for computing  $\partial \mathcal{K}_{a,n}/ \partial \vrho_{b,m}$ because there are only $N_\nu$ diagonal matter components instead of $N_\nu^2$ flavour components. Both Jacobian blocks are related thanks to
\begin{equation}\label{RelateJacob}
\frac{\partial \widetilde{\mathcal{K}}_{i,n}}{\partial \widetilde\vrho_{j,m}} = T^{a}_i(n) \frac{\partial {\mathcal{K}}_{a,n}}{\partial \vrho_{b,m}} T_{b}^j(m) \, . 
\end{equation}
Let us review the other blocks in \eqref{JacobNumATAOH}. First we use the chain rule
\begin{equation}\label{dzpdrhoi}
\frac{\partial z'}{\partial \widetilde \vrho_{j,m}} = \frac{\partial z'}{\partial \widetilde{\mathcal{K}}_{i,n}} \frac{\partial \widetilde{\mathcal{K}}_{i,n}}{\partial \widetilde\vrho_{j,m}}\,,
\end{equation}
where $\partial z' / \partial \widetilde{\mathcal{K}}_{i,n} $ is easily read from equation~\eqref{D1Froustey2020} since only the trace of the collision term sources $z'$. And finally $\partial \widetilde{\mathcal{K}}_{i,n} / \partial z $ is similar to the computation of a collision term, but with the contribution from the $e^\pm$ distribution varied upon $z$.

\subsection{\ATAOJH scheme}

The general \ATAOJH equation \eqref{BasicATAOJH}, when written explicitly in matter basis components and using the previous notation, is also of the form \eqref{QKEcompact}. However in the \ATAOJH we also solve at the same time the evolution of $\Anti_{\alpha\beta}$ given by \eqref{dJdx}. Note that it depends on the full collision term $\mathcal{K}_{\alpha \beta}(y_n)$ (or the $\mathcal{K}_{a,n}$ in serialized basis) and not just on the diagonal components in the matter basis $\widetilde{\mathcal{K}}_{i,n}$, contrary to the evolution of the $\widetilde \vrho_i$ in \eqref{QKEcompact}.

Since we supplement $z$ and  the $\widetilde\vrho_{i,n}$ with the $N_\nu^2$ variables $\Anti_a$, this extends the size of the Jacobian. We now show how the new blocks in the Jacobian can be computed, and that this preserves the $\mathcal{O}(N^3)$ complexity.  The general form of the Jacobian is 
\begin{equation}\label{JacobNum}
 \begin{pmatrix}
\displaystyle \frac{\partial z'}{\partial z}& \displaystyle \frac{\partial z'}{\partial \widetilde\vrho_{j,m}}& \displaystyle \frac{\partial z'}{ \partial \Anti_b}\\ \\
\displaystyle \frac{\partial \widetilde{\mathcal{K}}_{i,n}}{\partial z}& \displaystyle \frac{\partial \widetilde{\mathcal{K}}_{i,n}}{\partial \widetilde \vrho_{j,m}}& \displaystyle\frac{\partial \widetilde{\mathcal{K}}_{i,n}}{\partial \Anti_b}\\ \\
 \displaystyle \frac{\partial \Anti'_a}{\partial z}& \displaystyle \frac{\partial \Anti'_a}{\partial \widetilde\vrho_{j,m}}& \displaystyle \frac{\partial \Anti'_a}{\partial \Anti_b}
 \end{pmatrix} \, ,
\end{equation}
and only the blocks in the right column or the bottom line are specific to the \ATAOJH scheme. 
As detailed hereafter, in order to compute these new blocks we shall need the computation of $\partial \mathcal{K}_{a,n}/\partial \vrho_{b,m}$, which is what is needed when computing the Jacobian in the QKE method. The block $\partial \widetilde{\mathcal{K}}_{i,n}/\partial \widetilde \vrho_{j,m}$ is then deduced through \eqref{RelateJacob}.

We also need to know how the $U(n)$ vary when the components of $\Anti$ are varied. Let us define the set of anti-hermitian matrices
\begin{equation}\label{DefWan}
W^{a,n} \equiv \frac{\partial U(n)}{\partial \Anti_a} \cdot U^\dagger(n) = - U(n) \cdot \frac{\partial U^\dagger(n)}{\partial \Anti_a}\,.
\end{equation}
which allow to know how the flavour components vary when $\Anti$ varies, for fixed matter basis components. They are obtained thanks to 
\begin{equation}\label{CrypticW}
\left(W^{a,n}\right)_{ij}= \frac{\sqrt{2} G_F}{(xH)}\left(\frac{m_e}{x}\right)^3   \frac{\left[U^\dagger(n) \cdot P^a \cdot U(n)\right]_{ij}}{(\Hamil_{j,n} + \Hself_{j,n} -\Hamil_{i,n} - \Hself_{i,n})}\quad \text{for}\quad i\neq j\,, 
\end{equation}
where the $\left(W^{a,n}\right)_{ij}$ are the components of $W^{a,n}$ in the matter basis, that is they are defined as $\left[U^\dagger(n)\cdot W^{a,n} \cdot U(n)\right]_{ij}$\,. The $(\Hamil+\Hself)_{j,n}$ are the diagonal components of $\Hamil+\Hself$ in the matter basis, which are by definition its eigenvalues. The $W^{a,n}$ are then found by transforming~\eqref{CrypticW} to the flavour basis with $U(n)$. Using their definition~\eqref{DefWan}, one then finds 
\begin{equation}\label{drhoadAb}
\frac{\partial \vrho_{a,n}}{\partial \Anti_b} =\vrho_{c,n} [W^{b,n}, P^c]_{\alpha \beta} {P^{\alpha \beta}}_a(n)\,.
\end{equation}
We now have all the tools to compute the blocks in the Jacobian \eqref{JacobNum} that are specific to the presence of $\Anti$.

\begin{itemize}
\item $\partial \widetilde{\mathcal{K}}_{i,n}/\partial \Anti_b$ 

This is deduced from~\eqref{eq:relation_K_Ktilde}. Using the Leibniz rule, we deduce
\begin{equation}\label{dKidAb}
\frac{\partial \widetilde{\mathcal{K}}_{i,n}}{\partial \Anti_b} = \frac{\partial \mathcal{K}_{a,n}}{\partial \Anti_b}T^a_i(n) + \mathcal{K}_{a,n} \frac{\partial T^a_i(n)}{\partial \Anti_b}\quad \text{with} \quad \frac{\partial T^a_i(n)}{\partial \Anti_b} = - \left(U^\dagger(n) \cdot [ W^{b,n}, P^a ]\cdot U(n)\right)_{ii} \, .
\end{equation}

\item $\partial z'/\partial A_b$

We only need to apply the chain rule using equation \eqref{dKidAb} since
\begin{equation}
\frac{\partial z'}{\partial \Anti_b} = \frac{\partial z'}{\partial \widetilde{\mathcal{K}}_{i,n}} \frac{\partial \widetilde{\mathcal{K}}_{i,n}}{\partial \Anti_b}\,,
\end{equation}
with $\partial z'/\partial \widetilde{\mathcal{K}}_{i,n}$ already needed for equation \eqref{dzpdrhoi}.

\item $\partial \Anti'_a/\partial z$

This is similar to the treatment of equation~\eqref{dJdx} with the replacement $\Hamil \to \partial \Hamil/\partial z$ and $\mathcal{K} \to \partial \mathcal{K}/\partial z$.

\item $\partial \Anti'_a/\partial \widetilde \vrho_{j,n}$
 
Let us define the following derivative :
\begin{equation}\label{TotalStrangeDer}
\frac{\partial \Anti'_a}{\partial \vrho_{b,n}} \equiv \left(\left. \frac{\partial \Anti'_a}{\partial \vrho_{b,n}}\right|_\mathrm{mf} + \frac{\partial \Anti'_a}{\partial \mathcal{K}_{c,m}} \frac{\partial \mathcal{K}_{c,m}}{\partial \vrho_{b,n}}\right)\,,
\end{equation}
where it is understood that the first term corresponds to the mean-field term, that is the first term on the rhs of equation~\eqref{dJdx}, and the second term is the indirect contribution via the dependence of the collision term.  Since the integrals appearing in equation~\eqref{dJdx} are computed with a quadrature method, that is a sum on the grid of comoving momenta with appropriate weights, these derivatives select only one term in these sums (the one corresponding to the comoving momentum $y_n$). We then immediately get from equation~\eqref{Convertrhoarhoi} and the chain rule
 \begin{equation} 
\frac{\partial \Anti'_a}{\partial \widetilde\vrho_{j,n}} = \frac{\partial \Anti'_a}{\partial \vrho_{b,n}}T^j_b(n)\,.
\end{equation}

\item $\partial \Anti'_a/\partial \Anti_b$

Finally, using again the derivative \eqref{TotalStrangeDer}, we find a simple expression for the last block
\begin{equation}\label{dApdA}
\frac{\partial \Anti'_a}{\partial \Anti_b} = \frac{\partial \Anti'_a}{\partial \vrho_{c,n}} \frac{\partial \vrho_{c,n}}{\partial \Anti_b} \,.
\end{equation}
 
\end{itemize}
In practice, pairs of indices like ${i,n}$ or ${a,n}$ are also serialized (e.g. with $I = n N_\nu + i$ and $A = n N_\nu^2 +a$), such that all products with implicit summations in this section appear as matrix multiplications when implemented in the code.

\vspace{0.3cm}

To summarize we need to compute the Jacobian as in the QKE method, which gives $\partial \mathcal{K}_{a,n}/\partial \vrho_{b,m}$. It corresponds to the variation of all flavour components in the collision term with respect to variations in all flavour components of the density matrices. We also need to compute the $T^a_i(n)$ from equation~\eqref{DefPaij}, and the $W^{a,n}$ from equation~\eqref{CrypticW}. We then deduce from equation~\eqref{drhoadAb} the $\partial \vrho_{a,n}/\partial \Anti_b$. Finally, knowing $\partial z'/\partial \widetilde{\mathcal{K}}_i(n)$, $\left.\partial \Anti'_a/\partial \vrho_{b,n}\right|_\mathrm{mf}$ and $\partial \Anti'_a/\partial \mathcal{K}_{c,m}$ from the equations governing the evolution of $z$ and $\Anti_{\alpha\beta}$, we can compute the five new blocks of the Jacobian as described from equations~\eqref{dKidAb} to~\eqref{dApdA}. The step which is the most time-consuming is the first one, that is the computation of $\partial \mathcal{K}_{a,n}/\partial \vrho_{b,m}$, whose complexity is $\mathcal{O}(N^3)$. Since this is already the longest step in the direct computation of the Jacobian in the QKE scheme, we deduce that for large $N$ the direct computation of the Jacobian in the \ATAOJH method takes roughly the same time as the direct computation of the Jacobian in the QKE method.

\newpage

\renewcommand{\partmarkLOCAL}{Bibliography}
\bibintoc
\printbibliography

@article{Dolgov_2002PhysRep,
    author = "Dolgov, A. D.",
    title = "{Neutrinos in cosmology}",
    eprint = "hep-ph/0202122",
    archivePrefix = "arXiv",
    reportNumber = "INFN-FE",
    doi = "10.1016/S0370-1573(02)00139-4",
    journal = "Phys. Rept.",
    volume = "370",
    pages = "333--535",
    year = "2002"
}

@article{LesgourguesPastor,
    author = "Lesgourgues, Julien and Pastor, Sergio",
    title = "{Neutrino mass from Cosmology}",
    eprint = "1212.6154",
    archivePrefix = "arXiv",
    primaryClass = "hep-ph",
    doi = "10.1155/2012/608515",
    journal = "Adv. High Energy Phys.",
    volume = "2012",
    pages = "608515",
    year = "2012"
}

@article{PDG,
    author = "Zyla, P.A. and others",
    collaboration = "Particle Data Group",
    title = "{Review of Particle Physics}",
    doi = "10.1093/ptep/ptaa104",
    journal = "Prog. Theor. Exp. Phys.",
    volume = "2020",
    number = "8",
    pages = "083C01",
    year = "2021"
}

@article{deSalas_Mixing,
    author = "de Salas, P. F. and Forero, D. V. and Gariazzo, S. and Mart\'{i}nez-Mirav\'e, P. and Mena, O. and Ternes, C. A. and T\'ortola, M. and Valle, J. W. F.",
    title = "{2020 global reassessment of the neutrino oscillation picture}",
    eprint = "2006.11237",
    archivePrefix = "arXiv",
    primaryClass = "hep-ph",
    doi = "10.1007/JHEP02(2021)071",
    journal = "JHEP",
    volume = "02",
    pages = "071",
    year = "2021"
}

@article{Esteban2020,
    author = "Esteban, Ivan and Gonzalez-Garcia, M. C. and Maltoni, Michele and Schwetz, Thomas and Zhou, Albert",
    title = "{The fate of hints: updated global analysis of three-flavor neutrino oscillations}",
    eprint = "2007.14792",
    archivePrefix = "arXiv",
    primaryClass = "hep-ph",
    reportNumber = "IFT-UAM/CSIC-112, YITP-SB-2020-21",
    doi = "10.1007/JHEP09(2020)178",
    journal = "JHEP",
    volume = "09",
    pages = "178",
    year = "2020"
}

@article{Kelly2020,
    author = "Kelly, Kevin J. and Machado, Pedro A. N. and Parke, Stephen J. and Perez-Gonzalez, Yuber F. and Funchal, Renata Zukanovich",
    title = "{Neutrino mass ordering in light of recent data}",
    eprint = "2007.08526",
    archivePrefix = "arXiv",
    primaryClass = "hep-ph",
    reportNumber = "FERMILAB-PUB-20-330-T",
    doi = "10.1103/PhysRevD.103.013004",
    journal = "Phys. Rev. D",
    volume = "103",
    number = "1",
    pages = "013004",
    year = "2021"
}

@article{WMAP2012,
    author = "Hinshaw, G. and others",
    collaboration = "WMAP",
    title = "{Nine-Year Wilkinson Microwave Anisotropy Probe (WMAP) Observations: Cosmological Parameter Results}",
    eprint = "1212.5226",
    archivePrefix = "arXiv",
    primaryClass = "astro-ph.CO",
    doi = "10.1088/0067-0049/208/2/19",
    journal = "Astrophys. J. Suppl.",
    volume = "208",
    pages = "19",
    year = "2013"
}

@article{Planck18,
    author = "Aghanim, N. and others",
    collaboration = "Planck",
    title = "{Planck 2018 results. VI. Cosmological parameters}",
    eprint = "1807.06209",
    archivePrefix = "arXiv",
    primaryClass = "astro-ph.CO",
    doi = "10.1051/0004-6361/201833910",
    journal = "Astron. Astrophys.",
    volume = "641",
    pages = "A6",
    year = "2020"
}

@article{ACT:2020,
    author = "Aiola, Simone and others",
    collaboration = "ACT",
    title = "{The Atacama Cosmology Telescope: DR4 Maps and Cosmological Parameters}",
    eprint = "2007.07288",
    archivePrefix = "arXiv",
    primaryClass = "astro-ph.CO",
    doi = "10.1088/1475-7516/2020/12/047",
    journal = "JCAP",
    volume = "12",
    pages = "047",
    year = "2020"
}

@article{SPT-3G:2021,
    author = "Balkenhol, L. and others",
    collaboration = "SPT-3G",
    title = "{Constraints on \ensuremath{\Lambda}CDM extensions from the SPT-3G 2018 EE and TE power spectra}",
    eprint = "2103.13618",
    archivePrefix = "arXiv",
    primaryClass = "astro-ph.CO",
    reportNumber = "FERMILAB-PUB-21-565-PPD",
    doi = "10.1103/PhysRevD.104.083509",
    journal = "Phys. Rev. D",
    volume = "104",
    number = "8",
    pages = "083509",
    year = "2021"
}

@article{eBOSS,
    author = "Alam, Shadab and others",
    collaboration = "eBOSS",
    title = "{Completed SDSS-IV extended Baryon Oscillation Spectroscopic Survey: Cosmological implications from two decades of spectroscopic surveys at the Apache Point Observatory}",
    eprint = "2007.08991",
    archivePrefix = "arXiv",
    primaryClass = "astro-ph.CO",
    doi = "10.1103/PhysRevD.103.083533",
    journal = "Phys. Rev. D",
    volume = "103",
    number = "8",
    pages = "083533",
    year = "2021"
}

@ARTICLE{NotzoldRaffelt_NuPhB1988,
   author = {{N{\"o}tzold}, D. and {Raffelt}, G.},
    title = "{Neutrino dispersion at finite temperature and density}",
  journal = {Nucl. Phys. B},
     year = 1988,
    month = oct,
   volume = 307,
    pages = {924-936},
      doi = {10.1016/0550-3213(88)90113-7},
   adsurl = {http://adsabs.harvard.edu/abs/1988NuPhB.307..924N}
}

@ARTICLE{SiglRaffelt,
   author = {{Sigl}, G. and {Raffelt}, G.},
    title = "{General kinetic description of relativistic mixed neutrinos}",
  journal = {Nucl. Phys. B},
     year = 1993,
    month = sep,
   volume = "406",
    pages = {423-451},
      doi = {10.1016/0550-3213(93)90175-O},
   adsurl = {http://cdsads.u-strasbg.fr/abs/1993NuPhB.406..423S}
}

@article{Cirigliano2014,
    author = "Cirigliano, Vincenzo and Fuller, George M. and Vlasenko, Alexey",
    title = "{A New Spin on Neutrino Quantum Kinetics}",
    eprint = "1406.5558",
    archivePrefix = "arXiv",
    primaryClass = "hep-ph",
    reportNumber = "LA-UR-14-24564",
    doi = "10.1016/j.physletb.2015.04.066",
    journal = "Phys. Lett. B",
    volume = "747",
    pages = "27--35",
    year = "2015"
}

@article{Vlasenko_PhRevD2014,
    author = "Vlasenko, Alexey and Fuller, George M. and Cirigliano, Vincenzo",
    title = "{Neutrino Quantum Kinetics}",
    eprint = "1309.2628",
    archivePrefix = "arXiv",
    primaryClass = "hep-ph",
    reportNumber = "LA-UR-13-27035",
    doi = "10.1103/PhysRevD.89.105004",
    journal = "Phys. Rev. D",
    volume = "89",
    number = "10",
    pages = "105004",
    year = "2014"
}

@ARTICLE{Volpe_2013,
   author = {{Volpe}, C. and {V{\"a}{\"a}n{\"a}nen}, D. and {Espinoza}, C.
	},
    title = "{Extended evolution equations for neutrino propagation in astrophysical and cosmological environments}",
  journal = {Phys. Rev. D},
archivePrefix = "arXiv",
   eprint = {1302.2374},
 primaryClass = "hep-ph",
     year = 2013,
    month = jun,
   volume = 87,
   number = 11,
      eid = {113010},
    pages = {113010},
      doi = {10.1103/PhysRevD.87.113010},
   adsurl = {http://adsabs.harvard.edu/abs/2013PhRvD..87k3010V}
}

@article{SerreauVolpe,
    author = "Serreau, Julien and Volpe, Cristina",
    title = "{Neutrino-antineutrino correlations in dense anisotropic media}",
    eprint = "1409.3591",
    archivePrefix = "arXiv",
    primaryClass = "hep-ph",
    doi = "10.1103/PhysRevD.90.125040",
    journal = "Phys. Rev. D",
    volume = "90",
    number = "12",
    pages = "125040",
    year = "2014"
}

@ARTICLE{Volpe_2015,
   author = {{Volpe}, C.},
    title = "{Neutrino quantum kinetic equations}",
  journal = {Int. J. Mod. Phys. E},
archivePrefix = "arXiv",
   eprint = {1506.06222},
 primaryClass = "astro-ph.SR",
     year = 2015,
    month = sep,
   volume = 24,
      eid = {1541009},
    pages = {1541009},
      doi = {10.1142/S0218301315410098},
   adsurl = {http://adsabs.harvard.edu/abs/2015IJMPE..2441009V}
}

@article{KartavtsevRaffelt,
    author = "Kartavtsev, A. and Raffelt, G. and Vogel, H.",
    title = "{Neutrino propagation in media: Flavor, helicity, and pair correlations}",
    eprint = "1504.03230",
    archivePrefix = "arXiv",
    primaryClass = "hep-ph",
    reportNumber = "MPP-2015-45",
    doi = "10.1103/PhysRevD.91.125020",
    journal = "Phys. Rev. D",
    volume = "91",
    number = "12",
    pages = "125020",
    year = "2015"
}

@article{BlaschkeCirigliano,
    author = "Blaschke, Daniel N. and Cirigliano, Vincenzo",
    title = "{Neutrino Quantum Kinetic Equations: The collision term}",
    eprint = "1605.09383",
    archivePrefix = "arXiv",
    primaryClass = "hep-ph",
    reportNumber = "LA-UR-15-25029",
    doi = "10.1103/PhysRevD.94.033009",
    journal = "Phys. Rev. D",
    volume = "94",
    number = "3",
    pages = "033009",
    year = "2016"
}

@article{FidlerPitrou,
    author = "Fidler, Christian and Pitrou, Cyril",
    title = "{Kinetic theory of fermions in curved spacetime}",
    eprint = "1701.08844",
    archivePrefix = "arXiv",
    primaryClass = "cond-mat.stat-mech",
    doi = "10.1088/1475-7516/2017/06/013",
    journal = "JCAP",
    volume = "06",
    pages = "013",
    year = "2017"
}

@article{CalzettaHu,
    author = "Calzetta, E. and Hu, B. L.",
    title = "{Nonequilibrium Quantum Fields: Closed Time Path Effective Action, Wigner Function and Boltzmann Equation}",
    reportNumber = "MDDP-PP-87-104",
    doi = "10.1103/PhysRevD.37.2878",
    journal = "Phys. Rev. D",
    volume = "37",
    pages = "2878",
    year = "1988"
}

@ARTICLE{Simenel,
       author = {{Simenel}, C{\'e}dric and {Avez}, Beno{\^\i}t and {Lacroix}, Denis},
        title = "{Microscopic approaches for nuclear Many-Body dynamics: applications to nuclear reactions}",
         year = "2008",
        month = "6",
          eid = {arXiv:0806.2714},
        pages = {arXiv:0806.2714},
        doi={10.48550/arXiv.0806.2714},
archivePrefix = {arXiv},
       eprint = {0806.2714},
 primaryClass = {nucl-th},
       adsurl = {https://ui.adsabs.harvard.edu/abs/2008arXiv0806.2714S}
}

@article{Lac04,
title = "Nuclear collective vibrations in extended mean-field theory",
journal = "Prog. Part. Nucl. Phys.",
volume = "52",
pages = "497--563",
year = "2004",
issn = "0146-6410",
doi = "https://doi.org/10.1016/j.ppnp.2004.02.002",
url = "http://www.sciencedirect.com/science/article/pii/S0146641004000146",
author = "D. Lacroix and S. Ayik and Ph. Chomaz"
}

@ARTICLE{Lac14,
       author = {{Lacroix}, Denis and {Ayik}, Sakir},
        title = "{Stochastic quantum dynamics beyond mean field}",
      journal = {Eur. Phys. J. A},
         year = 2014,
        month = jun,
       volume = {50},
          eid = {95},
        pages = {95},
          doi = {10.1140/epja/i2014-14095-8},
archivePrefix = {arXiv},
       eprint = {1402.2393},
 primaryClass = {nucl-th},
       adsurl = {https://ui.adsabs.harvard.edu/abs/2014EPJA...50...95L}
}

@article{WangCassing1985,
    author = "Wang, S. J. and Cassing, W.",
    title = "{Explicit treatment of $N$-body correlations within a density-matrix formalism}",
    reportNumber = "GSI-PREPRINT-84-18",
    doi = "10.1016/0003-4916(85)90116-2",
    journal = "Annals Phys.",
    volume = "159",
    pages = "328--350",
    year = "1985"
}

@article{Cassing1990,
    author = "Cassing, W. and Mosel, U.",
    title = "{Many-body theory of high-energy heavy-ion reactions}",
    doi = "10.1016/0146-6410(90)90032-Y",
    journal = "Prog. Part. Nucl. Phys.",
    volume = "25",
    pages = "235--323",
    year = "1990"
}

@article{Reinhard1994,
  title={Correlations in nuclei and nuclear dynamics},
  author={Reinhard, P-G and Toepffer, C},
  journal={Int. J. Mod. Phys. E},
  volume={3},
  doi={https://doi.org/10.1142/S0218301394000139},
  %number={supp01},
  pages={435--521},
  year={1994},
  publisher={World Scientific}
}

@article{Berges:2004,
    author = "Berges, Juergen",
    editor = "Bracco, Mirian and Chiapparini, Marcelo and Ferreira, Erasmo and Kodama, Takeshi",
    title = "{Introduction to nonequilibrium quantum field theory}",
    eprint = "hep-ph/0409233",
    archivePrefix = "arXiv",
    doi = "10.1063/1.1843591",
    journal = "AIP Conf. Proc.",
    volume = "739",
    number = "1",
    pages = "3--62",
    year = "2004"
}

@article{Berges:2015,
    author = "Berges, Jurgen",
    title = "{Nonequilibrium Quantum Fields: From Cold Atoms to Cosmology}",
    eprint = "1503.02907",
    archivePrefix = "arXiv",
    primaryClass = "hep-ph",
    month = "3",
    year = "2015"
}

@article{Drewes:2017,
    author = "Drewes, M. and Garbrecht, B. and Hernandez, P. and Kekic, M. and Lopez-Pavon, J. and Racker, J. and Rius, N. and Salvado, J. and Teresi, D.",
    title = "{ARS Leptogenesis}",
    eprint = "1711.02862",
    archivePrefix = "arXiv",
    primaryClass = "hep-ph",
    doi = "10.1142/S0217751X18420022",
    journal = "Int. J. Mod. Phys. A",
    volume = "33",
    number = "05n06",
    pages = "1842002",
    year = "2018"
}

@article{Hannestad_PhRvD1995,
    author = "Hannestad, Steen and Madsen, Jes",
    title = "{Neutrino decoupling in the early universe}",
    eprint = "astro-ph/9506015",
    archivePrefix = "arXiv",
    reportNumber = "TAC-1995-006, IFA-95-14, AARHUS-ASTRO-1995-08",
    doi = "10.1103/PhysRevD.52.1764",
    journal = "Phys. Rev. D",
    volume = "52",
    pages = "1764--1769",
    year = "1995"
}

@article{Semikoz_Tkachev,
    author = "Semikoz, D.V. and Tkachev, I.I.",
    title = "{Condensation of bosons in kinetic regime}",
    eprint = "hep-ph/9507306",
    archivePrefix = "arXiv",
    reportNumber = "FERMILAB-PUB-95-220-A",
    doi = "10.1103/PhysRevD.55.489",
    journal = "Phys. Rev. D",
    volume = "55",
    pages = "489--502",
    year = "1997"
}

@article{Dolgov_NuPhB1997,
      author         = "Dolgov, A. D. and Hansen, S. H. and Semikoz, D. V.",
      title          = "{Nonequilibrium corrections to the spectra of massless
                        neutrinos in the early universe}",
      journal        = "Nucl. Phys. B",
      volume         = "503",
      year           = "1997",
      pages          = "426-444",
      doi            = "10.1016/S0550-3213(97)00479-3",
      eprint         = "hep-ph/9703315",
      archivePrefix  = "arXiv",
      primaryClass   = "hep-ph",
      reportNumber   = "TAC-1997-010",
      SLACcitation   = "%%CITATION = HEP-PH/9703315;%%"
}

@article{Dolgov_NuPhB1999,
    author = "Dolgov, A. D. and Hansen, S. H. and Semikoz, D. V.",
    title = "{Non-equilibrium corrections to the spectra of massless neutrinos in the early universe}",
    eprint = "hep-ph/9805467",
    archivePrefix = "arXiv",
    doi = "10.1016/S0550-3213(98)00818-9",
    journal = "Nucl. Phys. B",
    volume = "543",
    pages = "269--274",
    year = "1999"
}

@article{Esposito_NuPhB2000,
    author = "Esposito, S. and Miele, G. and Pastor, S. and Peloso, M. and Pisanti, O.",
    title = "{Nonequilibrium spectra of degenerate relic neutrinos}",
    eprint = "astro-ph/0005573",
    archivePrefix = "arXiv",
    reportNumber = "SISSA-51-2000-EP, DSF-16-00",
    doi = "10.1016/S0550-3213(00)00554-X",
    journal = "Nucl. Phys. B",
    volume = "590",
    pages = "539--561",
    year = "2000"
}

@article{Mangano2002,
    author = "Mangano, G. and Miele, G. and Pastor, S. and Peloso, M.",
    title = "{A precision calculation of the effective number of cosmological neutrinos}",
    eprint = "astro-ph/0111408",
    archivePrefix = "arXiv",
    reportNumber = "DSF-37-2001, MPI-PHT-2001-51",
    doi = "10.1016/S0370-2693(02)01622-2",
    journal = "Phys. Lett. B",
    volume = "534",
    pages = "8--16",
    year = "2002"
}

@ARTICLE{Mangano2005,
       author = {{Mangano}, Gianpiero and {Miele}, Gennaro and {Pastor}, Sergio and
         {Pinto}, Teguayco and {Pisanti}, Ofelia and {Serpico}, Pasquale D.},
        title = "{Relic neutrino decoupling including flavour oscillations}",
      journal = {Nucl. Phys. B},
         year = 2005,
        month = nov,
       volume = {729},
       number = {1-2},
        pages = {221-234},
          doi = {10.1016/j.nuclphysb.2005.09.041},
archivePrefix = {arXiv},
       eprint = {hep-ph/0506164},
 primaryClass = {hep-ph},
       adsurl = {https://ui.adsabs.harvard.edu/abs/2005NuPhB.729..221M}
}

@article{Grohs2015,
    author = "Grohs, E. and Fuller, G. M. and Kishimoto, C. T. and Paris, M. W. and Vlasenko, A.",
    title = "{Neutrino energy transport in weak decoupling and big bang nucleosynthesis}",
    eprint = "1512.02205",
    archivePrefix = "arXiv",
    primaryClass = "astro-ph.CO",
    doi = "10.1103/PhysRevD.93.083522",
    journal = "Phys. Rev. D",
    volume = "93",
    number = "8",
    pages = "083522",
    year = "2016"
}

@article{Relic2016_revisited,
    author = "de Salas, Pablo F. and Pastor, Sergio",
    title = "{Relic neutrino decoupling with flavour oscillations revisited}",
    eprint = "1606.06986",
    archivePrefix = "arXiv",
    primaryClass = "hep-ph",
    reportNumber = "IFIC-16-10, TTK-16-23",
    doi = "10.1088/1475-7516/2016/07/051",
    journal = "JCAP",
    volume = "07",
    pages = "051",
    year = "2016"
}

@article{Gariazzo_2019,
    author = "Gariazzo, S. and de Salas, P.F. and Pastor, S.",
    title = "{Thermalisation of sterile neutrinos in the early Universe in the 3+1 scheme with full mixing matrix}",
    eprint = "1905.11290",
    archivePrefix = "arXiv",
    primaryClass = "astro-ph.CO",
    doi = "10.1088/1475-7516/2019/07/014",
    journal = "JCAP",
    volume = "07",
    pages = "014",
    year = "2019"
}

@article{Froustey2019,
    author = "Froustey, Julien and Pitrou, Cyril",
    title = "{Incomplete neutrino decoupling effect on big bang nucleosynthesis}",
    eprint = "1912.09378",
    archivePrefix = "arXiv",
    primaryClass = "astro-ph.CO",
    doi = "10.1103/PhysRevD.101.043524",
    journal = "Phys. Rev. D",
    volume = "101",
    number = "4",
    pages = "043524",
    year = "2020"
}

@article{Escudero_2018,
    author = "Escudero, Miguel",
    title = "{Neutrino decoupling beyond the Standard Model: CMB constraints on the Dark Matter mass with a fast and precise $N_{\mathrm{eff}}$ evaluation}",
    eprint = "1812.05605",
    archivePrefix = "arXiv",
    primaryClass = "hep-ph",
    reportNumber = "KCL-2018-76",
    doi = "10.1088/1475-7516/2019/02/007",
    journal = "JCAP",
    volume = "02",
    pages = "007",
    year = "2019"
}

@article{Escudero_2020,
    author = "Escudero, Miguel",
    title = "{Precision early universe thermodynamics made simple: $N_{\mathrm{eff}}$ and neutrino decoupling in the Standard Model and beyond}",
    eprint = "2001.04466",
    archivePrefix = "arXiv",
    primaryClass = "hep-ph",
    reportNumber = "KCL-2019-85",
    doi = "10.1088/1475-7516/2020/05/048",
    journal = "JCAP",
    volume = "05",
    pages = "048",
    year = "2020"
}

@article{Akita2020,
    author = "Akita, Kensuke and Yamaguchi, Masahide",
    title = "{A precision calculation of relic neutrino decoupling}",
    eprint = "2005.07047",
    archivePrefix = "arXiv",
    primaryClass = "hep-ph",
    doi = "10.1088/1475-7516/2020/08/012",
    journal = "JCAP",
    volume = "08",
    pages = "012",
    year = "2020"
}

@article{Froustey2020,
    author = "Froustey, Julien and Pitrou, Cyril and Volpe, Maria Cristina",
    title = "{Neutrino decoupling including flavour oscillations and primordial nucleosynthesis}",
    eprint = "2008.01074",
    archivePrefix = "arXiv",
    primaryClass = "hep-ph",
    doi = "10.1088/1475-7516/2020/12/015",
    journal = "JCAP",
    volume = "12",
    pages = "015",
    year = "2020"
}

@article{Bennett2021,
    author = "Bennett, Jack J. and Buldgen, Gilles and De Salas, Pablo F. and Drewes, Marco and Gariazzo, Stefano and Pastor, Sergio and Wong, Yvonne Y. Y.",
    title = "{Towards a precision calculation of $N_{\mathrm{eff}}$ in the Standard Model II: Neutrino decoupling in the presence of flavour oscillations and finite-temperature QED}",
    eprint = "2012.02726",
    archivePrefix = "arXiv",
    primaryClass = "hep-ph",
    reportNumber = "CPPC-2020-10",
    doi = "10.1088/1475-7516/2021/04/073",
    journal = "JCAP",
    volume = "04",
    pages = "073",
    year = "2021"
}

@book{KapustaGale,
    author = "Kapusta, J. I. and Gale, Charles",
    title = "{Finite-temperature field theory: Principles and applications}",
    doi = "10.1017/CBO9780511535130",
    isbn = "978-0-521-17322-3",
    publisher = "Cambridge University Press",
    series = "Cambridge Monographs on Mathematical Physics",
    year = "2011"
}

@article{Dicus1982,
    author = "Dicus, Duane A. and Kolb, Edward W. and Gleeson, A. M. and Sudarshan, E. C. G. and Teplitz, Vigdor L. and Turner, Michael S.",
    title = "{Primordial Nucleosynthesis Including Radiative, Coulomb, and Finite Temperature Corrections to Weak Rates}",
    reportNumber = "LA-UR-82-1553, EFI-82-25-CHICAGO, DOE-ER-03992-494",
    doi = "10.1103/PhysRevD.26.2694",
    journal = "Phys. Rev. D",
    volume = "26",
    pages = "2694",
    year = "1982"
}

@article{Heckler_PhRvD1994,
    author = "Heckler, A. F.",
    title = "{Astrophysical applications of quantum corrections to the equation of state of a plasma}",
    doi = "10.1103/PhysRevD.49.611",
    journal = "Phys. Rev. D",
    volume = "49",
    pages = "611--617",
    year = "1994"
}

@article{Fornengo1997,
    author = "Fornengo, N. and Kim, C. W. and Song, J.",
    title = "{Finite temperature effects on the neutrino decoupling in the early universe}",
    eprint = "hep-ph/9702324",
    archivePrefix = "arXiv",
    reportNumber = "JHU-TIPAC-97003, DFTT-15-97, KIAS-P-97002",
    doi = "10.1103/PhysRevD.56.5123",
    journal = "Phys. Rev. D",
    volume = "56",
    pages = "5123--5134",
    year = "1997"
}

@article{LopezTurner1998,
    author = "Lopez, Robert E. and Turner, Michael S.",
    title = "{An Accurate Calculation of the Big Bang Prediction for the Abundance of Primordial Helium}",
    eprint = "astro-ph/9807279",
    archivePrefix = "arXiv",
    reportNumber = "FERMILAB-PUB-98-232-A",
    doi = "10.1103/PhysRevD.59.103502",
    journal = "Phys. Rev. D",
    volume = "59",
    pages = "103502",
    year = "1999"
}

@article{BrownSawyer,
    author = "Brown, Lowell S. and Sawyer, R. F.",
    title = "{Finite temperature corrections to weak rates prior to nucleosynthesis}",
    eprint = "astro-ph/0006370",
    archivePrefix = "arXiv",
    reportNumber = "UW-PT-00-03",
    doi = "10.1103/PhysRevD.63.083503",
    journal = "Phys. Rev. D",
    volume = "63",
    pages = "083503",
    year = "2001"
}

@article{Esposito_QED,
    author = "Esposito, S. and Mangano, G. and Miele, G. and Picardi, Ilenia and Pisanti, O.",
    title = "{Neutrino energy loss rate in a stellar plasma}",
    eprint = "astro-ph/0301438",
    archivePrefix = "arXiv",
    reportNumber = "DSF-21-2002",
    doi = "10.1016/S0550-3213(03)00151-2",
    journal = "Nucl. Phys. B",
    volume = "658",
    pages = "217--253",
    year = "2003"
}

@article{Tomalak2019,
    author = "Tomalak, Oleksandr and Hill, Richard J.",
    title = "{Theory of elastic neutrino-electron scattering}",
    eprint = "1907.03379",
    archivePrefix = "arXiv",
    primaryClass = "hep-ph",
    reportNumber = "FERMILAB-PUB-19-076-T",
    doi = "10.1103/PhysRevD.101.033006",
    journal = "Phys. Rev. D",
    volume = "101",
    number = "3",
    pages = "033006",
    year = "2020"
}

@article{Bennett2020,
    author = "Bennett, Jack J. and Buldgen, Gilles and Drewes, Marco and Wong, Yvonne Y.Y.",
    title = "{Towards a precision calculation of the effective number of neutrinos $N_{\mathrm{eff}}$ in the Standard Model I: The QED equation of state}",
    eprint = "1911.04504",
    archivePrefix = "arXiv",
    primaryClass = "hep-ph",
    doi = "10.1088/1475-7516/2020/03/003",
    journal = "JCAP",
    volume = "03",
    pages = "003",
    year = "2020"
}

@article{Grohs2019,
    author = "Thomas, Luke C. and Dezen, Ted and Grohs, Evan B. and Kishimoto, Chad T.",
    title = "{Electron-Positron Annihilation Freeze-Out in the Early Universe}",
    eprint = "1910.14050",
    archivePrefix = "arXiv",
    primaryClass = "hep-ph",
    doi = "10.1103/PhysRevD.101.063507",
    journal = "Phys. Rev. D",
    volume = "101",
    number = "6",
    pages = "063507",
    year = "2020"
}

@article{Mirizzi2012,
    author = "Mirizzi, Alessandro and Saviano, Ninetta and Miele, Gennaro and Serpico, Pasquale Dario",
    title = "{Light sterile neutrino production in the early universe with dynamical neutrino asymmetries}",
    eprint = "1206.1046",
    archivePrefix = "arXiv",
    primaryClass = "hep-ph",
    reportNumber = "LAPTH-026-12",
    doi = "10.1103/PhysRevD.86.053009",
    journal = "Phys. Rev. D",
    volume = "86",
    pages = "053009",
    year = "2012"
}

@article{Saviano2013,
    author = "Saviano, Ninetta and Mirizzi, Alessandro and Pisanti, Ofelia and Serpico, Pasquale Dario and Mangano, Gianpiero and Miele, Gennaro",
    title = "{Multi-momentum and multi-flavour active-sterile neutrino oscillations in the early universe: role of neutrino asymmetries and effects on nucleosynthesis}",
    eprint = "1302.1200",
    archivePrefix = "arXiv",
    primaryClass = "astro-ph.CO",
    reportNumber = "DF-2013-3, LAPTH-004-13",
    doi = "10.1103/PhysRevD.87.073006",
    journal = "Phys. Rev. D",
    volume = "87",
    pages = "073006",
    year = "2013"
}

@article{HannestadTamborra,
    author = "Hannestad, Steen and Tamborra, Irene and Tram, Thomas",
    title = "{Thermalisation of light sterile neutrinos in the early universe}",
    eprint = "1204.5861",
    archivePrefix = "arXiv",
    primaryClass = "astro-ph.CO",
    reportNumber = "MPP-2012-79",
    doi = "10.1088/1475-7516/2012/07/025",
    journal = "JCAP",
    volume = "07",
    pages = "025",
    year = "2012"
}

@article{Bogoliubov,
    author = "Bogoliubov, N.N.",
    title = "{Kinetic equations}",
    journal = "Journal of Physics USSR",
    volume = "10",
    number = "3",
    pages = "265--274",
    year = "1946"
}

@article{BornGreen,
    author = "Born, M. and Green, H.S.",
    title = "{A General Kinetic Theory of Liquids. I. The Molecular Distribution Functions}",
    doi = "10.1098/rspa.1946.0093",
    journal = "Proc. Roy. Soc. Lond. A",
    volume = "A188",
    number = "1012",
    pages = "10--18",
    year = "1946"
}

@ARTICLE{Kirkwood,
       author = {{Kirkwood}, John G.},
        title = "{The Statistical Mechanical Theory of Transport Processes I. General Theory}",
      journal = {J. Chem. Phys.},
         year = 1946,
        month = mar,
       volume = {14},
       number = {3},
        pages = {180-201},
          doi = {10.1063/1.1724117},
       adsurl = {https://ui.adsabs.harvard.edu/abs/1946JChPh..14..180K}
}

@book{Yvon,
  title={La th{\'e}orie statistique des fluides et l'{\'e}quation d'{\'e}tat},
  author={Yvon, Jacques},
  volume={203},
  year={1935},
  series = {Actualit{\'e}s scientifiques et industrielles},
  publisher={Hermann \& cie},
  pages={}
}

@article{Bell98,
    author = "Bell, Nicole F. and Volkas, Raymond R. and Wong, Yvonne Y. Y.",
    title = "{Relic neutrino asymmetry evolution from first principles}",
    eprint = "hep-ph/9809363",
    archivePrefix = "arXiv",
    reportNumber = "UM-P-98-43, RCHEP-98-09",
    doi = "10.1103/PhysRevD.59.113001",
    journal = "Phys. Rev. D",
    volume = "59",
    pages = "113001",
    year = "1999"
}

@article{Abazajian2002,
    author = "Abazajian, Kevork N. and Beacom, John F. and Bell, Nicole F.",
    title = "{Stringent Constraints on Cosmological Neutrino Antineutrino Asymmetries from Synchronized Flavor Transformation}",
    eprint = "astro-ph/0203442",
    archivePrefix = "arXiv",
    reportNumber = "FERMILAB-PUB-02-056-A",
    doi = "10.1103/PhysRevD.66.013008",
    journal = "Phys. Rev. D",
    volume = "66",
    pages = "013008",
    year = "2002"
}

@article{Wong2002,
    author = "Wong, Yvonne Y. Y.",
    title = "{Analytical treatment of neutrino asymmetry equilibration from flavor oscillations in the early universe}",
    eprint = "hep-ph/0203180",
    archivePrefix = "arXiv",
    doi = "10.1103/PhysRevD.66.025015",
    journal = "Phys. Rev. D",
    volume = "66",
    pages = "025015",
    year = "2002"
}

@article{Dolgov_NuPhB2002,
    author = "Dolgov, A. D. and Hansen, S. H. and Pastor, S. and Petcov, S. T. and Raffelt, G. G. and Semikoz, D. V.",
    title = "{Cosmological bounds on neutrino degeneracy improved by flavor oscillations}",
    eprint = "hep-ph/0201287",
    archivePrefix = "arXiv",
    reportNumber = "MPI-PHT-2001-52",
    doi = "10.1016/S0550-3213(02)00274-2",
    journal = "Nucl. Phys. B",
    volume = "632",
    pages = "363--382",
    year = "2002"
}

@article{Mangano:2010ei,
    author = "Mangano, Gianpiero and Miele, Gennaro and Pastor, Sergio and Pisanti, Ofelia and Sarikas, Srdjan",
    title = "{Constraining the cosmic radiation density due to lepton number with Big Bang Nucleosynthesis}",
    eprint = "1011.0916",
    archivePrefix = "arXiv",
    primaryClass = "astro-ph.CO",
    reportNumber = "IFIC-10-38",
    doi = "10.1088/1475-7516/2011/03/035",
    journal = "JCAP",
    volume = "03",
    pages = "035",
    year = "2011"
}

@article{Pastor:2008ti,
    author = "Pastor, Sergio and Pinto, Teguayco and Raffelt, Georg G.",
    title = "{Relic density of neutrinos with primordial asymmetries}",
    eprint = "0808.3137",
    archivePrefix = "arXiv",
    primaryClass = "astro-ph",
    reportNumber = "MPP-2008-105, IFIC-07-59",
    doi = "10.1103/PhysRevLett.102.241302",
    journal = "Phys. Rev. Lett.",
    volume = "102",
    pages = "241302",
    year = "2009"
}

@article{Grohs:2016cuu,
    author = "Grohs, E. and Fuller, George M. and Kishimoto, C. T. and Paris, Mark W.",
    title = "{Lepton asymmetry, neutrino spectral distortions, and big bang nucleosynthesis}",
    eprint = "1612.01986",
    archivePrefix = "arXiv",
    primaryClass = "astro-ph.CO",
    doi = "10.1103/PhysRevD.95.063503",
    journal = "Phys. Rev. D",
    volume = "95",
    number = "6",
    pages = "063503",
    year = "2017"
}

@article{Grohs:2020xxd,
    author = "Grohs, E. and Fuller, George M. and Sen, Manibrata",
    title = "{Consequences of neutrino self interactions for weak decoupling and big bang nucleosynthesis}",
    eprint = "2002.08557",
    archivePrefix = "arXiv",
    primaryClass = "astro-ph.CO",
    doi = "10.1088/1475-7516/2020/07/001",
    journal = "JCAP",
    volume = "07",
    pages = "001",
    year = "2020"
}

@article{Johns:2016enc,
    author = "Johns, Lucas and Mina, Mattia and Cirigliano, Vincenzo and Paris, Mark W. and Fuller, George M.",
    title = "{Neutrino flavor transformation in the lepton-asymmetric universe}",
    eprint = "1608.01336",
    archivePrefix = "arXiv",
    primaryClass = "hep-ph",
    reportNumber = "LA-UR-16-25128",
    doi = "10.1103/PhysRevD.94.083505",
    journal = "Phys. Rev. D",
    volume = "94",
    number = "8",
    pages = "083505",
    year = "2016"
}

@article{Froustey2021,
    author = "Froustey, Julien and Pitrou, Cyril",
    title = "{Primordial neutrino asymmetry evolution with full mean-field effects and collisions}",
    eprint = "2110.11889",
    archivePrefix = "arXiv",
    primaryClass = "hep-ph",
    doi = "10.1088/1475-7516/2022/03/065",
    journal = "JCAP",
    volume = "03",
    pages = "065",
    year = "2022"
}

@article{Barenboim:2016shh,
    author = "Barenboim, Gabriela and Kinney, William H. and Park, Wan-Il",
    title = "{Resurrection of large lepton number asymmetries from neutrino flavor oscillations}",
    eprint = "1609.01584",
    archivePrefix = "arXiv",
    primaryClass = "hep-ph",
    reportNumber = "FTUV-16-08-05, IFIC-16-60",
    doi = "10.1103/PhysRevD.95.043506",
    journal = "Phys. Rev. D",
    volume = "95",
    number = "4",
    pages = "043506",
    year = "2017"
}

@article{Pastor:2001iu,
    author = "Pastor, Sergio and Raffelt, Georg G. and Semikoz, Dmitry V.",
    title = "{Physics of synchronized neutrino oscillations caused by self-interactions}",
    eprint = "hep-ph/0109035",
    archivePrefix = "arXiv",
    reportNumber = "MPI-PHT-2001-36",
    doi = "10.1103/PhysRevD.65.053011",
    journal = "Phys. Rev. D",
    volume = "65",
    pages = "053011",
    year = "2002"
}

@article{Samuel:1993uw,
    author = "Samuel, Stuart",
    title = "{Neutrino oscillations in dense neutrino gases}",
    reportNumber = "IUHET-244",
    doi = "10.1103/PhysRevD.48.1462",
    journal = "Phys. Rev. D",
    volume = "48",
    pages = "1462--1477",
    year = "1993"
}

@article{Kostelecky:1993yt,
    author = "Kostelecky, V. Alan and Pantaleone, James T. and Samuel, Stuart",
    title = "{Neutrino oscillation in the early universe}",
    reportNumber = "IUHET-250",
    doi = "10.1016/0370-2693(93)90156-C",
    journal = "Phys. Lett. B",
    volume = "315",
    pages = "46--50",
    year = "1993"
}

@article{Kostelecky:1993dm,
    author = "Kostelecky, V. Alan and Samuel, Stuart",
    title = "{Neutrino oscillations in the early universe with an inverted neutrino mass hierarchy}",
    reportNumber = "IUHET-258, CCNY-HEP-93-3",
    doi = "10.1016/0370-2693(93)91795-O",
    journal = "Phys. Lett. B",
    volume = "318",
    pages = "127--133",
    year = "1993"
}

@article{Kostelecky:1993ys,
    author = "Kostelecky, V. Alan and Samuel, Stuart",
    title = "{Nonlinear neutrino oscillations in the expanding universe}",
    reportNumber = "IUHET-251",
    doi = "10.1103/PhysRevD.49.1740",
    journal = "Phys. Rev. D",
    volume = "49",
    pages = "1740--1757",
    year = "1994"
}

@article{Langacker1982a,
    author = "Langacker, Paul and Segre, Gino",
    title = "{Comment on `Does the Standard Hot Big Bang Model Explain the Primordial Abundances of Helium and Deuterium'?}",
    reportNumber = "UPR-0196T",
    doi = "10.1103/PhysRevLett.49.1363",
    journal = "Phys. Rev. Lett.",
    volume = "49",
    pages = "1363",
    year = "1982"
}

@article{Langacker1982b,
    author = "Langacker, Paul and Segre, Gino and Soni, Sanjeev",
    title = "{Majorana Neutrinos, Nucleosynthesis, and the Lepton Asymmetry of the Universe}",
    reportNumber = "UPR-0199T",
    doi = "10.1103/PhysRevD.26.3425",
    journal = "Phys. Rev. D",
    volume = "26",
    pages = "3425",
    year = "1982"
}

@article{Bernstein1982,
    author = "Bernstein, Jeremy",
    title = "{Comment on `Does the Standard Hot Big Bang Model Explain the Primordial Abundances of Helium and Deuterium'?}",
  journal = {Phys. Rev. Lett.},
  volume = {48},
  pages = {774--774},
  numpages = {0},
  year = {1982},
  month = {Mar},
  publisher = {American Physical Society},
  doi = {10.1103/PhysRevLett.48.774},
  url = {https://link.aps.org/doi/10.1103/PhysRevLett.48.774}
}

@article{Bernstein1984,
    author = "Bernstein, Jeremy",
    title = "{Neutrino Cosmology}",
    reportNumber = "CERN-84-06",
    doi = "10.5170/CERN-1984-006",
    month = "4",
    year = "1984"
}

@article{Shapiro1980,
    author = "Shapiro, S. L. and Teukolsky, S. A. and Wasserman, I.",
    title = "{Do neutrino rest masses affect cosmological Helium production?}",
    doi = "10.1103/PhysRevLett.45.669",
    journal = "Phys. Rev. Lett.",
    volume = "45",
    pages = "669--672",
    year = "1980"
}

@article{Kreisch:2019yzn,
    author = "Kreisch, Christina D. and Cyr-Racine, Francis-Yan and Dor{\'e}, Olivier",
    title = "{The Neutrino Puzzle: Anomalies, Interactions, and Cosmological Tensions}",
    eprint = "1902.00534",
    archivePrefix = "arXiv",
    primaryClass = "astro-ph.CO",
    doi = "10.1103/PhysRevD.101.123505",
    journal = "Phys. Rev. D",
    volume = "101",
    number = "12",
    pages = "123505",
    year = "2020"
}

@article{ODEPACK,
    author = "Hindmarsh, A. C.",
    title = "{ODEPACK, A Systematized Collection of ODE Solvers}",
    journal = "IMACS Transactions on Scientific Computation",
    volume = "1",
    pages = "55-64",
    year = "1983"
}

@article{AlpherBetheGamow,
    author = "Alpher, R. A. and Bethe, H. and Gamow, G.",
    title = "{The origin of chemical elements}",
    doi = "10.1103/PhysRev.73.803",
    journal = "Phys. Rev.",
    volume = "73",
    pages = "803--804",
    year = "1948"
}

@article{Wagoner,
    author = "Wagoner, Robert V. and Fowler, William A. and Hoyle, Fred",
    title = "{On the Synthesis of elements at very high temperatures}",
    doi = "10.1086/149126",
    journal = "Astrophys. J.",
    volume = "148",
    pages = "3--49",
    year = "1967"
}

@article{SmithBBN,
      author         = "Smith, Michael S. and Kawano, Lawrence H. and Malaney,
                        Robert A.",
      title          = "{Experimental, computational, and observational analysis
                        of primordial nucleosynthesis}",
      journal        = "Astrophys. J. Suppl.",
      volume         = "85",
      year           = "1993",
      pages          = "219-247",
      doi            = "10.1086/191763",
      reportNumber   = "OAP-716",
      SLACcitation   = "%%CITATION = APJSA,85,219;%%"
}

@article{Dodelson_Turner_PhRvD1992,
    author = "Dodelson, Scott and Turner, Michael S.",
    title = "{Nonequilibrium neutrino statistical mechanics in the expanding universe}",
    reportNumber = "FERMILAB-PUB-92-084-A",
    doi = "10.1103/PhysRevD.46.3372",
    journal = "Phys. Rev. D",
    volume = "46",
    pages = "3372--3387",
    year = "1992"
}

@article{Dolgov1992,
    author = "Dolgov, A.D. and Fukugita, M.",
    title = "{Nonequilibrium effect of the neutrino distribution on primordial helium synthesis}",
    reportNumber = "IASSNS-AST-92-12",
    doi = "10.1103/PhysRevD.46.5378",
    journal = "Phys. Rev. D",
    volume = "46",
    pages = "5378--5382",
    year = "1992"
}

@article{Fields_PhRvD1993,
    author = "Fields, Brian D. and Dodelson, Scott and Turner, Michael S.",
    title = "{Effect of neutrino heating on primordial nucleosynthesis}",
    eprint = "astro-ph/9210007",
    archivePrefix = "arXiv",
    reportNumber = "FERMILAB-PUB-92-298-A",
    doi = "10.1103/PhysRevD.47.4309",
    journal = "Phys. Rev. D",
    volume = "47",
    pages = "4309--4314",
    year = "1993"
}

@article{Pitrou_2018PhysRept,
title = "Precision big bang nucleosynthesis with improved Helium-4 predictions",
journal = "Phys. Rept.",
volume = "754",
pages = "1--66",
year = "2018",
issn = "0370-1573",
doi = "https://doi.org/10.1016/j.physrep.2018.04.005",
url = "http://www.sciencedirect.com/science/article/pii/S0370157318301054",
 author={Pitrou, Cyril and Coc, Alain and Uzan, Jean-Philippe and Vangioni, Elisabeth},
  archivePrefix = "arXiv",
  eprint = {1801.08023},
  publisher={Elsevier}
}

@article{Pitrou:2019nub,
    author = "Pitrou, Cyril and Coc, Alain and Uzan, Jean-Philippe and Vangioni, Elisabeth",
    editor = "Kawabata, T. and others",
    title = "{Precision Big Bang Nucleosynthesis with the New Code PRIMAT}",
    eprint = "1909.12046",
    archivePrefix = "arXiv",
    primaryClass = "astro-ph.CO",
    doi = "10.7566/JPSCP.31.011034",
    journal = "JPS Conf. Proc.",
    volume = "31",
    pages = "011034",
    year = "2020"
}

@article{Pitrou2020,
    author = "Pitrou, Cyril and Coc, Alain and Uzan, Jean-Philippe and Vangioni, Elisabeth",
    title = "{A new tension in the cosmological model from primordial deuterium?}",
    eprint = "2011.11320",
    archivePrefix = "arXiv",
    primaryClass = "astro-ph.CO",
    doi = "10.1093/mnras/stab135",
    journal = "Mon. Not. Roy. Astron. Soc.",
    volume = "502",
    number = "2",
    pages = "2474--2481",
    year = "2021"
}

@article{Pisanti2020,
    author = "Pisanti, Ofelia and Mangano, Gianpiero and Miele, Gennaro and Mazzella, Pierpaolo",
    title = "{Primordial Deuterium after LUNA: concordances and error budget}",
    eprint = "2011.11537",
    archivePrefix = "arXiv",
    primaryClass = "astro-ph.CO",
    doi = "10.1088/1475-7516/2021/04/020",
    journal = "JCAP",
    volume = "04",
    pages = "020",
    year = "2021"
}

@article{Yeh2020,
    author = "Yeh, Tsung-Han and Olive, Keith A. and Fields, Brian D.",
    title = "{The impact of new $d(p,\gamma){}^3$He rates on Big Bang Nucleosynthesis}",
    eprint = "2011.13874",
    archivePrefix = "arXiv",
    primaryClass = "astro-ph.CO",
    reportNumber = "UMN--TH--4004/20, FTPI--MINN--20/35",
    doi = "10.1088/1475-7516/2021/03/046",
    journal = "JCAP",
    volume = "03",
    pages = "046",
    year = "2021"
}

@article{Pitrou2021,
    author = "Pitrou, Cyril and Coc, Alain and Uzan, Jean-Philippe and Vangioni, Elisabeth",
    title = "{Resolving conclusions about the early Universe requires accurate nuclear measurements}",
    eprint = "2104.11148",
    archivePrefix = "arXiv",
    primaryClass = "astro-ph.CO",
    doi = "10.1038/s42254-021-00294-6",
    journal = "Nature Rev. Phys.",
    volume = "3",
    number = "4",
    pages = "231--232",
    year = "2021"
}

@article{Parthenope,
    author = "Pisanti, O. and Cirillo, A. and Esposito, S. and Iocco, F. and Mangano, G. and Miele, G. and Serpico, P. D.",
    title = "{PArthENoPE: Public Algorithm Evaluating the Nucleosynthesis of Primordial Elements}",
    eprint = "0705.0290",
    archivePrefix = "arXiv",
    primaryClass = "astro-ph",
    reportNumber = "DSF-13-07, FERMILAB-PUB-07-079-A, SLAC-PUB-12488",
    doi = "10.1016/j.cpc.2008.02.015",
    journal = "Comput. Phys. Commun.",
    volume = "178",
    pages = "956--971",
    year = "2008"
}

@article{Parthenope_reloaded,
    author = "Consiglio, R. and de Salas, P. F. and Mangano, G. and Miele, G. and Pastor, S. and Pisanti, O.",
    title = "{PArthENoPE reloaded}",
    eprint = "1712.04378",
    archivePrefix = "arXiv",
    primaryClass = "astro-ph.CO",
    doi = "10.1016/j.cpc.2018.06.022",
    journal = "Comput. Phys. Commun.",
    volume = "233",
    pages = "237--242",
    year = "2018"
}

@article{Parthenope_revolutions,
    author = "Gariazzo, S. and F. de Salas, P. and Pisanti, O. and Consiglio, R.",
    title = "{PArthENoPE revolutions}",
    eprint = "2103.05027",
    archivePrefix = "arXiv",
    primaryClass = "astro-ph.IM",
    doi = "10.1016/j.cpc.2021.108205",
    journal = "Comput. Phys. Commun.",
    volume = "271",
    pages = "108205",
    year = "2022"
}

@article{AlterBBN,
    author = "Arbey, Alexandre",
    title = "{AlterBBN: A program for calculating the BBN abundances of the elements in alternative cosmologies}",
    eprint = "1106.1363",
    archivePrefix = "arXiv",
    primaryClass = "astro-ph.CO",
    reportNumber = "CERN-PH-TH-2011-135, LYCEN-2011-06",
    doi = "10.1016/j.cpc.2012.03.018",
    journal = "Comput. Phys. Commun.",
    volume = "183",
    pages = "1822--1831",
    year = "2012"
}

@article{AlterBBNv2,
    author = "Arbey, A. and Auffinger, J. and Hickerson, K. P. and Jenssen, E. S.",
    title = "{AlterBBN v2: A public code for calculating Big-Bang nucleosynthesis constraints in alternative cosmologies}",
    eprint = "1806.11095",
    archivePrefix = "arXiv",
    primaryClass = "astro-ph.CO",
    reportNumber = "CERN-TH-2018-146",
    doi = "10.1016/j.cpc.2019.106982",
    journal = "Comput. Phys. Commun.",
    volume = "248",
    pages = "106982",
    year = "2020"
}

@article{Fields:2019pfx,
    author = "Fields, Brian D. and Olive, Keith A. and Yeh, Tsung-Han and Young, Charles",
    title = "{Big-Bang Nucleosynthesis after Planck}",
    eprint = "1912.01132",
    archivePrefix = "arXiv",
    primaryClass = "astro-ph.CO",
    reportNumber = "UMN--TH--3902/19, FTPI--MINN--19/25",
    doi = "10.1088/1475-7516/2020/03/010",
    journal = "JCAP",
    volume = "03",
    pages = "010",
    year = "2020",
    note = "[Erratum: JCAP 11, E02 (2020)]"
}

@article{bernstein1989,
    author = {Bernstein, Jeremy and Brown, Lowell S. and Feinberg, Gerald},
    title = "{Cosmological helium production simplified}",
    reportNumber = "DOE/ER/40423-02-P8",
    doi = "10.1103/RevModPhys.61.25",
    journal = "Rev. Mod. Phys.",
    volume = "61",
    pages = "25--39",
    year = "1989"
}

@article{Bania2002,
    author = "Bania, T. M. and Rood, Robert T. and Balser, Dana S.",
    title = "{The cosmological density of baryons from observations of $^{3}\mathrm{He}+$ in the Milky Way}",
    doi = "10.1038/415054a",
    journal = "Nature",
    volume = "415",
    pages = "54--57",
    year = "2002"
}

@article{Sbordone2010,
    author = "Sbordone, L. and others",
    title = "{The metal-poor end of the Spite plateau. 1: Stellar parameters, metallicities and lithium abundances}",
    eprint = "1003.4510",
    archivePrefix = "arXiv",
    primaryClass = "astro-ph.GA",
    doi = "10.1051/0004-6361/200913282",
    journal = "Astron. Astrophys.",
    volume = "522",
    pages = "A26",
    year = "2010"
}

@article{Cooke2017,
    author = "Cooke, Ryan J. and Pettini, Max and Steidel, Charles C.",
    title = "{One Percent Determination of the Primordial Deuterium Abundance}",
    eprint = "1710.11129",
    archivePrefix = "arXiv",
    primaryClass = "astro-ph.CO",
    doi = "10.3847/1538-4357/aaab53",
    journal = "Astrophys. J.",
    volume = "855",
    number = "2",
    pages = "102",
    year = "2018"
}

@article{Aver2020,
    author = "Aver, Erik and Berg, Danielle A. and Olive, Keith A. and Pogge, Richard W. and Salzer, John J. and Skillman, Evan D.",
    title = "{Improving helium abundance determinations with Leo P as a case study}",
    eprint = "2010.04180",
    archivePrefix = "arXiv",
    primaryClass = "astro-ph.CO",
    reportNumber = "UMN-TH-4001/20, FTPI-MINN-20/32",
    doi = "10.1088/1475-7516/2021/03/027",
    journal = "JCAP",
    volume = "03",
    pages = "027",
    year = "2021"
}

@article{Cooke2022,
    author = "Cooke, Ryan J. and Noterdaeme, Pasquier and Johnson, James W. and Pettini, Max and Welsh, Louise and Peroux, Celine and Murphy, Michael T. and Weinberg, David H.",
    title = "{Primordial Helium-3 Redux: The Helium Isotope Ratio of the Orion Nebula*}",
    eprint = "2203.11256",
    archivePrefix = "arXiv",
    primaryClass = "astro-ph.CO",
    doi = "10.3847/1538-4357/ac6503",
    journal = "Astrophys. J.",
    volume = "932",
    number = "1",
    pages = "60",
    year = "2022"
}

@article{Fields2011,
    author = "Fields, Brian D.",
    title = "{The primordial lithium problem}",
    eprint = "1203.3551",
    archivePrefix = "arXiv",
    primaryClass = "astro-ph.CO",
    doi = "10.1146/annurev-nucl-102010-130445",
    journal = "Ann. Rev. Nucl. Part. Sci.",
    volume = "61",
    pages = "47--68",
    year = "2011"
}

@article{Fields2022,
    author = "Fields, Brian D. and Olive, Keith A.",
    title = "{Implications of the Non-Observation of ${}^{6}{\mathrm{Li}}$ in Halo Stars for the Primordial ${}^{7}{\mathrm{Li}}$ Problem}",
    eprint = "2204.03167",
    archivePrefix = "arXiv",
    primaryClass = "astro-ph.GA",
    reportNumber = "UMN--TH--4118/22, FTPI--MINN--22/09",
    month = "4",
    year = "2022"
}

@article{Coc2006,
    author = "Coc, Alain and Olive, Keith A. and Uzan, Jean-Philippe and Vangioni, Elisabeth",
    title = "{Big bang nucleosynthesis constraints on scalar-tensor theories of gravity}",
    eprint = "astro-ph/0601299",
    archivePrefix = "arXiv",
    reportNumber = "UMN-TH-2424-05, FTPI-MINN-05-52",
    doi = "10.1103/PhysRevD.73.083525",
    journal = "Phys. Rev. D",
    volume = "73",
    pages = "083525",
    year = "2006"
}

@article{CocVangioni2010,
    author = "Coc, Alain and Vangioni, Elisabeth",
    editor = "Formicola, Alba and Gustavino, Carlo and Junker, Matthias",
    title = "{Big-Bang nucleosynthesis with updated nuclear data}",
    doi = "10.1088/1742-6596/202/1/012001",
    journal = "J. Phys. Conf. Ser.",
    volume = "202",
    pages = "012001",
    year = "2010"
}

@article{Coc2015,
    author = "Coc, Alain and Petitjean, Patrick and Uzan, Jean-Philippe and Vangioni, Elisabeth and Descouvemont, Pierre and Iliadis, Christian and Longland, Richard",
    title = "{New reaction rates for improved primordial D/H calculation and the cosmic evolution of deuterium}",
    eprint = "1511.03843",
    archivePrefix = "arXiv",
    primaryClass = "astro-ph.CO",
    doi = "10.1103/PhysRevD.92.123526",
    journal = "Phys. Rev. D",
    volume = "92",
    number = "12",
    pages = "123526",
    year = "2015"
}

@article{Mossa2020,
    author = "Mossa, V. and others",
    title = "{The baryon density of the Universe from an improved rate of deuterium burning}",
    doi = "10.1038/s41586-020-2878-4",
    journal = "Nature",
    volume = "587",
    number = "7833",
    pages = "210--213",
    year = "2020"
}

@article{Grohs:2016vef,
    author = "Grohs, E. and Fuller, G. M.",
    title = "{The surprising influence of late charged current weak interactions on Big Bang Nucleosynthesis}",
    eprint = "1607.02797",
    archivePrefix = "arXiv",
    primaryClass = "astro-ph.CO",
    doi = "10.1016/j.nuclphysb.2016.08.034",
    journal = "Nucl. Phys. B",
    volume = "911",
    pages = "955--973",
    year = "2016"
}

@article{Balantekin:2007es,
    author = "Balantekin, Akif Baha and Gava, J. and Volpe, C.",
    title = "{Possible CP-violation effects in core-collapse supernovae}",
    eprint = "0710.3112",
    archivePrefix = "arXiv",
    primaryClass = "astro-ph",
    doi = "10.1016/j.physletb.2008.03.038",
    journal = "Phys. Lett. B",
    volume = "662",
    pages = "396--404",
    year = "2008"
}

@article{Gava:2008rp,
    author = "Gava, Jerome and Volpe, Cristina",
    title = "{Collective neutrino oscillations in matter and CP violation}",
    eprint = "0807.3418",
    archivePrefix = "arXiv",
    primaryClass = "astro-ph",
    doi = "10.1103/PhysRevD.78.083007",
    journal = "Phys. Rev. D",
    volume = "78",
    pages = "083007",
    year = "2008"
}

@article{Gava:2010kz,
    author = "Gava, Jerome and Volpe, Cristina",
    title = "{CP violation effects on the neutrino degeneracy parameters in the Early Universe}",
    eprint = "1002.0981",
    archivePrefix = "arXiv",
    primaryClass = "hep-ph",
    doi = "10.1016/j.nuclphysb.2010.04.024",
    journal = "Nucl. Phys. B",
    volume = "837",
    pages = "50--60",
    year = "2010"
}

@article{Gava_corr,
title = "{Corrigendum to “CP violation effects on the neutrino degeneracy parameters in the Early Universe” [Nucl. Phys. B 837 (2010) 50–60)]}",
journal = "Nucl. Phys. B",
volume = "957",
pages = "115035",
year = "2020",
issn = "0550-3213",
doi = "https://doi.org/10.1016/j.nuclphysb.2020.115035",
url = "http://www.sciencedirect.com/science/article/pii/S0550321320301218",
author = "Maria Cristina Volpe"
}

@article{Cabibbo,
    author = "Cabibbo, Nicola",
    title = "{Unitary Symmetry and Leptonic Decays}",
    doi = "10.1103/PhysRevLett.10.531",
    journal = "Phys. Rev. Lett.",
    volume = "10",
    pages = "531--533",
    year = "1963"
}

@article{KobayashiMaskawa,
    author = "Kobayashi, Makoto and Maskawa, Toshihide",
    title = "{CP Violation in the Renormalizable Theory of Weak Interaction}",
    reportNumber = "KUNS-242",
    doi = "10.1143/PTP.49.652",
    journal = "Prog. Theor. Phys.",
    volume = "49",
    pages = "652--657",
    year = "1973"
}

@article{Pontecorvo1957,
    author = "Pontecorvo, B.",
    title = "{Mesonium and antimesonium}",
    journal = "Sov. Phys. JETP",
    volume = "6",
    pages = "429",
    url={"http://jetp.ras.ru/cgi-bin/dn/e_006_02_0429.pdf"},
    year = "1957"
}

@article{Pontecorvo1958,
    author = "Pontecorvo, B.",
    title = "{Inverse beta processes and nonconservation of lepton charge}",
    journal = "Sov. Phys. JETP",
    volume = "7",
    pages = "172--173",
    url="http://jetp.ras.ru/cgi-bin/dn/e_007_01_0172.pdf",
    year = "1958"
}

@article{MNS,
    author = "Maki, Ziro and Nakagawa, Masami and Sakata, Shoichi",
    title = "{Remarks on the unified model of elementary particles}",
    doi = "10.1143/PTP.28.870",
    journal = "Prog. Theor. Phys.",
    volume = "28",
    pages = "870--880",
    year = "1962"
}

@article{Eliezer1975,
    author = "Eliezer, Shalom and Swift, Arthur R.",
    title = "{Experimental Consequences of electron Neutrino-Muon-neutrino Mixing in Neutrino Beams}",
    reportNumber = "Print-75-0213 (MASS U.,AMHERST)",
    doi = "10.1016/0550-3213(76)90059-6",
    journal = "Nucl. Phys. B",
    volume = "105",
    pages = "45--51",
    year = "1976"
}

@article{MSW_W,
    author = "Wolfenstein, L.",
    title = "{Neutrino Oscillations in Matter}",
    reportNumber = "COO-3066-102",
    doi = "10.1103/PhysRevD.17.2369",
    journal = "Phys. Rev. D",
    volume = "17",
    pages = "2369--2374",
    year = "1978"
}

@article{MSW_MS,
    author = "Mikheyev, S.P. and Smirnov, A.Yu.",
    title = "{Resonance Amplification of Oscillations in Matter and Spectroscopy of Solar Neutrinos}",
    journal = "Sov. J. Nucl. Phys.",
    volume = "42",
    pages = "913--917",
    year = "1985"
}

@ARTICLE{Hoyle1967,
       author = {{Hoyle}, F.},
        title = "{Concluding Remarks}",
      journal = {Proc. Royal. Soc. (London) A},
         year = 1967,
       volume = {301},
       number = {1465},
        pages = {171},
          doi = {10.1098/rspa.1967.0199},
       adsurl = {https://ui.adsabs.harvard.edu/abs/1967RSPSA.301..171H}
}

@article{King2015,
    author = "King, Stephen F.",
    title = "{Models of Neutrino Mass, Mixing and CP Violation}",
    eprint = "1510.02091",
    archivePrefix = "arXiv",
    primaryClass = "hep-ph",
    doi = "10.1088/0954-3899/42/12/123001",
    journal = "J. Phys. G",
    volume = "42",
    pages = "123001",
    year = "2015"
}

@ARTICLE{Landau1932,
       author = {{Landau}, L.~D.},
        title = "{On the theory of energy transmission in collisions. II}",
      journal = {Phys. Zs. Sowjet},
         year = 1932,
        month = dec,
       volume = {2},
        pages = {46},
       adsurl = {https://ui.adsabs.harvard.edu/abs/1932PhyZS...2...46L}
}

@ARTICLE{Zener1932,
       author = {{Zener}, Clarence},
        title = "{Non-Adiabatic Crossing of Energy Levels}",
      journal = {Proceedings of the Royal Society of London Series A},
         year = 1932,
        month = sep,
       volume = {137},
       number = {833},
        pages = {696-702},
          doi = {10.1098/rspa.1932.0165},
       adsurl = {https://ui.adsabs.harvard.edu/abs/1932RSPSA.137..696Z}
}

@article{Wittig,
    author = "Wittig, C.",
    title = "{The Landau-Zener formula }",
    journal = "J. Phys. Chem. B",
    volume = "109 (17)",
    pages = "8428-8430",
    year = "2005"
}

@article{Haxton:1986bc,
    author = "Haxton, W. C.",
    title = "{Analytic Treatments of Matter Enhanced Solar Neutrino Oscillations}",
    reportNumber = "DOE-ER-40048-03-N7",
    doi = "10.1103/PhysRevD.35.2352",
    journal = "Phys. Rev. D",
    volume = "35",
    pages = "2352",
    year = "1987"
}

@article{Abazajian:2001nj,
    author = "Abazajian, Kevork and Fuller, George M. and Patel, Mitesh",
    title = "{Sterile neutrino hot, warm, and cold dark matter}",
    eprint = "astro-ph/0101524",
    archivePrefix = "arXiv",
    doi = "10.1103/PhysRevD.64.023501",
    journal = "Phys. Rev. D",
    volume = "64",
    pages = "023501",
    year = "2001"
}

@article{Malkus:2014iqa,
    author = "Malkus, A. and Friedland, A. and McLaughlin, G. C.",
    title = "{Matter-Neutrino Resonance Above Merging Compact Objects}",
    eprint = "1403.5797",
    archivePrefix = "arXiv",
    primaryClass = "hep-ph",
    month = "3",
    year = "2014"
}

@article{Hansen_Isotropy,
    author = "Hansen, Rasmus S. L. and Shalgar, Shashank and Tamborra, Irene",
    title = "{Neutrino flavor mixing breaks isotropy in the early universe}",
    eprint = "2012.03948",
    archivePrefix = "arXiv",
    primaryClass = "astro-ph.CO",
    doi = "10.1088/1475-7516/2021/07/017",
    journal = "JCAP",
    volume = "07",
    pages = "017",
    year = "2021"
}

@article{Simha:2008mt,
      author         = "Simha, Vimal and Steigman, Gary",
      title          = "{Constraining The Universal Lepton Asymmetry}",
      journal        = "JCAP",
      volume         = "0808",
      year           = "2008",
      pages          = "011",
      doi            = "10.1088/1475-7516/2008/08/011",
      eprint         = "0806.0179",
      archivePrefix  = "arXiv",
      primaryClass   = "hep-ph",
      SLACcitation   = "%%CITATION = ARXIV:0806.0179;%%"
}

@article{Oldengott:2017tzj,
      author         = "Oldengott, Isabel M. and Schwarz, Dominik J.",
      title          = "{Improved constraints on lepton asymmetry from the cosmic
                        microwave background}",
      journal        = "Europhys. Lett.",
      volume         = "119",
      year           = "2017",
      number         = "2",
      pages          = "29001",
      doi            = "10.1209/0295-5075/119/29001",
      eprint         = "1706.01705",
      archivePrefix  = "arXiv",
      primaryClass   = "astro-ph.CO",
      SLACcitation   = "%%CITATION = ARXIV:1706.01705;%%"
}

@article{Mangano:2011ip,
    author = "Mangano, Gianpiero and Miele, Gennaro and Pastor, Sergio and Pisanti, Ofelia and Sarikas, Srdjan",
    title = "{Updated BBN bounds on the cosmological lepton asymmetry for non-zero $\theta_{13}$}",
    eprint = "1110.4335",
    archivePrefix = "arXiv",
    primaryClass = "hep-ph",
    reportNumber = "IFIC-11-58, DSF-13-2011",
    doi = "10.1016/j.physletb.2012.01.015",
    journal = "Phys. Lett. B",
    volume = "708",
    pages = "1--5",
    year = "2012"
}

@article{Castorina2012,
    author = "Castorina, Emanuele and Franca, Urbano and Lattanzi, Massimiliano and Lesgourgues, Julien and Mangano, Gianpiero and Melchiorri, Alessandro and Pastor, Sergio",
    title = "{Cosmological lepton asymmetry with a nonzero mixing angle $\theta_{13}$}",
    eprint = "1204.2510",
    archivePrefix = "arXiv",
    primaryClass = "astro-ph.CO",
    reportNumber = "CERN-PH-TH-2012-089, IFIC-12-28, LAPTH-018-12",
    doi = "10.1103/PhysRevD.86.023517",
    journal = "Phys. Rev. D",
    volume = "86",
    pages = "023517",
    year = "2012"
}

@article{FrousteyTAUP2021,
    author = "Froustey, Julien",
    title = "{Precision calculation of neutrino evolution in the early Universe}",
    eprint = "2110.11296",
    archivePrefix = "arXiv",
    primaryClass = "hep-ph",
    %doi = "10.1088/1742-6596/2156/1/012013",
    url ={https://iopscience.iop.org/article/10.1088/1742-6596/2156/1/012013},
    journal = "J. Phys. Conf. Ser.",
    volume = "2156",
    number = "3",
    pages = "012013",
    year = "2021"
}

@article{Davidson_Leptogenesis,
    author = "Davidson, Sacha and Nardi, Enrico and Nir, Yosef",
    title = "{Leptogenesis}",
    eprint = "0802.2962",
    archivePrefix = "arXiv",
    primaryClass = "hep-ph",
    doi = "10.1016/j.physrep.2008.06.002",
    journal = "Phys. Rept.",
    volume = "466",
    pages = "105--177",
    year = "2008"
}

@article{March-Russell:1999hpw,
    author = "March-Russell, John and Murayama, Hitoshi and Riotto, Antonio",
    title = "{The Small observed baryon asymmetry from a large lepton asymmetry}",
    eprint = "hep-ph/9908396",
    archivePrefix = "arXiv",
    reportNumber = "CERN-TH-99-218, LBNL-44140, UCB-PTH-99-30, LBL-44140",
    doi = "10.1088/1126-6708/1999/11/015",
    journal = "JHEP",
    volume = "11",
    pages = "015",
    year = "1999"
}

@article{McDonald:1999in,
    author = "McDonald, John",
    title = "{Naturally large cosmological neutrino asymmetries in the MSSM}",
    eprint = "hep-ph/9908300",
    archivePrefix = "arXiv",
    doi = "10.1103/PhysRevLett.84.4798",
    journal = "Phys. Rev. Lett.",
    volume = "84",
    pages = "4798--4801",
    year = "2000"
}

@article{Gu:2010dg,
    author = "Gu, Pei-Hong",
    title = "{Large Lepton Asymmetry for Small Baryon Asymmetry and Warm Dark Matter}",
    eprint = "1005.1632",
    archivePrefix = "arXiv",
    primaryClass = "hep-ph",
    doi = "10.1103/PhysRevD.82.093009",
    journal = "Phys. Rev. D",
    volume = "82",
    pages = "093009",
    year = "2010"
}

@article{Snowmass_Abazajian,
    author = "Abazajian, Kevork N. and others",
    title = "{Synergy between cosmological and laboratory searches in neutrino physics: a white paper}",
    eprint = "2203.07377",
    archivePrefix = "arXiv",
    primaryClass = "hep-ph",
    month = "3",
    year = "2022"
}

@book{KolbTurner,
    author = "Kolb, Edward W. and Turner, Michael S.",
    title = "{The Early Universe}",
    doi = "10.1201/9780429492860",
    isbn = "978-0-201-62674-2",
    volume = "69",
    series="Frontiers in Physics",
    year = "1990",
    publisher="Addison-Wesley"
}

@book{PeterUzan,
    author = "Peter, Patrick and Uzan, Jean-Philippe",
    title = "{Primordial Cosmology}",
    isbn = "978-0-19-966515-0",
    publisher = "Oxford University Press",
    series = "Oxford Graduate Texts",
    month = "2",
    year = "2013"
}

@book{CohenIII,
      author         = "Cohen-Tannoudji, Claude and Diu, Bernard and Lalo{\"e}, Franck",
      title          = "{Mécanique quantique -- Tome III}",
      publisher      = "EDP Sciences",
      year           = "2017",
      url            = "https://laboutique.edpsciences.fr/produit/1061/9782759823369/Mecanique%20quantique%20-%20Tome%203",
      ISBN           = "9782759823352"
}

@book{Neutrino_Cosmology,
      author         = "Lesgourgues, Julien and Mangano, Gianpiero and Miele,
                        Gennaro and Pastor, Sergio",
      title          = "{Neutrino Cosmology}",
      publisher      = "Cambridge University Press",
      year           = "2013",
      url            = "https://www.cambridge.org/academic/subjects/physics/particle-physics-and-nuclear-physics/neutrino-cosmology?format=PB&isbn=9781108705011",
      ISBN           = "9781108705011",
      SLACcitation   = "%%CITATION = INSPIRE-1705305;%%",
      doi = {10.1017/CBO9781139012874}
}

@book{ModernCosmology,
  title={Modern Cosmology},
  author={Dodelson, Scott and Schmidt, Fabian},
  year={2020},
  isbn={9780128159484},
  publisher={Academic Press}
}

@book{PeskinSchroeder,
      author         = "Peskin, Michael E. and Schroeder, Daniel V.",
      title          = "{An Introduction to quantum field theory}",
      publisher      = "Addison-Wesley",
      address        = "Reading, USA",
      year           = "1995",
      url            = "http://www.slac.stanford.edu/~mpeskin/QFT.html",
      ISBN           = "9780201503975,",
      SLACcitation   = "%%CITATION = INSPIRE-407703;%%"
}

@book{SchwartzQFT,
      author         = "Schwartz, Matthew D.",
      title          = "{Quantum Field Theory and the Standard Model}",
      publisher      = "Cambridge University Press",
      year           = "2014",
      url            = "http://www.cambridge.org/us/academic/subjects/physics/theoretical-physics-and-mathematical-physics/quantum-field-theory-and-standard-model",
      ISBN           = "9781107034730",
      SLACcitation   = "%%CITATION = INSPIRE-1276589;%%"
}

@book{Srednicki,
    author = "Srednicki, M.",
    title = "{Quantum Field Theory}",
    isbn = "978-0-521-86449-7",
    publisher = "Cambridge University Press",
    month = "1",
    year = "2007"
}

@book{GiuntiKim,
      author         = "Giunti, Carlo and Kim, Chung W.",
      title          = "{Fundamentals of Neutrino Physics and Astrophysics}",
      publisher = "Oxford University Press",
      journal        = "Oxford, UK: Univ. Pr. (2007) 710 p",
      year           = "2007",
      url				= "https://global.oup.com/academic/product/fundamentals-of-neutrino-physics-and-astrophysics-9780198508717?cc=fr&lang=en&",
      ISBN           = "9780198508717",
      SLACcitation   = "%%CITATION = INSPIRE-747694;%%"
}

@BOOK{MohapatraPal,
       author = {{Mohapatra}, Rabindra N. and {Pal}, Palash B.},
        title = "{Massive Neutrinos in Physics and Astrophysics (Third Edition)}",
         year = 2004,
          doi = {10.1142/5024},
          publisher={World Scientific},
          series={Lecture Notes in Physics},
          volume=72,
       adsurl = {https://ui.adsabs.harvard.edu/abs/2004mnpa.book.....M}
}

@book{Wald,
    author = "Wald, Robert M.",
    title = "{General Relativity}",
    doi = "10.7208/chicago/9780226870373.001.0001",
    publisher = "Chicago Univ. Pr.",
    address = "Chicago, USA",
    year = "1984"
}

@article{Carroll_RG,
    author = "Carroll, Sean M.",
    title = "{Lecture notes on general relativity}",
    eprint = "gr-qc/9712019",
    archivePrefix = "arXiv",
    reportNumber = "NSF-ITP-97-147",
    month = "12",
    year = "1997"
}

@book{WeinbergCosmology,
    author = "Weinberg, Steven",
    title = "{Cosmology}",
    isbn = "978-0-19-852682-7",
    publisher={Oxford University Press},
    year = "2008"
}

@article{Long_PTOLEMY,
    author = "Long, Andrew J. and Lunardini, Cecilia and Sabancilar, Eray",
    title = "{Detecting non-relativistic cosmic neutrinos by capture on tritium: phenomenology and physics potential}",
    eprint = "1405.7654",
    archivePrefix = "arXiv",
    primaryClass = "hep-ph",
    doi = "10.1088/1475-7516/2014/08/038",
    journal = "JCAP",
    volume = "08",
    pages = "038",
    year = "2014"
}

@article{Vitagliano_Review,
    author = "Vitagliano, Edoardo and Tamborra, Irene and Raffelt, Georg",
    title = "{Grand Unified Neutrino Spectrum at Earth: Sources and Spectral Components}",
    eprint = "1910.11878",
    archivePrefix = "arXiv",
    primaryClass = "astro-ph.HE",
    reportNumber = "MPP-2019-205",
    doi = "10.1103/RevModPhys.92.045006",
    journal = "Rev. Mod. Phys.",
    volume = "92",
    pages = "45006",
    year = "2020"
}

@article{PTOLEMY2018,
    author = "Baracchini, E. and others",
    collaboration = "PTOLEMY",
    title = "{PTOLEMY: A Proposal for Thermal Relic Detection of Massive Neutrinos and Directional Detection of MeV Dark Matter}",
    eprint = "1808.01892",
    archivePrefix = "arXiv",
    primaryClass = "physics.ins-det",
    month = "8",
    year = "2018"
}

@article{PTOLEMY2019,
    author = "Betti, M. G. and others",
    collaboration = "PTOLEMY",
    title = "{Neutrino physics with the PTOLEMY project: active neutrino properties and the light sterile case}",
    eprint = "1902.05508",
    archivePrefix = "arXiv",
    primaryClass = "astro-ph.CO",
    doi = "10.1088/1475-7516/2019/07/047",
    journal = "JCAP",
    volume = "07",
    pages = "047",
    year = "2019"
}

@article{Cheipesh:2021fmg,
    author = "Cheipesh, Yevheniia and Cheianov, Vadim and Boyarsky, Alexey",
    title = "{Navigating the pitfalls of relic neutrino detection}",
    eprint = "2101.10069",
    archivePrefix = "arXiv",
    primaryClass = "hep-ph",
    doi = "10.1103/PhysRevD.104.116004",
    journal = "Phys. Rev. D",
    volume = "104",
    number = "11",
    pages = "116004",
    year = "2021"
}

@article{PTOLEMY2022,
    author = "Apponi, A. and others",
    collaboration = "PTOLEMY",
    title = "{Heisenberg's uncertainty principle in the PTOLEMY project: a theory update}",
    eprint = "2203.11228",
    archivePrefix = "arXiv",
    primaryClass = "hep-ph",
    doi = "10.1103/PhysRevD.106.053002",
    journal = "Phys. Rev. D",
    volume = "106",
    pages = "5",
    year = "2022"
}

@article{Bahcall,
    author = "Bahcall, John N. and Pinsonneault, M. H. and Basu, Sarbani",
    title = "{Solar models: Current epoch and time dependences, neutrinos, and helioseismological properties}",
    eprint = "astro-ph/0010346",
    archivePrefix = "arXiv",
    doi = "10.1086/321493",
    journal = "Astrophys. J.",
    volume = "555",
    pages = "990--1012",
    year = "2001"
}

@article{SNO_2002,
    author = "Ahmad, Q. R. and others",
    collaboration = "SNO",
    title = "{Direct evidence for neutrino flavor transformation from neutral current interactions in the Sudbury Neutrino Observatory}",
    eprint = "nucl-ex/0204008",
    archivePrefix = "arXiv",
    doi = "10.1103/PhysRevLett.89.011301",
    journal = "Phys. Rev. Lett.",
    volume = "89",
    pages = "011301",
    year = "2002"
}

@article{SNO_2016,
    author = "Bellerive, A. and Klein, J. R. and McDonald, A. B. and Noble, A. J. and Poon, A. W. P.",
    collaboration = "SNO",
    title = "{The Sudbury Neutrino Observatory}",
    eprint = "1602.02469",
    archivePrefix = "arXiv",
    primaryClass = "nucl-ex",
    doi = "10.1016/j.nuclphysb.2016.04.035",
    journal = "Nucl. Phys. B",
    volume = "908",
    pages = "30--51",
    year = "2016"
}

@article{Vissani2017a,
    author = "Vissani, Francesco",
    title = "{Solar neutrino physics on the beginning of 2017}",
    eprint = "1706.05435",
    archivePrefix = "arXiv",
    primaryClass = "nucl-th",
    doi = "10.15407/jnpae2017.01.005",
    journal = "Nucl. Phys. Atom. Energy",
    volume = "18",
    number = "1",
    pages = "5--12",
    year = "2017"
}

@article{Vissani2017b,
    author = "Vissani, Francesco",
    title = "{Joint analysis of Borexino and SNO solar neutrino data and reconstruction of the survival probability}",
    eprint = "1709.05813",
    archivePrefix = "arXiv",
    primaryClass = "hep-ph",
    doi = "10.15407/jnpae2017.04.303",
    journal = "Nucl. Phys. Atom. Energy",
    volume = "18",
    number = "4",
    pages = "303--312",
    year = "2017"
}

@article{Borexino,
    author = "Agostini, M. and others",
    collaboration = "BOREXINO",
    title = "{Comprehensive measurement of $pp$-chain solar neutrinos}",
    reportNumber = "FERMILAB-PUB-18-592-ND",
    doi = "10.1038/s41586-018-0624-y",
    journal = "Nature",
    volume = "562",
    number = "7728",
    pages = "505--510",
    year = "2018"
}

@article{Fixsen:2009,
    author = "Fixsen, D. J.",
    title = "{The Temperature of the Cosmic Microwave Background}",
    eprint = "0911.1955",
    archivePrefix = "arXiv",
    primaryClass = "astro-ph.CO",
    doi = "10.1088/0004-637X/707/2/916",
    journal = "Astrophys. J.",
    volume = "707",
    pages = "916--920",
    year = "2009"
}

\backcover

\end{document}